\documentclass[12pt,a4paper,twoside,sort&compress]{book}

\usepackage{epsfig}
\usepackage[skins]{tcolorbox}
\usepackage{slashed,verbatim,subfigure}
\usepackage[colorlinks=true,linktocpage=true,
linkcolor=blue,citecolor=blue]{hyperref}
\usepackage[a4paper]{geometry}
\usepackage{afterpage}
\newcommand\blankpage{%
    \null
    \thispagestyle{empty}%
    \addtocounter{page}{-1}%
    \newpage}
\usepackage{amsfonts}
\usepackage{amssymb}
\usepackage{amsmath}
\usepackage{setspace}
\usepackage{lscape}
\usepackage[latin1]{inputenc}
\usepackage[TS1,T1]{fontenc}
\usepackage{lmodern}
\usepackage{amsthm}
\usepackage{latexsym}
\usepackage{psfrag}
\usepackage{graphicx}
\usepackage{rotating}
\usepackage{graphics}
\usepackage{subfigure}
\usepackage{mathenv}
\usepackage{amsthm}
\usepackage[titletoc]{appendix}
\usepackage{etoolbox}
\usepackage{tikz}
\newrobustcmd*{\mysquare}[1]{\tikz{\filldraw[draw=#1,fill=#1] (0,0)
rectangle (0.2cm,0.2cm);}}
\newrobustcmd*{\mycircle}[1]{\tikz{\filldraw[draw=#1,fill=#1] (0,0) circle [radius=0.1cm];}}
\newrobustcmd*{\mytriangle}[1]{\tikz{\filldraw[draw=#1,fill=#1] (0,0) --
(0.2cm,0) -- (0.1cm,0.2cm);}}
\usepackage{amssymb}
\usepackage{amsmath}
\usepackage{tcolorbox}
\usepackage{graphics}
\usepackage[final]{pdfpages}
\usepackage{slashed}
\usepackage{amssymb}
\usepackage{epstopdf}

\newcommand{\ignore}[1]{}
\usepackage[latin1]{inputenc}
\usepackage{cite}
\usepackage{subfigure}
\usepackage[small,bf]{caption}
\usepackage{color}
\usepackage[Conny]{fncychap}
\ChNameVar{\LARGE\sffamily\bfseries} \ChNumVar{\Huge} \ChTitleVar{\Huge\sffamily\bfseries}
\ChRuleWidth{0.85pt} \ChNameUpperCase \ChTitleUpperCase 
\usepackage{fancyhdr}
\usepackage{amssymb}

\renewcommand{\vec}[1]{\mbox{\boldmath$#1$\unboldmath}}

\catcode`@=11
\def\seceqaa{\@addtoreset{equation}{section}
           \def\theequation{A\arabic{equation}}}
\def\seceqbb{\@addtoreset{equation}{section}
           \def\theequation{B\arabic{equation}}}
\def\seceqcc{\@addtoreset{equation}{section}
           \def\theequation{C\arabic{equation}}}
\def\seceqdd{\@addtoreset{equation}{section}
           \def\theequation{D\arabic{equation}}}
\def\seceqee{\@addtoreset{equation}{section}
           \def\theequation{E\arabic{equation}}}
\def\seceqff{\@addtoreset{equation}{section}
           \def\theequation{F\arabic{equation}}}
\def\seceqgg{\@addtoreset{equation}{section}
           \def\theequation{G\arabic{equation}}}
\def\seceqhh{\@addtoreset{equation}{section}
           \def\theequation{H\arabic{equation}}}
\def\seceqjj{\@addtoreset{equation}{section}
           \def\theequation{J\arabic{equation}}}
\def\seceqll{\@addtoreset{equation}{section}
           \def\theequation{L\arabic{equation}}}
\catcode`@=11

\begin{document}
\frontmatter
\pagestyle{plain}
    \pagestyle{fancy}                       
    \fancyfoot{}                            
    \renewcommand{\chaptermark}[1]{         
      \markboth{\chaptername\ \thechapter.\ #1}{}} %
    \renewcommand{\sectionmark}[1]{         
      \markright{\thesection.\ #1}}         %
    \fancyhead[LE,RO]{\bfseries\thepage}    
    \fancyhead[RE]{\bfseries\leftmark}      
    \fancyhead[LO]{\bfseries\rightmark}     
    \renewcommand{\headrulewidth}{0.3pt}    
    \makeatletter
    \def\cleardoublepage{\clearpage\if@twoside \ifodd\c@page\else%
        \hbox{}%
        \thispagestyle{empty}
        \newpage%
        \if@twocolumn\hbox{}\newpage\fi\fi\fi}
    \makeatother

  \newpage
\pagestyle{plain}
\begin{center}
{\Large \bf String/${\cal M}$-theory Dual of Large-$N$ Thermal QCD-Like Theories at Intermediate Gauge/'t Hooft Coupling and Holographic Phenomenology\footnote{Based on author's Ph.D thesis successfully defended on Nov 19, 2021}}
\end{center}
\begin{center}
Vikas Yadav\footnote{email:vyadav@ph.iitr.ac.in}
\end{center}
\begin{center}
{Department of Physics, Indian Institute of Technology, Roorkee- 247 667,\\
Uttarakhand, India}
\end{center}

{\bf Abstract:}Considering the setup of \cite{metrics} involving UV-complete top-down type IIB holographic dual of large-N thermal QCD with a fluxed resolved warped deformed conifold, in \cite{MQGP} delocalized type IIA S(trominger)-Y(au)-Z(aslow)mirror of the type IIB background of \cite{metrics} was constructed via three T dualities along a special Lagrangian $T^{3}$ fibered over a large base and then uplifted, locally, to the 11-dimensional ${\cal M}$-theory. Considering the aforementioned setup \cite{MQGP} in the ``MQGP'' limit, this thesis makes the following contributions.

In \cite{Sil+Yadav+Misra-glueball} we obtained the masses of the $0^{++},  0^{-+},0^{--}, 1^{++}, 2^{++}$ (`glueball') states which correspond to fluctuations in the dilaton or complexified two-forms or appropriate metric components in the aforementioned backgrounds in the `MQGP' limit of \cite{MQGP}. We used WKB quantization conditions and Neumann/Dirichlet boundary conditions at an IR cut-off (`$r_0$')/horizon radius (`$r_h$') on the solutions to the equations of motion. We found that the former technique produced results closer to the lattice results \cite{Teper98},\cite{Chen_et_al_latest_lattice_2006}. We also discussed the $r_h=0$-limits of all calculations which correspond to the thermal background.

This thesis fills the gap in the literature by computing the meson spectroscopy at finite gauge coupling and a finite number of colors, from a top-down holographic string model. In \cite{Yadav+Misra+Sil-Mesons}, we obtained analytical expressions for the vector and scalar meson spectra by using the delocalized type IIA SYZ mirror of the holographic type IIB dual of large-$N$ thermal QCD of \cite{metrics} as obtained in \cite{MQGP}. We verified via explicit computations that the spectra for both vector and scalar mesons obtained by the gravity dual with a black hole for all ranges of temperature are nearly isospectral with the spectra obtained by a thermal gravity dual valid only for low temperatures.

Subsequently, in \cite{mesons_0E++-to-mesons-decays} we obtained the interaction Lagrangian corresponding to exotic scalar glueball $\left( G_{E}\right)-\rho/\pi$- meson by using the pull-back of the perturbed type IIA metric corresponding to the perturbation of \cite{metrics}'s ${\cal M}$-theory uplift of\cite{MQGP}'s. The coupling constants in the interaction Lagrangian, linear in the exotic scalar glueball and up to quartic in the $\pi$ mesons, were obtained as (radial integrals of) \cite{MQGP}'s ${\cal M}$-theory uplift's metric components and six radial functions appearing in the ${\cal M}$-theory metric perturbations. Assuming $M_G>2M_\rho$, we then computed $\rho\rightarrow2\pi, G_E\rightarrow2\pi, 2\rho, \rho+2\pi$ decay widths as well as the direct and indirect (mediated via $\rho$ mesons) $G_E\rightarrow4\pi$ decays.

In \cite{OR4-Yadav+Misra} we obtained ${\cal O}\left(l_p^6\right)$ corrections to the MQGP background of \cite{MQGP} to study a top-down holographic dual of the thermal QCD-like theories at intermediate 't Hooft coupling  and in \cite{Vikas+Gopal+Aalok} we obtained the values of the coupling constants of the ${\cal O}(p^4)$ $\chi$PT Lagrangian in the chiral limit involving the NGBs and $\rho$ meson (and its flavor partners) from the  ${\cal M}$-theory/type IIA dual of large-$N$ thermal QCD \cite{MQGP}, inclusive of the ${\cal O}(R^4)$ corrections. We observed that there is a competition between non-conformal IR enhancement and Planckian and large-$N$ suppression and going to orders beyond the ${\cal O}(l_p^6)$ is necessitated if the former wins out. For the latter part \cite{Vikas+Gopal+Aalok}, we observed that ensuring compatibility with phenomenological/lattice results (the values ) as given in \cite{Ecker-2015}, required a relationship relating the ${\cal O}(R^4)$ corrections and large-$N$ suppression.




\afterpage{\blankpage}
\begin{spacing}{1.53}
\addcontentsline{toc}{chapter}{Table of Contents}\tableofcontents
\end{spacing}
\afterpage{\blankpage}
\addcontentsline{toc}{chapter}{List of Figures}
\begin{spacing}{1.34}
\listoffigures
\end{spacing}
\newpage
\addcontentsline{toc}{chapter}{List of Tables}
\begin{spacing}{1.34}
\listoftables
\end{spacing}
\mainmatter
\pagestyle{fancy}
\setcounter{chapter}{0}
\setcounter{section}{0}
\setcounter{subsection}{0}
\setcounter{tocdepth}{3}

\afterpage{\blankpage}
\chapter{Introduction}

\graphicspath{{Chapter5/}{Chapter5/}}

It is not incorrect to assert that the motivation for Physics is the human quest for a complete understanding of nature's fundamental laws. Quantum Electrodynamics, Weak Interactions, and Quantum Chromodynamics are the three fundamental theories that describe the interactions of elementary particles in a consistent manner. As we all know, the Standard Model of Elementary Particle Physics describes elementary particles. The term ``Standard Model'' refers to the theoretical framework of a quantum theory that incorporates the theory of strong interactions, abbreviated QCD, and the electroweak theory (unified theory of weak and electromagnetic interactions).
Although the Standard Model has been confirmed experimentally, it is incomplete because it does not include gravity. It treats gravity and the three forces independently, with gravity serving as a classical background. To overcome this shortcoming, we required a mathematically precise unified theory capable of explaining all forces. String theory, a physical theory that first emerged in the early 1960s, is currently the most promising candidate for unifying gravity and the other three fundamental forces.

The advantage of using String Theory as the theoretical framework is that it is not plagued by the lack of natural mechanisms for determining the values of parameters. Both the Standard Model and General Relativity lack natural mechanisms and rely on fine-tuning of their parameters for fixing their values. String Theory has only one unfixed parameter, the string scale, while the remaining parameters are treated as dynamical variables that are fixed via mechanisms provided by the theory's internal dynamics. This gives us reason to be optimistic about String Theory's ability to provide us with the values of observed ``real world'' parameters.
The identification of $D$-branes as fundamental objects in the string spectrum resulted in the discovery of AdS/CFT correspondence \cite{Maldacena}- see \cite{Minwalla} for the applications. This connection says that a quantum field theory existing in d-dimensional spacetime is equivalent to its dual quantum gravity theory existing in a one-dimensional higher (d+1) spacetime dimension. AdS/CFT serves as a bridge between Standard Model and String Theory. We will discuss holography, standard AdS/CFT correspondence, and its generalizations in the sections that follow.

\section{AdS-CFT}
Holography has remained a fascinating concept since the earliest days of Greek philosophy. The term holography (derived from the Greek words holo, which means whole, and graphy, which means writing) refers to the concept that Physics in a volume of any d-dimension can be described in terms of the degrees of freedom on the surface of the volume of dimensionality (d-1). This holography explanation is connected to the operation of a hologram, which records the image of a three-dimensional object on a two-dimensional surface. When we look back in time, we discover that Anaxagoras pre-Socratic philosophy bears a striking resemblance to holography. To address the issue of change, Anaxagoras proposed that the tiniest part of the world, obtained by infinitely dividing the universe into smaller parts, still contains all other things' information. Consider the case of holograms to illustrate his concept of ``everything in everything''. When a hologram is fragmented into smaller pieces, unlike normal photographs, each fragmented portion of the hologram depicts the original image in a smaller but complete form.
Recently, string theory has seen a revival of Anaxagora's philosophy with the discovery of the AdS/CFT correspondence \cite{Maldacena}. Correspondence can be viewed as the modern incarnation of holography.
\subsection{Why should QCD have a string dual$?$}
Heuristically, the urge to reformulate QCD as a string theory stems from the fact that QCD contains flux tubes between quark-antiquark couples that contribute to their confinement.  Using a string as a model for the flux tubes leads to Regge behaviour. The regge behaviour of the flux tube is defined by the relationship between its mass M and its mass J, which is defined as $M^2\propto J$. In the real world, the meson spectrum exhibits the same behaviour, i.e., the quark-antiquark bound state. This argument, however, does not apply to non-constricting gauge theories.
Consideration of 't Hooft's large-$N_c$ limit provides a compelling argument for the existence of a string dual of quantum gravity or any confining gauge theory. The gauge group of QCD is $SU(3)$ and, as a result of dimensional transmutation, it lacks an expansion parameter. 't Hooft generalised QCD by substituting $SU(3)$ for $SU(N_c)$, taking the limit $N_c\rightarrow\infty$, and expanding in $1/(N_c)$.
The quark and gluon fields $q^i_a$ and $A^i_\mu j$, where $i,j=1,...,N_c$ and $a=1,...,N_f$, respectively, represent the degrees of freedom of the generalised theory. Due to the fact that the gauge group is now $SU(N_c)$ rather than $U(N_c)$, the number of independent gauge fields is reduced to $N^2_c-1$, but since the working limit is $N_c\rightarrow\infty$, this difference can be ignored. As a result, the number of gluons is assumed to be $N^2_c$. In comparison to $N_{f}N_{c}$, which denotes the number of quark degrees of freedom, this $N^2_{c}$ is significantly larger. Thus, in the large-$N_{c}$ limit, the dynamics is governed by gluons.

 One-loop gluon self energy Feynman diagram scales as $g^{2}_{YM}N_{c}$. The Feynman diagram possess a smooth limit if along with limit $N_{c}\rightarrow\infty$ one simultaneously takes a limit on the gauge coupling given as $g_{YM}\rightarrow 0$ while keeping the 't Hooft coupling $\lambda\equiv g^{2}_{YM}N_{c}$, fixed. This is equivalent to demanding that in the large-$N_{c}$ limit the confinement scale, $\Lambda_{QCD}$, remain fixed. The one-loop $\beta-$function given as,
 $$\mu\frac{d}{d\mu}g^{2}_{YM}\propto-N_{c}g^{4}_{YM}$$,
when written in terms of $\lambda$, it becomes independent of $N c$.:
 $$\mu \frac{d}{d\mu}\lambda\propto-\lambda^{2}.$$

 The use of double-line notation makes determining the $N_c$-scaling of Feynman diagrams easier. This is accomplished by replacing the line associated with a gluon with a pair of lines associated with a quark and anti-quark. 
 As one can see, Feynman diagrams can be expanded in terms of the powers of $\lambda$ and $1/N_c$. Consider the examples of one-, two-, and three-loop vacuum diagrams to demonstrate this explicitly. They scale to the same power of $N_c$ but to a different power of $\lambda^{l-1}$, where $l$ denotes the loop count. 
 Topology plays a crucial role in classifying diagrams and in the large-$N_{c}$ limit non-planar diagrams are suppressed.

 The classification of diagrams according to their topology paves the way for their connection to string theory; the connection can be made more precise by associating with each Feynman diagram a Riemann surface. Each line representing a closed loop in a Feynman diagram in double-line notation can be thought of as the boundary of a two-dimensional surface. A Riemann surface is obtained by glueing these surfaces together along their boundaries. To create a compact surface, one adds the point at infinity to the face associated with the diagram's external line.
The Riemann surface of a planar diagram is a sphere, whereas the Riemann surface of a non-planar diagram is a torus. Scaling is as follows for a given Feynman diagram: $N^{\chi}_c$, where $\chi$ is the Euler number of the corresponding Riemann surface.
 $$\chi=2-2g,$$
 where $g$ is the genus of the orientable, compact surface with no boundaries. For sphere $\chi=2$ and for torus $\chi=0$. Therefore amplitude of Feynamnn diagrams can be written as
 $${\cal A}=\sum^{\infty}_{g=0}N^{\chi}_{c}\sum^{\infty}_{n=0}c_{g,n}\lambda^{n},$$

 where $c {g,n}$ are constants The analysis applies to any confining gauge theory with Yang-Mills fields and matter in adjoint representation. The introduction of matter or quarks into the theory results in the introduction of a boundary in the Riemann surface that is associated with a Feynmann diagram. The presence of matter has no effect on the power of $N_c$ associated with the Feynmann diagram, which remains $N^{\chi}_c$, but the Euler number changes to $\chi=2-2g-b$, where $b$ is the number of boundaries. The boundaries are associated with open strings. As a result, the summation over the number of boundaries is recognised as an expansion for a theory with both closed  and open strings. A gauge theory's large-$N_c$ expansion corresponds to a string theory's genus expansion. As a result, the gauge theory's planar limit corresponds to the string theory's classical limit. This duality is what gives rise to the AdS/CFT correspondence.

\subsection{Maldacena's conjecture}
The Maldacena's conjecture(also known as AdS/CFT correspondence)\cite{Maldacena} relates gravity theories in AdS to conformal field theories. Originally the conjecture postulated that the type IIB superstring theory on $AdS_{5}\times S^{5}$ is equivalent to $\cal{N}$=4 Super Yang-Mills theory in 3+1 dimensions. The strongest form of the $AdS_{5}/CFT_{4}$ states that $\cal{N}$=4 SYM theory with gauge group $SU(N_{c})$ and Yang-Mills coupling constant $g_{YM}$ is dynamically equivalent to type IIB superstring theory with string length $l_{s}=\sqrt{\alpha\prime}$ and coupling constant $g_{s}$ on $AdS_{5}\times S^{5}$ with radius of curvature R and N units of $F_{(5)}$ flux on $S^{5}$. The two free parameters on both side of the theory are related as:
$$g^{2}_{\rm YM}=2\pi g_{s}$$ and
$$2g^{2}_{\rm YM}N=R^{4}/\alpha^{\prime^{2}}$$
 The conjectured duality gives the equivalence of two theories which means that there exists a precise one-to-one mapping between the gauge invariants and local operators of the gauge side, and the states and fields of the string theory. However, a full quantum treatment of the superstring is not possible which prevents the use of duality in its strongest form, so one works with the weaker version of duality, obtained by taking suitable approximations. For weaker version of duality one considers the large-'t Hooft coupling limit: $\lambda=g^{2}_{YM}N_{c}$ fixed but very large, as $N_{c}\rightarrow \infty$ and $g_{\rm YM}^2=g_s\rightarrow 0$. This limit relates to a topological expansion of the Feynman diagrams on the gauge theory side, and to weakly coupled string theory on the string side.

 The weaker limit of the conjecture discussed above still presents some difficulty to work with and we have to dig even further to obtain a tractable setting.For the new limit, we are left with only one free parameter, $\lambda$; hence we can study the behaviour at both ends, $\lambda$ very small or very large. In D-brane picture which motivated the correspondence these limits arise naturally, and so in the next section we give a brief introduction to this duality from this perspective.

 \subsection{D-branes and strings}
 Strings are not the only extended objects that can be defined in superstring theory; it also contains a variety of non-perturbative higher dimensional objects known as D-branes. D-branes can be viewed from two different perspectives: {\it Open string} and {\it Closed string}. The appropriate perspective is determined by the value of the string coupling, $g_s$, which defines the strength of the interaction between open and closed strings.
 \begin{itemize}
\item{Open string perspective}:D-brans can be viewed in this light as higher-dimensional objects onto which open strings can terminate. This viewpoint is consistent for small coupling constant, e.g. $g_s<<1$ for both closed and open strings. The dynamics of the open strings at low energies $E<<\alpha^{\prime^{-1/2}}$, when massive string excitations are ignored, is described by a supersymmetric gauge theory based on the world volume of the D-branes. Gauge field $A_mu$ corresponds to open string excitations parallel to the D-brane, whereas scalar field $\phi$ corresponds to open string excitations transverse to the D-brane. The product $g_sN$ is the effective coupling constant for a stack of {\it N} coincident D-branes with gauge group $U(N)$, and the open string perspective is reliable for $g_{s}N<<1$.

\item{Closed string perspective}: In this perspective D-branes may be visualised as solitonic solutions of supergravity( low energy limit of superstring theory). D-branes act as the source of gravitational field responsible for curving the surrounding spacetime. To ensure the validity of supergravity approximation and small curvature, the length scale L should be large. For a stack of N coincident D-branes, $L^{4}/\alpha^{\prime^{ 2}}\propto g_{s}N$. The closed string perspective is valid only for $g_{s}N>>1$.
    \end{itemize}
Applying these two perspective to a stack of $N$ D3-branes in Minkowski spacetime give rise to $AdS_{5}/CFT_{4}$ correspondence. The stack of $N$ D3-branes extend along Minkowskian spacetime directions, and is transverse to rest of the six spatial direction.

\subsection{Open string perspective}
For $g_{s}N<<1$ type IIB perturbative string theory consists of two kinds of strings:
\begin{itemize}
\item Open strings beginning and ending on the D3-branes, where open strings may be viewed as excitation of (3+1)dimensional hyperplane.
\item Closed strings, which are considered as excitation of (9+1)dimensional flat spacetime.
\end{itemize}
One works with only massless excitations while ignoring all other stringy excitations for $N$ D3-branes in flat spacetime at energies $E<<\alpha^{\prime^{ -1/2}}$. The massless closed string states are grouped into ten dimensional $\cal{N}$=1 supergravity multiplet, while the massless open string excitations are grouped into a four dimensional $\cal{N}$=4 which consists of a gauge field $A_{\mu}$ and six real scalar fields $\phi^{i}$ along with fermionic superpartners. As per the transformation properties of D3-branes under their world volume rotations, gauge field $A_{\mu}$ arises from the bosonic massless open string excitationns along the D3-branes and six real scalar fields $\phi^{i}$ arises from excitations of bosonic massless open string transverse to the D3-branes. The complete effective actionn for all massless string mods is given as
$$S=S_{closed}+S_{open}+S_{int}$$
where $S_{closed}$ contains the closed string modes contribution, $S_{open}$ the open string modes and $S_{int}$ the interactions between closed and open string modes; $S_{closed}$ is the ten dimensional supergravity action.
The action for $S_{open}$ and $S_{int}$ can be derived from the DBI action and the Wess-Zumino term. The Dirac-Born-Infeld action for a single D3-brane is give as
$$S_{DBI}=-\frac{1}{(2\pi)^{3}\alpha^{\prime 2}g_{s}}\int d^{4}x e^{-\phi}\sqrt{-det({\cal P}[g]+2\pi \alpha^\prime F),}$$
 where ${\cal P}[g]$ is the pullback metric on the $D3$-brane. Expanding up to leading order in $\alpha^\prime$,
 $$S_{open}=-\frac{1}{2\pi g_{s}}\int d^{4}x(\frac{1}{4}F_{\mu\nu}F^{\mu\nu}+\frac{1}{2}\eta^{\mu\nu}\partial_{\mu}\phi^{i}\partial_{\nu}\phi^{i}+{\cal O}(\alpha^\prime))$$
 $$S_{int}=-\frac{1}{8\pi g_{s}}\int d^{4}x \phi F_{\mu\nu}F^{\mu\nu}+...$$

 For the case of $N$ coincident D3 branes tha gauge and scalar fields are $U(N)$ valued, $A_{\mu}=A^{a}_{\mu}T_{a}$, $\phi^{i}=\phi^{ia}T_{a}$. Taking a trace over the gauge group when generalising the low-energy-effective actions $S_{open}$ and $S_{int}$ for $N$ coincident D3-branes ensures gauge invariance. Partial derivatives are replaced by covariant derivatives and one needs to add a scalar potential V to the action $S_{open}$ with V given as
 $$V=\frac{1}{2\pi g_{s}}\sum_{i,j}Tr[\phi^{i},\phi^{j}]^{2}.$$

For the limit $\alpha^\prime\rightarrow 0$(also known as Maldacena limit) $S_{open}$ is just the bosonic part of the ${\cal N}=4$ SYM theory action provided
$$2\pi g_{s}=g^{2}_{YM}.$$
Since $\kappa\propto\alpha^\prime\rightarrow 0$, one observes that $S_{closed}$ is nothing but the free supergravity action in 10-dimensional Minkowski spacetime. Following a rescaling of the dilation $\phi$ by $\kappa$ for canonical normalisation on sees that $S_{int}$ also vanishes leaving open and closed strings decoupled.

\subsection{Closed string perpective}
The $N$ $D3$-branes in the strong coupling limit $g_{s}N\rightarrow\infty$, need to be looked at from closed string perspective. Wherein they are considered as massive charged objects which act as source for various type IIB supergravity fields and therefore type IIB string theory.

The ten dimensional supergravity solution of $N$ D3-branes preserving $SO(3,1)\times SO(6)$ isometries of spacetime, and half of the supercharges of type IIB supergravity is given by
\begin{eqnarray}
\label{Closed-metric}
ds^{2}&=&H(r)^{-1/2}\eta_{\mu\nu}dx^{\mu}d^{\nu}+H(r)^{1/2}\delta_{ij}dx^{i}dx^{j},
\end{eqnarray}
\begin{eqnarray}
\label{closed-exp}
e^{2\phi(r)}&=&g^{2}_{s},
\end{eqnarray}
\begin{eqnarray}
\label{closed-four-form}
C_{(4)}&=&(1-H(r)^{-1})dx^{0}\wedge dx^{1}\wedge dx^{2}\wedge dx^{3}+...
\end{eqnarray}
where $i,j=4,...,9$ and $\mu,\nu=0,...,3$, $r^{2}=\sum_{i=4}^{9}x^{2}_{i}$. Both the (\ref{closed-exp}) and the eoms of type IIB supergravity tell that
\begin{equation}
\label{closed-H}
H(r)=1+(\frac{L}{r})^{4}.
\end{equation}
Since the flux $F_{(5)}$ through the sphere $S^{5}$ is quantised because it counts the number of coincident $D3$-branes.
$$L^{4}=4\pi g_{s}N\alpha^{\prime^{2}}$$
The background can be divided into two regions depending on the value of r with respect to L:
\begin{itemize}
\item{r$\gg$ L}:In this region H(r)can be approximated by 1. For this value of H(r)the metric (\ref{Closed-metric})reduces to 10-dimensional flat spacetime.
\item{r$\ll$ L}: In this region H(r) can be approximated by $L^{4}/r^{4}$. The metric in this region can be approximated as
    \begin{eqnarray}
    \label{closed-metric-ii}
    ds^{2}&=&\frac{r^{2}}{L^{2}}\eta_{\mu\nu}dx^{\mu}dx^{\nu}+\frac{L^{2}}{r^{2}}\delta_{ij}dx^{i}dx^{j}\nonumber\\
    && =\frac{L^{2}}{z^{2}}(\eta_{\mu\nu}dx^{\mu}dx^{\nu}+dz^{2})+L^{2}ds^{2}_{S^{5}}.
    \end{eqnarray}
\end{itemize}
where the new coordinate z is given by $z=L^{2}/r$ and $\delta_{ij}dx^{i}dx^{j}=dr^{2}+r^{2}ds^{2}_{S^{5}}$. The first term occuring in the second line of equation (\ref{closed-metric-ii})is the $AdS_{5}$ metric.
Thus we have two different type of closed strings depending on the the region:
\begin{itemize}
\item closed strings propagating in flat 10-dimensional spacetime which corresponds to the region for the limit $r\gg L$.
\item closed strings propagating in the near horizon region($AdS_{5}\times S^{5}$) which corresponds to the limit $r\ll L$
\end{itemize}
For the low energy limit both type of closed strings mentioned above gets decoupled. To summarize, there are two regions : an asymptotically flat region and a near-horizon region. In asymptotically flat spacetime closed string dynamics is given by type IIB supergravity modes in 10-dimensional flat spacetime., while in the near-horizon region dynamics of the closed strings is given by string excitations about $AdS_{5}\times S^{5}$solution of type IIB supergravity. In the low energy limit both types of strings get decoupled.
\subsection{Combining Both Perspectives}
In the low-energy limit one sees that in both the open and the closed string perspectives, we get two decoupled effective theories,
\begin{itemize}
\item{Closed string perspective}: type IIB supergravity on $AdS_{5}\times S^{5}$ spacetime and type IIB supergravity on $\rm{R}^{9,1}$ spacetime.

\item{Open string perspective}:${\cal N}=4$ SYM theory on flat four dimensional spacetime and type IIB supergravity on $\rm{R}^{9,1}$spacetime.
\end{itemize}
The fact that physics remains unchanged irrespective of the choice of perspective and presence of type IIB supergravity modes on $\rm{R}^{9,1}$ in both the perspectives, leads to the conclusion that the two perspectives are dual to each other which is nothing but what the Maldacena conjectured.

 \section{Generalization of AdS/CFT Correspondence}
 The original Maldacena conjecture or AdS/CFT correspondence as discussed above talks about a maximally supersymmetric conformal gauge theory with 16 supercharges. But a realistic gauge theory for example QCD is non-conformal and non-supersymmetric. So to be able to use AdS/CFT correspondence for non-conformal gauge theories one has to generalize the correspondence. To generalize the correspondence one has to
 \begin{itemize}
 \item break the supersymmetry
 \item break the conformal invariance.
 \end{itemize}
 As we have seen in the previous section while explaining AdS/CFT correspondence open and closed string degrees of freedom decouple in the low energy limit. But the same is not true for a generalised version of AdS/CFT also known as gauge/gravity duality which has a non-AdS background. This decoupling was necessary in establishing an equivalence between closed superstring theory on AdS space and maximally supersymmetric gauge theory. However the AdS/CFT correspondence can also be obtained by taking supergravity approximation of the superstring theory (weaker version of the correspondence) without including any stringy correction. In the  supergravity limit decoupling of open and closed string sector is not possible for the the case of non-conformal gauge theories. One needs to include the string states and hence go beyond the supergravity limit which is a difficult task.  See \cite{S.Roy-1}, \cite{S.Roy-2}, \cite{S.Roy-3}, \cite{S.Roy-4}, \cite{S.Roy-5}, \cite{S.Roy-6},\cite{S.Roy-7} for more recent works on the aforementioned decoupling and extension of the AdS/CFT to non-supersymmetric cases. There are different ways to choose a target spacetime along with a particular $D$-brane configuration to break both the conformal invariance and supersymmetry at the very beginning. Depending on the choice of target spacetime  there exists many generalized versions of the AdS/CFT correspondence - a few of them are \cite{KW},\cite{KS},\cite{Nunez-6},\cite{Kruczenski:2003uq},\cite{Bergman:2001rw},\cite{Ballon-Bayona:2018ddm}. Placing $D$-branes at the point of conical singularity instead of any smooth point in a CY maifold breaks the supersymmetry.  In the next sections we have discussed about a few examples which include $D3$-branes placed at the tip of the conifold in Calabi Yau spaces.

 \subsection{Regular D3-branes at Conifold Singularity/Klebanov-Witten Solution }
 The Klebanov-Witten\cite{KW} solution contains only $D3$-branes, with less symmetric conifold $T^{1,1}$ internal manifold. The conifold gives rise to superpotential in the gauge theory by breaking the supersymmetry to ${\cal N}=1$. The branes are located at the tip of the conifold from gauge theory perspective. The resulting gauge theory on D3-branes is superconformal gauge theory with minimal supersymmetry ${\cal N}=1$ with $SU(N)\times SU(N)$ gauge group coupled to 4 chiral fields $A_{1}, A_{2}, B_{1}, B_{2}$. Fields $A_{1},A_{2}$ tranforms' as $(N,\bar N)$ of the gauge group while fields $B_{1}, B_{2}$ tranforms' as $(\bar{N},N)$ of the gauge group.
 On the gravity dual side, D3-branes are compactified on a warped conifold in the large radius limit of $AdS_{5}\times T^{1,1}$.  The conifold is a six dimensional cone whose base is $T^{1,1}$ with topology $S^{2}\times S^{3}$. Symmetry of the $T^{1,1}$ is $(SU(2)\times SU(2))/U(1)$. The full 10-dimensional metric is given as
 \begin{equation}
 ds^{2}_{10}=h^{-1/2}\eta_{\mu\nu}dx^{\mu}dx^{\nu}+h^{1/2}ds^{2}_{6}
 \end{equation}
  where $h$ is the warp factor given by
  \begin{equation}
  h(r)=1+\frac{L^{4}}{r^{4}}
  \end{equation}
  with $L^{4}=4\pi g_{s}N\alpha^{\prime^{2}}$ and $ds^{6}$ is the conifold metric:
  \begin{equation}
  ds^{2}_{conifold}=dr^{2}+r^{2}d^{2}_{T^{1,1}}
  \end{equation}
  with
  \begin{equation}
  ds^{2}_{T^{1,1}}=\frac{1}{9}(d\psi+\cos\theta_{1}d\phi_{1}+\cos\theta_{2}d\phi_{2})^{2}+\frac{1}{2}\sum^{2}_{i=1}(d\theta^{2}_{i}+\sin^{2}\theta_{i}d\phi^{2}_{i}).
  \end{equation}
  Combining both gravity and gauge perspective, one can see that the two couplings corresponding to the different gauge groups are related to the integral of $NS-NS$ $B_2$ and R-R $C_2$ over the $S^2$ of $T^{1,1}$. With only D3-brane in the picture, both dilaton and $B_2$ are constant, which renders both the couplings corresponding to both the gauge groups to be constant. Hence the gauge theory is conformal.
  \subsection{Fractional D3 branes at Conifold singularity/Klebanov-Tseytlin Solution }
   The Klebanov-Tsytlin \cite{KT}solution considered addition of M fractional D3-branes to the KW solution. M fractional D3-branes in reality are  D5-branes which are wrapping the vanishing two-cycle of $T^{1,1}$ at the tip of the conifold. Addition of these extra branes breaks the conformality of the gauge theory while still preserving ${\cal N}=1$ supersymmetry. These extra D5-branes modifies the gauge group to $SU(N+M)\times SU(N)$ by providing additional locations for the string endpoints. On the gauge theory side, the chiral superfields $A_i$ and $B_i$ now transforms' as  $(N+M,\bar N)$ represention of the product gauge group.
   Supergravity solution for the KT contains 3- and 5- from fluxes. D5-branes being charged under the Hodge dual of $F_3$ act as a source of M units of fluxes on $F_3$. The 3-form flux solutions are given as
   \begin{equation}
   F_3=M \omega_3,\ B_2=3g_s M\omega_2\ln(r/r_0),\ H_3=dB_2=3g_s M\frac{1}{r}dr\wedge\omega_2
   \end{equation}
    where $\omega_2$ and $\omega_3$ are the basis of the base of the conifold. The five form flux with the backreaction of $F_3$ is given as
    \begin{equation}
    F_{5}={\cal F}_{5}+*_{10}{\cal F}_{5},\ {\cal F}_{5}=(N+\frac{3}{2\pi}g_{s}M^{2}\ln(r/r_{0}))vol(T^{1,1})
    \end{equation}
   The number of effective D3-branes can be defined as,
   \begin{equation}
   N_{eff}(r)=N+\frac{3}{2\pi}g_{s}M^{2}\log(\frac{r}{r_{0}})
   \end{equation}
   one can see that number of effective branes decreases as $r\rightarrow 0$. The 10-dimensional warp factor in the near-horizon limit takes the form
   \begin{equation}
   h(r)=\frac{L^{4}}{r^{4}}(1+\frac{3g_{s}M^{2}}{2\pi N}\log r)
   \end{equation}
   The logarithmic dependence on the radial coordinate, also known as the RG scale, makes both ${\cal F}_5$ and h(r) tend to zero and eventually negative as r decreases, {\it i.e.}, towards the IR of the RG flow. Introduction of fractional $D3$-branes explicitly breaks the conformality by making the couplings run logarithmically.
   \subsection{Seiberg duality cascade}
   The Klebanov-Strassler \cite{KS} model gave a better understanding of singularity by providing a mechanism to resolve the singularity. It also gave a geometric realisation of confinement which is essential for QCD-like gauge theories. While the flux through the vanishing $S^2$ decreases by one unit when r decreases, the five form flux decreases by M units. In the same way the effective number of $D3$-branes also decreases:$N_{\rm eff}\rightarrow N_{\rm eff}-M$. In terms of gauge group this reduction is seen as:$SU(N+M)\times SU(N)\rightarrow SU(N-M)\times SU(N)$ as one goes down one step in the RG scale. This is known as the Seiberg duality. If one assumes N to be an integral multiple of M, this reduction can be repeated till only the gauge group $SU(M)$ remains in the IR. This series of dualities is called a duality cascade.
   From the gauge theory point of view at the end of the duality cascade $U(1)_R$ symmetry of the chiral fields get broken down to $Z_{2M}$ due to strong IR dynamics the presence of M fractional $D3$-branes. This symmetry is further broken down to $Z_2$ in the IR due to gaugino condensation resulting in desingularizing the singular conifold into a deformed conifold.
   The complete KS solution can be seen as an interpolation between the deformed solution in the IR and the KT solution in the UV. Due to the duality cascade in the IR the gauge theory is pure $SU(M)$ theory. If one can explicitly realize the UV completion of this setup then the string theory can provide a complete picture of the confinement. Hence it neccessitates modification of the UV sector of KS.

   \subsection{Resolved conifold and D7 brane embedding}
   As seen in the previous sub-section the singularity in the KT solution was removed via a deformed conifold which corresponds to a blown up $S^{3}$ in the IR. Another way to remove the singularity is via a $S^{2}$ blow-up in the deep IR at the tip of the conifold. This is known as the resolved conifold geometry and the metric for the resolved internal manifold is given as:
   \begin{eqnarray}
   ds^{2}_{res}&=&\kappa(r)^{-1}dr^{2}+\frac{\kappa(r)}{9}r^{2}(d\psi+\cos\theta_{1}d\phi_{1}+\cos\theta_{2}d\phi_{2})^{2}+\frac{r^{2}}{6}(d\theta_{1}^{2}+\sin^{2}\theta_{1}d\phi_{1}^{2})\nonumber\\
   &&+\frac{r^{2}+6a^{2}}{6}(d\theta_{2}^{2}+\sin^{2}\theta_{2}d\phi_{2}^{2})
   \end{eqnarray}
   where r is the suitable radial coordinate, $\kappa(r)=\frac{r^{2}+9 a^2}{r^{2}+6 a^2}$ and a is the resolution parameter. In the limit$\rho\rightarrow 0$ of newly defined radial coordinate the metric on $S^{2}(\theta_1,\phi_1)$ vanishes but the metric on the blown-up $S^2(\theta_2,\phi_2)$ does not. The sphere remains finite because of fintie a. For $a\rightarrow 0$ we get a singular conifold metric.

   \cite{Zayas+Tseytlin} explored a geometric arrangement of N $D3$-branes as well as fractional M $D$-branes. N $D3$-branes were put at the conifold's tip in the geometric arrangement, whereas fractional M $D3$-branes wrapped around the conifold's blown-up two cycles. The related supergravity solution included an RR three-form flux, an NS-NS three-form flux $H_3$, a self dual five-form flux $F_5$, and a dilaton $\phi$. But the arrangement lacked in terms of absence of fundamental quarks.
   The fundamental quarks were introduced into the picture via flavor branes. First attempt in this direction was done in \cite{KK}, where $D7$-branes were embedded in the AdS spacetime to include quarks. This inclusion was done in the probe limit to avoid the back reaction of the flavor branes. In \cite{ouyang} a stack of $D7$-branes were embedded in a conifold background  via Ouyang embedding. In \cite{metrics} $N_f$ D7-branes wrapping a non-trivial four cycle were embedded in a non-supersymmetric resolved conifold background via Ouyang embedding:
   \begin{equation}
   (r^{2}+9 a^{2}r^{4})^{1/4}e^{\frac{\iota}{2}\left(\psi-\phi_{1}-\phi_{2}\right)}\sin\frac{\theta_1}{2}\sin\frac{\theta_2}{2}=\mu
   \end{equation}
where $\mu $ is the embedding  parameter. In the presence of flavor branes the dilaton acquires a radial dependence as,
\begin{equation}
e^{-\phi}=\frac{1}{g_s}-\frac{N_f}{8\pi}\log(r^{6}+9 a^{2}r^{4})-\frac{N_{f}}{2\pi}\log(\sin\frac{\theta_1}{2}\sin\frac{\theta_2}{2})
\end{equation}
\section{Mia-Dasgupta-Gale-Jeon setup}
 In this section, via two subsections, we will:

\begin{itemize}
\item
discuss about the type IIB background of \cite{metrics},  a UV complete holographic dual of large-$N$ thermal QCD,
\item
discuss the 'MQGP' limit of \cite{MQGP} and the rationale behind it.
\end{itemize}

\subsection{Type IIB Dual of Large-$N$ Thermal QCD}
In this subsection, we will discuss a UV complete holographic dual of large-$N$ thermal QCD in this subsection, as described in \cite{metrics}. As partly mentioned in Sec. {\bf 1}, this was inspired by the zero-temperature Klebanov-Witten model \cite{KW}, the non-conformal Klebanov-Tseytlin model \cite{KT}, its IR completion as given in the Klebanov-Strassler model \cite{KS} and Ouyang's inclusion \cite{ouyang} of flavor in the same 
, as well as the non-zero temperature/non-extremal version of \cite{Buchel} ( the solution was not regular, since both the non-extremality/black hole function and the ten-dimensional warp factor disappeared concurrently at the horizon radius.), \cite{Gubser-et-al-finitetemp} (valid only at large temperatures) of the Klebanov-Tseytlin model  and \cite{Leo-ii} (addressing the IR), in the absence of flavors.  \\

To our knowledge, the following model proposed in \cite{metrics}, building up on previous work \cite{KW}, \cite{KS}, \cite{ouyang}, \cite{Nunez-2} -\cite{Nunez-4}, in the context of gauge/gravity duality, is the closest to a UV complete holographic dual of large-$N$ thermal QCD for including fundamental quarks at non-zero temperatures in type IIB string theory. With regard to gauge group $SU(M)$, both QCD and the KS model (after duality cascade) have similar behaviour in the infrared: IR confinement. Nevertheless, they differ dramatically in the ultraviolet (UV), where the former produces a logarithmically divergent gauge coupling (Landau pole). Radially dependent 3-form fluxes are responsible for this phenomenon. The UV sector of the KS model must be modified as a result. The type IIB holographic dual of \cite{metrics} was constructed with all of the above in mind. Below is a brief description of the set up of \cite{metrics}.

 \begin{itemize}
 \item
  From a gauge-theory perspective, the authors of \cite{metrics} considered  $N$ black $D3$-branes placed at the tip of six-dimensional conifold, $M\ D5$-branes wrapping the vanishing two-cycle and $M\ \overline{D5}$-branes  \cite{KS},\ \cite{Ganor:2014wua} (in the context of fractional particles) distributed along the resolved two-cycle and placed at the antipodal point relative to the location of $M\ D5$ branes on the blown-up $S^2$. Let ${\cal R}_{D5/\overline{D5}}$ denote the separation between $D5/\overline{D5}$ branes. The radial space, in \cite{metrics} is divided into three parts the IR, the IR-UV interpolating region and the UV depending on ${\cal R}_{D5/\overline{D5}}$. To summarize the above:
  \begin{itemize}
  \item
  $r_0/r_h<r<|\mu_{\rm Ouyang}|^{\frac{2}{3}}/{\cal R}_{D5/\overline{D5}}$: the IR/IR-UV interpolating regions with $r\sim\Lambda$: deep IR where the $SU(M)$ gauge theory confines

  \item
  $r>|\mu_{\rm Ouyang}|^{\frac{2}{3}}/ {\cal R}_{D5/\overline{D5}}$: the UV region.

  \end{itemize}

\item
$N_f\ D7$-branes, via Ouyang embedding,  are holomorphically embedded in the UV (asymptotically $AdS_5\times T^{1,1}$), the IR-UV interpolating region and dipping into the (confining) IR (up to a certain minimum value of $r$ corresponding to the lightest quark)  and $N_f\ \overline{D7}$-branes present in the UV and the UV-IR interpolating (not the confining IR). This is to ensure turning off of three-form fluxes, constancy of the axion-dilaton modulus and hence conformality and absence of Landau poles in the UV.
The resultant ten-dimensional geometry hence involves a resolved warped deformed conifold \cite{Panda:2009ji}, \cite{Ali:2008ij} in the context of $D-$brane inflation. Back-reactions are included, e.g., in the ten-dimensional warp factor(\ref{eq:h}), fluxes(\ref{three-form-fluxes}). Of course, the gravity dual, as in the Klebanov-Strassler construct, at the end of the Seiberg-duality cascade will have no $D3$-branes and the $D5$-branes are smeared/dissolved over the blown-up $S^3$ and thus replaced by fluxes in the IR.
\end{itemize}

 Due to the presence of $D3$-branes and $D5$-branes in the UV:$r\geq {\cal R}_{D5/\overline{D5}}$, one has $SU(N+M)\times SU(N+M)$ color gauge group and $SU(N_f)\times SU(N_f)$ flavor gauge group. It is expected that there will be a partial Higgsing of $SU(N+M)\times SU(N+M)$ to $SU(N+M)\times SU(N)$ at $r={\cal R}_{D5/\overline{D5}}$  \cite{K. Dasgupta et al [2012]}. The two gauge couplings, $g_{SU(N+M)}$ and $g_{SU(N)}$ flow  logarithmically  and oppositely in the IR:
\begin{equation}
\label{RG}
4\pi^2\left(\frac{1}{g_{SU(N+M)}^2} + \frac{1}{g_{SU(N)}^2}\right)e^\phi \sim \pi;\
 4\pi^2\left(\frac{1}{g_{SU(N+M)}^2} - \frac{1}{g_{SU(N)}^2}\right)e^\phi \sim \frac{1}{2\pi\alpha^\prime}\int_{S^2}B_2.
\end{equation}
  Had it not been for $\int_{S^2}B_2$, in the UV, one could have set $g_{SU(M+N)}^2=g_{SU(N)}^2=g_{YM}^2\sim g_s\equiv$ constant (implying conformality) which is the reason for inclusion of $M$ $\overline{D5}$-branes at the common boundary of the UV-IR interpolating and the UV regions, to annul this contribution. In fact, the running also receives a contribution from the $N_f$ flavor $D7$-branes which needs to be annulled via $N_f\ \overline{D7}$-branes. The gauge coupling $g_{SU(N+M)}$ flows towards strong coupling and the $SU(N)$ gauge coupling flows towards weak coupling. Upon application of Seiberg duality, $SU(N+M)_{\rm strong}\stackrel{\rm Seiberg\ Dual}{\longrightarrow}SU(N-(M - N_f))_{\rm weak}$ in the IR;  assuming after repeated Seiberg dualities or duality cascade, $N$ decreases to 0 and there is a finite $M$, {one will be left with $SU(M)$ gauge theory with $N_f$ flavors that confines in the IR - the finite temperature version of the same is what was looked at by \cite{metrics}}. Interesting work has been done in the context of large N gauge theories at finite temperature with quarks in electric and magnetic field in \cite{arnabkundu0709.1547}\cite{arnabkundu0709.1554}; also see \cite{Minwalla-3},\cite{Obers:2008pj} for a review on large-N gauge theoretic description of black holes in $AdS_5\times S^5$ in backgrounds dual to confining gauge theories.


In the gravity dual the working metric is given by :
\begin{equation}
\label{metric}
ds^2 = \frac{1}{\sqrt{h}}
\left(-g_1 dt^2+dx_1^2+dx_2^2+dx_3^2\right)+\sqrt{h}\biggl[g_2^{-1}dr^2+r^2 d{\cal M}_5^2\biggr],
\end{equation}
 where $g_i$'s are black hole functions in modified OKS(Ouyang-Klebanov-Strassler)-BH (Black Hole) background and are assumed to be:
$ g_{1,2}(r,\theta_1,\theta_2)= 1-\frac{r_h^4}{r^4} + {\cal O}\left(\frac{g_sM^2}{N}\right)$
where $r_h$ is the horizon, and the ($\theta_1, \theta_2$) dependence come from the
${\cal O}\left(\frac{g_sM^2}{N}\right)$ corrections. The compact five dimensional metric in (\ref{metric}), is given as:
\begin{eqnarray}
\label{RWDC}
& & d{\cal M}_5^2 =  h_1 (d\psi + {\rm cos}~\theta_1~d\phi_1 + {\rm cos}~\theta_2~d\phi_2)^2 +
h_2 (d\theta_1^2 + {\rm sin}^2 \theta_1 ~d\phi_1^2) +   \nonumber\\
&&  + h_4 (h_3 d\theta_2^2 + {\rm sin}^2 \theta_2 ~d\phi_2^2) + h_5~{\rm cos}~\psi \left(d\theta_1 d\theta_2 -
{\rm sin}~\theta_1 {\rm sin}~\theta_2 d\phi_1 d\phi_2\right) + \nonumber\\
&&  + h_5 ~{\rm sin}~\psi \left({\rm sin}~\theta_1~d\theta_2 d\phi_1 +
{\rm sin}~\theta_2~d\theta_1 d\phi_2\right),
\end{eqnarray}
$r\gg a, h_5\sim\frac{({\rm deformation\ parameter})^2}{r^3}\ll  1$ for $r \gg({\rm deformation\ parameter})^{\frac{2}{3}}$, i.e. in the UV/IR-UV interpolating region.  The $h_i$'s appearing in internal metric as well as $M, N_f$ are not constant and up to linear order depend on $g_s, M, N_f$ are given as below:
\begin{eqnarray}
\label{h_i}
& & \hskip -0.45in h_1 = \frac{1}{9} + {\cal O}\left(\frac{g_sM^2}{N}\right),\  h_2 = \frac{1}{6} + {\cal O}\left(\frac{g_sM^2}{N}\right),\ h_4 = h_2 + \frac{a^2}{r^2},\nonumber\\
& & h_3 = 1 + {\cal O}\left(\frac{g_sM^2}{N}\right),\ h_5\neq0,\
 L=\left(4\pi g_s N\right)^{\frac{1}{4}}.
\end{eqnarray}
One sees from (\ref{RWDC}) and (\ref{h_i}) that one has a non-extremal resolved warped deformed conifold involving
an $S^2$-blowup (as $h_4 - h_2 = \frac{a^2}{r^2}$), an $S^3$-blowup (as $h_5\neq0$) and squashing of an $S^2$ (as $h_3$ is not strictly unity). The horizon (being at a finite $r=r_h$) is warped squashed $S^2\times S^3$.
In the IR the warp factor that includes the back-reaction, is given as:
\begin{eqnarray}
\label{eq:h}
&& \hskip -0.45in h =\frac{L^4}{r^4}\Bigg[1+\frac{3g_sM_{\rm eff}^2}{2\pi N}{\rm log}r\left\{1+\frac{3g_sN^{\rm eff}_f}{2\pi}\left({\rm
log}r+\frac{1}{2}\right)+\frac{g_sN^{\rm eff}_f}{4\pi}{\rm log}\left({\rm sin}\frac{\theta_1}{2}
{\rm sin}\frac{\theta_2}{2}\right)\right\}\Biggr],
\end{eqnarray}
where, in principle, $M_{\rm eff}/N_f^{\rm eff}$ are not necessarily the same as $M/N_f$; we however will assume that up to ${\cal O}\left(\frac{g_sM^2}{N}\right)$, they are.

In the IR, up to ${\cal O}(g_s N_f)$ and setting $h_5=0$, the three-forms are as given in \cite{metrics}:
\begin{eqnarray}
\label{three-form-fluxes}
& & \hskip -0.4in (a) {\widetilde F}_3  =  2M { A_1} \left(1 + \frac{3g_sN_f}{2\pi}~{\rm log}~r\right) ~e_\psi \wedge
\frac{1}{2}\left({\rm sin}~\theta_1~ d\theta_1 \wedge d\phi_1-{ B_1}~{\rm sin}~\theta_2~ d\theta_2 \wedge
d\phi_2\right)\nonumber\\
&& \hskip -0.3in -\frac{3g_s MN_f}{4\pi} { A_2}~\frac{dr}{r}\wedge e_\psi \wedge \left({\rm cot}~\frac{\theta_2}{2}~{\rm sin}~\theta_2 ~d\phi_2
- { B_2}~ {\rm cot}~\frac{\theta_1}{2}~{\rm sin}~\theta_1 ~d\phi_1\right)\nonumber \\
&& \hskip -0.3in -\frac{3g_s MN_f}{8\pi}{ A_3} ~{\rm sin}~\theta_1 ~{\rm sin}~\theta_2 \left(
{\rm cot}~\frac{\theta_2}{2}~d\theta_1 +
{ B_3}~ {\rm cot}~\frac{\theta_1}{2}~d\theta_2\right)\wedge d\phi_1 \wedge d\phi_2, \nonumber\\
& & \hskip -0.4in (b) H_3 =  {6g_s { A_4} M}\Biggl(1+\frac{9g_s N_f}{4\pi}~{\rm log}~r+\frac{g_s N_f}{2\pi}
~{\rm log}~{\rm sin}\frac{\theta_1}{2}~
{\rm sin}\frac{\theta_2}{2}\Biggr)\frac{dr}{r}\nonumber \\
&& \hskip -0.3in \wedge \frac{1}{2}\Biggl({\rm sin}~\theta_1~ d\theta_1 \wedge d\phi_1
- { B_4}~{\rm sin}~\theta_2~ d\theta_2 \wedge d\phi_2\Biggr)
+ \frac{3g^2_s M N_f}{8\pi} { A_5} \Biggl(\frac{dr}{r}\wedge e_\psi -\frac{1}{2}de_\psi \Biggr)\nonumber  \\
&&  \wedge \Biggl({\rm cot}~\frac{\theta_2}{2}~d\theta_2
-{ B_5}~{\rm cot}~\frac{\theta_1}{2} ~d\theta_1\Biggr). \nonumber\\
\end{eqnarray}
The asymmetry factors in (\ref{three-form-fluxes}) are given by: $ A_i=1 +{\cal O}\left(\frac{a^2}{r^2}\ {\rm or}\ \frac{a^2\log r}{r}\ {\rm or}\ \frac{a^2\log r}{r^2}\right) + {\cal O}\left(\frac{{\rm deformation\ parameter }^2}{r^3}\right),$ $  B_i = 1 + {\cal O}\left(\frac{a^2\log r}{r}\ {\rm or}\ \frac{a^2\log r}{r^2}\ {\rm or}\ \frac{a^2\log r}{r^3}\right)+{\cal O}\left(\frac{({\rm deformation\ parameter})^2}{r^3}\right)$. To ensure UV conformality, it is important to ensure that the axion-dilaton modulus approaches a constant implying a vanishing beta function in the UV.
\subsection{The `MQGP Limit'}
In \cite{metrics}, the following limit is considered:
\begin{eqnarray}
\label{limits_Dasguptaetal-i}
&   & \hskip -0.17in {\rm weak string}(g_s){\rm coupling-large\ t'Hooft\ coupling\ limit}:\nonumber\\
& & \hskip -0.17in g_s\ll  1, g_sN_f\ll  1, \frac{g_sM^2}{N}\ll  1, g_sM\gg1, g_sN\gg 1.
 \end{eqnarray}
In \cite{MQGP}), the following limit was considered:
\begin{eqnarray}
\label{limits_Dasguptaetal-ii}
& & \hskip -0.17in {\rm MQGP\ limit}: \frac{g_sM^2}{N}\ll  1, g_sN\gg1, {\rm finite}\
 g_s, M.
\end{eqnarray}

 There are principally two reasons for the motivation for considering the MQGP limit.
\begin{enumerate}
\item
For strongly coupled thermal systems like sQGP, one works with $g_{\rm YM}\sim{\cal O}(1)$ and $N_c=3$ which differs from the case of the AdS/CFT limit wherein $g_{\rm YM}\rightarrow0, N\rightarrow\infty$ such that $g_{\rm YM}^2N$ is large. In the IR after the Seiberg duality cascade, effectively $N_c=M$ which in the MQGP limit of (\ref{limits_Dasguptaetal-ii})  can be tuned to 3. Further, in the same limit, the string coupling $g_s\stackrel{<}{\sim}1$. The finiteness of the string coupling (see \cite{A.Sinha-4} for holography at finite coupling)necessitates addressing the same from an M theory perspective. This is the reason for coining the name: `MQGP limit' \cite{MQGP}. In fact this is the reason why one is required to first construct a type IIA mirror, which was done in \cite{MQGP} a la delocalized Strominger-Yau-Zaslow mirror symmetry, and then take its M-theory uplift.

\item
From the perspective of calculational simplification in supergravity, the following are examples of the same and constitute therefore the second set of reasons for looking at the MQGP limit of (\ref{limits_Dasguptaetal-ii}):
\begin{itemize}
\item
In the UV-IR interpolating region and the UV,
$(M_{\rm eff}, N_{\rm eff}, N_f^{\rm eff})\stackrel{\rm MQGP}{\approx}(M, N, N_f)$
\item
Asymmetry Factors $A_i, B_j$(in three-form fluxes)$\stackrel{MQGP}{\rightarrow}1$  in the UV-IR interpolating region and the UV.

\item
Simplification of ten-dimensional warp factor and non-extremality function in MQGP limit
\end{itemize}
\end{enumerate}
\section{Summary}
Let us now summarize the rest of the thesis, chapter-wise, below.
In Chapter 2, using the delocalized type IIA SYZ mirror of the holographic type IIB dual of large-N thermal QCD of \cite{metrics} as constructed in \cite{MQGP} we calculated (pseudo-)vector and (pseudo-)scalar meson spectra at finite coupling , and compared their result with \cite{Sakai-Sugimoto-1,Dasgupta_et_al_Mesons} and PDG data. It was found that masses of the (pseudo-)vector $(\rho[770], a1[1260], \rho[1450], a1[1640])$ and (pseudo-)scalar $(f0[980]/a0[980], f0[1370], f0[1450)$ mesons were closer to the PDG data than previously obtained in the literature.

In Chapter 3, we have discussed specific portion of \cite{Sil+Yadav+Misra-glueball} in which we had contributed maximally. In \cite{Sil+Yadav+Misra-glueball} we obtained the spectrum for $0^{++},0^{-+}, 0^{--},1^{++},2^{++}$ glueball states in the supergravity dual of large-N thermal QCD as constructed in \cite{MQGP}. The thermal QCD that we have considered here is essentially $QCD_3$. The glueball states were obtained by considering fluctuations in one form, two form, dilaton or appropriate metric components. For the computations we considered two different type of background geometries. We considered a black-hole background as well as one without the black hole known as thermal background. In the thermal background black-hole horizon radius is replaced by an infrared cut-off and the IR cut-off value is a result of $D7$-brane embedding. Analytically it is taken to be proportional to the two-third power of the Ouyang embedding parameter.

In Chapter 4, we studied (exotic) scalar glueball $0^{++}_E$ which correspond to metric fluctuations of the M theory uplift  at finite coupling. Using the same (exotic) scalar glueball $0^{++}_E$-meson interaction Lagrangian linear in (exotic) scalar glueball and quartic in meson fields was derived. Decay widths of the processess  $G_E \rightarrow 2\pi, G_E \rightarrow 2\rho, \rho \rightarrow 2\pi, G_E \rightarrow 4\pi, G_E \rightarrow \rho + 2\pi$ as well as indirect four $\pi$ decay associated with $G_E \rightarrow \rho + 2\pi \rightarrow 4\pi$ and $G_E \rightarrow 2\rho \rightarrow 4\pi$, were also obtained. By appropriate choice of combination of constants of integration appearing in the solutions to the EOMs of the profile functions of the $\pi$ and $\rho$ mesons and six metric perturbation,s these decay widths were shown to match exactly with PDG data.

In Chapter 5, we worked out the ${\cal O}l^{6}_{p}$ corrections to the M theory metric of \cite{MQGP}arising from terms quartic in the eleven dimensional supergravity action.

In Chapter 6, we used the ${\cal O}\left( \frac{1}{N}\right )$-corrections in conjuction with the ${\cal O}(R^{4})$-corrections to match the experimental values of the coupling constants up to ${\cal O}(p^{4})$ appearing in the $SU(3)$ Chiral Perturbation theory Lagrangian for the $\pi$ and $\rho$ vector mesons as well as their flavor partners, in the chiral limit. 
\chapter{Glueball Spectrum}

\graphicspath{{Chapter2/}{Chapter2/}}

\section{Introduction}
In recent years, research on glueballs and computation of their mass spectrum has been instrumental in identifying resonance states observed in a variety of experiments like RHIC and future EIC experiment. Gauge fields serve as dynamical degrees of freedom in QCD, and the non-abelian nature of the theory enables gauge bosons to form color-neutral bound states of gluons dubbed glueballs (gg, ggg, etc.). To gain a better understanding of quantum gravity's non-perturbative regime, one needs to have a careful look at glueballs \cite{Ni_Kochelev}, \cite{Hernandez_glueball}. The quantum state of a glueball is denoted by $J^{PC}$, where J, P, and C denote the total angular momentum, parity, and charge conjugation respectively.

Different generalized versions of AdS/CFT correspondence has thus far been proposed to study non-supersymmetric field theories with a running gauge coupling constant. For non-supersymmetric theories  for example, $QCD_3$ and $QCD_4$, Witten proposed extending the AdS/CFT correspondence \cite{Maldacena} to obtain a gravity dual. $QCD_3$ is a non-supersymmetric three-dimensional theory that can be obtained by compactifying a four-dimensional theory on a stack of parallel $D3$-branes around a circle and then imposing anti-periodic boundary conditions on the fermions around the same circle. On the other hand, $QCD_4$ is a four-dimensional non-supersymmetric theory obtained by successively compactifying a six-dimensional superconformal theory on a stack of $M5$-branes on two circles while imposing anti-periodic boundary conditions on fermions along either of the two circles. In both the cases the compactification leads to a black hole in AdS geometry. The glueball masses on these supergravity backgrounds were studied in detail in \cite{Czaki_et_al-0-+}, \cite{Mathur_et_al}.  Apart from these top-down approaches, the glueball spectra have been obtained using a variety of bottom-up holographic setups based on soft-wall, hard-wall, and modified-soft-wall models \cite{BH_all_T,Nicotri,Forkel,ForkelStructure,FolcoCapossoli,Jugeau,Colangelolight}.

Correlation functions of gauge invariant local operators can be used to estimate the masses of glueballs. To obtain the glueball spectrum in $QCD_3$, the first step is to identify the gauge operators with the quantum numbers corresponding to the desired glueballs. Each supergravity mode corresponds to a gauge theory operator according to the gauge/gravity duality. At the AdS space boundary, this operator is coupled to the supergravity mode, for example, the lowest dimension operator with quantum numbers. $J^{PC}=0^{++}$ is Tr$F^{2}=$Tr$F^{\mu\nu}F_{\mu\nu}$, and this operator is coupled to the boundary dilaton mode. To determine the mass of a $0^{++}$ glueball, one must first evaluate the correlator $\left<TrF^{2}(x)TrF^{2}(y)\right>$ =$\Sigma_{i} c_{i} e^{-m{i}|x-y|}$, where $m_i$ denotes the mass of the glueball. However, masses can be determined by solving the wave equations for supergravity modes that are coupled to the gauge theory operators on the boundary. This project takes the latter approach. The 11D metric obtained by uplifting the delocalized SYZ type IIA metric can be interpreted as a black $M5$-brane wrapping a two-cycle, i.e. a black $M3$-brane \cite{transport-coefficients,NPB}. Using this as a starting point, compactifying again along the M-theory circle, we arrive at the type IIA metric. Then compactifying again along the periodic temporal circle (with a radius equal to the reciprocal of the temperature), we obtain ${\rm QCD}_3$, which corresponds to the three non-compact directions of the black $M3$-brane world volume. In principle, the type IIB background of \cite{ metrics} consists of $M_4\times$ RWDC($\equiv$ Resolved Warped Deformed Conifold); asymptotically, this becomes $AdS_5\times T^{1,1}$. To determine the gauge theory fields that would couple to suitable supergravity fields via gauge-gravity duality, one should ideally perform the same calculation for the $M_4\times$ RWDC background (which would also involve solving for the Laplace equation for the internal RWDC). We did not follow that approach. Motivated, on the other hand, by, for example,

\noindent {\bf (a)} asymptotically the type IIB background of \cite{metrics} and its delocalized type IIA mirror of \cite{MQGP} consist of $AdS_5$

\noindent and

\noindent {\bf (b)} terms of the type $Tr(F^2(AB)^k), (F^4(AB)^k)$ where $F^2 = F_{\mu\nu}F^{\mu\nu}, F^4 = F_{\mu_1}^{\ \mu_2}F_{\mu_2}^{\ \mu_3}F_{\mu_3}^{\ \mu_4}F_{\mu_4}^{\ \mu_1} - \frac{1}{4}\left(F_{\mu_1}^{\ \mu_2}F_{\mu_2}^{\ \mu_1}\right)^2$, $A, B$ being the bifundamental fields that appear in the gauge theory superpotential corresponding to $AdS_5\times T^{1,1}$ in \cite{KW}, form part of the gauge theory operators corresponding to the solution to the Laplace equation on $T^{1,1}$ \cite{Gubser-Tpq} (The operator $Tr F^2$, which shares the $0^{++}$ glueball's quantum numbers, couples to the dilaton, while $Tr F^4$ which also has the $0^{++}$ glueball's quantum number couples to the four-form potential and the metric fluctuations' trace.),

\noindent we calculated:\\
$\bullet$ type IIB dilaton fluctuations, which we referred to as $0^{++}$ glueball\\
$\bullet$ type IIB complexified two-form fluctuations that couple to
$d^{abc}Tr(F_{\mu\rho}^aF^{b\ \rho}_{\lambda}F^{c\ \lambda}_{\ \ \ \ \ \nu})$, which we referred to as $0^{--}$ glueball \\
$\bullet$ type IIA one-form fluctuations that couple to $Tr(F\wedge F)$, which we referred to as $0^{-+}$ glueball \\
$\bullet$  M-theory metric's scalar fluctuations which we referred to as another (lighter) $0^{++}$ glueball \\
$\bullet$ M-theory metric's vector fluctuations which we referred to as $1^{++}$ glueball, \\
and \\
$\bullet$ M-theory metric's tensor fluctuations which we referred to as $2^{++}$ glueball.

In this chapter, I have discussed only a specific portion of \cite{Sil+Yadav+Misra-glueball} which is the spectrum of $2^{++}$ glueball from the type IIB theory perspective in which I had contributed maximally. Now, for the glueball mass computation we have solved the supergravity equations by two different methods: (i) using WKB approximation, (ii) imposing Neumann/Dirichlet boundary condition at IR cut-off $r_h/r_0$.
\paragraph{WKB method:} To obtain the mass spectra for different glueballs one need to solve the differential equations involving appropriate field perturbation. For example, assuming a particular perturbation with structure $H(r)=\tilde{H}(r)e^{ikx}$ with $k^2=-m^2$, where $m$ is the corresponding glueball's mass, the equation of motion for the glueball has the following general form,
\begin{equation}
\label{geneom}
\tilde{H}^{\prime\prime}+f_1(r)\tilde{H}^{\prime}+m^2f_2(r)\tilde{H}=0.
\end{equation}
Next, following the redefinition of \cite{Minahan}, one can introduce new variables as given below,
\begin{itemize}
\item{\it{Background with a black hole}}: $r\rightarrow \sqrt{y}$, $r_h\rightarrow \sqrt{y_h}$ and then $y\rightarrow y_h\left(1+e^z\right)$
\end{itemize}
\begin{itemize}
\item{\it{Background without a black hole}}: $r\rightarrow \sqrt{y}$, $r_0\rightarrow \sqrt{y_0}$ and then $y\rightarrow y_0\left(1+e^z\right)$.
\end{itemize}
In terms of these newly defined variables equation (\ref{geneom}) can be rewritten as,
\begin{equation}
\label{geneominz}
\partial_z\left(f_3(z)\partial_z\tilde{H}\right)+m^2f_4(z)\tilde{H}=0.
\end{equation}
With a field redefinition: $\tilde{H}$ as $\psi(z)=\sqrt{f_3(z)}\tilde{H}(z)$, the above equation reduces to the following Schr\" {o}dinger like form:
\begin{equation}\label{pote}
\left(\frac{d^2}{dz^2} + V(z)\right)\psi(z)=0,
\end{equation}
where $V(z)$ is the potential term. Once one get the potential term, the mass can be found from the WKB quantization condition:
\begin{equation}
\int^{z_2}_{z_1}\sqrt{V(z)}dz=\left(n+\frac{1}{2}\right)\pi,
\end{equation}
where $z_1$ and $z_2$ are the turning points obtained by solving for the roots of the equation $V(z)=0$: $V(z)>0$ for $z\epsilon [z_1,z_2]$.

In the present work, we have considered the two regions namely, IR and IR-UV interpolating/UV region separately. Moreover, the potential, the corresponding turning points, and finally the spectrum were calculated for each of the above two regions.
\section{Tensor Glueball $2^{++}$ Mass From Type IIB}

In the low energy limit, the $10$-dimensional type IIB supergravity action is given by,
\begin{equation}\label{action}
\frac{1}{2k_{10}^2}\left\{\int d^{10}x~ e^{-2\phi}\sqrt{-g}\left(R-\frac{1}{2}H_3^2\right)-\frac{1}{2}\int d^{10}x
~\sqrt{-g}\left(F_1^2+\widetilde{F_3^2}+\frac{1}{2}\widetilde{F_5^2}\right)\right\},
\end{equation}
where $\phi$ and $g_{MN}$ represents the dilaton and $10$ dimensional metric respectively. $F_1$, $H_3$, $\widetilde{F_3}$, $\widetilde{F_5}$ are one form, three form and five form fluxes.\\
$\widetilde{F_5}$ and $\widetilde{F_3}$ are defined as follows:
\begin{equation}\label{F5F3tilde}
\widetilde{F_5}=F_5+\frac{1}{2}B_2\wedge F_3 ,~~~~~~~~~~~~~~~~~~~~            \widetilde{F_3}=F_3-C_0\wedge H_3,
\end{equation}
where the branes $D3$ and $D5$ sources $F_5$ and $F_3$ respectively. $B_2$ denotes the NS-NS two form, whereas $C_0$ denotes the axion.
Fluxes $\tilde{F_5}$, $\widetilde{F_3}$, $H_3$ and the axion $C_0$, were computed in \cite{metrics}. Now, by varying the action in (\ref{action}) with respect to the metric $g_{\mu\nu}$, one gets the following equation of motion:
\begin{equation}\label{ricci scalar}
\begin{split}
R_{\mu\nu} & =\left(\frac{5}{4}\right)e^{2\phi}\widetilde{F}_{\mu p_2p_3p_4p_5}\widetilde{F}_{\nu}^{p_2p_3p_4p_5}-\left(\frac{g_{\mu\nu}}{8}\right)e^{2\phi}\widetilde{F}_{5}^2+\left(\frac{3}{2}\right)H_{\mu\alpha_2\alpha_3}
H^{\alpha_2\alpha_3}_{\nu}\\ & -\left(\frac{g_{\mu\nu}}{8}\right)H_3^2
 +\left(\frac{3}{2}\right)e^{2\phi}\widetilde{F}_{\mu\alpha_2\alpha_3}
\widetilde{F}^{\alpha_2\alpha_3}_{\nu}
-\left(\frac{g_{\mu\nu}}{8}\right)e^{2\phi}\widetilde{F}_{3}^2+\left(\frac{1}{2}\right)e^{2\phi}F_{\mu}F_{\nu}.
\end{split}
\end{equation}
 we considered the following structure for the linear perturbation of the metric:
\begin{equation}
g_{\mu\nu}=g_{\mu\nu}^{(0)}+h_{\mu\nu},
\end{equation}
where $\mu,\nu=\{t,x_1,x_2,x_3,r,\theta_1,\theta_2,\phi_1,\phi_2,\psi\}$. As per the tensor mode of metric fluctuation, the only non-vanishing component is $h_{x_2x_3}$. The fluxes $F_1$, $\widetilde{F_3}$ and $H_3$ do not have non-vanishing components along the coordinates $\{x_2,x_3\}$, which further simplified final EOM, given as:
\begin{equation}\label{final EOM}
\begin{split}
R^{(1)}_{x_2x_3} & =\left(\frac{5}{4}\right)e^{2\phi}\left(4\widetilde{F}_{x_2 x_3p_3p_4p_5}\widetilde{F}_{x_2x_3 q_3q_4q_5}g^{p_3q_3}g^{p_4q_4}g^{p_5q_5}h^{x_2x_3}\right)-\left(\frac{h_{x_2x_3}}{8}\right)e^{2\phi}\widetilde{F}_{5}^2
\\&-\left(\frac{h_{yz}}{8}\right)H_3^2
-\left(\frac{h_{x_2x_3}}{8}\right)e^{2\phi}\widetilde{F}_{3}^2.
\end{split}
\end{equation}
for fixed $\theta_1$ and $\theta_2$ given as: $\theta_1=1/N^{1/5}$ and $\theta_2=1/N^{3/10}$, the squared fluxes appearing in (\ref{final EOM}) at the LO in $N$, can be read off from (\ref{fluxes-squared}).
We took an ansatz for the perturbation as $h_{x_2x_3}=\frac{r^2}{2 (g_{s}\pi N)^{1/2}} H(r)e^{i k x_1}$, which reduced (\ref{final EOM}) to the following second order differential equation in $H(r)$:
{\scriptsize
\begin{eqnarray}
\label{EOM}
& &  H^{\prime\prime}(r)+\Biggl(\frac{5 r^4-{r_h}^4}{r \left(r^4-{r_h}^4\right)}-\frac{9 a^2}{r^3}+\Biggl\{\frac{3}{256 \pi ^2 N^{2/5} r^3}\Biggl[-54 a^2 {g_s}^2 M^2 {N_f}-72 \pi  a^2 {g_s} M^2+768 \pi ^2 a^2+12 {g_s}^2 M^2 {N_f} r^2+\nonumber\\
& &  9 a^2 {g_s}^2 M^2 {N_f} \log (16)-2 {g_s}^2 M^2 {N_f} r^2 \log (16)+16 \pi  {g_s} M^2 r^2+{g_s}^2 M^2 {N_f} \left(9 a^2-2 r^2\right) \log (N)-24 {g_s}^2 M^2 {N_f}\nonumber\\
& & \left(9 a^2-2 r^2\right) \log (r)\Biggr]\Biggr\}\Biggr)H^{\prime}(r)+\Biggl(\frac{1}{4 \pi  r^4 \left(r^4-{r_h}^4\right)}\Biggr\{8 \pi  \left(a^2 \left(6 \pi  {g_s} N q^2 r^2-9 r^4+9 {r_h}^4\right)-2 \pi  {g_s} N q^2 r^4+4 r^6\right)\nonumber\\
& &+3 {g_s}^2 M^2 q^2 r^2 \left(r^2-3 a^2\right) \log (r) ({g_s} {N_f} \log (16 N)-6 {g_s} {N_f}-8 \pi )-36 {g_s}^3 M^2 {N_f} q^2 r^2 \left(r^2-3 a^2\right) \log ^2(r)\Bigg\}-\nonumber\\
& &\frac{{g_s}^{2}}{512\pi^{3}}\Biggl\{\frac{34992 a^2 {g_s} M^2 \left(\sqrt[5]{N}+3\right) {N_f}^2 \log (r)}{r^3}+9 a^2 {g_s} {N_f} \Biggl(\frac{7831552 \pi ^5}{\left(r^4-{r_h}^4\right) ({g_s} {N_f} \log (16 N)-3 {g_s} {N_f} \log (r)+4 \pi )^3}-\nonumber\\
&&\frac{81 M^2
  \left(7 \sqrt[5]{N}-1\right) {N_f}}{r^4}\Biggr )+\frac{2 \left(243 {g_s} M^2 \left(\sqrt[5]{N}+1\right) {N_f}^2+\frac{3915776 \pi ^5 r^4}{\left(r^4-{r_h}^4\right) ({g_s} {N_f} \log (16 N)-3 {g_s}
   {N_f} \log (r)+4 \pi )^2}\right)}{r^2}\Biggr\}\Biggr)H(r)=0.\nonumber\\
&&
\end{eqnarray}}

\subsubsection{$2^{++}$ glueball mass from Neumann BC at $r=r_h$}

Performing a small-$T$ expansion while rewriting and solving (\ref{EOM}) around $r=r_h$ gave a mass quantization condition:
\begin{equation}
\label{2++-IIB-i}
0.5 -\frac{0.174071 m^2 \left(1-\frac{3. \left({g_s} M^2 (2. \log (N)+11.8963)+4. {g_s} M^2 \log (T)+0.6
   N\right)^2}{N^2}\right)}{T \sqrt{-m^2 \left(4-7 \left(1-\frac{3. \left({g_s} M^2 (2. \log (N)+11.8963)+4. {g_s} M^2 \log (T)+0.6
   N\right)^2}{N^2}\right)\right)}}=-n,
\end{equation}
which for $g_s=0.8,N=g_s^{-39}\sim6000, {\rm and}\ M=3$ yields:
\begin{eqnarray}
\label{2++IIB-spectrum}
& & m_{2^{++}} = 4.975 T \nonumber\\
& & m_{2^{++}}^* = 14.925 T \nonumber\\
& & m_{2^{++}}^{**} = 24.876 T \nonumber\\
& & m_{2^{++}}^{***} = 34.826 T \nonumber\\
& & m_{2^{++}}^{****} = 44.776 T \nonumber\\
\end{eqnarray}
\subsubsection{ $2^{++}$ Glueball Mass From WKB Quantization Method}
$\bullet$ {$Black Hole\ Background\ $ ($r_h\neq0$)}\\
Using the variables of \cite{Minahan}, the `potential',  defining $m = \tilde{m}\frac{\sqrt{y_h}}{L^2}$, yields:
\begin{eqnarray}
\label{V_2++_IIB-i}
& & V(2^{++}, IIB, r_h\neq0) = \frac{1}{4 \left(e^z+1\right)^3 \left(e^z+2\right)^2}\Biggl\{e^z \Biggl(3 b^2 \left(e^z+2\right) \left(-\left(\tilde{m}^2-6\right) e^z-\tilde{m}^2+3 e^{2 z}+6\right)\nonumber\\
& & +\left(-e^z-1\right)
   \left(\left(25-3 \tilde{m}^2\right) e^z-\left(\tilde{m}^2-18\right) e^{2 z}-2 \left(\tilde{m}^2-6\right)+4 e^{3
   z}\right)\Biggr)\Biggr\} + {\cal O}\left(\frac{g_s M^2}{N}\right).
\end{eqnarray}
In the IR ($z<<0 $), the potential is given by:
\begin{eqnarray}
\label{V-IR}
& & V(IR,T) = e^z \left(\left(0.15 \tilde{m}^2-1.3375\right) e^z-0.01 \tilde{m}^2+0.06\right) + {\cal O}(e^{3z}),
\end{eqnarray}
in IR for $\tilde{m}>2.986$ and $z\in[\log\left(0.067 + {\cal O}\left(\frac{1}{\tilde{m}^2}\right)\right)\approx -2.708,-2.526]$ the potential follows a bound $V(IR<T)>0$ and for this interval WKB quantization condition
$$\int_{2.708}^{-2.526}\sqrt{V(IR,T)}\approx0$$
hence IR gave no contribution to the WKB quantization.

In UV ($z>>1$) the potential was obtained as:
\begin{eqnarray}
\label{V_UV_T}
& & V(UV,T) = \frac{1}{4} \left(\tilde{m}^2+9.24\right) e^{-z}-\frac{3}{4} \left(\tilde{m}^2+0.36 \left(\tilde{m}^2+9\right)+3\right) + {\cal O}\left(e^{-3z}\right),
   e^{-2 z}-1
\end{eqnarray}
 the bounds on the interval $\left(\log(4.08 + {\cal O}\left(\frac{1}{\tilde{m}^2}\right),0.25\tilde{m}^2 - 1.77)\right)$ gives the turning points for the potential for $\tilde{m}>7.141$.
  Thus,
\begin{eqnarray}
\label{WKB-2++_T_IIB}
& & \int_{1.406}^{\log(0.25\tilde{m}^2 - 1.77)}\sqrt{V(UV,T)} = \int_{1.406}^{\log(0.25\tilde{m}^2 - 1.77)} \frac{e^{-z}}{2}\sqrt{e^z - 4.08} \tilde{m} + {\cal O}\left(\frac{1}{\tilde{m}}\right)\nonumber\\
& & = 0.389 \tilde{m} - 2 = \left(n + \frac{1}{2}\right)\pi,
\end{eqnarray}
which gave:
\begin{equation}
\label{mn2++_T_IIB}
m_n^{2^{++}}(T) = (9.18 + 8.08 n)\frac{r_h}{L^2}.
\end{equation}

$\bullet${$Thermal\ Background\ $($r_h=0$)}\\
Using minahan's\cite{Minahan} coordinate redefinition for the thermal ($r_h=0$) background given by: $r=\sqrt{y},\ y = y_{0} \left(1 + e^z\right),\ r_h = \sqrt{y_0}$. The `potential' in the IR, took the form:
\begin{equation}
\label{V1++_ii}
 V(r_h=0) = -\frac{1}{4} + \frac{1}{4}\left(1 + \tilde{m}^2\right)e^{2z} + {\cal O}(e^{-3z}).
\end{equation}
 In IR  for the interval $[-\frac{1}{2}\log(5 + \tilde{m}^2),\log(\delta^2-1)]$ potential is $V(r_h=0)>0$. Thus writing WKB quantization condition with lower and upper bound of the interval as the turning points:
\begin{equation}
\label{WKB_1++_T=0}
\int_{-\frac{1}{2}\log(5 + \tilde{m}^2)}^{-2.526}\sqrt{-\frac{1}{4} + \frac{1}{4}\left(1 + \tilde{m}^2\right)e^{2z}} = \frac{(\delta^2-1)}{2}\tilde{m} - \frac{\pi}{4} = \left(n + \frac{1}{2}\right)\pi,
\end{equation}
yields :
\begin{equation}
\label{mn1++-IR}
m_n^{2^{++}}(IR,IIB,r_h=0) = m_n^{2^{++}}(IR,M\ {\rm theory},r_h=0).
\end{equation}
Zeros of the potential, as a function of $e^{Z}$, in UV,
\begin{equation}
\label{V1++-UV}
V(UV,r_h=0) = \frac{1}{4} \left(\tilde{m}^2-10\right) e^{-z}-\frac{3}{4} \left(\tilde{m}^2-5\right) e^{-2 z}+1 + {\cal O}\left(e^{-3z}\right),
\end{equation}
  are at $(\frac{1}{8} \left(-\tilde{m}^2\pm\sqrt{\tilde{m}^4+28 \tilde{m}^2-140}+10\right)$ $=(-\frac{\tilde{m}^2}{4} - \frac{1}{2}, 3 + {\cal O}\left(\frac{1}{\tilde{m}^2}\right))$. Thus for the interval $[\log 3,\infty)$ $V(UV,r_h=0)>0$. Therefore the WKB quantization condition is given as:
\begin{eqnarray}
\label{WKB1++-i}
& & \int_{\log 3}^\infty\sqrt{V(UV, r_h=0)} = \frac{1}{2}\int_{\log 3}^\infty e^{-z}\sqrt{e^z - 3}\tilde{m} + {\cal O}\left(\frac{1}{\tilde{m}}\right) = \frac{\tilde{m}\pi}{4\sqrt{3}} = \left(n + \frac{1}{2}\right)\pi,
\end{eqnarray}
yielding:
\begin{equation}
\label{mn1++-ii}
m_n^{2^{++}}(r_h=0) = (3.464 + 6.928 n)\frac{r_0}{L^2}.
\end{equation}
\subsubsection{NLO-in-$N$/Non-Conformal Corrections in the IR  in the $r_h=0$ Limit}
In IR the potential inclusive of the NLO-in-$N$ corrections in the $r_h=0$ limit, is as follows:
\begin{eqnarray}
\label{IIB_2++_IR_T=0}
& & V(IR,r_h=0) = e^{2 z} \Biggl(\frac{{g_s} M^2 \left(\frac{1}{N}\right)^{2/5} ({g_s} {N_f} (6 {\log N}-72+\log (16777216))-72 {g_s} {N_f} \log ({y_0})-48
   \pi )}{512 \pi ^2}\nonumber\\
   & & +\frac{1}{4} \left(\tilde{m}^2-3\right)\Biggr)-\frac{1}{4} + {\cal O}(e^{-3z}),
\end{eqnarray}
whose turning points, in the large-$\tilde{m}$ limit are: $[\log\left\{\frac{1}{\tilde{m}} + {\cal O}\left(\frac{1}{\tilde{m}^3}\right)\right\},\log(\delta^2-1)]$\\ $\approx
[-\log\tilde{m},\log\left(\delta^2-1\right)]$.
Now,
\begin{eqnarray}
\label{WKB_NLO_IIB_2++_T=0}
& & \int_{-\log\tilde{m}}^{\log(\delta^2-1)}\sqrt{V(IR,r_h=0)} = \frac{\left(\delta^2-1\right)}{2}\tilde{m} - \frac{\pi}{4} + {\cal O}\left(\frac{1}{\tilde{m}}\right) = \left(n + \frac{1}{2}\right),
\end{eqnarray}
which produced the same LO spectrum as the $0^{--}, 1^{++}$, and $2^{++}$ spectrums obtained through M theory. Thus, the type IIB failed to include the non-conformal NLO-in-$N$ corrections at $r_h=0$ in the $2^{++}$ corrections, as they either gets nullified or gets a  $\frac{1}{\tilde{m}}$-suppression in the large-$\tilde{m}$ limit.
\section{Summary and Discussion}
Although I focused on a specific section of \cite{Sil+Yadav+Misra-glueball} in this chapter to which I contributed maximally, in this section I present the results for all the glueballs considered in the project \cite{Sil+Yadav+Misra-glueball} to which Karunava Sil contributed maximally, for completeness and clarity. In \cite{Sil+Yadav+Misra-glueball} we obtained the spectra of $0^{++},0^{-+}, 0^{--}, 1^{++}$, $ 2^{++}$ glueballs in a type IIB/delocalized SYZ IIA mirror/(its) M-theory (uplift) model corresponding to the top-down holographic dual of \cite{metrics} in the MQGP limit introduced in \cite{MQGP}.
Table 2.1 and 2.3 summarizes the results of all the calculations- the former table summarizes the results obtained by WKB quantization condition using the coordinate/field redefinitions of \cite{Minahan} and the latter one summarizes the results obtained by imposing Neumann/Dirichlet boundary condition at $r_h$/IR cut-off $r_0$. Some of the salient features of the results are given as separate bullets.

\newpage
\begin{table}[h]
\vskip 0.5in
\begin{tabular}{|c|c|c|}\hline
 {\tiny Glueball} & {\tiny $\tilde{m}$  using WKB $r_h\neq0$} & {\tiny
$\tilde{m}$ using WKB  $r_h=0$}   \\
& {\tiny (units of $\pi T$, up to LO in $N$)} & {\tiny (units of $\frac{r_0}{L^2}$, up to NLO in $N$)}\\
& {\tiny (large-$\tilde{m}$ limit)} & {\tiny (large-$\tilde{m}$ limit)}\\ \hline
{\tiny $0^{++}$} & {\tiny (M theory)} & {\tiny (M theory)}  \\
 {\tiny (Fluctuations: $h_{00,rx_1,rr}$}& {\tiny$\frac{\sqrt{35 + 70 n}}{\sqrt{{\cal PL}(13.15 + 26.31 n)}}$} & {\tiny\rm No\ turning\ points} \\
{\tiny in M-theory metric)} &   & \\ \cline{1-3}
{\tiny $0^{++}$} & {\tiny (Type IIB)} & {\tiny (Type IIB)} \\
 {\tiny (Dilaton Fluctuations)}& {\tiny$9.18 + 8.08 n$}  & {\tiny$ \frac{(4.64 + 6.28 n)}{(\delta^2-1)}\left[1 - 0.01 \frac{g_s M^2 }{N}(g_s N_f)\log N\log r_0\right]$} \\  \hline
& {\tiny UV (redefinition of \cite{WKB-i}): $5.7n$} & \\
{\tiny$0^{-+}$} & {\tiny (Type IIA)} & {\tiny (Type IIA)} \\
 {\tiny(1-form fluctuation $a_{\theta_2}$) }&  {\tiny  $11.12\left(n + \frac{1}{2}\right), n=0$} & {\tiny $\frac{3.72 + 4.36 n}{(\delta^2-1)}, n=0$}  \\
& {\tiny  $(6.22 + 4.80 n), n\in\mathbb{Z}^+$}  & {\tiny $4.8\left(n + \frac{1}{2}\right), n\in\mathbb{Z}^+$} \\ 
{\tiny $0^{--}$} & {\tiny (Type IIB)} & {\tiny (Type IIB)} \\
{\tiny 2-form fluctuation $A_{23}$}& {\tiny $= m_n^{0^{++}}({\rm dilaton},T)$} & {\tiny  $ \frac{6.28 n+4.71}{(\delta ^2-1)}\left(1 + \frac{0.01 {g_s}^2 {\log N} M^2 {N_f} \log ({r_0})}{ N}\right), n=0$} \\
&  & {\tiny  $(7.87 + 6.93 n), n\in\mathbb{Z}^+$} \\ \hline
 {\tiny$1^{++}$} & {\tiny (M theory)} & {\tiny (M theory)} \\
{\tiny (Fluctuations: $h_{it}=h_{ti},i=x_{2,3}$}& {\tiny $8.08\left(n + \frac{1}{2}\right)$} & {\tiny $m_n^{1^{++}}(n=0,r_h=0) = m_n^{0^{--}}(n=0,r_h=0)$} \\
{\tiny in M-theory metric)}& &  {\tiny $(3.46 + 6.93 n),n\in\mathbb{Z}^+$} \\ \hline
$2^{++}$ & (M theory) & (M theory) \\
{\scriptsize (Fluctuations: $h_{x_2x_3}=h_{x_3x_2}, $}& {\scriptsize $8.08\left(n + \frac{1}{2}\right) = m_n^{1^{++}}(T)$} & {\scriptsize $=m_n^{1^{++}}(r_h=0)$} \\
{\scriptsize $h_{x_2x_2}=h_{x_3x_3}$ in M-theory metric)} & &  \\ \hline
 $2^{++}$& (Type IIB) & (Type IIB) \\
  {\scriptsize (Fluctuation $h_{x_2x_3}=h_{x_3x_2}$}& {\scriptsize $9.18 + 8.08 n = m_n^{0^{++}}(IIB,T)$} & {\scriptsize $=m_n^{1^{++}}(r_h=0)$} \\
{\scriptsize in type IIB metric)}&  &
$m_n^{2^{++}}(IR,M\ {\rm theory},LO,r_h=0)$
\\ \hline
\end{tabular}
\caption{ Mass spectrum of Glueballs from Type IIB, IIA and M Theory using WKB quantization condition (equalities in the fourth column, are valid up to NLO-in-$N$)}
\end{table}


\newpage

Some of the most salient features of Table 2.1:
\begin{enumerate}
\item
The higher excited $r_h\neq 0$ $2^{++}$ glueball states corresponding to M-theory metric fluctuations and those corresponding to type IIB metric fluctuations are isospectral. The $r_h=0$ $2^{++}$ glueball states corresponding to M-theory/type IIB string theory metric fluctuations are isospectral. Furthermore, unlike a M-theoretic computation, a type IIB $r_h=0$ $2^{++}$ glueball spectrum was unable to capture the NLO-in-$N$ corrections to the LO-in-$N$ type IIB $2^{++}$ glueball spectrum due to internal cancellation of terms and $\frac{1}{\tilde{m}}$-suppression.

\item
$m^{2^{++}}_n({\rm NLO},r_h=0) = m_n^{1^{++}}({\rm NLO},r_h=0)\stackrel{n\gg1}{\longrightarrow}m_n^{0^{--}}({\rm NLO},r_h=0)$, where the `NLO' implies equality with the inclusion of NLO-in-$N$ corrections.

\item
Contrastingly, the lightest $0^{++}$ glueball spectrum for $r_h\neq0$ obtained from M theory's scalar metric fluctuations compares favourably with the $N\rightarrow\infty$ lattice results of \cite{Teper98} (Table 2.2)... Like \cite{Mathur_et_al-0++-Mtheory}, the $0^{++}$ from M theory scalar fluctuations is lighter than the $0^{++}$ from type IIB dilaton fluctuations. The coordinate and field redefinitions of \cite{WKB-i} applied to the EOM for dilaton fluctuation yield a WKB quantization condition, for $a=0.6 r_h$ - as in \cite{EPJC-2} - matching the UV limit of the $0^{++}$ glueball spectrum obtained in \cite{Minahan}. The method based on \cite{WKB-i} coordinate/field redefinitions is not suitable for obtaining the $0^{++}$ glueball ground state and was not used for any other glueball calculations.
\begin{table}[h]
\begin{tabular}{|c|c|c|c|}\hline
{\scriptsize State} & $N\rightarrow\infty$ {\scriptsize Entry in Table 34} of \cite{Teper98} & {\scriptsize M-theory scalar metric perturbations}  & {\scriptsize Type IIB Dilaton  fluctuations of} \cite{Ouguri et al}\\
& {\scriptsize in units of square root of} & ({\bf 6.1.2} - {\scriptsize in units of} & {\scriptsize in units of reciprocal of} \\
& {\scriptsize string tension} & $\frac{r_h}{L^2}$) & {\scriptsize temporal circle's diameter} \\ \hline
$0^{++}$ & $4.065\pm0.055$ & 4.267 & 4.07 ({\scriptsize normalized to match lattice}) \\ \hline
$0^{++*}$ & $6.18\pm0.13$ & 6.251 & 7.02 \\ \hline
$0^{++**}$ & $7.99\pm0.22$ & 7.555 & 9.92 \\ \hline
$0^{++**}$ & - & 8.588 & 12.80 \\ \hline
$0^{++***}$ & - & 9.464 & 15.67 \\ \hline
\end{tabular}
\caption{Comparison of \cite{Teper98}'s $N\rightarrow\infty$ lattice results for $0^{++}$ glueball with our supergravity results obtained  using WKB quantization condition and redefinitions of \cite{Minahan} for M theory scalar metric fluctuations}
\end{table}

\item
The higher excited states of the type IIA $0^{-+}$ glueball are isospectral for both $r_{h}\neq 0$ and $r_{h}=0$. This is advantageous because large-$n$ corresponds to the UV, which deviates from the BH geometry, i.e. toward $r_{h}=0$. The non-conformal corrections up to NLO in $N$ have a semi-universal behaviour of $\frac{(g_{s}M^{2})(g_{s}N^{f})\log r_{0}}{N}$ and are multiplied by a numerical pre-factor of ${\cal O}(10^{-2})$; which can be disregarded in the MQGP limit.

\item
As per a more recent lattice calculation \cite{Chen_et_al_latest_lattice_2006}\footnote{We thank P.Majumdar for bringing this reference to our attention.}, the $0^{++}$-glueball has a mass $4.16\pm0.11\pm0.04$ (in units of the reciprocal of the `hadronic scale parameter' of \cite{Sommer-r0}), which compares rather well with $m_{n=0}^{0^{++}}=4.267$ (in units of $\frac{r_h}{L^2}$) of Table 2.2 coming from scalar fluctuations of the M theory metric.

\item
$1^{++}$ and $0^{--}$ glueballs ground state and $n\gg 1$ excited states are both isospectral.

\end{enumerate}
\newpage
\begin{table}[h]
\begin{tabular}{|c|c|c|c|}\hline
S. No. & Glueball & Spectrum Using   & Spectrum Using     \\
&& N(eumann)/D(irichlet) & N(eumann)/D(irichlet) \\
&& b.c., $r=r_h$(units of $\pi T$) & b.c., $r=r_0$(units of $\frac{r_0}{L^2}$)\\ \hline
1 & $0^{++}$ & (M theory) & (M theory)  \\
&& (N) {\scriptsize $12.25\sqrt{2+n}$} & (N) 4.1 \\
&& (D) {\scriptsize $12.25\sqrt{1+n}$} & \\ \hline
2 & $0^{-+}$ & (Type IIA) & (Type IIA) \\
& & (N/D) {\scriptsize $\frac{3.1}{\pi}\sqrt{n}$} & (N) {\scriptsize $m_{n=0}^{0^{-+}}=0, m_{n=1}^{0^{-+}}\approx 3.4, m_{n=2}^{0^{-+}}\approx 4.35$} \\
&&& (D) {\scriptsize $m_{n=0}^{0^{-+}}=0, m_{n=1}^{0^{-+}}\approx 3.06, m_{n=2}^{0^{-+}}\approx 4.81$} \\ \hline
3 & $0^{--}$ & (Type IIB) & (Type IIB) \\
&& (N/D) {\scriptsize $m_{n=0}^{0^{--}}(T)=0, m_{n=1}^{0^{--}}(T)=\frac{32.46}{\pi},$} & (large $n$)  \\
&& {\scriptsize $m_{n=2}^{0^{--}}(T)=\frac{32.88}{\pi}$} & (N/D)  \\
&&& {\scriptsize ${\scriptsize \frac{1}{2} 5^{3/4} \sqrt[4]{\frac{2 \left(\sqrt{6} \sqrt{\pi ^2 \left(16 n^2+22 n+7\right)+6}+6\right)+3 \pi ^2 (2 n+1)}{32 - 3 \pi ^2}}}$} \\ \hline
4 & $1^{++}$ & (M theory) & (M theory) \\
& & (N/D) {\scriptsize $m_{n=0}^{1^{++}}(T) = 2.6956, m_{n=1}^{1^{++}}(T)=2.8995$}  & (N) {\scriptsize $m_{n=0}^{1^{++}}(r_h=0)\approx1.137$} \\
& & {\scriptsize $m_{n=2}^{1^{++}}(T) = 2.9346$} & (D) {\scriptsize $m_{n=0}^{1^{++}}(r_h=0)\approx0.665$} \\ \hline
5 & $2^{++}$ & (M theory) & (M theory) \\
&& (N) {\scriptsize $m_{n=0}^{2^{++}}(T)=\frac{5.086}{\pi}, m_{n=1}^{2^{++}}(T)=\frac{5.269}{\pi}$} & $=m_n^{1^{++}}(r_h=0)$ \\
&& {\scriptsize $m_{n=2}^{2^{++}}(T)=\frac{5.318}{\pi}$} &  \\
&& {\scriptsize $m_{n=0}^{2^{++}}(D,T)=0, m_{n+1}^{2^{++}}(D,T)=m_n^{2^{++}}(N,T)$} & \\ \hline
   \end{tabular}
   \caption{Summary of Glueball Spectra from Type IIB, IIA and M Theory for $r_h\neq0/r_h=0$ using Neumann/Dirichlet boundary conditions at the horizon $r_h$/IR cut-off $r_0$}
   \end{table}
\newpage
\begin{itemize}

\item The following table compares the ratios of $0^{--}$ glueball masses determined in \cite{Sil+Yadav+Misra-glueball} using Neumann/Dirichlet boundary conditions at the horizon to those obtained in \cite{Ouguri et al}:

\begin{table}[h]
\begin{center}
\begin{tabular}{|c|c|c|}\hline
Ratio & Our Results & \cite{Ouguri et al}'s Results\\  \hline
$\frac{m_{0^{--}}^*}{m_{0^{--}}}$ & 1.0129 & 1.5311\\ \hline
$\frac{m_{0^{--}}^{**}}{m_{0^{--}}^*}$ & 1.0033 & 1.3244\\ \hline
$\frac{m_{0^{--}}^{***}}{m_{0^{--}}^{**}}$ & 1.0013 & 1.2393\\ \hline
$\frac{m_{0^{--}}^{****}}{m_{0^{--}}^{***}}$ & 1.0007 & 1.1588\\ \hline
 \end{tabular}
   \caption{Comparison of ratios of $0^{--}$ glueball masses obtained  from Neumann/Dirichlet boundary conditions at the horizon,  with \cite{Ouguri et al}}
   \end{center}
   \end{table}

The ratio of consecutive excited state masses approached unity quicker than \cite{Ouguri et al} for higher excited states.

 \item  Compared to the spectra obtained via imposing Neumann/Dirichlet boundary conditions at horizon/IR cut-off, the WKB quantization-based spectra are closer to $N\rightarrow\infty$ lattice findings. For the lightest $0^{++}$ glueball spectra, the WKB quantization technique outperformed the traditional calculations of \cite{Ouguri et al} for ground and lower excited states.

     \end{itemize}

\chapter{ Meson Spectrum at Finite \lowercase{g} and $N_c$ }

\graphicspath{{Chapter1/}{Chapter1/}}


\section{Introduction}
Mesons (and glueballs) have been examined intensively for more than a decade using various bottom up holographic soft/hard-wall holographic models to obtain fresh insights into QCD's non-perturbative regime \cite{Kim:2009bp,Domokos:2007kt,Alvares:2011wb,Bellantuono:2014lra,BH_all_T,Nicotri,Forkel,ForkelStructure,FolcoCapossoli,Jugeau,Colangelolight,Cui:2013xva}.
%

Most of existing literature on holographic meson spectroscopy is of the bottom-up variety based often on soft/hard wall AdS/QCD models. Here is a short summary of some of the relevant works.
Soft-wall holographic QCD model was used in \cite{Bellantuono:2014lra} and \cite{BH_all_T} to obtain spectrum and decay constants for $1^{-+}$ hybrid mesons and to study the scalar glueballs and scalar mesons at $T\neq 0$ respectively. In \cite{Bellantuono:2014lra} no states with exotic quantum numbers were observed in the heavy quark sector. Comparison of the computed mass with the experimental mass of the $1^{-+}$ candidates $\pi_{1}(1400)$, $\pi_{1}(1600)$ and $\pi_{1}(2015)$, favored $\pi_{1}(1400)$ as the lightest hybrid state.
%
In \cite{Cui:2013xva} an IR-improved soft-wall AdS/QCD model in good agreement with linear confinement and chiral symmetry breaking was constructed to study the mesonic spectrum. The model was constructed to rectify inconsistencies associated with both simple soft-wall and hard-wall models. The hard-wall model gave a good realization for the chiral symmetry breaking, but the mass spectra obtained for the excited mesons didn't match up with the experimental data well. The soft-wall model with a quadratic dilaton background field showed the Regge behaviour for excited vector mesons but chiral symmetry breaking phenomena cannot be realized consistently in the simple soft-wall AdS/QCD model.
%
A hard wall holographic model of QCD was used in \cite{Alvares:2011wb}, \cite{Domokos:2007kt} and \cite{Kim:2009bp} to analyze the mesons.
In \cite{Li:2013oda} a two-flavor quenched dynamical holographic QCD(hQCD) model was constructed in the graviton-dilaton framework by adding two light flavors.
%
In  \cite{Sakai-Sugimoto-1} the mesonic spectrum was obtained for a $D4/D8(-\overline{D8})$-brane configuration in type IIA string theory; in  \cite{Imoto:2010ef}  massive excited states in the open string spectrum were used to obtain the spectrum for higher spin mesons $J\geq 2$. NLO terms were obtained by taking into account the effect of the curved background perturbatively which led to corrections in formula  $J=\alpha_{0}+\alpha{'}M^{2}$. The results obtained for the meson spectrum were compared with the experimental data to identify $a_{2}(1320),b_{1}(1235),\pi(1300),a_{0}(1450)$ etc. first and second excited states. In \cite{Kruczenski:2004me} a holographic model was constructed with extremal $N_{c}$ $D4$-branes and $D6$-flavor branes in the probe approximation. The model gave a good approximation for Regge behaviour of glueballs but failed to explain mesonic spinning strings because the dual theory did not include quark in the fundamental representations.

 In \cite{Sakai-Sugimoto-1}, a top-down holographic setup consisting of $D4/D8(-\overline{D8})$-brane, was used to obtain mesonic spectrum in type IIA string theory. As far as we know, the only UV complete top-down holographic model for large-$N$ thermal QCD containing fundamental quarks and simultaneoulsy satisfying both, UV conformality and IR confinement, is given in \cite{metrics}. The spectrum for vector and scalar meson was obtained in \cite{Dasgupta_et_al_Mesons} (with some co-authors common with those of \cite{metrics} ) by T-dualizing the holographic type IIB background of \cite{metrics}. A reasonable agreement was found when the (pseudo-)vector mesons were compared with the PDG results. The focus of \cite{Yadav+Misra+Sil-Mesons} was to obtain mesonic spectra from a top-down holographic dual of large-$N$ thermal QCD like theories and to see if the results obtained in \cite{Yadav+Misra+Sil-Mesons} were an improvement upon previous results \cite{Sakai-Sugimoto-1,Dasgupta_et_al_Mesons} when comparing with P(article) D(ata) group) data. In this chapter we used delocalized Strominger-Yau-Zaslow's triple-T-duality prescription to obtain the type IIA mirror of the type IIB background of \cite{metrics} - at finite coupling and at finite $N_c$ - in the MQGP limit which is closer to realistic strongly coupled thermal QCD
We evaluated (pseudo-)vector and (pseudo-)scalar mesons spectra, and  compared the ratio of mass squares for both kinds with PDG results in this chapter to investigate top-down holographic large-$N$ thermal QCD phenomenology at {\it finite gauge coupling}.

The (pseudo-)vector mesons correspond to gauge variations about a background gauge field along the world volume of the $D6$ branes. Unlike \cite{Dasgupta_et_al_Mesons}, wherein the authors studied the low temperature thermal gravitational dual of large-$N$ QCD like theories we considered the low and high temperature gravitational dual wherein the former would involves a thermal gravitational dual while latter will involve a black hole gravitational dual. Unlike \cite{Dasgupta_et_al_Mesons}, wherein the authors looked at a single T-dual of type IIB we performed a SYZ triple T-dual of type IIB background and in the process we generated a new set of sub-dominant corrections in N which were independent of M as shown in (\ref{B-IIA-diag-nondiag}) resulting from mixing of type IIB metric and type IIB NS-NS B. The gravity dual unlike \cite{Dasgupta_et_al_Mesons} involves a black hole ($r_h\neq 0$), and thus, when factoring the gauge fluctuations along $\mathbb{R}^3\times \mathbb{S}^1\times \mathbb{R}_{\ge 0}$-radial direction into fluctuations along $\mathbb{R}^3\times \mathbb{S}^1$ and eigenmode fluctuations along the $\mathbb{R}_{\ge 0}$ radial direction, there are two types of eigenmodes along the radial direction: $\alpha^{i}_n(Z)$ along the space-like $\mathbb{R}^3$ and $\alpha^{0}_{n}(Z)$ coupled to the time-like $\mathbb{S}^{1}$. In the absence of any background gauge fields, the (pseudo-)scalar meson spectrum is derived by considering fluctuations of the $D6$-branes orthogonal to their world-volume.

\section{SYZ Mirror of Ouyang Embedding}
 We obtained the embedding of flavour $D6$-branes in the mirror (built in \cite{MQGP}) of the RWDC(resolved warped deformed conifold) of \cite{metrics}. In other words we obtained the delocalized SYZ mirror of the Ouyang embedding of the flavor
$D7$-branes in \cite{metrics}.
For constant $\theta_1=\frac{\alpha_{\theta_1}}{N^{\frac{1}{5}}}$, the $(\theta_2,T^3(x,y,z))$-subspace of the type IIA mirror metric of the RWDC as obtained in \cite{MQGP} in the vicinity $\theta
_2=\frac{\alpha_{\theta_2}}{N^{\frac{3}{10}}}$ takes the form:
\begin{eqnarray}
\label{metric4x4_i}
ds^2_{\rm IIA}(\theta_2,T^3(x,y,z)) & = & d\theta_2^2 N^{\frac{7}{10}}\left(\xi_{\theta_2\theta_2}\frac{\alpha_{\theta_1}^2}{\alpha_{\theta_2}^2}\sqrt{g_s} d\theta_2 + \xi_{\theta_2y}N^{-\frac{7}{20}}g_s^{\frac{1}{4}} dy - \xi_{\theta_2z}\frac{\log r M N_f}{\alpha_{\theta_2}} N^{-\frac{13}{20}} g_s^{\frac{7}{4}}dz\right)\nonumber\\
& &+ds^2(T^3(x,y,z))\nonumber\\
& &  \stackrel{N\gg1}{\longrightarrow} \xi_{\theta_2\theta_2} \frac{\alpha_{\theta_1}^2}{\alpha_{\theta_2}^2}\sqrt{g_s} d\theta_2^2  + ds^2(T^3(x,y,z)),\nonumber\\
&&
\end{eqnarray}
where $T^3(x,y,z)$ is a $3\times 3$ symmetric metric,
\begin{eqnarray}
\label{metric-T3-i}
& & g_{ij}(T^3(x,y,z)) = \nonumber\\
& & \left(
\begin{array}{ccc}
 \frac{3^{2/3} \left(\alpha_{\theta_1} ^2-\alpha_{\theta_2}^2 \sqrt[5]{\frac{1}{N}}\right)}{\alpha_{\theta_1} ^2} & \frac{2 \sqrt{2} \left(\alpha_{\theta_1} ^2 \alpha_{\theta_2} \sqrt{N}-2 \alpha_{\theta_2}^3
   N^{3/10}\right)}{3 \sqrt[6]{3} \alpha_{\theta_1} ^6} & \frac{2 \left(9 \sqrt{2} \sqrt[6]{3} \alpha_{\theta_1}  N^{4/5}-2\ 3^{2/3} N\right)}{27 \alpha_{\theta_1} ^2 \alpha_{\theta_2}^2} \\
 \frac{2 \sqrt{2} \left(\alpha_{\theta_1} ^2 \alpha_{\theta_2} \sqrt{N}-2 \alpha_{\theta_2}^3 N^{3/10}\right)}{3 \sqrt[6]{3} \alpha_{\theta_1} ^6} & 3^{2/3} & \frac{\sqrt{2}
   \left(\alpha_{\theta_2}^2-3 N^{3/5}\right)}{3 \sqrt[6]{3} \alpha_{\theta_2} N^{3/10}} \\
 \frac{2 \left(9 \sqrt{2} \sqrt[6]{3} \alpha_{\theta_1}  N^{4/5}-2\ 3^{2/3} N\right)}{27 \alpha_{\theta_1} ^2 \alpha_{\theta_2}^2} & \frac{\sqrt{2} \left(\alpha_{\theta_2}^2-3 N^{3/5}\right)}{3
   \sqrt[6]{3} \alpha_{\theta_2} N^{3/10}} & \frac{2 \left(\sqrt[5]{N} \alpha_{\theta_1} ^2+\alpha_{\theta_2}^2\right) N^{2/5}}{3 \sqrt[3]{3} \alpha_{\theta_1} ^2 \alpha_{\theta_2}^2}
\end{array}
\right).\nonumber\\
&&
\end{eqnarray}
The local $T^3$ metric is diagonalizable in a set of new basis $(d\tilde{x},d\tilde{y},d\tilde{z})$,
\begin{eqnarray}
\label{metric-T3-ii}
 ds^2_{\rm IIA}(T^3(x,y,z))& = & \frac{2 d\tilde{x}^2 \left(9 \sqrt{2} \sqrt[6]{3} \alpha  N^{4/5}-2\ 3^{2/3} N\right)}{27 \alpha_{\theta_1}^22 \alpha_{\theta_2}^2}+\frac{2 d\tilde{y}^2 \left(2\ 3^{2/3}
   N-9 \sqrt{2} \sqrt[6]{3} \alpha  N^{4/5}\right)}{27 \alpha_{\theta_1}^22 \alpha_{\theta_2}^2}\nonumber\\
   & & +\frac{2 d\tilde{z}^2 \left(3^{2/3} \alpha_{\theta_1}^22 N^{3/5}+3^{2/3} \alpha_{\theta_2}^2
   N^{2/5}\right)}{27 \alpha_{\theta_1}^22 \alpha_{\theta_2}^2},
\end{eqnarray}
where:
\begin{eqnarray}
\label{metric-T3-iii}
& & d\tilde{x} = \frac{{dx} \left(3 \alpha_{\theta_1} ^2 \left(\frac{1}{N}\right)^{2/5}+4\right)}{4 \sqrt{2}}+\frac{{dz} \left(4-3 \alpha_{\theta_1} ^2 \left(\frac{1}{N}\right)^{2/5}\right)}{4
   \sqrt{2}}\nonumber\\
   & & +\frac{{dy} \sqrt{\frac{1}{N}} \left(2 \alpha_{\theta_2}^2 \sqrt[5]{\frac{1}{N}} \left(\left(54-3^{2/3}\right) \alpha_{\theta_1} ^6-54 \sqrt{6} \alpha_{\theta_1} ^3
   \alpha_{\theta_2}^2+66 \alpha_{\theta_2}^4\right)-\alpha_{\theta_1} ^2 \left(3^{2/3} \alpha_{\theta_1} ^6+72 \alpha_{\theta_2}^4\right)\right)}{16\ 3^{5/6} \alpha_{\theta_1} ^4 \alpha_{\theta_2}}\nonumber\\
   & & d\tilde{y} = \frac{{dx} \left(3 \alpha_{\theta_1} ^2 \left(\frac{1}{N}\right)^{2/5}-4\right)}{4 \sqrt{2}}+\frac{{dz} \left(3 \alpha_{\theta_1} ^2 \left(\frac{1}{N}\right)^{2/5}+4\right)}{4
   \sqrt{2}}\nonumber\\
   & & +\frac{{dy} \sqrt{\frac{1}{N}} \left(\alpha_{\theta_1} ^2 \left(-\left(3^{2/3} \alpha_{\theta_1} ^6+72 \alpha_{\theta_2}^4\right)\right)-2 \alpha_{\theta_2}^2
   \sqrt[5]{\frac{1}{N}} \left(\left(54+3^{2/3}\right) \alpha_{\theta_1} ^6+54 \sqrt{6} \alpha_{\theta_1} ^3 \alpha_{\theta_2}^2-66 \alpha_{\theta_2}^4\right)\right)}{16\ 3^{5/6} \alpha_{\theta_1} ^4
   \alpha_{\theta_2}}\nonumber\\
   & & d\tilde{z} = -\frac{9 \sqrt[6]{3} \sqrt{\alpha_{\theta_2}^2} {dx} \left(\frac{1}{N}\right)^{7/10} \left(\alpha_{\theta_2}^2 \sqrt[5]{\frac{1}{N}}-2 \alpha_{\theta_1} ^2\right)}{4
   \sqrt{2}}+\frac{{dy} \left(-3^{2/3} \alpha_{\theta_1} ^{12}-768 \alpha_{\theta_1} ^4 \alpha_{\theta_2}^2 N+1728 \sqrt[3]{3} \alpha_{\theta_2}^8\right)}{768 \alpha_{\theta_1} ^4
   \sqrt{\alpha_{\theta_2}^2} \alpha_{\theta_2} N}\nonumber\\
   & & -\frac{3 \sqrt[6]{3} \alpha_{\theta_2}^4 {dz} \sqrt{\frac{1}{N}} \left(3 \sqrt{3} \alpha_{\theta_1} ^3
   \sqrt[5]{\frac{1}{N}}+\sqrt{2} \alpha_{\theta_1} ^2-2 \sqrt{2} \alpha_{\theta_2}^2 \sqrt[5]{\frac{1}{N}}\right)}{2 \alpha_{\theta_1} ^4 \sqrt{\alpha_{\theta_2}^2}}.
\end{eqnarray}

In the MQGP limit of M-theory uplift we worked with small $\phi_{1,2}$ which decoupled the compact space $M_6(\theta_{1,2},\phi_{1,2},\psi,x^{10})$ and the  non-compact space $M_5(\mathbb{R}^{1,3},r)$. For small values of  $\phi_{1,2}$ the NS-NS $B$ field obtained after the Buscher triple T-duality results as given in (\ref{B}) is as follow:
{\footnotesize
\begin{eqnarray}
\label{B-IIA-diag-nondiag}
& & B^{\rm IIA}\left(\theta_1=\frac{\alpha_{\theta_1}}{N^{\frac{1}{5}}},\theta_2\sim\frac{\alpha_{\theta_2}}{N^{\frac{3}{10}}}\right) = d\theta_2\wedge {dx} \left(-\frac{2 \sqrt{2} \sqrt[4]{\pi } \sqrt[4]{{g_s}} N^{3/4} \left(3 \sqrt{6} \alpha_{\theta_1}^3-2 \alpha_{\theta_1}^2 \sqrt[5]{N}+2 \alpha_{\theta_2}^2\right)}{27 \alpha_{\theta_1}^4
   \alpha_{\theta_2}}\right)\nonumber\\
   & & +d\theta_2\wedge {dz} \left(\frac{\sqrt[4]{\pi } \sqrt[4]{{g_s}} \left(5 \alpha_{\theta_2}^2 \sqrt[20]{\frac{1}{N}}-6 N^{11/20}\right)}{27 \sqrt{2} \alpha_{\theta_2}}\right) + d\theta_2\wedge {dy}\left(\frac{\sqrt[4]{\pi } \sqrt[4]{{g_s}} N^{3/20} \left(2 \alpha  \sqrt[10]{N}+\alpha_{\theta_2}\right)}{\sqrt{3} \alpha }\right)
  \nonumber\\
   & & = d\theta_2\wedge d\tilde{x} \left(-\frac{2 \sqrt[4]{\pi } \sqrt[4]{{g_s}} N^{3/4} \left(3 \sqrt{6} \alpha_{\theta_1}^3-2 \alpha_{\theta_1}^2 \sqrt[5]{N}+2 \alpha_{\theta_2}^2\right)}{27 \alpha_{\theta_1}^4 \alpha_{\theta_2}}\right)+d\theta_2\wedge d\tilde{y} \left(\frac{2 \sqrt[4]{\pi } \sqrt[4]{{g_s}} N^{3/4} \left(3 \sqrt{6} \alpha_{\theta_1}^3-2 \alpha_{\theta_1}^2 \sqrt[5]{N}+2 \alpha_{\theta_2}^2\right)}{27 \alpha_{\theta_1}^4 \alpha_{\theta_2}}\right)\nonumber\\
   & &  + d\theta_2\wedge d\tilde{z}\left(-\frac{\sqrt[4]{\pi } \alpha_{\theta_2} \sqrt[4]{{g_s}} N^{3/20} \left(2 \left(\sqrt[3]{3}-1\right) \alpha  \sqrt[10]{N}+\sqrt[3]{3}
   \alpha_{\theta_2}\right)}{3^{5/6} \alpha  \sqrt{\alpha_{\theta_2}^2}}\right)\nonumber\\
   & &
\end{eqnarray}}
From (\ref{B}) and (\ref{B-IIA-diag-nondiag}), we found that the even upto NLO in $N$ $B^{\rm IIA}$ has no dependence on $M$ even though $B^{IIB}$ is proportional to $M$. This result allows one to skip the step where one has to invoke order ${\cal O}\left(\frac{g_sM^2}{N}\right)$ corrections which was done in \cite{Dasgupta_et_al_Mesons} (which they fine tuned to a quite significant numerical value of $0.5$ to get a match with \cite{PDG}). This is crucial in order to obtain a good match between mesonic spectra and experimental values \cite{PDG}.
In large-$N$ limit, $10$-dimensional type IIA metric can be written as:
\begin{eqnarray}
\label{full IIA}
& & ds^{2}_{\rm IIA} \approx g^{\rm IIA}_{00} dx^{2}_{0}+g^{\rm IIA}_{11} dx^{2}_{1}+g^{\rm IIA}_{22} dx^{2}_{2}+g^{\rm IIA}_{33} dx^{2}_{3}+g^{\rm IIA}_{rr} dr^{2}+g^{\rm IIA}_{\theta_{1}\theta_{1}} d\theta_{1}^{2}+g^{\rm IIA}_{\theta_{1}\tilde{x}} d\theta_{1}d\tilde{x}\nonumber\\
& &+g^{\rm IIA}_{\theta_{1}\tilde{y}} d\theta_{1}d\tilde{y}+g^{\rm IIA}_{\theta_{1}\tilde{z}} d\theta_{1}d\tilde{z}+g^{\rm IIA}_{\theta_{2}\theta_{2}} d\theta_{2}^{2}+ds^{2}(T^{3}(\tilde{x},\tilde{y},\tilde{z}))
 \end{eqnarray}
We worked with the Ouyang embedding's first branch given by  $(\theta_1,\tilde{x})=(0,0)$ to obtain pullback metric on $D6$-brane. We took coordinate $\tilde{z}$ as a function of r, i.e. $\tilde{z}(r)$ whose explicit functional dependence was obtained by solving equation of motion of the field. The seven dimensional pullback metric with world volume coordinates $x^{\mu}={x^{(0,1,2,3)},r,\theta_2,\tilde{y}}$  on the $D6$-brane is given as:
\begin{eqnarray}
\label{pull-back}
& & g^{\rm IIA}_{6\mu \nu}dx^{\mu}dx^{\nu}= g^{\rm IIA}_{00} dx^{2}_{0}+g^{\rm IIA}_{11} dx^{2}_{1}+g^{\rm IIA}_{22} dx^{2}_{2}+g^{\rm IIA}_{33} dx^{2}_{3}+(g^{\rm IIA}_{rr}+g^{\rm IIA}_{\tilde{z}\tilde{z}}\tilde{z}^{\prime}(r)^{2}) dr^{2}\nonumber\\
 & & g^{\rm IIA}_{\theta_{2}\theta_{2}}d\theta_{2}^{2}+g^{\rm IIA}_{\tilde{y}\tilde{y}}d\tilde{y}^{2}
\end{eqnarray}
Components of type IIA metric near $\theta_1=\alpha_{\theta_1}N^{\frac{-1}{5}}$ and $\theta_2=\alpha_{\theta_2}N^{\frac{-3}{10}}$ upto NLO in large-$N$ expansion are given in (\ref{GIIA}). One will assume the $D6$-brane embedding effected by $\tilde{z}$ functional dependence, $\tilde{z}(r)$ by $i:\Sigma^{1,6}\left(\mathbb{R}^{1,3},r,\theta_2\sim\frac{\alpha_{\theta_2}}{N^{\frac{3}{10}}},\tilde{y}\right)
\hookrightarrow M^{1,9}$. The NS-NS $B$ field pull back on $D6$-brane is given as:
 \begin{eqnarray}
 \label{pull-back flux}
 & &  \delta\left(\theta_2 - \frac{\alpha_{\theta_2}}{N^{\frac{3}{10}}}\right) i^*B= \delta\left(\theta_2 - \frac{\alpha_{\theta_2}}{N^{\frac{3}{10}}}\right)\left[ -B_{\theta_{2}\tilde{z}}\tilde{z}^\prime(r)dr\wedge d\theta_{2}+B_{\theta_2\tilde{y}}d\theta_2\wedge d\tilde{y}+B_{\theta_{2}\tilde{x}}d\theta_{2}\wedge d\tilde{x}\right],\nonumber\\
 & &
 \end{eqnarray}
where $B_{\theta_2\tilde{x}}, B_{\theta_2\tilde{y}}, B_{\theta_2\tilde{z}}$ are given in (\ref{B-IIA-diag-nondiag}).
Determinent of the combination of pullback metric and pull back NS-NS $B$ field can be written as:
\begin{eqnarray}
\label{DBI-det_i}
& & {\rm det}\left(i^*(g+B)\right) = \Sigma_0(r;g_s,N_f,N,M) + \Sigma_1(r;g_s,N_f,M,N)\left(\tilde{z}^\prime(r)\right)^2,
\end{eqnarray}
(\ref{DBI-det_ii}) contain the expressions for $\Sigma_{0,1}(r;g_s,N_f,M,N)$.
Euler-Lagrange EOM obtained by minimizing the $D6-$brane DBI action is given as:
\begin{equation}
\label{EL-DBI-i}
\frac{d}{dr}\left(\frac{\tilde{z}^\prime(r)}{\sqrt{\Sigma_0(r;g_s,N_f,N,M) + \Sigma_1 (r;g_s,N_f,N,M) (\tilde{z}^\prime)^2}}\right) = 0.
\end{equation}
Both $\tilde{z}=$ constant like \cite{Dasgupta_et_al_Mesons} and $\frac{\tilde{z}^\prime(r)}{\sqrt{\Sigma_0(r;g_s,N_f,N,M) + \Sigma_1 (r;g_s,N_f,N,M) (\tilde{z}^\prime)^2}} = K$ provides a solution for (\ref{EL-DBI-i}). 
Thus, the equation (\ref{EL-DBI-i}) can be written as:
\begin{eqnarray}
\label{embedding eq}
& &z'(r)^2-\frac{24461180928 \pi ^{19/2} \alpha_{\theta_1}^{16} \alpha_{\theta_2}^8 {C_1} {g_s}^4 K N^{49/5}}{{C_2}^2-24461180928 \pi ^{19/2} \alpha_{\theta_1}^{16}
   \alpha_{\theta_2}^8 {C_3} {g_s}^4 K N^{49/5}}=0\nonumber\\
   & &
\end{eqnarray}
where $K$ is a random constant while $C_{1,2,3}$'s are function of $f(r;g_s,N_f,N)$. Doing a large-$N$ expansion and retaining terms upto NLO in $N$, $C_{1,2,3}(r;g_s,N_f,N)$ were obtained as :
\begin{eqnarray}
\label{coefficients}
& & C_1(r;g_s,N_f,N) \nonumber\\
& & = \frac{4194304 \pi ^{17/2} \sqrt[5]{N} r^6 \left(6 a^2+r^2\right) \left(27 \sqrt[3]{3} \alpha_{\theta_1}^6-12 \sqrt{6} \alpha_{\theta_1}^3 \alpha_{\theta_2}^2+4 \alpha_{\theta_1}^2
   \alpha_{\theta_2}^2 \sqrt[5]{N}-8 \alpha_{\theta_2}^4\right)}{\alpha_{\theta_1}^6 \alpha_{\theta_2}^4 {g_s} \left(9 a^2+r^2\right)}\nonumber\\
   & & C_2(r;g_s,N_f,N) \nonumber\\
   & & = \frac{8388608 \pi ^8 \alpha_{\theta_1}^2 \sqrt{{g_s}} N^{26/5} r^4 \left(r^4-{r_h}^4\right) \left(81 \alpha_{\theta_1}^6-36 \sqrt{2} \sqrt[6]{3} \alpha_{\theta_1}^3 \alpha_{\theta_2}^2+4
   3^{2/3} \alpha_{\theta_1}^2 \alpha_{\theta_2}^2 \sqrt[5]{N}-4 3^{2/3} \alpha_{\theta_2}^4\right)}{27 \alpha_{\theta_2}^2}\nonumber\\
   & & C_3(r;g_s,N_f,N)\nonumber\\
   & &  = \frac{4194304 \pi ^8 N^{3/10} r^4 \left(r^4-{r_h}^4\right) \left(81 \alpha_{\theta_1}^6-36 \sqrt{2} \sqrt[6]{3} \alpha_{\theta_1}^3 \alpha_{\theta_2}^2+4 3^{2/3} \alpha_{\theta_1}^2
   \alpha_{\theta_2}^2 \sqrt[5]{N}-4 3^{2/3} \alpha_{\theta_2}^4\right)}{27 \alpha_{\theta_1}^6 \alpha_{\theta_2}^6 {g_s}^{3/2}}.\nonumber\\
   & &
\end{eqnarray}
Plugging in the expressions for $C_{1,2,3}(r;g_s,N_f,N)$ in the equation (\ref{embedding eq}) followed by a large-$N$ expansion and truncating the series at NLO in $N$ the equation acquired the following form:
\begin{eqnarray}
\label{diff_eqn}
& & {\tilde{z}}'(r)^2 -\frac{59049 3^{2/3} \pi ^2 \alpha_{\theta_1}^4 \alpha_{\theta_2}^6 {g_s}^2 K \left(\frac{1}{N}\right)^{3/5} \left(6 a^2+r^2\right)}{2 r^2 \left(9 a^2+r^2\right)
   \left(r^4-{r_h}^4\right)^2}\nonumber\\
   & & -\frac{531441 \pi ^2 \alpha_{\theta_1}^5 \alpha_{\theta_2}^4 {g_s}^2 K \left(\frac{1}{N}\right)^{4/5} \left(6 a^2+r^2\right) \left(4
   \sqrt{2} \sqrt[6]{3} \alpha_{\theta_2}^2-9 \alpha_{\theta_1}^3\right)}{8 r^2 \left(9 a^2+r^2\right) \left(r^4-{r_h}^4\right)^2}=0\nonumber\\
   & &
\end{eqnarray}
As mentioned above for (\ref{EL-DBI-i}), one can verify that $\tilde{z}=$constant like \cite{Dasgupta_et_al_Mesons}, provides a solution. This solution allows one to fix the positions of $D6/\overline{D6}$-branes.  They can be placed at ``antipodal'' points with respect to each other by fixing $\tilde{z} = \pm{\cal C}\frac{\pi}{2}$. The constant embedding of $D6$-brane is independent of the choice of background. It is valid for both black hole($r_{h}\neq 0$) as well as a thermal background($r_{h}=0$).
\section{Spectroscopy of Vector Mesons in a Black-Hole Background for All Temperatures}
We obtained the spectra as the Kaluza-Klein modes of the massless sector of open strings in type IIA string theory at finite gauge coupling. The (pseudo-)vector mesons spectra is obtained by considering the gauge fluctuations of a background gauge field along the world volume of the embedded flavor $D6$-branes. For both high temperature and low temperature we have considered a background with a black hole without worrying about the issue that a themal background is suitable for low temperature. Meson spectra, for both (pseudo-)vector meson in  section {\bf 3.3} and (pseudo-)scalar meson in section {\bf 3.4} were shown to be isospectral with the spectra obtained with a thermal background in section {\bf 3.5}. 
Embedding condition of $D6$-branes can be further simplified by working with a new set of variables $(Y,Z)$ instead of $(r,\tilde{Z})$.
The coordinate transformation is given as  \cite{Dasgupta_et_al_Mesons}:
\begin{eqnarray}
\label{YZ}
& & r = r_h e^{\sqrt{Y^2+Z^2}}\nonumber\\
& & \tilde{z} = {\cal C} \arctan \frac{Z}{Y},
\end{eqnarray}
in the new variable the constant embedding of $D6(\overline{D6})$-branes can correspond to $\tilde{z} = \frac{\pi}{2}$ for ${\cal C}=1$ for $D6$-branes and $\tilde{z} = -\frac{\pi}{2}$ for ${\cal C}=-1$ for $\overline{D6}$-branes, both corresponding to $Y=0$.
In the background $i^*(g+B)$ consisting of a gauge field $F_{0}\frac{\sigma^3}{2}$ we activated  a gauge field fluctuation $F\frac{\sigma^3}{2}$ which implies:
{\footnotesize
\begin{eqnarray}
\label{DBI-i}
& & {\rm Str}\left.\sqrt{-{\rm det}_{\mathbb{R}^{1,3},Z,\theta_2,\tilde{y}}\left(i^*(g+B) + (2\pi\alpha\prime)(F_0 + F)\frac{\sigma^3}{2}\right)}\right|_{Y=0}\delta\left(\theta_2 - \frac{\alpha_{\theta_2}}{N^{\frac{3}{10}}}\right)\nonumber\\
& & = \sqrt{{\rm det}_{\theta_2,\tilde{y}}\left(i^*(g+B)\right)}\left. {\rm Str}\sqrt{-{\rm det}_{\mathbb{R}^{1,3},Z}\left(i^*(g+B) +(2\pi\alpha\prime) (F_0 + F)\frac{\sigma^3}{2}\right)}\right|_{Y=0}\delta\left(\theta_2 - \frac{\alpha_{\theta_2}}{N^{\frac{3}{10}}}\right)
\nonumber\\
& & = \left.\sqrt{{\rm det}_{\theta_2,\tilde{y}}\left(i^*(g+B)\right)}\sqrt{-{\rm det}_{\mathbb{R}^{1,3},Z}(i^*g)} {\rm Str}\left({\bf 1}_2 - \frac{1}{2}\left[(i^*g)^{-1}\left((2\pi\alpha\prime)(F_0 + F)\frac{\sigma^3}{2}\right)\right]^2 + ....\right)\right|_{Y=0}\delta\left(\theta_2 - \frac{\alpha_{\theta_2}}{N^{\frac{3}{10}}}\right).\nonumber\\
&&
\end{eqnarray}}
Picking up terms quadratic in $F$:
{\small
\begin{equation}
\label{DBI action}
\hskip -0.6in {\rm S}^{IIA}_{D6}=\frac{T_{D_{6}}(2\pi\alpha\prime)^{2}}{4}Str\int d^{4}xdZd\theta_{2}dy \delta\Bigg(\theta_{2}-\frac{\alpha_{\theta_{2}}}{N^{\frac{3}{10}}}\Bigg) e^{-\Phi^{IIA}}\sqrt{-{\rm det}_{\theta_{2}y}(\iota^*(g+B))}\sqrt{{\rm det}_{t,{\mathbb  R}^{1,2},Z}(\iota^*g)}g^{\mu\nu}F_{\nu\rho}g^{\rho\sigma}F_{\sigma\mu},
\end{equation}}
Here $\iota^*g$ and $\iota^*B$ are the pulled back metric and NS-NS $B$ on the $D6$-brane respectively. Plugging in the Klauza-Klein mode expansion for gauge fields:
\begin{eqnarray}
\label{AZ+Amu}
& & A_\mu(x^\nu,Z) = \sum_{n=1} B_\mu^{(n)}(x^\nu)\alpha_n(Z),\ \text{no summation w.r.t.}\ \mu,\nonumber\\
& & A_Z(x^\nu,Z) = \sum_{n=0} \phi^{(n)}(x^\nu)\beta_n(Z),
\end{eqnarray}
one obtains:
\begin{eqnarray}
\label{full-expansion}
& & -\frac{V}{4}\int d^4x dZ \sum_{nm}\Biggl({\cal V}_2(Z)\tilde{F}^{(n)}_{\mu\nu}\tilde{F}^{(m)\mu\nu}\alpha_m(Z)\alpha_n(Z) + {\cal V}_1(Z)B^{(m)}_\mu B^{(n)\mu }\dot{\alpha}_m\dot{\alpha}_n+ {\cal V}_1(Z)\partial_{\mu}\phi^{n}\partial^{\mu}\phi^{m}\beta_{n}\beta_{m}\nonumber\\
&&- {\cal V}_1(Z)\partial_{\mu}\phi^{n}{B^{(m)}}^{\mu}\beta_{n}\dot{\alpha}_{m}- {\cal V}_1(Z)\partial_{\mu}\phi^{m}{B^{n}}^{\mu}\beta_{m}\dot{\alpha}_{n}\Biggr).
\end{eqnarray}
The terms quadratic in $\alpha/\dot{\alpha}$ in (\ref{full-expansion}) are given as:
\begin{eqnarray}
\label{Ftildesq-i}
& & -\frac{V}{4}\int d^4x dZ \sum_{nm}\left({\cal V}_2(Z)\tilde{F}^{(n)}_{\mu\nu}\tilde{F}^{(m)\mu\nu}\alpha_m(Z)\alpha_n(Z) + {\cal V}_1(Z)B^{(m)}_\mu B^{(n)\mu }\dot{\alpha}_m\dot{\alpha}_n\right),
\end{eqnarray}
where:
\begin{eqnarray}
\label{V_V1_V2-defs}
&& V = -T_{D_{6}}2(2\pi\alpha')^2\int dy d\theta_{2}\delta\Bigg(\theta_{2}-\frac{\alpha_{\theta_{2}}}{N^{3/10}}\Bigg) \nonumber\\
&&{\cal V}_{1}(Z)=2\sqrt{h}g^{ZZ}e^{-\Phi^{IIA}}\sqrt{-{\rm det}_{\theta_2,y}\left(i^*(g+B)\right)}\sqrt{{\rm det}_{\mathbb{R}^{1,2},t,Z}(i^*g)}\nonumber\\
&&{\cal V}_{2}(Z)=h e^{-\Phi^{IIA}}\sqrt{-{\rm det}_{\theta_2,y}\left(i^*(g+B)\right)}\sqrt{{\rm det}_{\mathbb{R}^{1,2},t,Z}(i^*g)}.
\end{eqnarray}

Now, $F_{\mu\nu}(x^B,|Z|) = \sum_n\partial_{[\mu}B_{\nu]}^{(n)}\alpha_n(Z)\equiv \tilde{F}_{\mu\nu}^{(n)}\alpha_n(Z)$. The EOM satisfied by $B_\mu(x^\nu)^{(n)}$ is: $\partial_\mu \tilde{F}^{\mu\nu}_{(n)} + \partial_\mu\log\sqrt{g_{t,\mathbb{R}^{1,2},|Z|}}\tilde{F}^{\mu\nu}_{(n)} = \partial_\mu\tilde{F}^{\mu\nu}_{(n)} = {\cal M}_{(n)}^2B^\nu_{(n)}$. After integrating by parts once, and utilizing the EOM for $B^{(n)}_\mu$, one writes :
\begin{eqnarray}
\label{Ftildesq-ii}
& & \int d^4x dZ\  \left(-2 {\cal V}_2(Z) {\cal M}_{(m)}^2\alpha_n^{B_\mu}\alpha_m^{B_\mu} + {\cal V}_1(Z)\dot{\alpha}_n^{B_\mu}\dot{\alpha}_m^{B_\mu}\right)B^{\mu (n)}B_{\mu}^{(m)},
\end{eqnarray}
which yields the following equations of motion:
\begin{eqnarray}
\label{eoms_alpha_n_Bmu}
& & \alpha^{\left\{0\right\}}_{m}: \frac{d}{dZ\ }\left({\cal V}_1(Z) \tilde{g}^{00}(Z)\dot{\alpha}_{m}^{\left\{0\right\}}\right) + 2 {\cal V}_2(Z)\tilde{g}^{00}{\cal M}_{(m)}^2\alpha^{\left\{0\right\}}_m = 0,\nonumber\\
& & \alpha^{\left\{i\right\}}_{m}: \frac{d}{dZ\ }\left({\cal V}_1(Z) \dot{\alpha}_{m}^{\left\{i\right\}}\right) + 2 {\cal V}_2(Z){\cal M}_{(m)}^2\alpha^{\left\{i\right\}}_m = 0.
\end{eqnarray}

The normalization condition of $\alpha_{n}$ are given as
\begin{eqnarray}
\label{norm_psi}
&&V\int dZ\ {\cal V}_{2}(Z)\ \alpha_{n}\alpha_{m}=\delta_{nm}\nonumber\\
&&\frac{V}{2}\int dZ\ {\cal V}_{1}(Z)\ \partial_{Z}\alpha_{n} \partial_{Z}\alpha_{m}=m_{n}^2\delta_{nm}.
\end{eqnarray}
Thus the action for vector meson part for all $n\ge 1$ can be wriiten as
\begin{eqnarray}
& & -\int d^4x  \sum_{n}\left(\frac{1}{4}\tilde{F}^{(n)}_{\mu\nu}\tilde{F}^{(n)\mu\nu}+\frac{m_{n}^2}{2}B^{(n)}_\mu B^{(n)\mu }\right).
\end{eqnarray}

To normalize the kinetic term for $\phi ^{n}$, we impose the normalization condition for all n corresponding to $\phi ^{n}$ which ranges from 0 to $\infty $
\begin{eqnarray}
\label{norm_scalar}
&&\frac{V}{2}\int dZ\ {\cal V}_{1}(Z)\ \beta_{n}\beta_{m}=\delta_{nm}.
\end{eqnarray}
From (\ref{norm_psi}), it is seen that we can choose $\beta_{n}=m_{n}^{-1}\dot{\alpha}_{n}$ for all $n\ge 1$. For $n=0$ corresponding to $\phi_0$ we choose its form such as it is orthogonal to $\dot{\alpha}_{n}$ for all $n\ge 1$. By writing $\beta_{0}=\frac{C}{{\cal V}_{1}(Z)}$, we have
$$(\beta_{0},\beta_{n})\propto\int\ dZ\ C\partial_{Z}\alpha\ =0.$$
Thus the cross component in (\ref{full-expansion}) vanishes for $n=0$, and the remaining cross components can be absorbed in the $B_{\mu}^{n}$ by following a specific gauge transformation given as,
$$B_{\mu}^{n}\rightarrow B_{\mu}^{n}+m_{n}^{-1}\partial_{\mu}\phi^{n}.$$
Then the action becomes:
\begin{eqnarray}
& & -\int d^4x  \left[\frac{1}{2}\partial_{\mu}\phi^{0}\partial^{\mu}\phi^{0} + \sum_{n\ge 1}\left(\frac{1}{4}\tilde{F}^{(n)}_{\mu\nu}\tilde{F}^{(n)\mu\nu}+\frac{m_{n}^2}{2}B^{(n)}_\mu B^{(n)\mu }\right)\right].
\end{eqnarray}

Taking $a = r_h\left(0.6 + 4\frac{g_s M^2}{N}\left(1 + \log r_h\right)\right), m = \tilde{m}\frac{r_h}{\sqrt{4\pi g_s N}}$, from (\ref{Ftildesq-ii})one obtains :
\begin{eqnarray}
\label{EOMs-i}
& & \hskip -0.8in  \alpha_n^{\left\{i\right\}}\  ^{\prime\prime}(Z)+A_{0}(Z;g_{s},N_{f},N,M)\  \alpha_n^{\left\{i\right\}}\ ^{\prime}(Z) + A_{1}(Z;g_{s},N_{f},N,M)\  \alpha_n^{\left\{i\right\}}(Z)= 0,
\end{eqnarray}
and
\begin{eqnarray}
\label{EOMs-ii}
& & \hskip -0.8in {\alpha_n^{\left\{0\right\}}}\ ^{\prime\prime}(Z) + B_{0}(Z;g_{s},N_{f},N,M)\ \alpha_n^{\left\{0\right\}}\ ^{\prime}(Z)+ B_{1}(Z;g_{s},N_{f},N,M)\ {\alpha_n^{\left\{0\right\}}}(Z)=0.
\end{eqnarray}
Where the expressions for the coefficients $A_{0,1}(Z;g_{s},N_{f},N,M)$ and $B_{0,1}(Z;g_{s},N_{f},N,M)$ are given in (\ref{coeff-vector-eom}). We obtained the (pseudo-)vector meson spectrum via three different routes:
\begin{itemize}
\item{\it{First Route}}
\end{itemize}
The first route involved solving spatial mode $\alpha^{\left\{i\right\}}_n(Z)$ and temporal $\alpha^{\left\{0\right\}}_n(Z)$ EOMs near $r=r_{h}$(IR) and then imposing Neumann boundary conditions at the $r=r_{h}$(IR). This IR computation yields quantizated (pseudo-)vector meson masses and we found a near isospectrality in (pseudo-)vector meson masses up to LO in $N$, via both $\alpha^{\left\{0,i\right\}}_n(Z)$ EOMs. The temperature dependance for the spectrum was obtained via $N_f-$ and $M-$ dependent terms.
\begin{itemize}
\item{\it{Second Route}}
\end{itemize}
The second route involved converting $\alpha^{\left\{0,i\right\}}_n(Z)$ EOMs into Schr\"{o}dinger-like EOMs and solving the new EOMs in the IR and UV separately. Imposing Neumann BC at both the horizon(IR) as well in asymptotic boundary(UV) resulted in quantization of (pseudo-)vector meson mass. For IR, results up to LO in $N$, obtained via first route as well as second route were substantially equal up to LO in $N$. Imposing Neumann and/or Dirichlet boundary conditions in the UV failed to give any mass quantization conditions.
\begin{itemize}
\item{\it{Third Route}}
\end{itemize}
While the previous two routes caters to the both deep IR and UV region, third route addressed the IR-UV interpolating region. In this route WKB quantization condition was used to obtain our main results for (pseudo-) vector meson masses which were directly compared with PDG results. We also showed that  route one's (pseudo-)vector meson spectrum results were nearly isospectral with IR WKB quantization results.

\subsection{Vector Meson Spectrum from Solution of EOMs near $r=r_h$}

Near the horizon $Z=0 (Y=0)$ the EOM for $\alpha_n^i(Z)$ was obtained as :
\begin{equation}
\label{alpha_n^i_EOM-i}
\alpha_n^i\ ^{\prime\prime}(Z) + \left(\frac{1}{|Z|} + \alpha_1\right)\alpha_n^i\ ^\prime(Z) + \left(\frac{\beta_2}{|Z|} + \alpha_2\right)\alpha_n^i(Z) = 0,
\end{equation}
whose solution was given by:
\begin{eqnarray}
\label{alpha_n^i_EOM-ii}
& & \alpha_n^{\left\{i\right\}}(Z) = c_1 e^{\frac{1}{2} |Z| \left(-\sqrt{\alpha_1^2-4 \alpha_2}-\alpha_1\right)} U\left(-\frac{-\alpha_1+2 \beta_2-\sqrt{\alpha_1^2-4 \alpha_2}}{2
   \sqrt{\alpha_1^2-4 \alpha_2}},1,\sqrt{\alpha_1^2-4 \alpha_2} |Z|\right)\nonumber\\
   & & +c_2 e^{\frac{1}{2} |Z| \left(-\sqrt{\alpha_1^2-4 \alpha_2}-\alpha_1\right)}
   L_{\frac{-\sqrt{\alpha_1^2-4 \alpha_2}-\alpha_1+2 \beta_2}{2 \sqrt{\alpha_1^2-4 \alpha_2}}}\left(|Z| \sqrt{\alpha_1^2-4 \alpha_2}\right).
\end{eqnarray}
Since associate Laguerre function doesn't satisy Neumann BC, we set $c_2=0$. 
Under the transformation $Z\rightarrow-Z$, the diffrential operator in (\ref{eoms_alpha_n_Bmu}) is even -pertinent to parity and charge conjugation \cite{Sakai-Sugimoto-1}. Thus solutions are either odd or even under the transformation $Z\rightarrow-Z$. As given in \cite{Sakai-Sugimoto-1} that $\alpha^{\left\{i\right\}}_{2n}(-Z) = - \alpha^{\left\{i\right\}}_{2n}(Z)$ and $\alpha^{\left\{i\right\}}_{2n+1}(-Z) = \alpha^{\left\{i\right\}}_{2n+1}(Z)$, (\ref{alpha_n^i_EOM-ii}) with vanishing $c_2$ must be interpreted as:
\begin{eqnarray}
\label{vector-i_P_C_wavefunction}
& &\hskip -0.3in \alpha^{\left\{i\right\}}_n(Z) = \left(\delta_{n,2\mathbb{Z}^+}{\rm Sign}(Z) + \delta_{n,(2\mathbb{Z}^+\cup\left\{0\right\})+1}\right)e^{\frac{1}{2} |Z| \left(-\sqrt{\alpha_1^2-4 \alpha_2(n)}-\alpha_1\right)} \nonumber\\
 & & U\left(-\frac{-\alpha_1+2 \beta_2(n)-\sqrt{\alpha_1^2-4 \alpha_2(n)}}{2
   \sqrt{\alpha_1^2-4 \alpha_2(n)}},1,\sqrt{\alpha_1^2-4 \alpha_2(n)} |Z|\right).\nonumber\\
\end{eqnarray}
Setting $c_2=0$:
\begin{eqnarray}
\label{alpha_n^i_EOM-iii}
&&\alpha_n^{\left\{i\right\}}\ ^\prime(Z) = -\frac{1}{2} e^{-\frac{1}{2} |Z| \left(\sqrt{\alpha_1^2-4 \alpha_2}+\alpha_1\right)} \Biggl[\left(\sqrt{\alpha_1^2-4 \alpha_2}+\alpha_1\right)
   U\left(\frac{\alpha_1-2 \beta_2+\sqrt{\alpha_1^2-4 \alpha_2}}{2 \sqrt{\alpha_1^2-4 \alpha_2}},1,\sqrt{\alpha_1^2-4 \alpha_2}
   |Z|\right)\nonumber\\
& & +\left(\sqrt{\alpha_1^2-4 \alpha_2}+\alpha_1-2 \beta_2\right) U\left(\frac{\alpha_1-2 \beta_2+3 \sqrt{\alpha_1^2-4 \alpha_2}}{2
   \sqrt{\alpha_1^2-4 \alpha_2}},2,\sqrt{\alpha_1^2-4 \alpha_2} |Z|\right)\Biggr]\nonumber\\
   & & = -\frac{1}{|Z| \gamma \left(\frac{\alpha_1-2 \beta_2+\sqrt{\alpha_1^2-4 \alpha_2}}{2 \sqrt{\alpha_1^2-4 \alpha_2}}\right)}\nonumber\\
   & & +\frac{1}{2 \gamma \left(\frac{\alpha_1-2 \beta_2+\sqrt{\alpha_1^2-4 \alpha_2}}{2 \sqrt{\alpha_1^2-4
   \alpha_2}}\right)}\Biggl\{\beta_2 \log
   \left(\alpha_1^2-4 \alpha_2\right)+\left(\sqrt{\alpha_1^2-4 \alpha_2}+\alpha_1\right) \psi ^{(0)}\left(\frac{\alpha_1-2 \beta_2+\sqrt{\alpha_1^2-4
   \alpha_2}}{2 \sqrt{\alpha_1^2-4 \alpha_2}}\right)\nonumber\\
   & & -\left(\sqrt{\alpha_1^2-4 \alpha_2}+\alpha_1-2 \beta_2\right) \psi ^{(0)}\left(\frac{\alpha_1-2
   \beta_2+3 \sqrt{\alpha_1^2-4 \alpha_2}}{2 \sqrt{\alpha_1^2-4 \alpha_2}}\right)\nonumber\\
   & & +2 \sqrt{\alpha_1^2-4 \alpha_2}+2 \alpha_1+2 \beta_2 {\log |Z|}-2
   \beta_2+4 \gamma  \beta_2\Biggr\}+ {\cal O}\left(|Z|\right)
\end{eqnarray}
As a result the Neumann/Dirichlet BC $\alpha_n^i\ ^\prime({r = r_h})=0$ can be imposed so long as the following condition is met:
\begin{equation}
\label{alpha_n^i_EOM-iv}
\frac{\sqrt{\alpha_1^2-4 \alpha_2}+\alpha_1-2 \beta_2}{2 \sqrt{\alpha_1^2-4 \alpha_2}} = - n\in\mathbb{Z}^-.
\end{equation}
For $a = 0.6 r_h$, (\cite{EPJC-2,Sil+Yadav+Misra-glueball}) (\ref{alpha_n^i_EOM-iv}) with reference (\ref{EOMs-i}) can be shown:
\begin{eqnarray}
\label{alphas+betas}
& & \alpha_1 = -1.08 -\frac{9. {g_s} M^2 (4.8 \log ({r_h})+4.8)}{{\log N} N}+\frac{3. {g_s} M^2 (4.8 \log ({r_h})+4.8)}{N}+\frac{(-4.14
   {g_s} {N_f}-3.016) \log ({r_h})}{{g_s} {N_f} \log ^2(N)}\nonumber\\
   & & +\frac{1.5 M^2 (4.8 \log ({r_h})+4.8) (-18. {g_s} {N_f} \log
   ({r_h})+9 {g_s} {N_f}+75.398)}{N {N_f} \log ^2(N)}+\frac{0.24}{{\log N}},\nonumber\\
& & \alpha_2 = \frac{{g_s} M^2 \tilde{m}^2 (7.2 \log ({r_h})+7.2)}{N}+0.54 \tilde{m}^2,\nonumber\\
& & \beta_2 = \frac{{g_s} M^2 \tilde{m}^2 (-3.6 \log ({r_h})-3.6)}{N}-0.02 \tilde{m}^2.\nonumber\\
\end{eqnarray}
We performed a large-$N$, large $|\log r_h|$ and large $\log N$ expansion and retained terms up to leading order in both $N$ and $\log r_h$ and up to next-to-leading order in $\log N$ and obtained the following spectrum for mesons:
\begin{eqnarray}
\label{meson-spectroscopy-i}
& & \tilde{m}_n^{\alpha^{\left\{i\right\}}_n} = 0.5 \sqrt{-10800. n^2-10800. n+10800. \sqrt{(n+0.36) (n+0.5) (n+0.5)
   (n+0.64)}-2592.}\nonumber\\
   & & +\frac{0.25 \left(\frac{23.04 (n+0.5)^2}{\sqrt{(n+0.36) (n+0.5) (n+0.5) (n+0.64)}}-24.\right)}{{\log N} \sqrt{-10800. n^2-10800.
   n+10800. \sqrt{(n+0.36) (n+0.5) (n+0.5) (n+0.64)}-2592.}}\nonumber\\
   & & + \frac{1}{{g_s} (\log N)^2 \sqrt{-10800. n^2-10800. n+10800. \sqrt{(n+0.36) (n+0.5) (n+0.5) (n+0.64)}-2592.}
   {N_f}}\nonumber\\
   & & \Biggl\{0.25 \log ({r_h}) \Biggl(\frac{1}{((n+0.36) (n+0.5) (n+0.5) (n+0.64))^{3/2}}\nonumber\\
   & & \times\Biggl\{n^6 (-397.44 {g_s} {N_f}-289.529)+n^5 (-1192.32 {g_s} {N_f}-868.588)+n^4 (-1482.61
   {g_s} {N_f}-1080.06)\nonumber\\
   & & +n^3 (-978.02 {g_s} {N_f}-712.473)+n^2 (-360.915 {g_s} {N_f}-262.921)+n (-70.6251 {g_s}
   {N_f}-51.4493)\nonumber\\
   & & -5.72314 {g_s} {N_f}-4.16922\Biggr\}+414. {g_s}
   {N_f}+301.593\Biggr)\Biggr\}+ {\cal O}\left(\frac{1}{(\log N)^3},\frac{1}{N}\right)\nonumber\\
   \end{eqnarray}
For $n=0$ we obtained imagionary values with ${\cal O}\left(\frac{1}{N^2}\right)$-suppression. Hence we disregarded $n=0$ mode, for $n\geq 1$ we obtained:
\begin{eqnarray}
\label{meson-spectroscopy-ii}
& & \tilde{m}_{n=1}^{\alpha^{\left\{i\right\}}_0} = \frac{0.18 (414 {g_s} {N_f}+0.089 (-4487.65 {g_s} {N_f}-3269.19)+301.6) \log ({r_h})}{{g_s} {N_f} \log
   ^2(N)}-\frac{0.15}{{\log N}}+0.69,\nonumber\\
& & \tilde{m}_{n=2}^{\alpha^{\left\{i\right\}}_n} = \frac{1}{{g_s} {N_f} \log ^2(N)}\Biggl\{0.173 \log ({r_h}) (0.004 (-5.723 {g_s} {N_f}+16 (-1482.61 {g_s} {N_f}-1080.06)\nonumber\\
& & +32 (-1192.32 {g_s}
   {N_f}-868.588)+8 (-978.02 {g_s} {N_f}-712.473)+4 (-360.915 {g_s} {N_f}-262.92)\nonumber\\
   &&+2 (-70.625 {g_s} {N_f}-51.449)+64
   (-397.44 {gsNf}-289.529)-4.169)+414. {g_s} {N_f}+904.77)\Biggr\}\nonumber\\
 & &   -\frac{0.16}{{\log N}}+0.721.
\end{eqnarray}
(\ref{meson-spectroscopy-ii}) verifies the expectation that masses must be small since one is looking at near horizon solutions for the EOM.

Near $r=r_h$ EOM (\ref{EOMs-ii}) can be written as:
\begin{eqnarray}
\label{alpha_n^0_EOM-ii}
& & \alpha_n^{\left\{0\right\}}\ ^{\prime\prime}(Z) + \alpha_1\alpha_n^{\left\{0\right\}}\ ^{\prime}(Z) + \left(\frac{\beta_2}{|Z|} + \alpha_2\right)\alpha_n^{\left\{0\right\}}(Z) = 0,
\end{eqnarray}
with solution:
\begin{eqnarray}
\label{alpha_n^0_EOM-iii}
& & \alpha_n^{\left\{0\right\}}(Z) = c_1 |Z| e^{-\frac{1}{2} |Z| \left(\sqrt{\alpha_1^2-4 \alpha_2(n)}+\alpha_1\right)}
   U\left(1-\frac{\beta_2}{\sqrt{\alpha_1^2-4 \alpha_2(n)}},2,\sqrt{\alpha_1^2-4 \alpha_2(n)} |Z|\right)\nonumber\\
& & + c_2 |Z| e^{-\frac{1}{2} |Z| \left(\sqrt{\alpha_1^2-4 \alpha_2(n)}+\alpha_1\right)} \, _1F_1\left(1-\frac{\beta_2(n)}{\sqrt{\alpha_1^2-4
   \alpha_2(n)}};2;\sqrt{\alpha_1^2-4 \alpha_2(n)} |Z|\right).
\end{eqnarray}
At $|Z|\rightarrow0^+$ the Neumann BC for the second part of the solution \label{alpha_n^0_EOM-iii} given by \\
$\frac{d}{dZ\ }\left[c_2 |Z| e^{-\frac{1}{2} |Z| \left(\sqrt{\alpha_1^2-4 \alpha_2}+\alpha_1\right)} \, _1F_1\left(1-\frac{\beta_2}{\sqrt{\alpha_1^2-4
\alpha_2}};2;\sqrt{\alpha_1^2-4 \alpha_2} |Z|\right)\right]$  vanishes only for $c_2=0$. Hence similar to (\ref{vector-i_P_C_wavefunction}) after putting $c_2=0$ we obtain odd, even solutions:
\begin{eqnarray}
\label{vector-ii_P_C_wavefunction}
& &\hskip -0.3in \alpha^{\left\{0\right\}}_n(Z) = \left(\delta_{n,2\mathbb{Z}^+}{\rm Sign}(Z) + \delta_{n,(2\mathbb{Z}^+\cup\left\{0\right\})+1}\right)e^{-\frac{1}{2} |Z| \left(\sqrt{\alpha_1^2-4 \alpha_2(n)}+\alpha_1\right)}\nonumber\\
&& U\left(1-\frac{\beta_2(n)}{\sqrt{\alpha_1^2-4 \alpha_2(n)}},2,\sqrt{\alpha_1^2-4 \alpha_2(n)} |Z|\right).
\end{eqnarray}
Now:
\begin{eqnarray}
\label{alpha_n^0_EOM-iv}
   & & \left.\alpha_n^{\left\{0\right\}}\ ^{\prime}(Z)\right|_{c_2=0} = -\frac{1}{2} c_1 e^{-\frac{1}{2} |Z| \left(\sqrt{\alpha_1^2-4 \alpha_2}+\alpha_1\right)}\nonumber\\
   & & \times \Biggl[\left(|Z| \sqrt{\alpha_1^2-4 \alpha_2}+\alpha_1 |Z|-2\right)
   U\left(1-\frac{\beta_2}{\sqrt{\alpha_1^2-4 \alpha_2}},2,\sqrt{\alpha_1^2-4 \alpha_2} |Z|\right)\nonumber\\
   & & +2 |Z| \left(\sqrt{\alpha_1^2-4 \alpha_2}-\beta_2\right)
   U\left(2-\frac{\beta_2}{\sqrt{\alpha_1^2-4 \alpha_2}},3,\sqrt{\alpha_1^2-4 \alpha_2} |Z|\right)\Biggr]\nonumber\\
   & & = \frac{c_1 \left(\beta_2 \log \left(\alpha_1^2-4 \alpha_2\right)+2 \beta_2 \psi ^{(0)}\left(1-\frac{\beta_2}{\sqrt{\alpha_1^2-4
   \alpha_2}}\right)+\sqrt{\alpha_1^2-4 \alpha_2}+\alpha_1+2 \beta_2 {\log |Z|}+4 \gamma  \beta_2\right)}{2 \beta_2 \gamma
   \left(-\frac{\beta_2}{\sqrt{\alpha_1^2-4 \alpha_2}}\right)} + {\cal O}\left(|Z|\right).\nonumber\\
   & &
\end{eqnarray}
As a result the Neumann/Dirichlet BC  $\alpha_n^{\left\{0\right\}}\ ^{\prime}({r = r_h})=0$ are successfully imposed as long as the following condition is met:
\begin{equation}
   \label{Neumann_f0}
   \frac{\beta_2}{\sqrt{\alpha_1^2-4 \alpha_2}} = n \in \mathbb{Z}^+.
   \end{equation}
In relevance to (\ref{EOMs-ii}):
\begin{eqnarray}
   \label{alphas+betas}
    \alpha_1 & = & \frac{-\frac{3.016}{{g_s} {N_f}}+0.72 \log ({r_h})-4.86}{{\log N}^2}+\frac{43.2 {g_s} M^2 {N_f} \log ({r_h})+43.2
   {g_s} M^2 {N_f}}{{\log N} N {N_f}}\nonumber\\
   & & +\frac{{g_s} M^2 (-14.4 \log ({r_h})-14.4)}{N}+\frac{0.24}{{\log N}}+0.92,\nonumber\\
    \alpha_2 & = & \frac{{g_s} M^2 \tilde{m}^2 (7.2 \log ({r_h})+7.2)}{N}+0.54 \tilde{m}^2,\nonumber\\
   \beta_2  & = & \frac{{g_s} M^2 \tilde{m}^2 (-3.6 \log ({r_h})-3.6)}{N}-0.02 \tilde{m}^2.
\end{eqnarray}
Only for a single value of mode `n' the conditon (\ref{Neumann_f0}) was satisified. Hence we considered that mode corresponding to the ground state (primarily due to the fact that its value was very close to (\ref{meson-spectroscopy-i}))- which satisfied:
\begin{eqnarray}
\label{condition-single-mtilde}
&& \frac{1}{{g_s}^2
   {\log N}^4 {N_f}^2}\Biggl\{\Biggl({g_s}^2 {\log N} M^2 {N_f} (43.2-14.4 {\log N})\nonumber\\
   & & +{g_s} {N_f} \log ({r_h}) \left({g_s} {\log N} M^2
   (43.2-14.4 {\log N})+0.72 N\right)\nonumber\\
   & & +N (0.92 {g_s} ({\log N}-2.17166) ({\log N}+2.43253) {N_f}-3.016)\Biggr)^2\Biggr\}\nonumber\\
   & & +\tilde{m}^2 N \left(-28.8 {g_s} M^2 \log ({r_h})-28.8 {g_s} M^2-2.16 N\right)=0.\nonumber\\
   & &
\end{eqnarray}
The solution following a large-$N$ and large-$\log N$ expansion were obtained as:
   \begin{eqnarray}
   \label{meson-spectroscopy-iii}
  & &   \tilde{m}_{n=0}^{\alpha^{\left\{0\right\}}_n} = \frac{0.163}{{\log N}} + 0.626 + \frac{-\frac{2.052}{{g_s} {N_f}}+0.49 \log ({r_h})-3.307}{{\log N}^2}
   \nonumber\\
   & & +\frac{{g_s} M^2 (28.305 \log
   ({r_h})+28.305)}{N{\log N} }+\frac{{g_s} M^2 (-13.971 \log ({r_h})-13.971)}{N} + ...
   \end{eqnarray}
  Now, from (\ref{meson-spectroscopy-ii}) and (\ref{meson-spectroscopy-iii}), disregarding
  ${\cal O}\left(\frac{\log r_h}{(\log N)^2}\right)$ terms, one sees that
\begin{equation}
\label{match-ground-states}
 \tilde{m}_{n=1}^{\alpha^{\left\{i\right\}}} = \tilde{m}_{n=0}^{\alpha^{\left\{0\right\}}},\ {\rm for}\ N=105.
\end{equation}
Hence, from (\ref{meson-spectroscopy-ii}) and (\ref{meson-spectroscopy-iii}), one observes that $\alpha_{n=1}^i$ and $\alpha_{n=0}^0$ spectra are isospectral in IR. The conformal ($N\rightarrow\infty$) and the non-conformal
($N_f,M$-dependence) are captured by (\ref{meson-spectroscopy-iii}). Along with it the temperature dependence for vector mesons introduced via  $\log r_h$ occuring at ${\cal O}\left(\frac{1}{N}\right)$ is captured by both, (\ref{meson-spectroscopy-ii}) and (\ref{meson-spectroscopy-iii}).

\subsection{Vector Meson Spectroscopy from Conversion of $\alpha^{\left\{i\right\}}_n(Z)$'s EOM to Schr\"{o}dinger-like Equations}
Following a field redefinition given as: $\psi_n^i(Z) = \sqrt{{\cal C}_1(Z)}\alpha_n^i(Z)$ the $\alpha^{\left\{i\right\}}_n$ EOM (\ref{EOMs-i})
\begin{eqnarray*}
   & & \alpha^{\left\{i\right\}}_n\ ^{\prime\prime}(Z) + A_{0}(Z)\alpha^{\left\{i\right\}}_n\ ^{\prime}(Z) + A_{1}(Z)\alpha^{\left\{i\right\}}_n(Z) = 0
\end{eqnarray*}
takes the Schr\"{o}dinger-like equation form:
\begin{eqnarray}
\label{WKB-i-1}
& & \psi_n^{\left\{i\right\}}\ ^{\prime\prime}(Z) + V(\alpha^{\left\{i\right\}}_n)\psi_n^{\left\{i\right\}}(Z) = 0,
\end{eqnarray}
where:
$V = \frac{{\cal C}_1^{\prime\prime}}{2{\cal C}_1} - \frac{1}{4}\left(\frac{{\cal C}_1^\prime}{{\cal C}_1}\right)^2 + A_1$.
For $\alpha^{\left\{i\right\}}_n$ ,
\begin{eqnarray}
\label{C1_WKB}
& & {\cal C}_1 = -\frac{1}{2 {r_h}^2}\Biggl\{e^{-4 |Z|} \left(e^{4 |Z|}-1\right) \Biggl({g_s} {N_f} {\log N} \left(3 a^2+2 {r_h}^2 e^{2 |Z|}\right)-3 {g_s}
   {N_f} \log ({r_h}) \left(3 a^2+2 {r_h}^2 e^{2 |Z|}\right)\nonumber\\
   & & -9 a^2 {g_s} {N_f} |Z|-9 a^2 {g_s} {N_f}+12 \pi
   a^2-6 {g_s} {N_f} {r_h}^2 e^{2 |Z|} |Z|+8 \pi  {r_h}^2 e^{2 |Z|}\Biggr)\Biggr\}.
\end{eqnarray}

\subsubsection{IR}
Analytical expression after two consecutive: large-$N$ and small-$|Z|$ expansion of the potential $V(\alpha^{\left\{i\right\}}_n)$, for $a=r_h
\left(0.6 + 4 \frac{g_s M^2}{N}\left(1 + \log r_h\right)\right)$ \cite{EPJC-2}, is as follow:
\begin{eqnarray}
\label{V-WKB}
& & \hskip -0.4in V(\alpha^{\left\{i\right\}}_n) = \nonumber\\
& & \hskip -0.4in \frac{1}{\left(e^{4
   |Z|}-1.\right)^2}\Biggl\{e^{-2 |Z|} \left(e^{6 |Z|} \left(6.-1.08 \tilde{m}^2\right)+e^{4 |Z|} \left(2.16- \tilde{m}^2\right)+
   \tilde{m}^2 e^{8 |Z|}+\left(1.08 \tilde{m}^2-1\right) e^{2 |Z|}- e^{10 |Z|}-2.16\right)\Biggr\}
   \nonumber\\
   & & \hskip -0.4in +\frac{e^{-2 |Z|} \left({g_s}^3 {N_f}^3 \left(4.86-3. e^{2 |Z|}-3.24 e^{4 |Z|}-1.62 e^{8 |Z|}+3. e^{10
   |Z|}\right)\right)}{{g_s}^3 {\log N} {N_f}^3 \left(e^{4
   |Z|}-1.\right)^2} + {\cal O}\left(\frac{1}{{(\log N)^2}},\frac{g_sM^2}{N}\right).
\end{eqnarray}
IR form of, (\ref{V-WKB}) is:
\begin{eqnarray}
\label{V-WKB-alphai-IR}
& & V(\alpha^{\left\{i\right\}}_n; IR) = \frac{-0.02 \tilde{m}^2-\frac{0.12}{{\log N}}+0.54}{|Z|}+0.54 \tilde{m}^2+\frac{4.86}{{\log N}}+\frac{0.25}{Z^2}-3.49333 + {\cal O}\left(|Z|,\frac{1}{{(\log N)^2}},\frac{g_sM^2}{N}\right).\nonumber\\
& &
\end{eqnarray}
The solution to (\ref{V-WKB}) were obtained in terms of Whittaker functions:
\begin{eqnarray}
\label{solution-Schrodinger}
& & \psi_n^{\left\{i\right\}}(Z) = c_1 M_{\frac{{\log N} \left(0.27-0.01 \tilde{m}^2\right)-0.06}{\sqrt{{\log N}} \sqrt{-0.54 {\log N}
   \tilde{m}^2+3.49333 {\log N}-4.86}},0}\left(\frac{2 \sqrt{-0.54 {\log N} \tilde{m}^2+3.49333
   {\log N}-4.86} |Z|}{\sqrt{{\log N}}}\right)\nonumber\\
   & & +c_2 W_{\frac{{\log N} \left(0.27-0.01
   \tilde{m}^2\right)-0.06}{\sqrt{{\log N}} \sqrt{-0.54 {\log N} \tilde{m}^2+3.49333
   {\log N}-4.86}},0}\left(\frac{2 \sqrt{-0.54 {\log N} \tilde{m}^2+3.49333 {\log N}-4.86} |Z|}{\sqrt{{\log N}}}\right).\nonumber\\
   & &
\end{eqnarray}
One can show that:
\begin{eqnarray}
\label{inf_der_horizon}
& & \left.\frac{d}{dZ\ }\left(\frac{M_{\frac{{\log N} \left(0.27-0.01 \tilde{m}^2\right)-0.06}{\sqrt{{\log N}} \sqrt{-0.54 {\log N}
   \tilde{m}^2+3.49333 {\log N}-4.86}},0}\left(\frac{2 \sqrt{-0.54 {\log N} \tilde{m}^2+3.49333
   {\log N}-4.86} |Z|}{\sqrt{{\log N}}}\right)}{\sqrt{C_1}}\right)\right|_{{r = r_h}}=0,\nonumber\\
   & &
\end{eqnarray}
implies
\begin{equation}
\label{alpha-Sch-IR}
\tilde{m} = \left(2.543-\frac{1.769}{{\log N}}\right) + {\cal O}\left(\left(\frac{1}{{\log N}}\right)^{3/2}\right)
\end{equation}
One can also show that
\begin{eqnarray}
\label{Neumann_horizon}
& & \left.\frac{d}{dZ\ }\left(\frac{W_{\frac{{\log N} \left(0.27-0.01 \tilde{m}^2\right)-0.06}{\sqrt{{\log N}} \sqrt{-0.54 {\log N}
   \tilde{m}^2+3.49333 {\log N}-4.86}},0}\left(\frac{2 \sqrt{-0.54 {\log N} \tilde{m}^2+3.49333 {\log N}-4.86}
   |Z|}{\sqrt{{\log N}}}\right)}{\sqrt{{C_1}}}\right)\right|_{{r = r_h}}=0\nonumber\\
   & &
\end{eqnarray}
implies:
\begin{equation}
\label{Neumann-Schrodinger}
\tilde{m}_n^{\alpha_n^i} = 0.5 \sqrt{-10800. n^2+10800. \sqrt{(n+0.376679) (n+0.623321) \left(n^2+n+0.25\right)}-10800. n-2592}.
   \end{equation}
 The $n=0$ result of  (\ref{alpha-Sch-IR}) and $n=0$ result- 2.479 of (\ref{Neumann-Schrodinger})- are approximately close at LO. Similar to
 (\ref{vector-i_P_C_wavefunction}) and (\ref{vector-ii_P_C_wavefunction}) we have even or odd solutions hence,
{\footnotesize \begin{eqnarray}
 \label{alphai-wavefunction-Sch-IR}
 & & \hskip -0.5in \alpha^{\left\{i\right\}}_{n=0}(Z) ={\rm Sign}(Z) \frac{ {M\ {or}\ W}_{\frac{{\log N} \left(0.27-0.01 \tilde{m}^2\right)-0.06}{\sqrt{{\log N}} \sqrt{-0.54 {\log N}
   \tilde{m}^2+3.49333 {\log N}-4.86}},0}\left(\frac{2 \sqrt{-0.54 {\log N} \tilde{m}^2+3.49333
   {\log N}-4.86} |Z|}{\sqrt{{\log N}}}\right)}{\sqrt{{\cal C}_1(Z)}}.\nonumber\\
   & &
 \end{eqnarray}}
\subsubsection{UV}
In the UV, $N_f, M$ are very small which allowed us to drop the terms dependent on both paramters. Hence we obtained simplified analytical expression for the potential:
\begin{eqnarray}
\label{V_alphai_UV}
& & V(\alpha^{\left\{i\right\}}_n;UV)  \nonumber\\
& & = \frac{e^{-2 |Z|} \left(e^{6 |Z|} \left(6.-1.08 \tilde{m}^2\right)+e^{4 |Z|} \left(2.16-1\tilde{m}^2\right)+\tilde{m}^2
   e^{8 |Z|}+\left(1.08 \tilde{m}^2-1\right) e^{2 |Z|}- e^{10 |Z|}-2.16\right)}{\left(e^{4 |Z|}-1\right)^2}\nonumber\\
   & & = -1 + \left(2.16 + \tilde{m}^2\right)e^{-2 |Z|} + {\cal O}\left(e^{-4 |Z|}\right).
\end{eqnarray}
Schr\"{o}dinger-like EOM solution were obtained as:
\begin{eqnarray}
\label{solution-Sch-UV}
& & \psi_n^{\left\{i\right\}}(|Z|\in{\rm UV}) =\left(\delta_{n,2\mathbb{Z}^+}{\rm Sign}(Z) + \delta_{n,(2\mathbb{Z}^+\cup\left\{0\right\})+1}\right)\nonumber\\
& & \times \left[c_1 I_{1}\left(0.2i e^{-|Z|} \sqrt{25. \tilde{m}^2(n)+54.}\right)+c_2 K_{1}\left(0.2i e^{-|Z|}
   \sqrt{25 \tilde{m}^2(n)+54}\right)\right].\nonumber\\
   & &
\end{eqnarray}
One can show that both $I_1$ and $K_1$ satisfies Neumann BC while only the former one satisfies Dirichlet BC:
\begin{equation}
\label{dI}
\lim_{Z\rightarrow\infty}\frac{d}{dZ\ }\left(\frac{c_1 I_{1}\left(0.2 i e^{-|Z|} \sqrt{25. \tilde{m}^2+54.}\right)}{\sqrt{{C_1}}}\right)=0
\end{equation}

\begin{equation}
\label{dK}
\lim_{Z\rightarrow\infty}\frac{d}{dZ\ }\left(\frac{c_2 K_{1}\left(0.2 i e^{-|Z|} \sqrt{25. \tilde{m}^2+54.}\right)}{\sqrt{{C_1}}}\right)=0,
\end{equation}
Which left us without any quantization conditon for the mass of vector mesons therefore not providing values for $\tilde{m}$.

\subsection{Vector Meson Spectrum from Conversion of $\alpha^{\left\{0\right\}}_n(Z)$'s EOM to Schr\"{o}dinger-like Equations}
\subsubsection{IR}
The potential was obtained as:
\begin{eqnarray}
\label{V-alpha0-IR}
& & V(\alpha^{\left\{0\right\}}_n;IR) = -1 + |Z| \left(\frac{3.24}{{\log N}}-0.526667 \tilde{m}^2\right)-\frac{0.02 \tilde{m}^2}{|Z|}+0.54 \tilde{m}^2-\frac{3.24 Z^2}{\log
   (N)}+\frac{1.38}{{\log N}} \nonumber\\
   & &  + {\cal O}\left(\frac{1}{(\log N)^2},\frac{g_s M^2}{N},Z^3\right).
\end{eqnarray}
The solution to Schr\"{o}dinger-like EOM:
\begin{equation}
\psi^0_n\ ^{\prime\prime} (Z) + \left(\frac{a_1}{|Z|} + b_1\right)\psi^0_n(Z) = 0,
\end{equation}
was obtained as:
\begin{eqnarray}
\label{solution-alpha0-IR-i}
& & \psi^0_n(Z) = c_2 |Z| e^{-\sqrt{-{b_1}} |Z|}\ _1F_1 \left(\frac{- {a_1}}{2 \sqrt{-{b_1}}}+1;2;2 \sqrt{-{b_1}} |Z|\right)+c_1 |Z| e^{-\sqrt{-{b_1}} |Z|} U\left(\frac{- {a_1}}{2 \sqrt{-{b_1}}}+1,2,2 \sqrt{-{b_1}}
   |Z|\right).\nonumber\\
   & &
\end{eqnarray}
Imposing Neumann BC at ${r = r_h}$:
\begin{eqnarray}
\label{dU}
& & \frac{d}{dZ\ }\left(\frac{c_1 |Z| e^{-\sqrt{-{b_1}}|Z|} U\left(\frac{- {a_1}}{2 \sqrt{-{b_1}}}+1,2,2 \sqrt{-{b_1}} |Z|\right)}{{\sqrt{C_1}}(Z)}\right)\nonumber\\
   & & = \frac{1}{2 {a_1} {C_1}(0)^{3/2} \left({a_1} \sqrt{-{b_1}}+2
   {b_1}\right) \gamma \left(-\frac{{a_1}}{2 \sqrt{-{b_1}}}\right)}\nonumber\\
   & & \times\Biggl\{c_1 \Biggl(2 {a_1}^2 \sqrt{-{b_1}} {C_1}(0) \log |Z|+4 \gamma  {a_1}^2 \sqrt{-{b_1}} {C_1}(0)+{a_1}
   \sqrt{-{b_1}} {C_1}'(0)+4 {a_1} {b_1} {C_1}(0) \log |Z|\nonumber\\
   & & -2 {a_1} {b_1} {C_1}(0)+8 \gamma  {a_1} {b_1}
   {C_1}(0)+2 {a_1} {C_1}(0) \left({a_1} \sqrt{-{b_1}}+2 {b_1}\right) \log \left(2 \sqrt{-{b_1}}\right)
   \nonumber\\
   & & +2 {a_1}
   {C_1}(0) \left({a_1} \sqrt{-{b_1}}+2 {b_1}\right) \psi ^{(0)}\left(1-\frac{{a_1}}{2 \sqrt{-{b_1}}}\right)+2 {b_1}
   {C_1}'(0)+4 \sqrt{-{b_1}} {b_1} {C_1}(0)\Biggr)\Biggr\}+ {\cal O}(Z).\nonumber\\
& & \end{eqnarray}
 satisfied if:
\begin{equation}
\label{alpha0-IR-quantization}
\frac{a_1}{2\sqrt{-b_1}} = n \in\mathbb{Z}^+\cup\left\{0\right\},
\end{equation}
with:
\begin{eqnarray}
\label{a1+b1}
& & a_1 = -0.02 \tilde{m}^2\nonumber\\
& & b_1 =  -1 + 0.54 \tilde{m}^2+\frac{1.38}{{\log N}},
\end{eqnarray}
This gave:
\begin{eqnarray}
\label{m-alpha0-IR}
& & m^{\alpha^{\left\{0\right\}}_n}_n = 0.5 \sqrt{10800. \sqrt{n^4+0.001 n^2}-10800 n^2}-\frac{2.56 n^2}{{\log N} \sqrt{n^4+0.001 n^2} \sqrt{10800. \sqrt{n^4+0.001
   n^2}-10800 n^2}} \nonumber\\
   & & + {\cal O}\left(\frac{1}{(\log N)^2}\right)\nonumber\\
   & & = \left\{1.36059-\frac{0.938489}{{\log N}},1.36077-\frac{0.93885}{{\log N}},1.3608-\frac{0.938917}{{\log N}},1.36081-\frac{0.938941}{{\log N}},...\right\}.
\end{eqnarray}

\subsubsection{UV}
Disregarding the $M$ and $N_f$ dependent terms we obtained
\begin{eqnarray}
\label{V-UV-alpha-0}
& & V(\alpha^{\left\{0\right\}};UV) = \frac{e^{-2 |Z|} \left(3. e^{2 |Z|}-1.62\right)}{{\log N}}+\frac{\tilde{m}^2 e^{2 |Z|}}{e^{4 |Z|}-1.}-\frac{1.08 \tilde{m}^2}{e^{4 |Z|}-1.}-1\nonumber\\
& & = -1 + e^{-2 |Z|} \left(\tilde{m}^2-\frac{1.62}{{\log N}}\right)+\frac{3}{{\log N}} +
{\cal O}(e^{-4|Z|}).
\end{eqnarray}
The solution to  Schr\"{o}dinger-like EOM:
\begin{equation}
\psi^{\left\{0\right\}}_n "(Z) + \left(B_{0} + B_{1} e^{-|Z|}\right)\psi^{\left\{0\right\}}_n (Z) = 0
\end{equation}
were obtained as:
\begin{equation}
\psi^{\left\{0\right\}}_n(Z) = c_1 J_{-i \sqrt{B_{0}}}\left(\sqrt{B_{1}} \sqrt{e^{-2 |Z|}}\right)+c_2 J_{i \sqrt{B_{0}}}\left(\sqrt{B_{1}} \sqrt{e^{-2 |Z|}}\right).
\end{equation}
In the UV $\psi^0_n(Z)$ satisfied the Neumann BC but failed to satisfy the Dirichlet BC:
\begin{equation}
\lim_{Z\rightarrow\infty}\frac{d}{dZ\ }\left(\frac{\psi^{\left\{0\right\}}_n(Z)}{\sqrt{{\cal C}_1(Z)}}\right)=0,
\end{equation}
 hence one does not obtain any quantization condition on the masses $\tilde{m}$.

\subsection{$\alpha^{\left\{i\right\}}_n(Z)$ Meson Spectroscopy from WKB Quantization}
The potential is same as given in (\ref{WKB-i-1}) obtained after converting  $\alpha^{\left\{i\right\}}_n(Z)$ into Schr\"{o}dinger-like EOM. After performing a large-$N$ expansion and retaining terms up to LO in $N$ of the potential, we again performed a large-$\log N$ expansion of $\sqrt{V(\alpha^{\left\{i\right\}}_n)}$ and retained terms up to NLO in $\log N$. Such that,
{\footnotesize
\begin{eqnarray}
\label{V_f_large_LogN}
& & \sqrt{V^{\alpha^{\left\{i\right\}}_n}(\tilde{m},N)} \nonumber\\
& & = \sqrt{\frac{e^{-2 |Z|} \left(e^{6 |Z|} \left(6.-1.08 \tilde{m}^2\right)+e^{4 |Z|} \left(2.16-
   \tilde{m}^2\right)+\tilde{m}^2 e^{8 |Z|}+\left(1.08 \tilde{m}^2-1.\right) e^{2 |Z|}-e^{10
   |Z|}-2.16\right)}{\left(e^{4 |Z|}-1.\right)^2}}\nonumber\\
   & & -\frac{0.75 \left(e^{4 |Z|}-1\right) \left(2. e^{2 |Z|}-1.08 e^{4 |Z|}+2. e^{6
   |Z|}-3.24 \right)}{{\log N} \left(-\tilde{m}^2 e^{8 |Z|}+\left(1-1.08
   \tilde{m}^2\right) e^{2 |Z|}+\left(\tilde{m}^2-2.16\right) e^{4 |Z|}+\left(1.08 \tilde{m}^2-6.\right) e^{6 |Z|}+e^{10
   |Z|}+2.16\right)}\nonumber\\
   & & \times\sqrt{\frac{e^{-2 |Z|} \left(e^{6 |Z|} \left(6.-1.08 \tilde{m}^2\right)+e^{4 |Z|} \left(2.16-1.
   \tilde{m}^2\right)+\tilde{m}^2 e^{8 |Z|}+\left(1.08 \tilde{m}^2-1.\right) e^{2 |Z|}-e^{10
   |Z|}-2.16\right)}{\left(e^{4 |Z|}-1\right)^2}} \nonumber\\
   & & + {\cal O}\left(\left(\frac{1}{{\log N}}\right)^2\right).
\end{eqnarray}}
\paragraph{(a) Large-$\tilde{m}$ expansion : In the UV sector, i.e., $r>0.6\sqrt{3}r_h$ or $|Z|>0.04$}
From (\ref{V_f_large_LogN}) one observes that for large $\tilde{m}$ in the UV, $\sqrt{V}\in\mathbb{R}$ :
\begin{eqnarray}
\label{range-Exp|Z|}
& & \hskip -0.7in \sqrt{0.5 \tilde{m}^2-0.1 \sqrt{25. \tilde{m}^4-108. \tilde{m}^2}}<e^{|Z|}<\sqrt{0.5 \tilde{m}^2+0.1 \sqrt{25.
   \tilde{m}^4-108. \tilde{m}^2}},
\end{eqnarray}
or
\begin{eqnarray}
\label{range-|Z|}
& & |Z|\in\left[\log\left(1.039 + \frac{0.561}{\tilde{m}^2} + {\cal O}\left(\frac{1}{\tilde{m}^3}\right)\right), \log\left(\tilde{m} - \frac{0.54}{\tilde{m}} + {\cal O}\left(\frac{1}{\tilde{m}^3}\right)\right)\right].
\end{eqnarray}
we performed three consecutive large expansions in the parameters: $N$, $\tilde{m}$ and $|Z|$, and obtained:
\begin{eqnarray}
\label{large-mtilde}
& & \sqrt{V^{\alpha^{\left\{i\right\}}_n}(\tilde{m},N)} = \left(e^{-|Z|} - 0.54 e^{-3|Z|}\right)\tilde{m} + \frac{1}{\tilde{m}}\left(-{0.5}e^{-|Z|} - {0.27}e^{-|Z|} + {2.03}e^{-3|Z|}\right) \nonumber\\
& &  + \frac{1}{\tilde{m}N}\left(1.5 e^{-|Z|} + 2.47 e^{-3|Z|}\right)  + {\cal O}\left(\frac{1}{\tilde{m}^2},\frac{1}{(\log N)^2},e^{-5|Z|}\right).
\end{eqnarray}
Finally, the WKB quantization condition:
\begin{eqnarray}
\label{WKB-alphai-1}
& & \int_{\log\left(1.039 + \frac{0.561}{\tilde{m}^2}\right)}^{\log\left(\tilde{m} - \frac{0.54}{\tilde{m}}\right)}\sqrt{V} = \left(n + \frac{1}{2}\right)\pi
\end{eqnarray}
up to ${\cal O}\left(\frac{1}{\log N}\right)$ gave:
\begin{table}[h]
\begin{center}
\begin{tabular}{|c|c|c|c|c|}\hline
& (Pseudo-)Vector Meson Name & $J^{PC}$ & $m_{n>0}$ & PDG Mass \cite{PDG} (MeV) \\
&&&(units of $\frac{r_h}{\sqrt{4\pi g_s N}}$) & \\ \hline
$B^{(1)}_{\mu}$& $\rho[770]$ & $1^{++}$ & 7.649 - $\frac{1.759}{\log N}$ &775.49 \\ \hline
$B^{(2)}_{\mu}$ &$a_1[1260]$ & $1^{--}$ & 11.60 - $\frac{1.792}{\log N}$ & 1230 \\ \hline
$B^{(3)}_{\mu}$ &$\rho[1450]$ &$1^{++}$ & 15.535 - $\frac{1.81}{\log N}$ & 1465\\  \hline
$B^{(4)}_{\mu}$ & $a_1[1640]$ & $1^{--}$ & 19.462 - $\frac{1.821}{\log N}$ & 1647 \\ \hline
\end{tabular}
\end{center}
\caption{(Pseudo-)Vector Meeson masses from WKB Quantization applied to $V(\alpha^{\left\{i\right\}}_n)$}
\end{table}
\paragraph{(b) Small-$\tilde{m}$ Expansion} We expanded  $\sqrt{V(\alpha^{\left\{i\right\}}_n)}$ and retained terms up to ${\cal O}\left(\frac{1}{\log N},\tilde{m}^4\right)$,  $\sqrt{V}\in\mathbb{R}$ for
$Z\in[0.01,0.47]$; given that $Z=0.0385$ corresponding to $r={\cal R}_{D5/\overline{D5}}$. Where ${\cal R}_{D5/\overline{D5}}$ denotes the separation between $D5-\overline{D5}$ branes. Fixing 0.01 to be the lower limit by hand we used the WKB quantization condition:
\begin{equation}
\label{WKB-quantization-alphain}
\int_{0.01}^{0.47}dZ\sqrt{V\left(\alpha^{\left\{i\right\}}_n;Z\right)} = \left(n + \frac{1}{2}\right)\pi,
\end{equation}
to obtain the following vector meson spectrum in IR:
{\scriptsize
\begin{eqnarray}
\label{mn_IR}
& & m_n (IR) = 0.5 \sqrt{\frac{3.036-0.1136 {\log N}}{0.068 {\log N}+56.946}+2. \sqrt{\frac{(0.854513 n-0.0765252) \log ^2(N)+(715.605 n-67.1225)
   {\log N}-134.138}{(0.068 {\log N}+56.946)^2}}}.\nonumber\\
& &
\end{eqnarray}}
We obtained a non-zero ground state which was $0.81$ for $N=6000$, and was close to the result obtained in (\ref{meson-spectroscopy-ii}), $0.694 - \frac{0.155}{\log N}$ - which for $N=6000$  yields $0.677$.

\subsection{$\alpha^{\left\{0\right\}}_n$ Spectroscopy  WKB Quantization}
Writing mass of the vector mesons as $m=\tilde{m}\frac{r_h}{\sqrt{4 \pi g_s N}s},\ a = r_h\left(0.6 + 4 \frac{g_sM^2}{N}\left(1 + \log r_h\right)\right)$, one can obtain the Schr\"{o}dinger-like potential for $\alpha^{\left\{0\right\}}_n(Z)$.
After performing a large-$N$ expansion for the potential and retaining terms upto LO in $N$ , followed by a large-$\log N$ expansion of the   square root of the Schr\"{o}dinger-like potential we obtained:
\begin{eqnarray}
\label{sqrtValpha0n}
& & \sqrt{V^{\alpha^{\left\{0\right\}}_n}(|Z|,N,\tilde{m})} = \sqrt{\frac{\tilde{m}^2 e^{2 |Z|}}{e^{4 |Z|}-1}-\frac{1.08 \tilde{m}^2}{e^{4 |Z|}-1.}+0. e^{-2 |Z|}-1.}+\frac{e^{-2 |Z|} \left(1.5 e^{2
   |Z|}-0.81\right)}{{\log N} \sqrt{\frac{\tilde{m}^2 e^{2 |Z|}}{e^{4 |Z|}-1.}-\frac{1.08 \tilde{m}^2}{e^{4 |Z|}-1}-1.}} \nonumber\\
   & & + {\cal O}\left(\left(\frac{1}{{\log N}}\right)^2\right).
\end{eqnarray}
\paragraph{(a)Large-$\tilde{m}$ Expansion}
 For the range :
 {\footnotesize
\begin{eqnarray}
\label{range_e^|Z|_alpha0}
& & \hskip -0.5in 0.5 \log \left(0.1 \left(5 \tilde{m}^2-\sqrt{25. \tilde{m}^4-108\tilde{m}^2+100}\right)\right)<|Z|<0.5 \log \left(0.1 \left(5
   \tilde{m}^2+\sqrt{25 \tilde{m}^4-108 \tilde{m}^2+100}\right)\right),\nonumber\\
   & &
\end{eqnarray}}
or
\begin{eqnarray}
\label{range_|Z|_alpha0}
& & |Z|\in\left[0.0385 + {\cal O}\left(\frac{1}{\tilde{m}^2}\right),\log \tilde{m} - \frac{0.54}{\tilde{m}^2} + {\cal O}\left(\frac{1}{\tilde{m}^3}\right)\right]\nonumber\\
& & \approx \left[0.0385,\log \tilde{m}\right],
\end{eqnarray}
$V(\alpha^{\left\{0\right\}}_n)\in\mathbb{R}$. For the implementation of WKB quantization condition the upper and lower bound of (\ref{range_|Z|_alpha0}) acted as the turning points.

We performed a large-$\tilde{m}$ expansion for (\ref{sqrtValpha0n}):
\begin{eqnarray}
\label{sqrtVlargemtilde}
& &  \sqrt{V^{\alpha^{\left\{0\right\}}_n}(|Z|,N,\tilde{m})} = \tilde{m}\left(e^{-|Z|} - 0.54 e^{-3|Z|}\right) + \frac{1}{\tilde{m}}\left(-\frac{e^{-|Z|}}{2} - 0.27 e^{-|Z|} + 0.03 e^{-3|Z|}\right)\nonumber\\
& &  + \frac{1}{\tilde{m}\log N}\left(1.5 e^{|Z|}- 0.531 e^{-3|Z|}\right)  + {\cal O}\left(\frac{1}{(\log N)^2},\frac{1}{\tilde{m}^2},e^{-5|Z|}\right).
\end{eqnarray}
The WKB quantization condition:
\begin{equation}
\label{WKB-alpha0_largem}
\int_{0.0385}^{\log\tilde{m}} dZ\sqrt{V^{\alpha^{\left\{0\right\}}_n}(Z)} = \left(n + \frac{1}{2}\right)\pi
\end{equation}
gave a cubic of the form: $a + b \tilde{m} + \frac{c}{\tilde{m}} + \frac{d}{\tilde{m}^2} = g$ where:
\begin{eqnarray}
\label{abcdg}
& & a = -1.5 + \frac{1.5}{\log N}\nonumber\\
& & b = 0.802\nonumber\\
& & c = 0.269 - \frac{1.717}{\log N}\nonumber\\
& & d = 0.27\nonumber\\
& & g = \left(n + \frac{1}{2}\right)\pi.
\end{eqnarray}
The only real root up to ${\cal O}\left(\frac{1}{\log N}\right)$ gave the following vector meson spectrum (disregarding $n=0$ as it did not satisfy the large-$\tilde{m}$ assumption):
\begin{table}[h]
\begin{center}
\begin{tabular}{|c|c|c|c|c|}\hline
& (Pseudo-)Vector Meson Name & $J^{PC}$ & $m_{n>0}$ & PDG Mass \cite{PDG} (MeV) \\
&&&(units of $\frac{r_h}{\sqrt{4\pi g_s N}}$) & \\ \hline
$B^{(1)}_{\mu}$& $\rho[770]$ & $1^{++}$ & 7.698 - $\frac{1.604}{\log N}$ &775.49 \\ \hline
$B^{(2)}_{\mu}$ &$a_1[1260]$ & $1^{--}$ & 11.634 - $\frac{1.692}{\log N}$ & 1230 \\ \hline
$B^{(3)}_{\mu}$ &$\rho[1450]$ &$1^{++}$ & 15.56 - $\frac{1.736}{\log N}$ & 1465\\  \hline
$B^{(4)}_{\mu}$ & $a_1[1640]$ & $1^{--}$ & 19.483 - $\frac{1.762}{\log N}$ & 1647 \\ \hline
\end{tabular}
\end{center}
\caption{(Pseudo-)Vector Meson masses from WKB Quantization applied to $V(\alpha^{\left\{0\right\}}_n)$}
\end{table}
One hence notes a near isospectrality between Tables 3.1 and 3.2 results for (pseudo-)vector masses.

\paragraph{(b)Small-$\tilde{m}$ Expansion}
The WKB quantization failed to give a result because of the unavailability of turning points in the large-$N$ limit.
\section{Scalar Meson Spectroscopy using a Black-Hole Background for All Temperatures}
The spectrum for scalar mesons was obtained by turning off any $D6$-brane world-volume fluxes as in \cite{Dasgupta_et_al_Mesons} and then considering  fluctuation along the direction Y of the world volume $D6$-brane.
For $Y\neq0$ the metric of $D6$-brane (\ref{pull-back}), using(\ref{YZ}) and the embedding $ Y = Y(x^\mu,Z)$ was given as:
\begin{eqnarray}
\label{metric_IIA_pull_back}
 g^{\rm IIA}_{6\mu \nu}dx^{\mu}dx^{\nu} &=& g^{\rm IIA}_{\mu\nu}\left(1 + {\cal C}_1(x^\kappa,Z) g^{{\rm IIA} \rho\lambda}\partial_\rho{Y} \partial_\lambda{Y} \right)dx^\mu dx^\nu + {\cal C}_2(x^\kappa,Z,\dot{Y}) dZ^2 + {\cal C}_3(x^\kappa,Z,\dot{Y})dx^\mu dZ \partial_\mu Y
\nonumber\\
 & & g^{\rm IIA}_{\theta_{2}\theta_{2}}d\theta_{2}^{2}+g^{\rm IIA}_{\tilde{y}\tilde{y}}d\tilde{y}^{2},
\end{eqnarray}
where:
\begin{eqnarray}
\label{Cs}
& & {\cal C}_1(x^\kappa,Z) = {\cal A} Y^2 + {\cal B} Z^2,\nonumber\\
& & {\cal C}_2(x^\kappa,Z,\dot{Y}) = \left({\cal A}Y^2 + {\cal B}Z^2\right)\dot{Y}^2
 + \left({\cal A}Y^2 + {\cal B}Z^2\right) + 2 Y Z\left({\cal A} - {\cal B}\right)\dot{Y},\nonumber\\
 & & {\cal C}_3(x^\kappa,Z,\dot{Y}) = 2 \left({\cal A}Y^2 + {\cal B}Z^2\right)\dot{Y} + 2 Y Z \left({\cal A} - {\cal B}\right),
\end{eqnarray}
wherein:
\begin{eqnarray}
\label{AB}
& & {\cal A} = \frac{g^{\rm IIA}_{rr}r_h^2 e^{2\sqrt{Y^2 + Z^2}}}{\left(Y^2 + Z^2\right)},\nonumber\\
& & {\cal B} = \frac{g^{\rm IIA}_{\tilde{z}\tilde{z}}}{\left(Y^2 + Z^2\right)^2}.
\end{eqnarray}
  $B^{\rm IIA}_{NS-NS}$\cite{MQGP} in diagonal basis $(\theta_2,\tilde{x},\tilde{y},\tilde{z})$ was given by:
\begin{eqnarray}
\label{B-IIA}
B^{IIA} &=& B_{\theta_2\tilde{y}}d\theta_2\wedge d\tilde{y} + B_{\theta_2\tilde{z}}d\theta_2\wedge d\tilde{z}+ B_{\theta_2\tilde{x}}d\theta_{2}\wedge d\tilde{x}.
\end{eqnarray}
Thus, its pull-back on $D6$-brane was given by:
\begin{eqnarray}
\label{i*B}
i^* B^{\rm IIA} &=& B_{\theta_2\tilde{y}}d\theta_2\wedge d\tilde{y} + {\cal C}_4(x^\kappa,Z,\dot{Y})dZ\wedge d\theta_2 + {\cal C}_5(x^\kappa,Z)\partial_\mu Y dx^\mu\wedge d\theta_2
\end{eqnarray}
where:
\begin{eqnarray}
\label{BIIA-Cs}
& &   {\cal C}_4(x^\kappa,Z,\dot{Y}) = \left(\frac{B^{\rm IIA}_{\theta_2\tilde{z}}}{Y^2 + Z^2}\right)
\left(\dot{Y} Z - Y\right),\nonumber\\
& & {\cal C}_5(x^\kappa,Z) = \left(\frac{B^{\rm IIA}_{\theta_2\tilde{z}}}{Y^2 + Z^2}\right)Z.
\end{eqnarray}
One can write:
\begin{equation}
\label{i*(g+B)_7x7-i}
i^*(g+ B)^{\rm IIA} = \left(\begin{array}{cc} \mathbb{A}_{4\times4} & \mathbb{B}_{4\times3}\\
\mathbb{C}_{3\times4} & \mathbb{D}_{3\times3}\end{array} \right),
\end{equation}
where:
\begin{eqnarray}
& & \mathbb{A} = \left(\begin{array}{cccc} g^{\rm IIA}_{00}{\cal T} & 0 & 0 & 0 \\
0 & g^{\rm IIA}_{x^1x^1}{\cal T} & 0 & 0 \\
0 & 0 & g^{\rm IIA}_{x^2x^2}{\cal T}& 0 \\
0 & 0 & 0 & g^{\rm IIA}_{x^3x^3}{\cal T}\end{array}\right),\nonumber\\
& & {\cal T}=\left(1 + {\cal C}_1 g^{{\rm IIA} \rho\lambda}\partial_\rho{Y} \partial_\lambda{Y} \right),\nonumber\\
& & \mathbb{B}_{4\times3} = \left(\begin{array}{ccc} i^* g^{\rm IIA}_{x^0 Z} & i^* B^{\rm IIA}_{x^0\theta_2} &
0 \\
 i^* g^{\rm IIA}_{x^1 Z} & i^* B^{\rm IIA}_{x^1\theta_2} &
0 \\
 i^* g^{\rm IIA}_{x^2 Z} & i^* B^{\rm IIA}_{x^2\theta_2} &
0 \\
 i^* g^{\rm IIA}_{x^3 Z} & i^* B^{\rm IIA}_{x^3\theta_2} &
0
\end{array}\right),\nonumber\\
& & \mathbb{D}_{3\times3} = \left(\begin{array}{ccc} i^* g^{\rm IIA}_{Z Z} & i^* B^{\rm IIA}_{Z\theta_2} & 0
\nonumber\\
- i^* B^{\rm IIA}_{Z\theta_2}& i^* g^{\rm IIA}_{\theta_2 \theta_2} & i^* B^{\rm IIA}_{\theta_2\tilde{y}}\\
0 & -i^* B^{\rm IIA}_{\theta_2\tilde{y}} & i^* g^{\rm IIA}_{\tilde y \tilde y}
\end{array}\right).
\end{eqnarray}
Now, using the relation ${\rm det}\left(i^*(g+ B)^{\rm IIA} \right) = {\rm det}\mathbb{A}\ {\rm det}\left(\mathbb{D} - \mathbb{C}\mathbb{A}^{-1}\mathbb{B}\right)$, and keeping terms in the relation only up to ${\cal O}\left(Y^2,\dot{Y}^2,\partial_\mu Y\partial_\nu Y\right)$ , we obtained:
\begin{eqnarray}
\label{DBI-quad-Ys}
& & \sqrt{{\rm det}\mathbb{A}} \sim \sqrt{-g^{\rm IIA}}_{\mathbb{R}^{1,3}}\left(1 + \frac{{\cal C}_1(Y=0)}{2}g^{\rm IIA \mu\nu}\partial_\mu Y\partial_\nu Y\right),\nonumber\\
& & {\rm det}\left(\mathbb{D} - \mathbb{C}\mathbb{A}^{-1}\mathbb{B}\right) \sim \dot{Y}^2\Omega_1 + Y^2\Omega_2 + \Omega_3 g^{\rm IIA \mu\nu}\partial_\mu Y\partial_\nu Y + \Omega_4,
\nonumber\\
& & {\rm implying}:\nonumber\\
& & \sqrt{{\rm det}\left(\mathbb{D} - \mathbb{C}\mathbb{A}^{-1}\mathbb{B}\right)}\sim
\sqrt{\Omega_4}\left(1 + \frac{\Omega_1}{\Omega_4}\frac{\dot{Y}^2}{2} + \frac{\Omega_2}{\Omega_4}\frac{\dot{Y}^2}{2} + \frac{\Omega_3}{\Omega_4}\frac{g^{\rm IIA \mu\nu}}{2}\partial_\mu Y\partial_\nu Y\right);\ \cdot\equiv\frac{d}{dZ}.
\end{eqnarray}
Thus the $N_f$ $D6$-branes DBI action (setting the tension to unity) is:
{\scriptsize
\begin{eqnarray}
\label{DBI-scalar}
& &S_{D6}\nonumber\\
& &= N_f\left.\int d^4x dZ d\theta_2 d\tilde{y} \delta\left(\theta_2 - \frac{\alpha_{\theta_2}}{N^{\frac{3}{10}}}\right)e^{-\phi^{\rm IIA}}\sqrt{\Omega_4}\sqrt{-g^{\rm IIA}}_{\mathbb{R}^{1,3}}\left[1
+ \frac{{\cal C}_1  + \frac{\Omega_3}{\Omega_4}}{2}g^{\rm IIA \mu\nu}\partial_\mu Y\partial_\nu Y  + \frac{\Omega_1}{\Omega_4}\frac{\dot{Y}^2}{2} + \frac{\Omega_2}{\Omega_4}\frac{\dot{Y}^2}{2}\right]\right|_{\theta_1=\frac{\alpha_{\theta_1}}{N^{\frac{1}{5}}},\tilde{x}=0} \nonumber\\
& &= N_f\left.\int d^4x dZ d\theta_2 d\tilde{y}\delta\left(\theta_2 - \frac{\alpha_{\theta_2}}{N^{\frac{3}{10}}}\right)\left[{\cal S}_1(Z)g^{\rm IIA \mu\nu}\partial_\mu Y\partial_\nu Y  + {\cal S}_2(Z)\dot{Y}^2 + {\cal S}_3(Z)Y^2 \right]\right|_{\theta_1=\frac{\alpha_{\theta_1}}{N^{\frac{1}{5}}},\tilde{x}=0},
\end{eqnarray}}
expressions for ${\cal S}_{1,2,3}$ are given in (\ref{S123}).
Taking a Klauza-klien ansatz for field $Y(x^\mu,Z)$ similar to \cite{Dasgupta_et_al_Mesons}:
\begin{equation}
\label{KK-Y}
Y(x^\mu,Z) = \sum_{n=1} {\cal Y}^{(n)}(x^\mu){\cal Z}_n(Z),
\end{equation}
  together with  normalization condition and EOM:
\begin{eqnarray}
\label{mass-term-identification+EOM}
& & \int dZ\left({\cal S}_2(Z)\dot{\cal Z}_m(Z)\dot{\cal Z}_n(Z) + {\cal S}_3(Z){\cal Z}_m(Z){\cal Z}_n(Z)\right) = \frac{m_n^2}{2}\delta_{mn},\nonumber\\
& & \int dZ {\cal S}_1(Z) {\cal Z}_m(Z) {\cal Z}_n(Z) = \frac{1}{2}\delta_{mn};\nonumber\\
& & -\partial_Z\left({\cal S}_2(Z)\partial_Z{\cal Z}_n\right) + {\cal S}_3(Z){\cal Z}_n(Z)
 = {\cal S}_1(Z)m_n^2{\cal Z}_n.
\end{eqnarray}
The EOM following a field redefinition: ${\cal Z}_n(Z) = |Z| {\cal G}_n(Z)$, in terms of ${\cal G}(Z)$ was obtained  as:
\begin{eqnarray}
\label{G_EOM}
& &{\cal G}_n''(Z) + N_0\ {\cal G}_n'(Z)+ N_1\ {\cal G}_n(Z) = 0.\nonumber\\
   & &
\end{eqnarray}

Expressions for $N_0$ and $N_1$ are given in (\ref{Scalar_EOM_Coefficients}). Analogous to the three different routes used to obtain the (pseudo-)vector meson spectrum, (pseudo-)scalar meson spectrum was also obtained via same routes.

\subsection{Scalar Meson Spectrum from Solution to EOM near ${r = r_h}$}
Similar to (\ref{alpha_n^i_EOM-i}) - (\ref{alpha_n^i_EOM-iv}), one can rewrite and solve the (\ref{G_EOM}) near ${r = r_h}$, impose Neumann BC at ${r = r_h}$ with the following identifications:
\begin{eqnarray}
\label{alphas+betas}
& & \alpha_1 = 0.92+\frac{0.24}{{\log N}},\nonumber\\
& & \alpha_2 = \frac{0.02 \alpha_{\theta_2}^2 \tilde{m}^2}{\alpha_{\theta_1}^22 \sqrt[5]{N}}+\frac{{g_s} M^2 \tilde{m}^2 (-3.6 \log ({r_h})-3.6)}{N}-0.02
   \tilde{m}^2,\nonumber\\
& & \beta_2 = -\frac{0.54 \alpha_{\theta_2}^2 \tilde{m}^2}{\alpha_{\theta_1}^22 \sqrt[5]{N}}+\frac{{g_s} M^2 \tilde{m}^2 (7.2 \log ({r_h})+7.2)}{N}+0.54
   \tilde{m}^2.
\end{eqnarray}
For scalar mesons the analog of (\ref{alpha_n^i_EOM-iv}) up to ${\cal O}\left(\frac{1}{\log N}\right)$, gives:
\begin{eqnarray}
\label{mn-Dirichlet}
& & \tilde{m}_n = 0.5 \sqrt{0.548697 n^2+0.548697 \sqrt{\left(n^2+n+0.25\right) \left(n^2+n+166.926\right)}+0.548697 n+3.54458}\nonumber\\
& & +\frac{0.25 \left(\frac{22.9689
   (n+0.5)^2}{\sqrt{\left(n^2+n+0.25\right) \left(n^2+n+166.926\right)}}+0.888889\right)}{{\log N} \sqrt{0.548697 n^2+0.548697
   \sqrt{\left(n^2+n+0.25\right) \left(n^2+n+166.926\right)}+0.548697 n+3.54458}} \nonumber \\ & & + {\cal O}\left(\frac{1}{(\log N)^2}\right).
\end{eqnarray}
The masses of lightest scalar mesons were obtained as:
\begin{table}[h]
\begin{center}
\begin{tabular}{|c|c|}\hline
$m_{n=1}$ & 1.331 - $\frac{0.167}{\log N}$\\ \hline
$m_{n=2}$ & 1.958 - $\frac{0.226}{\log N}$ \\ \hline
\end{tabular}
\end{center}
\caption{The lightest Sector Meson masses}
\end{table}

We get ratio of masses as $\frac{m_{n=1}^2}{m_{n=0}^2} = 2.16$ after disregarding the ${\cal O}\left(\frac{1}{N}\right)$ corrections.
Looking at the PDG table if one assumes $f0[980]/a0[980], f0[1370]$ as lightest scalar mesons then the ratio of square of their masses is 1.95 - not too far from our result.

\subsection{Scalar Mass Spectrum from Solution of the Schr\"{o}dinger-Like Equation}
\subsubsection{IR}
Potential for Schr\"{o}dinger-like EOM in the IR , simplifies to:
\begin{eqnarray}
\label{V_IR_scalar}
& & V(IR) = \frac{\frac{-0.36 {g_s}^2 {N_f}^2 \log ({r_h})+2.43 {g_s}^2 {N_f}^2+1.50796 {g_s} {N_f}}{{g_s}^2 {\log N}^2
   {N_f}^2}-\frac{0.12}{{\log N}}-0.02 \tilde{m}^2-0.46}{|Z|}\nonumber\\
   & & +\frac{13.86 {g_s}^4 {N_f}^4 \log ({r_h})+{g_s}^2
   {N_f}^2 \left(-2.97 {g_s}^2 {N_f}^2-58.0566 {g_s} {N_f}\right)}{{g_s}^4 {\log N}^2 {N_f}^4}+0.54
   \tilde{m}^2+\frac{0.25}{Z^2}-3.413.\nonumber\\
& & \end{eqnarray}
Following a field redefinition $\Phi_n(Z) = \sqrt{{\cal C}_1}{\cal G}_n(Z)$ the Schr\"{o}dinger-like EOM: \\
$\Phi^{\prime\prime}_n(Z) + V(IR)(Z)\Phi_n(Z) = 0$, with potential  $V(IR)(Z) = \frac{c_1}{Z^2}+\frac{a_1}{|Z|}+b_1$ where:
\begin{eqnarray}
\label{c1a1b1}
& & c_1 = 0.25,\nonumber\\,
& & a_1 = \frac{-0.36 {g_s}^2 {N_f}^2 \log ({r_h})+2.43 {g_s}^2 {N_f}^2+1.50796 {g_s} {N_f}}{{g_s}^2 {N_f}^2 \log
   ^2(N)}-0.02 \tilde{m}^2-\frac{0.12}{\log (N)}-0.46,\nonumber\\
   & & b_1 = 0.54 \tilde{m}^2 - 3.413,
\end{eqnarray}
was solved with solution given as:
\begin{eqnarray}
\label{solution_Sch_IR}
& & \Phi_n(Z) = c_1 M_{\frac{{g_s} \left(\left(-0.01 \tilde{m}^2(n)-0.23\right) {\log N}^2-0.06 {\log N}+1.215\right) {N_f}-0.18 {g_s}
   \log ({r_h}) {N_f}+0.753982}{{g_s} {\log N}^2 \sqrt{3.41333-0.54 \tilde{m}^2(n)} {N_f}},0}\left(2 \sqrt{3.41333-0.54
   \tilde{m}^2(n)} |Z|\right)\nonumber\\
   & & +c_2 W_{\frac{{g_s} \left(\left(-0.01 \tilde{m}^2(n)-0.23\right) {\log N}^2-0.06 {\log N}+1.215\right)
   {N_f}-0.18 {g_s} \log ({r_h}) {N_f}+0.753982}{{g_s} {\log N}^2 \sqrt{3.41333-0.54 \tilde{m}^2(n)}
   {N_f}},0}\left(2 \sqrt{3.41333-0.54 \tilde{m}^2(n)} |Z|\right).
\end{eqnarray}
Now,
\begin{eqnarray}
\label{C1_scalar}
& & {\cal C}_1(Z) = -\frac{1}{2 {r_h}^2}\Biggl\{e^{-2 |Z|} \left(e^{4 |Z|}-1\right)\nonumber\\
& & \times \Biggl(2 {g_s} {N_f} \log (N) \left(3 {r_h}^2 \left(\frac{4 {g_s} M^2 \log
   ({r_h})}{N}+\frac{4 {g_s} M^2}{N}+0.6\right)^2+2 {r_h}^2 e^{2 |Z|}\right)\nonumber\\
   & & -6 {g_s} {N_f} (\log ({r_h})+|Z|) \left(3
   {r_h}^2 \left(\frac{4 {g_s} M^2 \log ({r_h})}{N}+\frac{4 {g_s} M^2}{N}+0.6\right)^2+2 {r_h}^2 e^{2 |Z|}\right)\nonumber\\
   & & +2
   \left({r_h}^2 (12 \pi -9 {g_s} {N_f}) \left(\frac{4 {g_s} M^2 \log ({r_h})}{N}+\frac{4 {g_s} M^2}{N}+0.6\right)^2+8
   \pi  {r_h}^2 e^{2 |Z|}\right)\Biggr)\Biggr\}.
\end{eqnarray}
Imposing Neumann boundary condition on ${\cal G}_n(Z) =  \frac{\Phi_n(Z)}{\sqrt{{\cal C}_1(Z)}}$. One sees that:
\begin{eqnarray}
\label{DMoversqrtC1}
& & \left.\frac{d}{dZ}\left(\frac{M_{\frac{{g_s} \left(\left(-0.01 \tilde{m}^2-0.23\right) {\log N}^2-0.06 {\log N}+1.215\right) {N_f}-0.18 {g_s}
   \log ({r_h}) {N_f}+0.753982}{{g_s} {\log N}^2 \sqrt{3.41333-0.54 \tilde{m}^2} {N_f}},0}\left(2 \sqrt{3.41333-0.54
   \tilde{m}^2} |Z|\right)}{\sqrt{{\cal C}_1(Z)}}\right)\right|_{|Z|\sim0}\nonumber\\
   & & = \frac{1}{N^2}\left(0.54 \tilde{m}^2-3.41333\right)^{\frac{1}{4}}\left(\omega_1(g_s, M, N_f)+ \omega_2(g_s, M, N_f; \log r_h)\tilde{m}^2\right) + {\cal O}(Z).
\end{eqnarray}
Hence, either:
\begin{equation}
\label{mtilde-i-Sch-scalar-IR}
0.54 \tilde{m}^2-3.41333 = 0,\ {\rm implying}\  \tilde{m} = 2.514,
\end{equation}
or
\begin{eqnarray}
\label{mtilde-ii-Sch-scalar-IR}
& & \omega_1(g_s, M, N_f)+ \omega_2(g_s, M, N_f; \log r_h)\tilde{m}^2 = 0,\ {\rm implying}\nonumber\\
& & \tilde{m} = 3.07694 +\frac{\frac{95.6605}{{g_s} {N_f}}-22.8373 \log ({r_h})+3.34479}{\log ^2(N)}-\frac{7.61242}{\log (N)}+ {\cal O}\left(\frac{1}{(\log N)^2}\right).
\end{eqnarray}

One can show that:
\begin{eqnarray}
\label{DWoversqrtC1}
& & \left.\frac{W_{-\frac{i {a_1}}{2 \sqrt{{b_1}}},0}\left(2 i \sqrt{{b_1}} |Z|\right)}{\sqrt{{\cal C}_1(Z)}}\right|_{|Z|\sim0}
\nonumber\\
& & = \frac{\frac{2 i \sqrt{{b_1}} \left(\psi ^{(0)}\left(\frac{i
   {a_1}}{2 \sqrt{{b_1}}}-\frac{1}{2}\right)+\log \left(2 i
   \sqrt{{b_1}}\right)+\log |Z|+2 \gamma \right)}{\Gamma
   \left(\frac{i {a_1}}{2
   \sqrt{{b_1}}}-\frac{1}{2}\right)}+\frac{{a_1} \left(\psi
   ^{(0)}\left(\frac{i {a_1}}{2
   \sqrt{{b_1}}}+\frac{1}{2}\right)+\log \left(2 i
   \sqrt{{b_1}}\right)+\log |Z|+2 \gamma \right)}{\Gamma
   \left(\frac{i {a_1}}{2
   \sqrt{{b_1}}}+\frac{1}{2}\right)}}{\sqrt{2} \sqrt{i
   \sqrt{{b_1}}} \sqrt{{\cal C}_1(0)} \sqrt{|Z|}}\nonumber\\
   & & +\frac{\sqrt{|Z|}
   \left(\frac{2 {b_1} {C1}'(0) \log |Z|}{\Gamma \left(\frac{i
   {a_1}}{2 \sqrt{{b_1}}}-\frac{1}{2}\right)}-\frac{2 {b_1}
   {C1}'(0) \log |Z|}{\Gamma \left(\frac{i {a_1}}{2
   \sqrt{{b_1}}}+\frac{1}{2}\right)}-\frac{i {a_1}
   \sqrt{{b_1}} {C1}'(0) \log |Z|}{\Gamma \left(\frac{i
   {a_1}}{2 \sqrt{{b_1}}}+\frac{1}{2}\right)}\right)}{2 \sqrt{2}
   \left(i \sqrt{{b_1}}\right)^{3/2}
   {\cal C}_1(0)^{3/2}}+O\left(Z^{3/2}\right).
\end{eqnarray}
Hence Neumann BC at ${r = r_h}$ can be satisfied  if:
\begin{equation}
\label{Neumann_scalar_W_IR}
\frac{1}{2} -\frac{i {a_1}}{2 \sqrt{{b_1}}} = - n\in\mathbb{Z}^-\cup\left\{0\right\},
\end{equation}
which gave a $\tilde{m}$ quantization condition:
{\scriptsize
\begin{eqnarray}
\label{mtilde_W}
& &  m_n = 0.5 \sqrt{-10800. n^2+10800. \sqrt{\left(n^2+n+0.25\right) \left(n^2+n+0.271719\right)}-10800. n-2792}\nonumber\\
& &  + \frac{6. n^2-6. \sqrt{\left(n^2+n+0.25\right) \left(n^2+n+0.271719\right)}+6. n+1.5}{{\log N} \sqrt{\left(n^2+ n+0.25\right)
   \left(n^2+n+0.271719\right)} \sqrt{-10800. n^2+10800. \sqrt{\left(n^2+n+0.25\right) \left(n^2+n+0.271719\right)}-10800.
   n-2792.}}\nonumber\\
   & & + {\cal O}\left(\frac{1}{(\log N)^2}\right).
\end{eqnarray}}
mass value for $n=0$ mode $2.39 - \frac{0.51}{\log N}$ is close to $2.51$ of (\ref{mtilde-i-Sch-scalar-IR}), and not very far from
$3.08 - \frac{7.61}{\log N}$ of (\ref{mtilde-ii-Sch-scalar-IR}).

\subsubsection{UV}

The potential after dropping the terms involving $M$ and $N_f$ in UV, was obtained as:
{\begin{eqnarray}
\label{V_scalar_UV}
& &  V(UV) = \frac{e^{-2 |Z|} \left(e^{6 |Z|} \left(8-1.08 \tilde{m}^2\right)+e^{4 |Z|} \left(- \tilde{m}^2\right)+1.08 \tilde{m}^2 e^{2
   |Z|}+\left(\tilde{m}^2+1.08\right) e^{8 |Z|}-4 e^{10 |Z|}-1.08\right)}{\left(e^{4 |Z|}-1\right)^2}
   \nonumber\\
   & &  +\frac{\alpha_{\theta_2}^2 \tilde{m}^2
   \left(e^{2 |Z|}+1.08 e^{4 |Z|}- e^{6 |Z|}-1.08\right)}{\alpha_{\theta_1}^2 \sqrt[5]{N} \left(e^{4 |Z|}-1\right)^2}.
\end{eqnarray}}
 Schr\"{o}dinger-like EOM $\Phi^{\prime\prime}(Z) + V(Z)\Phi(Z) = 0$ solution was obtained as:
\begin{eqnarray}
\label{solution_UV_scalar}
& & \Phi(Z) = c_1 J_{-i \sqrt{{x_2}}}\left(e^{|Z|} \sqrt{{x_1}}\right)+c_2 J_{i \sqrt{{x_2}}}\left(e^{|Z|} \sqrt{{x_1}}\right),
\end{eqnarray}
where:
\begin{eqnarray}
\label{x1-x2}
& & x_1 = 1.08 + \tilde{m}^2,\nonumber\\
& & x_2 = 8 - 1.08 \tilde{m}^2.
\end{eqnarray}
In the UV:
\begin{eqnarray}
\label{Neumann_UV_scalar}
& & \frac{d}{dZ\ }\left(\frac{J_{\pm i \sqrt{{x_2}}}\left(e^{|Z|}\sqrt{{x_1}}\right)}{\sqrt{{\cal C}_1(Z)}}\right)\sim \cos\left(e^{|Z|}\sqrt{x_1}\right)\times {\cal O}\left(e^{-\frac{3|Z|}{2}}\right),
\end{eqnarray}
one fails to obtain a mass quantization condition because Neumann BC is identically satisfied.

\subsection{Scalar Mass Spectrum via WKB Quantization Condition}
WKB quantization condition integral fails to converge for
IR-valued $\tilde{m}$. In IR $N_f$ and $M$ are non zero while quite small in UV. Thus, ignoring $N_f$ and $M$ in the large-$\tilde{m}$/UV limit from (\ref{V_scalar_UV}), one obtains:
{\begin{eqnarray}
\label{Sqrt[V]_scalar}
 & & \sqrt{V}  =  \tilde{m} \sqrt{\frac{1.08-e^{2 |Z|}-1.08 e^{4 |Z|}+e^{6 |Z|}}{1.-2e^{4 |Z|}+e^{8 |Z|}}}\nonumber\\
 & & +\frac{0.5 \left(8e^{6 |Z|}+1.08 e^{8 |Z|}-4e^{10
   |Z|}-1.08\right) \sqrt{\frac{1.08-e^{2 |Z|}-1.08 e^{4 |Z|}+e^{6 |Z|}}{1.-2e^{4 |Z|}+e^{8 |Z|}}}}{\tilde{m} \left(1.08 e^{2 |Z|}-e^{4
   |Z|}-1.08 e^{6 |Z|}+e^{8 |Z|}\right)}\nonumber\\
   & & + \frac{0.5 \alpha_{\theta_2}^2 \tilde{m} \left(e^{2 |Z|}+1.08 e^{4 |Z|}-e^{6 |Z|}-1.08\right)}{\alpha_{\theta_1}^2 \sqrt[5]{N} \left(e^{4 |Z|}-1.\right)^2
   \sqrt{\frac{1.08-e^{2 |Z|}-1.08 e^{4 |Z|}+e^{6 |Z|}}{1.-2e^{4 |Z|}+e^{8 |Z|}}}} + {\cal O}\left(\frac{1}{\tilde{m}N^{\frac{1}{5}}},\frac{1}{\tilde{m}^2}\right).
\end{eqnarray}}
 For $|Z|\in\left[0.0385, 0.5\log\left(0.54 + \frac{1.013\times 10^9}{\tilde{m}^2}\right)\right]$ the square root of the potential $\sqrt{V}\in\mathbb{R}$. To apply WKB method we approximated the interval for $|Z|$ by $|Z|\in\left[0.0385, 0.5\log(1.013\times 10^9)\approx10.368\right]$:
\begin{equation}
\label{WKB-scalar}
\int_{0.0385}^{10.368}d Z\sqrt{V(Z)} = \left(n + \frac{1}{2}\right)\pi,
\end{equation}
which gave scalar meson spectrum:
\begin{eqnarray}
\label{scalar-meson-WKB-i}
& & \hskip -0.3in \tilde{m}_n = \frac{\alpha_{\theta_1}^2 (157.08 n+78.5398)+0.5 \sqrt{\alpha_{\theta_1}^4 \left(98696n^2+98696 n+2.08747\times 10^9\right)-1.04372\times 10^9 \alpha_{\theta_1}^2
   \alpha_{\theta_2}^2 \sqrt[5]{N}}}{82\alpha_{\theta_1}^2-41\alpha_{\theta_2}^2 \sqrt[5]{N}}.
   \nonumber\\
   & &
\end{eqnarray}
One can argue that $Y(x^{\mu},Z)$ is even under parity: $(x^{1,2,3},Z)\rightarrow (-x^{1,2,3},-Z)$. The idea is the following. The type IIB setup of \cite{metrics} includes
$D3$-branes with world-volume coordinates $x^{0,1,2,3}$ and $D7$-branes with world-volume coordinates $(x^{0,1,2,3},r,\tilde{x},\theta_1,\tilde{z})$\footnote{There are also $D5$-branes with world volume coordinates $(x^{0,1,2,3},\theta_1,\tilde{x})$ and $\overline{D5}$-branes with
world volume coordinates $(x^{0,1,2,3},\theta_1,\tilde{x})$ which, relative to the $D5$-branes are at the antipodal point of the resolved $S^2_a(\theta_2,\phi_2)$; their bound state however is equivalent to producing a net $D3$-brane charge provided a certain topological condition is satisfied (See \cite{EPJC-2} and references therein).}, which after three T-dualities along $\tilde{x},\tilde{y},\tilde{z}$ yield two sets of $D6$-branes, one set with world-volume coordinates $(x^{0,1,2,3},\tilde{x},\tilde{y},\tilde{z})$ (obtained from a triple T-dual of the $D3$-branes) and the other set with world-volume coordinates $(x^{0,1,2,3},r,\theta_2,\tilde{y})$ (obtained from a triple T-dual of the $D7$-branes). One hence sees that the two sets of $D6$-branes are separated in $r$
or $Y$. In the type IIB setup of \cite{metrics}, the flavor $D7$-branes never touch the $D3$-branes which in the SYZ or triple T-dual picture implies that the two sets of $D6$-branes never touch each other. This, like \cite{Sakai-Sugimoto-1}, \cite{Dasgupta_et_al_Mesons}, implies one can construct a $C_5\sim Y dx^0\wedge dx^1\wedge dx^2\wedge dx^3\wedge dZ$  which vanishes precisely when the two sets of $D6$-branes touch. From this $C_5$, one can construct a Chern-Simons action: $\int_{{w.v. of }D6}F_2\wedge C_5$ where $F_2=dA_1$ correponds to a gauge field on the $D6$-brane world volume. If one demands the Cherns-Simons action be invariant under parity - which includes $Z\rightarrow-Z$ - given that $F_2$ is even, one sees that $Y$ is even under parity. Similarly, under charge conjugation - which includes $Z\rightarrow - Z$ - and noting that $F_2$ is charge-conjugation odd implies that $Y$ is charge-conjugation even. From (\ref{KK-Y}), under 5D parity, ${\cal Z}_n(-Z) = (-)^{n+1}{\cal Z}_n(Z),
{\cal Y}^{(n)}(-x^{\mu}) = (-)^{n+1}{\cal Y}^{(n)}(x^{\mu}),  n\in\mathbb{Z}^+$ \cite{Sakai-Sugimoto-1}.\\
We assume that the three lightest scalar mesons from the PDG  are $f0[980]/a0[980], f0[1370]$ and $f0[1450]$. We could choose $\alpha_{\theta_1}$ and $\alpha_{\theta_2}$ to match $\frac{m_{n=3}}{m_{n=1}}$ with PDG exactly (this is not normalizing our $\frac{m_{n=3}}{m_{n=1}}$ result to match PDG values)! This is effected by imposing the following condition on
$\alpha_{\theta_1}, \alpha_{\theta_2}$:
\begin{eqnarray}
\label{match_Dasgupta_et_al}
\frac{m_{n=3}}{m_{n=1}} = \frac{549.7787 \alpha_{\theta_1}^2+0.5 \sqrt{2.08767\times 10^9 \alpha_{\theta_1}^4-1.04372\times 10^9 \alpha_{\theta_1}^2 \alpha_{\theta_2}^2 \sqrt[5]{N}}}{235.6194 \alpha
   ^2+0.5 \sqrt{2.08747\times 10^9 \alpha_{\theta_1}^4-1.04372\times 10^9 \alpha_{\theta_1}^2 \alpha_{\theta_2}^2 \sqrt[5]{N}}} = \frac{1350}{980},
\end{eqnarray}
which is satisfied by $\alpha_{\theta_1} = 0.70765N^{\frac{1}{10}}\alpha_{\theta_2}$. - From Table 3.8 in Section {\bf 3.6}, one sees that the ratio $\frac{m_{n=5}}{m_{n=1}}$ is very close to the PDG value . Table 3.4 below summarizes the mass of ($0^{--}$)$0^{++}$ (pseudo-)scalar mesons. (The entries for $0^{--}$ are unfilled , as there are no known candidates for this particular $J, P, C$ assignment at the moment.)
\begin{table}[h]
\begin{center}
\begin{tabular}{|c|c|c|c|c|}\hline
& (Pseudo-)Scalar Meson Name & $J^{PC}$ & $m_{n>0}$ & PDG Mean Mass \cite{PDG} (MeV) \\
&&&(units of $\frac{r_h}{\sqrt{4\pi g_s N}}$) & \\ \hline
${\cal Y}^{(1)}$& $f0[980]/a0[980$ & $0^{++}$ & 9207.44 & 980 \\ \hline
${\cal Y}^{(2)}$ & -- & $0^{--}$ & 10861.9 & -- \\ \hline
${\cal Y}^{(3)}$ &f0[1370] &$0^{++}$ & 12683.7 & 1350\\  \hline
${\cal Y}^{(4)}$ & -- & $0^{--}$ & 14640.8 & -- \\ \hline
${\cal Y}^{(5)}$ & f0[1450] & $0^{++}$ & 16704 & 1474 \\ \hline
\end{tabular}
\end{center}
\caption{(Pseudo-)Scalar Meson masses from WKB Quantization}
\end{table}

\section{Meson Spectroscopy in a Thermal Background and Near Isospectrality with Black-Hole Background}
Previous sections used a background with a black-hole for all range temperatures. But in principle for low temperatures one should consider a thermal background.
Here, we show that the (pseudo-) vector meson spectrum and (pseudo-)scalar meson spectrum obtained via a thermal background are nearly isospectral with the results obtained via a black hole background. 
Here, we followed the same three routes of calculation techniques used in above sections. Following section  presents the main results to validate our claim.

\subsection{Vector Meson Spectroscopy in a Thermal Background}
\subsubsection{Solving the EOM near an IR cut-off $r=r_0$}
Writing $r = r_0 e^{\sqrt{Y^2+Z^2}}$ - $r_0$ being an IR cut-off \footnote{This is not actually a parameter put in by hand. In the spirit of a top-down approach, one can show that a Hawking-Page transition occurs at a temperature at an $r_0$ given in terms of  $r_h$ and an ${\cal O}(1)$ constant of proportionality relating the modulus of the Ouyang embedding parameter corresponding to the holomorphic embedding of type IIB flavor branes to $r_0$  - see \cite{EPJC-2}.} - and defining $m = \tilde{m}\frac{r_0}{\sqrt{4\pi g_s N}}$, setting $r_h=0$ and introducing a bare resolution parameter $a = \gamma r_0$ (to ensure that ${\cal R}_{D5/\overline{D5}}=\sqrt{3}a\neq0$), one can show that the $\alpha_N(Z)$ EOM - there is no need to attach a superscript to $\alpha_N$ anymore as $r_h=0$ - near the horizon can be written in the form:
\begin{equation}
\label{alphaEOM-rh=0}
\alpha_n^{\prime\prime}(Z) + (a_1 + b_1 |Z|)\alpha_n^\prime(Z) + (a_2 + b_2 |Z|)\alpha_n(Z) = 0,
\end{equation}
where up to ${\cal O}\left(\frac{1}{\log N}\right)$:
\begin{eqnarray}
\label{a1-b1-c1}
& & a_1 = 2 - 3\gamma^2 - \frac{3}{\log N} + \frac{9\gamma^2}{\log N},\nonumber\\
& & b_1 = 6\gamma^2 - 18 \frac{\gamma^2}{\log N},\nonumber\\
& & a_2 = \left(1 - 3\gamma^2\right)\tilde{m}^2,
\end{eqnarray}
whose solution is:
\begin{eqnarray}
\label{solution_EOM_alpha-rh=0}
& & \alpha_n(Z) = e^{-{a_1} |Z|+\frac{{b_2} |Z|}{{b_1}}-\frac{{b_1} Z^2}{2}} \Biggl(c_2 \, _1F_1\left(\frac{{b_1}^3-{a_2} {b_1}^2+{a_1} {b_2}
   {b_1}-{b_2}^2}{2 {b_1}^3};\frac{1}{2};\frac{\left(|Z| {b_1}^2+{a_1} {b_1}-2 {b_2}\right)^2}{2 {b_1}^3}\right)\nonumber\\
   & & +c_1
   H_{\frac{-{a_1} {b_1} {b_2}+{a_2} {b_1}^2-{b_1}^3+{b_2}^2}{{b_1}^3}}\left(\frac{{a_1} {b_1}+{b_1}^2 |Z|-2
   {b_2}}{\sqrt{2} {b_1}^{3/2}}\right)\Biggr).
\end{eqnarray}
Imposing Neumann BC at $r=r_0$  required setting $c_2=0$, numerically for lightest vector meson, e.g., for $N=6000, \gamma=0.6$ ,  one obtained a root - $\tilde{m}\approx 1.04$ - identical to the LO value in (\ref{meson-spectroscopy-ii}). While for $c_1=0$ the only root was $\tilde{m}=0$.

\subsubsection{Schr\"{o}dinger-like EOM}
Using a field redefinition  $\psi_n(Z) = \sqrt{{\cal C}_1(Z)}\alpha_n(Z)$ one can write the Schr\"{o}dinger-like EOM where $\psi_n(Z)$ satisfies: $\psi^{\prime\prime}_n(Z) + V(Z)\psi_n(Z) = 0$,
where:
\begin{eqnarray}
\label{C1-alpha-rh=0}
&& {\cal C}_1(Z) = \frac{1}{2} \Biggl[3 {\gamma}^2 ({g_s} {N_f} (\log (N)-3 |Z|-3)+4 \pi )-3 {g_s} {N_f} \left(3 {\gamma}^2+2 e^{2 |Z|}\right)
   \log ({r_0})\nonumber\\
   & & +2 e^{2 |Z|} ({g_s} {N_f} (\log (N)-3 |Z|)+4 \pi )\Biggr],
\end{eqnarray}
\begin{equation}
\label{V_vectormesons_rh=0}
V(Z) = -1 + e^{-2|Z|}\tilde{m}^2 - 3 e^{-4|Z|}\gamma^2\tilde{m}^2 + \frac{3 - \frac{9}{2}e^{-2|Z|}\gamma^2}{\log N} + {\cal O}\left(\frac{1}{(\log N)^2}\right).
\end{equation}
In IR ($r=r_0$)-the EOM took the form $\psi_n^{\prime\prime}(Z) + (a + b|Z|)\psi_n(Z)$,
where:
\begin{eqnarray}
\label{abc}
& & a = - 1 + \frac{3}{\log N} - \frac{9 \gamma^2}{2\log N} + \tilde{m}^2(1 - 3 \gamma^2),\nonumber\\
& & b = \frac{9\gamma^2}{\log N} + \tilde{m}^2(12\gamma^2 - 2),
\end{eqnarray}
 whose solution was obtained as:
\begin{equation}
\label{solution_psi-rh=0}
{\psi}(Z) = c_1 {Ai}\left(-\frac{a+b |Z|}{(-b)^{2/3}}\right)+c_2 {Bi}\left(-\frac{a+b |Z|}{(-b)^{2/3}}\right).
\end{equation}
For large $|\log r_0|$-limit and $\gamma=0.6$, one obtains:
$\tilde{m} = 0.36$.

In the UV, potential was given by $V(Z) = -1 + e^{-2|Z|}\tilde{m}^2 + {\cal O}\left(e^{-4|Z|}\right)$, and EOM's solution were obtained in terms of Bessel functions $J_1\left(e^{-|Z|}\tilde{m}\right)$ and $Y_1\left(e^{-|Z|}\tilde{m}\right)$. One doesn't get any mass quantization condition for $\tilde{m}$ in UV as
Neumann/Dirichlet BC were identically satisfied.

\subsubsection{WKB Quantization Condition}
In the UV:
\begin{eqnarray}
\label{sqrtV-alpha-rh=0}
& & \sqrt{V(\alpha_n)} = \frac{3-\frac{9}{2} {\gamma}^2 e^{-2 |Z|}}{2 \log (N) \sqrt{-3 {\gamma}^2 {\tilde{m}}^2 e^{-4 |Z|}+{\tilde{m}}^2 e^{-2
   |Z|}-1}}+\sqrt{-3 {\gamma}^2 {\tilde{m}}^2 e^{-4 |Z|}+{\tilde{m}}^2 e^{-2 |Z|}-1}\nonumber\\
   & & +{\cal O}\left(\left(\frac{1}{{\log N}}\right)^2\right).
\end{eqnarray}
One can see that $\sqrt{V}\in\mathbb{R}$ for $|Z|\in\left[\log\left(\sqrt{3}\gamma + {\cal O}\left(\frac{1}{\tilde{m}^2}\right)\right),\log\left(\tilde{m} - \frac{3}{2\tilde{m}} + {\cal O}\left(\frac{1}{\tilde{m}^2}\right)\right)\right]$. One can then show that:
\begin{equation}
\label{WKB_alpha_rh=0}
\int_{\log\left(\sqrt{3}\gamma\right)}^{\log\left(\tilde{m} - \frac{3}{2\tilde{m}} \right)}\sqrt{V} = \left(n + \frac{1}{2}\right)\pi,
\end{equation}
yields:
\begin{eqnarray}
\label{mn-alpha-rh=0}
& & m_n^{\alpha_n} = \frac{3}{20} \sqrt{3} \left(\sqrt{2} \sqrt{{\gamma}^2 \left(2 (2 \pi  n+\pi )^2+12
   \pi  (2 n+1)+13\right)}+2 {\gamma} (2 \pi  n+\pi +3)\right)\nonumber\\
& & \frac{\sqrt{3} \left(\frac{{\gamma}^2 (7-36 \pi  (2 n+1))}{\sqrt{2} \sqrt{{\gamma}^2 \left(2 (2 \pi  n+\pi )^2+12 \pi  (2
   n+1)+13\right)}}-18 {\gamma}\right)}{20 \log (N)}+ {\cal O}\left(\left(\frac{1}{\log N}\right)^2\right).
   \end{eqnarray}
Table 3.5, nearly isospectral with Table 3.1(black-hole results), summarizes the results for vector mesons obtained using a thermal background.
\begin{table}[h]
\begin{center}
\begin{tabular}{|c|c|c|c|c|}\hline
& (Pseudo-)Vector Meson Name & $J^{PC}$ & $m_{n>0}$ & PDG Mass \cite{PDG} (MeV) \\
&&&(units of $\frac{r_0}{\sqrt{4\pi g_s N}}$) & \\ \hline
$B^{(1)}_{\mu}$& $\rho[770]$ & $1^{++}$ & 7.716 - $\frac{1.636}{\log N}$ &775.49 \\ \hline
$B^{(2)}_{\mu}$ &$a_1[1260]$ & $1^{--}$ & 11.644 - $\frac{1.714}{\log N}$ & 1230 \\ \hline
$B^{(3)}_{\mu}$ &$\rho[1450]$ &$1^{++}$ & 15.567 - $\frac{1.753}{\log N}$ & 1465\\  \hline
$B^{(4)}_{\mu}$ & $a_1[1640]$ & $1^{--}$ & 19.488 - $\frac{1.776}{\log N}$ & 1647 \\ \hline
\end{tabular}
\end{center}
\caption{(Pseudo-)Vector Meson masses from WKB Quantization applied to $V(\alpha^{\left\{0\right\}}_n)$}
\end{table}

\subsection{Scalar Meson Spectroscopy in a Thermal Background}

\subsubsection{Solving the EOM near an IR cut-off $r = r_0$}

 Near horizon ${\cal G}_n(Z)$'s EOM has same form as (\ref{alphaEOM-rh=0}) where:
\begin{eqnarray}
\label{a1_b1_a2_b2-scalar}
& & a_1 = 4 - 3\gamma^2 +\frac{(9\gamma^2 - 3)}{\log N} + \frac{(27\gamma^2 - 9)\log r_0}{(\log N)^2},\nonumber\\
& & b_1 = 6\gamma^2 - \frac{18\gamma^2}{\log N} - \frac{54\gamma^2\log r_0}{(\log N)^2},
\nonumber\\
& & a_2 = \tilde{m}^2(1 - 3 \gamma^2),\nonumber\\
& & b_2 = \tilde{m}^2(12\gamma^2 - 2).
\end{eqnarray}
Quite interestingly, this IR computation is able to resolve $f0[980](m_{f0[980]}=990 MeV)$ and $a0[980](m_{a0[980]}=980 MeV)$ because, for
$\gamma=0.6$, numerically one can show that the two smallest roots of the equation obtained by imposing Neumann boundary condition on ${|Z|\cal G}_n(r=r_0)$ by setting $c_2=0$ are:
$1.83$ and $1.94$ - the second in particular not far off of the results of Table 3.3 gotten using a black-hole gravity dual - and $\frac{1.94}{1.83} = 1.06$ and $\frac{m_{f0[980]}}{m_{a0[980]}}=1.01$ - very close indeed! A black-hole computation could not do so.

\subsubsection{Schr\"{o}dinger-like EOM}

For:
\begin{eqnarray}
\label{C1-scalar}
& & {\cal C}_1(Z) = e^{2 |Z|} \Biggl(3 {\gamma}^2 ({g_s} {N_f} (\log (N)-3 |Z|-3)+4 \pi )-3 {g_s} {N_f} \left(3
   {\gamma}^2+2 e^{2 |Z|}\right) \log ({r_0})\nonumber\\
   & & +2 e^{2 |Z|} ({g_s} {N_f} (\log (N)-3 |Z|)+4 \pi
   )\Biggr),
\end{eqnarray}
and  Schr\"{o}dinger-like EOM's potential (similar to (\ref{V_vectormesons_rh=0})) was obtained as:
\begin{eqnarray}
\label{V-scalar-rh=0}
& & V({\cal G}_n) = -4 - 3 e^{-4|Z|}\gamma^2\tilde{m}^2 + e^{-2|Z|}(3\gamma^2 + \tilde{m}^2)
\nonumber\\
& & + \frac{6 - \frac{27}{2}e^{-2|Z|}\gamma^2}{\log N} + {\cal O}\left(\frac{1}{(\log N)^2}\right),
\end{eqnarray}
the $a, b$, similar to (\ref{abc}), are as follows:
\begin{eqnarray}
\label{ab-rh=0}
& & a = -4 + 3\gamma^2 + \tilde{m}^2(1 - 3 \gamma^2) + \frac{(6 - 27 \gamma^2)}{\log N},\nonumber\\
& & b = - 6 \gamma^2 + \tilde{m}^2(12\gamma^2 - 2) + \frac{27 \gamma^2}{\log N}.
\end{eqnarray}
One gets a solution similar to (\ref{solution_psi-rh=0}); numerically for $\gamma=0.6$, $\tilde{m}\approx 0.85$ - very close to the smallest root in {\bf 3.5.2.1} and identical in order as the results of {\bf 3.4.2.1} - imposing Neumann BC at $r=r_0$ requires $c_2=0$ .

\subsubsection{WKB Quantization Condition}

Addressing the UV-IR interpolating/UV region for very small $M, N_f$, lead to:
\begin{eqnarray}
\label{sqrtV_scalar_rh=0}
& & \sqrt{V({\cal G}_n)} = \sqrt{-4 - 3 e^{-2|Z|}\gamma^2\tilde{m}^2 + e^{-2|Z|}\left(3\gamma^2 + \tilde{m}^2\right)} + {\cal O}\left(\frac{1}{N^{\frac{1}{5}}}\right).
\end{eqnarray}
One sees that $|Z|\in\left[\log\left(\sqrt{3}\gamma\right),\log\left(\frac{\tilde{m}}{2} - \frac{9\gamma^2}{4\tilde{m}^2}\right)\right]$, $\sqrt{V}\in\mathbb{R}$ and the WKB quantization condition:
\begin{equation}
\label{QKB_scalar_rh=0}
\int_{\log\left(\sqrt{3}\gamma\right)}^{\log\left(\frac{\tilde{m}}{2} - \frac{9\gamma^2}{4\tilde{m^2}}\right)}\sqrt{V\left({\cal G}_n(Z)\right)} = \left(n + \frac{1}{2}\right)\pi,
\end{equation}
yields:
\begin{equation}
\label{mn-scalar-rh=0}
m_n = \frac{1}{20} \sqrt{3} {\gamma} \left(6 (2 \pi  n+\pi +6)+\sqrt{2} \sqrt{18 (2 \pi  n+\pi )^2+216 \pi  (2
   n+1)+349}\right).
\end{equation}
Table 3.6 summarizes the results for scalar mesons spectrum disregarding $n=0$ mode:
\begin{table}[h]
\begin{center}
\begin{tabular}{|c|c|c|c|c|}\hline
& (Pseudo-)Scalar Meson Name & $J^{PC}$ & $m_{n>0}$ & PDG Mean Mass \cite{PDG} (MeV) \\
&&&(units of $\frac{r_0}{\sqrt{4\pi g_s N}}$) & \\ \hline
${\cal Y}^{(1)}$& $f0[980]/a0[980]$ & $0^{++}$ & 15.745  & 980 \\ \hline
${\cal Y}^{(2)}$ & -- & $0^{--}$ & 22.359 & -- \\ \hline
${\cal Y}^{(3)}$ &f0[1370] &$0^{++}$ & 28.934 & 1350\\  \hline
\end{tabular}
\end{center}
\caption{(Pseudo-)Scalar Meson masses from WKB Quantization}
\end{table}
For a low-temperature thermal gravity dual, the results for modes greater than 3, $n>3$, are not reliable.

\section{Summary and New Insights into Thermal QCD }
In this chapter, we calculated (pseudo-)vector and (pseudo-)scalar meson spectra from the delocalized type IIA SYZ mirror (built in \cite{MQGP}) of the UV-complete top-down type IIB holographic dual of large-$N$ thermal QCD (built in \cite{metrics}), and compared our results with \cite{Sakai-Sugimoto-1}, \cite{Dasgupta_et_al_Mesons} and \cite{PDG}. We first computed the spectra with a black hole background, assuming it is valid for all temperatures, low and high (similar to \cite{ BH_all_T}). Then we repeated the computation in a thermal background without a black hole. We observed that the mirror of \cite{metrics}, when considered in the `MQGP limit', can generate the low-lying vector and scalar meson spectra from the massless string sector almost without fine tuning. Computation of the perturbative (thermal) finite-gauge coupling is difficult or impossible. In our setup, this can be done as a classical supergravity computation.

\begin{itemize}
\item
{\bf Summary of New Results Obtained (Points 1. - 6.) and the New Insights Gained into Thermal QCD (Point 7.)}
\begin{enumerate}
\item
In tables 3.1 and 3.2, even if we  drop the ${\cal O}\left(\frac{1}{\log N}\right)$ terms appearing in  the  vector meson masses (BH/thermal background) obtained by a WKB quantization condition, and assume $n=1,2,3,4$ to correspond respectively to $\rho[770], a_1[1260], \rho[1450], a_1[1640]$, then the following table compares mass ratios from our results at LO in $N$ (obtained from a WKB quantization condition) with those from \cite{Sakai-Sugimoto-1}, \cite{Dasgupta_et_al_Mesons} (up to first order in $\delta=\frac{g_sM^2}{N}<1$) and \cite{PDG}:

\begin{table}[h]
\begin{center}
\begin{tabular}{|c|c|c|c|c|c|c|}\hline
                      & ratio ($\alpha_{n}^{\left\{i\right\}}$)&ratio ($\alpha_{n}^{\left\{0\right\}}$)&ratio  &Sakai-Sugimoto & Best value & Exp. value PDG\\
                      &(BH)&(BH)&(Thermal)&{\scriptsize (as given in \cite{Dasgupta_et_al_Mesons})}& in \cite{Dasgupta_et_al_Mesons}:$\delta=0.5$ &{\scriptsize (as given in \cite{Dasgupta_et_al_Mesons})}\\ \hline
$\frac{m_{a_1[1260]}^{2}}{m_{\rho[770]}^{2}}$ & 2.30& 2.28 &2.28 & 2.32 &2.31& 2.52\\ \hline
$\frac{m_{\rho[1450]}^{2}}{m_{\rho[770]}^{2}}$ & 4.12 & 4.09 & 4.07& 4.22 &4.09& 3.57\\ \hline
$\frac{m_{a_1[1640]}^{2}}{m_{\rho[770]}^{2}}$ & 6.47 & 6.41& 6.38& 6.62 &5.93& 4.51\\ \hline
\end{tabular}\begin{center}
\end{center}
\end{center}
\caption{Comparison of  Mesons masses ratio }
\end{table}

The authors of \cite{Dasgupta_et_al_Mesons} obtained a variety of values by adjusting the values of and working up to first order in $\delta$, as well as a constant appearing in a `squashing factor' in the metric. Table 3.7 column 5 lists their best (pseudo-)vector meson mass ratios. They choose $\delta=0.5$, which in fact does not justify disregarding terms of higher powers of $\delta$, as $\delta=0.5$ is not small enough to justify the same. We found that the WKB quantization condition applied to $V(\alpha^{\left\{0\right\}}_n)$ for the BH gravitational dual or $V(\alpha_n)$ for the thermal gravitational background works even up to LO in $N$ without having to explicitly compute the ${\cal O}(\delta)$ (we used $\delta\sim0.001$ for our calculations, justifying dropping higher powers of $\delta$) contribution, display the following features:
\begin{itemize}
\item
our $m_{a_1[1260]}^{2}/m_{\rho[770]}^{2}$ is close to \cite{Sakai-Sugimoto-1} and\cite{Dasgupta_et_al_Mesons}, and not too far off of the PDG value

\item
our $m_{\rho[1450]}^{2}/m_{\rho[770]}^{2}$ is the same as (for BH background)/very close to (for thermal background) \cite{Dasgupta_et_al_Mesons} (but without any fine tuning and already at LO in $N$) - within $\approx$ 15$\%$ of the PDG value

\item
our $m_{a_1[1640]}^{2}/m_{\rho[770]}^{2}$ is closer to the PDG value than \cite{Sakai-Sugimoto-1}
\end{itemize}

\item
spectrum for lightest (pseudo-)vector meson is almost isospectral for both black hole and thermal backgrounds

\item
At low temperatures, the thermal background fails to provide a temperature dependence of $\tilde{m}$, which is in accordance with one's expectations \cite{BH_all_T}. Motivated by the aforementioned isospectrality the explicit temperature dependence of the lowest lying vector meson mass is captured explicitly by solving the EOMs in a black hole gravitational dual close to the horizon $r=r_h $, with the temperature dependence appearing at order ${\cal O}\left(\frac{1}{(\log N)^2}, \frac{g_s M^2}{N}\right)$. The temperature-dependent meson mass $\tilde{m}$ will have the following form:
\begin{equation}
\label{rh_T_meson_mass}
 \tilde{m}_{\rm lightest} = \alpha + \frac{\beta}{\log N} + \frac{\left(\delta_1 + \delta_2 \log r_h\right)}{\left(\log N\right)^2} + \frac{g_sM^2\left(\kappa_1 + \kappa_2\log r_h\right)}{N} + {\cal O}\left(\frac{g_sM^2}{N}\frac{\log r_h}{\log N}\right),
\end{equation}
$\delta_2>0$.
The temperature, assuming the resolution to be larger than the deformation in the resolved warped deformed conifold in the type IIB background of \cite{metrics} in the MQGP limit, and utilizing the IR-valued warp factor $h(r,\theta_1\sim N^{-\frac{1}{5}},\theta_2\sim N^{-\frac{3}{10}})$, is \cite{EPJC-2}:
\begin{eqnarray}
\label{T}
& & T = \frac{\partial_rG^{\cal M}_{00}}{4\pi\sqrt{G^{\cal M}_{00}G^{\cal M}_{rr}}}\nonumber\\
& &= {r_h} \left[\frac{1}{2 \pi ^{3/2} \sqrt{{g_s} N}}-\frac{3 {g_s}^{\frac{3}{2}} M^2 {N_f} \log ({r_h}) \left(-\log
   {N}+12 \log ({r_h})+\frac{8 \pi}{g_s N_f} +6-\log (16)\right)}{64 \pi ^{7/2} N^{3/2}} \right]\nonumber\\
   & &+ a^2 \left(\frac{3}{4 \pi ^{3/2} \sqrt{{g_s}} \sqrt{N} {r_h}}-\frac{9 {g_s}^{3/2} M^2 {N_f} \log ({r_h})
   \left(\frac{8 \pi }{{g_s} {N_f}}-\log (N)+12 \log ({r_h})+6-2 \log (4)\right)}{128 \pi ^{7/2} N^{3/2}
   {r_h}}\right).\nonumber\\
   & &
\end{eqnarray}
Using (\ref{T}) and the arguments of \cite{EPJC-2}, one can  invert (\ref{T}) and express $r_h$ in terms of $T$ \cite{Misra+Gale_Conformal_Anomaly} in the MQGP limit. Assuming $\log r_h$ in (\ref{rh_T_meson_mass}) to be in fact
$\log\left(\frac{r_h}{\Lambda}\right), \Lambda>r_h$ being the scale at which confinement occurs, one sees that, as per expectations, the vector meson masses decrease with temperature
\cite{BH_all_T} with the same being large-$N$ suppressed \cite{Herzog_Tc} (and references therein).

\item
On comparing scalar meson mass ratios obtained from (\ref{scalar-meson-WKB-i}) using a black hole gravitational dual WKB quantization and PDG values, we obtained Table 3.8:
\begin{table}[h]
\begin{center}
\begin{tabular}{|c|c|}\hline
Our results & PDG values \\ \hline\hline
$\frac{m_{n=3}}{m_{n=1}}$ & $\frac{m_{{f0}[1370]}}{m_{{f0}[980]}}$\\ \hline
1.38 & 1.38 \\ \hline\hline
$\frac{m_{n=5}}{m_{n=1}}$ & $\frac{m_{{f0}[1450]}}{m_{{f0}[980]}}$\\ \hline
1.81 & 1.50 \\ \hline\hline
\end{tabular}
\end{center}
\caption{The lightest Scalar Meson mass ratios}
\end{table}

The agreement with the PDG values for the lightest three scalar meson candidates (if assumed to be $f0[980], f0[1370], f0[1450]$) is quite nice. We do not expect the agreement for more massive scalar mesons. This is for the following reason. As discussed in \cite{Imoto:2010ef,Dasgupta_et_al_Mesons}\footnote{One of us (AM) thanks K. Dasgupta for emphasizing this point to him.}  massive open string excitations can contribute to the  meson (specially scalar) masses (as scalar mesons are typically heavier than (pseudo-)vector mesons). We do not attempt to estimate the same as open string quantization in a curved RR background is a hard problem, and in this chapter, like \cite{Dasgupta_et_al_Mesons}, we have assumed that the mesons arise from the massive KK modes of the massless open string sector. The difference between our results and the PDG results for the mass ratios of vector and scalar mesons, for heavier mesons, could be attributed to the contributions coming to meson masses from the massive  open string sector (which we have not calculated) in addition to the ones coming from the massless open string sector (which we have calculated in this chapter).

\item
Using a thermal background, though on one hand, it appears to be possible to  in fact resolve $a0[980]$ (average mass of $980 MeV$) and $f0[980]$ (average mass $990 MeV$), on the other hand assuming $f0[980]/a0[980], f0[1370], f0[1450]$ to be the lightest scalar mesons,  a WKB quantization condition yields a mass ratio of the first two as 1.83, the mass ratio of $f0[1370]$ and $a0[980]$ being 1.38; as $f0[1370]$ is expected to have mass range of $1200 - 1500 MeV$ \cite{PDG}, with $1500 MeV$ the ratio - as per PDG values - increases to 1.53.

The thermal background is not able to correctly account for $f0[1470]$. The black hole background, as is evident from Table 3.8, does a good job though.

\item
The $0^{--}$ pseudo-scalar mesons in Table 3.4 do not, thus far, have corresponding entries in the PDG tables. This is one point of difference between our results and PDG tables.

\item
There are {\bf three main insights} we gain into thermal QCD by evaluating mesonic (vector/scalar) spectra and comparing with PDG values. \\
(a) The first is the realization that from a type IIA perspective, meson spectroscopy in the mirror of top-down holographic type IIB duals of large-$N$ thermal QCD at {\it finite coupling and number of colors} \footnote{In the IR, $N_c=M$ which can be ${\cal O}(1)$ in the MQGP limit of \cite{MQGP} - taken to be three in this chapter - and not $N$.} (closer to a realistic QCD calculation) which are UV complete - we know of only \cite{metrics} that falls in this category for which the mirror was worked out in \cite{MQGP} - will give results closer to PDG values rather than a single T-dual of the same. Even though obtaining the mirror requires a lot of work, but once obtained, one can obtain very good agreement with PDG tables already at ${\cal O}\left(\left(\frac{g_sM^2}{N}\right)^0\right)$ (for vector mesons, {\it without any fine tuning}). This is a major lesson we learned from our computation. There are two major reasons for the same.
\begin{itemize}
\item
 As noted in section {\bf 3.2}, the mirror of \cite{metrics} picks up sub-dominant terms in $N$ of ${\cal O}(N^0)$ in $B^{\rm IIA}$ which are therefore bigger than the ${\cal O}(\frac{g_sM^2}{N})$ contributions, and were missed in \cite{Dasgupta_et_al_Mesons}. This is the reason why the authors of \cite{Dasgupta_et_al_Mesons} had to set $\frac{g_sM^2}{N}=0.5$ - not a small enough fraction to warrant disregarding ${\cal O}\left(\left(\frac{g_sM^2}{N}\right)^2\right)$ contributions which they did - to obtain a reasonable match with \cite{PDG}.

\item
 In the context of \cite{metrics}, this is expected to be related to the following \footnote{One of us (AM) thanks K. Dasgupta for discussion on this point.}. The brane construct of \cite{metrics} involves $N\ D3$-branes, $M\ D5$-branes wrapping the vanishing $S^2$, $M\ \overline{D5}$-branes wrapping the same $S^2$ but placed at the antipodal points of the resolved $S^2(a)$ relative to the $D3, D5$-branes, $N_f$ flavor $D7$-branes wrapping an $S^3$ and radially extending all the way into the IR starting from the UV and $N_f\ \overline{D7}$-branes wrapping the same $S^3$ but going only up to the IR-UV interpolating region. The mirror of this results in $D6$ branes in a deformed conifold. Now, the gravity dual of the above picture - which is what we work with - involves a resolved warped deformed conifold with a black hole and $\overline{D5}$, $D7$ branes and $\overline{D7}$ branes (plus fluxes) on the type IIB side and the delocalized mirror yields a non-K\"{a}hler warped resolved conifold with a black hole and $D6, \overline{D6}$ branes (plus fluxes) on the type IIA side.   The latter (warped resolved conifolds) are more easier to deal with computationally than  the former(resolved warped deformed conifolds).

\end{itemize}

(b) (related to (a) above) Thermal QCD at strong coupling and finite $N_c$ is intimately related to the SYZ mirror of RWD conifolds.; hence, {\it delocalized Strominger-Yau-Zaslow mirror construction is an entirely new technique  used  for hadron spectroscopy}.

(c) In addition, BH gravitational duals as well as thermal gravitational duals produce nearly isospectral spectra for the lightest vector mesons; this is also true to a certain extent for the lightest scalar mesons, but not completely. The results of explicit computations show that the type IIA gravity dual obtained by delocalized SYZ mirror transform of the type IIB holographic dual of \cite{metrics} is not only capable of providing a good match with PDG values for the lightest vector and scalar mesons, but it is also capable of obtaining an explicit temperature-dependence of the (pseudo-)vector masses as a bonus and realising the $\log r_h$--dependence in the same appears at the sub-dominant ${\cal O}\left(\frac{1}{(\log N)^2}\right)$ with the desired feature of a small large-$N$ suppressed  decrease with increase of temperature.

\end{enumerate}

\end{itemize}


\chapter{Glueball decay}

\graphicspath{{Chapter3/}{Chapter3/}}

\section{Introduction}
The study of glueballs with gauge fields in dynamical role is important to understand the nonperturbative behaviour of QCD. Existence of glueballs is well established, but their experimental identification continues to be a difficult task to accomplish. This lack of clarity in identifying glueballs arises from the fact that the theoretical developments are not able to distinguish how the production and decay properties of glueballs are distinct from those of mesons \cite{Mathieu:2008me}.

Glueballs and mesons have been extensively studied over the last decade and more to gain new insights into QCD's non-perturbative regime. Various holographic setups, including soft-wall models, hard-wall models, modified-soft-wall models \cite{Kim:2009bp,Domokos:2007kt,Alvares:2011wb,Bellantuono:2014lra,BH_all_T,Nicotri,Forkel,ForkelStructure,FolcoCapossoli,Jugeau,Colangelolight,Cui:2013xva}, and the Sakai-Sugimoto model \cite{Sakai-Sugimoto-1}, have been used to obtain the glueball and meson spectra and study their interactions. Let us now summarise our recent work in this area.

In chapter 2, we initiated top-down holographic glueball spectroscopy obtained from the M theory uplift of the SYZ type IIA mirror of the top-down UV complete holographic dual of large-$N$ thermal QCD of \cite{metrics}, as constructed in \cite{MQGP}. Therein we computed masses of $0^{++}, 0^{--},\ 0^{-+},\ 1^{++}$ and $2^{++}$ glueballs up to NLO in $N$, and found very good agreement with several lattice results. As discussed in chapter 3, we continued our work in \cite{Yadav+Misra+Sil-Mesons} by evaluating the spectra of (pseudo-)vector and (pseudo-)scalar mesons and compared our results to those of \cite{Sakai-Sugimoto-1}, \cite{Dasgupta_et_al_Mesons}, and \cite{PDG-2018}. In this project, we examined two-, three-, and four-body (`exotic' scalar) glueball decays into $\rho$ and $\pi$ mesons and demonstrated that our results allowed for an exact match with PDG data, including any future improvements.
Lagrangian consisting mixed terms representing interaction of glueball and mesons were obtained by performing a Kaluza-Klein reduction for gauge fields as done in  \cite{Hashimoto-glueball}: $A_Z(x^\nu,Z) = \sum_{n=0} \pi^{(n)}(x^\nu)\phi_n(Z), A_\mu(x^\nu,Z) = \sum_{n=1} \rho_\mu^{(n)}(x^\nu)\psi_n(Z)$, which introduced scalar and vector meson into the picture and similar to \cite{Myers}, M-theory metric perturbations $h_{MN}(M,N=0,...,10;\mu=t,a, a=1,2,3)$ were taken into account to introduce glueballs into the picture:
\begin{eqnarray}
\label{M-theory-metric-perturbations}
& & h_{tt}(r,x^\mu) = q_1(r) G(x^\mu) g^{\cal M}_{tt}\nonumber\\
& & h_{rr}(r,x^\mu) = q_2(r)G(x^\mu) g^{\cal M}_{rr}\nonumber\\
& & h_{ra}(r,x^\mu) = q_3(r)\partial_a G(x^\mu) g^{\cal M}_{aa}\nonumber\\
& & h_{ab}(r,x^\mu) = g^{\cal M}_{ab}\left(q_4(r) + q_5 \frac{\partial_a\partial_b}{m^2}\right)G(x^\mu)\ {\rm no\ summation\ on\ a\ and\ b}\nonumber\\
& & h_{10\ 10}(r,x^\mu) = q_6(r) G(x^\mu)g^{\cal M}_{1010}.
\end{eqnarray}
In accordance with Witten's prescription:
\begin{equation}
\label{Witten_IIA_to_M+vice_versa}
ds_{ M}^2 = e^{-\frac{2\phi^{IIA}}{3}}ds_{\rm IIA}^2 + e^{\frac{4\phi^{IIA}}{3}}\left(dx^{10} + A\right)^2,
\end{equation}
$A$ being the type $IIA$ RR one-form,
we worked back the type IIA metric perturbations which gave(in the following $\tilde{g}^{\rm IIA}_{\alpha\beta} = g^{\rm IIA}_{\alpha\beta} + h_{\alpha\beta}; \alpha,\beta=0,1...,9$ and $h_{\alpha\beta}$ being type IIA metric perturbations):
\begin{eqnarray*}
& & e^{\frac{4\phi^{\rm IIA}}{3}} = g^{\cal M}_{10\ 10} + h_{10\ 10},\nonumber\\
& & \frac{\tilde{g}^{\rm IIA}_{rr,tt}}{\sqrt{g^{\cal M}_{10\ 10} + h_{10\ 10}}} = g^{\cal M}_{rr,tt} + h_{rr,tt},\nonumber\\
& &  \frac{\tilde{g}^{\rm IIA}_{ra}}{\sqrt{g^{\cal M}_{10\ 10} + h_{10\ 10}}} = h_{ra},\nonumber\\
& & \frac{\tilde{g}^{\rm IIA}_{ab}}{\sqrt{g^{\cal M}_{10\ 10} + h_{10\ 10}}} = g^{\cal M}_{ab} + h_{ab}.
\end{eqnarray*}
Using Einstein's EOM, we solved the first order perturbation of the M-theory (assuming the flux term providing a cosmological constant):
$R_{MN}^{(1)} \sim G_{PQRS}G^{PQRS}h_{MN}$,
 for $q_{1,...6}$,we obtained the glueball-meson interaction Lagrangian density (metric perturbation corresponding to glueballs and gauge field fluctuations corresponding to mesons).As previously mentioned, similar considerations were made in the context of the type IIA Sakai-Sugimoto model, first in \cite{Hashimoto-glueball} and more recently in \cite{Brunner_Hashimoto-results}. According to our understanding, a top-down investigation of holographic glueball-to-meson decays in the context of a M-theory uplift of a UV-complete type IIB holographic counterpart at finite coupling has hitherto been completely absent from the scientific literature. This effort fills in the gaps left by the previous work.\\
 In the MQGP limit as explained in the chapter 1,  apart from the gluon-bound states, i.e. glueballs, and the light ($\rho/\pi$) mesons, all other scalar mesons are integrated out. As per \cite{witten}, supersymmetry can be broken by imposing anti-periodic boundary conditions for fermions along the $x^0$-circle (which at finite temperature has periodicity given by the reciprocal of the Hawking temperature). This is expected to generate fermionic masses of the order of the reciprocal of the $S^1_{t}$ radius $R_{r_h}$ and scalar masses of the order of $\frac{g_s^2N}{R_{r_h}}$. We will now argue that $R_{r_h}$ is very small implying scalar mesons (apart from the lightest $\rho$-vector and pionic pseudo-scalar mesons) are very heavy and are hence integrated out, and effectively the 3+1-dimensional QCD-like theory thus reduces to 2+1 dimensions. From (\ref{M-th components}), one sees that working with a near-horizon coordinate $\chi: r = r_h + \chi, \frac{\chi}{r_h}\ll1$, $g^{\cal M}_{rr}dr^2 = \frac{d\chi^2}{\chi}F_{rr}(r_h) \equiv d\rho^2$ or
$\chi = \frac{\rho^2}{4F_{rr}(r_h)}$. Thus:
\begin{equation}
\label{radius-x0}
-g^{\cal M}_{tt}dt^2 + g^{\cal M}_{rr}dr^2 = -g^{\cal M}_{tt}\ ^\prime(r_h)\frac{\rho^2}{4F_{rr}(r_h)}dt^2 + d\rho^2 \equiv - 4\pi^2 R_{r_h}^2 dt^2 + d\rho^2.
\end{equation}
As a result, we obtained the radius of the temporal direction as follows:
\begin{eqnarray}
\label{radius-t}
& & R_{r_h} = \sqrt{\frac{1+9b^2}{1+6b^2}}\frac{r_h}{\pi L^2}\rho \sim T\rho\sim T \sqrt{\chi}\sqrt{F_{rr}(r_h)}\sim\sqrt{\chi}T^{\frac{3}{2}}.
\end{eqnarray}
As a result, one can observe that $R_{r_h}$ is very tiny, which supports the claim.
\section{M-theory metric perturbations for `exotic' scalar glueball}

 The general M-theory metric fluctuations corresponding to `exotic' scalar glueball $0^{++}$ in terms of 2+1 dimensional spacetime $x^{1},x^{2},x^{3}$ can be written following \cite{Myers}, \cite{Hashimoto-glueball}as:
\begin{eqnarray}
\label{M-th perturbation}
h_{tt}&=&-q_{1}(r)g^{\cal M}_{tt}G_{E}(x^{1},x^{2},x^{3})\nonumber\\
h_{rr}&=&-q_{2}(r)g^{\cal M}_{rr}G_{E}(x^{1},x^{2},x^{3})\nonumber\\
h_{ra}&=&q_{3}(r)g^{\cal M}_{aa}\frac{\partial_{a}G_{E}(x^{1},x^{2},x^{3})}{M_g^2}, \ \ a=1,2,3\nonumber\\
h_{ab}&=&g^{\cal M}_{ab}\Bigg(q_{4}(r)\eta_{ab}-q_{5}(r)\frac{\partial_{a}\partial_{b}}{M_g^2}\Bigg)G_{E}(x^{1},x^{2},x^{3}),\ \ b=1,2,3\nonumber\\
h_{11,11}&=&q_{6}(r)g^{\cal M}_{11,11}G_{E}(x^{1}x^{2}x^{3})\nonumber\\
\end{eqnarray}
where $G_E(x^{1},x^{2},x^{3})$ and $M_g$ represents the three dimensional glueball and mass of the glueball respectively. The M-theory metric components up to NLO in $N$ near $\theta_{1}=\alpha_{\theta_{1}}N^{-\frac{1}{5}}$, $\theta_{2}=\alpha_{\theta_{2}}N^{-\frac{3}{10}}$, $\phi_{1,2}=0/2\pi$ are given in (\ref{M-th components}). The explicit expression for functions $q_{i={1,2,...,6}}$ were obtained by solving their EOM's derived from 11-D action. The 11-d action, using $\int C_3\wedge G_4\wedge G_4=0$ \cite{MQGP}, is given as:
$${\rm S}_{11}=\int d^{11}x\sqrt{-{\rm det}{\rm g}}\ \Bigg(R-\frac{1}{2 \times 4!}|G_4|^2\Bigg),$$
minimizing the action yields:
\begin{equation}
\label{Einstein-fluxes}
R_{{M}{N}}^{(1)}=\frac{1}{12}\left(-3{G_{{M}}^{\ \ {P}_2}}_{{Q}{R}}G_{{N}}^{\ \ {{P}_{3}}{Q}{R}}h_{{{P}_{2}}{{P}_{3}}}+\frac{1}{3}G^{{P}_{2}}_{\ \ {N_1}{P}{Q}}G^{{{P}_{3}}{N_1}{P}{Q}}h_{{P}_{2}{P}_{3}}g^{\cal M}_{{M}{N}}-\frac{G^2}{12}
h_{{M}{N}}\right).
\end{equation}
Here $R_{{M}{N}}^{(1)}$ denotes the perturbed part of the Ricci tensor while, M, N, etc ranges from 0 to 10. Following coupled eom's were produced by plugging in the expressions for each of the components.{\footnote{ EOM for the coordinates ${M},{N}$ is represented by $\delta R[{M},{N}]$.}},

\noindent $\bullet\delta {\rm R[t,t]}$\\
$q_1(r)$ equation of motion (\ref{EOM11}), near $r=r_h$, can be written as:
\begin{equation}
\label{q1-near_rh}
{q_1}''(r) + \left({a_1}+\frac{1}{ (r-{r_h})}\right) {q_1}'(r)+{a_2}\ {q_1}(r)=0,
\end{equation}
where,
\begin{eqnarray}
\label{a1-b1-a2-defs}
& & {a_1}=\frac{243 \sqrt{\frac{3}{2}} b^2 \left(9 b^2-1\right) {g_s}^{3/2} M \left(\frac{1}{N}\right)^{2/5} {N_f} \alpha _{\theta _1}^4 \log ^2({r_h})}{\pi ^{3/2} \left(3
   b^2-1\right) \alpha _{\theta _2}^3}+\frac{\left(\frac{12}{1-3 b^2}-\frac{6}{54 b^4+15 b^2+1}\right) b^2+5}{2 {r_h}}\nonumber\\
& & {a_2}=\frac{\left(6 b^2+1\right) {g_s} {K^1}^2 \left(16 \pi ^2 N-3{g_s}^2 {\log (N)} M^2 {N_f} \log ({r_h})\right)}{16 \pi  \left(9 b^2+1\right) {r_h}^3}.
\end{eqnarray}
Solution of (\ref{q1-near_rh}) was obtained as:
\begin{eqnarray}
\label{solution-q1}
& & \hskip -0.3in q_1(r) =  e^{- {a_1}r} \Biggl(c_{1\ q_1}
U\left(1-\frac{a_2}{a_1},1,a_1(r-{r_h})\right)+c_{2\ q_1}L_{\frac{{a_2}}{{a_1}}-1}({a_1} (r-{r_h}))\Biggr).
\end{eqnarray}
Since
\begin{eqnarray}
\label{Neumann-q1(r)-1}
& &  q_1'(r) = -\frac{{c_{1\ {q_1}}} e^{-{a_1} {r_h}}}{(r-{r_h}) \Gamma \left(1-\frac{{a_2}}{{a_1}}\right)}+\frac{e^{-{a_1} {r_h}}}{\Gamma \left(1-\frac{{a_2}}{{a_1}}\right)} \Biggl({c_{1\ {q_1}}} \Biggl({a_2} \log
   ({a_1})+{a_1} \psi ^{(0)}\left(1-\frac{{a_2}}{{a_1}}\right)+({a_2}-{a_1}) \psi ^{(0)}\left(2-\frac{{a_2}}{{a_1}}\right)\nonumber\\
   &&  +2 {a_1}+{a_2}
   \log (r-{r_h})-{a_2}+2 \gamma  {a_2}\Biggr)-{a_1} c_2 \left(L_{\frac{{a_2}}{{a_1}}-1}(0)+L_{\frac{{a_2}}{{a_1}}-2}^1(0)\right) \Gamma
   \left(1-\frac{{a_2}}{{a_1}}\right)\Biggr)+O\left((r-{r_h})^1\right),
\end{eqnarray}
we found that in order to impose Neumann boundary condition at $r=r_h$: $q_1'(r_h)=0$, one must set $c_{2\ q_1}=0$ and
\begin{equation}
\label{Neumann-q1(r)-2}
 \left(-\frac{{a_2}}{a_1}+1\right)=-n, n\in\mathbb{Z}^+\cup\left\{0\right\}.
\end{equation}
We fixed $n=1$, which implied $a_1 = \frac{a_2}{2}$,\\
So:
\begin{eqnarray}
\label{q1-EOM-Nbc_n=1}
& & q_1(r\sim r_h) = \frac{1}{2} e^{1-\frac{{a_2} r}{2}} \left(c_2 e^{\frac{{a_2} {r_h}}{2}} ({a_2} ({r_h}-r)+2) {Ei}\left(\frac{1}{2} {a_2} (r-{r_h})\right)+4 c_1
   e^{\frac{{a_2} {r_h}}{2}} (-{a_2} r+{a_2} {r_h}+2)+2 c_2 e^{\frac{{a_2} r}{2}}\right).\nonumber\\
& &
\end{eqnarray}
For $c_{2\ q_1}=0$:
\begin{equation}
\label{q1_solution_near_rh}
q_1(r \sim r_h) = -\frac{1}{3} a_2^3 c_{1_{{q1}}} (r-{r_h})^3+\frac{3}{2} a_2^2 c_{1_{{q1}}} (r-{r_h})^2-4 a_2 c_{1_{{q1}}} (r-{r_h})+4 c_{1_{{q1}}}+O
   (r-{r_h})^4.
\end{equation}
Further, using (\ref{EOM55}),  $c_{1\ {q_1}}=0$, i.e.:
\begin{equation}
\label{q1_solution_near_rh1}
q_1(r \sim r_h) = 0.
\end{equation}
In the UV, :
\begin{eqnarray}
&& \alpha=5-\frac{27 \sqrt{\frac{3}{2}} {g^{UV}_s}^{3/2} M^{UV} \left(\frac{1}{N}\right)^{2/5} {N^{UV}_f} \alpha _{\theta _1}^4}{2 \pi ^{3/2} \alpha _{\theta _2}^3}\nonumber\\
&&\beta=\frac{1}{16} {m_0}^2 {r_h}^2 \left(16-\frac{3 {g^{UV}_s}^2 {\log(N)} {\log(r)} {M^{UV}}^2 {N^{UV}_f}}{\pi ^2 N}\right),
\end{eqnarray}
(\ref{EOM11}) reduced to:
\begin{eqnarray}
\label{EOM11UV}
{q_1}''(r)+\frac{\left| \alpha \right|  {q_1}'(r)}{r}+\frac{\left| \beta \right|  {q_1}(r)}{r^4}=0,
\end{eqnarray}
whose solution was obtained as:
{\small
\begin{eqnarray}
\label{q1solution_UV}
&&q_1(r)=\left(\frac{1}{r}\right)^{\frac{1}{2} (\left| \alpha \right| +\left| \left| \alpha \right| -1\right| -1)} \Biggl(c_2 e^{-\frac{i \sqrt{\left| \beta \right| }}{r}} \,
   _1F_1\left(\frac{1}{2} (\left| \left| \alpha \right| -1\right| +1);\left| \left| \alpha \right| -1\right| +1;\frac{2 i \sqrt{\left| \beta \right| }}{r}\right)\nonumber\\
   &&+\frac{c_1
   2^{-\frac{\left| \left| \alpha \right| -1\right| }{2}} \left| \beta \right| ^{-\frac{\left| \left| \alpha \right| -1\right| }{4}} \left(\frac{i}{r}\right)^{-\frac{\left|
   \left| \alpha \right| -1\right| }{2}} K_{\frac{\left| \left| \alpha \right| -1\right| }{2}}\left(\frac{i \sqrt{\left| \beta \right| }}{r}\right)}{\sqrt{\pi }}\Biggr).
\end{eqnarray}
}
For a vanishing solution in the UV region one must set $c_1=0$, then the solution can be approximated as:
\begin{eqnarray}
\label{q1solution_r_rUV}
&&q_1(r\rightarrow r_{UV})={c^{UV}_{2_{q_1}}} \left(\frac{-\frac{3 \sqrt{\frac{3}{2}} {g^{UV}_s}^{3/2} {m_0}^2 {M^{UV}} {N^{UV}_f} {r_h}^2 \alpha _{\theta _1}^4}{16 \pi ^{3/2}
   N^{2/5} \alpha _{\theta _2}^3}-\frac{{m_0}^2 {r_h}^2}{12}}{r^6}+\frac{1}{r^4}\right).
\end{eqnarray}\\\\\\
\noindent$\bullet\delta {\rm R[x^1,x^1]}$\\\\\\
EOM(\ref{EOM22}) near $r=r_h$ was written as:
\begin{eqnarray}
\label{q5-near-rh}
& & \hskip -0.3in{q_5}''(r)+\frac{{q_5}'(r)}{ (r-{r_h})}+\frac{2{q_5}(r)}{{r_h} (r-{r_h})} +\gamma_{52}+\gamma_{56}+\gamma_{33}+\frac{\gamma_{51}+\gamma_{55}+\gamma_{32}}{r-{r_h}}=0,
\end{eqnarray}
where:
{\small
\begin{eqnarray}
\label{gamma51---gamma56_defs}
& & \gamma_{32} \equiv -\frac{2 {\alpha_4}}{{\beta_3}}\nonumber\\
& & \gamma_{33} \equiv -\frac{{\alpha_4} \left(\frac{243 \sqrt{6} b^2 \left(9 b^2-1\right) {g_s}^{3/2} M \left(\frac{1}{N}\right)^{2/5} {N_f} \alpha _{\theta _1}^4 \log ^2({r_h})}{\pi
   ^{3/2} \left(3 b^2-1\right) \alpha _{\theta _2}^3}+\frac{-486 b^6+261 b^4+90 b^2+7}{-162 b^6 {r_h}+9 b^4 {r_h}+12 b^2 {r_h}+{r_h}}\right)}{{\beta_3}}\nonumber\\
& & \gamma_{51} \equiv \frac{100 {a_2} {c_{1_{q4}}}}{{K^1}^2}\nonumber\\
& & \gamma_{52} \equiv  \frac{25 {a_2} {c_{1_{q4}}} \left(-3 {a_2}-\frac{2 \left(\frac{-1134 b^6+297 b^4+138 b^2+11}{54 b^4 {r_h}+15 b^2 {r_h}+{r_h}}-\frac{243 \sqrt{6}
   b^2 \left(9 b^2-1\right) {g_s}^{3/2} M \left(\frac{1}{N}\right)^{2/5} {N_f} \alpha _{\theta _1}^4 \log ^2({r_h})}{\pi ^{3/2} \alpha _{\theta _2}^3}\right)}{3
   b^2-1}\right)}{{K^1}^2}\nonumber\\
& & \gamma_{55} \equiv \frac{25 \left(6 b^2+1\right) {g_s} {C_{1_{q4}}} \left(3 {g_s}^2 {\log (N)} M^2 {N_f} \log ({r_h})-16 \pi ^2 N\right)}{2 \left(9 \pi  b^2+\pi \right)
   {r_h}^3}\nonumber\\
& & \gamma_{56} \equiv \frac{25 {g_s} {c_{1_{q4}}} \left(54 b^4 (2 {a_2} {r_h}+3)+b^2 (30 {a_2} {r_h}+33)+2 {a_2} {r_h}+3\right) \left(16 \pi ^2 N-3 {g_s}^2
   {\log (N)} M^2 {N_f} \log ({r_h})\right)}{4 \pi  \left(9 b^2+1\right)^2 {r_h}^4},\nonumber
\end{eqnarray}}

solution of (\ref{q5-near-rh})was obtained as:
{\small
\begin{eqnarray}
\label{q5-solution-near-rh}
 q_5(r\sim r_h)& =& \frac{1}{2} \left(2 \sqrt{2} {c_{1_{q5}}}-{\gamma_{51}} {r_h}-{\gamma_{55}} {r_h}-{\gamma_{32}} {r_h}\right)\nonumber\\
& & +\frac{1}{4} (r-{r_h})^2 \left(\frac{4
   \sqrt{2} {c_{1_{q5}}}}{{r_h}^2}-{\gamma_{52}}-{\gamma_{56}}-{\gamma_{33}}\right)-\frac{2 \sqrt{2} {c_{1_{q5}}} (r-{r_h})}{{r_h}}.\nonumber\\
   &&
\end{eqnarray}}
 Imposing Neumann boundary condition $q_5'(r=r_h)=0$, required setting $c_2=0$ and $c_{1_{q5}}=N^{-\alpha_5}$, $\alpha_5>0$.
In the UV region($r>r_h$), (\ref{EOM22}) was approximated as:
{\small
\begin{eqnarray}
\label{EOM22_UV}
&&{q_5}''(r)+\frac{0.75}{r}{q_5}'(r)-\frac{192019. b^4 {g^{UV}_s}^{5/2} {M^{UV}}^3 N^{4/5} {r_h}^2}{{\log(N)}^5 {m_0}^2 r^2}{q_5}(r)+\frac {12. {c^{UV} _ {1 _ {q_ 3}}}} {r} - \frac {1256.64 {g^{UV} _s} \
N {c^{UV} _ {2 _ {q_ 1}}}} {{m_ 0}^2 r^6 {r_h}^2} \nonumber\\
&&+ \frac {3769.91 \
{g^{UV} _s} N {c^{UV}_ {2 _ {q_ 4}}}} {{m_ 0}^2 r^6 {r_h}^2}=0
\end{eqnarray}}
after two consecutive expansion in `large-r' and 'large-$N$' the solution of preceding equation was obtained as:
{\small
\begin{eqnarray}
\label{q5solution_UV}
&&q_5(r\rightarrow r_{UV})=\frac{{\log(N)}^5 \sqrt[5]{N} (0.0196\  {c^{UV}_{2_{q_4}}}-0.0065\  {c^{UV}_{2_{q_1}}})}{b^4 {g^{UV}_s}^{3/2} {M^{UV}}^3 r^4 {r_h}^4}
\end{eqnarray}}
\noindent$\bullet\delta {\rm R[x^1,r]}$
(\ref{EOM25}) around $r=r_h$ was written as:
\begin{equation}
\label{q3-EOM-near-rh}
{q_3}'(r) + \frac{({\alpha_1} + {\alpha_4}+\gamma_1)}{ (r-{r_h})}+\frac{{\beta_1}}{(r-{r_h})^2}+\frac{{\beta_3}
	{q_3}(r)}{r-{r_h}}=0,
\end{equation}
where:
\begin{eqnarray}
\label{alpha1+alpha4+beta1+beta3-defs}
& & \alpha_1 \equiv  -\frac{25 {a_2} \left(6 b^2+1\right) {g_s} {c_{1_{q_1}}} \left(16 \pi ^2 N-3 {g_s}^2 {\log (N)} M^2 {N_f} \log ({r_h})\right)}{42 \pi  \left(9
   b^2+1\right) {r_h}^2}\nonumber\\
& & \alpha_4 \equiv-\frac{25 {a_2} \left(6 b^2+1\right) {g_s} {c_{1_{q4}}} \left(3 {g_s}^2 {\log (N)} M^2 {N_f} \log ({r_h})-16 \pi ^2 N\right)}{21 \pi  \left(9
   b^2+1\right) {r_h}^2}\nonumber\\
& & \beta_1 \equiv \frac{25 \left(6 b^2+1\right) {g_s} {c_{1\ {q_1}}} \left(16 \pi ^2 N-3 {g_s}^2 {\log (N)} M^2 {N_f} \log ({r_h})\right)}{84 \left(9 \pi  b^2+\pi
   \right) {r_h}^2}\nonumber\\
   &&\gamma_1 \equiv \frac{25 {g_s} {c_{1\ {q_1}}} \left(54 b^4 ({a_2} {r_h}+3)+3 b^2 (5 {a_2} {r_h}+13)+{a_2} {r_h}+3\right) \left(3 {g_s}^2 {\log (N)}
   M^2 {N_f} \log ({r_h})-16 \pi ^2 N\right)}{84 \pi  \left(9 b^2+1\right)^2 {r_h}^3}\nonumber\\
& & \beta_3 = \frac{13}{42},
\end{eqnarray}

solution of (\ref{q3-EOM-near-rh})was obtained as:
\begin{equation}
\label{q3-solution-near-rh}
q_3(r\sim r_h) = -\frac{({\alpha_1}+{\alpha_4}+{\gamma_1})}{\beta_3}-\frac{{\beta_1}}{(1-{\beta_3})(r-{r_h})}+c_{q_3}
(r-{r_h})^{-{\beta_3}}.
\end{equation}
 Imposing Neumann boundary condition at $r=r_h$, required setting $c_{q_3}=0$.
For ${c_{1\ {q_1}}}$ only $\alpha_4$, $\beta_3$, and $\gamma_3$ gave a non-zero value,
\begin{equation}
\label{q3-solution-near-rh-1}
q_3(r\sim r_h) = -\frac{{\alpha_4}}{\beta_3}.
\end{equation}
(\ref{EOM25}) in the UV region ($r>r_h $) was approximated as:
{\small
\begin{eqnarray}
\label{q3EOM_UV}
&&r^4 \Biggl(r^7 \alpha _{\theta _2}^3 \left(18287.5 b^4 {g^{UV}_s}^{5/2} {M^{UV}}^3 N^{4/5} {r_h}^2 {q_3}(r)+1. {\log(N)}^5 {m_0}^2 r {q_3}'(r)\right)+{g_s}
   {\log(N)}^5 {m_0}^2 N {c^{UV}_{2_{q_4}}}\nonumber\\
   && \left(239.359 \alpha _{\theta _2}^3-177.683 {g^{UV}_s}^{3/2} M \left(\frac{1}{N}\right)^{2/5} {N^{UV}_f} \alpha
   _{\theta _1}^4\right) r^{\frac{27 \sqrt{\frac{3}{2}} {g^{UV}_s}^{3/2} M \left(\frac{1}{N}\right)^{2/5} {N^{UV}_f} \alpha _{\theta _1}^4}{2 \pi ^{3/2} \alpha _{\theta
   _2}^3}}\Biggr)\nonumber\\
   && +{g^{UV}_s} {\log(N)}^5 {m_0}^2 N {c^{UV}_{2_{q_1}}} r^{\frac{27 \sqrt{\frac{3}{2}} {g^{UV}_s}^{3/2} M \left(\frac{1}{N}\right)^{2/5}
   {N^{UV}_f} \alpha _{\theta _1}^4}{2 \pi ^{3/2} \alpha _{\theta _2}^3}} \Biggl(88.8414 {g^{UV}_s}^{3/2} M \left(\frac{1}{N}\right)^{2/5} {N^{UV}_f} r^4 \alpha _{\theta
   _1}^4\nonumber\\
   &&+\alpha _{\theta _2}^3 \left(59.8399 {r_h}^4-119.68 r^4\right)\Biggr)\nonumber\\
   &&
\end{eqnarray}}

after two consecutive large expansion in `large-r' and `large-$N$' the solution of preceding equation was obtained as:
\begin{eqnarray}
\label{q3solution_UV}
&&q_3(r\rightarrow r_{UV})=\frac{0.00654434\  {\log(N)}^5 {m_0}^2 \sqrt[5]{N} ({c^{UV}_{2_{q_1}}} -2 {c^{UV}_{2_{q_4}}} )}{b^4 {g^{UV}_s}^{3/2} {M^{UV}}^3 r^7
   {r_h}^2}+{c^{UV}_{1_{q_3}}}
\end{eqnarray}
\noindent$\bullet\delta {\rm R[x^3,x^3]}$\\
This resulted in an EOM for $q_4(r)$ which was identical to $q_1(r)$ for both UV and IR region.

%

\noindent $\bullet\delta {\rm R[r,r]}$\\
\begin{eqnarray}
\label{EOM55}
& & {q_1}'(r)=0.
\end{eqnarray}
This along with the $\delta {\rm R[t,t]}$ EOM implies that $c_{1\ q_1}$ is vanishingly small. In the calculations for exotic scalar glueball decay widths we set it to zero.

\noindent$\bullet\delta {\rm R[\theta_1,\theta_1]}$
\begin{eqnarray}
\label{EOM66}
& & {q_2}(r)-\frac{49 \pi ^3 N^{3/5} r^2 \alpha _{\theta _2}^2 \left(6 a^2+r^2\right)\left(1-\frac{{r_h}^4}{r^4}\right)}{216 {g_s}^3 M^2 {N_f}^2 \left(9 a^2+r^2\right) \log ^2(r) \left(108 a^2+r\right)^2}q_6(r) = 0.
\end{eqnarray}




\noindent$\bullet\delta {\rm R[\theta_2,\theta_2]}$\\
\begin{eqnarray}
\label{EOM77}
& & {q_6}(r)\Bigg( -\frac{3 {g_s}^2 {\log N} M^2 {N_f} \log (r)\left(1-\frac{{r_h}^4}{r^4}\right)}{32 \pi ^2 N}+\frac{{r_h}^4}{r^4}-1\Bigg)+{q_2}(r)=0.
\end{eqnarray}

\noindent$\bullet\delta {\rm R[\theta_1,\theta_2]}$\\
\begin{eqnarray}
\label{EOM67}
& & -\frac{49 \sqrt{3} \pi ^{3/2} \sqrt[5]{N} r \left(6 a^2+r^2\right) \left(1-\frac{{r_h}^4}{r^4}\right)\left(36 a^2 \log (r)+r\right){q_6}(r)}{\sqrt{2}32 {g_s}^{3/2} M {N_f} \alpha _{\theta _2} \left(9
   a^2+r^2\right) \log (r) \left(108 a^2 +r\right)^2}+{q_2}(r) = 0.
\end{eqnarray}



We see that (\ref{EOM66}) - (\ref{EOM67}) are identically satisfied by setting $q_2(Z) = q_6(Z) = 0$.

\noindent $\bullet$ All other remaining equations $\delta {\rm R[m,n]}$ for $(m,n)\in\left\{\theta_{1,2},x,y,z,x^{11}\right\}$ are automatically satisfied provided:
\begin{equation}
\label{EOMS-allothercompact}
\frac{1}{2} (K^1)\ ^2 {q_3}(r)+\frac{1}{4} (K^1)\ ^2 {q_5}'(r)+{q_1}'(r)-3 {q_4}'(r)=0.
\end{equation}
In the IR, near $r=r_h$, by plugging in solutions for functions $q_{3,4,5}(Z)$, one sees that (\ref{EOMS-allothercompact}) is identically satisfied.

\section{Meson Sector}

To start off our study of glueball-meson interaction in the type IIA background we first have to understand how the mesons are obtained in the theory. As explained in {\bf Chapter 3} meson sector in the type IIA dual background of top-down holographic type IIB setup\cite{MQGP} is given by the flavor D6-branes action. We explained the embedding of D6 branes in the mirror(constructed in \cite{MQGP}) of the resolved warped deformed conifold of \cite{metrics}. To obtain the pullback metric and the pullback NS-NS flux on the D6 branes, we choose the first branch of the Ouyang embedding where $(\theta_{1},x)=(0,0)$ and we considered the `z' coordinate as a function of r, i.e z(r)\cite{Dasgupta_et_al_Mesons}. In \cite{Yadav+Misra+Sil-Mesons} a diagonal metric $\{t,x^1,x^2,x^3,r,\theta_1,\theta_2,\tilde{x},\tilde{y},\tilde{z}\}$was used to obtain the mirror of the Ouyang embedding, but it turns out that the embedding conditions remains same even with the nondiagonal basis $\{t,x^1,x^2,x^3,r,\theta_1,\theta_2,x,y,z\}$ . For $\theta_{1}=\alpha_{\theta_{1}}N^{-\frac{1}{5}}$ and $\theta_{2}=\alpha_{\theta_{2}}N^{-\frac{3}{10}}$ one will assume that the embedding of the D6-brane will be given by $\iota :\Sigma ^{1,6}\Bigg(t, R^{1,2},r,\theta_{2}\sim\frac{\alpha_{\theta_{2}}}{N^{\frac{3}{10}}},y\Bigg)\hookrightarrow M^{1,9}$, effected by: $z=z(r)$. As obtained in \cite{Yadav+Misra+Sil-Mesons} one sees that z=constant is still a solution and by choosing $z=\pm {\cal C}\frac{\pi}{2}$, one can choose the $D6/\bar{D6}$-branes to be at ``antipodal" points along the z coordinate.\\
\subsection{Radial Profile Function $\psi_1(Z)$ for $\rho$-Meson}

Up to NLO in $N$:
{\footnotesize
\begin{eqnarray}
\label{V1NLO}
& & {\cal V}_1(Z) = \frac{1}{108 \pi ^2 {\log N} \alpha _{\theta _1}^3 \alpha _{\theta _2}^2}\Biggl\{M \sqrt[5]{\frac{1}{N}} {N_f} e^{-4 Z} {e^{4 Z}-1} \left(2 \sqrt[5]{\frac{1}{N}} \alpha _{\theta _2}^2+81 \alpha _{\theta
		_1}^2\right)\nonumber\\
	& & \log \left({r_h} e^Z\right) \Biggl(3 \log \left({r_h} e^Z\right) \left(3 a^2 \left({g_s} {N_f} \left(8 {\log (N)} {r_h} e^Z-1\right)+32 \pi  {r_h} e^Z\right)-2 {g_s} {N_f}
   {r_h}^2 e^{2 Z}\right)+3 a^2 ({g_s} ({\log (N)}-3) {N_f}+4 \pi )\nonumber\\
   &&-216 a^2 {g_s} {N_f} {r_h} e^Z \log ^2\left({r_h} e^Z\right)+2
   {r_h}^2 e^{2 Z} ({g_s} {\log (N)} {N_f}+4 \pi )\Biggr)\Biggr\},
\end{eqnarray}}
and
\begin{eqnarray}
\label{V2NLO}
& & {\cal V}_2(Z) = \frac{1}{54 \pi  {\log N} {r_h}^2 \alpha _{\theta _1}^3 \alpha _{\theta _2}^2}\Biggl\{{g_s} M N^{3/5} {N_f} \left(81 \sqrt[5]{N} \alpha _{\theta _1}^2+2 \alpha _{\theta _2}^2\right)\nonumber\\
& & \log
	\left({r_h} e^Z\right) \Biggl(3 a^2 e^{-2 Z} \Biggl((3 \log \left({r_h} e^Z\right) \left({g_s} {N_f} \left(8 {\log (N)} {r_h} e^Z+1\right)+32 \pi  {r_h} e^Z\right)-{g_s}
   ({\log (N)}+3) {N_f}\nonumber\\
   &&-72 {g_s} {N_f} {r_h} e^Z \log ^2\left({r_h} e^Z\right)-4 \pi \Biggr)+2 {r_h}^2 \left({g_s} {\log (N)} {N_f}-3
   {g_s} {N_f} \log \left({r_h} e^Z\right)+4 \pi \right)\Biggr)\Biggr\}.
\end{eqnarray}
For $g(Z)\equiv\sqrt{{\cal V}_1(Z)}\psi_1(Z)$ the EOM for $\psi(Z)$ takes the form of Schr\"{o}dinger-like equation with a potential given by (\ref{V-psi1}). The aforementioned Schr\"{o}dinger-like equation near the horizon, $Z=0$, can be written as:
\begin{equation}
\label{near_Z=0-EOM-redefined_psi1}
g''(Z)+g(Z) \left(\frac{\omega_1}{Z}+\omega_2+\frac{1}{4 Z^2}\right)=0,
\end{equation}
wherein:
{\footnotesize
\begin{eqnarray}
\label{omega_1-and-2_defs}
& & \omega_1\equiv \frac{1}{4} \left({m_0}^2-3 b^2 \left({m_0}^2-2\right)\right)+18 b^2 {r_h} \log
   ({r_h})-\frac{3 b \gamma  {g_s} M^2 \left({m_0}^2-2\right) \log ({r_h})}{2 N}+\frac{36 b
   \gamma  {g_s} M^2 {r_h} \log ^2({r_h})}{N},\nonumber\\
& & \omega_2\equiv -\frac{4}{3}+\frac{3}{2} b^2 \left({m_0}^2+72 {r_h}-4\right)-36 b^2 {r_h} \log ({r_h})+\frac{3 b \gamma
   {g_s} M^2 \left({m_0}^2-4\right) \log ({r_h})}{N}-\frac{72 b \gamma  {g_s} M^2 {r_h}
   \log ^2({r_h})}{N}.\nonumber\\
& &
\end{eqnarray}}
The solution to (\ref{near_Z=0-EOM-redefined_psi1}) was obtained as:
\begin{eqnarray}
\label{solution-redefined-psi1-near_Z=0}
 & & g(Z) = \tilde{c}_{1\ \psi_1} M_{-\frac{i {\omega_1}}{2 \sqrt{{\omega_2}}},0}\left(2 i \sqrt{{\omega_2}} Z\right) + \tilde{c}_{2\ \psi_1}
   W_{-\frac{i {\omega_1}}{2 \sqrt{{\omega_2}}},0}\left(2 i \sqrt{{\omega_2}} Z\right).
\end{eqnarray}
Now,
{\footnotesize
\begin{eqnarray}
\label{1oversqrtV1}
&&\frac{1}{\sqrt{{\cal V}_1}} = \frac{2 \pi  \sqrt[10]{N}}{\sqrt{3} \sqrt{{\cal D}}} + {\cal O}\left(N^{-\frac{1}{10}}\right)\nonumber\\
		& & \hskip-0.2in=\frac{\pi  \sqrt[10]{N}}{\sqrt{3} \sqrt{Z} \sqrt{\frac{M {N_f} {r_h}^2 \log ({r_h}) (3 \log
   ({r_h}) ({g_s} {N_f} (3 b^2 (8 {\log (N)} {r_h}-1)-2)+96 \pi  b^2
   {r_h})+3 b^2 ({g_s} ({\log (N)}-3) {N_f}+4 \pi )-216 b^2 {g_s} {N_f} {r_h}
   \log ^2({r_h})+2 {g_s} {\log (N)} {N_f}+8 \pi )}{{\log (N)} \alpha _{\theta _1} \alpha
   _{\theta _2}^2}}} \nonumber\\
				& &  +	{\cal O}\left(N^{-\frac{1}{10}},Z^{\frac{3}{2}}\right),
\end{eqnarray}}
where:
{\small
\begin{eqnarray}
\label{cal D-def}
& &  {\cal D} \equiv  \frac{1}{{\log N} \alpha _{\theta _1} \alpha _{\theta _2}^2}\Biggl\{M {N_f} {r_h}^2 e^{-4 Z} \sqrt{e^{4 Z}-1} (\log(e^Zr_h))\nonumber\\
& & \Biggl(3 (\log ({r_h})+Z) \left(3 b^2 \left({g_s} {N_f} \left(8 {\log (N)} {r_h} e^Z-1\right)+32 \pi
    {r_h} e^Z\right)-2 {g_s} {N_f} e^{2 Z}\right)+3 b^2 ({g_s} ({\log (N)}-3) {N_f}+4 \pi
   )\nonumber\\
   &&-216 b^2 {g_s} {N_f} {r_h} e^Z (\log ({r_h})+Z)^2+2 e^{2 Z} ({g_s} {\log (N)}
   {N_f}+4 \pi )\Biggr)\Biggr\}
\end{eqnarray}}
Thus:
\begin{eqnarray}
\label{solution-redefinied-psi1-near-Z=0}
& & \psi_1(Z) = Z^{-\frac{1}{2}}\Biggl[c_{1\ \psi_1} M_{-\frac{i {\omega_1}}{2 \sqrt{{\omega_2}}},0}\left(2 i \sqrt{{\omega_2}} Z\right)+c_{2\ \psi_1}
   W_{-\frac{i {\omega_1}}{2 \sqrt{{\omega_2}}},0}\left(2 i \sqrt{{\omega_2}} Z\right)\Biggr],
\end{eqnarray}
which yields:
\begin{eqnarray}
\label{psi1'}
 \psi_1^\prime(Z) &=&\frac{1}{{\sqrt{2} \sqrt{i \sqrt{{\omega_2}}} Z \Gamma \left(\frac{i {\omega_1}}{2
   \sqrt{{\omega_2}}}-\frac{1}{2}\right) \Gamma \left(\frac{i {\omega_1}}{2
   \sqrt{{\omega_2}}}+\frac{1}{2}\right)}}\nonumber\\
   &&\times\Biggl\{i c_2 \Biggl(2 \sqrt{{\omega_2}} \Gamma \left(\frac{i {\omega_1}}{2
   \sqrt{{\omega_2}}}+\frac{1}{2}\right) \Biggl(\psi ^{(0)}\left(\frac{i {\omega_1}}{2
   \sqrt{{\omega_2}}}-\frac{1}{2}\right)\nonumber\\
   &&+\log \left(2 i \sqrt{{\omega_2}}\right)+\log (Z)+2 \gamma
   \Biggr)\nonumber\\
   &&+\left(\sqrt{{\omega_2}}-i {\omega_1}\right) \Gamma \left(\frac{i {\omega_1}}{2
   \sqrt{{\omega_2}}}-\frac{1}{2}\right) \Biggl(\psi ^{(0)}\left(\frac{i {\omega_1}}{2
   \sqrt{{\omega_2}}}+\frac{1}{2}\right)+\log \left(2 i \sqrt{{\omega_2}}\right)+\log (Z)\nonumber\\
   &&+2 \gamma
   \Biggr)\Biggr)\Biggr\}+ ....
\end{eqnarray}
In order to ensure that the $\frac{1}{Z}$ term's coefficient in $\psi_1^\prime(Z\sim0)$ vanishes, we set:
\begin{equation}
\label{quantization}
-\frac{1}{2}+\frac{i \omega_1}{2 \sqrt{\omega_2}}=-1,
\end{equation}
which implies $\omega_1= i \sqrt{\omega_2}$, and:
\begin{eqnarray}
\label{m_0}
& & m_0 = 2.479 +2.911 {r_h} \log ({r_h}) -\frac{0.289 \gamma  {g_s} M^2 \log ({r_h})}{N}.
\end{eqnarray}
Near $Z=0$ one requires to set $c_{2\ \psi_1}=0$ for well-behaved $\psi_1'(Z)$. Therefore:
\begin{eqnarray}
\label{psi1-quant}
& & \psi_1(Z) = -\frac{{c_{\psi_1}} \sqrt{i \sqrt{{\omega_2}}} {\omega_2} Z^2}{\sqrt{2}}-\sqrt{2} {c_{\psi_1}} \left(i
   \sqrt{{\omega_2}}\right)^{3/2} Z+\sqrt{2} {c_{\psi_1}} \sqrt{i \sqrt{{\omega_2}}},
\end{eqnarray}
and
\begin{eqnarray}
\label{psi1'-quant}
& & \psi_1^\prime(Z) = -\sqrt{2} {c_{\psi_1}} {\omega_2} \sqrt{i \sqrt{{\omega_2}}} Z-\sqrt{2} {c_{\psi_1}} \left(i
   \sqrt{{\omega_2}}\right)^{3/2}.
\end{eqnarray}
Imposing Neumann boundary condition at $Z=0$, required  setting: $ c_{1_{\ \psi_1}} = c_{\psi_1} = N^{-\Omega_\psi}, \Omega_\psi>1$. Also, for $b=0.57,$ $|\omega_2|={\cal O}\left(r_h\log r_h, \frac{g_s M^2}{N}r_h(\log r_h)^2\right)<<1$.

\subsection{Radial Profile Function $\phi_0(Z)$ for $\pi$-Meson}

Near $Z=0$:
{\footnotesize
\begin{eqnarray}
\label{phi0-1}
& & \hskip -0.2in \phi_0(Z) = \frac{{\cal C}_{\phi_0}}{{\cal V}_1(Z)} = \frac{\sqrt[5]{N} \left(\frac{\pi ^2 {\cal C}_{\phi_0} \alpha _{\theta _1} \alpha _{\theta _2}^2}{3 \left(3 b^2+2\right) {g_s} M
   {N_f}^2 {r_h}^2 \log ({r_h})}-\frac{\pi ^2 {\cal C}_{\phi_0} \alpha _{\theta _1} \alpha _{\theta _2}^2}{\left(3
   b^2+2\right) {g_s} {\log (N)} M {N_f}^2 {r_h}^2}\right)}{Z}\nonumber\\
   & & \hskip -0.2in +\frac{\frac{2 \pi ^2 {\cal C}_{\phi_0} \alpha _{\theta _2}^4}{81 \left(3 b^2+2\right) {g_s} {\log(N)} M {N_f}^2 {r_h}^2
   \alpha _{\theta _1}}+\frac{2 \pi ^2 {\cal C}_{\phi_0} \alpha _{\theta _2}^4}{243 \left(3 b^2+2\right) {g_s} M {N_f}^2
   {r_h}^2 \alpha _{\theta _1} \log ({r_h})}}{Z}\nonumber\\
& & \hskip -0.2in +\sqrt[5]{N} \left(\frac{\pi ^2 {\cal C}_{\phi_0} \alpha _{\theta _1} \alpha _{\theta _2}^2}{72 b^2 {g_s}^2 {\log(N)} M {N_f}^2
   {r_h}^3 \log ({r_h})}+\frac{2 \pi ^2 b^2 {\cal C}_{\phi_0} \alpha _{\theta _1} \alpha _{\theta _2}^2}{\left(3 b^2+2\right)^2
   {g_s} M {N_f}^2 {r_h}^2 \log ({r_h})}\right)\nonumber\\
   & & \hskip -0.2in +\Bigg(-\frac{\pi ^2 {\cal C}_{\phi_0} \alpha _{\theta _2}^4}{2916 b^2{g_s}^2 {\log(N)} M {N_f}^2 {r_h}^3 \alpha _{\theta _1} \log
   ({r_h})} -\frac{4 \pi ^2 b^2 {\cal C}_{\phi_0} \alpha _{\theta _2}^4}{81 \left(3 b^2+2\right)^2 {g_s} M {N_f}^2 {r_h}^2
   \alpha _{\theta _1} \log ({r_h})}\Bigg)\nonumber\\
   & & \hskip -0.2in +\frac{\pi ^2 {\cal C}_{\phi_0} \sqrt[5]{N} Z \alpha _{\theta _1} \alpha _{\theta _2}^2 \left(9 b^4 \left(4 {\log(r_h)}^2-6
   {\log(r_h)}+3\right)-12 b^2 \left(5 {\log(r_h)}^2+3 {\log(r_h)}-3\right)-8 {\log(r_h)}^2+12\right)}{9 \left(3 b^2+2\right)^3
   {g_s}^{7/2} M {N_f}^2 \log ^3({r_h})}\nonumber\\
   && \hskip -0.2in -\frac{\pi ^2 {\cal C}_{\phi_0}\sqrt[5]{N} Z^2 \alpha _{\theta _1} \alpha _{\theta _2}^2 \left(432 b^4 {\log(r_h)}^3-18 \left(3
   b^2+2\right)^2 b^2 {\log(r_h)}+3 \left(3 b^2+2\right)^3+4 \left(3 b^2+2\right) \left(9 b^4-15 b^2-2\right) {\log(r_h)}^2\right)}{9
   \left(3 b^2+2\right)^4 {g_s} {\log(r_h)}^4 M {N_f}^2 {r_h}^2}.\nonumber\\
   & &
\end{eqnarray}}
Let $Z\rightarrow\epsilon\rightarrow0$. Then, in (\ref{phi0-1}), one can demonstrate that the summation of all terms, with the exception of ${\cal O}(N^{\frac{1}{5}}Z^{2})$, assuming $\log N > 3 |\log r_h|$, vanishes provided:
\begin{eqnarray}
\label{alpha_theta_2-2}
& & \alpha _{\theta _2}=\frac{9 \sqrt[10]{N} \alpha _{\theta _1} \sqrt{\log ^2({r_h}) ({\log(N)}+3 \log ({r_h}))}}{\sqrt{2} \sqrt{\log ^2({r_h})
   ({\log(N)}-3 \log ({r_h}))}}.
\end{eqnarray}
Therefore:
\begin{equation}
\label{phi0-3}
\phi_0(Z\sim0) =-\frac{0.682249 {\cal C}_{\phi _0} \sqrt[5]{N} Z^2 \alpha _{\theta _1} \alpha _{\theta _2}^2}{{g_s} {|\log(r_h)|} M {N_f}^2
   {r_h}^2}.
\end{equation}

\section{Glueball-Meson Interaction Lagrangian}

After ignoring derivatives and possible indices, the flavour structure of interaction terms for single glueball (as well as multi glueball) cases appearing in the DBI action can be written as follows:
\begin{equation}
\label{vertices}
G_E Tr(\pi^{2}),\ \ G_E Tr(\pi,[\pi,\rho]),\ \ G_E Tr([\pi,\rho]^{2}),\ \ G_E Tr(\rho^{2}),\ \  G_E Tr(\rho,[\rho,\rho]),\ \ G_E Tr([\rho,\rho]^{2}).
\end{equation}
The interaction terms mentioned above are general outcomes for a single glueball case. In the case of multiple-glueballs, the flavour structure remains essentially unchanged. The following sections will discuss the $n=1,0$ modes in the KK expansion of $A_\mu, A_Z$. We obtained the glueball-meson interaction \ref{M-th perturbation} by substituting all the fluctuations for the metric in the $D6$-brane action. We considered only linear interaction terms in the glueball field $G_E$, as we are interested in single glueball decays.
The DBI action on the $D6$-brane is expressed in terms of a ten-dimensional type IIA metric and dilaton field. The glueball modes and dilaton field for type-IIA background were calculated using witten's relation in terms of 11-dimensional M theory metric perturbations. The perturbed type IIA field components and dilaton in terms of M-theory metric components were obtained as:
\begin{eqnarray}
\label{type IIA components}
&& g^{IIA}_{tt}=\sqrt{g^{\cal M}_{11,11}}\left[\left(1+\frac{h_{11,11}}{2g^{\cal M}_{11,11}}\right)g^{\cal M}_{tt}+h_{tt}\right]\nonumber\\
&& g^{IIA}_{rr}=\sqrt{g^{\cal M}_{11,11}}\left[\left(1+\frac{h_{11,11}}{2g^{\cal M}_{11,11}}\right)g^{\cal M}_{rr}+h_{rr}\right]\nonumber\\
&& g^{IIA}_{ab}=\sqrt{g^{\cal M}_{11,11}}\left[\left(1+\frac{h_{11,11}}{2g^{\cal M}_{11,11}}\right)g^{\cal M}_{ab}+h_{ab}\right]\nonumber\\
&& g^{IIA}_{ra}=\sqrt{g^{\cal M}_{11,11}}\left[\left(1+\frac{h_{11,11}}{2g^{\cal M}_{11,11}}\right)g^{\cal M}_{ra}+h_{ra}\right]\nonumber\\
&& g^{IIA}_{yy}=\sqrt{g^{\cal M}_{11,11}}\left[\left(1+\frac{h_{11,11}}{2g^{\cal M}_{11,11}}\right)g^{\cal M}_{yy}\right]\nonumber\\
&& g^{IIA}_{\theta_{2}y}=\sqrt{g^{\cal M}_{11,11}}\left[\left(1+\frac{h_{11,11}}{2g^{\cal M}_{11,11}}\right)g^{\cal M}_{\theta_{2}y}\right]\nonumber\\
&& g^{IIA}_{\theta_{2}\theta_{2}}=g^{\cal M}_{11,11}\sqrt{g^{\cal M}_{11,11}}\left[\left(1+\frac{3h_{11,11}}{2g^{\cal M}_{11,11}}\right){\cal A}_{\theta_{2}\theta_{2}}\right],
\end{eqnarray}
where $a, b$ run from 1 to 3 corresponding to the spatial components of the metric. Plugging type IIA metric components and the M-theory  perturbations into the $D6$-brane DBI action and, working only upto linear order in perturbation we obtained three different type of terms as:
\begin{equation}
\label{action-types-of-terms}
{\cal L}_{{\cal O}_d(h^0){\cal O}_\phi (h^0){\cal O}_F(h)} + {\cal L}_{{\cal O}_\phi (h^0){\cal O}_F(h^0){\cal O}_d(h)} + {\cal L}_{{\cal O}_d(h^0){\cal O}_F(h^0){\cal O}_\phi (h)}.
\end{equation}
Here ${\cal O}(h^0)$ represent term without any perturbation while ${\cal O}(h)$ represents term with linear order in perturbation. The subscripts d,F, and $\phi$ correspond to part of the integrand of the DBI action from which they are obtained.  ${\cal O}_d$,\ ${\cal O}_\phi$\  and ${\cal O}_F$ corresponds to term obtained from $\sqrt{-{\rm det}(\iota^*(g+B))}$,  $e^{-\phi}$ and,  $g^{-1}Fg^{-1}F$ respectively. Contributions to the interaction lagrangian from these three different terms were obtained as:
\begin{itemize}
\item
${\cal O}_d(h^0){\cal O}_F(h^0){\cal O}_\phi (h):$\\
\end{itemize}
{\small
\begin{eqnarray}
&& {\cal L}_{{\cal O}_d(h^0){\cal O}_F(h^0){\cal O}_\phi (h)}\nonumber\\
&&=\sqrt{-{\cal A}_{\theta_2\theta_2} {g^{\cal M}_{11,11}}^{2} g^{\cal M}_{yy}-B^{IIA}_{\theta_2y}\ ^2+g^{\cal M}_{11,11} {g^{\cal M}_{\theta_2 y}}^2} \sqrt{g^{\cal M}_{11} {g^{\cal M}_{11,11}}^{5/2} g^{\cal M}_{22} g^{\cal M}_{33}
   g^{\cal M}_{rr} g^{\cal M}_{tt} {r}_h^2 e^{2 Z}} \nonumber\\
 &&\Bigg(-\frac{3 e^{-2 Z} \phi_{0}(Z)^2 q_{6}(Z)\partial_\mu   \pi \partial^{\mu }\pi  G_E(x^{1},x^{2},x^{3})}{2
  g^{\cal M}_{11} g^{\cal M}_{rr} {g^{\cal M}_{11,11}}^{7/4} {r}_h^2}-\frac{3 \iota e^{-2 Z}  \psi_{1}(Z)   \phi_{0}(Z)^2 q_{6}(Z) \partial_{\mu }\pi\left[\pi ,\rho ^{\mu }\right]
   G_E(x^{1},x^{2},x^{3})}{g^{\cal M}_{11} g^{\cal M}_{rr} {g^{\cal M}_{11,11}}^{7/4} {r}_h^2}\nonumber\\
   &&-\frac{3 e^{-2 Z}q_{6}(Z) \psi_{1}'(Z)^2 \rho ^{\mu } \rho _{\mu }
   G_E(x^{1},x^{2},x^{3})}{2 g^{\cal M}_{11}g^{\cal M}_{rr} {g^{\cal M}_{11,11}}^{7/4} {r}_h^2}-\frac{3 \psi_{1}(Z)^2
   q_{6}(Z) \tilde{F}_{\mu\nu} \tilde{F}^{\mu\nu} }{4 {g^{\cal M}_{11}}^2{g^{\cal M}_{11,11}}^{7/4}}\Bigg).
\end{eqnarray}}

\begin{itemize}
\item
${\cal O}_\phi (h^0){\cal O}_F(h^0){\cal O}_d (h):$\\
\end{itemize}
{\small
\begin{eqnarray}
&&{\cal L}_{{\cal O}_\phi (h^0){\cal O}_F(h^0){\cal O}_d (h)}=\nonumber\\
& &\Bigg(\frac{1}{2} \sqrt{-{\cal A}_{\theta_{2}\theta_{2}} {g^{\cal M}_{11,11}}^{2}  g^{\cal M}_{yy}-B^{IIA}_{\theta_2y}\ ^2+g^{\cal M}_{11,11} {g^{\cal M}_{\theta_{2}y}}^2} \sqrt{{g^{\cal M}_{x^1x^1}}^3 {g^{\cal M}_{11,11}}^{5/2} g^{\cal M}_{rr}
   g^{\cal M}_{tt} {r}_h^2 e^{2 Z}}\nonumber\\
& & \times \left(3q_{4}(Z)-q_{1}(Z)-q_{2}(Z)-q_{5}(Z)\frac{\partial_{\mu}\partial^{\mu}}{M_g^2}\right)\nonumber\\
   &&-\frac{q_{6}(Z) \left(9 {\cal A}_{\theta_{2}\theta_{2}}
   {g^{\cal M}_{11,11}}^{2} g^{\cal M}_{yy}+5 B^{IIA}_{\theta_2y}\ ^2-7 g^{\cal M}_{11,11} {g^{\cal M}_{\theta_{2}y}}^2\right) \sqrt{{g^{\cal M}_{x^1x^1}}^3 {g^{\cal M}_{11,11}}^{5/2} g^{\cal M}_{rr} g^{\cal M}_{tt} {r}_h^2 e^{2
   Z}}}{4 \sqrt{g^{\cal M}_{11,11} \left({g^{\cal M}_{\theta_{2}y}}^2-g^{\cal M}_{\theta_{2}\theta_{2}}g^{\cal M}_{11,11} g^{\cal M}_{yy}\right)-B^{IIA}_{\theta_2y}\ ^2}}\Bigg)\nonumber\\
   &&\Bigg(\frac{2 e^{-2 Z}}{ {g^{\cal M}_{11,11}}^{7/4}g^{\cal M}_{11} g^{\cal M}_{rr} {r}_h^2}\Bigg) \Bigg(\phi_{0}(Z)^2 \partial_{\mu }\pi \partial^{\mu }\pi +\psi_{1}'(Z)^2\rho ^{\mu } \rho _{\mu }+2\iota \phi_{0}^{2}(Z)\psi_{1}(Z)\partial_{\mu}\pi[\pi,\rho^{\mu}] \Bigg)G_E(x^1,x^2,x^3)\nonumber
   \end{eqnarray}}
   {\small
   \begin{eqnarray}
   &&+\Bigg(\frac{1}{2} \sqrt{-{\cal A}_{\theta_{2}\theta_{2}} {g^{\cal M}_{11,11}}^{2} g^{\cal M}_{yy}-B^{IIA}_{\theta_2y}\ ^2+g^{\cal M}_{11,11} {g^{\cal M}_{\theta_{2}y}}^2} \sqrt{{g^{\cal M}_{x^1x^1}}^3 {g^{\cal M}_{11,11}}^{5/2} g^{\cal M}_{rr}
   g^{\cal M}_{tt} {r}_h^2 e^{2 Z}}\nonumber\\
& & \times  \left(3q_{4}(Z)-q_{1}(Z)-q_{2}(Z)-g^{\cal M}_{11}\sqrt{g^{\cal M}_{11,11}}q_5(Z)\frac{\partial_{\mu}\partial^{\mu}}{M_g^2}\right)\nonumber\\
   && -\frac{q_{6}(Z) \left(9 g^{\cal M}_{\theta_{2}\theta_{2}}
   g^{\cal M}_{11,11} g^{\cal M}_{yy}+5 B^{IIA}_{\theta_2y}\ ^2-7 g^{\cal M}_{11,11} {g^{\cal M}_{\theta_{2}y}}^2\right) \sqrt{{g^{\cal M}_{x^1x^1}}^3 {g^{\cal M}_{11,11}}^{5/2} g^{\cal M}_{rr} g^{\cal M}_{tt} {r}_h^2 e^{2
   Z}}}{4 \sqrt{g^{\cal M}_{11,11} \left({g^{\cal M}_{\theta_{2}y}}^2-g^{\cal M}_{\theta_{2}\theta_{2}}g^{\cal M}_{11,11} g^{\cal M}_{yy}\right)-B^{IIA}_{\theta_2y}\ ^2}}\Bigg)\nonumber\\
   &&\Bigg(\frac{1}{{g^{\cal M}_{11}}^2 g^{\cal M}_{11,11} {g^{\cal M}_{11,11}}^{7/4} }\Bigg) \Bigg(\psi_{1}(Z)^2\tilde{F}_{\mu\nu}\tilde{F}^{\mu\nu}\Bigg)G_E(x^1,x^2,x^3).\nonumber\\
   & &
\end{eqnarray}}
\begin{itemize}
\item
${\cal O}_d(h^0){\cal O}_\phi(h^0){\cal O}_F (h):$\\
\end{itemize}
{\small
\begin{eqnarray}
&& {\cal L}_{{\cal O}_d(h^0){\cal O}_\phi(h^0){\cal O}_F (h)}\nonumber\\
&&=\sqrt{-{\cal A}_{\theta_{2}\theta_{2}} {g^{\cal M}_{11,11}}^2 g^{\cal M}_{yy}-B^{IIA}_{\theta_2y}\ ^2+g^{\cal M}_{11,11} {g^{\cal M}_{\theta_{2}y}}^2} \sqrt{g^{\cal M}_{x^1x^1} {g^{\cal M}_{11,11}}^{5/2} g^{\cal M}_{x^2x^2} g^{\cal M}_{x^3x^3}
   g^{\cal M}_{rr} g^{\cal M}_{tt} {r}_h^2 e^{2 Z}} {g^{\cal M}_{11,11}}^{-3/4}\nonumber\\
   &&\Bigg(\frac{2 e^{-2 Z} \psi_{1}'(Z)^2\rho _{\mu }^2  G_E(x^{1},x^{2},x^{3}) \left(q_{2}(Z)-q_{4}(Z)-q_{6}(Z)\right)}{g^{\cal M}_{rr}g^{\cal M}_{11} {g^{\cal M}_{11,11}} {r}_h^2}+\frac{2 e^{-2 Z}  \psi_{1}'(Z)^2\rho _{\mu } \rho _{\nu } q_{5}(Z)
   \partial^{\mu} \partial^{\nu } G_E(x^{1},x^{2},x^{3})}{ g^{\cal M}_{rr} g^{\cal M}_{11}g^{\cal M}_{11,11}M_g^2 {r}_h^2}\nonumber\\
   &&+\frac{2 e^{-2 Z} {\partial_{\mu }\pi}^2 \phi_{0}(Z)^2 G_E(x^{1},x^{2},x^{3}) \left(q_{2}(Z)-q_{4}(Z)-q_{6}(Z)\right)}{ g^{\cal M}_{11,11} g^{\cal M}_{rr} g^{\cal M}_{11}{r}_h^2}+\frac{2 e^{-2 Z} g^{\cal M}_{11}\phi_{0}(Z)^2 q_{5}(Z) \partial_{\mu }\pi \partial_{\nu }\pi
   \partial^{\mu} \partial^{\nu } G_E(x^{1},x^{2},x^{3})}{ g^{\cal M}_{11,11} g^{\cal M}_{rr}g^{\cal M}_{11} M_g^2 {r}_h^2}\nonumber\\
   &&+\iota {\phi_{0}{Z}}^2 \psi_{1}(Z)\partial_{\mu}\pi\left[\pi,\rho_{\nu}\right]\frac{4e^{-2Z}}{g^{\cal M}_{11}g^{\cal M}_{11,11}g^{\cal M}_{rr}r_{h}^2}q_5(Z)\frac{\partial^\mu\partial^\nu G_{E}(x^1,x^2,x^3)}{M_{g}^2}\nonumber\\
   &&+\iota {\phi_{0}{Z}}^2 \psi_{1}(Z)\partial_{\mu}\pi\left[\pi,\rho^{\mu}\right]\frac{4e^{-2Z}}{g^{\cal M}_{11}g^{\cal M}_{11,11}g^{\cal M}_{rr}r_{h}^2}(-q_4(Z)) \nonumber\\
   &&+\tilde{F}^{\mu\nu} \tilde{F}_{\mu\nu} \psi_{1}(Z)^2 G_E(x^{1},x^{2},x^{3})
   \left(-2 q_{4}(Z)-q_{6}(Z)\right)\nonumber\\
   &&+\frac{2  \psi_{1}(Z)^2 q_{5}(Z)}{{g^{\cal M}_{11}}^2 g^{\cal M}_{11,11}}
   \tilde{F}_{\mu l} {\tilde{F}_{\nu}\ ^ l} \frac{\partial^{\mu}\partial^{\nu}G_E(x^{1},x^{2},x^{3})}{M_g^2}\nonumber\\
  &&-\frac{4 e^{-Z}\psi_{1}(Z)\psi_{1}'(Z)q_{3}(Z)}{ g^{\cal M}_{11}g^{\cal M}_{rr}g^{\cal M}_{11,11}r_{h} } \rho _{\mu }  F_{\nu }\ ^{\mu }  \frac{\partial^{\nu}G_E(x^{1},x^{2},x^{3})}{M_g^2}\Bigg).
\end{eqnarray}}
By combining all of the expressions, we obtained the following  glueball-meson interaction lagrangian :
\begin{eqnarray}
\label{interaction action full}
&&\hskip -0.5in{\rm S}_{int}= {\cal T}{Str}\int \left(\frac{1}{T_h}\right) d^{3}x\Bigg[c_{1}(\partial_\mu\pi)^{2}G_E
+c_{2}\partial_\mu\pi\partial_\nu\pi \frac{\partial^{\mu}\partial^{\nu}}{M^2}G_E\nonumber\\
&& +c_{3}\rho_{\mu}^{2}G_E+c_{4}\rho_\mu\rho_\nu \frac{\partial^{\mu}\partial^{\nu}}{M^2}G_E + c_{5}\tilde{F}_{\mu\nu}\tilde{F}^{\mu\nu}G_E
+c_{6}\tilde{F}_{\mu\rho}\tilde{F}_\nu^{\ \rho}\frac{\partial^{\mu}\partial^{\nu}}{M^2}G_E\nonumber\\
&&+\iota c_{7}\partial_{\mu}\pi [\pi,\rho^\mu]G_E+\iota c_{8}\partial_{\mu}\pi [\pi,\rho_\nu]\frac{\partial^{\mu}\partial^{\nu}}{M^2}G_E + c_{9}\rho_\mu \tilde{F}_{\nu}^{\ \mu}\frac{\partial^\nu G_E}{M^2}\nonumber\\
&&+c_{10}\tilde{F}_{\mu\nu}\tilde{F}^{\mu\nu}G_E +c_{11}\partial_{\mu}\pi\partial^{\mu}\pi G_E +c_{12}\rho_{\mu}\rho^{\mu} G_E +\iota c_{13}\partial_{\mu}\pi [\pi,\rho^\mu]G_E \Bigg],
\end{eqnarray}
where:
\begin{equation}
{\cal T}=\frac{-T_{D_{6}}(2\pi\alpha\prime)^2}{4}\int dyd\theta_{2}\delta\Bigg(\theta_{2}-\frac{\alpha_{\theta_{2}}}{N^{3/10}}\Bigg).
\end{equation}

These are the only interaction terms in the field strength tensor at quadratic order.  Higher order terms can be generated by retaining higher order terms of F in the DBI action. 
 Coefficient expressions $c_{i}'s$ are listed in (\ref{interaction coefficients}).

  \section{Decay widths}
Using established approaches in scattering theory (specifically with regard to multi-particle phase-space integrals: see \cite{Savage-Phase-Space-Lecture-Notes}, \cite{Murayama-Phase-Space}\footnote{We would like to thank M.Dhuria for bringing \cite{Murayama-Phase-Space} to our attention.}), in the following sub-sections, we calculated decay widths for $G_E\rightarrow2\pi,\ G_E\rightarrow2\rho,\ \rho\rightarrow2\pi,\ G_E\rightarrow 4\pi^0,\ G_E\rightarrow\rho+2\pi$ as well as indirect four-$\pi$ decay width associated with $G_E\rightarrow\rho+2\pi\rightarrow 4\pi$ as well as $G_E\rightarrow 2\rho\rightarrow 4\pi$ assuming $M_G>2M_\rho$ for definiteness and specifically concentrating on the potential glueball candidate $f0[1710]$.
\subsection{$G_E\rightarrow2\pi$}
For two body decay, the decay width is given as:
\begin{eqnarray}
 &&\Gamma=\frac{S}{8 m^{2}}|{\cal M}|^{2}
\end{eqnarray}
where ${\cal M}$ is the amplitude for the decay, and $\bold{p}$ is the final momentum of one of the identical particles in the decay product. For $2\pi$ decay in rest frame of glueball, following terms in interaction lagranjian give relevant coupling.
\begin{eqnarray}
{\cal T}\left(\frac{1}{T_h}\right){\rm Str}\left( c_{1}(\partial_\mu\pi)^{2}G_E
+ c_{2}\partial_\mu\pi\partial_\nu\pi \frac{\partial^{\mu}\partial^{\nu}}{M_g^2}G_E + c_{11}\partial_{\mu}\pi\partial^{\mu}\pi G_E\right)
\end{eqnarray}

Considering a specific adjoint index for the pion $\pi^{a}$(a=1,2,3), ${\cal M}$ for two pions $\pi^{1}$ and $\pi^{2}$ as final state particles in glueball's rest frame was obtained as,
\begin{eqnarray}
\iota{\cal M}=-\iota 2\ {\cal T}\left(\frac{1}{T_h}\right)\Bigg( 2c_{1}\iota k_{1\mu}\iota k_{2}^{\mu}+2c_{2}\iota k_{1\mu}\iota k_{2\nu}\frac{\iota k_{g}^{\mu}\iota k_{g}^{\nu}}{{M_g}^{2}}\Bigg)
\end{eqnarray}
when exchanging two final state particles, this factor of 2 takes into account their symmetry. For massless pions $k^{0}=|\bold{k}|=m/2$ for both the particles, which gives:
\begin{eqnarray}
\iota{\cal M}&&=-\iota {\cal T}\left(\frac{1}{T_h}\right)\Bigg(-2c_{1}(k_{10}k_{2}^{0}+k_{1i}k_{2}^{i})+2c_{2}k_{10}k_{20}\frac{k_{g}^{0}k_{g}^{0}}{{M_g}^{2}}\Bigg)\nonumber\\
&&=-\iota {\cal T}\left(\frac{1}{T_h}\right)(k_{1}^{0}k_{2}^{0}(2\eta_{00}^{2}c_{2}-2c_{1}\eta_{00})-2c_{1}\bold{k}_{1}.\bold{k}_{2})\nonumber\\
&&=\frac{-\iota M_g^2}{4} {\cal T}\left(\frac{1}{T_h}\right)(2\eta_{00}^{2}c_{2}-2\eta_{00}c_{1}+2c_{1})\nonumber\\
&&= -\iota \frac{{\cal T}}{2}\left(\frac{1}{T_h}\right){M_g}^{2}\left(2c_{1}+c_{2}\right)
\end{eqnarray}
The decay width over the range $a=1,2,3$ was obtained as follows:
{\small
\begin{eqnarray}
\label{Gto2pi-decay-width}
& & \Gamma_{G_E\rightarrow\pi\pi}=\frac{|2c_{1}+c_{2}|^{2}{M_g}^{2}}{32}{\cal T}^{2}{\left(\frac{1}{T_h}\right)}^{2}\times3\times\frac{1}{2} \approx \frac{3}{64}c_2^2m_0^2\pi^2{\cal T}^2\nonumber\\
& &  {\rm which\ for\ } b\sim0.6:\nonumber\\
& &  = 0.003m_0\Biggl(\frac{1.834 \times 10^{-4} {{\cal C}^{2}_{\phi_0}} N^{7/5} \alpha _{\theta _1}^3 {c_{1_{q_4}}}}{M {N_f}^2 {r_h}^3}+\frac{15.379 {{\cal C}^{UV}_{\phi_0}}^2 {\log(N)}^5 {N^{UV}_f}^2
   \log ({r_h}) ({c^{UV}_{2_{q_1}}}-3.015 {c^{UV}_{2_{q_4}}})}{\sqrt{{g^{UV}_s}} {M^{UV}}^2 {r_h}^6 \alpha _{\theta _1}^3}\Biggl)\nonumber\\
   & &  \equiv 0.003m_0\times \Lambda_{G_E\rightarrow2\pi}.
\end{eqnarray}}
we assumed $|\log r_h| = \frac{f_{r_h}}{3}\log N, 0<f_{r_h}<1$, or equivalently $r_h=N^{-\frac{f_{r_h}}{3}}$.
From \cite{PDG-2018}, $\Gamma_{G_E\rightarrow\pi\pi}/m$ associated with $f0[1710]$  is $\sim 10^{-2}$. Therefore by a suitable choice of ${\cal C}_{\phi_0}, c_{1\ q_4}, {\cal C}^{UV}_{\phi_0}, {c^{UV}_{2_{q_1}}}-3.015 {c^{UV}_{2_{q_4}}}: \Lambda_{G_E\rightarrow2\pi}\sim10$ - implying a constraint on a linear combination of ${\cal C}_{\phi_0}^2c_{1\ q_4}$ and ${{\cal C}^{UV}_{\phi_0}}^2({c^{UV}_{2_{q_1}}}-3.015 {c^{UV}_{2_{q_4}}})$ - we obtained: $\frac{\Gamma_{G_E\rightarrow2\pi}}{m_0}=10^{-2}$ - clearly an exact match with the PDG-2018 results is also similarly possible.

\subsection{$G_E\rightarrow2\rho$}
We considered the onshell decay for $G_E\rightarrow\rho\rho$.
The differential width was given by
$$d\Gamma=\frac{1}{16\pi}\Sigma_{pol}|{\cal M}|^{2}\frac{S}{m^{2}}d\Omega_{k_{1}}$$
where
$${\cal M}={\cal T}\frac{1}{T_h}\epsilon_{\alpha}(k_{1})\epsilon_{\beta}(k_{2})(A\eta^{\alpha\beta}+B^{\alpha\beta})$$
where expression for A and $B^{\alpha\beta}$ were obtained as
{\small
\begin{eqnarray}
\label{A-B}
& & A =
\frac{c_6\left(k_1.k_{\rm g}\right)\left(k_2.k_{\rm g}\right)}{M_g^2}
 -\frac{c_9\left(k_1+k_2\right).k_{\rm g}}{2M_g^2}
 -2 \left(c_5 + c_{10}\right) k_1.k_2+c_3+c_{12}\nonumber\\
 & &  {\rm which\ for}\ q_6(Z)=c_{1\ q_1}=0\ {\rm yields}:\nonumber\\
 & & = c_3 - c_5\left(m_\rho^2 - M_g^2\right) + \frac{c_9}{2} + \frac{c_6}{4}M_g^2\nonumber\\
 & & {\rm For}\ b\sim0.6\ {\rm dominated\ by\ }\frac{c_6}{4}M_g^2;\nonumber\\
& &  B^{\alpha\beta}= \frac{1}{2} c_6 \delta _0^{\beta } k_2^{\alpha } k_g^{\beta }+\frac{1}{2} c_6 \delta _0^{\alpha } k_1^{\beta } k_g^{\alpha }+\frac{c_9 \delta _0^{\beta } k_2^{\alpha } k_g^{\beta
   }}{2 M^2}+\frac{c_9 \delta _0^{\alpha } k_1^{\beta } k_g^{\alpha }}{2 M^2}-\frac{c_4 \delta _0^{\alpha }\delta _0^{\beta } k_g^{\alpha}k_g^{\beta }}{M^2}+\frac{c_6 k_{1}.k_{2}\delta _0^{\alpha }\delta _0^{\beta } k_g^{\alpha}k_g^{\beta }}{M^2}\nonumber\\
& &\hskip 0.5in+2 c_5 k_2^{\alpha } k_1^{\beta }
\end{eqnarray}}
 Now using:
\begin{eqnarray}
   k_{1}.k_{2}&&=\frac{1}{2}(2M_{\rho}^{2}-m^{2})\nonumber\\
   &&=\frac{1}{2}\sqrt{m^{4}\lambda(M_{\rho}^{2},M_{\rho}^{2};m^{2})+4M_{\rho}^{4}}\nonumber\\
   &&|\bold{k_{1}}|=\frac{m}{2}\sqrt{\lambda(M_{\rho}^{2},M_{\rho}^{2};m^{2})}
\end{eqnarray}
we were able to write
\begin{eqnarray}
&&\sum_{\rm pol}={\cal T}^2\Bigg(\frac{1}{T_h}\Bigg)^2\frac{m^{4}}{4M_{\rho}^{4}}\left(A^2\lambda(M_{\rho}^{2},M_{\rho}^{2};m^{2})+8 A^2\frac{M_{\rho}^{4}}{m^{4}}+4\frac{M_{\rho}^{4}}{m^{4}}X(k_{1},k_{2},m,B)\right),
\end{eqnarray}
where:
\begin{eqnarray}
\label{X-1}
& & X=\frac{A c_4 \left(-6 M_g^2 M_{\rho }^2+M_g^4+8 M_{\rho }^4\right)}{4 M_{\rho }^4}-\frac{A c_5
	\left(M_g^2-4 M_{\rho }^2\right) \left(-6 M_g^2 M_{\rho }^2+M_g^4+8 M_{\rho }^4\right)}{2 M_{\rho }^4}\nonumber\\
& & -\frac{A
	c_6 \left(M_g^2+2 M_{\rho }^2\right) \left(-6 M_g^2 M_{\rho }^2+M_g^4+8 M_{\rho }^4\right)}{8 M_{\rho
	}^4}-\frac{A c_9 \left(-6 M_g^2 M_{\rho }^2+M_g^4+8 M_{\rho }^4\right)}{4 M_{\rho }^4}\nonumber\\
& & +\frac{c_4^2
	\left(M_g^2-4 M_{\rho }^2\right){}^2}{16 M_{\rho }^4}-\frac{c_4 c_5 \left(M_g^3-4 M_g M_{\rho
	}^2\right){}^2}{4 M_{\rho }^4}\nonumber\\
& & -\frac{c_4 c_6 \left(-6 M_g^4 M_{\rho }^2+M_g^6+32 M_{\rho }^6\right)}{16
	M_{\rho }^4}-\frac{c_4 c_9 \left(M_g^2-4 M_{\rho }^2\right){}^2}{8 M_{\rho }^4}\nonumber\\
& & +\frac{c_5^2 M_g^4
	\left(M_g^2-4 M_{\rho }^2\right){}^2}{4 M_{\rho }^4}+\frac{c_5 c_6 \left(-5 M_g^6 M_{\rho }^2-4 M_g^4
	M_{\rho }^4+36 M_g^2 M_{\rho }^6+M_g^8\right)}{8 M_{\rho }^4}\nonumber\\
& & +\frac{c_5 c_9 \left(M_g^3-4 M_g M_{\rho
	}^2\right){}^2}{4 M_{\rho }^4}+\frac{c_6^2 \left(-2 M_g^2 M_{\rho }^2+M_g^4-8 M_{\rho }^4\right){}^2}{64
	M_{\rho }^4}\nonumber\\
& & +\frac{c_6 c_9 \left(-6 M_g^4 M_{\rho }^2+M_g^6+32 M_{\rho }^6\right)}{16 M_{\rho
	}^4}+\frac{c_9^2 \left(M_g^2-4 M_{\rho }^2\right){}^2}{16 M_{\rho }^4}.
\end{eqnarray}
For $b\sim0.6$, (\ref{X-1}) recieved most dominant contribution from the terms involving coupling constants, $c_6^2, A c_6$ and $c_4c_6$ terms (if $M_g = 2 M_\rho + \epsilon, 0<\epsilon\ll M_\rho$ then the $c_4^2$ term were further suppressed). Demanding $\Gamma_{G_E\rightarrow2\rho} = \Gamma_{G_E\rightarrow4\pi}$ for $M_g>2M_\rho$ \cite{Brunner_Hashimoto-results}, would require $c_4 = \frac{3}{4}c_6 M_g^2$; so for $M_g = m_0$ MeV $\equiv m_0 \left(\frac{r_h}{\pi\sqrt{4\pi g_s N}}\right)$, $c_4 = \frac{3 m_0^2}{4}c_6\left(\frac{r_h}{\pi\sqrt{4\pi g_s N}}\right)^2$.
\subsection{$\rho\rightarrow2\pi$}
The action's relevant interaction term is given by :
\begin{equation}
\label{rho-pi^2-c_16}
c_{16}{\cal T}\left(\frac{1}{T_h}\right)\int d^3x\partial_\mu\pi[\pi,\rho^\mu],
\end{equation}
where:
\begin{eqnarray}
\label{c_16}
& & c_{16} = \frac{5.61\times 10^{-9} {{\cal C}_{\phi _0}}^2 \sqrt[4]{{\omega_2}} \alpha _{\theta _1} \alpha _{\theta _2}^2 c_{ \psi_1}
   N^{\frac{f_{r_h}}{3}+\frac{1}{5}}}{{g_s} M {N_f}^2}\nonumber\\
	& & -\frac{43017.7 {{{\cal C}^{UV}_{\phi _0}}}^2 {f_{r_h}} {g_s}^{UV} {M^{UV}} \sqrt[5]{\frac{1}{N}} {N_f^{UV}}^2 \log (N) {c^{UV}_{2_{\psi_1}}} N^{-\frac{2 {fr}
   h}{3}}}{\alpha _{\theta _1} \alpha _{\theta _2}^2},
\end{eqnarray}
\begin{eqnarray}
&&\omega_2\equiv \upsilon_2+\frac{\upsilon_1
   {g_s} M^2 \left({m_0}^2-4\right) \log ({r_h})}{N};\nonumber\\
\end{eqnarray}
 $M_{UV}\ll M$ and $N_{f\ UV}\ll N_f$ are the tiny values of the number of fractional $D3$-branes and flavor branes in the UV. The $\rho\rightarrow2\pi$ decay width was hence obtained as under:
\begin{equation}
\label{rho-to-2-rho-Gamma}
\Gamma_{\rho\rightarrow2\pi} = {\cal T}^2\left(\frac{1}{T_h}\right)^2\frac{c_{16}^2}{2}.
\end{equation}
 We demanded $\Gamma_{\rho\rightarrow2\pi} = 149 MeV$ (\cite{PDG-2018}); exchanging MeV by $\frac{r_h}{\pi \sqrt{4 \pi g_s N}}$, this implied a constraint on ${\cal C}_{\phi_0}^2 \left(c_{\psi_1} \right)$ and  ${\cal C}_{\phi_0}^{UV}\ ^2c_{2\ \psi_1}^{UV}$:
\begin{eqnarray}
\label{constraint-rho-2pi-decay}
& & \Biggl[\frac{5.61\times 10^{-9} {{\cal C}_{\phi _0}}^2 \sqrt[4]{{\omega_2}} \alpha _{\theta _1} \alpha _{\theta _2}^2 c_{\psi_1}
   N^{\frac{f_{r_h}}{3}+\frac{1}{5}}}{{g_s} M {N_f}^2}\nonumber\\
	& & -\frac{43017.7 {{{\cal C}^{UV}_{\phi _0}}}^2 {f_{r_h}} {g_s}^{UV} {M^{UV}} \sqrt[5]{\frac{1}{N}} {N_f^{UV}}^2 \log (N) {c^{UV}_{2_{ \psi_1}}} N^{-\frac{2 {fr}
   h}{3}}}{\alpha _{\theta _1} \alpha _{\theta _2}^2}\Biggr]^2 = \frac{298}{{\cal T}^2}\left(\frac{r_h}{\sqrt{4\pi g_s N}}\right)^3.
\end{eqnarray}

\subsection{Direct Glueball Decay to 4$\pi^{0}$s }
To study the glueball decay to four pions, the DBI action must be expanded to quartic order in $F_{\mu \nu}$. Restricted to quartic order, the action reads
\begin{eqnarray}
\label{NLO dbi action}
&&\hskip -0.5in S=-T_{D_{6}}(2\pi\alpha^{\prime})^{4}{\rm Str}\int d^{4}x dZ d\theta_{2} dy \delta \left(\theta_{2}-\frac{\alpha_{\theta_{2}}}{N^{3/10}}\right) e^{-\Phi}\sqrt{-det(\iota^{ *}(g+B))}\nonumber\\
& & \times
\Bigg\{\frac{1}{32}STr\left(g^{-1}Fg^{-1}F)Tr(g^{-1}Fg^{-1}F\right)-\frac{1}{8}STr\left(g^{-1}Fg^{-1}Fg^{-1}Fg^{-1}F\right)\Bigg\}
\end{eqnarray}
By incorporating the metric fluctuations associated with the glueball and retaining the quartic terms in $\phi_{0}(Z)$, the following interaction term were obtained:
\begin{itemize}
\item
${\cal O}_d(h^0){\cal O}_F(h^0){\cal O}_\phi (h):$\\
\end{itemize}

{\small
\begin{eqnarray}
&& {\cal L}_{{\cal O}_d(h^0){\cal O}_F(h^0){\cal O}_\phi (h)} = \sqrt{-{\cal A}^M_{\theta_2\theta_2} {g^{\cal M}_{11,11}}^{2} g^{\cal M}_{yy}-B^{IIA}_{\theta_2y}\ ^2+g^{\cal M}_{11,11} {g^{\cal M}_{\theta_2 y}}^2} \sqrt{g^{\cal M}_{11} {g^{\cal M}_{11,11}}^{5/2} g^{\cal M}_{22} g^{\cal M}_{33}
   g^{\cal M}_{rr} g^{\cal M}_{tt} {r}_h^2 e^{2 Z}}\nonumber\\
& & \times{g^{\cal M}_{11,11}}^{-3/4}\Bigg(\frac{3 e^{-4 Z} \phi_{0}(Z)^4 q_{6}(Z)\partial_\mu   \pi \partial^{\mu }\pi \partial_\nu   \pi \partial^{\nu }\pi  G_E(x^{1},x^{2},x^{3})}{16
   {g^{\cal M}_{rr}}^{2} {g^{\cal M}_{11,11}}^{2} {g^{\cal M}_{11}}^2{r}_h^4}\Bigg)
\end{eqnarray}}
\begin{itemize}
\item
${\cal O}_\phi (h^0){\cal O}_F(h^0){\cal O}_d (h):$\\
\end{itemize}
{\small
\begin{eqnarray}
&&{\cal L}_{{\cal O}_\phi (h^0){\cal O}_F(h^0){\cal O}_d (h)} = \Bigg(\frac{1}{2} \sqrt{-{\cal A}^{M}_{\theta_{2}\theta_{2}} {g^{\cal M}_{11,11}}^{2}  g^{\cal M}_{yy}-B^{IIA}_{\theta_2y}\ ^2+g^{\cal M}_{11,11} {g^{\cal M}_{\theta_{2}y}}^2} \sqrt{{g^{\cal M}_{x^1x^1}}^3 {g^{\cal M}_{11,11}}^{5/2} g^{\cal M}_{rr}
   g^{\cal M}_{tt} {r}_h^2 e^{2 Z}}\nonumber\\
& &  \times \left(3q_{4}(Z)-q_{1}(Z)-q_{2}(Z)-q_{5}(Z)\right)\nonumber\\
   && -\frac{q_{6}(Z) \left(9 {\cal A}^{M}_{\theta_{2}\theta_{2}}
   {g^{\cal M}_{11,11}}^{2} g^{\cal M}_{yy}+5 B^{IIA}_{\theta_2y}\ ^2-7 g^{\cal M}_{11,11} {g^{\cal M}_{\theta_{2}y}}^2\right) \sqrt{{g^{\cal M}_{x^1x^1}}^3 {g^{\cal M}_{11,11}}^{5/2} g^{\cal M}_{rr} g^{\cal M}_{tt} {r}_h^2 e^{2
   Z}}}{4 \sqrt{g^{\cal M}_{11,11} \left({g^{\cal M}_{\theta_{2}y}}^2-{\cal A}^{M}_{\theta_{2}\theta_{2}}g^{\cal M}_{11,11} g^{\cal M}_{yy}\right)-B^{IIA}_{\theta_2y}\ ^2}}\Bigg)\nonumber\\
& & \times{g^{\cal M}_{11,11}}^{-3/4}\Bigg(\frac{- e^{-4 Z}}{ {g^{\cal M}_{11,11}}^{2} {g^{\cal M}_{rr}}^{2}{g^{\cal M}_{11}}^2 {r}_h^4}\Bigg)\Bigg(\phi_{0}(Z)^4 \partial_{\nu }\pi \partial^{\nu }\pi \partial_{\mu }\pi \partial^{\mu }\pi  \Bigg)G_E(x^1,x^2,x^3)
\end{eqnarray}}
\begin{itemize}
\item
${\cal O}_d(h^0){\cal O}_\phi(h^0){\cal O}_F (h):$\\
\end{itemize}
{\small
\begin{eqnarray}
   && {\cal L}_{{\cal O}_d(h^0){\cal O}_\phi(h^0){\cal O}_F (h)} =\nonumber\\
    &&\sqrt{-{\cal A}^{M}_{\theta_{2}\theta_{2}} {g^{\cal M}_{11,11}}^2 g^{\cal M}_{yy}-B^{IIA}_{\theta_2y}\ ^2+g^{\cal M}_{11,11} {g^{\cal M}_{\theta_{2}y}}^2} \sqrt{g^{\cal M}_{x^1x^1} {g^{\cal M}_{11,11}}^{5/2} g^{\cal M}_{x^2x^2} g^{\cal M}_{x^3x^3}g^{\cal M}_{rr} g^{\cal M}_{tt} {r}_h^2 e^{2 Z}} \nonumber\\
   &&{g^{\cal M}_{11,11}}^{-3/4}\Bigg(\frac{e^{-4 Z} {\partial_{\nu }\pi}^2{\partial_{\mu }\pi}^2 \phi_{0}(Z)^4 G_E(x^{1},x^{2},x^{3}) \left(-q_{2}(Z)+q_{4}(Z)+q_{6}(Z)\right)}{ 4{g^{\cal M}_{11,11}}^{2} {g^{\cal M}_{rr}}^{2} {r}_h^4{g^{\cal M}_{11}}^2}\nonumber\\
   & & -\frac{ e^{-4 Z} \partial_{\sigma }\pi\partial^{\sigma }\pi \partial_{\nu }\pi\partial_{\mu }\pi
   \phi_{0}(Z)^4 q_{5}(Z) \partial^{\mu} \partial^{\nu } G_E(x^{1},x^{2},x^{3})}{ 4{g^{\cal M}_{11,11}}^{2} {g^{\cal M}_{rr}}^{2} M_g^2 {r}_h^4{g^{\cal M}_{11}}^2}\Bigg)
\end{eqnarray}}

By combining all of the above expressions and setting $q_2(Z)=q_6(Z)=0$, we obtained the following interaction Lagranjian corresponding to the direct $G_{E}\rightarrow 4\pi$ decay:
\begin{eqnarray}
\label{interaction-G-4pi_direct}
S_{\rm int}^{G_E\rightarrow4\pi} = {\cal T}\left(\frac{1}{T_h}\right){\rm Str}\int d^3x\left(c_{14}\partial_\mu\pi\partial^\mu\pi \partial_\nu\pi\partial^\nu\pi G_E(x^{1,2,3})
+ c_{15}\partial_\sigma\pi\partial^\sigma\pi \partial_\mu\pi\partial_\nu\pi\frac{\partial^\mu\partial^\nu}{M_g^2}G_E(x^{1,2,3})\right),\nonumber\\
& &
\end{eqnarray}
where:
\begin{eqnarray}
\label{c14_c15-IR}
& &  c_{14}^{\rm IR} = \int dZ\Biggl[ -\frac{{\cal G} e^{-4 Z} \phi_0(Z)^4 \left(-\frac{{q_1}(Z)}{2}-\frac{{q_2}(Z)}{2}+\frac{3
		{q_4}(Z)}{2}-\frac{{q_5}(Z)}{2}\right)}{8 g^{\cal M}_{x^1x^1}\ ^2 g^{\cal M}_{11\ 11}\ ^2 g^{\cal M}_{rr}\ ^2 {r_h}^4}\nonumber\\
&&-\frac{{\cal G} e^{-4 Z}
	\phi_0(Z)^4 ({q_2}(Z)-{q_4}(Z)-{q_6}(Z))}{4 g^{\cal M}_{x^1x^1}\ ^2 g^{\cal M}_{11\ 11}\ ^2 g^{\cal M}_{rr}\ ^2 {r_h}^4}\Biggr]\nonumber\\
& &{\rm which\ for}\ b\sim0.6\ {\rm yields}:\nonumber\\
& &  =\frac{6.219\times 10^{-16} {{\cal C}_{\phi _0}}^4 N^{8/5} \alpha _{\theta _1}^3 \alpha _{\theta _2}^6 c_{1_{q4}}}{{g_s}^2 M^3
   {N_f}^6 {r_h}^9 \log ^2({r_h})} \nonumber\\
& &c_{15}^{\rm IR} = \int dZ \Biggl[-\frac{{\cal G} e^{-4 Z} \phi_0(Z)^4 {q_5}(Z)}{4 g^{\cal M}_{x^1x^1}\ ^2 g^{\cal M}_{11\ 11}\ ^2 g^{\cal M}_{rr}\ ^2 {r_h}^4}\Biggr]\nonumber\\
& & = -\frac{11.224\  {{\cal C}_{\phi _0}}^4 N^{8/5}  \alpha _{\theta _1}^3 \alpha _{\theta _2}^6 c_{1_{q4}}}{{g_s}^2 M^3 {N_f}^6 {r_h}^9 \log ^2({r_h})}.
\end{eqnarray}
From (\ref{c14_c15-IR}), $c_{15}^{\rm IR}>c_{14}^{\rm IR}$. which justified dropping $c_{14}$ in the direct $4\pi^0$-decay of the glueball decay. One can show that the contribution from the UV: $Z\in\left[\log\left(\sqrt{3}b\right),\infty\right]$ yields:
\begin{equation}
\label{c15-UV}
c_{15}^{\rm UV} = \frac{638116. {{\cal C}_{\phi _0}^{UV}}^4 {\log (N)}^5 \sqrt[5]{N} {{N_{f}}^{UV}}^2 \log ({r_h}) (  {c_{2_{q1}}}^{UV}-3.015 {c_{2_{q4}}}^{UV})}{\sqrt{{g^{UV}_s}}
   {M^{UV}}^2 {r_h}^8 \alpha _{\theta _1} \alpha _{\theta _2}^2}.
\end{equation}
From (\ref{c14_c15-IR}) and (\ref{c15-UV}):
\begin{eqnarray}
\label{c15-IR+UV}
& & c_{15} = {1.35\times 10^{-13}} N^{21/20} \Biggl(\frac{4.72\times 10^{18} {{\cal C}_{\phi _0}^{UV}}^4 {\log (N)}^5  {{N^{UV}_{f}}}^2 \log ({r_h}) (  {c_{2_{q1}}}^{UV}-3.01538 {c_{2_{q4}}}^{UV})}{\sqrt{{g^{UV}_s}}N^{17/20}
   {M^{UV}}^2 {r_h}^8 \alpha _{\theta _1} \alpha _{\theta _2}^2}\nonumber\\
& &  -\frac{8.31\times 10^{13}\  {{\cal C}_{\phi _0}}^4 N^{11/20}  \alpha _{\theta _1}^3 \alpha _{\theta _2}^6 c_{1_{q4}}}{{g_s}^2 M^3 {N_f}^6 {r_h}^9 \log ^2({r_h})}\Biggr).
\end{eqnarray}
For $f0[1710](M_g>2 M_\rho)$:
\begin{eqnarray}
\label{Gamma-direct_4pi0_i}
& & \frac{\Gamma_{G_E\rightarrow4\pi^0}}{m_0} \sim 10^{17} c_{15}^2 \sim 10^{-5}\Biggl(\frac{4.72\times 10^{18} {{\cal C}_{\phi _0}^{UV}}^4 {\log (N)}^5  {{N^{UV}_{f}}}^2 \log ({r_h}) (  {c_{2_{q1}}}^{UV}-3.015 {c_{2_{q4}}}^{UV})}{\sqrt{{g^{UV}_s}}N^{17/20}
   {M^{UV}}^2 {r_h}^8 \alpha _{\theta _1} \alpha _{\theta _2}^2}\nonumber\\
& &  -\frac{8.31\times 10^{13}\  {{\cal C}_{\phi _0}}^4 N^{11/20}  \alpha _{\theta _1}^3 \alpha _{\theta _2}^6 c_{1_{q4}}}{{g_s}^2 M^3 {N_f}^6 {r_h}^9 \log ^2({r_h})}\Biggr)^2.
\end{eqnarray}
Currently \cite{PDG-2018} does not contain a reference to the experimental value of $\frac{\Gamma_{G_E\rightarrow4\pi^0}}{m_0}$. We assumed it is $\sim 10^{-5 +{\rm required}_1},\ {\rm where} {\rm required}_1$ could be positive or negative \footnote{$\mu\times 10^n = 10^{n + \log_{10}\mu} \equiv 10^{\rm required}$.}.  As a result, the following constraint was obtained:
\begin{eqnarray}
\label{Constraint_direct_4pi0_decay}
& &  \Biggl. \Biggl(\frac{4.72\times 10^{18} {{\cal C}_{\phi _0}^{UV}}^4 {\log (N)}^5  {{N^{UV}_{f}}}^2 \log ({r_h}) (  {c_{2_{q1}}}^{UV}-3.015 {c_{2_{q4}}}^{UV})}{\sqrt{{g^{UV}_s}}N^{17/20}
   {M^{UV}}^2 {r_h}^8 \alpha _{\theta _1} \alpha _{\theta _2}^2}\nonumber\\
& &  -\frac{8.31\times 10^{13}\  {{\cal C}_{\phi _0}}^4 N^{11/20}  \alpha _{\theta _1}^3 \alpha _{\theta _2}^6 c_{1_{q4}}}{{g_s}^2 M^3 {N_f}^6 {r_h}^9 \log ^2({r_h})}\Biggr)^2\Biggr| _{N={10^2}}\sim 10^{{\rm required}_1}.
\end{eqnarray}
\begin{figure}[t]
	\begin{center}
		\begin{tabular}{c c}
			\includegraphics[width=5.5cm]{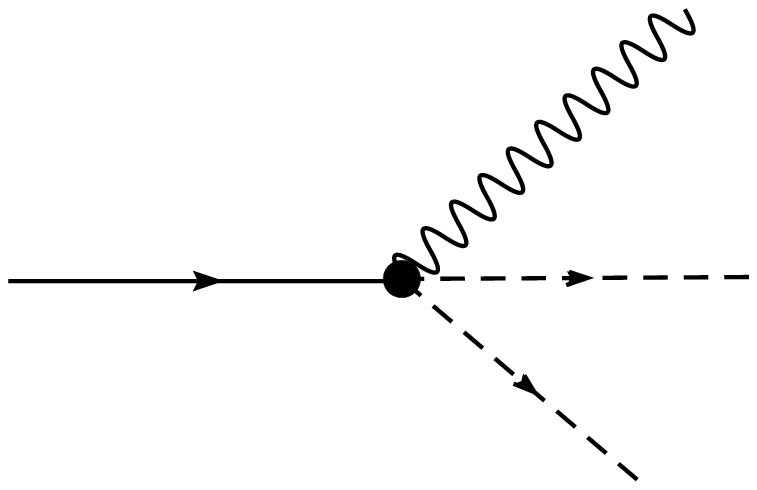}&
			\hspace{1 cm}
			\includegraphics[width=5.5cm]{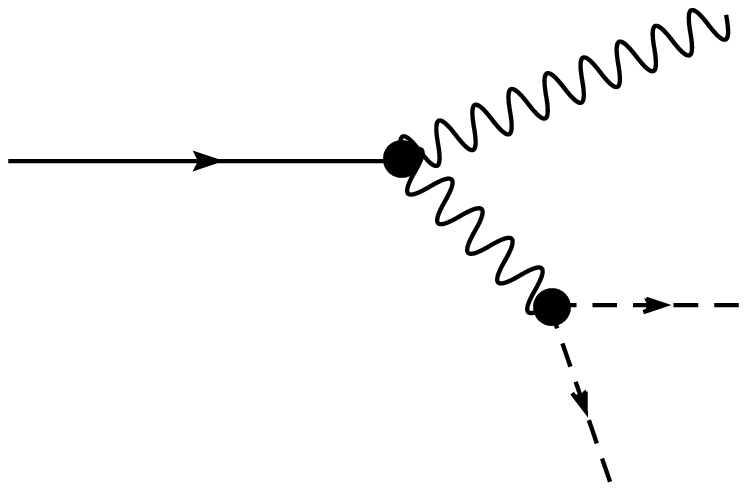}\\
			(a) & (b)
		\end{tabular}
		\caption{$G_E\rightarrow\rho+2\pi$}
	\end{center}
	
\end{figure}
\subsection{$G_E\rightarrow\rho+2\pi$}
Referring to fig. 4.1 One obtains:
{\footnotesize
\begin{eqnarray}
\label{Gamma-both_and_cross_terms}
& & \Gamma_{(a)} = -3 c_8^2  L^4 {\cal T}^2\int_{k_1=0}^{\frac{\left(M_g^2 - M_\rho^2\right)}{2M_g}}\int_{\frac{\left(M_g^2 - M_\rho^2\right)}{2M_g} - k_1}^{\frac{\left(M_g^2 - M_\rho^2\right)}{2M_g}}{dk_1} {dk_2}\frac{ (k_1-k_2)^2 \left(\frac{\left(M_g-k_1-k_2\right){}^2}{M_{\rho }^2}-1\right)}{4
	k_1 k_2 {r_h}^2 M_g \sqrt{1-\frac{\left(\left(M_g-k_1-k_2\right){}^2-k_1^2-k_2^2-M_{\rho }^2\right){}^2}{4
			k_1^2 k_2^2}}}\nonumber\\
		& & \times\theta\left(1-\frac{\left[\left(M_g-k_1-k_2\right){}^2-k_1^2-k_2^2-M_{\rho }^2\right]{}^2}{4 k_1^2 k_2^2}\right)\nonumber\\
		& & \Gamma_{(b)} = -\frac{3 \pi ^2 c_6^2 c_{16}^2  L^8}{4M_g M_{\rho }^4} {\cal T}^4\int_{k_1=0}^{\frac{\left(M_g^2 - M_\rho^2\right)}{2M_g}}\int_{\frac{\left(M_g^2 - M_\rho^2\right)}{2M_g} - k_1}^{\frac{\left(M_g^2 - M_\rho^2\right)}{2M_g}}{dk_1} {dk_2}\frac{\theta\left(1-\frac{\left[\left(M_g-k_1-k_2\right){}^2-k_1^2-k_2^2-M_{\rho }^2\right]{}^2}{4 k_1^2 k_2^2}\right)}{\sqrt{1-\frac{\left(\left(M_g-k_1-k_2\right){}^2-k_1^2-k_2^2-M_{\rho }^2\right){}^2}{4 k_1^2 k_2^2}}}(k_1-k_2)^2\nonumber\\
		& & \times\frac{  \left((k_1+k_2)^2-M_{\rho }^2\right){}^2
			\left(\frac{\left(M_g-k_1-k_2\right){}^2}{M_{\rho }^2}-1\right) \left(\left(M_g-k_1-k_2\right){}^2+4 k_1
			\left(M_g-k_1-k_2\right)+4 k_2 \left(M_g-k_1-k_2\right)-M_{\rho }^2\right){}^2}{ k_1 k_2
			\left(\left(-\left(M_g-k_1-k_2\right){}^2+k_1^2+2 k_1 k_2+k_2^2+2 M_{\rho }^2\right){}^2+M_{\rho }^2 \Gamma _{\rho
			}^2\right)}\nonumber\\
		& &\Gamma_{(a)(b)^*+(a)^*(b)} = \frac{3 \pi  c_6 c_8 c_{16}  L^6}{2 M_g
			M_{\rho }^2 } {\cal T}^3\int_{k_1=0}^{\frac{\left(M_g^2 - M_\rho^2\right)}{2M_g}}\int_{\frac{\left(M_g^2 - M_\rho^2\right)}{2M_g} - k_1}^{\frac{\left(M_g^2 - M_\rho^2\right)}{2M_g}}{dk_1} {dk_2}\nonumber\\
&&\frac{\theta\left(1-\frac{\left[\left(M_g-k_1-k_2\right){}^2-k_1^2-k_2^2-M_{\rho }^2\right]{}^2}{4 k_1^2 k_2^2}\right)}{\sqrt{1-\frac{\left(\left(M_g-k_1-k_2\right){}^2-k_1^2-k_2^2-M_{\rho }^2\right){}^2}{4 k_1^2 k_2^2}}}\frac{(k_1-k_2)^2\left((k_1+k_2)^2-M_{\rho }^2\right)}{k_1 k_2}\nonumber\\
		& &\hskip -0.8in \times\frac{
			\left(\frac{\left(M_g-k_1-k_2\right){}^2}{M_{\rho }^2}-1\right) \left(\left(M_g-k_1-k_2\right){}^2+k_1
			\left(M_g-k_1-k_2\right)+k_2 \left(M_g-k_1-k_2\right)+3 (k_1+k_2)
			\left(M_g-k_1-k_2\right)-M_{\rho }^2\right) }{
			\left(-4 M_{\rho }^2 \left(\frac{1}{2} \left(\left(M_g-k_1-k_2\right){}^2-k_1^2-k_2^2-M_{\rho }^2\right)-k_1
			k_2\right)+4 \left(\frac{1}{2} \left(\left(M_g-k_1-k_2\right){}^2-k_1^2-k_2^2-M_{\rho }^2\right)-k_1
			k_2\right){}^2+M_{\rho }^2 \Gamma _{\rho }^2+M_{\rho }^4\right)}\nonumber\\
		& & \times \left(M_{\rho }^2-2 \left(\frac{1}{2}
		\left(\left(M_g-k_1-k_2\right){}^2-k_1^2-k_2^2-M_{\rho }^2\right)-k_1 k_2\right)\right).
\end{eqnarray}}
Writing:
\begin{eqnarray}
\label{c_6-IR+UV}
& &  c_6 = 10^{-5}N^{\frac{13}{10}}\Biggl(-\frac{2.23\times 10^7 {f_{r_h}}^2 {g_s}^3 M {N_f}^2 \sqrt{{\omega_2}} {c_{\psi_1}}^2 {c_{1_{\ q4}}} \log ^2\left(\sqrt[3]{N}\right)
   N^{\frac{fr_ h}{3}}}{\alpha _{\theta _1} \alpha _{\theta _2}^2}\nonumber\\
& &  -\frac{460099 {f_{r_h}} \sqrt{{g_s}} {\log (N)}^5 {N_f^{UV}}^2 \log \left(\sqrt[3]{N}\right) {{c_{2_{\ \psi_1}}}^{UV}}^2 (  {c_{2_{\ q1}}}^{UV}-3.015
   {c_{2_{\ q4}}}^{UV}) N^{\frac{8 {fr}_h}{3}-\frac{1}{10}}}{{M^{UV}}^2 \alpha _{\theta _1} \alpha _{\theta _2}^2}\Biggr),
\end{eqnarray}
working with $f0[1710]: M_g>2M_\rho$ having dropping $\upsilon_2$:
\begin{eqnarray}
\label{Gamma-3body-over-m0}
& & \hskip -0.7in \frac{\Gamma_{G_E\rightarrow\rho+2\pi}}{m_0}\sim \frac{\Gamma_{(b)}}{m_0} \sim c_6^2c_{16}^2\nonumber\\
& & \hskip -0.7in\sim 10^{-10}N^{\frac{13}{5}}\Biggl(-\frac{2.23\times 10^7 {f_{r_h}}^2 {g_s}^3 M {N_f}^2 \sqrt{{\omega_2}} {c_{\psi_1}}^2 {c_{1_{ q4}}} \log ^2\left(\sqrt[3]{N}\right)
   N^{\frac{fr_ h}{3}}}{\alpha _{\theta _1} \alpha _{\theta _2}^2}\nonumber\\
& & \hskip -0.7in -\frac{460099 {f_{r_h}} \sqrt{{g^{UV}_s}} {\log (N)}^5 {N_f^{UV}}^2 \log \left(\sqrt[3]{N}\right) {{c^{UV}_{2_{ \psi_1}}}}^2 (  {c_{2_{q1}}}^{UV}-3.0153
   {c_{2_{ q4}}}^{UV}) N^{\frac{8 {fr}_h}{3}-\frac{1}{10}}}{{M^{UV}}^2 \alpha _{\theta _1} \alpha _{\theta _2}^2}\Biggr)^2.
\end{eqnarray}
We assumed the experimental value for $\frac{\Gamma_{G_E\rightarrow\rho+2\pi}}{m_0}$ - not yet known in \cite{PDG-2018} - to be $10^{-5 + {\rm required}_2}$, (\ref{Gamma-3body-over-m0}) for $N=10^2$, implied the following constaint:
\begin{eqnarray}
\label{constraint_3body_decay-glueball}
& & \Biggl.\Biggl(-\frac{2.23\times 10^7 {f_{r_h}}^2 {g_s}^3 M {N_f}^2 \sqrt{{\omega_2}} {c_{\psi_1}}^2 {c_{1_{ q4}}} \log ^2\left(\sqrt[3]{N}\right)
   N^{\frac{fr_ h}{3}}}{\alpha _{\theta _1} \alpha _{\theta _2}^2}\nonumber\\
& &  -\frac{460099 {f_{r_h}} \sqrt{{g^{UV}_s}} {\log (N)}^5 {N_f^{UV}}^2 \log \left(\sqrt[3]{N}\right) {{c^{UV}_{2_{ \psi_1}}}}^2 (  {c_{2_{q1}}}^{UV}-3.0153
   {c_{2_{ q4}}}^{UV}) N^{\frac{8 {fr}_h}{3}-\frac{1}{10}}}{{M^{UV}}^2 \alpha _{\theta _1} \alpha _{\theta _2}^2}\Biggr|_{N=10^2} = 10^{{\rm required}_2}.\nonumber\\
   &&
\end{eqnarray}

\subsection{Indirect Decay of Glueball to $4\pi$ }
The relevant interaction lagrangian was given as follows:
\begin{eqnarray}
\label{interaction action}
&&\hskip -0.5in{\rm S}_{int}= {\cal T}{\rm Str}\int \left(\frac{1}{T_h}\right) d^{3}x\Bigg[c_{3}\rho_{\mu}^{2}G_E + c_{4}\rho_\mu\rho_\nu \frac{\partial^{\mu}\partial^{\nu}}{M^2}G_E + c_{5}\tilde{F}_{\mu\nu}\tilde{F}^{\mu\nu}G_E
+ c_{6}\tilde{F}_{\mu\rho}\tilde{F}_\nu^{\ \rho}\frac{\partial^{\mu}\partial^{\nu}}{M^2}G_E\nonumber\\
&& + \iota c_{7}\partial_{\mu}\pi [\pi,\rho^\mu]G_E + \iota c_{8}\partial_{\mu}\pi [\pi,\rho_\nu]\frac{\partial^{\mu}\partial^{\nu}}{M^2}G_E + c_{9}\rho_\mu \tilde{F}_{\nu}^{\ \mu}\frac{\partial^\nu G_E}{M^2}\nonumber\\
&&+c_{10}\tilde{F}_{\mu\nu}\tilde{F}^{\mu\nu}G_E +c_{11}\partial_{\mu}\pi\partial^{\mu}\pi G_E +c_{12}\rho_{\mu}\rho^{\mu} G_E +\iota c_{13}\partial_{\mu}\pi [\pi,\rho^\mu]G_E \Bigg].
\end{eqnarray}
 Glueball decay into four pions is a combination of two process: $G_E\rightarrow \rho \rho$ and $G_E\rightarrow \pi\pi\rho$. In the first process each
$\rho$ meson further decays into two $\pi$ each and, in the later one involves $\rho$ meson further decaying into two $\pi$ mesons. Two pairs of pions with different isospin indexes determined the amplitude of a glueball's decay into four pions. If  {\cal M} is the amplitude for $G_E\rightarrow 2\pi^{a}2\pi^{b}$ where $a\neq b$ then one can set $a=1$ and $b=2$ without any loss of generality. The total decay rate was calculated as follows:
\begin{eqnarray}
&&\Gamma =\frac{3}{4}\frac{1}{2M}\int d\Phi_{4}|M|^{2}
\end{eqnarray}
 the three different pairs of isospin introduce a factor of 3, and symmetry factor of two pairs of identical particles gives 4. Hence a factor of $\frac{3}{4}$ appears in the decay width.
The full four body phase space in 2+1 dimension spacetime is given by
\begin{eqnarray}
\int d\Phi_{4}=\prod_{i=1}^{4} \frac{d^{2}\vec{k_i}}{(2\pi)^2 2E_i}(2\pi)^3\delta(k^\mu -k_1^\mu -k_2^\mu -k_3^\mu -k_4^\mu)
\end{eqnarray}

The amplitude corresponding to process $G_E\rightarrow \rho\pi\pi\rightarrow\pi\pi\pi\pi$ (Fig. 1.2(a)) is given as :
\begin{eqnarray}
 {\cal M}_{(a)}&=&{\cal T}^{2}\Bigg(\frac{\pi L^{2}}{r_h} \Bigg)^{2}8 c_{8}c_{16} \Bigg( \Delta_{\rho}^{\mu\mu\prime}(k_{2}+k_{4})(-k_{(1)\mu}k_{(2)\mu\prime}+k_{(1)\mu}k_{(4)\mu\prime}+k_{(3)\mu}k_{(2)\mu\prime}-k_{(3)\mu}k_{(4)\mu\prime})\nonumber\\
&&+\Delta_{\rho}^{\mu\mu\prime}(k_{3}+k_{4})(-k_{(1)\mu}k_{(3)\mu\prime}+k_{(1)\mu}k_{(4)\mu\prime}+k_{(2)\mu}k_{(3)\mu\prime}-k_{(2)\mu}k_{(4)\mu\prime})\nonumber\\
&&+\Delta_{\rho}^{\mu\mu\prime}(k_{1}+k_{2})(-k_{(1)\mu}k_{(3)\mu\prime}+k_{(4)\mu}k_{(1)\mu\prime}+k_{(2)\mu}k_{(3)\mu\prime}-k_{(2)\mu}k_{(4)\mu\prime})\nonumber\\
&&+\Delta_{\rho}^{\mu\mu\prime}(k_{1}+k_{3})(-k_{(1)\mu}k_{(2)\mu\prime}+k_{(1)\mu}k_{(4)\mu\prime}+k_{(3)\mu}k_{(2)\mu\prime}-k_{(3)\mu}k_{(4)\mu\prime})\Bigg),
\end{eqnarray}
where $\Delta_{\rho}^{\mu\mu\prime}$ represents the vector meson propagator  given as:
\begin{eqnarray}
& & \Delta_{\rho}^{\mu\mu\prime}(k_{i}+k_{j})=\frac{\delta _0^{\mu  }\delta _0^{\nu } \left(-\delta _{\nu }^{\mu '}-\frac{\left(k_i+k_j\right){}^{\mu '} \left(k_{{i}}+k_{{j}}\right)_{\nu}}{M_{\rho
   }^2}\right)}{\left(k_i+k_j\right){}^2+i \Gamma _{\rho } M_{\rho }^2-M_{\rho }^2}.
\end{eqnarray}

 The amplitude for the second process $G_E\rightarrow \rho\rho \rightarrow \pi\pi\pi\pi$ (Fig. 4.2(b)) is given as under:
 \begin{eqnarray}
 &&{\cal M}_{(b)}=-\frac{16 \pi ^3 {c_{16}}^2 L^6 {\cal T}^3}{r_h ^3}
 \Biggl(\left({A\eta }^{\sigma \gamma}+B^{\sigma \gamma }\right) \Delta_{\rho}\left(k_3+k_4\right)_{\sigma }^{\mu } \Delta_{\rho }\nonumber\\
 & & \times\left(k_1+k_2\right)_{\gamma}^{\mu '} \left(-k_{{\mu 1}} k_{{\mu 3}'}+k_{{\mu 1}}
   k_{{\mu 4}'}+k_{{\mu 2}} k_{{\mu 3}'}-k_{{\mu 2}} k_{{\mu 4}'}\right)+k_2\longleftrightarrow k_3\Biggr),
\end{eqnarray}
where the expression for $A$ and $B^{\mu\nu}$ are same as given in the section for $G_E\rightarrow \rho\rho$ decay with appropriate momentum substitution. For $f0[1710]:M_g>2M_\rho$,  (b) dominates.

As a result, the total width of decay can be approximated by
\begin{eqnarray}
\label{Gamma-indirect-4pi-decay-glueball-i}
&& \hskip -0.7in \frac{\Gamma_{G_E\rightarrow2\rho\rightarrow4\pi}}{m_0} \sim c_6^2c_{16}^4\nonumber\\
& & \hskip -0.7in \sim 10^{-10}N^{\frac{13}{5}}\Biggl(-\frac{2.23007\times 10^7 {f_{r_h}}^2 {g_s}^3 M {N_f}^2 \sqrt{{\omega_2}} {c_{\psi_1}}^2 {c_{1_{ q4}}} \log ^2\left(\sqrt[3]{N}\right)
   N^{\frac{fr_ h}{3}}}{\alpha _{\theta _1} \alpha _{\theta _2}^2}\nonumber\\
& & \hskip -0.7in -\frac{460099. {fr_h} \sqrt{{g^{UV}_s}} {\log (N)}^5 {N_f^{UV}}^2 \log \left(\sqrt[3]{N}\right) {{c^{UV}_{2_{ \psi_1}}}}^2 (  {c_{2_{q1}}}^{UV}-3.015
   {c_{2_{ q4}}}^{UV}) N^{\frac{8 {fr}_h}{3}-\frac{1}{10}}}{{M^{UV}}^2 \alpha _{\theta _1} \alpha _{\theta _2}^2}\Biggr)^2.
\end{eqnarray}
Assuming the experimental value of $\frac{\Gamma_{G_E\rightarrow2\rho\rightarrow4\pi}}{m_0}$ - not currently known \cite{PDG-2018} - is $10^{-5 + {\rm required}_3}$, one obtains, for $N=10^2$ the same constraint as (\ref{Constraint_direct_4pi0_decay}) with ${\rm required}_2$ replaced by ${\rm required}_3$; we expect ${\rm required}_2\sim{\rm required}_3$.


\begin{figure}[t]
	\begin{center}
		\begin{tabular}{c c}
			\includegraphics[width=5.5cm]{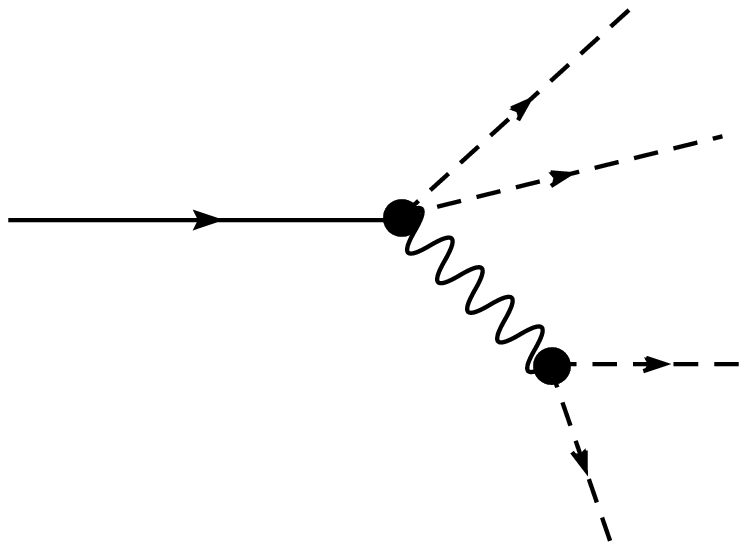}&
			\hspace{1 cm}
			\includegraphics[width=5.5cm]{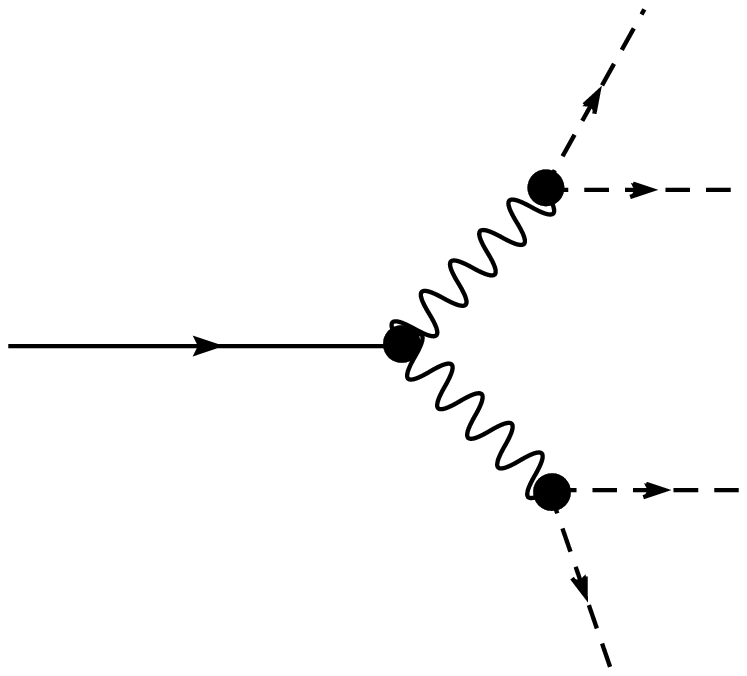}\\
			(a) & (b)
		\end{tabular}
		\caption{$G_E\rightarrow4\pi$}
	\end{center}
	
\end{figure}

\section{Summary and Discussion}

We studied (exotic) scalar glueball $0_E^{++}$-meson interaction and (exotic) scalar glueball decays at tree level wherein the glueballs corresponded to metric fluctuations of the M theory uplift of \cite{metrics}'s UV-complete type IIB holographic dual of large-$N$ thermal QCD at finite coupling - MQGP limit of \cite{MQGP} - and the mesons corresponded to gauge fluctuations on the world-volume of type IIA flavor $D6$-branes. The following is a summary of the main results.
\begin{itemize}
	\item
	We obtained $0_E^{++}-\rho,\pi$ interaction Lagrangian linear in the exotic scalar glueball and up to quartic in $\pi$ mesons wherein the coefficients are given by radial integrals of components of the M theory metric that corresponds to the uplift of \cite{MQGP}'s SYZ type IIA mirror of \cite{metrics}, and perturbations thereof. This is quite satisfying because the coupling constants may be obtained directly from the underlying fundamental M theory.
	
	\item
	Using the solutions to the metric fluctuations and meson radial profile functions, assuming $M_G > 2M_\rho$ , we can perform calculations that, when appropriate choices of integration constants are made, exactly reproduce the PDG results on scalar glueball decay widths.
	\begin{enumerate}

\item
The normalization condition for $\psi_1(Z)$ implies the following quadratic constaint on $c_{\psi_1}$ and $c_{2\ \psi_1}^{\rm UV}$:
\begin{eqnarray}
\label{psi_1_normalization}
& & V\Biggl(\frac{5\times10^{-5}}{\alpha_{\theta_1}\alpha_{\theta_2}^2}g_s^2MN^{\frac{4}{5}}N_f^2\sqrt{|\omega_2|}\left(1433.4 + b^2(-2067.37 + \omega_2)\right)(c_{\psi_1})^2\log r_h\nonumber\\
 & & + \frac{244.91 \log r_hg_s^{\rm UV 2}M^{\rm UV}N^{\frac{4}{5}}N_f^{\rm UV}\ ^2c_{2\ \psi_1}^{\rm UV}\ ^2}{\alpha_{\theta_1}\alpha_{\theta_2}^2}\Biggr)=1.
\end{eqnarray}

\item
The normalization condition for $\phi_0(Z)$ implies the following quadratic constraint on ${\cal C}_{\phi_0}$ and ${\cal C}_{\phi_0}^{\rm UV}$:
\begin{eqnarray}
\label{phi_0_normalization}
& & \frac{V}{2}\Biggl(\frac{5.51\times10^{-9}{\cal C}_{\phi_0}^2N^{\frac{1}{5}}(0.03 + 0.042 b^2)\alpha_{\theta_1}\alpha_{\theta_2}^2}{g_s r_h^2\log r_h M N_f^2}\nonumber\\
& & + \frac{793.58 {\cal C}_{\phi_0}^{\rm UV\ 2}g_s^{\rm UV}M^{\rm UV}N_f^{\rm UV}r_h^2\log r_h}{N^{\frac{1}{5}}\alpha_{\theta_1}\alpha_{\theta_2}^2}\Biggr)=1.
\end{eqnarray}

		\item
	From (\ref{Gto2pi-decay-width}),	${\cal C}_{\phi_0}^2 c_{1\ q_4}$ and $\left({\cal C}_{\phi_0}\ ^{\rm UV}({\cal C}_{\phi_0})\right)^2 \left({c_{2_{q1}}}^{\rm UV}-3.015{c_{2_{q4}}}^{\rm UV}\right)$  can be fine tuned to reproduce the PDG value of $\Gamma_{G_E\rightarrow2\pi}$ {\it exactly}.
		
		\item
		Requiring $\Gamma_{G_E\rightarrow2\rho} = \Gamma_{G_E\rightarrow4\pi}$ yields: $c_4 \approx \frac{3}{4}c_6 m_0^2\left(\frac{r_h}{\pi\sqrt{4\pi g_s N}}\right)^2$, the glueball mass written as
		$M_g = m_0 \frac{r_h}{\pi\sqrt{4 \pi g_s N}}$ \cite{Sil+Yadav+Misra-glueball}. Assuming $\omega_2\equiv {\cal O}(1)$, writing $c_{1\ q_4} = N^{-\alpha_4} (\alpha_4\geq1), M_g = m_\rho \frac{r_h}{\pi\sqrt{4 \pi g_s N}}$ and setting $m_0=1710, m_\rho=775$ one sees that the aforementioned relation between $c_4$ and $c_6$ implies:
\begin{equation}
g_s = \frac{N^{\frac{2(-9+35f_{r_h}+15\alpha_4)}{75}}\left(\frac{N_f^{\rm UV}}{M^{\rm UV}}\right)^{\frac{4}{5}}\left(\frac{\left({c_{2_{\ \psi_1}}}^{UV}\ \right)^4}{\left({\cal O}(1){c_{2_{\ q1}}}^{UV} - {\cal O}(1){c_{2_{\ q4}}}^{UV}\right)^2}\right)^{\frac{1}{5}}}{{\cal O}(1)\left(f_{r_h}^2M^2N_f^4\omega_2c_{\psi_1}^4\right)^{\frac{1}{5}}},
\end{equation}
which can be made to be finite as part of the MQGP limit.

		\item
		The combination of constants of integration appearing in the solutions to the EOMS of $\phi_0(Z), \psi_1(Z)$ in the IR and UV, using (\ref{psi_1_normalization}): ${\cal C}_{\phi_0}^2c_{\ \psi_1}$ and $\left({\cal C}_{\phi_0}\ ^{\rm UV}({\cal C}_{\phi_0})\right)^2 c_{2_{\ \psi_1}}^{\rm UV}(c_{\ \psi_1})$, can be adjusted to reproduce the PDG value of $\Gamma_{\rho\rightarrow2\pi}$ {\it exactly}.

		\item
	From (\ref{constraint-rho-2pi-decay}), (\ref{Constraint_direct_4pi0_decay}) and (\ref{Gamma-indirect-4pi-decay-glueball-i}), and also using (\ref{psi_1_normalization}) as well as (\ref{phi_0_normalization}), we note that	the combination of constants of integration appearing in the solutions to the EOMS of $\phi_0(Z), \psi_1(Z)$ and $q_{1,2,3,4,5,6}(Z)$ in the IR and UV:\\
		-- involving  ${\cal C}_{\phi_0}^4 c_{1\ q_4}$ and $\left({\cal C}_{\phi_0}\ ^{\rm UV}({\cal C}_{\phi_0})\right)^4 \left({c_{2_{q1}}}^{\rm UV}-3.015{c_{2_{q4}}}^{\rm UV}\right)$ 	appearing in $\Gamma_{G_E\rightarrow4\pi^0}$ \\
		-- involving $c_{1\ q_4}\left(c_{\psi_1}\right)^2$ and $c_{2\ \psi_1}\ ^{\rm UV}(c_{\psi_1})\left({c_{2_{q1}}}^{\rm UV}-3.015{c_{2_{q4}}}^{\rm UV}\right)$ appearing in
		$\Gamma_{G_E\rightarrow\rho+2\pi} \approx \Gamma_{G_E\rightarrow2\rho\rightarrow4\pi}$\\
can be tuned and equality of these two combinations can be effected such that one can reproduce the PDG value of $\Gamma_{G_E\rightarrow4\pi^0} = \Gamma_{G_E\rightarrow\rho+2\pi} \approx \Gamma_{G_E\rightarrow2\rho\rightarrow4\pi}$ {\it exactly}.
		
	\end{enumerate}	
\end{itemize}


\chapter{Towards Thermal QCD from M theory at Intermediate '\lowercase{t} Hooft Coupling }
\graphicspath{{Chapter4/}{Chapter4/}}

\section{Introduction}


In recent years, AdS/CFT correspondence \cite{Maldacena} and its generalization has provided a simple, classical computational tool for understanding strongly coupled systems and overcoming theoretical limitations in studying the plasma phase of non-Abelian gauge theories. In its simplest form for maximally supersymmetric $SU(N_c)$ Yang-Mills theory (${\cal N} = 4$ SYM), in the $N_c\rightarrow\infty$ limit, the gauge/gravity duality provides a tool for analysing its properties in the large `t Hooft coupling limit. Numerous studies of strongly coupled dynamics relevant to heavy ion collisions have been conducted using gauge/gravity duality; see, for example, the \cite{Krishna-Rajagopal-et-al-book} and its references. Additionally, the gauge/gravity duality enables us to investigate corrections to the limit of infinite coupling. On the gravity side, these corrections appear as higher order derivative corrections to the Einstein-Hilbert action. These corrections to gravity's action are regarded as minor perturbations to the second order equation of motions. Our starting point will include terms admitting terms quartic in the eleven-dimensional Riemann curvature $R$, i.e. schematically of the form $R^4$. We focus on leading order corrections to supergravity that begin at the order of $l^6_p$, where $l_p$ is the 11D Planckian length.

The importance of higher order derivative corrections is not limited to corrections to the limit of infinite coupling. Additionally, they serve as the leading quantum gravity corrections to the ${\cal M}$-Theory action, allowing for the study of ${\cal M}$-Theory compactifications on compact eight-dimensional manifolds.
It is interesting to study the warped compactification of ${\cal M}$-Theory on eight-dimensional compact manifolds. On the one hand, this compactification enables the study of three-dimensional effective theories that contain only a small amount of supersymmetry. On the other hand, it enables us to investigate the extension of three-dimensional theories to four space-time dimensions for a class of eight-dimensional manifolds via the ${\cal M}$-Theory to F-theory limit. Previously, the vacuum for warped compactifications of ${\cal M}$-Theory on compact eight-dimensional manifolds was investigated by including the action's higher derivative terms. The leading quantum gravity corrections to ${\cal M}$-Theory actions are of fourth order, $R^4$, and third order, $R^3G^2$, in the eleven-dimensional Riemann curvature $R$, where $G$ denotes the field strength of the ${\cal M}$-Theory three form. The terms of ${\cal O}(R^4)$ have been used previously in \cite{O(R^4)}, while terms of third order have been analysed recently in \cite{O(R^3G^2)}. Previously, higher derivative terms were used to investigate the effects of intermediate coupling in thermal gauge theories \cite{previous-higher-ders}. However, the literature lacks a top-down holographic dual of thermal QCD with intermediate 't Hooft coupling. This work closes this gap by examining the M theory dual of large-$N$ thermal QCD at intermediate gauge and 't Hooft couplings and obtaining the ${\cal O}(l^6_p)$ corrections to the ${\cal M}$-Theory uplift of \cite{metrics} as constructed in \cite{MQGP}.

\section{${\cal O}(l_p^6)$ Corrections to the Background of \cite{MQGP} in the MQGP Limit}

In this section, we discuss how the equations of motion (EOMs) starting from $D=11$ supergravity action inclusive of the ${\cal O}(R^4)$ terms in the same (which provide the ${\cal O}(l_p^6)$ corrections to the leading order terms in the action), were obtained and how the same were solved. The actual EOMs are given in Appendix {\bf D.1} - EOMs in {\bf D.1.1} were obtained in the $\psi=0, 2\pi, 4\pi$ coordinate patches (wherein $g^{M}_{rM}, M\neq r$  and $g^{M}_{x^{10}N}, N\neq x^{10}$ vanish) and EOMs in {\bf D.1.2} were obtained away from the same. The solutions of the EOMs  are similarly split across subsections {\bf 5.2.1} and {\bf 5.2.2}.

The ${\cal N}=1, D=11$ supergravity action inclusive of ${\cal O}(l_p^6)$ terms, is given by:
\begin{eqnarray}
\label{D=11_O(l_p^6)}
& & \hskip -0.8inS = \frac{1}{2\kappa_{11}^2}\int_M\left[  R *_{11}1 - \frac{1}{2}G_4\wedge *_{11}G_4 -
\frac{1}{6}C\wedge G\wedge G\right] + \frac{1}{\kappa_{11}^2}\int_{\partial M} d^{10}x \sqrt{h} K \nonumber\\
& & \hskip -0.8in+ \frac{1}{(2\pi)^43^22^{13}}\left(\frac{2\pi^2}{\kappa_{11}^2}\right)^{\frac{1}{3}}\int d^{11}x\sqrt{-g}\left( J_0 - \frac{1}{2}E_8\right) + \left(\frac{2\pi^2}{\kappa_{11}^2}\right)\int C_3\wedge X_8,
\end{eqnarray}
where:
\begin{eqnarray}
& & J_0  =3\cdot 2^8 (R^{HMNK}R_{PMNQ}{R_H}^{RSP}{R^Q}_{RSK}+
{1\over 2} R^{HKMN}R_{PQMN}{R_H}^{RSP}{R^Q}_{RSK})+O(R_{MN}) \nonumber\\
& & E_8  ={ 1\over 3!} \epsilon^{ABCM_1 N_1 \dots M_4 N_4}
\epsilon_{ABCM_1' N_1' \dots M_4' N_4' }{R^{M_1'N_1'}}_{M_1 N_1} \dots
{R^{M_4' N_4'}}_{M_4 N_4},\nonumber\\
& & X_8={ 1\over 192 (2 \pi)^4} \left[ {\rm tr} R^4 -{1 \over 4}
({\rm tr}R^2)^2 \right],\nonumber\\
& & \kappa_{11}^2 = \frac{(2\pi)^8 l_p^{9}}{2}.
\end{eqnarray}
$\kappa_{11}^2$ being related to the eleven-dimensional Newtonian coupling constant, and $G=dC$ with $C$ being the ${\cal M}$-theory three-form potential with the four-form $G$ being the associated four-form field strength.  The EOMS are:
\begin{eqnarray}
\label{eoms}
& &  R_{MN} - \frac{1}{2}g_{MN} R - \frac{1}{12}\left(G_{MPQR}G_N^{\ PQR} - \frac{g_{MN}}{8}G_{PQRS}G^{PQRS} \right)\nonumber\\
 & &  = - \beta\left[\frac{g_{MN}}{2}\left( J_0 - \frac{1}{2}E_8\right) + \frac{\delta}{\delta g^{MN}}\left( J_0 - \frac{1}{2}E_8\right)\right],\nonumber\\
& &  \partial_M\left(\sqrt{-g}G^{M M_1 M_2 M_3}\right) + \frac{1}{2 4!}\epsilon^{M_1 \dots M_{11}}G_{M_4 \dots M_7} G_{M_8 \dots M_{11}} + 2\kappa_{11}^2\left(\frac{2\pi^2}{\kappa_{11}^2}\right)^{\frac{1}{3}}\epsilon^{M_1 \dots M_{11}}\left(X_8\right)_{M_4 \dots M_{11}},\nonumber\\
& &
\end{eqnarray}
where \cite{Becker-sisters-O(R^4)}:
\begin{equation}
\label{beta-def}
\beta \equiv \frac{\left(2\pi^2\right)^{\frac{1}{3}}\left(\kappa_{11}^2\right)^{\frac{2}{3}}}{\left(2\pi\right)^43^22^{12}} \sim l_p^6.
\end{equation}
$R_{MNPQ}, R_{MN}, {\cal R}$  in  (\ref{D=11_O(l_p^6)})/(\ref{eoms}) being respectively the elven-dimensional Riemann curvature tensor, Ricci tensor and the Ricci scalar.

As  shown in \cite{MQGP}, $C_3\wedge X_8=0$, one sees that if one makes the ansatz:
\begin{eqnarray}
\label{ansaetze}
& & \hskip -0.8ing_{MN} = g_{MN}^{(0)} +\beta g_{MN}^{(1)},\nonumber\\
& & \hskip -0.8inC_{MNP} = C^{(0)}_{MNP} + \beta C_{MNP}^{(1)},
\end{eqnarray}
then symbolically, one obtains:
\begin{eqnarray}
\label{deltaC=0consistent}
& & \beta \partial\left(\sqrt{-g}\partial C^{(1)}\right) + \beta \partial\left[\left(\sqrt{-g}\right)^{(1)}\partial C^{(0)}\right] + \beta\epsilon_{11}\partial C^{(0)} \partial C^{(1)} = {\cal O}(\beta^2) \sim 0 [{\rm up\ to}\ {\cal O}(\beta)].
\nonumber\\
& & \end{eqnarray}
One can see that one can find a consistent set of solutions to (\ref{deltaC=0consistent}) wherein $C^{(1)}_{MNP}=0$ up to ${\cal O}(\beta)$. Assuming that one can do so, henceforth we will  define:
\begin{eqnarray}
\label{fMN-definitions}
\delta g_{MN} =\beta g^{(1)}_{MN} = g_{MN}^{\rm MQGP} f_{MN}(r),
\end{eqnarray}
no summation implied.

To evaluate the variation of $J_0$ with respect to the metric, schematically, one will need to evaluate, e.g., $(\delta R^\bullet_{\ \bullet\bullet\bullet})\chi = \delta\left( \left(g^{-1}\right)^3R^{\bullet\bullet\bullet\bullet}\right) \chi$ where $\chi$ is cubic in the curvature tensor to be defined below. This will yield, schematically -$\left(g^{-1}\right)^4\delta g R^{\bullet\bullet\bullet\bullet}\chi + \left(g^{-1}\right)^3\left(D\delta\Gamma\right)\chi =
-\left(g^{-1}\right)^4\delta g R^{\bullet\bullet\bullet\bullet}\chi + \left(g^{-1}\right)^3\left(D^2\delta g\right)\chi \equiv  -\left(g^{-1}\right)^4\delta g R^{\bullet\bullet\bullet\bullet}\chi + \delta g \left(g^{-1}\right)^3 D^2\chi$. Now, in the MQGP limit and in the IR, one can show that the contribution of the derivative in $D$ is sub-dominant as compared to the affine connection $\Gamma$ contained therein. Hence, the first term is sub-dominant as compared to the second term ($\sim\delta g \left(g^{-1}\right)^3\Gamma^2\chi$) and will be disregarded henceforth. Hence, in the MQGP limit and in the IR,
$R\left(\left\{{\alpha_i}\right\}\right)\delta R\left(\left\{{\beta_i}\right\}\right)\prod_{i=1}^k \delta(\alpha_i,\beta_i) =
R\left(\left\{\overset{\sim}{\alpha_i}\right\}\right)\delta R\left(\left\{\overset{\sim}{\beta_i}\right\}\right)\prod_{i=1}^k\delta(\beta_i,\alpha_i) $ which implies terms with either  one or up to four indices corresponding to $k=1$ or $2$ or $3$ or $4$, $ \left\{\alpha_i\right\}$ being contracted with respectively one or up to four indices $\left\{\beta_i\right\}$ with no restriction on the remaining indices that are not contracted in both sets ($\left\{\alpha_i\right\},\left\{\beta_j\right\})$) to be contravariant, covariant or a mixture of the same; the tildes imply that the covariance/contravariance of $k$ of the $\alpha_i$s and $\beta_i$s, has been swapped. So, as an example, in the MQGP limit and in the IR, $R^\times_{\ \bullet\bullet\bullet}\delta R_\times^{\ \bullet\bullet\bullet}
\sim R^{\times\bullet\bullet\bullet}\delta R_{\times\bullet\bullet\bullet}$, etc.

Hence, one can show:
{\footnotesize
\begin{eqnarray}
\label{delta J_0}
 \delta J_0 & \stackrel{\rm MQGP,\ IR}{\xrightarrow{\hspace*{1.5cm}}} & 3\times 2^8 \delta R^{HMNK} R_H^{\ RSP}\Biggl(R_{PQNK}R^Q_{\ RSM} + R_{PSQK}R^Q_{\ MNR}  \nonumber\\
 & & + 2\left[R_{PMNQ}R^Q_{\ RSK} + R_{PNMQ}R^Q_{\ SRK}\right]\Biggr) \nonumber\\
& &  \equiv 3\times 2^8 \delta R^{HMNK}\chi_{HMNK}\nonumber\\
& & = -\delta g_{\tilde{M}\tilde{N}}\Biggl[ g^{M\tilde{N}} R^{H\tilde{N}NK}\chi_{HMNK}
+ g^{N\tilde{N}} R^{HM\tilde{M}K}\chi_{HMNK} + g^{K\tilde{M}}R^{HMN\tilde{N}}\chi_{HMNK}
\nonumber\\
& & + \frac{1}{2}\Biggl(g^{H\tilde{N}}[D_{K_1},D_{N_1}]\chi_H^{\tilde{M}N_1K_1} +
g^{H\tilde{N}}D_{M_1}D_{N_1} \chi_H^{M_1[\tilde{N}_1\tilde{M}]}  - g^{H\tilde{H}} D_{\tilde{H}}D_{N_1}\chi_H^{\tilde{N}[N_1\tilde{M}]}\Biggr)\Biggr],\nonumber\\
& &
\end{eqnarray}
}
where:
{\footnotesize
\begin{eqnarray}
\label{chi-def}
& & \chi_{HMNK} \equiv R_H^{\ \ RSP}\left[R_{PQNK} R^Q_{\ \ RSM}
+ R_{PSQK} R^Q_{\ \ MNR} + 2\left(R_{PMNQ} R^Q_{\ \ RSK} + R_{PNMQ}R^Q_{\ \ SRK}\right)\right].
\end{eqnarray}
}
Further:
{\footnotesize
\begin{eqnarray}
\label{deltaE8}
& & \delta E_8 \sim -\frac{2}{3}\delta g_{\tilde{M}\tilde{N}} g^{N_1^\prime\tilde{N}} \epsilon^{ABCM_1N_1,,,M_4N_4}\epsilon_{ABCM_1^\prime N_1^\prime...M_4^\prime N_4^\prime} R^{M_1^\prime\tilde{M}}_{\ \ \ \ \ M_1N_1}R^{M_2^\prime N_2^\prime}_{\ \ \ \ \ M_2N_2}R^{M_3^\prime N_3^\prime}_{\ \ \ \ \ M_2N_2}R^{M_4^\prime N_4^\prime}_{\ \ \ \ \ M_4N_4}\nonumber\\
& & + \frac{\delta g_{\tilde{M}\tilde{N}}}{3}\Biggl[ 2 \epsilon^{ABCM_1\tilde{N},,,M_4N_4}\epsilon_{ABCM_1^\prime N_1^\prime...M_4^\prime N_4^\prime}
g^{N_1^\prime\tilde{N}_1}g^{M_1^\prime\tilde{M}}D_{\tilde{N}_1}D_{M_1}\left(R^{M_2^\prime N_2^\prime}_{\ \ \ \ \ M_2N_2}R^{M_3^\prime N_3^\prime}_{\ \ \ \ \ M_2N_2}R^{M_4^\prime N_4^\prime}_{\ \ \ \ \ M_4N_4}\right)\nonumber\\
& & + \epsilon^{ABCM_1N_1,,,M_4N_4}\epsilon_{ABCM_1^\prime N_1^\prime...M_4^\prime N_4^\prime}
g^{N_1^\prime\tilde{N}_1}g^{M_1^\prime\tilde{M}}[D_{\tilde{N}_1},D_{M_1}]\left(R^{M_2^\prime N_2^\prime}_{\ \ \ \ \ M_2N_2}R^{M_3^\prime N_3^\prime}_{M_2N_2}R^{M_4^\prime N_4^\prime}_{\ \ \ \ \ M_4N_4}\right)\nonumber\\
& & - 2 \epsilon^{ABCM_1\tilde{M},,,M_4N_4}\epsilon_{ABCM_1^\prime N_1^\prime...M_4^\prime N_4^\prime}
g^{N_1^\prime\tilde{N}}g^{M_1^\prime\tilde{L}}D_{\tilde{L}_1}D_{M_1}\left(R^{M_2^\prime N_2^\prime}_{\ \ \ \ \ M_2N_2}R^{M_3^\prime N_3^\prime}_{\ \ \ \ \ M_2N_2}R^{M_4^\prime N_4^\prime}_{\ \ \ \ \ M_4N_4}\right)\Biggr],
\end{eqnarray}
}
where, e.g.,  \cite{Tseytlin-epsilonD^2R^4-kroneckerdeltaR^4}
{\footnotesize
\begin{eqnarray}
\label{epsilonD^2R^4}
& &  \epsilon^{ABCM_1M_2,,,M_8}\epsilon_{ABCM_1^\prime M_2^\prime...M_8^\prime}R^{M_1^\prime M_2^\prime}_{\ \ \ \ \ M_1M_2}R^{M_3^\prime M_4^\prime}_{\ \ \ \ \ M_3M_4}R^{M_5^\prime M_6^\prime}_{M_5M_6}R^{M_7^\prime M_8^\prime}_{\ \ \ \ \ M_7M_8}\nonumber\\
& &  = -3!8!
\delta^{M_1}_{N_1]}...\delta^{M_8}_{M_8^\prime]}R^{M_1^\prime M_2^\prime}_{\ \ \ \ \ M_1M_2}R^{M_3^\prime M_4^\prime}_{\ \ \ \ \ M_3M_4}R^{M_5^\prime M_6^\prime}_{M_5M_6}R^{M_7^\prime M_8^\prime}_{\ \ \ \ \ M_7M_8}.
\end{eqnarray}
}

Writing: $T_{MN} \equiv G_M^{\ \ PQR}G_{NPQR} - \frac{g_{MN}}{8}G^2$, the ${\cal O}(l_p^6)$ ``perturbations"  $T_{MN}^{(1)}$ therein will be given by:
\begin{eqnarray}
\label{TMN-first-order-in-beta-1}
& & T_{MN}^{(1)} = {\cal T}^{(1)}_{MN} + {\cal T}^{(2)}_{MN} -\frac{g_{MN}}{2}\delta g^{PP^\prime} {\cal T}^{(3)}_{PP^\prime},
\end{eqnarray}
where:
\begin{eqnarray}
\label{TMN-first-order-in-beta-2}
& & {\cal T}^{(1)}_{MN} \equiv 3 \delta g^{PP^\prime}g^{QQ^\prime}G_{MPQR}G_{NP^\prime Q^\prime R^\prime} \equiv \delta g^{PP^\prime} {\cal C}_{MNPP^\prime},\nonumber\\
& & {\cal T}^{(2)}_{MN} \equiv -\frac{\delta g_{MN}}{8} G^2,\nonumber\\
& & {\cal T}^{(3)}_{PQ} \equiv g^{QQ^\prime}g^{RR^\prime}g^{SS^\prime}G_{PQRS}G_{P^\prime Q^\prime R^\prime S^\prime}.
\end{eqnarray}
The non-zero components of ${\cal T}^{(3)}_{PQ}$ are given in appendix {\bf D.2} via two sub-sections: {\bf D.2.1} near the $\psi=0,2\pi,4\pi$-coordinate patches and {\bf D.2.2} away from the same. These along with the ${\cal T}^{(1),(2)}_{MN}$ were used in obtaining the EOMs obtained as the ${\cal O}(l_p^6)$ variation of (\ref{eoms}) (via the substitution of (\ref{ansaetze}) into (\ref{eoms})).

\subsection{$\psi=0, 2\pi, 4\pi$-Coordinate Patches Near $r=r_h$}

In this sub-section, we obtained the EOMs and their solutions, in the IR, near the $\psi=0, 2\pi, 4\pi$-coordinate patches. For simplicity, we restricted to the Ouyang embedding $\left(r^6 + 9 a^2 r^4\right)^{\frac{1}{4}}e^{\frac{i}{2}\left(\psi - \phi_1 - \phi_2\right)}\sin \frac{\theta_1}{2}\sin\frac{\theta_2}{2} = \mu,\ \mu$ being the Ouyang embedding parameter assuming $|\mu|\ll r^{\frac{3}{2}}$, effected, e.g., by working near the $\theta_1=\frac{\alpha_{\theta_1}}{N^{\frac{1}{5}}}, \theta_2 = \frac{\alpha_{\theta_2}}{N^{\frac{3}{10}}}$-coordinate patch. Note, using 
arguments similar to the ones given in \cite{SYZ 3 Ts}, one can show that our results are indepedent of any delocalization in $\theta_{1,2}$.   

Using:
{\footnotesize
\begin{eqnarray}
\label{Gsquared}
& & G^2 = \frac{2 6^{5/6} a^4 {g_s} M^3 \left(\frac{68260644 \left(54 a^2+5\right) \left({r_h}^4-10^4\right) \log (10)}{\left(100-3 a^2\right)^4}-\frac{30876125 \left(12 a^2+1\right)
   \left({r_h}^4-6561\right) \log 9}{9 \left(a^2-27\right)^4}\right)}{125 \pi ^{4/3} N^{7/10} {N_f}^{2/3} \log ^{\frac{17}{3}}(N)},
\end{eqnarray}
}
and:
\begin{eqnarray}
\label{J0-1}
J_0 = \frac{1}{2} R^{\phi_2r \theta_1r}  R_{r \psi\theta_1r}  R_{\phi_2}^{\ \ r \phi_1r}  R^{\psi}_{r \phi_1r} -R^{\phi_2r \theta_1r}
   R_{\phi_1r\theta_1r}  R_{\phi_2}^{\ \ r \phi_1r}  R^{\theta_1}_{\ \ r\theta_1r} ,
\end{eqnarray}
where:
{\footnotesize
\begin{eqnarray}
R^{\phi_2r \theta_1r}  & = & \frac{12 \sqrt{2} \pi ^{3/4} \left(\frac{1}{N}\right)^{5/4} \left(9 a^2+r^2\right) \left(r^4-{r_h}^4\right) \alpha _{\theta _1}^3}{{g_s}^{5/4}
   {N_f}^2 r^4 \left(r^2-3 a^2\right)^3 \left(6 a^2+r^2\right)^3 \log ^3(N) \alpha _{\theta _2}^2}\nonumber\\
& & \times  \left(-81 a^8
   r^4+243 a^8 {r_h}^4+27 a^6 r^6-36 a^6 r^2 {r_h}^4+15 a^4 r^8-27 a^4 r^4 {r_h}^4+a^2 r^{10}-2 a^2 r^6 {r_h}^4\right)
\nonumber\\
   R_{\phi_1r \theta_1r} & = & -\frac{3^{2/3} {g_s}^{7/4} M N^{11/20} {N_f}^{5/3} \log ^{\frac{5}{3}}(N) \log (r)}{2^{5/6} \pi ^{23/12} r^4 \left(r^2-3 a^2\right) \left(6
   a^2+r^2\right) \left(9 a^2+r^2\right) \left(r^4-{r_h}^4\right) \alpha _{\theta _1} \alpha _{\theta _2}^2}\nonumber\\
& & \times  \left(-81 a^8 r^4+243 a^8 {r_h}^4+27 a^6 r^6-36 a^6 r^2
   {r_h}^4+15 a^4 r^8-27 a^4 r^4 {r_h}^4+a^2 r^{10}-2 a^2 r^6 {r_h}^4\right)\nonumber\\
 R_{r \psi\theta_1r} & = & -\frac{3 3^{2/3} {g_s}^{7/4} M N^{3/20} {N_f}^{5/3} \log ^{\frac{5}{3}}(N) \log (r) \alpha _{\theta _1} }{2 2^{5/6} \pi ^{23/12} r^4 \left(r^2-3
   a^2\right) \left(6 a^2+r^2\right) \left(9 a^2+r^2\right) \left(r^4-{r_h}^4\right) \alpha _{\theta _2}^2}\nonumber\\
& & \times \left(-81 a^8 r^4+243 a^8 {r_h}^4+27
   a^6 r^6-36 a^6 r^2 {r_h}^4+15 a^4 r^8-27 a^4 r^4 {r_h}^4+a^2 r^{10}-2 a^2 r^6 {r_h}^4\right)\nonumber\\
  R_{\phi_2}^{\ \ r \phi_1r} & = & -\frac{\sqrt[3]{\frac{\pi }{6}} a^4 \left(\frac{1}{N}\right)^{21/20} \left(9 a^2+r^2\right)^2 \left(\frac{1}{\log (N)}\right)^{4/3}
   \left(r^4-{r_h}^4\right)^2 \alpha _{\theta _2}}{54 {g_s} {N_f}^{4/3} r^6 \left(r^2-3 a^2\right)^2 \left(6 a^2+r^2\right)^2 \alpha _{\theta _1}^2}\nonumber\\
& & \times  \left(19683 \sqrt{6} \alpha _{\theta _1}^6+6642 \alpha _{\theta _2}^2 \alpha _{\theta _1}^3-40
   \sqrt{6} \alpha _{\theta _2}^4\right)\nonumber\\
   R^{\theta_1}_{\ \ r \theta_1r} & = & \frac{6 \left(-81 a^8 r^4+243 a^8 {r_h}^4+27 a^6 r^6-36 a^6 r^2 {r_h}^4+15 a^4 r^8-27 a^4 r^4 {r_h}^4+a^2 r^{10}-2 a^2 r^6
   {r_h}^4\right)}{r^2 \left(r^2-3 a^2\right)^2 \left(6 a^2+r^2\right) \left(9 a^2+r^2\right) \left(r^4-{r_h}^4\right)},\nonumber\\
R^{\psi}_{\ \ r \phi_1r} &  = & -\frac{4 N^{2/5}}{r^2 \left(r^2-3 a^2\right)^2 \left(6 a^2+r^2\right) \left(9 a^2+r^2\right) \left(r^4-{r_h}^4\right) \alpha _{\theta _1}^2}\nonumber\\
& & \times  \left(-81 a^8 r^4+243 a^8 {r_h}^4+27 a^6 r^6-36 a^6 r^2 {r_h}^4+15 a^4 r^8-27 a^4 r^4 {r_h}^4+a^2 r^{10}-2 a^2 r^6
   {r_h}^4\right),\nonumber\\
& &
\end{eqnarray}
}
and therefore one obtains:
{\footnotesize
\begin{eqnarray}
\label{J0-2}
& &J_0 = -\frac{\sqrt[3]{6} a^{10} \left(\frac{1}{{\log N}}\right)^{8/3} M \left(\frac{1}{N}\right)^{7/4} \left(9 a^2+r^2\right) \left(r^4-{r_h}^4\right)
   \log (r) \left(-19683 \sqrt{6} \alpha _{\theta _1}^6-6642 \alpha _{\theta _2}^2 \alpha _{\theta _1}^3+40 \sqrt{6} \alpha _{\theta _2}^4\right)
  }{\pi
   ^{5/6} \sqrt{{g_s}} {N_f}^{5/3} r^{16} \left(3 a^2-r^2\right)^8 \left(6 a^2+r^2\right)^7 \alpha _{\theta _2}^3}\nonumber\\
& &  \times  \left(81 a^6 r^4-243 a^6 {r_h}^4-27 a^4 r^6+36 a^4 r^2 {r_h}^4-15 a^2 r^8+27 a^2 r^4 {r_h}^4-r^{10}+2 r^6 {r_h}^4\right)^3.
\end{eqnarray}
}

Before we proceed, let us provide an estimate of $E_8$. Using (\ref{epsilonD^2R^4}), we picked out eight of the eleven space-time indices (and anti-symmetrize appropriately). Let us estimate $R^{M_1N_1}_{\ \ \ \ \ M_1N_1} R^{M_2N_2}_{\ \ \ \ \ M_2N_2}R^{M_3N_3}_{\ \ \ \ \ M_3N_3}R^{M_4N_4}_{\ \ \ \ \ M_4N_4}$ which will be one of the kinds of terms one will obtain using (\ref{epsilonD^2R^4}).
After a very long and careful computation, we showed that the above contributes a $\frac{1}{N^2}$ via the following most dominant term in the MQGP limit:
\begin{eqnarray}
\label{E8-dominant-large-N}
E_8 \ni R^{tx^1}_{\ \ \ tx^1}R^{x^2x^3}_{\ \ \ x^2x^3}R^{r\theta_1}_{\ \ \ r\theta_1}\left(R^{\psi x^{10}}_{\ \ \ \psi x^{10}}
+ R^{\phi_1\psi}_{\ \ \ \phi_1\psi} +  R^{yz}_{\ \ \ \phi_2\psi}\right) \left(\sim {\cal O}\left(\frac{1}{N^2}\right)\right),
\end{eqnarray}
(and its variation) was disregarded as compared to $J_0$ (and its variation) in the MQGP limit.

The following EOMs' solutions were obtained assuming $f_{\theta_1y}''(r)=0$. We have listed all 15 non-trivial simplified ${\rm EOM}_{MN}$ listed out expanding the coefficients of $f^{(n)}_{MN}, n=0,1,2$ near $r=r_h$ and retaining the LO terms in the powers of $(r-r_h)$ in the same, and then performing a large-$N$-large-$|\log r_h|$-$\log N$ expansion the resulting LO terms are written out and four EOMs which serve as consistency checks.

(i) ${\rm EOM}_{tt}$:

{\footnotesize
\begin{eqnarray}
& & \frac{4 \left(9 b^2+1\right)^3 \left(4374 b^6+1035 b^4+9 b^2-4\right) \beta  b^8 M \left(\frac{1}{N}\right)^{9/4} \Sigma_1
   \left(6 a^2+  {r_h}^2\right) \log (  {r_h})}{27 \pi  \left(18 b^4-3 b^2-1\right)^5  \log N ^2   {N_f}   {r_h}^2
   \alpha _{\theta _2}^3 \left(9 a^2+  {r_h}^2\right)}-\frac{6 \left(  {r_h}^2-2 a^2\right)  {f_t}(r)}{  {r_h}
   \left(  {r_h}^2-3 a^2\right) (r-  {r_h})}\nonumber\\
    & & -\frac{32 \sqrt{2} \left(9 b^2+1\right)^4 \beta  b^{12}
   \left(\frac{1}{N}\right)^{3/20}\Sigma_1 (r-  {r_h})}{81 \pi ^3 \left(1-3 b^2\right)^{10} \left(6 b^2+1\right)^8
     {g_s}^{9/4}  \log N ^4 N^{61/60}   {N_f}^3   {r_h}^4 \alpha _{\theta _1}^7 \alpha _{\theta _2}^6 \left(-27 a^4+6
  a^2   {r_h}^2+  {r_h}^4\right)}+2  {f_t}''(r)=0,\nonumber\\
  &&
\end{eqnarray}

}
where $\Sigma_1$ is defined in (\ref{Sigma_1-def}).

As, the solution to the differential equation:
\begin{equation}
2  {f_t}''(r)+\frac{ {\Gamma_{f_{t1}}}  {f_t}(r)}{r-  {r_h}}+ \Gamma_{f_{t2}} (r-  {r_h})+ \Gamma_{f_{t3}}=0,
\end{equation}
is given by:
{\scriptsize
\begin{eqnarray}
& & \frac{1}{2} \Biggl[-\frac{1}{ {\Gamma_{f_{t1}}}}\Biggl\{(r-  {r_h})^2 \Biggl( {\Gamma_{f_{t1}}}  {\Gamma_{f_{t2}}} (r-  {r_h}) I_1\left(\sqrt{2}
   \sqrt{ {\Gamma_{f_{t1}}} (  {r_h}-r)}\right) G_{1,3}^{2,1}\left(\left.\frac{\sqrt{ {\Gamma_{f_{t1}}}
   (  {r_h}-r)}}{\sqrt{2}},\frac{1}{2}\right|
\begin{array}{c}
 -\frac{3}{2} \\
 -\frac{1}{2},\frac{1}{2},-\frac{5}{2} \\
\end{array}
\right)\nonumber\\
& & + {\Gamma_{f_{t1}}}  {\Gamma_{f_{t3}}} I_1\left(\sqrt{2} \sqrt{ {\Gamma_{f_{t1}}} (  {r_h}-r)}\right)
   G_{1,3}^{2,1}\left(\left.\frac{\sqrt{ {\Gamma_{f_{t1}}} (  {r_h}-r)}}{\sqrt{2}},\frac{1}{2}\right|
\begin{array}{c}
 -\frac{1}{2} \\
 -\frac{1}{2},\frac{1}{2},-\frac{3}{2} \\
\end{array}
\right)\nonumber\\
& & +K_1\left(\sqrt{2} \sqrt{ {\Gamma_{f_{t1}}} (  {r_h}-r)}\right) \Biggl[\sqrt{2} \sqrt{ {\Gamma_{f_{t1}}} (  {r_h}-r)}
   \left(2  {\Gamma_{f_{t2}}} I_4\left(\sqrt{2} \sqrt{ {\Gamma_{f_{t1}}} (  {r_h}-r)}\right)- {\Gamma_{f_{t1}}}  {\Gamma_{f_{t3}}} \,
   _0\tilde{F}_1\left(;3;\frac{1}{2}  {\Gamma_{f_{t1}}} (  {r_h}-r)\right)\right)\nonumber\\
   & & +8  {\Gamma_{f_{t2}}} I_3\left(\sqrt{2}
   \sqrt{ {\Gamma_{f_{t1}}} (  {r_h}-r)}\right)\Biggr]\Biggr)\Biggr\} \nonumber\\
& &    +\sqrt{2} c_1 \sqrt{ {\Gamma_{f_{t1}}} (  {r_h}-r)}
   I_1\left(\sqrt{2} \sqrt{ {\Gamma_{f_{t1}}} (  {r_h}-r)}\right)-\sqrt{2} c_2 \sqrt{ {\Gamma_{f_{t1}}} (  {r_h}-r)}
   K_1\left(\sqrt{2} \sqrt{ {\Gamma_{f_{t1}}} (  {r_h}-r)}\right)\Biggr].\nonumber\\
   &&
\end{eqnarray}
}
To prevent the occurency of a (logarithmic) singularity at $r=r_h$, one sets: $c_2=0$ which yields:

\begin{equation}
f_t(r) = \frac{1}{4}  {\Gamma_{f_{t3}}} (r-  {r_h})^2 + {\cal O}\left((r-r_h)^2\right),
\end{equation}
where:
{\footnotesize
\begin{eqnarray}
&&\Gamma_{f_{t3}} \equiv \frac{4 b^8 \left(9 b^2+1\right)^3 \left(4374 b^6+1035 b^4+9 b^2-4\right) \beta  M \left(\frac{1}{N}\right)^{9/4} \Sigma_1
   \left(6 a^2+  {r_h}^2\right) \log (  {r_h})}{27 \pi  \left(18 b^4-3 b^2-1\right)^5  \log N ^2   {N_f}   {r_h}^2
   \alpha _{\theta _2}^3 \left(9 a^2+  {r_h}^2\right)},\nonumber\\
& &
\end{eqnarray}
}
where $\Sigma_1$ is defined in (\ref{Sigma_1-def}).

(ii) ${\rm EOM}_{x^1x^1}$:

{\footnotesize
\begin{eqnarray}
& &-\frac{6   {r_h} \left(57 a^4+14 a^2   {r_h}^2+  {r_h}^4\right)
   f(r)}{\left(  {r_h}^2-3 a^2\right) \left(6 a^2+  {r_h}^2\right) \left(9 a^2+  {r_h}^2\right) (r-  {r_h})}+2 f''(r)
\nonumber\\
& & -\frac{4 \left(9 b^2+1\right)^4 \left(39 b^2-4\right) \beta  b^8 M \left(\frac{1}{N}\right)^{9/4} \Sigma_1\left(6
   a^2+  {r_h}^2\right) \log (  {r_h})}{9 \pi  \left(3 b^2-1\right)^5 \left(6 b^2+1\right)^4  \log N ^2   {N_f}
     {r_h}^2 \alpha _{\theta _2}^3 \left(9 a^2+  {r_h}^2\right)}\nonumber\\
     & & -\frac{32 \sqrt{2} \left(9 b^2+1\right)^4 \beta  b^{12}
   \left(\frac{1}{N}\right)^{3/20} \Sigma_1 (r-  {r_h})}{81 \pi ^3 \left(1-3 b^2\right)^{10} \left(6 b^2+1\right)^8
     {g_s}^{9/4}  \log N ^4 N^{61/60}   {N_f}^3   {r_h}^4 \alpha _{\theta _1}^7 \alpha _{\theta _2}^6 \left(-27 a^4+6
   a^2   {r_h}^2+  {r_h}^4\right)}=0.
\nonumber\\
& &
\end{eqnarray}
}
This yields:

\begin{equation}
f(r) = \frac{1}{4}  \gamma_{f_2} (r - {r_h})^2 + {\cal O}\left((r-r_h)^3\right),
\end{equation}
where:
\begin{equation}
\gamma_{f_2} \equiv -\frac{4 b^8 \left(9 b^2+1\right)^4 \left(39 b^2-4\right) M \left(\frac{1}{N}\right)^{9/4} \beta  \left(6 a^2+{r_h}^2\right) \log
   ({r_h})\Sigma_1}{9 \pi  \left(3 b^2-1\right)^5 \left(6 b^2+1\right)^4 \log N ^2 {N_f} {r_h}^2 \left(9 a^2+{r_h}^2\right) \alpha
   _{\theta _2}^3}.
\end{equation}

(iii) ${\rm EOM}_{\theta_1x}$:

{\footnotesize
\begin{eqnarray}
\label{f-theta1phi1-i}
& &  -3
    {f_{\theta_1z}}''(r)+2  {f_{\theta_1x}}''(r)-3  {f_{\theta_2y}}''(r) -\frac{4 \left(9 b^2+1\right)^4 \beta  b^{10} M \sqrt[5]{\frac{1}{N}} \Sigma_1 \left(6
   a^2+  {r_h}^2\right)}{3 \pi  \left(-18 b^4+3 b^2+1\right)^4  \log N  \sqrt[3]{N}   {N_f} \alpha
   _{\theta _2}^3 \left(  {r_h}^2-3 a^2\right) \left(9 a^2+  {r_h}^2\right)}\nonumber\\
   & & -\frac{32 \sqrt{2} \left(9
   b^2+1\right)^4 \beta  b^{12} \left(\frac{1}{N}\right)^{3/20} \Sigma_1 (r-  {r_h})}{81 \pi ^3
   \left(1-3 b^2\right)^{10} \left(6 b^2+1\right)^8   {g_s}^{9/4}  \log N ^4 N^{61/60}   {N_f}^3
     {r_h}^4 \alpha _{\theta _1}^7 \alpha _{\theta _2}^6 \left(-27 a^4+6 a^2   {r_h}^2+  {r_h}^4\right)}
     =0.\nonumber\\
     &&
\end{eqnarray}
}
Choosing the two constants of integration obtained by solving (\ref{f-theta1phi1-i}) in such a way that the Neumann b.c. at $r=r_h: f_{\theta_1x}^\prime(r=r_h)=0$, one obtains:

{\footnotesize
\begin{eqnarray}
\label{f-theta1phi1-ii}
&&f_{\theta_1x}(r) = - \frac{\left(9 b^2+1\right)^4 b^{10} M \beta  \left(6 a^2+{r_h}^2\right) \left((r-{r_h})^2+{r_h}^2\right)
   \Sigma_1}{3 \pi  \left(-18 b^4+3 b^2+1\right)^4 \log N  N^{8/15} {N_f} \left(-27 a^4+6 a^2
   {r_h}^2+{r_h}^4\right) \alpha _{\theta _2}^3}+{C_{\theta_1x}}^{(1)} + {\cal O}(r-r_h)^3\nonumber\\
   &&
\end{eqnarray}
}

(iv) ${\rm EOM}_{\theta_1y}$:

{\footnotesize
\begin{eqnarray}
\label{EOM-theta1phi2}
& & \hskip -0.8in -6 f''(r)+2  {f_{zz}}''(r)-2
    {f_{x^{10} x^{10}}}''(r)-3  {f_{\theta_1z}}''(r)-3  {f_{\theta_2y}}''(r)-3
    {f_{xz}}''(r)-2  {f_t}''(r) \nonumber\\
& & \hskip -0.8in\frac{32 \sqrt{2} \left(9 b^2+1\right)^4 \beta  b^{12}
   \left(\frac{1}{N}\right)^{3/20}\Sigma_1
   (r-  {r_h})}{81 \pi ^3 \left(3 b^2-1\right)^{10} \left(6
   b^2+1\right)^8   {g_s}^{9/4}  \log N ^4 N^{61/60}
     {N_f}^3   {r_h}^4 \alpha _{\theta _1}^7 \alpha _{\theta
   _2}^6 \left(  {r_h}^2-3 a^2\right) \left(9
   a^2+  {r_h}^2\right)}=0.\nonumber\\
   &&
\end{eqnarray}
}
The equation (\ref{EOM-theta1phi2}) can be shown to be equivalent to a decoupled second order EOM for $f_{xz}$. Then, expanding the solution around the horizon and requiring the constant of integration $C_{xz}^{(1)}$ appearing in the ${\cal O}(r-r_h)^0$ term to satisfy:
\begin{eqnarray}
\label{fphi1psi-i}
& & -\frac{32 \left(9 b^2+1\right)^4 b^{12} \beta  \left(19683 \sqrt{3} \alpha _{\theta _1}^6+3321 \sqrt{2} \alpha
   _{\theta _2}^2 \alpha _{\theta _1}^3-40 \sqrt{3} \alpha _{\theta _2}^4\right)}{729 \pi ^3 \left(1-3
   b^2\right)^{10} \left(6 b^2+1\right)^8 {g_s}^{9/4} \log N ^4 N^{7/6} {N_f}^3 \left(-27 a^4 {r_h}+6
   a^2 {r_h}^3+{r_h}^5\right) \alpha _{\theta _1}^7 \alpha _{\theta _2}^6}\nonumber\\
   & & -\frac{4 \left(9 b^2+1\right)^4
   b^{10} M {r_h}^2 \beta  \log ({r_h}) \Sigma_1}{81 \pi ^{3/2} \left(3 b^2-1\right)^5 \left(6
   b^2+1\right)^4 \sqrt{{g_s}} \log N ^2 N^{23/20} {N_f} \alpha _{\theta _2}^5}+C_{xz}^{(1)} = 0,\nonumber\\
   &&
\end{eqnarray}
one obtains:

\begin{eqnarray}
& & f_{xz}(r) = \frac{18 b^{10} \left(9 b^2+1\right)^4 M \beta  \left(6 a^2+{r_h}^2\right)
   \left(\frac{(r-{r_h})^2}{{r_h}^2}+1\right) \log ^3({r_h}) \Sigma_1}{\pi  \left(3
   b^2-1\right)^5 \left(6 b^2+1\right)^4 \log N ^4 N^{5/4} {N_f} \left(9 a^2+{r_h}^2\right) \alpha
   _{\theta _2}^3}.\nonumber\\
   &&
\end{eqnarray}

(v) ${\rm EOM}_{\theta_1z}$:

{\footnotesize
\begin{eqnarray}
\label{f-theta1psi-i}
&&2  {f_{\theta_1z}}''(r)-\frac{32 \sqrt{2} b^{12} \left(9 b^2+1\right)^4 \beta  \left(\frac{1}{N}\right)^{3/20} \Sigma_1
   (r-  {r_h})}{81 \pi ^3 \left(1-3 b^2\right)^{10} \left(6 b^2+1\right)^8   {g_s}^{9/4}  \log N ^4 N^{61/60}
     {N_f}^3   {r_h}^4 \alpha _{\theta _1}^7 \alpha _{\theta _2}^6 \left(-27 a^4+6 a^2   {r_h}^2+  {r_h}^4\right)}=0.\nonumber\\
     &&
\end{eqnarray}
}
Choosing the two constants of integration obtained by solving (\ref{f-theta1psi-i}) in such a way that the Neumann b.c. at $r=r_h: f_{\theta_1z}^\prime(r=r_h)=0$, one obtains:

{\footnotesize
\begin{eqnarray}
\label{f-theta1psi-ii}
& & f_{\theta_1z}(r) = \frac{16 \left(9 b^2+1\right)^4 b^{12} \left(\frac{1}{N}\right)^{3/20} \beta  \left(\frac{(r-{r_h})^3}{{r_h}^3}+1\right) \left(19683
   \sqrt{3} \alpha _{\theta _1}^6+3321 \sqrt{2} \alpha _{\theta _2}^2 \alpha _{\theta _1}^3-40 \sqrt{3} \alpha _{\theta _2}^4\right)}{243
   \pi ^3 \left(1-3 b^2\right)^{10} \left(6 b^2+1\right)^8 {g_s}^{9/4} \log N ^4 N^{61/60} {N_f}^3 \left(-27 a^4 {r_h}+6 a^2
   {r_h}^3+{r_h}^5\right) \alpha _{\theta _1}^7 \alpha _{\theta _2}^6}+C_{\theta_1z}^{(1)}+ {\cal O}(r-r_h)^3.\nonumber\\
& &
\end{eqnarray}
}

(vi) ${\rm EOM}_{\theta_2x}$:

{\footnotesize
\begin{eqnarray}
&&2  {f_{\theta_2x}}''(r)-\frac{32 \sqrt{2} b^{12} \left(9 b^2+1\right)^4 \beta  \left(\frac{1}{N}\right)^{3/20} \Sigma_1 (r-  {r_h})}{81 \pi ^3
   \left(1-3 b^2\right)^{10} \left(6 b^2+1\right)^8   {g_s}^{9/4}  \log N ^4 N^{61/60}   {N_f}^3   {r_h}^4 \alpha _{\theta
   _1}^7 \alpha _{\theta _2}^6 \left(-27 a^4+6 a^2   {r_h}^2+  {r_h}^4\right)} = 0.\nonumber\\
&&
\end{eqnarray}
}

{\footnotesize
\begin{eqnarray}
\label{f-theta2phi1-i}
& & \hskip -0.3in f_{\theta_2x}(r) = \frac{16 \left(9 b^2+1\right)^4 b^{12} \left(\frac{1}{N}\right)^{3/20} \beta  \left(\frac{(r-{r_h})^3}{{r_h}^3}+1\right) \left(19683 \sqrt{3}
   \alpha _{\theta _1}^6+3321 \sqrt{2} \alpha _{\theta _2}^2 \alpha _{\theta _1}^3-40 \sqrt{3} \alpha _{\theta _2}^4\right)}{243 \pi ^3 \left(1-3
   b^2\right)^{10} \left(6 b^2+1\right)^8 {g_s}^{9/4} \log N ^4 N^{61/60} {N_f}^3 \left(-27 a^4 {r_h}+6 a^2
   {r_h}^3+{r_h}^5\right) \alpha _{\theta _1}^7 \alpha _{\theta _2}^6}+C_{\theta_2x}^{((1)}\nonumber\\
& &
\end{eqnarray}
}

(vii) ${\rm EOM}_{\theta_2y}$:

{\footnotesize
\begin{eqnarray}
\label{f-theta2phi2-i}
& & 2  {f_{\theta_2y}}''(r) + \frac{12 \left(9 b^2+1\right)^4 \beta  b^{10} M \left(\frac{1}{N}\right)^{7/5} \Sigma_1 \left(6
   a^2+  {r_h}^2\right) \log (  {r_h})}{\pi  \left(3 b^2-1\right)^5 \left(6 b^2+1\right)^4  \log N ^2   {N_f}
     {r_h}^2 \alpha _{\theta _2}^3 \left(9 a^2+  {r_h}^2\right)}\nonumber\\
     & & -\frac{32 \sqrt{2} \left(9 b^2+1\right)^4 \beta  b^{12}
   \left(\frac{1}{N}\right)^{3/20} \Sigma_1 (r-  {r_h})}{81 \pi ^3 \left(1-3 b^2\right)^{10} \left(6 b^2+1\right)^8
     {g_s}^{9/4}  \log N ^4 N^{61/60}   {N_f}^3   {r_h}^4 \alpha _{\theta _1}^7 \alpha _{\theta _2}^6 \left(-27
   a^4+6 a^2   {r_h}^2+  {r_h}^4\right)}=0.\nonumber\\
   &&
\end{eqnarray}
}

Choosing the two constants of integration obtained by solving (\ref{f-theta2phi2-i}) in such a way that the Neumann b.c. at $r=r_h: f_{\theta_2y}^\prime(r=r_h)=0$, and requiring the constant of integration $C_{\theta_2y}^{(1)}$ that figures in the ${\cal O}(r-r_h)^0$-term to satisfy:
{\footnotesize
\begin{eqnarray}
\label{Ctheta2phi2}
\frac{16 \left(9 b^2+1\right)^4 b^{12} \beta  \left(19683 \sqrt{3} \alpha _{\theta _1}^6+3321 \sqrt{2} \alpha _{\theta _2}^2 \alpha
   _{\theta _1}^3-40 \sqrt{3} \alpha _{\theta _2}^4\right)}{243 \pi ^3 \left(1-3 b^2\right)^{10} \left(6 b^2+1\right)^8 {g_s}^{9/4}
   \log N ^4 N^{7/6} {N_f}^3 \left(-27 a^4 {r_h}+6 a^2 {r_h}^3+{r_h}^5\right) \alpha _{\theta _1}^7 \alpha _{\theta
   _2}^6}+C_{\theta_2y}^{(1)} = 0,
\end{eqnarray}
}
 one obtains:

{\footnotesize
\begin{eqnarray}
\label{f-theta2phi2-ii}
f_{\theta_2y} = \frac{3 b^{10} \left(9 b^2+1\right)^4 M \beta  \left(6 a^2+{r_h}^2\right) \left(1-\frac{(r-{r_h})^2}{{r_h}^2}\right) \log
   ({r_h}) \Sigma_1}{\pi  \left(3 b^2-1\right)^5 \left(6 b^2+1\right)^4 \log N ^2 N^{7/5} {N_f} \left(9 a^2+{r_h}^2\right) \alpha
   _{\theta _2}^3}+ {\cal O}\left((r-r_h)^3\right).
\end{eqnarray}
}

(viii) ${\rm EOM}_{\theta_2z}$

{\footnotesize
\begin{eqnarray}
\label{f-theta2z-i}
& & \frac{12 \left(9 b^2+1\right)^4 \beta  b^{10} M \sqrt{\frac{1}{N}} \Sigma_1 \left(6 a^2+{r_h}^2\right) \log
   ({r_h})}{\pi  \left(3 b^2-1\right)^5 \left(6 b^2+1\right)^4 {\log N}^2 N^{2/3} {N_f} {r_h}^2 \alpha
   _{\theta _2}^3 \left(9 a^2+{r_h}^2\right)}\nonumber\\
& & -\frac{32 \sqrt{2} \left(9 b^2+1\right)^4 \beta  b^{12}
   \left(\frac{1}{N}\right)^{3/20} \Sigma_1 (r-{r_h})}{81 \pi ^3 \left(1-3 b^2\right)^{10} \left(6 b^2+1\right)^8
   {g_s}^{9/4} {\log N}^4 N^{61/60} {N_f}^3 {r_h}^4 \alpha _{\theta _1}^7 \alpha _{\theta _2}^6 \left(-27
   a^4+6 a^2 {r_h}^2+{r_h}^4\right)}+2 {f_{\theta_2z}}''(r)=0.\nonumber\\
& &
\end{eqnarray}
}

Choosing the two constants of integration obtained by solving (\ref{f-theta2z-i}) in such a way that the Neumann b.c. at $r=r_h: f_{\theta_2z}^\prime(r=r_h)=0$, one obtains:

{\footnotesize
\begin{eqnarray}
\label{f-theta2z-ii}
& &  f_{\theta_2z} = \frac{3 \left(9 b^2+1\right)^4 b^{10} M \beta  \left(6 a^2+{r_h}^2\right) \left(1-\frac{(r-{r_h})^2}{{r_h}^2}\right) \log
   ({r_h}) \left(19683 \sqrt{6} \alpha _{\theta _1}^6+6642 \alpha _{\theta _2}^2 \alpha _{\theta _1}^3-40 \sqrt{6} \alpha _{\theta
   _2}^4\right)}{\pi  \left(3 b^2-1\right)^5 \left(6 b^2+1\right)^4 {\log N}^2 N^{7/6} {N_f} \left(9 a^2+{r_h}^2\right) \alpha
   _{\theta _2}^3}+{C_{\theta_2z}}^{(1)}.\nonumber\\
& &
\end{eqnarray}
}

(ix) ${\rm EOM}_{xx}$:

{\footnotesize
\begin{eqnarray}
\label{f-phi1phi1-i}
& & {f_{zz}}(r)-2  {f_{\theta_1z}}(r)+2  {f_{\theta_1\phi_{1}}}(r)- {f_r}(r)\nonumber\\
& & +\frac{81 \left(9 b^2+1\right)^4 \beta  b^{10} M \left(\frac{1}{N}\right)^{53/20} \alpha _{\theta _1}^4 \left(19683
   \sqrt{6} \alpha _{\theta _1}^6+6642 \alpha _{\theta _2}^2 \alpha _{\theta _1}^3-40 \sqrt{6} \alpha _{\theta
   _2}^4\right) \left(  {r_h}^2-3 a^2\right)^2 \left(6 a^2+  {r_h}^2\right) \log (  {r_h})}{16 \pi  \left(3
   b^2-1\right)^5  \log N ^2   {N_f} \left(6 a b^2+a\right)^4 \alpha _{\theta _2} \left(9
   a^2+  {r_h}^2\right)}=0.\nonumber\\
   & &
\end{eqnarray}
}
Substituting (\ref{f-theta1phi1-ii}), (\ref{f-theta1psi-ii}) and (\ref{f-psipsi-ii}) into (\ref{f-phi1phi1-i}), one obtains:

{\footnotesize
\begin{eqnarray}
\label{f-phi1phi1-ii}
f_r(r) & = & - \frac{2 \left(9 b^2+1\right)^4 b^{10} M \beta  \left(6 a^2+{r_h}^2\right) \left((r-{r_h})^2+{r_h}^2\right)\Sigma_1}{3 \pi
   \left(-18 b^4+3 b^2+1\right)^4 \log N  N^{8/15} {N_f} \left(-27 a^4+6 a^2 {r_h}^2+{r_h}^4\right) \alpha _{\theta
   _2}^3}\nonumber\\
& & +{C_{zz}}^{(1)}-2 {C_{\theta_1z}}^{(1)}+2 {C_{\theta_1x}}^{(1)} + {\cal O}(r-r_h)^3.
\end{eqnarray}
}

(x) ${\rm EOM}_{xy}$:

{\footnotesize
\begin{eqnarray}
& & 2  {f_{xy}}''(r) + \frac{12 \left(9 b^2+1\right)^4 \beta  b^{10} M \left(\frac{1}{N}\right)^{21/20} \Sigma_1 \left(6 a^2+  {r_h}^2\right) \log
   (  {r_h})}{\pi  \left(3 b^2-1\right)^5 \left(6 b^2+1\right)^4  \log N ^2   {N_f}   {r_h}^2 \alpha _{\theta _2}^3
   \left(9 a^2+  {r_h}^2\right)}\nonumber\\
   & & -\frac{32 \sqrt{2} \left(9 b^2+1\right)^4 \beta  b^{12} \left(\frac{1}{N}\right)^{3/20}
  \Sigma_1 (r-  {r_h})}{81 \pi ^3 \left(1-3 b^2\right)^{10} \left(6 b^2+1\right)^8   {g_s}^{9/4}  \log N ^4
   N^{61/60}   {N_f}^3   {r_h}^4 \alpha _{\theta _1}^7 \alpha _{\theta _2}^6 \left(-27 a^4+6 a^2
     {r_h}^2+  {r_h}^4\right)}= 0.\nonumber\\
     & &
\end{eqnarray}
}
Choosing the two constants of integration obtained by solving (\ref{f-phi1phi2-i}) in such a way that the Neumann b.c. at $r=r_h: f_{xy}^\prime(r=r_h)=0$, one obtains:

{\footnotesize
\begin{eqnarray}
\label{f-phi1phi2-i}
&&f_{xy}(r) =\frac{3 \left(9 b^2+1\right)^4 b^{10} M \beta  \left(6 a^2+{r_h}^2\right) \left(\frac{(r-{r_h})^2}{{r_h}^2}+1\right) \log
   ({r_h}) \alpha _{\theta _2}^3\Sigma_1}{\pi  \left(3 b^2-1\right)^5 \left(6 b^2+1\right)^4 \log N ^2 N^{21/20} {N_f} \left(9
   a^2+{r_h}^2\right) \alpha _{\theta _{2 l}}^6}+C_{xy}^{(1)}+ {\cal O}\left(r-r_h\right)^3.\nonumber\\
   &&
\end{eqnarray}
}

(xi) EOM$\  _{xz}$:

{\footnotesize
\begin{eqnarray}
& & -8  {f_t}''(r)+\frac{24 \left(9 b^2+1\right)^4 \beta  b^{10} M \left(\frac{1}{N}\right)^{3/4} \Sigma_1 \left(9
   a^2+  {r_h}^2\right) \log (  {r_h})}{\pi ^{3/2} \left(3 b^2-1\right)^5 \left(6 b^2+1\right)^4 \sqrt{  {g_s}}
    \log N ^2   {N_f} \alpha _{\theta _1}^2 \alpha _{\theta _2}^5}\nonumber\\
    & & -\frac{64 \sqrt{2} \left(9 b^2+1\right)^4 \beta
   b^{12} \left(\frac{1}{N}\right)^{3/20} \Sigma_1 (r-  {r_h})}{81 \pi ^3 \left(1-3 b^2\right)^{10} \left(6
   b^2+1\right)^8   {g_s}^{9/4}  \log N ^4 N^{61/60}   {N_f}^3   {r_h}^4 \alpha _{\theta _1}^7 \alpha _{\theta
   _2}^6 \left(-27 a^4+6 a^2   {r_h}^2+  {r_h}^4\right)}=0.\nonumber\\
   &&
\end{eqnarray}
}
The solution is given as under:
\begin{equation}
f_t(r) = \frac{1}{16}  \gamma_{f_{t2}}  (r-  {r_h})^2 + {\cal O}\left((r-  {r_h})^3\right),
\end{equation}
where:
\begin{eqnarray}
\gamma_{f_{t2}} \equiv \frac{24 b^{10} \left(9 b^2+1\right)^4 \beta  M \left(\frac{1}{N}\right)^{3/4} \Sigma_1 \left(9
   a^2+  {r_h}^2\right) \log (  {r_h})}{\pi ^{3/2} \left(3 b^2-1\right)^5 \left(6 b^2+1\right)^4 \sqrt{  {g_s}}
    \log N ^2   {N_f} \alpha _{\theta _1}^2 \alpha _{\theta _2}^5}.
\end{eqnarray}

(xii) ${\rm EOM}_{yy}$:

{\footnotesize
\begin{eqnarray}
\label{f-phi2phi2-i}
& & 2
    {f_{\phi_{2}\phi_{2}}}''(r) + \frac{12 \left(9 b^2+1\right)^4 \beta  b^{10} M \left(\frac{1}{N}\right)^{7/4} \Sigma_1 \left(6 a^2+  {r_h}^2\right) \log
   (  {r_h})}{\pi  \left(3 b^2-1\right)^5 \left(6 b^2+1\right)^4  \log N ^2   {N_f}   {r_h}^2 \alpha _{\theta _2}^3
   \left(9 a^2+  {r_h}^2\right)}\nonumber\\
   & & -\frac{32 \sqrt{2} \left(9 b^2+1\right)^4 \beta  b^{12} \left(\frac{1}{N}\right)^{3/20}
  \Sigma_1 (r-  {r_h})}{81 \pi ^3 \left(1-3 b^2\right)^{10} \left(6 b^2+1\right)^8   {g_s}^{9/4}  \log N ^4 N^{61/60}
     {N_f}^3   {r_h}^4 \alpha _{\theta _1}^7 \alpha _{\theta _2}^6 \left(-27 a^4+6 a^2   {r_h}^2+  {r_h}^4\right)}=0\nonumber\\
     &&
\end{eqnarray}
}
Choosing the two constants of integration obtained by solving (\ref{f-phi2phi2-i}) in such a way that the Neumann b.c. at $r=r_h: f_{yy}^\prime(r=r_h)=0$, and choosing the constant of integration $C_{yy}^{(1)}$ appearing in the ${\cal O}(r-r_h)^0$-term  to satisfy:
{\footnotesize
\begin{eqnarray}
\label{Cphi2phi2[1]}
\frac{16 \left(9 b^2+1\right)^4 b^{12} \beta  \left(19683 \sqrt{3} \alpha _{\theta _1}^6+3321 \sqrt{2} \alpha _{\theta _2}^2 \alpha _{\theta
   _1}^3-40 \sqrt{3} \alpha _{\theta _2}^4\right)}{243 \pi ^3 \left(1-3 b^2\right)^{10} \left(6 b^2+1\right)^8 {g_s}^{9/4} \log N ^4
   N^{7/6} {N_f}^3 \left(-27 a^4 {r_h}+6 a^2 {r_h}^3+{r_h}^5\right) \alpha _{\theta _1}^7 \alpha _{\theta _2}^6}+C_{yy}^{(1)} = 0,
\end{eqnarray}
}
one obtains:

{\footnotesize
\begin{eqnarray}
\label{f-phi2phi2-ii}
& & \hskip -0.8in f_{yy}(r) = - \frac{3 b^{10} \left(9 b^2+1\right)^4 M \left(\frac{1}{N}\right)^{7/4} \beta  \left(6 a^2+{r_h}^2\right) \log ({r_h})\Sigma_1
   \left(\frac{(r-{r_h})^2}{h^2 r^2}+1\right)}{\pi  \left(3 b^2-1\right)^5 \left(6 b^2+1\right)^4 \log N ^2 {N_f} {r_h}^2 \left(9
   a^2+{r_h}^2\right) \alpha _{\theta _2}^3}+ {\cal O}\left(\left(r-r_h\right)^3\right).\nonumber\\
   & &
\end{eqnarray}
}

(xiii) ${\rm EOM}_{yz}$:
{\footnotesize
\begin{eqnarray}
\label{f-phi2psi-i}
&&2  {f_{\phi_{2}\psi}}''(r)-\frac{128 \sqrt{2} b^{22} \left(9 b^2+1\right)^8 \beta ^2 M \left(\frac{1}{N}\right)^{3/5} \Sigma_1{}^2
   \left(6 a^2+  {r_h}^2\right) (r-  {r_h}) \log (  {r_h})}{27 \pi ^4 \left(3 b^2-1\right)^{15} \left(6 b^2+1\right)^{12}
     {g_s}^{9/4}  \log N ^6 N^{109/60}   {N_f}^4   {r_h}^6 \alpha _{\theta _1}^7 \alpha _{\theta _2}^9
   \left(  {r_h}^2-3 a^2\right) \left(9 a^2+  {r_h}^2\right)^2}=0.\nonumber\\
   &&
\end{eqnarray}
}
Choosing the two constants of integration obtained by solving (\ref{f-x10x10-i}) in such a way that the Neumann b.c. at $r=r_h: f_{x^{10}x^{10}}^\prime(r=r_h)=0$, one obtains:

{\footnotesize
\begin{eqnarray}
\label{f-phi2psi-ii}
& & f_{yz}(r) = \frac{64 \left(9 b^2+1\right)^8 b^{22} M \left(\frac{1}{N}\right)^{3/5} \beta ^2 \left(6 a^2+{r_h}^2\right)
   \left(\frac{(r-{r_h})^3}{{r_h}^3}+1\right) \log ({r_h}) }{27 \pi ^4 \left(3 b^2-1\right)^{15} \left(6 b^2+1\right)^{12}
   {g_s}^{9/4} \log N ^6 N^{109/60} {N_f}^4 {r_h}^3 \left({r_h}^2-3 a^2\right) \left(9 a^2+{r_h}^2\right)^2 \alpha
   _{\theta _1}^7 \alpha _{\theta _2}^9}\nonumber\\
& &  \times \left(387420489 \sqrt{2} \alpha _{\theta _1}^{12}+87156324 \sqrt{3}
   \alpha _{\theta _2}^2 \alpha _{\theta _1}^9+5778054 \sqrt{2} \alpha _{\theta _2}^4 \alpha _{\theta _1}^6-177120 \sqrt{3} \alpha _{\theta
   _2}^6 \alpha _{\theta _1}^3+1600 \sqrt{2} \alpha _{\theta _2}^8\right)\nonumber\\
   &&+C_{yz}^{(1)}+ {\cal O}(r-r_h)^3.\nonumber\\
& &
   \end{eqnarray}
}

(xiv) ${\rm EOM}_{zz}$:

{\footnotesize
\begin{eqnarray}
\label{f-psipsi-i}
& & 2  {f_{zz}}''(r)-\frac{32 \sqrt{2} \left(9 b^2+1\right)^4 \beta  b^{12} \left(\frac{1}{N}\right)^{3/20} \Sigma_1
   (r-  {r_h})}{81 \pi ^3 \left(1-3 b^2\right)^{10} \left(6 b^2+1\right)^8   {g_s}^{9/4}  \log N ^4 N^{61/60}
     {N_f}^3   {r_h}^4 \alpha _{\theta _1}^7 \alpha _{\theta _2}^6 \left(-27 a^4+6 a^2
     {r_h}^2+  {r_h}^4\right)}\nonumber\\
 & &   +\frac{4 \left(9 b^2+1\right)^4 \beta  b^{10} M \left(\frac{1}{N}\right)^{23/20}
  \Sigma_1 (r-  {r_h}) \log (  {r_h})}{9 \pi ^{3/2} \left(3 b^2-1\right)^5 \left(6 b^2+1\right)^4
   \sqrt{  {g_s}}  \log N ^2   {N_f}   {r_h} \alpha _{\theta _2}^5}=0.\nonumber\\
   &&
\end{eqnarray}
}
Choosing the two constants of integration obtained by solving (\ref{f-psipsi-i}) in such a way that the Neumann b.c. at $r=r_h: f_{zz}^\prime(r=r_h)=0$, one obtains:

{\footnotesize
\begin{eqnarray}
\label{f-psipsi-ii}
&& f_{zz}(r) = C_{zz}^{(1)}-\frac{b^{10} \left(9 b^2+1\right)^4 M \beta  \left({r_h}^2-\frac{(r-{r_h})^3}{{r_h}}\right) \log ({r_h})
   \Sigma_1}{27 \pi ^{3/2} \left(3 b^2-1\right)^5 \left(6 b^2+1\right)^4 \sqrt{{g_s}} \log N ^2 N^{23/20} {N_f} \alpha
   _{\theta _2}^5}+ {\cal O}(r-r_h)^3.\nonumber\\
   &&
\end{eqnarray}
}

(xv) ${\rm EOM}_{x^{10}x^{10}}$:
{\footnotesize
\begin{eqnarray*}
\label{f-x10x10-i}
& & \frac{4 \left(9 b^2+1\right)^3 \beta  b^8 M \left(\frac{1}{N}\right)^{5/4} \Sigma_1 \left(6 a^2+  {r_h}^2\right) \left(9
   b^4 (27  \log N +16)+3 b^2 (9  \log N -8)-8\right) \log ^3(  {r_h})}{\pi  \left(3 b^2-1\right)^5 \left(6
   b^2+1\right)^4  \log N ^5   {N_f}   {r_h}^2 \alpha _{\theta _2}^3 \left(9 a^2+  {r_h}^2\right)}\nonumber\\
   & & - \frac{4 \left(9
   b^2+1\right)^4 \beta  b^{10} M \left(\frac{1}{N}\right)^{23/20} \Sigma_1 (r-  {r_h}) \log (  {r_h})}{9 \pi ^{3/2}
   \left(3 b^2-1\right)^5 \left(6 b^2+1\right)^4 \sqrt{  {g_s}}  (\log N) ^2   {N_f}   {r_h} \alpha _{\theta _2}^5}+2
    {f_{x^{10} x^{10}}}''(r)=0.\nonumber\\
    &&
\end{eqnarray*}
}
Choosing the two constants of integration obtained by solving (\ref{f-x10x10-i}) in such a way that the Neumann b.c. at $r=r_h: f_{x^{10}x^{10}}^\prime(r=r_h)=0$, and requiring the constant of integration $C_{x^{10}x^{10}}^{(1)}$ appearing in the ${\cal O}(r-r_h)^0$ to satisfy:
\begin{eqnarray}
\frac{\left(9 b^2+1\right)^4 b^{10} M {r_h}^2 \beta  \log ({r_h}) \Sigma_1}{27 \pi ^{3/2} \left(3 b^2-1\right)^5 \left(6 b^2+1\right)^4
   \sqrt{{g_s}} \log N ^2 N^{23/20} {N_f} \alpha _{\theta _2}^5}+C_{x^{10}x^{10}}^{(1)} = 0,
\end{eqnarray}
 one obtains:

{\footnotesize
\begin{eqnarray}
\label{f-x10x10-ii}
&&f_{x^{10}x^{10}} = -\frac{27 b^{10} \left(9 b^2+1\right)^4 M \left(\frac{1}{N}\right)^{5/4} \beta  \left(6 a^2+{r_h}^2\right)
   \left(1-\frac{(r-{r_h})^2}{{r_h}^2}\right) \log ^3({r_h}) \Sigma_1}{\pi  \left(3 b^2-1\right)^5 \left(6 b^2+1\right)^4 \log N ^4
   {N_f} {r_h}^2 \left(9 a^2+{r_h}^2\right) \alpha _{\theta _2}^3} + {\cal O}(r-r_h)^3.\nonumber\\
   &&
\end{eqnarray}
}

The remaining EOMS provide consistency checks and are listed below:

\begin{itemize}
\item ${\rm EOM}_{rr}$:

\begin{eqnarray}
\label{consistency-EOMrr-i}
& &  \frac{3  {\alpha_h} \left(9 b^2+1\right)^3 \beta  b^{10} M \Sigma_1}{\pi
   \left(3 b^2-1\right)^5 \left(6 b^2+1\right)^3  \log N ^2 N^{11/12}   {N_f}   {r_h}^2 \alpha
   _{\theta _2}^3}-\frac{ {f_{\theta_1z}}''(r)}{4}-\frac{ {f_{\theta_2y}}''(r)}{4}-\frac{ {f_{xz}}''(r)}{4} = 0.\nonumber\\
& &
\end{eqnarray}

\item ${\rm EOM}_{\theta_1\theta_1}$
{\footnotesize
\begin{eqnarray}
\label{consistency-EOMtheta1theta1-i}
& &  \frac{ {f_{zz}}(r)}{2}- {f_{yz}}(r)+\frac{ {f_{yy}}(r)}{2}\nonumber\\
& & -\frac{32 \sqrt{2} \sqrt[4]{\pi } \left(9 b^2+1\right)^3 \beta  b^{12} \left(\frac{1}{N}\right)^{7/10}
   \left(-19683 \alpha _{\theta _1}^6+216 \sqrt{6} \alpha _{\theta _2}^2 \alpha _{\theta _1}^3+530 \alpha
   _{\theta _2}^4\right)\Sigma_1 \left(6 a^2   {r_h}+  {r_h}^3\right)
   (r-  {r_h})^2}{14348907 \left(1-3 b^2\right)^4   {g_s}^{7/4}  \log N  N^{4/5} \alpha _{\theta _1}^8
   \left(9 a^2+  {r_h}^2\right) \left(6 b^2   {r_h}+  {r_h}\right)^3 \left(108 b^2   {N_f}
     {r_h}^2+  {N_f}\right)^2}=0.\nonumber\\
     & &
\end{eqnarray}
}

\item ${\rm EOM}_{\theta_1\theta_2}$
{\scriptsize
\begin{eqnarray}
\label{consistency-EOMtheta1theta2-i}
&&-\frac{441 N^{3/10} \left(2   {r_h}^2 \alpha _{\theta _1}^3  {f_{x^{10} x^{10}}}(r)+  {r_h}^2 \alpha
   _{\theta _1}^3  {f_{\theta_2y}}(r)\right)}{512 \alpha _{\theta _2}^3 \left(  {r_h}^2-3 a^2\right) \log
   (  {r_h})}-\frac{3 \sqrt{\frac{3}{2}}   {g_s}^{3/2} M \sqrt[10]{N}   {N_f}   {r_h} \left(9
   a^2+  {r_h}^2\right) \left(108 b^2   {r_h}^2+1\right)^2  {f_r}(r) (r-  {r_h})}{\pi ^{3/2}
   \alpha _{\theta _1}^3 \left(-18 a^4+3 a^2   {r_h}^2+  {r_h}^4\right)} = 0.\nonumber\\
   &&
\end{eqnarray}
}
\item ${\rm EOM}_{\theta_2\theta_2}$:
\begin{equation}
\label{consistency-EOMtheta2theta2-i}
 {f_{zz}}(r)- {f_{x^{10} x^{10}}}(r)-2  {f_{\theta_1z}}(r)- {f_r}(r) = 0.
\end{equation}
One can show that by requiring:
\begin{eqnarray}
\label{consistency-EOMtheta1theta1-ii}
& & C_{zz}^{(1)} - 2 C_{\theta_1z}^{(1)} + 2 C_{\theta_1x}^{(1)} =0,\nonumber\\
& & C_{zz}^{(1)} - 2 C_{yz} = 0,\nonumber\\
& & \left|\Sigma_1\right|\ll1,\nonumber\\
& & \frac{2 b^{10} \left(9 b^2+1\right)^4 M {r_h}^2 \beta  \left(6 a^2+{r_h}^2\right) \Sigma_1}{3 \pi
   \left(-18 b^4+3 b^2+1\right)^4 \log N  N^{8/15} {N_f} \left(-27 a^4+6 a^2 {r_h}^2+{r_h}^4\right) \alpha
   _{\theta _2}^3}-2 {C_{\theta_1x}}^{(1)} = 0,\nonumber\\
& &
\end{eqnarray}
 (\ref{consistency-EOMrr-i}) -(\ref{consistency-EOMtheta2theta2-i}) will automatically be satisfied.
\end{itemize}
The ${\cal O}(\beta)$-corrected ${\cal M}$-theory metric of \cite{MQGP} in the MQGP limit near the $\psi=2n\pi, n=0, 1, 2$-branches up to ${\cal O}((r-r_h)^2)$ [and up to ${\cal O}((r-r_h)^3)$ for some of the off-diagonal components along the delocalized $T^3(x,y,z)$] - the components which do not receive an ${\cal O}(\beta)$ corrections, are not listed in (\ref{M-theory-metric-psi=2npi-patch_1}) - is given below:
{\scriptsize
\begin{eqnarray}
\label{M-theory-metric-psi=2npi-patch_1}
 \hskip -0.5in g_{tt} & = & g^{\rm MQGP}_{tt}\Biggl[1 + \frac{1}{4}  \frac{4 b^8 \left(9 b^2+1\right)^3 \left(4374 b^6+1035 b^4+9 b^2-4\right) \beta  M \left(\frac{1}{N}\right)^{9/4} \Sigma_1
   \left(6 a^2+  {r_h}^2\right) \log (  {r_h})}{27 \pi  \left(18 b^4-3 b^2-1\right)^5  \log N ^2   {N_f}   {r_h}^2
   \alpha _{\theta _2}^3 \left(9 a^2+  {r_h}^2\right)} (r-  {r_h})^2\Biggr]
\nonumber\\
g_{x^{1,2,3}x^{1,2,3}} & = &  g^{\rm MQGP}_{x^{1,2,3}x^{1,2,3}}
\Biggl[1 - \frac{1}{4} \frac{4 b^8 \left(9 b^2+1\right)^4 \left(39 b^2-4\right) M \left(\frac{1}{N}\right)^{9/4} \beta  \left(6 a^2+{r_h}^2\right) \log
   ({r_h})\Sigma_1}{9 \pi  \left(3 b^2-1\right)^5 \left(6 b^2+1\right)^4 \log N ^2 {N_f} {r_h}^2 \left(9 a^2+{r_h}^2\right) \alpha
   _{\theta _2}^3} (r - {r_h})^2\Biggr]\nonumber\\
g_{rr} & = & g^{\rm MQGP}_{rr}\Biggl[1 + \Biggl(- \frac{2 \left(9 b^2+1\right)^4 b^{10} M   \left(6 a^2+{r_h}^2\right) \left((r-{r_h})^2+{r_h}^2\right)\Sigma_1}{3 \pi
   \left(-18 b^4+3 b^2+1\right)^4 \log N  N^{8/15} {N_f} \left(-27 a^4+6 a^2 {r_h}^2+{r_h}^4\right) \alpha _{\theta
   _2}^3}\nonumber\\
& & +{C_{zz}}^{(1)}-2 {C_{\theta_1z}}^{(1)}+2 {C_{\theta_1x}}^{(1)}\Biggr)\beta\Biggr]\nonumber\\
 g_{\theta_1x} & = & g^{\rm MQGP}_{\theta_1x}\Biggl[1 + \Biggl(
- \frac{\left(9 b^2+1\right)^4 b^{10} M  \left(6 a^2+{r_h}^2\right) \left((r-{r_h})^2+{r_h}^2\right)
   \Sigma_1}{3 \pi  \left(-18 b^4+3 b^2+1\right)^4 \log N  N^{8/15} {N_f} \left(-27 a^4+6 a^2
   {r_h}^2+{r_h}^4\right) \alpha _{\theta _2}^3}+{C_{\theta_1x}}^{(1)}
\Biggr)\beta\Biggr]\nonumber\\
g_{\theta_1z} & = & g^{\rm MQGP}_{\theta_1z}\Biggl[1 + \Biggl(\frac{16 \left(9 b^2+1\right)^4 b^{12}  \beta  \left(\frac{(r-{r_h})^3}{{r_h}^3}+1\right) \left(19683
   \sqrt{3} \alpha _{\theta _1}^6+3321 \sqrt{2} \alpha _{\theta _2}^2 \alpha _{\theta _1}^3-40 \sqrt{3} \alpha _{\theta _2}^4\right)}{243
   \pi ^3 \left(1-3 b^2\right)^{10} \left(6 b^2+1\right)^8 {g_s}^{9/4} \log N ^4 N^{7/6} {N_f}^3 \left(-27 a^4 {r_h}+6 a^2
   {r_h}^3+{r_h}^5\right) \alpha _{\theta _1}^7 \alpha _{\theta _2}^6}+C_{\theta_1z}^{(1)}\Biggr)\Biggr]\nonumber\\
   g_{\theta_2x} & = & g^{\rm MQGP}_{\theta_2x}\Biggl[1 + \Biggl(
   \frac{16 \left(9 b^2+1\right)^4 b^{12} \left(\frac{(r-{r_h})^3}{{r_h}^3}+1\right) \left(19683 \sqrt{3}
   \alpha _{\theta _1}^6+3321 \sqrt{2} \alpha _{\theta _2}^2 \alpha _{\theta _1}^3-40 \sqrt{3} \alpha _{\theta _2}^4\right)}{243 \pi ^3 \left(1-3
   b^2\right)^{10} \left(6 b^2+1\right)^8 {g_s}^{9/4} \log N ^4 N^{7/6} {N_f}^3 \left(-27 a^4 {r_h}+6 a^2
   {r_h}^3+{r_h}^5\right) \alpha _{\theta _1}^7 \alpha _{\theta _2}^6}+C_{\theta_2x}^{((1)}\Biggr)\beta\Biggr]\nonumber\\
g_{\theta_2y} & = & g^{\rm MQGP}_{\theta_2y}\Biggl[1 +  \frac{3 b^{10} \left(9 b^2+1\right)^4 M \beta \left(6 a^2+{r_h}^2\right) \left(1-\frac{(r-{r_h})^2}{{r_h}^2}\right) \log
   ({r_h}) \Sigma_1}{\pi  \left(3 b^2-1\right)^5 \left(6 b^2+1\right)^4 \log N ^2 N^{7/5} {N_f} \left(9 a^2+{r_h}^2\right) \alpha
   _{\theta _2}^3}\Biggr]\nonumber\\
g_{\theta_2z} & = & g^{\rm MQGP}_{\theta_2z}\Biggl[1 + \Biggl(\frac{3 \left(9 b^2+1\right)^4 b^{10} M  \left(6 a^2+{r_h}^2\right) \left(1-\frac{(r-{r_h})^2}{{r_h}^2}\right) \log
   ({r_h}) \left(19683 \sqrt{6} \alpha _{\theta _1}^6+6642 \alpha _{\theta _2}^2 \alpha _{\theta _1}^3-40 \sqrt{6} \alpha _{\theta
   _2}^4\right)}{\pi  \left(3 b^2-1\right)^5 \left(6 b^2+1\right)^4 {\log N}^2 N^{7/6} {N_f} \left(9 a^2+{r_h}^2\right) \alpha
   _{\theta _2}^3}\nonumber\\
& & +{C_{\theta_2z}}^{(1)}\Biggr)\beta\Biggr]\nonumber\\
g_{xy} & = & g^{\rm MQGP}_{xy}\Biggl[1 + \Biggl(\frac{3 \left(9 b^2+1\right)^4 b^{10} M  \left(6 a^2+{r_h}^2\right) \left(\frac{(r-{r_h})^2}{{r_h}^2}+1\right) \log
   ({r_h}) \alpha _{\theta _2}^3\Sigma_1}{\pi  \left(3 b^2-1\right)^5 \left(6 b^2+1\right)^4 \log N ^2 N^{21/20} {N_f} \left(9
   a^2+{r_h}^2\right) \alpha _{\theta _{2 l}}^6}+C_{xy}^{(1)}\Biggr)\beta\Biggr]\nonumber\\
g_{xz}  & = & g^{\rm MQGP}_{xz}\Biggl[1 + \frac{18 b^{10} \left(9 b^2+1\right)^4 M \beta  \left(6 a^2+{r_h}^2\right)
   \left(\frac{(r-{r_h})^2}{{r_h}^2}+1\right) \log ^3({r_h}) \Sigma_1}{\pi  \left(3b^2-1\right)^5 \left(6 b^2+1\right)^4 \log N ^4 N^{5/4} {N_f} \left(9 a^2+{r_h}^2\right) \alpha
   _{\theta _2}^3}\Biggr]\nonumber\\
g_{yy} & = & g^{\rm MQGP}_{yy}\Biggl[1  - \frac{3 b^{10} \left(9 b^2+1\right)^4 M \left(\frac{1}{N}\right)^{7/4} \beta  \left(6 a^2+{r_h}^2\right) \log ({r_h})\Sigma_1
   \left(\frac{(r-{r_h})^2}{r_h^2}+1\right)}{\pi  \left(3 b^2-1\right)^5 \left(6 b^2+1\right)^4 \log N ^2 {N_f} {r_h}^2 \left(9
   a^2+{r_h}^2\right) \alpha _{\theta _2}^3}\Biggr]\nonumber\\
 g_{yz} & = & g^{\rm MQGP}_{yz}\Biggl[1 + \Biggl(\frac{64 \left(9 b^2+1\right)^8 b^{22} M \left(\frac{1}{N}\right)^{29/12}  \left(6 a^2+{r_h}^2\right)
   \left(\frac{(r-{r_h})^3}{{r_h}^3}+1\right) \log ({r_h}) }{27 \pi ^4 \left(3 b^2-1\right)^{15} \left(6 b^2+1\right)^{12}
   {g_s}^{9/4} \log N ^6  {N_f}^4 {r_h}^3 \left({r_h}^2-3 a^2\right) \left(9 a^2+{r_h}^2\right)^2 \alpha
   _{\theta _1}^7 \alpha _{\theta _2}^9}\nonumber\\
& & \hskip -0.3in \times \left(387420489 \sqrt{2} \alpha _{\theta _1}^{12}+87156324 \sqrt{3}
   \alpha _{\theta _2}^2 \alpha _{\theta _1}^9+5778054 \sqrt{2} \alpha _{\theta _2}^4 \alpha _{\theta _1}^6-177120 \sqrt{3} \alpha _{\theta
   _2}^6 \alpha _{\theta _1}^3+1600 \sqrt{2} \alpha _{\theta _2}^8\right)+C_{yz}^{(1)}\Biggr)\beta\Biggr]\nonumber\\
g_{zz} & = & g^{\rm MQGP}_{zz}\Biggl[1 + \Biggl(C_{zz}^{(1)}-\frac{b^{10} \left(9 b^2+1\right)^4 M \left({r_h}^2-\frac{(r-{r_h})^3}{{r_h}}\right) \log ({r_h})
   \Sigma_1}{27 \pi ^{3/2} \left(3 b^2-1\right)^5 \left(6 b^2+1\right)^4 \sqrt{{g_s}} \log N ^2 N^{23/20} {N_f} \alpha
   _{\theta _2}^5}\Biggr)\beta\Biggr]\nonumber\\
g_{x^{10}x^{10}} & = & g^{\rm MQGP}_{x^{10}x^{10}}\Biggl[1 -\frac{27 b^{10} \left(9 b^2+1\right)^4 M \left(\frac{1}{N}\right)^{5/4} \beta  \left(6 a^2+{r_h}^2\right)
   \left(1-\frac{(r-{r_h})^2}{{r_h}^2}\right) \log ^3({r_h}) \Sigma_1}{\pi  \left(3 b^2-1\right)^5 \left(6 b^2+1\right)^4 \log N ^4
   {N_f} {r_h}^2 \left(9 a^2+{r_h}^2\right) \alpha _{\theta _2}^3}\Biggr],
\end{eqnarray}
   }
   where $\Sigma_1$ is defined in (\ref{Sigma_1-def}), and $g^{\rm MQGP}_{MN}$ are the ${\cal M}$ theory metric components in the MQGP limit at ${\cal O}(\beta^0)$ \cite{mesons_0E++-to-mesons-decays}. The explicit dependence on $\theta_{1,2}$ of the ${\cal M}$-theory metric components up to ${\cal O}(\beta)$ is effected by the replacemements:
$\alpha_{\theta_1}\rightarrow N^{\frac{1}{5}}\sin\theta_{1},\ \alpha_{\theta_2}\rightarrow N^{\frac{3}{1}}\sin\theta_{2}$.

\subsection{$\psi\neq0$ near $r=r_h$}

In this sub-section we looked at the EOMs and their solutions in the $\psi=\psi_0\neq0$-coordinate patch (wherein some $g^{M}_{rM}, M\neq r$ and $g^{M}_{x^{10}N}, N\neq x^{10}$ components are non-zero) near $r=r_h$.

Using:
\begin{eqnarray}
\label{Gsquared-psineq0}
& & G^2 = \frac{49 \pi ^{37/6} \left(\frac{1}{N}\right)^{3/10} \log ^{\frac{10}{3}}(N) \alpha _{\theta _1}^6 \alpha _{\theta _2}^2}{12\times 6^{2/3} ({g_s}-1)^2 {g_s}^{13/2} M^4 {N_f}^{14/3}
   \sin^2\phi_{20} \sin^2\left(\frac{\psi_0}{2}\right) \log ^4(r)},
\end{eqnarray}
and:
\begin{eqnarray}
\label{J0-psineq0}
& & J_0 \sim \frac{1}{2} R^{x^{10}t\theta_2t}R_{t\theta_1\theta_2t}R_{x^{10}}^{\ \ tx^{10}t}R^{\theta_1}_{\ \ tx^{10}t},
\end{eqnarray}
where:
{\footnotesize
\begin{eqnarray}
& & R^{x^{10}t\theta_2t}  =  \frac{512 \sqrt{2} \pi ^{19/4} a^2 \alpha _{\theta _1} \alpha _{\theta _2}^2 \left(9 a^2 r^4+9 a^2 {r_h}^4+r^6+r^2 {r_h}^4\right)}{6
   ({g_s}-1) {g_s}^{5/4} {\log N}^2 M \sqrt[20]{\frac{1}{N}} {N_f}^4 \sin\left(\frac{\psi_0}{2}\right)  r \left(6 a^2+r^2\right)
   \left(r^4-{r_h}^4\right)^2 \log (r)}\nonumber\\
& & R_{t\theta_1\theta_2t} = \frac{3^{2/3}\times4 {g_s}^{5/2} \left(\frac{1}{{\log N}}\right)^{4/3} M^2 {N_f}^{8/3} \sin^2\left(\frac{\psi_0}{2}\right)  \left(9 a^2+r^2\right)
   \left(r^4-{r_h}^4\right) \left(r^4+{r_h}^4\right) \log (r)}{16 \sqrt[3]{2} \pi ^{25/6} r^6 \alpha _{\theta _1}^3 \alpha _{\theta
   _2}^3 \left(6 a^2+r^2\right)}\nonumber\\
& & R_{x^{10}}^{\ \ tx^{10}t} = -\frac{2^4\pi ^{4/3} \left(\frac{1}{{\log N}}\right)^{4/3} \sin^4\left(\frac{\psi_0}{2}\right)  \left(9 a^2+r^2\right) \left(r^4+{r_h}^4\right)}{27 \sqrt[3]{6}
   \left(\frac{1}{N}\right)^{6/5} {N_f}^{4/3} \sin^4\phi_{20} r^2 \alpha _{\theta _2}^4 \left(6 a^2+r^2\right) \left(r^4-{r_h}^4\right)
   \log (r)}\nonumber\\
& & R^{\theta_1}_{\ \ tx^{10}t} = \frac{2\pi ^{5/4} \left(\frac{1}{N}\right)^{17/20} \sin\left(\frac{\psi_0}{2}\right)  \alpha _{\theta _1}^2 \left(9 a^2+r^2\right) \left(r^4-{r_h}^4\right)
   \left(r^4+{r_h}^4\right)}{18 \sqrt{2} {g_s}^{11/4} M {N_f}^2 \sin^2\phi_{20} r^6 \alpha _{\theta _2} \left(6 a^2+r^2\right) \log
   ^2(r)}.
\end{eqnarray}
}
yields:
\begin{eqnarray}
\label{J0}
& & \hskip -0.8in J_0 = -\frac{8\times2^6 \sqrt[3]{2} \pi ^{19/6} a^2 \left(\frac{1}{{\log N}}\right)^{14/3} N^{2/5} \sin^6\left(\frac{\psi_0}{2}\right) \left(9 a^2+r^2\right)^4
   \left(r^4+{r_h}^4\right)^4}{2\times 3^{17/3} ({g_s}-1) {g_s}^{3/2} {N_f}^{14/3} \sin^6\phi_{20} r^{15} \alpha _{\theta _2}^6
   \left(6 a^2+r^2\right)^4 \left(r^4-{r_h}^4\right) \log ^3(r)}.
\end{eqnarray}

From the appendix {\bf D.1.2}, one can show that the set of linearly independent EOMs for the ${\cal O}(l_p^6 R^4)$ corrections to the MQGP metric, with the simplifying assumtion ${f_{\theta_1\theta_1}} = {f_{\theta_1x^{10}}} = {f_{x^{10}x^{10}}} =0$,  reduce to the following set of equations:

(a) ${\rm EOM}_{tt}$:

\begin{eqnarray}
\label{EOMttII-i}
\alpha_{tt}^{f_{\theta_1\theta_2}^\prime} (r-{r_h})^2 f_{\theta_1\theta_2}'(r)+\frac{{\alpha_{tt}^\beta} \beta }{r-{r_h}}=0,
\end{eqnarray}
where:
{\footnotesize
\begin{eqnarray}
\label{defs-67}
& & \alpha_{tt}^{f_{\theta_1\theta_2}^\prime} \equiv \frac{12\times4 a^2 \left(\frac{1}{N}\right)^{2/5} \sin^2\left(\frac{\psi_0}{2}\right) \left(9 a^2+{r_h}^2\right) \log ({r_h})}{\pi  ({g_s}-1) {g_s}
   \sin^2\phi_{20} \left(6 a^2+{r_h}^2\right) \alpha _{\theta _2}^2}\nonumber\\
   & & \alpha_{tt}^\beta \equiv \frac{8192\times16 \pi ^{9/2} a^2 \sqrt[10]{\frac{1}{N}} \sin^4\left(\frac{\psi_0}{2}\right) \beta  \left(9 a^2+{r_h}^2\right)^2 (\log ({r_h})-1) \left(\left(9
   a^2+{r_h}^2\right) \log \left(9 a^2 {r_h}^4+{r_h}^6\right)-8 \left(6 a^2+{r_h}^2\right) \log ({r_h})\right)^2}{729\times16
   ({g_s}-1) {g_s}^{3/2} {N_f}^6 \sin^4\phi_{20} {r_h}^2 \left(6 a^2+{r_h}^2\right)^4 \log ^3({r_h}) \alpha _{\theta
   _2}^4 \log ^8\left(9 a^2 {r_h}^4+{r_h}^6\right)}.\nonumber\\
& &
\end{eqnarray}
}
whose solution is given by:

{\footnotesize
\begin{eqnarray}
\label{EOMttII-ii}
& & \hskip -0.4in f_{\theta_1\theta_2}(r) = \frac{1024\times4 \pi ^{11/2} \beta  N^{3/10} \sin^2\left(\frac{\psi_0}{2}\right) \beta  \left(9 a^2+{r_h}^2\right) (\log ({r_h})-1) \left(\left(9 a^2+{r_h}^2\right) \log
   \left(9 a^2 {r_h}^4+{r_h}^6\right)-8 \left(6 a^2+{r_h}^2\right) \log ({r_h})\right)^2}{2187 \sqrt{{g_s}} {N_f}^6
   \sin^2\phi_{20} {r_h}^2 \left(6 a^2+{r_h}^2\right)^3 (r-{r_h})^2 \log ^4({r_h}) \alpha _{\theta _2}^2 \log ^8\left(9 a^2
   {r_h}^4+{r_h}^6\right)}\nonumber\\
& & \hskip -0.4in +{C_{\theta_1\theta_2}}^{(1)}  = \frac{64 \pi ^{11/2} \left(1-3 b^2\right)^2 \left(9 b^2+1\right) \beta  N^{3/10}\sin^2\left(\frac{\psi_0}{2}\right) \beta }{14348907 \left(6 b^2+1\right)^3 \sqrt{{g_s}}
   {N_f}^6 \sin^2\phi_{20} {r_h}^2 (r-{r_h})^2 \log ^9({r_h}) \alpha _{\theta
   _2}^2}+{C_{\theta_1\theta_2}}^{(1)} + {\cal O}\left(\frac{1}{N^{7/10}}\right).
\end{eqnarray}
}

Assuming:
\begin{equation}
\label{ansatz-b}
b = \frac{1}{\sqrt{3}} - \kappa_b r_h^2\left(\log r_h\right)^{\frac{9}{2}} N^{-\frac{9}{10} - \alpha },
\end{equation}
one obtains:

\begin{eqnarray}
\label{f67II}
f_{\theta_1\theta_2}(r) = \frac{256\times4 \pi ^{11/2} \beta  \kappa_b ^2 \sin^2\left(\frac{\psi_0}{2}\right) {r_h}^2 \beta  \left(\frac{1}{N}\right)^{2 \alpha +\frac{3}{2}}}{129140163
   \sqrt{{g_s}} {N_f}^6 \sin^2\phi_{20} (r-{r_h})^2 \alpha _{\theta _2}^2}+{C_{\theta_1\theta_2}}^{(1)}.
\end{eqnarray}

(b) ${\rm EOM}_{ry}$

\begin{equation}
\label{EOM59II-i}
\alpha_{ry}^{\theta_1\theta_2} f_{\theta_1\theta_2}(r)+\alpha_{ry}^{yy} f_{yy}(r)=0,
\end{equation}
where:
{\footnotesize
\begin{eqnarray}
\label{defs-59}
& & \alpha_{ry}^{\theta_1\theta_2} \equiv \frac{7 \pi ^{17/4} \left(108 a^2+{r_h}\right) \alpha _{\theta _1}^4 \alpha _{\theta _2} \log ^4\left(9 a^2
   {r_h}^4+{r_h}^6\right)}{768 \sqrt{3} ({g_s}-1)^2 {g_s}^{19/4} M^3 \left(\frac{1}{N}\right)^{3/20} {N_f}^3
   \sin^4\phi_{20} {r_h}^2 \log ^3({r_h})},\nonumber\\
& & \alpha_{ry}^{yy} \equiv \frac{7 \sqrt{3} \pi ^{17/4} \log N ^2 \left(\frac{1}{N}\right)^{9/20} \left(108 a^2+{r_h}\right) \left({r_h}^2-3 a^2\right)^2
   (2 \log ({r_h})+1)^2 \alpha _{\theta _1}^{10} \log ^6\left(9 a^2 {r_h}^4+{r_h}^6\right)}{65536 ({g_s}-1)
   {g_s}^{19/4} M^3 {N_f}^3 \sin^6\phi_{20} {r_h}^6 \log ^5({r_h}) \alpha _{\theta _2}}, \nonumber\\
& &
\end{eqnarray}
}
and one obtains:
{\footnotesize
\begin{eqnarray}
\label{f99}
& & \hskip -0.3in f_{yy}(r) = -\frac{64\times4 N^{3/5} \sin^2\phi_{20} {r_h}^4 \alpha _{\theta _2}^2}{9 ({g_s}-1)
   \left({r_h}^2-3 a^2\right)^2 \log ^2(N) \alpha _{\theta _1}^6 \log ^2\left(9 a^2 {r_h}^4+{r_h}^6\right)}\nonumber\\
& & \hskip -0.3in\left(\frac{1024 \pi ^{11/2} \beta  N^{3/10} \sin^2\left(\frac{\psi_0}{2}\right) \beta  \left(9
   a^2+{r_h}^2\right) (\log ({r_h})-1) \left(\left(9 a^2+{r_h}^2\right) \log \left(9 a^2 {r_h}^4+{r_h}^6\right)-8 \left(6
   a^2+{r_h}^2\right) \log ({r_h})\right)^2}{2187 \sqrt{{g_s}} {N_f}^6 \sin^2\phi_{20} {r_h}^2 \left(6 a^2+{r_h}^2\right)^3
   (r-{r_h})^2 \log ^4({r_h}) \alpha _{\theta _2}^2 \log ^8\left(9 a^2 {r_h}^4+{r_h}^6\right)}+{C_{\theta_1\theta_2}}^{(1)}\right).\nonumber\\
& &
\end{eqnarray}
}

Even though $f_{yy}(r)$ is numerically suppressed as the same is ${\cal O}\left(10^{-7}\right)$ apart from an ${\cal O}\left(l_p^6\right)$-suppression - the latter of course common to most $f_{MN}$s -  $f_{yy}(r)$, near $r=r_h$ for ${\cal O}(1)$
${C_{\theta_1\theta_2}}^{(1)}$, goes like $\frac{\frac{N^{\frac{9}{10}}}{r_h^2\log^{11} r_h}}{(r-r_h)^2}$.  To ensure $f_{yy}$ remains finite one has to forego the assumption that ${C_{\theta_1\theta_2}}^{(1)}$ is ${\cal O}(1)$. Around a chosen $(\psi_0,\phi_{20})$, writing $r = r_h + \epsilon_r, \epsilon<<r_h$ close to the horizon, by assuming ${C_{\theta_1\theta_2}}^{(1)} = {C_{\theta_1\theta_2}}^{(1)}(\psi_0,\phi_{20})$:
{\footnotesize
\begin{eqnarray}
\label{finite-f99}
& & \frac{4 \pi ^{11/2} \beta  \left(9 a^2+{r_h}^2\right) \left({r_h}^2-3 a^2\right)^2\sin^2\left(\frac{\psi_0}{2}\right)}{14348907 {\epsilon_r}^3 \sqrt{{g_s}}
   \left(\frac{1}{N}\right)^{3/10} {N_f}^6 {r_h}^2 \left(6 a^2+{r_h}^2\right)^3 \log ^9({r_h}) \alpha _{\theta _2}^2\sin^2\phi_{20}}+{C_{\theta_1\theta_2}}^{(1)}(\psi_0,\phi_{20})=0,
\end{eqnarray}
}
which would imply one can consistently set $f_{yy}(r)=0$ up to ${\cal O}(\beta)$. The idea is that for every chosen value of
$(\psi_0,\phi_{20})$, once upgraded to a local uplift, using the ideas similar to \cite{SYZ 3 Ts}, one can show that the same will correspond to a $G_2$ structure.

(c) ${\rm EOM}_{x^1x^1}$

\begin{eqnarray}
\label{EOM22II-i}
\frac{\alpha_{tt}^\beta \beta }{4 \left(1-\frac{r}{{r_h}}\right) (r-{r_h})}+\alpha_{x^1x^1}^{\theta_1\phi_2}  f_{\theta_1y}(r) (r-{r_h})=0,
\end{eqnarray}
where:
\begin{eqnarray}
\label{defs-EOM22II}
\alpha_{x^1x^1}^{\theta_1\phi_2} \equiv -\frac{\sqrt{\frac{3 \pi }{2}} \left({r_h}^2-3 a^2\right) \left(9 a^2+{r_h}^2\right) \alpha _{\theta _1}^4 \log ^2\left(9 a^2
   {r_h}^4+{r_h}^6\right)}{32 ({g_s}-1) {g_s}^{5/2} M N {N_f} \sin^2\phi_{20} \left(6 a^2+{r_h}^2\right) \log
   ({r_h}) \alpha _{\theta _2}},
\end{eqnarray}
and obtain:
{\footnotesize
\begin{eqnarray}
\label{EOM22II-ii}
& & \hskip -0.4in f_{\theta_1y}(r) \nonumber\\
& & \hskip -0.4in = -\frac{65536 \sqrt{\frac{2}{3}} \pi ^4 a^2 \beta  {g_s} M N^{9/10} 16 \sin^4\left(\frac{\psi_0}{2}\right)\beta  \left(9 a^2+{r_h}^2\right) (\log ({r_h})-1)
   \left(\left(9 a^2+{r_h}^2\right) \log \left(9 a^2 {r_h}^4+{r_h}^6\right)-8 \left(6 a^2+{r_h}^2\right) \log
   ({r_h})\right)^2}{729 {N_f}^5 \sin^2\phi_{20} \left(6 a^2+{r_h}^2\right)^3 \left({r_h}^3-3 a^2 {r_h}\right) (r-{r_h})^3
   \log ^2({r_h}) \alpha _{\theta _1}^4 \alpha _{\theta _2}^3 \log ^{10}\left(9 a^2 {r_h}^4+{r_h}^6\right)}
   \nonumber\\
   & &  \hskip -0.4in = \frac{256\times16 \sqrt{\frac{2}{3}} \pi ^4 b^2 \left(3 b^2-1\right) \left(9 b^2+1\right) \beta  {g_s} M N^{9/10} \sin^4\left(\frac{\psi_0}{2}\right) \beta }{43046721 \left(6
   b^2+1\right)^3 {N_f}^5 \sin^2\phi_{20} {r_h} (r-{r_h})^3 \log ^9({r_h}) \alpha _{\theta _1}^4 \alpha _{\theta _2}^3},
\end{eqnarray}
}
yielding:

\begin{eqnarray}
\label{EOM22II-iii}
& & f_{\theta_1y}(r) = \frac{2048 \sqrt{2} \pi ^4 \beta  {g_s} \kappa_b  M 16\sin^4\left(\frac{\psi_0}{2}\right) {r_h} \beta  N^{-\alpha }}{3486784401 {N_f}^5 \sin^2\phi_{20}
   (r-{r_h})^3 \log ^{\frac{9}{2}}({r_h}) \alpha _{\theta _1}^4 \alpha _{\theta _2}^3}.
\end{eqnarray}

(d) ${\rm EOM}_{\theta_1z}$

\begin{equation}
\label{EOM610II-i}
\alpha_{y x^{10}}^{\theta_1\phi_2}  f_{\theta_1y}(r) + \alpha_{y x^{10}}^{\theta_1\psi}  f_{\theta_1z}(r) + \alpha_{y x^{10}}^{yy}  f_{yy}(r) = 0,
\end{equation}
where:
\begin{eqnarray}
\label{defs-EOM610II}
& & \alpha_{y x^{10}}^{\theta_1\phi_2}  \equiv -\frac{49 \pi ^{13/2} \log N  \left(\frac{1}{N}\right)^{3/10} \left({r_h}^2-3 a^2\right) (2 \log ({r_h})+1) \alpha _{\theta
   _1}^9 \log ^4\left(9 a^2 {r_h}^4+{r_h}^6\right)}{9216 \sqrt{6} {g_s}^{13/2} ({g_s}-1) M^4 {N_f}^5 \sin^4\phi_{20}
   2\sin\left(\frac{\psi_0}{2}\right)  {r_h}^2 \log ^5({r_h})}\nonumber\\
   & & \alpha_{y x^{10}}^{\theta_1\psi}  = - \alpha_{y x^{10}}^{\theta_1\phi_2}  = \alpha_{y x^{10}}^{yy} ,
\end{eqnarray}
and obtain:
\begin{eqnarray}
\label{EOM610II-ii}
& & f_{\theta_1z}(r) = \frac{49 \pi ^{13/2} \log N  \left(\frac{1}{N}\right)^{3/10} \left({r_h}^2-3 a^2\right) (2 \log ({r_h})+1) \alpha _{\theta _1}^9 \log
   ^4\left(9 a^2 {r_h}^4+{r_h}^6\right)}{9216 \sqrt{6} ({g_s}-1) {g_s}^{13/2} M^4 {N_f}^5 \sin^4\phi_{20} 2\sin\left(\frac{\psi_0}{2}\right)  {r_h}^2 \log
   ^5({r_h})}\nonumber\\
\end{eqnarray}
yielding:

\begin{eqnarray}
   & & f_{\theta_1z}(r) = \frac{2048\times16 \sqrt{2} \pi ^4 \beta  {g_s}^{15/2} \kappa_b  M \sin^4\left(\frac{\psi_0}{2}\right) {r_h} \beta  N^{-\alpha }}{3486784401 {g_s}^{13/2} {N_f}^5
   \sin^2\phi_{20} (r-{r_h})^3 \log ^{\frac{9}{2}}({r_h}) \alpha _{\theta _1}^4 \alpha _{\theta _2}^3}
\end{eqnarray}

(e1) ${\rm EOM}_{xy}$

\begin{eqnarray}
\label{EOM89II-i}
\frac{\alpha_{tt}^\beta \beta }{{R_{\frac{tt}{r\phi_1}}} (r-{r_h})^2}+\alpha_{r\phi_1}^{f_{xy}^{\prime\prime}} (r-{r_h}) f_{xy}''(r)+\alpha_{r\phi_1}^{f_{xy}^\prime}
   (r-{r_h}) f_{xy}'(r)=0,
\end{eqnarray}
where:
{\scriptsize
\begin{eqnarray}
\label{defs-EOM89II}
& & {R_{\frac{tt}{r\phi_1}}} \equiv \frac{3 \left(\frac{1}{N}\right)^{3/5} {r_h} \left(9 a^2+{r_h}^2\right) \alpha _{\theta _2} \log ^3\left(9 a^2
   {r_h}^4+{r_h}^6\right)}{160 \sqrt{\pi } \sqrt{{g_s}} \sin\phi_{10} \sin\left(\frac{\psi_0}{2}\right)  \alpha _{\theta _1} \left(24 ({g_s}-1)^2 \left(6
   a^2+{r_h}^2\right) \log ({r_h})+(2 {g_s}-3) \left(9 a^2+{r_h}^2\right) \log \left(9 a^2 {r_h}^4+{r_h}^6\right)\right)},\nonumber\\
   & &  \alpha_{r\phi_1}^{f_{xy}^{\prime\prime}}\equiv \frac{5 ({g_s}-1) \sin\phi_1 2\sin\left(\frac{\psi_0}{2}\right)  {r_h} \left(9 a^2+{r_h}^2\right) }{729 \sqrt{6 \pi } {g_s}^{3/2} {N_f} \sin^2\phi_{20} \left(6 a^2+{r_h}^2\right) \log
   ({r_h}) \alpha _{\theta _1} \alpha _{\theta _2}^3 \log ^2\left(9 a^2 {r_h}^4+{r_h}^6\right)} \Biggl(112 ({g_s}-1) {g_s} {N_f} \psi ^2 \log
   ({r_h}) \left(81 \alpha _{\theta _1}^3+5 \sqrt{6} \alpha _{\theta _2}^2\right)\nonumber\\
& & -\frac{1}{{r_h}^2}\Biggl\{243 \alpha _{\theta _1}^3 \log ^2\left(9 a^2
   {r_h}^4+{r_h}^6\right)\nonumber\\
& &  \times \Biggl[\log ({r_h}) \left(6 a^2 ({g_s} \log N  {N_f}-4 \pi )+4 {g_s} {N_f}
   \left({r_h}^2-3 a^2\right) \log \left(\frac{1}{4} \alpha _{\theta _1} \alpha _{\theta _2}\right)+{r_h}^2 (8 \pi -{g_s} (2
   \log N +3) {N_f})\right)\nonumber\\
   &&+{g_s} {N_f} \left(3 a^2-{r_h}^2\right) \left(\log N -2 \log \left(\frac{1}{4} \alpha
   _{\theta _1} \alpha _{\theta _2}\right)\right)\nonumber\\
& &  +18 {g_s} {N_f} \left({r_h}^2-3 a^2 (6 {r_h}+1)\right) \log
   ^2({r_h})\Biggr]\Biggr\}\Biggr),
\end{eqnarray}
}
which yields:
{\footnotesize
\begin{eqnarray}
\label{EOM89II-ii}
& &  f_{xy}(r) = \frac{e^{-\frac{\alpha_{r\phi_1}^{f_{xy}^\prime} r}{\alpha_{r\phi_1}^{f_{xy}^{\prime\prime}}}} \left(\alpha_{tt}^\beta \alpha_{r\phi_1}^{f_{xy}^\prime}\ ^2 \beta  (r-{r_h})
   e^{\frac{\alpha_{r\phi_1}^{f_{xy}^\prime} {r_h}}{\alpha_{r\phi_1}^{f_{xy}^{\prime\prime}}}} {Ei}\left(\frac{\alpha_{r\phi_1}^{f\ _{xy}^\prime}
   (r-{r_h})}{\alpha_{r\phi_1}^{f_{xy}^{\prime\prime}}}\right)-\alpha_{tt}^\beta \alpha_{r\phi_1}^{f_{xy}^{\prime\prime}} \alpha_{r\phi_1}^{f_{xy}^\prime} \beta  e^{\frac{\alpha_{r\phi_1}^{f_{xy}^\prime}
   r}{\alpha_{r\phi_1}^{f_{xy}^{\prime\prime}}}}+2 \alpha_{r\phi_1}^{f_{xy}^{\prime\prime}}\ ^3 c_1^{(89)} {R_{\frac{tt}{r\phi_1}}} ({r_h}-r)\right)}{2 \alpha_{r\phi_1}^{f_{xy}^{\prime\prime}}\ ^2
   \alpha_{r\phi_1}^{f\ _{xy}^\prime} {R_{\frac{tt}{r\phi_1}}} (r-{r_h})}+c_2^{(89)}\nonumber\\
& & = -\frac{\alpha_{tt}^\beta \beta }{2 (\alpha_{r\phi_1}^{f_{xy}^{\prime\prime}} {R_{\frac{tt}{r\phi_1}}}) (r-{r_h})}+\Biggl(\frac{\alpha_{tt}^\beta \alpha_{r\phi_1}^{f_{xy}^\prime} \beta  \log
   (r-{r_h})}{2 \alpha_{r\phi_1}^{f_{xy}^{\prime\prime}}\ ^2 {R_{\frac{tt}{r\phi_1}}}}+\frac{\gamma
\alpha_{tt}^\beta \alpha_{r\phi_1}^{f_{xy}^\prime} \beta }{2 \alpha_{r\phi_1}^{f_{xy}^{\prime\prime}}\ ^2
   {R_{\frac{tt}{r\phi_1}}}}-\frac{\alpha_{tt}^\beta \alpha_{r\phi_1}^{f_{xy}^\prime} \beta  \log \left(\frac{\alpha_{r\phi_1}^{f_{xy}^{\prime\prime}}}{\alpha_{r\phi_1}^{f_{xy}^\prime}}\right)}{4
   \alpha_{r\phi_1}^{f_{xy}^{\prime\prime}}\ ^2 {R_{\frac{tt}{r\phi_1}}}}+\frac{\alpha_{tt}^\beta \alpha_{r\phi_1}^{f_{xy}^\prime} \beta  \log
   \left(\frac{\alpha_{r\phi_1}^{f_{xy}^\prime}}{\alpha_{r\phi_1}^{f_{xy}^{\prime\prime}}}\right)}{4 \alpha_{r\phi_1}^{f_{xy}^{\prime\prime}}\ ^2 {R_{\frac{tt}{r\phi_1}}}}\nonumber\\
& & -\frac{\alpha_{r\phi_1}^{f_{xy}^{\prime\prime}} c_1
   e^{-\frac{\alpha_{r\phi_1}^{f_{xy}^\prime} {r_h}}{\alpha_{r\phi_1}^{f_{xy}^{\prime\prime}}}}}{\alpha_{r\phi_1}^{f_{xy}^\prime}}+c_2\Biggr)+O\left(r-{r_h}\right) \nonumber\\
    &&= \frac{128\times16 \sqrt{\frac{2}{3}} \pi ^{11/2} b^2 \left(3 b^2-1\right) \beta  \sqrt{N} \sin^4\left(\frac{\psi_0}{2}\right) \beta  \left(3 b^2 \left(8 {g_s}^2-10
   {g_s}-1\right)+4 {g_s}^2-6 {g_s}+1\right)}{14348907 \left(6 b^2+1\right)^3 ({g_s}-1)^2 \sqrt{{g_s}} \log N  {N_f}^6
   \sin^2\phi_{20} {r_h}^2 (r-{r_h}) \log ^{10}({r_h}) \alpha _{\theta _1} \alpha _{\theta _2}^2},\nonumber\\
   &&
\end{eqnarray}
}
implying:

\begin{eqnarray}
\label{EOM89II-iii}
 & & \hskip -0.8in f_{xy}(r) = \frac{16384 \sqrt{2} \pi ^{11/2} \beta  \sqrt{{g_s}} (3 {g_s}-4) \kappa_b  \left(\frac{1}{N}\right)^{2/5} \sin^4\left(\frac{\psi_0}{2}\right)  \beta  N^{-\alpha
   }}{1162261467 ({g_s}-1)^2 {N_f}^6 \sin^2\phi_{20} \log (N) (r-{r_h}) \log ^{\frac{11}{2}}({r_h}) \alpha _{\theta _1} \alpha
   _{\theta _2}^2}.
\end{eqnarray}

(e2) ${\rm EOM}_{\theta_1x}$ (consistency)

\begin{equation}
\label{EOM68II-i}
\alpha_{\theta_1x}^{f_{xy}^\prime}\  f_{xy}'(r)+\frac{\alpha_{\theta_1x}^\beta  \beta }{(r-{r_h})^2}=0,
\end{equation}
where:
{\scriptsize
\begin{eqnarray}
\label{defs-EOM68II}
& &   \alpha_{\theta_1x}^{f_{xy}^\prime}\  \equiv -\frac{4{g_s}^{5/4} \log N  M {N_f} \sin^2\left(\frac{\psi_0}{2}\right) \left({r_h}^2-3 a^2\right) \left(9 a^2+{r_h}^2\right) (2 \log ({r_h})+1)}{36
   \sqrt{2} \pi ^{7/4} \left(\frac{1}{N}\right)^{13/20} \sin^2\phi_{20} {r_h}^2 \left(6 a^2+{r_h}^2\right) \log ({r_h}) \alpha _{\theta
   _1} \alpha _{\theta _2}^4}\nonumber\\
& &   \alpha_{\theta_1x}^\beta  \equiv \frac{32768\times64 \pi ^{15/4} a^2 {g_s}^{3/4} M N^{23/20} \sin^6\left(\frac{\psi_0}{2}\right)  \beta  \left(9 a^2+{r_h}^2\right)^2 (\log ({r_h})-1) \left(\left(9
   a^2+{r_h}^2\right) \log \left(9 a^2 {r_h}^4+{r_h}^6\right)-8 \left(6 a^2+{r_h}^2\right) \log ({r_h})\right)^2}{2187 \sqrt{3}
   ({g_s}-1) {N_f}^5 \sin^4\phi_{20} {r_h}^3 \left(6 a^2+{r_h}^2\right)^4 \log ^2({r_h}) \alpha _{\theta _1}^2 \alpha _{\theta
   _2}^6 \log ^{10}\left(9 a^2 {r_h}^4+{r_h}^6\right)},\nonumber\\
& &
\end{eqnarray}
}
which obtains as its LHS:
\begin{eqnarray}
\label{EOM68II-ii}
& & \frac{128\times64 \pi ^{15/4} b^2 \left(1-3 b^2\right)^2 \left(9 b^2+1\right)^2 \beta  {g_s}^{3/4} M N^{23/20} \sin\left(\frac{\psi_0}{2}\right) \beta }{129140163 \sqrt{3}
   ({g_s}-1) {N_f}^5 {r_h} \left(6 b^2 +1\right)^4\sin^4\phi_{20} (r-{r_h})^2 \log ^9({r_h}) \alpha _{\theta _1}^2 \alpha
   _{\theta _2}^6} \nonumber\\
   & & = \frac{8192\times64 \pi ^{15/4} \beta  {g_s}^{3/4} \kappa_b ^2 M \sin^6\left(\frac{\psi_0}{2}\right)  {r_h}^3 \beta  \left(\frac{1}{N}\right)^{2 \alpha
   +\frac{13}{20}}}{10460353203 \sqrt{3} ({g_s}-1) {N_f}^5 \sin^4\phi_{20} (r-{r_h})^2 \alpha _{\theta _1}^2 \alpha _{\theta _2}^6},
\end{eqnarray}
that in the MQGP limit, is vanishingly small.

(f) ${\rm EOM}_{r\theta_1}$

\begin{eqnarray}
\label{EOM56II-i}
& & a_{r\theta_1}^{\theta_1y} f_{\theta_1y}(r)+a_{r\theta_1}^{xy} f_{xy}(r)+a_{r\theta_1}^{yz}  f_{yz}(r)+\frac{a_{r\theta_1}^\beta  \beta }{(r-{r_h})^2}=0,
\end{eqnarray}
where:
{\scriptsize
\begin{eqnarray}
\label{EOM56II-i}
& &  \hskip -0.4in a_{r\theta_1}^{\theta_1y} \equiv -\frac{7 \pi ^3 \log N  \left(108 a^2+{r_h}\right) \left({r_h}^2-3 a^2\right) (2 \log ({r_h})+1) \alpha _{\theta _1}^7 \log
   ^4\left(9 a^2 {r_h}^4+{r_h}^6\right)}{2048 ({g_s}-1) {g_s}^3 M^2 \sqrt[5]{\frac{1}{N}} {N_f}^2 \sin^4\phi_{20} {r_h}^4 \log
   ^3({r_h}) \alpha _{\theta _2}^2},\nonumber\\
   & & \hskip -0.4in a_{r\theta_1}^{xy} \equiv \frac{28 \pi ^3 \sin^2\left(\frac{\psi_0}{2}\right) \left(108 a^2+{r_h}\right) \alpha _{\theta _1}^6 \log ^2\left(9 a^2 {r_h}^4+{r_h}^6\right)}{64 \sqrt{6}
   {g_s}^3 M^2 \left(\frac{1}{N}\right)^{2/5} {N_f}^2 \sin^4\phi_{20} {r_h}^2 \log ^2({r_h}) \alpha _{\theta _2}^2},\nonumber\\
   & & \hskip -0.4in a_{r\theta_1}^{yz}  \equiv -\frac{28 \pi ^3  \sin^2\left(\frac{\psi_0}{2}\right) \left(108 a^2+{r_h}\right) \alpha _{\theta _1}^6 \log ^2\left(9 a^2 {r_h}^4+{r_h}^6\right)}{64 \sqrt{6}
   {g_s}^3 M^2 \left(\frac{1}{N}\right)^{2/5} {N_f}^2 \sin^4\phi_{20} {r_h}^2 \log ^2({r_h}) \alpha _{\theta _2}^2},\nonumber\\
   & & \hskip -0.4in a_{r\theta_1}^\beta  \equiv -\frac{163840\times32 \pi ^{15/4} a^2 {g_s}^{3/4} M N^{13/20} \sin\phi_{10}  \sin^5\left(\frac{\psi_0}{2}\right) \beta  \left(9 a^2+{r_h}^2\right)^2 (\log ({r_h})-1)
   \left(\left(9 a^2+{r_h}^2\right) \log \left(9 a^2 {r_h}^4+{r_h}^6\right)-8 \left(6 a^2+{r_h}^2\right) \log
   ({r_h})\right)^2}{243 \sqrt{3} ({g_s}-1) {N_f}^5 \sin^4\phi_{20} {r_h}^3 \left(6 a^2+{r_h}^2\right)^4 (r-{r_h})^2
   \log ^2({r_h}) \alpha _{\theta _1} \alpha _{\theta _2}^5 \log ^{10}\left(9 a^2 {r_h}^4+{r_h}^6\right)},\nonumber\\
& &
\end{eqnarray}
}
that yields:
\begin{eqnarray}
\label{EOM56II-ii}
& & f_{yz}(r) = \frac{512 \pi ^4 b^2 \left(1-3 b^2\right)^2 \left(9 b^2+1\right) \beta  {g_s} \log N  M N^{7/10}  \sin^2\left(\frac{\psi_0}{2}\right) \beta }{4782969 \left(6
   b^2+1\right)^3 ({g_s}-1) {N_f}^5 \sin^2\phi_{20} {r_h} (r-{r_h})^3 \log ^7({r_h}) \alpha _{\theta _1}^3 \alpha _{\theta _2}^3}\nonumber\\
\end{eqnarray}
implying:

\begin{eqnarray}
\label{EOM56II-iii}
   & & f_{yz}(r) = \frac{8192 \pi ^4 \beta  {g_s} \kappa_b ^2 \log N  M  \sin^2\left(\frac{\psi_0}{2}\right) {r_h}^3 \beta  N^{-2 \alpha -\frac{11}{10}} \log
   ^2({r_h})}{129140163 ({g_s}-1) {N_f}^5 \sin^2\phi_{20} (r-{r_h})^3 \alpha _{\theta _1}^3 \alpha _{\theta _2}^3}.
\end{eqnarray}

(g) ${\rm EOM}_{xx}$

\begin{equation}
\label{EOM88II-i}
a_{xx}^{f_{xy}^\prime}  f_{xy}'(r) + a_{xx}^{f_{xz}^\prime}    f_{xz}'(r) = 0,
\end{equation}
where:
{\scriptsize
\begin{eqnarray}
& &   a_{xx}^{f_{xy}^\prime}  \equiv \nonumber\\
& &   -\frac{224 \left(2 \pi ^3 \alpha _{\theta _1}^2 \alpha _{\theta _2}^4 \log ^2\left(9 a^2 {r_h}^4+{r_h}^6\right)+360 ({g_s}-1)
   {g_s}^3 M^2 {N_f}^2 \sin^2\left(\frac{\psi_0}{2}\right) \log ^2({r_h}) \alpha _{\theta _2}^2+243\times4 \sqrt{6} ({g_s}-1) {g_s}^3 M^2 {N_f}^2
   \sin^2\left(\frac{\psi_0}{2}\right) \log ^2({r_h}) \alpha _{\theta _1}^3\right)}{177147\times4 \sqrt{\pi } ({g_s}-1) {g_s}^{7/2} M^2
   \left(\frac{1}{N}\right)^{7/10} {N_f}^2 \sin^2\left(\frac{\psi_0}{2}\right) \log ^2({r_h}) \alpha _{\theta _1}^6 \alpha _{\theta _2}^2 \log \left(9 a^2
   {r_h}^4+{r_h}^6\right)},\nonumber\\
   & &   a_{xx}^{f_{xz}^\prime}    \equiv \nonumber\\
& &  -\frac{32 {r_h} \left(9 a^2+{r_h}^2\right) \alpha _{\theta _2}^2 }{177147 \sqrt{\pi } ({g_s}-1) {g_s}^{7/2} M^2
   \left(\frac{1}{N}\right)^{7/10} {N_f}^2 \sin^2\phi_{20} \psi ^2 \left(6 a^2+{r_h}^2\right) \log ^2({r_h}) \alpha _{\theta
   _1}^6 \log ^3\left(9 a^2 {r_h}^4+{r_h}^6\right)}\nonumber\\
& &   \times \Biggl(\frac{63\times4 ({g_s}-1) {g_s}^3 M^2 {N_f}^2
   \sin^2\phi_{20} \sin^2\left(\frac{\psi_0}{2}\right) \left(6 a^2+{r_h}^2\right) \log ^2({r_h}) \left(4 \alpha _{\theta _2}^2-27 \sqrt{6} \alpha _{\theta
   _1}^3\right) \log ^2\left(9 a^2 {r_h}^4+{r_h}^6\right)}{\left(9 a^2 {r_h}+{r_h}^3\right) \alpha _{\theta
   _2}^4}\nonumber\\
& & +\frac{2187\times64 ({g_s}-1)^2 {g_s}^3 M^2 {N_f}^2 \sin^6\left(\frac{\psi_0}{2}\right) \log ({r_h}) \alpha _{\theta _1}^6 \log \left(9 a^2
   {r_h}^4+{r_h}^6\right)}{{r_h} \alpha _{\theta _2}^6}+\frac{17496\times64 ({g_s}-1)^3 {g_s}^3 M^2 {N_f}^2 \sin^6\left(\frac{\psi_0}{2}\right)
   \left(6 a^2+{r_h}^2\right) \log ^2({r_h}) \alpha _{\theta _1}^6}{\left(9 a^2 {r_h}+{r_h}^3\right) \alpha _{\theta
   _2}^6}\nonumber\\
& &  -\frac{14 \pi ^3 \sin^2\phi_{20} \left(6 a^2+{r_h}^2\right) \alpha _{\theta _1}^2 \log ^4\left(9 a^2
   {r_h}^4+{r_h}^6\right)}{9 a^2 {r_h}+{r_h}^3}\Biggr),
\end{eqnarray}
}
which yields:

{\footnotesize
\begin{eqnarray}
\label{EOM88II-ii}
& &  f_{xz}(r) = C_{xz}^{(1)} -\frac{1024\times16 \Sigma_2\sqrt{2} \pi ^{11/2} \beta  \sqrt{{g_s}} (3 {g_s}-4) \kappa_b  \sin^4\left(\frac{\psi_0}{2}\right) \beta  N^{-\alpha -\frac{2}{5}}
  }{10460353203 ({g_s}-1)^2 {N_f}^6 \sin^2\phi_{20}
   (r-{r_h}) \log ^{\frac{13}{2}}({r_h}) \alpha _{\theta _1} \alpha _{\theta _2}^2}.\nonumber\\
& &
\end{eqnarray}
}
where:
{\footnotesize
\begin{eqnarray}
& & \Sigma_2 \equiv \frac{ \left(40 ({g_s}-1) {g_s}^3 M^2 {N_f}^2 \sin^2\left(\frac{\psi_0}{2}\right) \alpha _{\theta _2}^2+108 \sqrt{6} ({g_s}-1) {g_s}^3 M^2 {N_f}^2 \sin^2\left(\frac{\psi_0}{2}\right)
   \alpha _{\theta _1}^3+8 \pi ^3 \alpha _{\theta _1}^2 \alpha _{\theta _2}^4\right)}{ \left(-4 ({g_s}-1) {g_s}^3 M^2 {N_f}^2 \psi
   ^2 \alpha _{\theta _2}^2+108 \sqrt{6} ({g_s}-1) {g_s}^3 M^2 {N_f}^2 \sin^2\left(\frac{\psi_0}{2}\right) \alpha _{\theta _1}^3+8 \pi ^3 \alpha _{\theta _1}^2
   \alpha _{\theta _2}^4\right)}
\end{eqnarray}
}
 The following is the final result as regards the ${\cal O}(\beta)$-corrected ${\cal M}$-theory metric of \cite{MQGP} in the MQGP limit in the
$\psi\neq 2n\pi, n=0, 1, 2$-branches up to ${\cal O}((r-r_h)^2)$ - the components which do not receive an ${\cal O}(\beta)$ corrections, are not listed in (\ref{ M-theory-metric-psi-neq-2npi-patch}):
{\scriptsize
\begin{eqnarray}
\label{ M-theory-metric-psi-neq-2npi-patch}
g_{\theta_1\theta_2} & = & g_{\theta_1\theta_2}^{\rm MQGP}\Biggl[1 + \Biggl(\frac{256\times4 \pi ^{11/2}   \kappa_b ^2 \sin^2\left(\frac{\psi_0}{2}\right) {r_h}^2 \beta  \left(\frac{1}{N}\right)^{2 \alpha +\frac{3}{2}}}{129140163
   \sqrt{{g_s}} {N_f}^6 \sin^2\phi_{20} (r-{r_h})^2 \alpha _{\theta _2}^2}+{C_{\theta_1\theta_2}}^{(1)}\Biggr)\beta\Biggr]\nonumber\\
g_{yy} & = & g_{yy}^{\rm MQGP}\Biggl[1 + \Biggl(-\frac{64\times4 N^{3/5} \sin^2\phi_{20} {r_h}^4 \alpha _{\theta _2}^2}{9 ({g_s}-1)
   \left({r_h}^2-3 a^2\right)^2 \log ^2(N) \alpha _{\theta _1}^6 \log ^2\left(9 a^2 {r_h}^4+{r_h}^6\right)}\nonumber\\
& &\times  \Biggl\{\frac{1024 \pi ^{11/2}   N^{3/10} \sin^2\left(\frac{\psi_0}{2}\right) \beta  \left(9
   a^2+{r_h}^2\right) (\log ({r_h})-1) \left(\left(9 a^2+{r_h}^2\right) \log \left(9 a^2 {r_h}^4+{r_h}^6\right)-8 \left(6
   a^2+{r_h}^2\right) \log ({r_h})\right)^2}{2187 \sqrt{{g_s}} {N_f}^6 \sin^2\phi_{20} {r_h}^2 \left(6 a^2+{r_h}^2\right)^3
   (r-{r_h})^2 \log ^4({r_h}) \alpha _{\theta _2}^2 \log ^8\left(9 a^2 {r_h}^4+{r_h}^6\right)}\nonumber\\
& & + {C_{\theta_1\theta_2}}^{(1)}\Biggr\}\Biggr)\beta\Biggr]\nonumber\\
g_{\theta_1y} & = & g_{\theta_1y}^{\rm MQGP}\Biggl[1 + \frac{2048 \sqrt{2} \pi ^4 \beta  {g_s} \kappa_b  M 16\sin^4\left(\frac{\psi_0}{2}\right) {r_h} \beta  N^{-\alpha }}{3486784401 {N_f}^5 \sin^2\phi_{20}
   (r-{r_h})^3 \log ^{\frac{9}{2}}({r_h}) \alpha _{\theta _1}^4 \alpha _{\theta _2}^3}\Biggr]\nonumber\\
g_{\theta_1z} & = & g_{\theta_1z}^{\rm MQGP}\Biggl[1 + \frac{2048\times16 \sqrt{2} \pi ^4 \beta  {g_s}^{15/2} \kappa_b  M \sin^4\left(\frac{\psi_0}{2}\right) {r_h} \beta  N^{-\alpha }}{3486784401 {g_s}^{13/2} {N_f}^5
   \sin^2\phi_{20} (r-{r_h})^3 \log ^{\frac{9}{2}}({r_h}) \alpha _{\theta _1}^4 \alpha _{\theta _2}^3}\Biggr]\nonumber\\
g_{xy} & = & g_{xy}^{\rm MQGP}[1 + \frac{16384 \sqrt{2} \pi ^{11/2} \beta  \sqrt{{g_s}} (3 {g_s}-4) \kappa_b  \left(\frac{1}{N}\right)^{2/5} \sin^4\left(\frac{\psi_0}{2}\right)  \beta  N^{-\alpha
   }}{1162261467 ({g_s}-1)^2 {N_f}^6 \sin^2\phi_{20} \log (N) (r-{r_h}) \log ^{\frac{11}{2}}({r_h}) \alpha _{\theta _1} \alpha
   _{\theta _2}^2}]\nonumber\\
 g_{yz}  & = & g_{yz}^{\rm MQGP}\Biggl[1 + \frac{8192 \pi ^4 \beta  {g_s} \kappa_b ^2 \log N  M  \sin^2\left(\frac{\psi_0}{2}\right) {r_h}^3 \beta  N^{-2 \alpha -\frac{11}{10}} \log
   ^2({r_h})}{129140163 ({g_s}-1) {N_f}^5 \sin^2\phi_{20} (r-{r_h})^3 \alpha _{\theta _1}^3 \alpha _{\theta _2}^3}\Biggr]\nonumber\\
g_{xz} & = & g_{xz}^{\rm MQGP}\Biggl[1 + \Biggl(C_{xz}^{(1)} -\frac{1024\times16 \Sigma_2\sqrt{2} \pi ^{11/2}  \sqrt{{g_s}} (3 {g_s}-4) \kappa_b  \sin^4\left(\frac{\psi_0}{2}\right) \beta  N^{-\alpha -\frac{2}{5}}
  }{10460353203 ({g_s}-1)^2 {N_f}^6 \sin^2\phi_{20}
   (r-{r_h}) \log ^{\frac{13}{2}}({r_h}) \alpha _{\theta _1} \alpha _{\theta _2}^2}\Biggr)\beta\Biggr],\nonumber\\
& &
\end{eqnarray}
}
where $\Sigma_2$ is defined in (\ref{Sigma_2-def}) and $\alpha\in\mathbb{Z}^+$ appearing via (\ref{ansatz-b}). Analogous to working near the $\psi=2n\pi$- the explicit dependence on $\theta_{1,2}$ of the ${\cal M}$-theory metric components up to ${\cal O}(\beta)$ is effected by the replacemements:
$\alpha_{\theta_1}\rightarrow N^{\frac{1}{5}}\sin\theta_{1},\ \alpha_{\theta_2}\rightarrow N^{\frac{3}{10}}\sin\theta_{2}$.

\section{Physics Lessons Learnt}
Physics lessons learnt as a consequence of working out the ${\cal O}(R^4)
/{\cal  O}(l_p^6)$ corrections to the ${\cal M}$-theory dual of large-$N$ thermal QCD-like theories.

{\it IR-enhancement large-$N$/Planckian-suppresion competition}:  From (\ref{M-theory-metric-psi=2npi-patch_1}), one constructs Table 5.1.
\begin{table}[h]
\begin{center}
\begin{tabular}{|c|c|c|c|} \hline
S. No. & $g^{M}_{MN}$ & IR Enhancement Factor & $N$ Suppression \\
& &  $\frac{\left(\log {\cal R}_h\right)^m}{{\cal R}_h^n}, m,n\in\mathbb{Z}^+$ & Factor \\
& & in the ${\cal O}(R^4)$ Correction & in the ${\cal O}(R^4)$ Correction   \\ \hline
1 & $g^{M}_{\mathbb{R}^{1,3}}$ & $\log {\cal R}_h$ & ${N^{-\frac{9}{4}}}$ \\  \hline
2 & $g^{M}_{rr, \theta_1x}$ & 1 & ${N^{-\frac{8}{15}}}$ \\  \hline
3 & $g^{M}_{\theta_1z,\theta_2x}$ & ${{\cal R}_h^{-5}}$ & ${N^{-\frac{7}{6}}}$ \\  \hline
4 &  $g^{M}_{\theta_2y}$ & $\log {\cal R}_h$ & ${N^{-\frac{7}{5}}}$ \\ \hline
5 &  $g^{M}_{\theta_2z}$ & $\log {\cal R}_h$ & ${N^{-\frac{7}{6}}}$
\\ \hline
6 &  $g^{M}_{xy}$ & $\log {\cal R}_h$ & ${N^{-\frac{21}{20}}}$ \\  \hline
7 &  $g^{M}_{xz}$ &  $\left(\log {\cal R}_h\right)^3$ & ${N^{-\frac{5}{4}}}$ \\ \hline
8 &  $g^{M}_{yy}$ & $\log {\cal R}_h$ & ${N^{-\frac{7}{4}}}$ \\  \hline
9 &  $g^{M}_{yz}$ & $\frac{\log {\cal R}_h}{{\cal R}_h^7}$ & ${N^{-\frac{29}{12}}}$ \\ \hline
10 &  $g^{M}_{zz}$ & $\log {\cal R}_h$ & ${N^{-\frac{23}{20}}}$ \\ \hline
11  &  $g^{M}_{x^{10}x^{10}}$ & $\frac{\log {\cal R}_h^3}{{\cal R}_h^2}$ & ${N^{-\frac{5}{4}}}$ \\ \hline
\end{tabular}
\end{center}
\caption{IR Enhancement vs. large-$N$ Suppression in ${\cal O}(R^4)$-Corrections in the M-theory Metric in the $\psi=2n\pi, n=0,1,2$ Patches; ${\cal R}_h \equiv \frac{r_h}{{\cal R}_{D5/\overline{D5}}}\ll1$, ${\cal R}_{D5/\overline{D5}}$ being the $D5-\overline{D5}$ separation}
\end{table}
The main Physics-related take-away is the following. One notes that in the IR: $r = \chi r_h, \chi\equiv {\cal O}(1)$, and up to ${\cal O}(\beta)$:
\begin{equation}
\label{IR-beta-N-suppressed-logrh-rh-neg-exp-enhanced}
f_{MN} \sim \beta\frac{\left(\log {\cal R}_h\right)^{m}}{{\cal R}_h^n N^{\beta_N}},\ m\in\left\{0,1,3\right\},\ n\in\left\{0,2,5,7\right\},\
\beta_N>0.
\end{equation}
Now, $|{\cal R}_h|\ll1$. As estimated in \cite{Bulk-Viscosity}, $|\log {\cal R}_h|\sim N^{\frac{1}{3}}$, implying there is a competition between Planckian and large-$N$ suppression and infra-red enhancement arising from $m,n\neq0$ in (\ref{IR-beta-N-suppressed-logrh-rh-neg-exp-enhanced}). One could choose a heirarchy: $\beta\sim e^{-\gamma_\beta N^{\gamma_N}}, \gamma_\beta,\gamma_N>0: \gamma_\beta N^{\gamma_N}>7N^{\frac{1}{3}} + \left(\frac{m}{3} - \beta_N\right)\log N$ (ensuring that the IR-enhancement does not overpower Planckian suppression - we took the ${\cal O}(\beta)$ correction to $g^{M}_{yz}$, which had the largest IR enhancement, to set a lower bound on $\gamma_{\beta,N}$/Planckian suppression). If $\gamma_\beta N^{\gamma_N}\sim7N^{\frac{1}{3}}$, then one will be required to go to a higher order in $\beta$. This hence answers the question, when one can truncate at ${\cal O}(\beta)$.

\section{Summary }

In this work we obtained the M theory metric at  intermediate 't Hooft coupling in the infrared.

The following is  a summary of the important results obtained in this work.

\begin{enumerate}
\item
We work out the ${\cal O}(l_p^6)$ corrections to the ${\cal M}$-Theory metric worked out in \cite{MQGP,NPB} arising from the ${\cal O}(R^4)$ terms in $D=11$ supergravity. We realize that in the MQGP limit of \cite{MQGP}, the contribution from the $J_0$ (and its variation) dominate over the contribution from $E_8$ and its variation as a consequence of which $E_8$ has been disregarded. The computations have been partitioned into two portions - one near the $\psi=2n\pi, n=0, 1, 2$ patches and the other away from the same (wherein there is no decoupling of the radial direction, the six angles and the ${\cal M}$-Theory circle).

\item
We also note that there is a close connection between finite ('t Hooft) coupling effects in the IR and non-conformality (which being effected via the effective number of the fractional $D3$-branes, vanishes in the UV) as almost all corrections to the ${\cal M}$-Theory metric components of \cite{MQGP,NPB} in the IR arising from the aforementioned ${\cal O}(R^4)$ terms in $D=11$ supergravity action, vanish when the number of fractional $D3$-branes is set to zero.

\item
The importance of the higher derivative corrections arises from the competition between the non-conformal Infra-Red enhancement $\frac{\left(\log {\cal R}_h\right)^m}{{\cal R}_h^n}, {\cal R}_h\equiv\frac{r_h}{{\cal R}_{D5/\overline{D5}}}, m=0,1,2,3, n=0,$ and the Planckian and large-$N$ suppression $\frac{l_p^6}{N^{\beta_N}}, \beta_N>0$ in the ${\cal O}(l_p^6)$ corrections to the ${\cal M}$-theory dual \cite{MQGP,NPB} of thermal QCD-like theories. As $|\log {\cal R}_h|\sim N^{\frac{1}{3}}$ \cite{Bulk-Viscosity}, for appropriate values of $N$, it may turn out that this correction may become of ${\cal O}(1)$, and thereby very significant. This would also then imply that one will need to consider higher order corrections beyond ${\cal O}(l_p^6)$.
\end{enumerate}

%
\chapter{(Phenomenology/Lattice-Compatible) $SU(3)$ M$\chi$PT HD up to ${\cal O}(\lowercase{p}^4)$ and the  ${\cal O}\left(R^4\right)$-Large-$N$ Connection }
\graphicspath{{Chapter6/}{Chapter6/}}

\section{Introduction}

Chiral Perturbation Theory ($\chi$PT) is an effective field theory of Quantum Chromodynamics (QCD) which describes the low energy regime (IR) of Quantum Chromodynamics (QCD). In $\chi$PT, the degrees of freedom are hadrons (which include mesons and baryons and hyperons but we will study only mesonic $\chi$PT). $\chi$PT Lagrangian consists of various terms which are invariant under chiral symmetry $SU(N_f)_L\times SU(N_f)_R$, charge conjugation and parity symmetry of QCD. Chiral symmetry $SU(N_f)_L\times SU(N_f)_R$ is spontaneously broken to $SU(N_f)_V$ yielding $(N_f^2 -1)$ (pseudo-)Goldstone bosons where $N_f$ is the number of flavors. As an effective field theory, it is renormalizable  order-by-order in momentum.

The $\rho$ vector meson can also be incorporated by augmenting the chiral symmetry with the inclusion of a HLS (Hidden Local Symmetry) whose gauge group is $G_{global}\times H_{local}$, where $G_{global} = SU(N_f)_L\times SU(N_f)_R$ and $H = SU(N_f)_V$ is the HLS \cite{HARADA}. The $\rho$ meson and its flavor partners  are identified with gauge boson of HLS. Chiral symmetry breaking scale is $\Lambda_\chi \sim 4\pi F_\pi \sim 1.1 GeV$ \cite{MG} in the chiral limit\cite{GL}. Since $\Lambda_\chi$  is much larger than the $\rho$ mass scale and $M_\rho < \Lambda_\chi$ , therefore we can expand the generating functional of QCD in terms of $p/\Lambda_\chi$ or $m/\Lambda_\chi$. At scale  $\mu = \Lambda_\chi$, perturbative expansion in $\mu/\Lambda_\chi$ breaks down. One can construct the most general form of the Lagrangian order by order in the derivative expansion consistently with the chiral symmetry. The ``universal" leading order Lagrangian is constructed from the terms of $O(p^2)$ .
J. Gasser and H. Leutwyler worked out the $SU(2)$\cite{GLF}$/SU(3)$ \cite{GL}  chiral perturbation theory lagrangian up to $O(p^4)$ and the renormalization of the coupling constants at scale $\mu = m_\eta$ \cite{GL}. At NLO, i.e. $O(p^4)$, there are 12 coupling constants "$(L_{i = 1,2,..10}, H_1, H_2)$". Some of the low energy constants $L_4, L_5, L_6, L_8$ have been calculated from Lattice simulation from the MILC collaboration \cite{MILC}. In \cite{Pich}, the ten low energy constants $L_{i = 1,2,..10}$ were evaluated at scale $\mu = M_\rho$; \cite{EJP} also contains the low energy constants at $\mu = M_\rho$.  Updated values are given in \cite{Ecker-2015}.

As regards to a top-down approach, Kruczenski et al \cite{KMMW} considered an interesting probe $D6$-branes in a $D4$-brane type IIA supergravity background, which they used to explore various aspects of low energy phenomena in QCD. An important ingredient which was still missing from their model, however, were the massless pions as Nambu-Goldstone bosons associated with the spontaneous breaking of the $U(N_f)_L \times U(N_f)_R$ chiral symmetry in QCD. Following \cite{KMMW}, Sakai-Sugimoto (SS) in \cite{Sakai-Sugimoto-1,SS2} considered a nice modification by looking at a $D4/D8/\overline{D8}$ system in type IIA supergravity background with  anti-periodic boundary condition for fermions along a circle to break supersymmetry. This model exhibited chiral symmetry breaking  with $D8-\overline{D8}$ pairs merging into $D8$-branes. This model also yields massless pions which are identified with Nambu-Goldstone bosons associated with chiral symmetry breaking, and the lightest vector meson($\rho$ meson). The Sakai Sugimoto model is closely related to the HLS formalism which produces Kawarabayashi-Suzuki-Riazuddin-Fayyazuddin-type relation among the couplings. Chern-Simons term on the probe brane leads to the Wess-Zumino-Witten term. In \cite{SS2}, Sakai Sugimoto   obtained few low energy constants ($L_1, L_2, L_3$) of $SU(3)$ chiral perturbation theory at ${\cal O}(p^4)$ which were close to the values given in \cite{Pich} for suitable choice of a parameter $\kappa$. Although Sakai Sugimoto model reproduces various physical quantities of low energy QCD but this model deviates from realistic QCD above the energy scale  of the vector mesons because they obtain a four-dimensional theory by compactifying $D4$-branes on a circle of radius $\tilde{M}_{KK}^{-1}$ with an infinite tower of Kaluza-Klein modes of mass scale $\tilde{M}_{KK}$ arises. These Kaluza-Klein modes do not appear in realistic QCD. Further, the SS model caters  to the IR and is not UV complete. This was taken care of by the (only) UV-complete $D3, D5/\overline{D5}, D7/\overline{D7}$ holographic dual of large-$N$ thermal QCD of the McGill group \cite{metrics}, its type IIA SYZ mirror and the  ${\cal M}$-theory  uplift of the same (in particular in the `MQGP limit' ) as constructed in \cite{MQGP} (with one of the co-authors [AM]).

In\cite{HARADA}, the  authors considered the Sakai Sugimoto model \cite{Sakai-Sugimoto-1} as holographic QCD model and  proposed a method to integrate out infinite number of higher KK modes appearing in the expansion of five dimensional gauge field which consists of infinite number of vector and axial vector fields including pion as Nambu Goldstone boson arising due to spontaneous chiral symmetry breaking. First, they truncated the spectrum at certain level so that number of fields ``integrated in" becomes finite in the theory and then integrated out all the KK modes except pion and lowest lying vector mode($\rho$ meson and the flavor partners). Using this method they obtained the effective lagrangian up to $O(p^4)$. In this appendix, we summarize the HLS formalism and the arguments of how to obtain the $SU(3)$ $\chi$PT Lagrangian of \cite{GL} up to (NLO in momentum below the chiral symmetry breaking scale) ${\cal O}(p^4)$ by integrating out the vector mesons from the HLS Lagrangian (both as discussed in detail in  \cite{HLS-Physics-Reports}). The authors in \cite{HMM} derived relations between the $SU(3)$ low energy coupling constants of \cite{GL} and $O(p^4)$ couplings in \cite{HARADA}.

The main takeaway from this work is that QCD imposes a relationship between the higher derivative corrections and large-$N$ suppression when comparing our ${\cal M}$-theory/type IIA holographic computational results for the low energy coupling constants of $\chi$PT Lagrangian up to ${\cal O}(p^4)$ and corresponding experimental values of these coupling constants.

\section{Obtaining ${\cal L}_{\chi PT}^{(4)}[\pi,\rho]$ from DBI action on Flavor $D6$ Branes}

In this section, similar to the discussion in \cite{HARADA} using the ``HLS formalism",  starting from the DBI action on flavor $D6$-branes (obtained from the SYZ type IIA mirror of the type IIB holographic dual of large-$N$ thermal QCD as constructed in \cite{metrics}) we reviewed obtaining the $\chi$PT Lagrangian for $\pi, \rho$ mesons and their flavor partners, up to ${\cal O}(p^4)$ wherein the coupling constants were obtained as appropriate radial integrals.

The type IIB dual corresponding to high temperatures, i.e., $T>T_c$, involves a black hole with the metric given by \cite{metrics}:
\begin{eqnarray}
\label{BH-T>Tc}
& & \hskip -0.5in ds^2 = \frac{1}{\sqrt{h(r,\theta_{1,2})}}\left(-g(r)dt^2 + \left(dx^1\right)^2 +  \left(dx^2\right)^2 + \left(dx^3\right)^2 \right)
+ \sqrt{h(r,\theta_{1,2})}\left(\frac{dr^2}{g(r)} + r^2 ds^2(\theta_{1,2},\phi_{1,2},\psi)\right),\nonumber\\
& &
\end{eqnarray}
where $g(r) = 1 - \frac{r_h^4}{r^4}$, and for low temperatures, i.e., $T<T_c$, is given by the thermal gravitational dual:
\begin{eqnarray}
\label{Thermal-T<Tc}
& & \hskip -0.5in ds^2 = \frac{1}{\sqrt{h(r,\theta_{1,2})}}\left(-dt^2 + \left(dx^1\right)^2 +  \left(dx^2\right)^2 + \tilde{g}(r)\left(dx^3\right)^2 \right)
+ \sqrt{h(r,\theta_{1,2})}\left(\frac{dr^2}{\tilde{g}(r)} + r^2 ds^2(\theta_{1,2},\phi_{1,2},\psi)\right)\nonumber\\
& &
\end{eqnarray}
where $\tilde{g}(r) = 1 - \frac{r_0^4}{r^4}$. One notes that $t\rightarrow x^3,\ x^3\rightarrow t$ in (\ref{BH-T>Tc}) following by a Double Wick rotation in the new $x^3,\ t$ coordinates obtains (\ref{Thermal-T<Tc}); $h(r,\theta_{1,2})$ is the ten-dimensional warp factor \cite{metrics, MQGP}. This amounts to:  $-g_{tt}^{\rm BH}(r_h\rightarrow r_0) = g_{x^3x^3}\ ^{\rm Thermal}(r_0),$ $ g_{x^3x^3}^{\rm BH}(r_h\rightarrow r_0) = -g_{tt}\ ^{\rm Themal}(r_0)$ in the results of
\cite{VA-Glueball-decay, OR4-Yadav+Misra} (See \cite{KMMW} in the context of Euclidean/black $D4$-branes in type IIA).

In (\ref{Thermal-T<Tc}), we assumed the spatial part of the solitonic $D3$ brane world volume to be given by $\mathbb{R}^2(x^{1,2})\times S^1(x^3)$ where the period of $S^1(x^3)$ is given by a very large: $\frac{2\pi}{M_{\rm KK}}$, where the very small $M_{\rm KK} = \frac{2r_0}{ L^2}\left[1 + {\cal O}\left(\frac{g_sM^2}{N}\right)\right]$, $r_0$ being the very small IR cut-off in the thermal background (See also \cite{Armoni et al-2020}) and $L = \left( 4\pi g_s N\right)^{\frac{1}{4}}$. So, $\lim_{M_{\rm KK}\rightarrow0}\mathbb{R}^2(x^{1,2})\times S^1(x^3) = \mathbb{R}^3(x^{1,2,3})$, thereby recovering 4D Physics.

As explained in \cite{Knauf-thesis}, the $T^3$-valued $(x, y, z)$ (used for effecting SYZ mirror via a triple T-dual in \cite{MQGP, NPB}) are defined via:
\begin{eqnarray}
\label{xyz-definitions}
& & \phi_1 = \phi_{10} + \frac{x}{\sqrt{h_2}\left[h(r_0,\theta_{10,20})\right]^{\frac{1}{4}} \sin\theta_{10}\ r_0},\nonumber\\
& & \phi_2 = \phi_{20} + \frac{y}{ \sqrt{h_4}
\left[h( r_0,\theta_{10,20})\right]^{\frac{1}{4}}\sin\theta_{20}\ r_0}\nonumber\\
& & \psi = \psi_0 + \frac{z}{\sqrt{h_1} \left[h( r_0,\theta_{10,20})
\right]^{\frac{1}{4}}\ r_0},
\end{eqnarray}
$h_{1,2,4}$ defined in \cite{metrics}, and one works up to linear order in $(x, y, z)$. Up to linear order in $r$, i.e., in the IR, it can be shown \cite{theta0-theta} that $\theta_{10,20}$ can be promoted to global coordinates $\theta_{1,2}$ in all the results in this work.
 The meson sector in the type IIA dual background of top-down holographic type IIB setup is given by the flavor $D6$-branes action.
 For
\begin{equation}
\label{theta12-deloc}
\theta_{1}=\frac{\alpha_{\theta_{1}}}{N^{\frac{1}{5}}}, \ \theta_{2}=\frac{\alpha_{\theta_{2}}}{N^{\frac{3}{10}}},
\end{equation}
i.e., restricting to the Ouyang embedding for a vanishingly small $|\mu_{\rm Ouyang}|$,
 one will assume that the embedding of the $D6$-brane will be given by $\iota :\Sigma ^{1,6}\Bigg( R^{1,3},r,\theta_{2}\sim\frac{\alpha_{\theta_{2}}}{N^{\frac{3}{10}}},y\Bigg)\hookrightarrow M^{1,9}$, effected by: $z=z(r)$.  As obtained in \cite{Yadav+Misra+Sil-Mesons} one sees that $z$=constant is a solution and by choosing $z=\pm {\cal C}\frac{\pi}{2}$, one can choose the $D6/\overline{D6}$-branes to be at ``antipodal" points along the z coordinate. As in \cite{Yadav+Misra+Sil-Mesons}, we worked with redefined $(r,z)$ in terms of new variables $(Y,Z)$:
\[r=r_{0}e^{\sqrt{Y^{2}+Z^{2}}}\]
\[z={\cal C}\arctan\frac{Z}{Y}.\]
 Vector mesons were obtained by considering gauge fluctuations of a background gauge field along the world volume of the embedded flavor $D6$-branes (with world volume ${\Sigma}_7(x^{0,1,2,3},Z,\theta_2,\tilde{y}) = {\Sigma}_2(\theta_2,\tilde{y})\times{\Sigma}_5(x^{0,1,2,3},Z)$). Turning on a gauge field fluctuation $\tilde{F}$ about a small background gauge field $F_0$ and the backround $i^*(g+B) [i:\Sigma_7\hookrightarrow M_{10}$, $M_{10}$ being the ten-dimensional ambient space-time]. Following {\bf Chapter 3} this implies that the DBI action for $D6-$ brane is:

{\footnotesize
\begin{equation}
\label{DBI action}
{\rm S}^{IIA}_{D6}=\frac{T_{D_6}(2\pi\alpha^\prime)^{2}}{4} \left(\frac{\pi L^2}{r_0}\right){\rm Str}\int \prod_{i=0}^3dx^i dZd\theta_{2}dy \delta\Bigg(\theta_{2}-\frac{\alpha_{\theta_{2}}}{N^{\frac{3}{10}}}\Bigg) e^{-\Phi} \sqrt{-{\rm det}_{\theta_{2}y}(\iota^*(g+B))}\sqrt{{\rm det}_{{\mathbb  R}^{1,3},Z}(\iota^*g)}g^{\tilde{\mu}\tilde{\nu}}F_{\tilde{\nu}\tilde{\rho}}g^{\tilde{\rho}\tilde{\sigma}}F_{\tilde{\sigma}\tilde{\mu}},
\end{equation}
}
where $\tilde{\mu}=i(=0,1,2,3),Z$

Making a Klauza-Klein ansatz $A_\mu(x^\nu,Z) = \sum_{n=1}^\infty \rho^{(n)}_\mu(x^\nu)\psi_n(Z), A_Z(x^\nu,Z) = \sum_{n=0}^\infty\pi^{(n)}(x^\nu)\phi_n(Z) $ and using appropriate normalization conditions for the meson profile functions appearing in the action leads to(as done in chapter 2):
\begin{eqnarray}
& & -\int d^3x\ \ \left[\frac{1}{2}\partial_{\mu}\pi^{(0)}\partial^{\mu}\pi^{(0)} + \sum_{n\ge 1}\left(\frac{1}{4}\tilde{F}^{(n)}_{\mu\nu}\tilde{F}^{(n)\mu\nu}+\frac{m_{n}^2}{2}\rho^{(n)}_\mu \rho^{(n)\mu }\right)\right].
\end{eqnarray}

Working in the $A_Z(x^\mu,Z)=0$-gauge, integrating out all higher order vector and axial vector meson fields except keeping only the lowest vector meson field \cite{HARADA} \\ $V_\mu^{(1)}(x^\mu) = g \rho_\mu(x^\mu) = \left(\begin{array}{ccc}
\frac{1}{\sqrt{2}}\left(\rho_\mu^0 + \omega_\mu\right) & \rho_\mu^+ & K_\mu^{*+} \\
\rho_\mu^- & -\frac{1}{\sqrt{2}}\left(\rho_\mu^0 - \omega_\mu\right) & K_\mu^{*0} \\
K_\mu^{*-} & {\bar K}_\mu^{*0}  & \phi_\mu
\end{array} \right)$ and lightest pseudo-scalar meson field i.e. $\pi = \frac{1}{\sqrt{2}}\left(\begin{array}{ccc}
\frac{1}{\sqrt{2}}\pi^0 + \frac{1}{\sqrt{6}}\eta_8 + \frac{1}{\sqrt{3}}\eta_0 & \pi^+ & K^+ \\
\pi^- & -\frac{1}{\sqrt{2}}\pi^0 + \frac{1}{\sqrt{6}}\eta_8 + \frac{1}{\sqrt{3}}\eta_0 & K^0 \\
K^- & {\bar K}^0 & -\frac{2}{\sqrt{6}}\eta_8 + \frac{1}{\sqrt{3}}\eta_0
\end{array}\right)$ meson, the gauge field $A_\mu(x^\nu,Z)$ up to ${\cal O}(\pi)$ is given by:
\begin{eqnarray}
\label{Amu-exp}
& & A_\mu(x^\nu,Z) = \frac{\partial_\mu\pi}{F_\pi}\psi_0(Z) - V_\mu(x^\nu)\psi_1(Z),
\end{eqnarray}
where $\psi_0(z) = \int^Z_0 dZ^\prime \phi_0(Z^\prime), V_\mu^{(1)}(x^\nu) = \rho^{(1)}_\mu - \frac{1}{{\cal M}_{(1)}}\partial_\mu\pi^{(1)}$.

To introduce external vector ${\cal V}_\mu$ and axial vector fields ${\cal A}_\mu$, one could use the Hidden Local Symmetry (HLS) formalism of \cite{HARADA} and references therein, wherein  $\frac{1}{F_\pi} \partial_\mu \pi\rightarrow\hat{\alpha}_{\mu \perp}= \frac{1}{F_\pi} \partial_\mu \pi + {\cal A}_\mu - \frac{i}{F_\pi}[{\cal V}_\mu,\pi] + \cdots$ (refer to (\ref{al perp exp})), and one also works with $\hat{\alpha}_{\mu ||} \equiv -V_\mu + {\cal V}_\mu - \frac{i}{2F_\pi^2}[\partial_\mu\pi,\pi] + \cdots$ (refer to (\ref{al para exp})).
 To obtain the low energy effective theory of QCD, we truncated the KK spectrum at certain level because mode expansion of the gauge field contains infinite number of vector meson fields $V_\mu^{(n)}(x^\mu)$ and axial vector meson fields $A_\mu^{(n)}(x^\mu)$ \cite{HARADA},
\begin{equation}
A_\mu(x^\mu,z) = \hat{\alpha}_{\mu \perp}(x^\mu) \psi_0(z)
  + (\hat{\alpha}_{\mu ||}(x^\mu) + V_\mu^{(1)}(x^\mu)  )
  + \hat{\alpha}_{\mu ||}(x^\mu)  \psi_1(z),
  \end{equation}
implying therefore
\begin{eqnarray}
\label{F mu nu}
& &   F_{\mu\nu} = -V_{\mu\nu} \psi_1 +v_{\mu\nu}(1+\psi_1) + a_{\mu\nu} \psi_0 - i[\hat{\alpha}_{\mu ||},\hat{\alpha}_{\nu ||}]\psi_1(1+\psi_1)  + i[\hat{\alpha}_{\mu \perp},\hat{\alpha}_{\nu \perp}](1+\psi_1-\psi_0^{2}) \nonumber\\
  & &
  -i([\hat{\alpha}_{\mu \perp},\hat{\alpha}_{\nu ||}]+[\hat{\alpha}_{\mu ||},\hat{\alpha}_{\nu \perp}])\psi_1 \psi_0.
 \end{eqnarray}
From appendix {\bf E.3} (based on \cite{HLS-Physics-Reports}), as regards a chiral power counting, one notes that $M_\rho\equiv{\cal O}(p)$ implying $\hat{\alpha}_{\nu ||}\equiv \frac{{\cal O}(p^3)}{M_\rho^2}\equiv{\cal O}(p),
\hat{\alpha}_{\nu\perp}\equiv {\cal O}(p)$. Further, $V_{\mu\nu}, a_{\mu\nu}$ and $v_{\mu\nu}$ are of ${\cal O}(p^2)$. Hence, using (\ref{F mu nu}), $\left(F_{\mu\nu}F^{\mu\nu}\right)^m$ is of ${\cal O}(p^{4m}), m\in\mathbb{Z}^+$. Therefore, one considers the kinetic term ($m=1$) at ${\cal O}(p^4)$, which yields the  expression for $ F_{\mu\nu}F^{\mu\nu}$.
Defining  parity as $x^i\rightarrow-x^i$, $i$ indexing the conformally Minkowskian spatial directions and $Z\rightarrow-Z$, given that $A_\mu(x, Z)$ will be odd, $\alpha_\perp$ will be even, $\alpha_{||}$ will be odd and $V_\mu$ will be odd implies $\psi_0(Z)$  will be odd and $\psi_1(Z)$ will be even. As coupling constants are assumed to be scalars and they are given by $Z$-integrals, the $Z$-dependent terms in the action must be separately of even-$Z$ parity. As  $\psi_0$ has odd $Z$-parity and $\psi_1$ has even $Z$-parity, therefore at ${\cal O}(p^4)$, terms with $(3\hat{\alpha} _{||}s\ ,\ 1\hat{\alpha}_\perp\ {\rm or}\ 3\hat{\alpha}_\perp s\ ,\ 1\hat{\alpha} _{||} )$, were dropped as they involve coefficients of the type $\psi_0^{2m+1}\psi_1^{2n}(Z)$ for appropriate postive integral values of $n, m$. Similarly, at ${\cal O}(p^2)$, $tr(\hat{\alpha}_{\mu \perp}\hat{\alpha}_{||}^{\mu})$  accompanied by $\dot{\psi_0}\dot{\psi_1}(\ ^. \equiv \frac{d}{dZ})$ of odd-$Z$ parity, was dropped.
At ${\cal O}(p^4)$, one hence obtains \cite{HARADA}:
\begin{eqnarray}
\label{Lagrangian-Op4}
 & & {\mathcal{L}}_{(4)}
\ni
y_1 \,
{\rm tr}[{\hat{\alpha}}_{\mu\perp}{\hat{\alpha}}^{\mu}_{\perp}
{\hat{\alpha}}_{\nu\perp}{\hat{\alpha}}^{\nu}_{\perp}]
+
y_2 \,
{\rm tr}[{\hat{\alpha}}_{\mu\perp}{\hat{\alpha}}_{\nu\perp}
{\hat{\alpha}}^{\mu}_{\perp}{\hat{\alpha}}^{\nu}_{\perp}]
+y_3 \,
{\rm tr}[{\hat{\alpha}}_{\mu||}{\hat{\alpha}}^{\mu}_{||}
{\hat{\alpha}}_{\nu||}{\hat{\alpha}}^{\nu}_{||}]
+y_4 \,
{\rm tr}[{\hat{\alpha}}_{\mu||}{\hat{\alpha}}_{\nu||}
{\hat{\alpha}}^{\mu}_{||}{\hat{\alpha}}^{\nu}_{||}] \nonumber\\
& &
+y_5 \,
{\rm tr}[{\hat{\alpha}}_{\mu\perp}{\hat{\alpha}}^{\mu}_{\perp}
{\hat{\alpha}}_{\nu||}{\hat{\alpha}}^{\nu}_{||}]
+y_6 \,
{\rm tr}[{\hat{\alpha}}_{\mu\perp}{\hat{\alpha}}_{\nu\perp}
{\hat{\alpha}}^{\mu}_{||}{\hat{\alpha}}^{\nu}_{||}]
+y_7 \,
{\rm tr}[{\hat{\alpha}}_{\mu\perp}{\hat{\alpha}}_{\nu\perp}
{\hat{\alpha}}^{\nu}_{||}{\hat{\alpha}}^{\mu}_{||}] \nonumber\\
& &
+y_8 \,
\left\{ {\rm tr}[{\hat{\alpha}}_{\mu\perp}{\hat{\alpha}}^{\nu}_{||}
{\hat{\alpha}}_{\nu\perp}{\hat{\alpha}}^{\mu}_{||}]
+
{\rm tr}[{\hat{\alpha}}_{\mu\perp}{\hat{\alpha}}^{\mu}_{||}
{\hat{\alpha}}_{\nu\perp}{\hat{\alpha}}^{\nu}_{||}\right\}
+y_9 \,
{\rm tr}[{\hat{\alpha}}_{\mu\perp}{\hat{\alpha}}_{\nu ||}
{\hat{\alpha}}^{\mu}_{\perp}{\hat{\alpha}}^{\nu}_{||}]\nonumber\\
& &
+z_1 \,
{\rm tr}[v_{\mu\nu}v^{\mu\nu}]
+z_2 \,
{\rm tr}[a_{\mu\nu}a^{\mu\nu}]
+z_3 \,
{\rm tr}[v_{\mu\nu}V^{\mu\nu}]
+iz_4 \,
{\rm tr}[V_{\mu\nu}{\hat{\alpha}}^{\mu}_{\perp}
{\hat{\alpha}}^{\nu}_{\perp}]\nonumber\\
& &
+iz_5 \,
{\rm tr}[V_{\mu\nu}{\hat{\alpha}}^{\mu}_{||}
{\hat{\alpha}}^{\nu}_{||}]
+iz_6 \,
{\rm tr}[v_{\mu\nu}
{\hat{\alpha}}^{\mu}_{\perp}{\hat{\alpha}}^{\nu}_{\perp}]
+iz_7 \,
{\rm tr}[v_{\mu\nu}
{\hat{\alpha}}^{\mu}_{||}{\hat{\alpha}}^{\nu}_{||}]
-iz_8 \,
{\rm tr}\left[a_{\mu\nu}
\left({\hat{\alpha}}^{\mu}_{\perp}{\hat{\alpha}}^{\nu}_{||}
+{\hat{\alpha}}^{\mu}_{||}{\hat{\alpha}}^{\nu}_{\perp}
\right)\right]
\end{eqnarray}
where:
\begin{equation}
 v_{\mu\nu}
= \frac{1}{2} \left(
\xi_R {\cal R}_{\mu\nu} \xi^\dag_R + \xi_L {\cal L}_{\mu\nu} \xi_L^\dag \right)  \hspace{0.5cm}
{\rm and} \hspace{0.5cm}
a_{\mu\nu}
=  \frac{1}{2} \left(
\xi_R {\cal R}_{\mu\nu} \xi^\dag_R - \xi_L {\cal L}_{\mu\nu} \xi_L^\dag
\right),
\end{equation}
${\cal L}_{\mu\nu} = \partial_{[\mu}{\cal L}_{\nu]} - i[{\cal L}_\mu,{\cal L}_\nu]$ and ${\cal R}_{\mu\nu} = \partial_{[\mu}{\cal R}_{\nu]} - i[{\cal R}_\mu,{\cal R}_\nu]$ and ${\cal L}_\mu = {\cal V}_\mu - {\cal A}_\mu$ where ${\cal R}_\mu =  {\cal V}_\mu + {\cal A}_\mu $, and $\xi_L^\dagger(x^\mu) = \xi_R(x^\mu) = e^{\frac{i \pi(x^\mu)}{F_\pi}}$; also, $V_{\mu\nu} = \partial_{[\mu}V_{\nu]}-i[V_\mu,V_\nu]$.

The various  couplings, using $F_{\mu\nu} F^{\mu\nu}$, were given by the following expressions \cite{HARADA}:
 \begin{eqnarray}
\label{y_i+z_i}
& &  F_\pi^{2} = -\frac{{\cal V}_{\Sigma_2}}{4}<<\dot\psi_0^{2}>>
\nonumber\\
& &
 aF_\pi^{2} = -\frac{{\cal V}_{\Sigma_2}}{4}<<\dot\psi_1^{2}>>
 \nonumber\\
& &
 \frac{1}{g^2} = \frac{{\cal V}_{\Sigma_2}}{2}<\psi_1^{2}>
 \nonumber\\
& &
  y_1 = -y_2 = -\frac{{\cal V}_{\Sigma_2}}{2}<(1+\psi_1-\psi_0)^2>
  \nonumber\\
& &
  y_3 = -y_4 = -\frac{{\cal V}_{\Sigma_2}}{2}<\psi_1^{2}(1+\psi_1)^2>
  \nonumber\\
& &
  y_5 = -{\cal V}_{\Sigma_2}<\psi_0^{2}\psi_1^{2}>
  \nonumber\\
& &
  y_6 = -y_7=-{\cal V}_{\Sigma_2}<\psi_1(1+\psi_1)(1+\psi_1-\psi_0^{2})>
 \nonumber\\
& &
  y_8 = -y_9= -{\cal V}_{\Sigma_2}<\psi_0^{2}\psi_1^{2}>
  \nonumber\\
& &
  z_1 = -\frac{{\cal V}_{\Sigma_2}}{4}<(1+\psi_1)^2>
  \nonumber\\
& &
  z_2 = -\frac{{\cal V}_{\Sigma_2}}{4}<(\psi_0)^2>
  \nonumber\\
& &
  z_3 = \frac{{\cal V}_{\Sigma_2}}{2}<\psi_1(1+\psi_1)>
 \nonumber\\
& &
  z_4 = {\cal V}_{\Sigma_2}<\psi_1(1+\psi_1-\psi_0^{2})>
  \nonumber\\
& &
  z_5 = -{\cal V}_{\Sigma_2}<\psi_1^{2}(1+\psi_1)>
  \nonumber\\
& &
  z_6 = -{\cal V}_{\Sigma_2}<(1+\psi_1)(1+\psi_1-\psi_0^{2})>
  \nonumber\\
& &
  z_7 = {\cal V}_{\Sigma_2}<\psi_1(1+\psi_1)^{2}>
  \nonumber\\
& &
  z_8 = {\cal V}_{\Sigma_2}<\psi_0^{2}\psi_1>
  \end{eqnarray}
  where:
  \begin{equation}
\label{<<A>>}
  << A >> =  \int_{0}^{+\infty}{\mathcal{V}_1(z)} A dZ
\end{equation}
and
\begin{equation}
\label{<A>}
 < A > =  \int_{0}^{+\infty}{\mathcal{V}_2(z)} A dZ.
\end{equation}
\section{The Coupling Constants in ${\cal L}_{\chi PT}^{[4]}(\pi,\rho)$ from $S_{\rm DBI}^{D6}$ Incorporating ${\cal O}(R^4)$-Corrections from M Theory}
This section has the core of the main results. Inclusive of the ${\cal O}(R^4)$ corrections to the ${\cal M}$-theory uplift of large-$N$ thermal QCD as worked out in \cite{OR4-Yadav+Misra}, we showed how to obtain lattice-compatible values of the coupling constants up to ${\cal O}(p^4)$ of the $\chi$PT Lagrangian of \cite{GL} in the chiral limit. The ${\cal O}(R^4)$-corrections - indicated by a $\tilde{}$ (e.g., the  ${\cal M}$-theory  metric: $\tilde{g}_{MN}^{\cal M}=g^{\rm MQGP}_{MN}\left(1+f_{MN}\right)$ \cite{OR4-Yadav+Misra}) in (\ref{V_12}) - to the $D6$-brane DBI action is described in Appendix E.4.

The ${\cal O}(\beta)$-corrected ${\cal M}$-theory metric of \cite{MQGP} in the MQGP limit adapted to the thermal background (\ref{Thermal-T<Tc}) near the $\psi=2n\pi, n=0, 1, 2$-branches up to ${\cal O}((r-r_0)^2)$ [and up to ${\cal O}((r-r_0)^3)$ for some of the off-diagonal components along the delocalized $T^3(x,y,z)$] - the components which do not receive an ${\cal O}(\beta)$ corrections, are not listed in (\ref{ M-theory-metric-psi=2npi-patch}) - is given below \cite{OR4-Yadav+Misra}:
{\scriptsize
\begin{eqnarray}
 \label{ M-theory-metric-psi=2npi-patch}
 \hskip -0.5in \tilde{g}^{\cal M}_{x^3x^3} & = & g^{\rm MQGP}_{x^3x^3}\Biggl[1 + \frac{1}{4}  \frac{4 b^8 \left(9 b^2+1\right)^3 \left(4374 b^6+1035 b^4+9 b^2-4\right) \beta  M \left(\frac{1}{N}\right)^{9/4} \Sigma_1
   \left(6 a^2+  {r_0}^2\right) \log (  {r_0})}{27 \pi  \left(18 b^4-3 b^2-1\right)^5  \log N ^2   {N_f}   {r_0}^2
   \alpha _{\theta _2}^3 \left(9 a^2+  {r_0}^2\right)} (r-  {r_0})^2\Biggr]
\nonumber\\
\tilde{g}^{\cal M}_{x^{0,1,2}x^{0,1,2}} & = &  g^{\rm MQGP}_{x^{0,1,2}x^{0,1,2}}
\Biggl[1 - \frac{1}{4} \frac{4 b^8 \left(9 b^2+1\right)^4 \left(39 b^2-4\right) M \left(\frac{1}{N}\right)^{9/4} \beta  \left(6 a^2+{r_0}^2\right) \log
   ({r_0})\Sigma_1}{9 \pi  \left(3 b^2-1\right)^5 \left(6 b^2+1\right)^4 \log N ^2 {N_f} {r_0}^2 \left(9 a^2+{r_0}^2\right) \alpha
   _{\theta _2}^3} (r - {r_0})^2\Biggr]\nonumber\\
\tilde{g}^{\cal M}_{rr} & = & g^{\rm MQGP}_{rr}\Biggl[1 + \Biggl(- \frac{2 \left(9 b^2+1\right)^4 b^{10} M   \left(6 a^2+{r_0}^2\right) \left((r-{r_0})^2+{r_0}^2\right)\Sigma_1}{3 \pi
   \left(-18 b^4+3 b^2+1\right)^4 \log N  N^{8/15} {N_f} \left(-27 a^4+6 a^2 {r_0}^2+{r_0}^4\right) \alpha _{\theta
   _2}^3}\nonumber\\
& & +{C_{zz}}^{(1)}-2 {C_{\theta_1z}}^{(1)}+2 {C_{\theta_1x}}^{(1)}\Biggr)\beta\Biggr]\nonumber\\
 \tilde{g}^{\cal M}_{\theta_1x} & = & g^{\rm MQGP}_{\theta_1x}\Biggl[1 + \Biggl(
- \frac{\left(9 b^2+1\right)^4 b^{10} M  \left(6 a^2+{r_0}^2\right) \left((r-{r_0})^2+{r_0}^2\right)
   \Sigma_1}{3 \pi  \left(-18 b^4+3 b^2+1\right)^4 \log N  N^{8/15} {N_f} \left(-27 a^4+6 a^2
   {r_0}^2+{r_0}^4\right) \alpha _{\theta _2}^3}+{C_{\theta_1x}}^{(1)}
\Biggr)\beta\Biggr]\nonumber\\
\tilde{g}^{\cal M}_{\theta_1z} & = & g^{\rm MQGP}_{\theta_1z}\Biggl[1 + \Biggl(\frac{16 \left(9 b^2+1\right)^4 b^{12}  \beta  \left(\frac{(r-{r_0})^3}{{r_0}^3}+1\right) \left(19683
   \sqrt{3} \alpha _{\theta _1}^6+3321 \sqrt{2} \alpha _{\theta _2}^2 \alpha _{\theta _1}^3-40 \sqrt{3} \alpha _{\theta _2}^4\right)}{243
   \pi ^3 \left(1-3 b^2\right)^{10} \left(6 b^2+1\right)^8 {g_s}^{9/4} \log N ^4 N^{7/6} {N_f}^3 \left(-27 a^4 {r_0}+6 a^2
   {r_0}^3+{r_0}^5\right) \alpha _{\theta _1}^7 \alpha _{\theta _2}^6}+C_{\theta_1z}^{(1)}\Biggr)\Biggr]\nonumber\\
   \tilde{g}^{\cal M}_{\theta_2x} & = & g^{\rm MQGP}_{\theta_2x}\Biggl[1 + \Biggl(
   \frac{16 \left(9 b^2+1\right)^4 b^{12} \left(\frac{(r-{r_0})^3}{{r_0}^3}+1\right) \left(19683 \sqrt{3}
   \alpha _{\theta _1}^6+3321 \sqrt{2} \alpha _{\theta _2}^2 \alpha _{\theta _1}^3-40 \sqrt{3} \alpha _{\theta _2}^4\right)}{243 \pi ^3 \left(1-3
   b^2\right)^{10} \left(6 b^2+1\right)^8 {g_s}^{9/4} \log N ^4 N^{7/6} {N_f}^3 \left(-27 a^4 {r_0}+6 a^2
   {r_0}^3+{r_0}^5\right) \alpha _{\theta _1}^7 \alpha _{\theta _2}^6}+C_{\theta_2x}^{((1)}\Biggr)\beta\Biggr]\nonumber\\
\tilde{g}_{\theta_2y} & = & g^{\rm MQGP}_{\theta_2y}\Biggl[1 +  \frac{3 b^{10} \left(9 b^2+1\right)^4 M \beta \left(6 a^2+{r_0}^2\right) \left(1-\frac{(r-{r_0})^2}{{r_0}^2}\right) \log
   ({r_0}) \Sigma_1}{\pi  \left(3 b^2-1\right)^5 \left(6 b^2+1\right)^4 \log N ^2 N^{7/5} {N_f} \left(9 a^2+{r_0}^2\right) \alpha
   _{\theta _2}^3}\Biggr]\nonumber\\
\tilde{g}^{\cal M}_{\theta_2z} & = & g^{\rm MQGP}_{\theta_2z}\Biggl[1 + \Biggl(\frac{3 \left(9 b^2+1\right)^4 b^{10} M  \left(6 a^2+{r_0}^2\right) \left(1-\frac{(r-{r_0})^2}{{r_0}^2}\right) \log
   ({r_0}) \left(19683 \sqrt{6} \alpha _{\theta _1}^6+6642 \alpha _{\theta _2}^2 \alpha _{\theta _1}^3-40 \sqrt{6} \alpha _{\theta
   _2}^4\right)}{\pi  \left(3 b^2-1\right)^5 \left(6 b^2+1\right)^4 {\log N}^2 N^{7/6} {N_f} \left(9 a^2+{r_0}^2\right) \alpha
   _{\theta _2}^3}\nonumber\\
& & +{C_{\theta_2z}}^{(1)}\Biggr)\beta\Biggr]\nonumber\\
\tilde{g}^{\cal M}_{xy} & = & g^{\rm MQGP}_{xy}\Biggl[1 + \Biggl(\frac{3 \left(9 b^2+1\right)^4 b^{10} M  \left(6 a^2+{r_0}^2\right) \left(\frac{(r-{r_0})^2}{{r_0}^2}+1\right) \log
   ({r_0}) \alpha _{\theta _2}^3\Sigma_1}{\pi  \left(3 b^2-1\right)^5 \left(6 b^2+1\right)^4 \log N ^2 N^{21/20} {N_f} \left(9
   a^2+{r_0}^2\right) \alpha _{\theta _{2 l}}^6}+C_{xy}^{(1)}\Biggr)\beta\Biggr]\nonumber\\
\tilde{g}^{\cal M}_{xz}  & = & g^{\rm MQGP}_{xz}\Biggl[1 + \frac{18 b^{10} \left(9 b^2+1\right)^4 M \beta  \left(6 a^2+{r_0}^2\right)
   \left(\frac{(r-{r_0})^2}{{r_0}^2}+1\right) \log ^3({r_0}) \Sigma_1}{\pi  \left(3b^2-1\right)^5 \left(6 b^2+1\right)^4 \log N ^4 N^{5/4} {N_f} \left(9 a^2+{r_0}^2\right) \alpha
   _{\theta _2}^3}\Biggr]\nonumber\\
\tilde{g}^{\cal M}_{yy} & = & g^{\rm MQGP}_{yy}\Biggl[1  - \frac{3 b^{10} \left(9 b^2+1\right)^4 M \left(\frac{1}{N}\right)^{7/4} \beta  \left(6 a^2+{r_0}^2\right) \log ({r_0})\Sigma_1
   \left(\frac{(r-{r_0})^2}{r_0^2}+1\right)}{\pi  \left(3 b^2-1\right)^5 \left(6 b^2+1\right)^4 \log N ^2 {N_f} {r_0}^2 \left(9
   a^2+{r_0}^2\right) \alpha _{\theta _2}^3}\Biggr]\nonumber\\
 \tilde{g}^{\cal M}_{yz} & = & g^{\rm MQGP}_{yz}\Biggl[1 + \Biggl(\frac{64 \left(9 b^2+1\right)^8 b^{22} M \left(\frac{1}{N}\right)^{29/12}  \left(6 a^2+{r_0}^2\right)
   \left(\frac{(r-{r_0})^3}{{r_0}^3}+1\right) \log ({r_0}) }{27 \pi ^4 \left(3 b^2-1\right)^{15} \left(6 b^2+1\right)^{12}
   {g_s}^{9/4} \log N ^6  {N_f}^4 {r_0}^3 \left({r_0}^2-3 a^2\right) \left(9 a^2+{r_0}^2\right)^2 \alpha
   _{\theta _1}^7 \alpha _{\theta _2}^9}\nonumber\\
& & \hskip -0.3in \times \left(387420489 \sqrt{2} \alpha _{\theta _1}^{12}+87156324 \sqrt{3}
   \alpha _{\theta _2}^2 \alpha _{\theta _1}^9+5778054 \sqrt{2} \alpha _{\theta _2}^4 \alpha _{\theta _1}^6-177120 \sqrt{3} \alpha _{\theta
   _2}^6 \alpha _{\theta _1}^3+1600 \sqrt{2} \alpha _{\theta _2}^8\right)+C_{yz}^{(1)}\Biggr)\beta\Biggr]\nonumber\\
\tilde{g}^{\cal M}_{zz} & = & g^{\rm MQGP}_{zz}\Biggl[1 + \Biggl(C_{zz}^{(1)}-\frac{b^{10} \left(9 b^2+1\right)^4 M \left({r_0}^2-\frac{(r-{r_0})^3}{{r_0}}\right) \log ({r_0})
   \Sigma_1}{27 \pi ^{3/2} \left(3 b^2-1\right)^5 \left(6 b^2+1\right)^4 \sqrt{{g_s}} \log N ^2 N^{23/20} {N_f} \alpha
   _{\theta _2}^5}\Biggr)\beta\Biggr]\nonumber\\
\tilde{g}^{\cal M}_{x^{10}x^{10}} & = & g^{\rm MQGP}_{x^{10}x^{10}}\Biggl[1 -\frac{27 b^{10} \left(9 b^2+1\right)^4 M \left(\frac{1}{N}\right)^{5/4} \beta  \left(6 a^2+{r_0}^2\right)
   \left(1-\frac{(r-{r_0})^2}{{r_0}^2}\right) \log ^3({r_0}) \Sigma_1}{\pi  \left(3 b^2-1\right)^5 \left(6 b^2+1\right)^4 \log N ^4
   {N_f} {r_0}^2 \left(9 a^2+{r_0}^2\right) \alpha _{\theta _2}^3}\Biggr],
\end{eqnarray}
   }
   where:
\begin{eqnarray}
\label{Sigma_1-def}
& & \hskip -0.8in\Sigma_1 \equiv 19683
   \sqrt{6} \alpha _{\theta _1}^6+6642 \alpha _{\theta _2}^2 \alpha _{\theta _1}^3-40 \sqrt{6} \alpha _{\theta _2}^4\nonumber\\
   & & \hskip -0.8in\stackrel{\rm Global}{\longrightarrow} N^{\frac{6}{5}}\left(19683
   \sqrt{6} \sin^6\theta_1+6642 \sin^2{\theta _2} \sin^3{\theta _1}-40 \sqrt{6} \sin^4{\theta _2}\right),
\end{eqnarray}
${\cal C}^{(1)}_{MN}$ are constants of integration that figure in (\ref{ M-theory-metric-psi=2npi-patch}) after solving the EOMs for the ${\cal O}(\beta)$ metric perturbations $f_{MN}$, and $g^{\rm MQGP}_{MN}$ are the ${\cal M}$ theory metric components in the MQGP limit at ${\cal O}(\beta^0)$
\cite{mesons_0E++-to-mesons-decays}. The explicit dependence on $\theta_{10,20}$ of the ${\cal M}$-theory metric components up to ${\cal O}(\beta)$, using (\ref{theta12-deloc}), is effected by the replacemements:
$\alpha_{\theta_1}\rightarrow N^{\frac{1}{5}}\sin\theta_{10},\ \alpha_{\theta_2}\rightarrow N^{\frac{3}{10}}\sin\theta_{20}$ in (\ref{ M-theory-metric-psi=2npi-patch}). Also, see (\ref{xyz-definitions}).  The main Physics-related take-away  is the following. From (\ref{ M-theory-metric-psi=2npi-patch}), one notes that in the IR: $r = \chi r_0, \chi\equiv {\cal O}(1)$, up to ${\cal O}(\beta)$:
\begin{equation}
\label{IR-beta-N-suppressed-logrh-rh-neg-exp-enhanced}
f_{MN} \sim \beta\frac{\left(\log r_0\right)^{m}}{r_0^n N^{\beta_N}},\ m\in\left\{0,1,3\right\},\ n\in\left\{0,2,5,7\right\},\
\beta_N>0.
\end{equation}
As estimated in \cite{Bulk-Viscosity}, $|\log \left(\frac{r_0}{{\cal R}_{D5/\overline{D5}}}\right)|\sim N^{\frac{1}{3}}$, implying there is a competition between Planckian and large-$N$ suppression and infra-red enhancement arising from $m,n\neq0$ in (\ref{IR-beta-N-suppressed-logrh-rh-neg-exp-enhanced}).

Now, using the standard Witten's prescription of reading off the type IIA metric (inclusive of the ${\cal O}(R^4)$ corrections):
\begin{eqnarray}
\label{TypeIIA-from-M-theory-Witten-prescription}
\hskip -0.1in ds_{11}^2 & = & e^{-\frac{2\phi^{\rm IIA}}{3}}\Biggl[\frac{1}{\sqrt{h(r,\theta_{1,2})}}\left(-dt^2 + \left(dx^1\right)^2 +  \left(dx^2\right)^2 + \tilde{g}(r)\left(dx^3\right)^2 \right)
\nonumber\\
& & \hskip -0.1in+ \sqrt{h(r,\theta_{1,2})}\left(\frac{dr^2}{\tilde{g}(r)} + ds^2_{\rm IIA}(r,\theta_{1,2},\phi_{1,2},\psi)\right)
\Biggr] + e^{\frac{4\phi^{\rm IIA}}{3}}\left(dx^{11} + A_{\rm IIA}^{F_1^{\rm IIB} + F_3^{\rm IIB} + F_5^{\rm IIB}}\right)^2,
\end{eqnarray}
where $A_{\rm IIA}^{F^{\rm IIB}_{i=1,3,5}}$ are the type IIA RR 1-forms obtained from the triple T/SYZ-dual of the type IIB $F_{1,3,5}^{\rm IIB}$ fluxes in the type IIB holographic dual of \cite{metrics}.

Using ${\cal V}_1{\cal V}_2$-definition(obtained as in chapter 3) and (\ref{O4-corrections}), (\ref{CMN}) and (\ref{E1010F1010}), we first obtained:
\begin{eqnarray}
\label{expressions-O(R^4)-corrections}
& & \hskip -0.5in {\cal V}_1 = {\cal V}_1^{\rm LO} + {\cal V}_1^{{\cal O}(R^4)}\ {\rm where}:\nonumber\\
& & \hskip -0.5in {\cal V}_1^{{\cal O}(R^4)}=\frac{\sqrt{h} e^{-2 Z} (2 e^{-\phi} g^{\cal M}_{x^{10}x^{10}} g^{\cal M}_{rr}  {Tr} C.F+e^{-\phi^{\rm IIA}}\sqrt{g^{\rm IIA}}
   (- {{\cal F}_{x^{10}x^{10}} } g^{\cal M}_{rr}-2  {{\cal F}_{rr} } g^{\cal M}_{x^{10}x^{10}}+2 g^{\cal M}_{x^{10}x^{10}} g^{\cal M}_{rr}))}{g^{\cal M}_{x^{10}x^{10}}\ ^{3/2} g^{\cal M}_{rr}\ ^2 r_0 ^2}\nonumber\\
& & \hskip -0.5in = -\frac{3  {g_s} M \sqrt[5]{\frac{1}{N}}  {N_f}^2 e^{-4 Z} \left(e^{4 Z}-1\right) (  {\cal C}_{zz}^{(1)}-2
     {\cal C}_{\theta_1z}^{(1)}+2   {\cal C}_{\theta_1x}^{(1)}) \log \left(r_0  e^Z\right) \left(72 a^2 r_0  e^Z \log \left(r_0  e^Z\right)+3 a^2+2
   r_0 ^2 e^{2 Z}\right)}{8 \pi ^2 \alpha _{\theta _1} \alpha _{\theta _2}^2};\nonumber\\
& & \hskip -0.5in {\cal V}_2 = {\cal V}_2^{\rm LO} + {\cal V}_2^{{\cal O}(R^4)}\ {\rm where}:\nonumber\\
& & \hskip -0.5in {\cal V}_2^{{\cal O}(R^4)} = e^{-\phi^{\rm IIA}}h Tr({\cal C}.{\cal F}) + {\cal E}_{x^{10}x^{10}} h \sqrt{g^{\rm IIA}}
\nonumber\\
& & \hskip -0.5in = \frac{3  {g_s}^2 M N^{4/5}  {N_f}^2 e^{-2 Z} (  {\cal C}_{zz}^{(1)}-2   {\cal C}_{\theta_1z}^{(1)}+2   {\cal C}_{\theta_1x}^{(1)}) \log \left(r_0  e^Z\right) \left(72 a^2 r_0  e^Z \log
   \left(r_0  e^Z\right)-3 a^2+2 r_0 ^2 e^{2 Z}\right)}{4 \pi  r_0 ^2 \alpha _{\theta _1} \alpha _{\theta _2}^2},
\end{eqnarray}
where ${\cal V}_{1,2}^{\rm LO}$ are the LO terms as obtained in \cite{mesons_0E++-to-mesons-decays}. The equation of motion satisfied by the profile function of the vector meson $\psi_1(z)$, using ideas similar to \cite{mesons_0E++-to-mesons-decays}, was rewritten as a Sch\"{o}dinger-like equation with a potential ${\cal V} = {\cal V}^{\rm LO} + {\cal V}^{{\cal O}(R^4)}$ where (${\cal M}_{(1)} = m_0\frac{r_0}{\sqrt{4\pi g_sN}}$) and $ {\cal V}^{\rm LO}$ is the LO potential as given in  \cite{mesons_0E++-to-mesons-decays}. Further,
{
\begin{eqnarray}
\label{Schroedinger-like-psi}
& & \hskip -0.6in  {\cal V}^{{\cal O}(R^4)} = -2{\cal M}_{(1)}^2\frac{{\cal V}_1^{{\cal O}(R^4)}{\cal V}_2}{{\cal V}_1^2} + 2{\cal M}_{(1)}^2\frac{{\cal V}_2^{{\cal O}(R^4)}}{{\cal V}_1} - \frac{{\cal V}_1^{{\cal O}(R^4)}\left({\cal V}_1^\prime\right)\ ^2}{{\cal V}_1^3}
+ \frac{{\cal V}_1^\prime\left({\cal V}_1^{{\cal O}(R^4)}\right)^\prime}{2{\cal V}_1^2} + \frac{{\cal V}_1^{{\cal O}(R^4)}{\cal V}_1^{\prime\prime}}{2{\cal V}_1^2} - \frac{\left({\cal V}_1^{{\cal O}(R^4)}\right)^{\prime\prime}}{2{\cal V}_1}  \nonumber\\
& & \hskip -0.6in  =\beta  N \Biggl(\frac{2 \pi  {g_s}^{7/2} \log N  {\cal M}_{(1)}^2 {N_f} e^{-4 Z} ({\cal C}_{zz}^{(1)}-2   {\cal C}_{\theta_1z}^{(1)}+2   {\cal C}_{\theta_1x}^{(1)})
   \left(72 a^2 {r_0} e^Z \log \left({r_0} e^Z\right)-3 a^2+2 {r_0}^2 e^{2 Z}\right)}{{r_0}^8 \left(e^{4 Z}-1\right)
   \left(\frac{{g_s} e^{-4 Z}}{{r_0}^4}\right)^{3/2} \tilde{\Omega}(Z)}\nonumber\\
& & \hskip -0.6in +\frac{2 \pi  {g_s}^2 \log N  {\cal M}_{(1)}^2 {N_f} e^{4 Z}
   ({\cal C}_{zz}^{(1)}-2   {\cal C}_{\theta_1z}^{(1)}+2   {\cal C}_{\theta_1x}^{(1)}) \left(72 a^2 {r_0} e^Z \log \left({r_0} e^Z\right)+3 a^2+2 {r_0}^2 e^{2
   Z}\right)}{81 {r_0}^2 \left(e^{4 Z}-1\right) \alpha _{\theta _1}^2
   \tilde{\Omega}(Z)\ ^2}\Biggr)\nonumber\\
& & \hskip -0.6in \times  \Biggl[243 a^2 e^{-2 Z} \alpha _{\theta _1}^2 \Biggl(3 \log \left({r_0} e^Z\right) \biggl({g_s} {N_f} \left(8
   \log N  {r_0} e^Z+1\right)+32 \pi  {r_0} e^Z\biggr)-{g_s} (\log N +3) {N_f}\nonumber\\
& & \hskip -0.6in-72 {g_s} {N_f} {r_0}
   e^Z \log ^2\left({r_0} e^Z\right)-4 \pi \Biggr)+162 {r_0}^2 \alpha _{\theta _1}^2 \left({g_s} \log N  {N_f}-3
   {g_s} {N_f} \log \left({r_0} e^Z\right)+4 \pi \right)\Biggr]\nonumber\\
& & \hskip -0.6in  = -\beta\frac{\left(3 b^2-2\right) \log N  {m_0}^2 ({\cal C}_{zz}^{(1)}-2   {\cal C}_{\theta_1z}^{(1)}+2   {\cal C}_{\theta_1x}^{(1)})}{4 \left(3
   b^2+2\right) (\log N -3 \log ({r_0}))Z} + {\cal O}(Z^0),
\end{eqnarray}
}
where $a = \left(b + \gamma \frac{g_sM^2}{N}\left(1+\log r_0\right)\right)r_0$, and,
{
\begin{eqnarray}
\label{tildeOmegaZ}
& &\tilde{\Omega}(Z) \equiv 3 \log \left({r_0} e^Z\right) \left(3 a^2 \left({g_s}
   {N_f} \left(8 \log N  {r_0} e^Z-1\right)+32 \pi  {r_0} e^Z\right)-2 {g_s} {N_f} {r_0}^2 e^{2 Z}\right)\nonumber\\
& & +3
   a^2 ({g_s} (\log N -3) {N_f}+4 \pi )-216 a^2 {g_s} {N_f} {r_0} e^Z \log ^2\left({r_0} e^Z\right)+2
   {r_0}^2 e^{2 Z} ({g_s} \log N  {N_f}+4 \pi ).\nonumber\\
& &
\end{eqnarray}
}
As in \cite{mesons_0E++-to-mesons-decays}, we defined $g(Z) \equiv \sqrt{{\cal V}_1(Z)}\psi_1(Z)$ where $g(Z)$ satisfies the following Schr\"{o}dinger-like equation :
\begin{equation}
\label{g(Z)-EOM}
g^{\prime\prime}(Z) +  \left(\frac{\omega_1 + \beta  {\cal C}^{zz}_{\ \ \theta_1z\ \theta_1x}}{Z}+\omega_2+\frac{1}{4 Z^2}\right)g(Z) = 0,
\end{equation}
wherein:
{\footnotesize
\begin{eqnarray}
\label{omega_1-and-2_defs}
& & \omega_1\equiv \frac{1}{4} \left({m_0}^2-3 b^2 \left({m_0}^2-2\right)\right)+18 b^2 {r_h} \log
   ({r_h})-\frac{3 b \gamma  {g_s} M^2 \left({m_0}^2-2\right) \log ({r_h})}{2 N}+\frac{36 b
   \gamma  {g_s} M^2 {r_h} \log ^2({r_h})}{N},\nonumber\\
& & \omega_2\equiv -\frac{4}{3}+\frac{3}{2} b^2 \left({m_0}^2+72 {r_h}-4\right)-36 b^2 {r_h} \log ({r_h})+\frac{3 b \gamma
   {g_s} M^2 \left({m_0}^2-4\right) \log ({r_h})}{N}-\frac{72 b \gamma  {g_s} M^2 {r_h}
   \log ^2({r_h})}{N},\nonumber\\
& &
 {\cal C}^{zz}_{\ \ \theta_1z\ \theta_1x} = -\frac{\left(3 b^2-2\right) \log N  {m_0}^2 ({\cal C}_{zz}^{(1)}-2 {\cal C}_{\theta_1z}^{(1)}+2 {\cal C}_{\theta_1x}^{(1)})}{4 \left(3
   b^2+2\right) (\log N -3 \log ({r_0}))},
\end{eqnarray}}
and whose solution (using arguments similar to the ones in  \cite{mesons_0E++-to-mesons-decays}) was given in terms of Whittaker functions:
\begin{equation}
\label{Whittaker-solution}
g(Z) = c_{\psi_1}^{(1)}  M_{-\frac{i (\omega_1 +  \beta {\cal C}^{zz}_{\ \ \theta_1z\ \theta_1x} )}{2 \sqrt{\omega_2}},0}\left(2 i \sqrt{\omega_2} Z\right)+c_{\psi_1}^{(2)}
   W_{-\frac{i (\omega_1 +  \beta {\cal C}^{zz}_{\ \ \theta_1z\ \theta_1x}  )}{2 \sqrt{\omega_2}},0}\left(2 i \sqrt{\omega_2} Z\right).
\end{equation}
We note that the effect of the inclusion of the ${\cal O}(R^4)$ corrections into the EOM for the radial profile function $\psi_1(Z)$  for the $\rho$ meson is a shift in the residue of the simple pole in the potential of the Schr\"{o}dinger-like equation satisfied by the redefined $\rho$ meson profile function $g(Z)$.

Using arguments similar to the ones in \cite{mesons_0E++-to-mesons-decays}, implementing Neumann boundary condition ($\psi_1^\prime(Z=0)=0$) one sees that in the IR (i.e., near $Z=0$):
\begin{equation}
\label{psi1(Z)}
\psi_1(Z) = \sqrt{2} {\cal C}_{\psi_1}^{(1)\ {\rm IR}} \sqrt{i \sqrt{\omega_2}} \left[1-Z \left(\beta  {\cal C}^{zz}_{\ \ \theta_1z\ \theta_1x}+\omega_1\right)\right]; {\cal C}_{\psi_1}^{(1)\ {\rm IR}} \equiv {\cal C}_{\psi_1}^{{\rm IR}}=N^{-\Omega_{\psi_1}},\ \Omega_{\psi_1}>0.
\end{equation}
Now, as explained in chapter 3,
\begin{eqnarray}
\label{phi0}
& &  \phi_0(Z) = \frac{{\cal C}_{\phi_0}^{\rm IR}}{{\cal V}_1(Z)} = \frac{\phi_0^{(-1)}}{Z} + \phi_0^{({\rm constant})} + \phi_0^{(1)}Z + \phi_0^{(2)}Z^2 + {\cal O}(Z^3).
\end{eqnarray}
By requiring:
\begin{eqnarray}
\label{alphatheta2-alphatheta1}
& & \alpha _{\theta _2} = \frac{9 \sqrt{\log N -3 \log r_0 }}{\sqrt{2} \sqrt{\log N +3 \log r_0 }}N^{\frac{1}{10}} \alpha _{\theta _1},\ \log N > |\log r_0|,\nonumber\\
& & b = \frac{1}{\sqrt{3}} + \epsilon,
\end{eqnarray}
for a very tiny $\epsilon$ to be ascertained later, one can set: $\phi^{(-1)}_0 = \phi_0^{(1)} = 0$ and one obtains:
\begin{eqnarray}
\label{phi0(Z)}
& & \phi_0(Z) =\frac{\pi ^2 {\cal C}_{\phi_0}^{\rm IR}\  N^{2/5} \alpha _{\theta _1}^3 (\log N -3 \log r_0 ) \left(\frac{27}{8 b^2 {g_s}
   \log N  (\log N +3 \log r_0 )}-\frac{81 b^2 \beta  ({\cal C}_{zz}^{(1)}-2   {\cal C}_{\theta_1z}^{(1)}+2   {\cal C}_{\theta_1x}^{(1)})}{8 \log
   ({r_0})}\right)}{{g_s} M {N_f}^2 {r_0}^3 (\log N +3 \log r_0 )}\nonumber\\
& & -\frac{\pi ^2 {\cal C}_{\phi_0}^{IR} N^{2/5}
   \alpha _{\theta _1}^3 (\log N -3 \log r_0 ) \left(\frac{9 \left(3 b^2+1\right) \beta  ({\cal C}_{zz}^{(1)}-2   {\cal C}_{\theta_1z}^{(1)}+2   {\cal C}_{\theta_1x}^{(1)})}{4 \log r_0 }+\frac{1944 b^4}{\left(3 b^2+2\right)^4}\right)}{{g_s} \log r_0  M {N_f}^2 {r_0}^2
   (\log N +3 \log r_0 )} Z^2 + {\cal O}(Z^3).\nonumber\\
\end{eqnarray}
One can similarly show that one obtains the following profile functions in the UV:
\begin{eqnarray}
\label{profile-functions-UV}
& & \psi_1^{\rm UV}(Z) = {\cal C}_{\psi_1}^{\rm UV}\frac{e^{-2Z}}{Z^{\frac{3}{2}}},\nonumber\\
& & \phi_0^{\rm UV}(Z) = {\cal C}_{\phi_0}^{\rm UV}\frac{e^{-2Z}}{Z^2}.
\end{eqnarray}

To evaluate $y_{1,3,5,7}$ and $z_{1,...,8}$ using (\ref{y_i+z_i}) along with (\ref{<<A>>}) and (\ref{<A>}), one will be splitting the radial integral into the IR and the UV, e.g., $\langle A\rangle[\tilde{g}_{MN}^{\rm IR}] + \langle A\rangle[g_{MN}^{\rm UV}]$, where using the results of Appendix {\bf E.1}, $f_{MN}^{\rm UV}$'s are vanishingly small (impling $\tilde{g}_{MN}^{\rm UV} = g_{MN}^{\rm UV}$). Using (\ref{psi1(Z)}) and (\ref{phi0(Z)}), one arrives at the  expressions for the
coupling constants $y_{1,3,5,7}$ and $z_{1,...,8}$ as explained in Appendix {\bf E.2}:
\begin{eqnarray}
\label{y_i-z_j}
& & \left.y_{1,...,7},\ z_{1,...,8}\right|_{\sim\rm IIB\ Ouyang}\nonumber\\
& & = {\cal V}_{\Sigma_2} \left({\cal C}_{\psi_1}^{\rm IR}\right)^{n_{y_i/z_j}}
 \left({\cal C}_{\phi_0}^{\rm IR}\right)^{m_{y_i/z_j}}\nonumber\\
& & \times\Biggl({\cal F}_{y_i/z_j}(r_0; M, N, N_f ) + \beta \left({\cal C}_{zz}^{(1)}-2 {\cal C}_{\theta_1z}^{(1)}+2 {\cal C}_{\theta_1x}^{(1)}\right){\cal H}_{y_i/z_j}(r_0; M, N, N_f) \Biggr).
\end{eqnarray}
Further,
{\footnotesize
\begin{eqnarray}
\label{Fpisq}
& & \hskip -0.4in  F_\pi^2 = {\cal V}_{\Sigma_2}\Biggl(\frac{243 \pi ^2 \beta  \left({\cal C}_{zz}^{(1)} - 2 {\cal C}_{\theta_1z}^{(1)} + 2 {\cal C}_{\theta_1x}^{(1)}\right) {\cal C}_{\phi_0}^{\rm IR}\ ^2 {f_{r_0}} ({f_{r_0}}+1) \log ^2(3) \alpha _{\theta _1}^3 N^{\frac{4
   {f_{r_0}}}{3}+\frac{2}{5}}}{8192 ({f_{r_0}}-1)^3 {g_s}^3 \left(\log N\right) ^3 M N_f ^2}\nonumber\\
& & \hskip -0.4in +\frac{81 \sqrt{3} \pi ^2 {\cal C}_{\phi_0}^{\rm IR}\ ^2
   \epsilon  {f_{r_0}} ({f_{r_0}}+1)^2 \log (3) (\log (243)-6) \alpha _{\theta _1}^3 N^{\frac{4 {f_{r_0}}}{3}+\frac{2}{5}}}{2048
   ({f_{r_0}}-1)^3 {g_s}^3 \left(\log N\right) ^3 M N_f ^2} -\frac{243 \pi ^2 {\cal C}_{\phi_0}^{\rm IR}\ ^2 {f_{r_0}} ({f_{r_0}}+1)^2 \log ^2(3) \alpha
   _{\theta _1}^3 N^{\frac{4 {f_{r_0}}}{3}+\frac{2}{5}}}{4096 ({f_{r_0}}-1)^3 {g_s}^3 \left(\log N\right) ^3 M N_f ^2}\Biggr)\nonumber\\
& &
\end{eqnarray}
}
and
\begin{eqnarray}
\label{gsq}
& & g_{\rm YM}^2 =\frac{{\log N}   N \left(7 {({\cal C}_{zz}^{(1)}-2 {\cal C}_{\theta_1z}^{(1)}+2 {\cal C}_{\theta_1x}^{(1)})} {f_{r_0}}^2 \gamma ^2 {g_s}^2 M^4 \log ^2(N)+3456 ({f_{r_0}}+1)
   \lambda_{\epsilon}^2\right)}{288 \lambda_{\epsilon}^2 \alpha _{\theta _1}^2 \log N  \left(\sqrt{3} \beta ^{3/2} {({\cal C}_{zz}^{(1)}-2 {\cal C}_{\theta_1z}^{(1)}+2 {\cal C}_{\theta_1x}^{(1)})}
    \lambda_{\epsilon} m_0^2-12 ({f_{r_0}}+1)  N\right)}.
\end{eqnarray}

In the chiral limit, the ${\cal O}(p^4)$ $SU(3)\ \chi$PT Lagrangian is given by \cite{GL}:
\begin{eqnarray}
\label{ChPT-Op4}
& & L_1 \left({\rm Tr}(\nabla_\mu U^\dagger \nabla^\mu U)\right)^2 + L_2\left({\rm Tr}(\nabla_\mu U^\dagger \nabla_\nu U)\right)^2 + L_3{\rm Tr} \left(\nabla_\mu U^\dagger \nabla^\mu U\right)^2\nonumber\\
& & - i L_9 Tr\left({\cal L}_{\mu\nu}\nabla^\mu U \nabla^\nu U^\dagger + {\cal R}_{\mu\nu}\nabla^\mu U \nabla^\nu U^\dagger\right) + L_{10} Tr\left(U^\dagger {\cal L}_{\mu\nu}U{\cal R}^{\mu\nu}\right) + H_1 Tr\left({\cal L}_{\mu\nu}^2 + {\cal R}_{\mu\nu}^2\right),\nonumber\\
& &
\end{eqnarray}
where $\nabla_\mu U\equiv \partial_\mu U - i {\cal L}_\mu U + i U {\cal R}_\mu,\ U=e^{\frac{2i\pi}{F_\pi}}$.

\section{Summary and Physics Lessons Learnt}

\begin{enumerate}
\item
The low energy coupling constants (LECs) of the $SU(3)\ \chi$PT Lagrangian in the chiral limit at ${\cal O}(p^4)$,  is given by \cite{GLF}:
\begin{eqnarray}
\label{ChPT-Op4}
& & L_1 \left({\rm Tr}(D_\mu U^\dagger D^\mu U)\right)^2 + L_2\left({\rm Tr}(D_\mu U^\dagger D_\nu U)\right)^2 + L_3{\rm Tr} \left(D_\mu U^\dagger D^\mu U\right)^2\nonumber\\
& & - i L_9 Tr\left({\cal L}_{\mu\nu}D^\mu U D^\nu U^\dagger + {\cal R}_{\mu\nu}D^\mu U D^\nu U^\dagger\right) + L_{10} Tr\left(U^\dagger {\cal L}_{\mu\nu}U{\cal R}^{\mu\nu}\right) + H_1 Tr\left({\cal L}_{\mu\nu}^2 + {\cal R}_{\mu\nu}^2\right),\nonumber\\
& &
\end{eqnarray}
where $D_\mu U\equiv \partial_\mu U - i {\cal L}_\mu U + i U {\cal R}_\mu,\ U=e^{\frac{2i\pi}{F_\pi}}$ and ${\cal L}_{\mu\nu} = \partial_{[\mu}{\cal L}_{\nu]} - i[{\cal L}_\mu,{\cal L}_\nu]$ and ${\cal R}_{\mu\nu} = \partial_{[\mu}{\cal R}_{\nu]} - i[{\cal R}_\mu,{\cal R}_\nu]$ and ${\cal L}_\mu = {\cal V}_\mu - {\cal A}_\mu$, ${\cal R}_\mu =  {\cal V}_\mu + {\cal A}_\mu $, ${\cal V}_\mu$ and ${\cal A}_\mu$ being the external vector and axial-vector fields;{\small $\pi = \frac{1}{\sqrt{2}}\left(\begin{array}{ccc}
\frac{1}{\sqrt{2}}\pi^0 + \frac{1}{\sqrt{6}}\eta_8 + \frac{1}{\sqrt{3}}\eta_0 & \pi^+ & K^+ \\
\pi^- & -\frac{1}{\sqrt{2}}\pi^0 + \frac{1}{\sqrt{6}}\eta_8 + \frac{1}{\sqrt{3}}\eta_0 & K^0 \\
K^- & {\bar K}^0 & -\frac{2}{\sqrt{6}}\eta_8 + \frac{1}{\sqrt{3}}\eta_0
\end{array}\right)$.}
Inclusive of the ${\cal O}(R^4)$ corrections to the ${\cal M}$-theory dual of large-$N$ QCD-like theories as worked out in this project, the values of the 1-loop renormalized LECs were obtained in \cite{Vikas+Gopal+Aalok} holographically as appropriate radial integrals.

\item
\begin{itemize}
\item
There is a particular combination of the constants of integration appearing in the solutions to the ${\cal O}(R^4)$ corrections to the  ${\cal M}$-theory  dual of thermal QCD that will appear in all the coupling constants of $\chi$PT at least up to ${\cal O}(p^4)$.

\item
Matching the result obtained from our  ${\cal M}$-theory  ${\cal O}(R^4)$-corrected holographic computation with the experimental values of one-loop renormalized $L_{1,2,3}^r$, one sees one that one is required to do two things. One, the ${\cal O}\left(\frac{1}{N}\right)$ correction to the leading order (in $N$) result in expressing the resolution parameter in terms of the IR cut-off, also must involve a term proportional to $\frac{l_p^3}{N}$, i.e., $a = \left(b + \gamma \frac{g_sM^2}{N}\left(1+\log r_0\right)\right)r_0\rightarrow \left(\tilde{b} + \lambda \frac{l_p^3}{N} + \gamma\frac{g_sM^2}{N}(1+\log r_0)\right)r_0$. The second, the value of the aforementioned linear combination of integration constants figuring in the solutions to the ${\cal O}(R^4)$ corrections to the MQGP metric of \cite{MQGP,NPB}, gets fixed in terms of $\lambda, \gamma, M, g_s$ and $N$. This is the first evidence of the relationship between the ${\cal O}\left(\frac{1}{N}\right)$ and ${\cal O}(\beta)$ corrections.

\item
Matching the experimental values of $F_\pi^2$ and the one-loop renormalized $L_9^r$ and internal consistency, determine the angular delocalization in the polar angles $\theta_{1,2}$ (\ref{theta12-deloc}) consistent with the type IIB Ouyang embedding of the flavor $D7$-branes in the type IIB holographic dual and its SYZ type IIA mirror \cite{Yadav+Misra+Sil-Mesons}. Note, similar to as explained in \cite{SYZ 3 Ts}, the SYZ type IIA mirror (and hence its M theory uplift) is independent of the angular delocalization. In the context of obtaining the values of the $\chi$PT Lagrangian's coupling constants this is encoded in the fact even though the aforementioned angular delocalization  parameters $\alpha_{\theta_{1,2}}$ would change depending on the values of $\theta_{10,20}$ of (\ref{xyz-definitions}), the corresponding values of $\alpha_{\theta_{1,2}}$ can always be found.
\item
Matching the experimental value of $g_{\rm YM}^2$ at $\Lambda_{\rm QCD}=0.4$GeV, the HLS-QCD matching scale $\Lambda=1.1$GeV and renormalization scale $\mu=M_\rho$ with the value obtained from our setup, one obtains an upper bound on the constant of integration appearing in the expression for the profile function of the $\pi$ meson (and its flavor partners) in the IR.
\item
Upon matching with the experimental value of the one-loop renormalized $L_{10}^r$, one obtains a non-linear relation between the constants of integration appearing in the radial profile functions for the $\pi$ and $\rho$ mesons in the IR, as well as the coefficient of the $\frac{l_p^3}{N}$ term required to exist in the resolution-parameter-IR-cut-off relation (discussed in the second bullet above) upon matching with the experimental values of $L_{1,2,3}$. For numerical clarity, we explicitly wrote down the same for $N=10^2, g_s=0.1, M=N_f=3$.
\end{itemize}
\end{enumerate} 

\chapter{Conclusion and Future outlook }
\graphicspath{{Chapter6/}{Chapter6/}}

Over the course of this thesis we have used a top-down approach  to study the properties of strongly coupled thermal QCD-like theories by constructing gravitational duals of brane constructs. In a top-down approach one can achieve UV completion of the gauge theory by choosing a particular configuration of D-branes in a suitable background. Using the delocalized type IIA SYZ mirror and M-theory uplift constructed in \cite{MQGP} of the holograhic type IIB dual of large-N thermal QCD as constructed  in \cite{metrics} we obtained the analytical expressions for the vector and scalar meson spectra and compared our results with previous calculations of \cite{Sakai-Sugimoto-1,Dasgupta_et_al_Mesons} and obtained a closer match with the PDG results. In parallel to this  we also studied the mass spectra for various glueball states in the aforementioned background. For both cases we used two different type of background geometries, a black-hole(high temperature) background and a thermal(low temperature) background. For both vector mesons and scalar mesons spectra obtained for black hole gravity dual showed near isospectrality with the thermal gravity dual upto ${\cal O}\left (\frac{1}{\log N}\right )$.  For vector meson mass the black hole gravity dual gave a large-N suppression.

We used the results of glueball spectroscopy and meson spectroscopy to study the scalar glueball-meson interaction and two- three-,and four body glueball decay into scalar and vector mesons at tree level. Glueball in our case corresponds to the metric fluctuation of the M-theory uplift of \cite{MQGP} and mesons corresponds to the gauge fluctuations on or orthogonal to the world volume of type IIA flavor D6-Branes. Using Witten's prescription of relating M-theory and type IIA metric components, and DBI action for D6-branes we obtained the $G_{E}-\rho/\pi$ interaction Lagrangian, linear in exotic scalar glueball and up to quartic order in mesons. The expressions for the decay widths consisted of various constants of integration coming from solutions of M-theory metric perturbations  and mesons radial profile functions. Presence of these integration constants permitted a flexibility to our calculations allowing us to fine tune them and get bounds on them by comparing them with the PDG data. Further, in this thesis we tried to construct a top-down holographic dual of thermal QCD-like theories at intermedite 't Hooft coupling. In doing so we obtained ${\cal O}(l^{6}_{p})$ corrections to the MQGP background of \cite{MQGP}.  As an application of chapter 5 in chapter 6 we used the ${\cal O}\left( \frac{1}{N}\right )$-corrections in conjuction with the ${\cal O}(R^{4})$-corrections to match the experimental values of the coupling constants up to ${\cal O}(p^{4})$ appearing in the $SU(3)$ Chiral Perturbation theory Lagrangian for the $\pi$ and $\rho$ vector mesons as well as their flavor partners, in the chiral limit.

{\bf Future Physics outlook}: In the context of intermediate 't Hooft coupling top-down holography, there is no known literature on applying gauge-gravity duality techniques to studying the perturbative regime of thermal QCD-like theories  so as to be able to explain, e.g., low-frequency peaks expected to occur in spectral functions associated with transport coefficients, from ${\cal M}$ theory. In higher dimensional (Gauss-Bonnet or quartic in the Weyl tensor) holography, in the past couple of
years using previously known results, it has been shown by the (Leiden-)MIT-Oxford collaboration  \cite{previous-higher-ders}
that one obtains low frequency peaks in correlation/spectral functions of energy momentum tensor, per unit frequency, obtained
from the dissipative (i.e. purely imaginary) quasi-normal modes. As an extremely crucial application of the results of our work,
 for the first time,  spectral/correlation functions involving the energy momentum
tensor with the inclusion of the ${\cal O}(l_p^6)$ corrections in the ${\cal M}$ theory (uplift) metric of \cite{MQGP} can be evaluated and hence one would be able to make direct connection between previous results in perturbative thermal QCD-like theories  (e.g., \cite{Moore+Saremi}) as well as QCD plasma in RHIC
experiments, and ${\cal M}$ theory. Further, the temperature dependence of the speed of sound, the attenuation constant and bulk viscosity can also be obtained from its
solution, as well as the ${\cal O}(l_p^6)$ and the non-conformal corrections to the conformal results thereof. One could see if one could reproduce the known weak-coupling result from ${\cal M}$ theory that the ratio of the bulk and
shear viscosities has a lower bound that goes like the square of the deviation of the square of the speed of sound from its conformal
value (the last reference in \cite{EPJC-2}). Generically, the dissipative quasi-normal modes in the spectral functions at low frequencies can be investigated to study the existence of peaks at low frequencies in transport coefficients, thus making direct contact with perturbative QCD results as well as (QCD plasma in) RHIC experiments. 
\renewcommand{\chaptermark}[1]{         
\markboth{ \thechapter.\ #1}{}} %

\appendix

\chapter{}

\section{$\widetilde{F^2_5}, \widetilde{F^2_3}, H_3^2$}

 In the IR, up to ${\cal O}(g_s N_f)$ and setting $h_5=0$, the three-forms are as given in \cite{metrics}:
\begin{eqnarray}
\label{three-form-fluxes}
& & \hskip -0.4in (a) {\widetilde F}_3  =  2M { A_1} \left(1 + \frac{3g_sN_f}{2\pi}~{\rm log}~r\right) ~e_\psi \wedge
\frac{1}{2}\left({\rm sin}~\theta_1~ d\theta_1 \wedge d\phi_1-{ B_1}~{\rm sin}~\theta_2~ d\theta_2 \wedge
d\phi_2\right)\nonumber\\
&& \hskip -0.3in -\frac{3g_s MN_f}{4\pi} { A_2}~\frac{dr}{r}\wedge e_\psi \wedge \left({\rm cot}~\frac{\theta_2}{2}~{\rm sin}~\theta_2 ~d\phi_2
- { B_2}~ {\rm cot}~\frac{\theta_1}{2}~{\rm sin}~\theta_1 ~d\phi_1\right)\nonumber \\
&& \hskip -0.3in -\frac{3g_s MN_f}{8\pi}{ A_3} ~{\rm sin}~\theta_1 ~{\rm sin}~\theta_2 \left(
{\rm cot}~\frac{\theta_2}{2}~d\theta_1 +
{ B_3}~ {\rm cot}~\frac{\theta_1}{2}~d\theta_2\right)\wedge d\phi_1 \wedge d\phi_2, \nonumber\\
& & \hskip -0.4in (b) H_3 =  {6g_s { A_4} M}\Biggl(1+\frac{9g_s N_f}{4\pi}~{\rm log}~r+\frac{g_s N_f}{2\pi}
~{\rm log}~{\rm sin}\frac{\theta_1}{2}~
{\rm sin}\frac{\theta_2}{2}\Biggr)\frac{dr}{r}\nonumber \\
&& \hskip -0.3in \wedge \frac{1}{2}\Biggl({\rm sin}~\theta_1~ d\theta_1 \wedge d\phi_1
- { B_4}~{\rm sin}~\theta_2~ d\theta_2 \wedge d\phi_2\Biggr)
+ \frac{3g^2_s M N_f}{8\pi} { A_5} \Biggl(\frac{dr}{r}\wedge e_\psi -\frac{1}{2}de_\psi \Biggr)\nonumber  \\
&&  \wedge \Biggl({\rm cot}~\frac{\theta_2}{2}~d\theta_2
-{ B_5}~{\rm cot}~\frac{\theta_1}{2} ~d\theta_1\Biggr). \nonumber\\
\end{eqnarray}
The asymmetry factors in (\ref{three-form-fluxes}) are given by: $ A_i=1 +{\cal O}\left(\frac{a^2}{r^2}\ {\rm or}\ \frac{a^2\log r}{r}\ {\rm or}\ \frac{a^2\log r}{r^2}\right) + {\cal O}\left(\frac{{\rm deformation\ parameter }^2}{r^3}\right),$ $  B_i = 1 + {\cal O}\left(\frac{a^2\log r}{r}\ {\rm or}\ \frac{a^2\log r}{r^2}\ {\rm or}\ \frac{a^2\log r}{r^3}\right)+{\cal O}\left(\frac{({\rm deformation\ parameter})^2}{r^3}\right)$.    As in the UV, $\frac{({\rm deformation\ parameter})^2}{r^3}\ll  \frac{({\rm resolution\ parameter})^2}{r^2}$, we will assume the same three-form fluxes for $h_5\neq0$.

The expressions of squares of various fluxes that figure in the EOM (\ref{final EOM}) are given below for ready reference:
{\
\begin{eqnarray}
\label{fluxes-squared}
& & \widetilde{F^2_5}=-\frac{8}{\sqrt{\pi } \sqrt{N} \sqrt{g_s}}-\frac{1}{254803968 \pi ^{13/2} N^{7/10}}
\nonumber\\
&&\times\Biggl[M^4 r^6 N_f^4 g_s^{11/2} \left(r^4-r_h^4\right) \left(\phi _1+\phi_2-\psi \right){}^2\Biggl(N_f g_s \log (N) (2 (r+1) \log (r)+1)
+ 2 \Biggl\{-9 (r+1) N_f g_s \log ^2(r)\nonumber\\
&&-2 (r+1) \log (r) \left(2 \pi -\log (4) N_f g_s\right)+\log (4) N_f g_s\biggr\}\Biggr){}^2\Biggr]
\nonumber\\
&&+a^2\biggl[+\frac{\pi ^{3/2} r^{10} g_s^{3/2}}{956593800 N^{7/10}}-\frac{24}{\sqrt{\pi } \sqrt{N} \sqrt{g_s}r^2}-\frac{1}{84934656 \pi ^{13/2} N^{7/10}} \nonumber\\
&&\times\Biggl\{M^4 r^6 N_f^4 g_s^{11/2} \left(r^4-r_h^4\right) \left(\phi _1+\phi_2-\psi \right){}^2(24 r\log{r}-1)\Biggl(N_f g_s \log (N) (2 (r+1) \log (r)+1)\nonumber\\
&&+ 2 \Biggl\{-9 (r+1) N_f g_s \log ^2(r)-2 (r+1) \log (r) \left(2 \pi -\log (4) N_f g_s\right)+\log (4) N_f g_s\biggr\}\Biggr){}^2\Biggr\}\Biggr];\nonumber \\
&&\widetilde{F^2_3}=\frac{729 M^2 N_f^2 \sqrt{g_s} \left(r^4-r_h^4\right) \left(72 a^2 N^{2/5} \log (r)+a^2 \left(-\left(3 N^{2/5}+4\right)\right)+2 N^{2/5} r^2\right)}{128
   \pi ^{7/2} N^{11/10} r^6};\nonumber\\
&&   H_3^2 =\frac{243 M^2 N_f^2 g_s^{5/2} \left(r^4-r_h^4\right) \left(144 a^2 \left(\sqrt[5]{N}+3\right) r \log (r)+a^2 \left(9-15 \sqrt[5]{N}\right)+2
   \left(\sqrt[5]{N}+1\right) r^2\right)}{256 \pi ^{7/2} \sqrt{N} r^6};\nonumber\\
&&\widetilde{F}_{x_2 x_3p_3p_4p_5}\widetilde{F}_{x_2x_3 q_3q_4q_5}g^{p_3q_3}g^{p_4q_4}g^{p_5q_5}h^{x_2x_3}=\frac{60 r^4}{\pi ^{3/2} N^{3/2} g_s^{3/2}}-\frac{1}{169869312 \pi ^{13/2} N^{7/10}}\nonumber\\
&&\times\Biggl[5 M^4 r^{10} N_f^4 g_s^{9/2} \left(r^4-r_h^4\right) \left(\phi _1+\phi _2-\psi \right){}^2\Biggl(N_f g_s \log (N) (2 (r+1) \log (r)+1)
+ 2 \Biggl\{-9 (r+1) N_f g_s \log ^2(r)\nonumber\\
& &-2 (r+1) \log (r) \left(2 \pi -\log (4) N_f g_s\right)+\log (4) N_f g_s\biggr\}\Biggr){}^2\Biggr]+a^2\Biggl[\frac{180 r^2}{\pi ^{3/2} N^{3/2} g_s^{3/2}}+\frac{1}{56623104 \pi ^{15/2} N^{17/10}}\nonumber\\
&&\times\Biggl\{5 M^4 r^{10} N_f^4 g_s^{9/2} \left(r^4-r_h^4\right) \left(\phi _1+\phi _2-\psi \right){}^2(24 r\log{r}-1)\Biggl(N_f g_s \log (N) (2 (r+1) \log (r)+1)
\nonumber \\
&& + 2 \Biggl\{-9 (r+1) N_f g_s \log ^2(r)-2 (r+1) \log (r) \left(2 \pi -\log (4) N_f g_s\right)+\log (4) N_f g_s\biggr\}\Biggr){}^2\Biggr\}\Biggr].
\end{eqnarray}}

\chapter{}

\section{Triple-T Duality Rules}

In this section, we summarize the Buscher triple T-duality rules for T-dualizing first along $x$, then along $y$ followed by along $z$. The starting metric in the type IIB theory has the following components
\begin{eqnarray}
ds^2_{\rm IIB} & = &
g^{\rm IIB}_{\mu \nu}dx^\mu ~dx^\nu + g^{\rm IIB}_{x\mu} dx ~dx^\mu +  g^{\rm IIB}_{y \mu} dy~
dx^\mu +  g^{\rm IIB}_{z\mu} dz ~ dx^\mu +  g^{\rm IIB}_{xy} dx ~dy
  + g^{\rm IIB}_{xz} dx ~dz +  g^{\rm IIB}_{zy} dz ~ dy  \nonumber\\
 &&  +  g^{\rm IIB}_{xx} dx^2 +  g^{\rm IIB}_{yy}dy^2
 +  g^{\rm IIB}_{zz}~dz^2,
  \end{eqnarray}
where $\mu, \nu \neq x, y, z$.
As shown in \cite{SYZ 3 Ts}, the form of the metric of the mirror manifold after performing three T-dualities, first along $x$, then along $y$ and finally along $z$:
\begin{eqnarray}
\label{mirror_metric}
& & ds^2 =
\left( G_{\mu\nu} - {G_{z\mu}G_{z\nu} - {\cal B}_{z\mu} {\cal
B}_{z\nu} \over G_{zz}} \right) dx^\mu~dx^\nu +2 \left( G_{x\nu} -
{G_{zx}G_{z\nu} - {\cal B}_{zx} {\cal B}_{z\nu}
 \over G_{zz}} \right) dx~dx^\mu  \nonumber\\
& &  + 2\left( G_{y\nu} - {G_{zy}G_{z\nu} - {\cal B}_{zy} {\cal B}_{z\nu}
 \over G_{zz}}\right) dy~dx^\nu +
2\left( G_{xy} - {G_{zx}G_{zy} - {\cal B}_{zx} {\cal B}_{zy} \over
G_{zz}}\right) dx~dy  \nonumber\\
& &  + {dz^2\over G_{zz}} + 2{{\cal
B}_{\mu z} \over G_{zz}} dx^\mu~dz + 2{{\cal B}_{xz} \over G_{zz}}
dx~dz + 2{{\cal B}_{yz} \over G_{zz}} dy~dz \nonumber\\
& & +  \left( G_{xx}
- {G^2_{zx} - {\cal B}^2_{zx} \over G_{zz}} \right) dx^2 + \left(
G_{yy} - {G^2_{zy} - {\cal B}^2_{zy} \over G_{zz}} \right)
dy^2.
\end{eqnarray}
 The various components of the metric after three successive T-dualities along $x, y$ and $z$ respectively, can be written as \cite{SYZ 3 Ts}:
 {\small
\begin{eqnarray}
\label{G_munu}
& & G_{\mu\nu} = {g^{\rm IIB}_{\mu\nu}g^{\rm IIB}_{xx} -
g^{\rm IIB}_{x\mu}g^{\rm IIB}_{x\nu} + b^{\rm IIB}_{x\mu}b^{\rm IIB}_{x\nu} \over g^{\rm IIB}_{xx}} -
{(g^{\rm IIB}_{y\mu}g^{\rm IIB}_{xx} - g^{\rm IIB}_{xy} g^{\rm IIB}_{x \mu} + b^{\rm IIB}_{xy} b^{\rm IIB}_{x\mu})
(g^{\rm IIB}_{y\nu}g^{\rm IIB}_{xx}
 - g^{\rm IIB}_{xy} g^{\rm IIB}_{x \nu} + b^{\rm IIB}_{xy} b^{\rm IIB}_{x\nu}) \over
g^{\rm IIB}_{xx}(g^{\rm IIB}_{yy}g^{\rm IIB}_{xx}- g^{\rm IIB}_{xy}\ ^2 + b^{\rm IIB}_{xy}\ ^2)} \nonumber\\
&& + {(b^{\rm IIB}_{y\mu}g^{\rm IIB}_{xx} - g^{\rm IIB}_{xy} b^{\rm IIB}_{x \mu} + b^{\rm IIB}_{xy}
g^{\rm IIB}_{x\mu})(b^{\rm IIB}_{y\nu}g^{\rm IIB}_{xx}
 - g^{\rm IIB}_{xy} b^{\rm IIB}_{x \nu} + b^{\rm IIB}_{xy} g^{\rm IIB}_{x\nu})\over
g^{\rm IIB}_{xx}(g^{\rm IIB}_{yy}g^{\rm IIB}_{xx}- g^{\rm IIB}_{xy}\ ^2 + b^{\rm IIB}_{xy}\ ^2)},
\end{eqnarray}

\begin{eqnarray}
\label{Gmuz}
& & G_{\mu z} = {g^{\rm IIB}_{\mu z}g^{\rm IIB}_{xx} -
g^{\rm IIB}_{x\mu}g^{\rm IIB}_{xz} + b^{\rm IIB}_{x \mu}b^{\rm IIB}_{xz} \over g^{\rm IIB}_{xx}} - {(g^{\rm IIB}_{y\mu}g^{\rm IIB}_{xx}
- g^{\rm IIB}_{xy} g^{\rm IIB}_{x \mu} + b^{\rm IIB}_{xy} b^{\rm IIB}_{x\mu}) (g^{\rm IIB}_{yz}g^{\rm IIB}_{xx} - g^{\rm IIB}_{xy} g^{\rm IIB}_{x
z} + b^{\rm IIB}_{xy} b^{\rm IIB}_{xz}) \over g^{\rm IIB}_{xx}(g^{\rm IIB}_{yy}g^{\rm IIB}_{xx}- g^{\rm IIB}_{xy}\ ^2 +
b^{\rm IIB}_{xy}\ ^2)} \nonumber\\
& &  + {(b^{\rm IIB}_{y\mu}g^{\rm IIB}_{xx} - g^{\rm IIB}_{xy} b^{\rm IIB}_{x \mu} +
b^{\rm IIB}_{xy} g^{\rm IIB}_{x\mu})(b^{\rm IIB}_{yz}g^{\rm IIB}_{xx} - g^{\rm IIB}_{xy} b^{\rm IIB}_{x z} + b^{\rm IIB}_{xy}
g^{\rm IIB}_{xz})\over g^{\rm IIB}_{xx}(g^{\rm IIB}_{yy}g^{\rm IIB}_{xx}- g^{\rm IIB}_{xy}\ ^2 + b^{\rm IIB}_{xy}\ ^2)},
\end{eqnarray}

\begin{eqnarray}
\label{Gzz}
& & G_{zz} =  {g^{\rm IIB}_{zz}g^{\rm IIB}_{xx} - j^2_{xz} +
b^2_{xz}\over g^{\rm IIB}_{xx}} - {(g^{\rm IIB}_{yz}g^{\rm IIB}_{xx} - g^{\rm IIB}_{xy} g^{\rm IIB}_{xz} + b^{\rm IIB}_{xy}
b^{\rm IIB}_{xz})^2 \over g^{\rm IIB}_{xx}(g^{\rm IIB}_{yy}g^{\rm IIB}_{xx}- g^{\rm IIB}_{xy}\ ^2 + b^{\rm IIB}_{xy}\ ^2)} + {(b^{\rm IIB}_{yz}g^{\rm IIB}_{xx} - g^{\rm IIB}_{xy} b^{\rm IIB}_{x z} + b^{\rm IIB}_{xy} g^{\rm IIB}_{xz})^2 \over
g^{\rm IIB}_{xx}(g^{\rm IIB}_{yy}g^{\rm IIB}_{xx}- g^{\rm IIB}_{xy}\ ^2 + b^{\rm IIB}_{xy}\ ^2)},
\end{eqnarray}

\begin{eqnarray}
\label{Gymu}
& & G_{y \mu} = -{b^{\rm IIB}_{y \mu} g^{\rm IIB}_{xx} - b^{\rm IIB}_{x \mu} g^{\rm IIB}_{xy} + b^{\rm IIB}_{xy}
 g^{\rm IIB}_{\mu x} \over g^{\rm IIB}_{yy}g^{\rm IIB}_{xx}- g^{\rm IIB}_{xy}\ ^2 + b^{\rm IIB}_{xy}\ ^2},
~ G_{y z} = -{b^{\rm IIB}_{y z} g^{\rm IIB}_{xx} - b^{\rm IIB}_{x z} g^{\rm IIB}_{xy} + b^{\rm IIB}_{xy} g^{\rm IIB}_{z x}
\over g^{\rm IIB}_{yy}g^{\rm IIB}_{xx}- g^{\rm IIB}_{xy}\ ^2 + b^{\rm IIB}_{xy}\ ^2},
\end{eqnarray}

\begin{eqnarray}
\label{Gyy}
& & G_{yy} = {g^{\rm IIB}_{xx} \over g^{\rm IIB}_{yy}g^{\rm IIB}_{xx}- g^{\rm IIB}_{xy}\ ^2 +
b^{\rm IIB}_{xy}\ ^2},~ G_{xx} = {g^{\rm IIB}_{yy} \over g^{\rm IIB}_{yy}g^{\rm IIB}_{xx}- g^{\rm IIB}_{xy}\ ^2 +
b^{\rm IIB}_{xy}\ ^2}, ~G_{xy} = {-g^{\rm IIB}_{xy} \over g^{\rm IIB}_{yy}g^{\rm IIB}_{xx}- g^{\rm IIB}_{xy}\ ^2 +
b^{\rm IIB}_{xy}\ ^2},
\end{eqnarray}

\begin{eqnarray}
\label{Gmux}
& & G_{\mu x} = {b^{\rm IIB}_{\mu x} \over g^{\rm IIB}_{xx}} + {(g^{\rm IIB}_{\mu y} g^{\rm IIB}_{xx} -
 g^{\rm IIB}_{xy} g^{\rm IIB}_{x \mu} + b^{\rm IIB}_{xy} b^{\rm IIB}_{x \mu}) b^{\rm IIB}_{xy} \over
g^{\rm IIB}_{xx}(g^{\rm IIB}_{yy}g^{\rm IIB}_{xx}- g^{\rm IIB}_{xy}\ ^2 + b^{\rm IIB}_{xy}\ ^2)}
+ {(b^{\rm IIB}_{y \mu} g^{\rm IIB}_{xx} - g^{\rm IIB}_{xy} b^{\rm IIB}_{x \mu} + b^{\rm IIB}_{xy} g^{\rm IIB}_{x \mu}) g^{\rm IIB}_{xy}
 \over g^{\rm IIB}_{xx}(g^{\rm IIB}_{yy}g^{\rm IIB}_{xx}- g^{\rm IIB}_{xy}\ ^2 + b^{\rm IIB}_{xy}\ ^2)},
 \end{eqnarray}

\begin{eqnarray}
\label{Gzx}
& & G_{z x} = {b^{\rm IIB}_{z x} \over g^{\rm IIB}_{xx}} + {(g^{\rm IIB}_{z y} g^{\rm IIB}_{xx} -
g^{\rm IIB}_{xy} g^{\rm IIB}_{x z} + b^{\rm IIB}_{xy} b^{\rm IIB}_{x z}) b^{\rm IIB}_{xy} \over
g^{\rm IIB}_{xx}(g^{\rm IIB}_{yy}g^{\rm IIB}_{xx}- g^{\rm IIB}_{xy}\ ^2 + b^{\rm IIB}_{xy}\ ^2)}
 + {(b^{\rm IIB}_{y z} g^{\rm IIB}_{xx} - g^{\rm IIB}_{xy} b^{\rm IIB}_{x z} + b^{\rm IIB}_{xy} g^{\rm IIB}_{xz}) g^{\rm IIB}_{xy}
  \over g^{\rm IIB}_{xx}(g^{\rm IIB}_{yy}g^{\rm IIB}_{xx}- g^{\rm IIB}_{xy}\ ^2 + b^{\rm IIB}_{xy}\ ^2)}.
\end{eqnarray}}
In the above formulae we have denoted the type IIB
$B$ fields as $b^{\rm IIB}_{mn}$.  For the generic case we will switch on all the
components of the $B$ field:
{\small
\begin{eqnarray}
B^{\rm IIB} & = &
b^{\rm IIB}_{\mu\nu} ~ dx^\mu \wedge dx^\nu + b^{\rm IIB}_{x \mu} dx \wedge dx^\mu +  b^{\rm IIB}_{y \mu}
~ dy~\wedge dx^\mu + b^{\rm IIB}_{z \mu} ~ dz \wedge dx^\mu \nonumber\\
& & + ~ b^{\rm IIB}_{xy}
~ dx \wedge dy +
 b^{\rm IIB}_{xz} ~ dx  \wedge dz +  b^{\rm IIB}_{zy}~ dz  \wedge dy.
 \end{eqnarray}}
\noindent After applying again the T-dualities, the type IIA NS-NS $B$ field in the mirror set-up will take the form:
{\small
\begin{eqnarray}
\label{B}
B^{IIA} & = & \left( {\cal B}_{\mu\nu}
+ {2 {\cal B}_{z[\mu} G_{\nu]z} \over G_{zz}} \right) dx^\mu
\wedge dx^\nu + \left( {\cal B}_{\mu x} + {2 {\cal B}_{z[\mu}
G_{x]z} \over G_{zz}}\right)
 dx^\mu \wedge dx  \nonumber\\
& & \left( {\cal B}_{\mu y} + {2 {\cal B}_{z[\mu} G_{y]z} \over G_{zz}}
 \right) dx^\mu \wedge dy
+ \left( {\cal B}_{xy} + {2 {\cal B}_{z[x} G_{y]z} \over G_{zz}}
\right) dx \wedge dy \nonumber\\
& &  + {G_{z \mu} \over G_{zz}} dx^\mu
\wedge dz + {G_{z x} \over G_{zz}} dx \wedge dz + {G_{z y} \over
G_{zz}} dy \wedge dz.
\end{eqnarray}}
 Here the $G_{mn}$ components have been
given above, and the various ${\cal B}$  components can  be
written as:
{\small
\begin{eqnarray}
\label{Bmunu}
{\cal B}_{\mu\nu} & = &
{b^{\rm IIB}_{\mu\nu} g^{\rm IIB}_{xx} + b^{\rm IIB}_{x \mu} g^{\rm IIB}_{\nu x} - b^{\rm IIB}_{x \nu} g^{\rm IIB}_{\mu x}
\over g^{\rm IIB}_{xx}} \nonumber\\
& & +  {2 (g^{\rm IIB}_{y[\mu}g^{\rm IIB}_{xx} - g^{\rm IIB}_{xy}g^{\rm IIB}_{x[\mu} +
b^{\rm IIB}_{xy} b^{\rm IIB}_{x[\mu}) (b^{\rm IIB}_{\nu]y}g^{\rm IIB}_{xx} - b^{\rm IIB}_{\nu]x}g^{\rm IIB}_{xy} - b^{\rm IIB}_{xy}
g^{\rm IIB}_{\nu]x}) \over g^{\rm IIB}_{xx}(g^{\rm IIB}_{yy}g^{\rm IIB}_{xx}- g^{\rm IIB}_{xy}\ ^2 + b^{\rm IIB}_{xy}\ ^2)},
\end{eqnarray}

\begin{eqnarray}
\label{Bmuz}
{\cal B}_{\mu z} & = &  {b^{\rm IIB}_{\mu z} g^{\rm IIB}_{xx} +
b^{\rm IIB}_{x \mu} g^{\rm IIB}_{z x} - b^{\rm IIB}_{x z} g^{\rm IIB}_{\mu x} \over g^{\rm IIB}_{xx}} \nonumber\\
& & +  {2
(g^{\rm IIB}_{y[\mu}g^{\rm IIB}_{xx} - g^{\rm IIB}_{xy}g^{\rm IIB}_{x[\mu} + b^{\rm IIB}_{xy} b^{\rm IIB}_{x[\mu})
(b^{\rm IIB}_{z]y}g^{\rm IIB}_{xx} - b^{\rm IIB}_{z]x}g^{\rm IIB}_{xy} - b^{\rm IIB}_{xy} g^{\rm IIB}_{z]x}) \over
g^{\rm IIB}_{xx}(g^{\rm IIB}_{yy}g^{\rm IIB}_{xx}- g^{\rm IIB}_{xy}\ ^2 + b^{\rm IIB}_{xy}\ ^2)},
\end{eqnarray}

\begin{eqnarray}
\label{Bmuy}
{\cal B}_{\mu y} & = & {g^{\rm IIB}_{\mu y} g^{\rm IIB}_{xx} - g^{\rm IIB}_{xy} g^{\rm IIB}_{x \mu} + b^{\rm IIB}_{xy} b^{\rm IIB}_{x \mu}
 \over g^{\rm IIB}_{yy}g^{\rm IIB}_{xx}- g^{\rm IIB}_{xy}\ ^2 + b^{\rm IIB}_{xy}\ ^2},\nonumber\\
{\cal B}_{z y} & = & {g^{\rm IIB}_{z y} g^{\rm IIB}_{xx} - g^{\rm IIB}_{xy} g^{\rm IIB}_{x z} + b^{\rm IIB}_{xy} b^{\rm IIB}_{x z} \over g^{\rm IIB}_{yy}g^{\rm IIB}_{xx}-
 g^{\rm IIB}_{xy}\ ^2 + b^{\rm IIB}_{xy}\ ^2},
 \end{eqnarray}

\begin{eqnarray}
\label{Bmux}
{\cal B}_{\mu x} & = & {g^{\rm IIB}_{\mu x} \over g^{\rm IIB}_{xx}} - {g^{\rm IIB}_{xy} (g^{\rm IIB}_{\mu y} g^{\rm IIB}_{xx} -
g^{\rm IIB}_{xy} g^{\rm IIB}_{x \mu} + b^{\rm IIB}_{xy} b^{\rm IIB}_{x \mu}) \over
g^{\rm IIB}_{xx}(g^{\rm IIB}_{yy}g^{\rm IIB}_{xx}- g^{\rm IIB}_{xy}\ ^2 + b^{\rm IIB}_{xy}\ ^2)} + {b^{\rm IIB}_{xy} (b^{\rm IIB}_{x\mu}g^{\rm IIB}_{xy} - b^{\rm IIB}_{y\mu}g^{\rm IIB}_{xx} -
b^{\rm IIB}_{xy}g^{\rm IIB}_{xz})
 \over g^{\rm IIB}_{xx}(g^{\rm IIB}_{yy}g^{\rm IIB}_{xx}- g^{\rm IIB}_{xy}\ ^2 + b^{\rm IIB}_{xy}\ ^2)},
 \end{eqnarray}

\begin{eqnarray}
\label{Bzx}
{\cal B}_{z x} & = & {g^{\rm IIB}_{z x} \over g^{\rm IIB}_{xx}} - {g^{\rm IIB}_{xy} (g^{\rm IIB}_{z y} g^{\rm IIB}_{xx} -
g^{\rm IIB}_{xy} g^{\rm IIB}_{xz} + b^{\rm IIB}_{xy} b^{\rm IIB}_{x z}) \over
g^{\rm IIB}_{xx}(g^{\rm IIB}_{yy}g^{\rm IIB}_{xx}- g^{\rm IIB}_{xy}\ ^2 + b^{\rm IIB}_{xy}\ ^2)} + {b^{\rm IIB}_{xy} (b^{\rm IIB}_{xz}g^{\rm IIB}_{xy} - b^{\rm IIB}_{yz}g^{\rm IIB}_{xx} -
 b^{\rm IIB}_{xy}g^{\rm IIB}_{xz})
\over g^{\rm IIB}_{xx}(g^{\rm IIB}_{yy}g^{\rm IIB}_{xx}- g^{\rm IIB}_{xy}\ ^2 + b^{\rm IIB}_{xy}\ ^2)},
\end{eqnarray}

\begin{eqnarray}
\label{Bxy}
{\cal B}_{xy} & = & {-b^{\rm IIB}_{xy} \over g^{\rm IIB}_{yy}g^{\rm IIB}_{xx}- g^{\rm IIB}_{xy}\ ^2 + b^{\rm IIB}_{xy}\ ^2}.
\end{eqnarray}}

{\scriptsize
\begin{eqnarray}
\label{GIIA}
& &g^{\rm IIA}_{00}=-\frac{\left(r^4-{r_h}^4\right) \left(3 {g_s} M^2 \log (r) (-2 {g_s} {N_f} \log (\alpha_{\theta 1} \alpha_{\theta 2})+{g_s} {N_f} \log(N)-6 {g_s} {N_f}+{g_s} {N_f} \log (16)-8 \pi )-36 {g_s}^2 M^2 {N_f} \log ^2(r)+32 \pi ^2 N\right)}{64 \pi ^{5/2}\sqrt{{g_s}} N^{3/2} r^2}\nonumber\\
& &g^{\rm IIA}_{11}=\frac{r^2 \left(3 {g_s} M^2 \log (r) (-2 {g_s} {N_f} \log (\alpha_{\theta 1} \alpha_{\theta 2})+{g_s} {N_f} {\log N}-6 {g_s}{N_f}+{g_s} {N_f} \log (16)-8 \pi )-36 {g_s}^2 M^2 {N_f} \log ^2(r)+32 \pi ^2 N\right)}{64 \pi ^{5/2} \sqrt{{g_s}}
N^{3/2}}\nonumber\\
& &g^{\rm IIA}_{22}=\frac{r^2 \left(3 {g_s} M^2 \log (r) (-2 {g_s} {N_f} \log (\alpha_{\theta 1} \alpha_{\theta 2})+{g_s} {N_f} {\log N}-6 {g_s}
   {N_f}+{g_s} {N_f} \log (16)-8 \pi )-36 {g_s}^2 M^2 {N_f} \log ^2(r)+32 \pi ^2 N\right)}{64 \pi ^{5/2} \sqrt{{g_s}}
   N^{3/2}}\nonumber\\
& &g^{\rm IIA}_{33}=\frac{r^2 \left(3 {g_s} M^2 \log (r) (-2 {g_s} {N_f} \log (\alpha_{\theta 1} \alpha_{\theta 2})+{g_s} {N_f} {\log N}-6 {g_s}
   {N_f}+{g_s} {N_f} \log (16)-8 \pi )-36 {g_s}^2 M^2 {N_f} \log ^2(r)+32 \pi ^2 N\right)}{64 \pi ^{5/2} \sqrt{{g_s}}
   N^{3/2}}\nonumber\\
& &g^{\rm IIA}_{rr}=\frac{\sqrt{{g_s}} r^2 \left(6 a^2+r^2\right) \left(3 {g_s} M^2 \log (r) (2 {g_s} {N_f} \log (\alpha_{\theta 1} \alpha_{\theta 2})-{g_s}
   {N_f} {\log N}+6 {g_s} {N_f}-2 {g_s} {N_f} \log (4)+8 \pi )+36 {g_s}^2 M^2 {N_f} \log ^2(r)+32 \pi ^2
   N\right)}{16 \pi ^{3/2} \sqrt{N} \left(9 a^2+r^2\right) \left(r^4-{r_h}^4\right)}\nonumber\\
 & & g^{\rm IIA}_{\tilde{y}\tilde{y}}=-\frac{2 \left(9 \sqrt{2} \sqrt[6]{3} \alpha  N^{4/5}-2\ 3^{2/3} N\right)}{27 \alpha_{\theta_1}^22 \alpha_{\theta_2}^2}\nonumber\\
& &g^{\rm IIA}_{\theta_{2}\theta_{2}}=\frac{\sqrt{\pi } \sqrt{{g_s}} \sqrt{N} \left(\alpha_{\theta_1}^22 \sqrt[5]{N}+\alpha_{\theta_2}^2\right)}{\sqrt[3]{3} \alpha_{\theta_2}^2}.
\end{eqnarray}}

\section{Vector Meson Embedding Functions appearing in the DBI action for $D6$-branes}

In this appendix we give the embedding functions $\Sigma_{0,1}(r;g_s,N_f,M,N)$ relevant to the embedding of $D6$-branes in the delocalized SYZ type IIA mirror of the type IIB construct of \cite{metrics} that appear in (\ref{DBI-det_i}) in Section {\bf 3.2}.
The same are given as under:
{\footnotesize
\begin{eqnarray}
\label{DBI-det_ii}
& &  \Sigma_0(r;g_s,N_f,M,N) \equiv -\frac{1}{97844723712 \pi
   ^{11} \alpha_{\theta_1}^8 \alpha_{\theta_2}^4 {g_s} N^{26/5} \left(9 a^2+r^2\right)}\Biggl\{r^6 \left(6 a^2+r^2\right)\nonumber\\
   & & \times \Biggl[81 \sqrt{2} 3^{5/6} \alpha_{\theta_1}^9 \sqrt[5]{N}-54 \sqrt[3]{3} \alpha_{\theta_1}^8 N^{2/5}+81 \sqrt{2} 3^{5/6} \alpha_{\theta_1}^7
   \alpha_{\theta_2}^2\nonumber\\
   & & -54 \left(2+\sqrt[3]{3}\right) \alpha_{\theta_1}^6 \alpha_{\theta_2}^2 \sqrt[5]{N}+24 \sqrt{6} \alpha_{\theta_1}^5 \alpha_{\theta_2}^2 N^{2/5}-8 \alpha_{\theta_1}^4 \alpha_{\theta_2}^2
   N^{3/5}\nonumber\\
   & & -24 \sqrt{6} \alpha_{\theta_1}^3 \alpha_{\theta_2}^4 \sqrt[5]{N}+16 \alpha_{\theta_1}^2 \alpha_{\theta_2}^4 N^{2/5}-8 \alpha_{\theta_2}^6 \sqrt[5]{N}\Biggr]\nonumber\\
   & & \times \left(3 {g_s} M^2
   \log (r) (-2 {g_s} {N_f} \log (\alpha_{\theta_1}  \alpha_{\theta_2})+{g_s} {N_f} {\log N}-6 {g_s} {N_f}+{g_s} {N_f} \log (16)-8 \pi )-36
   {g_s}^2 M^2 {N_f} \log ^2(r)+32 \pi ^2 N\right)^4\nonumber\\
   & & \times \left(3 {g_s} M^2 \log (r) (2 {g_s} {N_f} \log (\alpha  \alpha_{\theta_2})-{g_s}
   {N_f} {\log N}+6 {g_s} {N_f}-2 {g_s} {N_f} \log (4)+8 \pi )+36 {g_s}^2 M^2 {N_f} \log ^2(r)+32 \pi ^2 N\right)\Biggr\}\nonumber\\
& &\Sigma_1(r;g_s,N_f,M,N) \equiv \frac{1}{165112971264 \pi ^{19/2} \alpha_{\theta_1}^{10} \alpha_{\theta_2}^6 {g_s}^{3/2} N^{49/10}}\Biggl\{r^4 \left(r^4-{r_h}^4\right)
\nonumber\\
& & \times \Bigg[-486 \sqrt{6} \alpha_{\theta_1}^{11} N+324 \alpha_{\theta_1}^{10} N^{6/5}+972 \sqrt{2} \sqrt[6]{3} \alpha_{\theta_1}^9 \alpha_{\theta_2}^2
   \sqrt[5]{N} \left(\left(-9-3 \sqrt[3]{3}+6 3^{2/3}\right) \alpha_{\theta_2}^2-\sqrt[3]{3} N^{3/5}\right)
 \nonumber\\
& &    +108 \alpha_{\theta_1}^8 \Biggl(27 \sqrt{2} \sqrt[6]{3}
   \left(3^{2/3}-3\right) \alpha_{\theta_2}^5 \sqrt[10]{N}+18 \left(\sqrt[3]{3}-1\right)^2 \alpha_{\theta_2}^4 N^{2/5}\nonumber\\
   & & +2 \left(3+3^{2/3}\right) \alpha_{\theta_2}^2
   N\Biggr)-9 \alpha_{\theta_1}^7 \Biggl(243 \sqrt{2} \sqrt[6]{3} \alpha_{\theta_2}^6-216 \sqrt[3]{3} \left(\sqrt[3]{3}-1\right) \alpha_{\theta_2}^5 N^{3/10}+54 \sqrt{6}
   \alpha_{\theta_2}^4 N^{3/5}\nonumber\\
   & & +16 \sqrt{2} \sqrt[6]{3} \alpha_{\theta_2}^2 N^{6/5}\Biggr)+2 \alpha_{\theta_1}^6 \Biggl(243 3^{2/3} \alpha_{\theta_2}^6 \sqrt[5]{N}+54 \left(3+2
   3^{2/3}\right) \alpha_{\theta_2}^4 N^{4/5}\nonumber\\
   & & +8 3^{2/3} \alpha_{\theta_2}^2 N^{7/5}\Biggr)-16 3^{2/3} \alpha_{\theta_1}^4 \alpha_{\theta_2}^4 N^{6/5}+144 \sqrt{2} \sqrt[6]{3} \alpha
   ^3 \alpha_{\theta_2}^6 N^{4/5} -16 3^{2/3} \alpha_{\theta_1}^2 \alpha_{\theta_2}^6 N+16 3^{2/3} \alpha_{\theta_2}^8 N^{4/5}\Biggr]\nonumber\\
   & & \times \left(3 {g_s} M^2 \log (r) (-2 {g_s}
   {N_f} \log (\alpha  \alpha_{\theta_2})+{g_s} {N_f} {\log N}-6 {g_s} {N_f}+{g_s} {N_f} \log (16)-8 \pi )-36 {g_s}^2 M^2 {N_f}
   \log ^2(r)+32 \pi ^2 N\right)^4\Biggr\}.\nonumber\\
   & &
\end{eqnarray}}

\section{Functions appearing in the vector meson EOMs'}

The functions $A_{0}(Z,g_{s},N_{f},N,M)$, $A_{1}(Z,g_{s},N_{f},N,M)$, $B_{0}(Z,g_{s},N_{f},N,M)$ and $B_{1}(Z,g_{s},N_{f},N,M)$ appearing in the equations of motion ({\ref{EOMs-i}}) and (\ref{EOMs-ii}) of vector meson modes $\alpha^{\left\{i\right\}}_{n}(Z)$ and $\alpha^{\left\{0\right\}}_{n}(Z)$ respectively in section {\bf 3.3} are given as:
{\scriptsize
\begin{eqnarray}
\label{coeff-vector-eom}
&&A_{0}=\alpha_n^{\left\{i\right\}}\ ^{\prime}(Z) \Biggl(\frac{-{g_s} {N_f} \left(e^{4 |Z|} (-2 {\log N}+6 |Z|+3)-2 {\log N}+6 |Z|-3\right)-6 {g_s} {N_f} \left(e^{4
   |Z|}+1\right) \log ({r_h})+8 \pi  \left(e^{4 |Z|}+1\right)}{\left(e^{4 |Z|}-1\right) ({g_s} {N_f} ({\log N}-3 |Z|)-3 {g_s}
   {N_f} \log ({r_h})+4 \pi )}\nonumber\\
   & &-\frac{1}{N^2 ({g_s} {N_f} ({\log N}-3. |Z|)-3. {g_s} {N_f} \log
   ({r_h})+12.5664)^2}\Biggl\{1.5 e^{-2 |Z|} \left(4. {g_s} M^2 \log ({r_h})+4. {g_s} M^2+0.6 N\right)^2
  \nonumber\\
 & &\times   \Biggl[{g_s}^2 {N_f}^2 \left(2. {\log N}^2-12. {\log N} |Z|-6. {\log N}+18. Z^2+18. |Z|+9.\right)+18. {g_s}^2 {N_f}^2
   \log ^2({r_h})\nonumber\\
   & &+{g_s} {N_f} \log ({r_h}) ({g_s} {N_f} (-12. {\log N}+36. |Z|+18.)-150.796)+{g_s} {N_f}
   (50.2655 {\log N}-150.796 |Z|-75.3982)\nonumber\\
   & & +315.827\Biggr]\Biggr\}\Biggr)\nonumber\\
 & & A_{1}= \frac{\tilde{m}^2  \left(e^{2 |Z|}-\frac{3. \left(4. {g_s} M^2 \log ({r_h})+4. {g_s} M^2+0.6
   N\right)^2}{N^2}\right)}{e^{4 |Z|}-1}\nonumber\\
 & &B_{0}=\frac{1}{2 ({g_s} {N_f} {\log N}-3 {g_s} {N_f} \log ({r_h})-3 {g_s}
   {N_f} |Z|+4 \pi )^2}\nonumber\\
   & &\times\Biggl\{ \Biggl(\frac{1}{N^2}\Biggl\{e^{-2 |Z|} \left(4. {g_s} M^2 \log ({r_h})+4. {g_s} M^2+0.6 N\right)^2 (9 {g_s}
   {N_f} (-{g_s} {N_f} {\log N}+3 {g_s} {N_f} \log ({r_h})+3 {g_s} {N_f} |Z|-4 \pi )\nonumber\\
   & &+(2 {g_s} {N_f} \log
   (N)-6 {g_s} {N_f} \log ({r_h})-6 {g_s} {N_f} |Z|-3 {g_s} {N_f}+8 \pi ) (-3 {g_s} {N_f} {\log N}+9 {g_s}
   {N_f} \log ({r_h})+9 {g_s} {N_f} |Z|+9 {g_s} {N_f}-12 \pi ))\Biggr\}\nonumber\\
   & & +2 (2 {g_s} {N_f} {\log N}-6 {g_s}
   {N_f} \log ({r_h})-6 {g_s} {N_f} |Z|-3 {g_s} {N_f}+8 \pi ) ({g_s} {N_f} {\log N}-3 {g_s} {N_f} \log
   ({r_h})-3 {g_s} {N_f} |Z|+4 \pi )\Biggr)\Biggr\}\nonumber\\
 & & B_{1}=\frac{\tilde{m}^2  \left({r_h}^2 e^{2 |Z|}-\frac{3. {r_h}^2 \left(4. {g_s} M^2 \log
   ({r_h})+4. {g_s} M^2+0.6 N\right)^2}{N^2}\right)}{{r_h}^2 \left(e^{4 |Z|}-1\right)}.
\end{eqnarray}}

\section{Functions Appearing in the Vector and Scalar Mesons' Actions }

\subsection{Vector Meson Action Functions}
The functions ${\cal V}_{1,2}$ appearing in equation (\ref{Ftildesq-i}) in Section {\bf 3.3} in the context of the vector meson action obtained by substituting a KK ansatz (\ref{AZ+Amu}) into the DBI action for $N_f=2$ $D6$-branes of (\ref{DBI action}) are given as under:
\begin{eqnarray*}
\label{V1}
& &  {\cal V}_1(Z) = e^{-\Phi^{IIA}}\sqrt{h}g^{ZZ}\sqrt{{\rm det}_{\theta_2,\tilde{y}}\left(i^*(g+B)\right)}\sqrt{{\rm det}_{\mathbb{R}^{1,3},Z}(i^*g)},\nonumber\\
& &  = \frac{1}{36 \sqrt{2} \pi ^{3/2} \alpha_{\theta_1}^2 \alpha_{\theta_2} {g_s}^{3/2}}\Biggl\{\sqrt[5]{N} e^{-4 |Z|} \left(e^{4 |Z|}-1\right)\nonumber\\
& &  \times\Biggl(3 a^2 {g_s} {N_f} \log N-9 a^2 {g_s} {N_f} \log
   ({r_h})-9 a^2 {g_s} {N_f} |Z|-9 a^2 {g_s} {N_f}+12 \pi  a^2+2 {g_s} {N_f} {r_h}^2 e^{2 |Z|} \log N
   \nonumber\\
& &    -6 {g_s} {N_f} {r_h}^2 e^{2 |Z|} |Z|-6 {g_s} {N_f} {r_h}^2 e^{2 |Z|} \log ({r_h})+8 \pi
   {r_h}^2 e^{2 |Z|}\Biggr)\Biggr\}\nonumber\\
   & & +\frac{1}{576 \pi ^{3/2} \alpha_{\theta_1}^4 \alpha_{\theta_2}^3
   {g_s}^{3/2}}\Biggl\{e^{-4 |Z|} \left(e^{4 |Z|}-1\right) \left(27
   \sqrt{2} \sqrt[3]{3} \alpha_{\theta_1}^6-24 \sqrt{3} \alpha_{\theta_1}^3 \alpha_{\theta_2}^2-8 \sqrt{2} \alpha_{\theta_2}^4\right)\nonumber\\
  & & \times  \Biggl(-3 a^2 {g_s} {N_f} \log
   \left(\frac{1}{N}\right)-9 a^2 {g_s} {N_f} \log ({r_h})-9 a^2 {g_s} {N_f} |Z|-9 a^2 {g_s} {N_f}+12 \pi  a^2-2
   {g_s} {N_f} {r_h}^2 e^{2 |Z|} \log \left(\frac{1}{N}\right)\nonumber\\
   & & -6 {g_s} {N_f} {r_h}^2 e^{2 |Z|} |Z|-6 {g_s} {N_f}
   {r_h}^2 e^{2 |Z|} \log ({r_h})+8 \pi  {r_h}^2 e^{2 |Z|}\Biggr)\Biggr\} + {\cal O}\left(\frac{1}{N^{6/5}}\right),
   \end{eqnarray*}
   \begin{eqnarray}
   \label{v1v2}
& & {\cal V}_2(Z) =  e^{-\Phi^{IIA}}\frac{h}{2}\sqrt{{\rm det}_{\theta_2,\tilde{y}}\left(i^*(g+B)\right)}\sqrt{{\rm det}_{\mathbb{R}^{1,3},|Z|}(i^*g)}\nonumber\\
& & =\frac{1}{18 \sqrt{2 \pi
   } \alpha_{\theta_1}^2 \alpha_{\theta_2} \sqrt{{g_s}} {r_h}^2}\Biggl\{N^{6/5} e^{-2 |Z|} \Biggl(3 a^2 {g_s} {N_f} \log \left(\frac{1}{N}\right)+9 a^2 {g_s} {N_f} \log ({r_h})+9 a^2 {g_s}
   {N_f} |Z|\nonumber\\
   & & -9 a^2 {g_s} {N_f}-12 \pi  a^2+2 {g_s} {N_f} {r_h}^2 e^{2 |Z|} \log N-6 {g_s}
   {N_f} {r_h}^2 e^{2 |Z|} |Z|-6 {g_s} {N_f} {r_h}^2 e^{2 |Z|} \log ({r_h})+8 \pi  {r_h}^2 e^{2 |Z|}\Biggr)\Biggr\}\nonumber\\
   & & +\frac{1}{288 \sqrt{\pi } \alpha_{\theta_1}^4 \alpha_{\theta_2}^3 \sqrt{{g_s}}
   {r_h}^2}\Biggl\{N e^{-2 |Z|} \Biggl(27 \sqrt{2} \sqrt[3]{3} \alpha_{\theta_1}^6-24 \sqrt{3} \alpha_{\theta_1}^3
   \alpha_{\theta_2}^2-8 \sqrt{2} \alpha_{\theta_2}^4\Biggr)\nonumber\\
   & & \times \Biggl(-3 a^2 {g_s} {N_f} \log N+9 a^2 {g_s} {N_f}
   \log ({r_h})+9 a^2 {g_s} {N_f} |Z|-9 a^2 {g_s} {N_f}-12 \pi  a^2+2 {g_s} {N_f} {r_h}^2 e^{2 |Z|} \log
  N \nonumber\\
  & & - 6 {g_s} {N_f} {r_h}^2 e^{2 |Z|} |Z| - 6 {g_s} {N_f} {r_h}^2 e^{2 |Z|} \log ({r_h})+8 \pi
   {r_h}^2 e^{2 |Z|}\Biggr)\Biggr\}  + {\cal O}\left(\frac{1}{N^{6/5}}\right).
\end{eqnarray}

\subsection{Scalar Meson Action Functions}

The scalar meson functions ${\cal S}_{1,2,3}$ appearing  in the DBI action (\ref{DBI-scalar}) for $N_f$ $D6$-branes and (\ref{mass-term-identification+EOM}) are given as under:
{\scriptsize
\begin{eqnarray}
\label{S123}
& &{\cal S}_1 = \frac{1}{1296 \sqrt{2} \sqrt[3]{3}
   \pi  \alpha_{\theta_1}^24 \alpha_{\theta_2}^5 {g_s} Z^2}\Biggl\{{\cal C}^2 N^{11/10} {r_h}^2
    \left(27 \sqrt[3]{3} \alpha_{\theta_1}^26-12 \sqrt{6} \alpha_{\theta_1}^23 \alpha_{\theta_2}^2+8 \alpha_{\theta_1}^22 \alpha_{\theta_2}^2 \sqrt[5]{N}-8
   \alpha_{\theta_2}^4\right)\nonumber\\
   & &\times \Bigg[-{g_s} {N_f} \log (N) \left(3 \left(\frac{4 {g_s} M^2 (\log ({r_h})+1)}{N}+0.6\right)^2-2 e^{2
   |Z|}\right)+3 {g_s} {N_f} (\log ({r_h})+|Z|) \left(3 \left(\frac{4 {g_s} M^2 (\log ({r_h})+1)}{N}+0.6\right)^2-2 e^{2 |Z|}\right)\nonumber\\
   & & -3
   (3 {g_s} {N_f}+4 \pi ) \left(\frac{4 {g_s} M^2 (\log ({r_h})+1)}{N}+0.6\right)^2+8 \pi  e^{2 |Z|}\Biggr]\Biggr\},\nonumber\\
   & &{\cal S}_2 = \frac{1}{15552 \sqrt{2}
   \pi ^2 \alpha_{\theta_1}^22 \alpha_{\theta_2}^5 {g_s}^2 Z^2}\Biggl\{{\cal C}^2 \sqrt[10]{N} {r_h}^4 e^{-2 |Z|} \left(e^{4 |Z|}-1\right) \left(81 \alpha_{\theta_1}^24-36 \sqrt{2} \sqrt[6]{3} \alpha  \alpha_{\theta_2}^2+8 3^{2/3}
   \alpha_{\theta_2}^2 \sqrt[5]{N}\right)\nonumber\\
   & &\times \Biggl[{g_s} {N_f} \log (N) \left(3 \left(\frac{4 {g_s} M^2 (\log ({r_h})+1)}{N}+0.6\right)^2+2
   e^{2 |Z|}\right)-3 {g_s} {N_f} (\log ({r_h})+|Z|) \left(3 \left(\frac{4 {g_s} M^2 (\log ({r_h})+1)}{N}+0.6\right)^2+2 e^{2
   |Z|}\right)\nonumber\\
   & &+(12 \pi -9 {g_s} {N_f}) \left(\frac{4 {g_s} M^2 (\log ({r_h})+1)}{N}+0.6\right)^2+8 \pi  e^{2 |Z|}\Biggr]\Biggr\}\nonumber\\
   & &{\cal S}_3 = \frac{1}{1944 \sqrt{2} \pi ^2 \alpha_{\theta_1}^22 \alpha_{\theta_2}^3 {g_s}^2
   \left(e^{4 |Z|}-1\right) Z^4}\Biggl\{\sqrt[5]{N} {r_h}^4 e^{-2 |Z|} \Biggl({g_s} {N_f} \log (N)\nonumber\\
   & &\times \Biggl(3 \left(3^{2/3} {\cal C}^2 \sqrt[10]{N} \left(e^{4 |Z|}-1\right)^2-27
   \sqrt{\pi } \alpha_{\theta_2}^2 \sqrt{{g_s}} e^{4 |Z|} Z^2 \left(-2 |Z|+e^{4 |Z|} (6 |Z|+1)-1\right)\right) \nonumber\\
   & &\times\left(\frac{4 {g_s} M^2 (\log
   ({r_h})+1)}{N}+0.6\right)^2+2 e^{2 |Z|} \left(27 \sqrt{\pi } \alpha_{\theta_2}^2 \sqrt{{g_s}} e^{4 |Z|} Z^2 \left(e^{4 |Z|} (4 |Z|+1)-1\right)+3^{2/3}
   {\cal C}^2 \sqrt[10]{N} \left(e^{4 |Z|}-1\right)^2\right)\Biggr)\nonumber\\
   & &-3 {g_s} {N_f} (\log ({r_h})+|Z|) \Biggl[3 \left(3^{2/3} {\cal C}^2 \sqrt[10]{N}
   \left(e^{4 |Z|}-1\right)^2-27 \sqrt{\pi } \alpha_{\theta_2}^2 \sqrt{{g_s}} e^{4 |Z|} Z^2 \left(-2 |Z|+e^{4 |Z|} (6 |Z|+1)-1\right)\right)\nonumber\\
   & &\times \left(\frac{4
   {g_s} M^2 (\log ({r_h})+1)}{N}+0.6\right)^2\nonumber\\
   & &+2 e^{2 |Z|} \left(27 \sqrt{\pi } \alpha_{\theta_2}^2 \sqrt{{g_s}} e^{4 |Z|} Z^2 \left(e^{4 |Z|} (4
   |Z|+1)-1\right)+3^{2/3} {\cal C}^2 \sqrt[10]{N} \left(e^{4 |Z|}-1\right)^2\right)\Biggr]\nonumber\\
   & &+3 \left(\frac{4 {g_s} M^2 (\log ({r_h})+1)}{N}+0.6\right)^2
   \Biggl[(3^{2/3} {\cal C}^2 \sqrt[10]{N} \left(e^{4 |Z|}-1\right)^2 (4 \pi -3 {g_s} {N_f})-27 \sqrt{\pi } \alpha_{\theta_2}^2 \sqrt{{g_s}} e^{4 |Z|} |Z|\nonumber\\
& &\times   \left(3 {g_s} {N_f} |Z| \left(3 |Z|+e^{4 |Z|} (|Z|+1)-1\right)+4 \pi  |Z| \left(-2 |Z|+e^{4 |Z|} (6 |Z|+1)-1\right)\right)\Biggr]+2 \sqrt{\pi } e^{2 |Z|}\nonumber\\
& &\times   \left(27 \alpha_{\theta_2}^2 \sqrt{{g_s}} e^{4 |Z|} Z^2 \left(4 \pi  \left(e^{4 |Z|} (4 |Z|+1)-1\right)-3 {g_s} {N_f} \left(e^{4 |Z|}-1\right)
   |Z|\right)+4 3^{2/3} \sqrt{\pi } {\cal C}^2 \sqrt[10]{N} \left(e^{4 |Z|}-1\right)^2\right)\Biggr)\Biggr\}.
\end{eqnarray}}

\subsection{Functions appearing in the Scalar Meson EOM}
Coeffiecients appearing in the (\ref{G_EOM}) are given as:
{\tiny
\begin{eqnarray}
\label{Scalar_EOM_Coefficients}
& &\hskip -0.8in N_0=\frac{1}{(2 {g_s} {N_f} {\log N}-6 {g_s} {N_f} (\log ({r_h})+|Z|)+8 \pi )^2}\nonumber\\
& & \hskip -0.8in\times\Biggl\{ \Biggl(\frac{2 (2 {g_s} {N_f}
   {\log N}-6 {g_s} {N_f} (\log ({r_h})+|Z|)+8 \pi ) \left(4 {g_s} {N_f} e^{4 |Z|} {\log N}-12 {g_s} {N_f} e^{4 |Z|} (\log
   ({r_h})+|Z|)-3 {g_s} {N_f} e^{4 |Z|}+3 {g_s} {N_f}+16 \pi  e^{4 |Z|}\right)}{e^{4 |Z|}-1}
 \nonumber\\
& & \hskip -0.8in   -3 e^{-2 |Z|} \left(\frac{4 {g_s} M^2 \log
   ({r_h})}{N}+\frac{4 {g_s} M^2}{N}+0.6\right)^2\nonumber\\
   & & \hskip -0.8in\times \Biggl[-24 {g_s}^2 {N_f}^2 {\log N} (\log ({r_h})+|Z|)+4 {g_s}^2 {N_f}^2
   \log ^2(N)-12 {g_s}^2 {N_f}^2 {\log N}+36 {g_s}^2 {N_f}^2 (\log ({r_h})+|Z|)^2+36 {g_s}^2 {N_f}^2 (\log
   ({r_h})+|Z|)\nonumber\\
   & & \hskip -0.8in+18 {g_s}^2 {N_f}^2+32 \pi  {g_s} {N_f} {\log N}-96 \pi  {g_s} {N_f} (\log ({r_h})+|Z|)-48 \pi  {g_s}
   {N_f}+64 \pi ^2\Biggr]\Biggr)\Biggr\}\nonumber\\
   & & \hskip -0.8in  N_1 =\frac{\tilde{m}^2  \left(\alpha_{\theta_1}^22 \sqrt[5]{N}-\alpha_{\theta_2}^2\right) \left(e^{2 |Z|}-\frac{3. \left(4. {g_s} M^2 \log ({r_h})+4.
   {g_s} M^2+0.6 N\right)^2}{N^2}\right)}{\alpha_{\theta_1}^22 \sqrt[5]{N} \left(e^{4 |Z|}-1\right)}
   \end{eqnarray}} 
\chapter{}
\section{M theory Metric Components}

Near  $\theta_{1}=\alpha_{\theta_{1}}N^{\frac{-1}{5}}$, $\theta_{2}=\alpha_{\theta_{2}}N^{\frac{-3}{10}}$, $\phi_{1,2}=0/2\pi$ and $\psi=0/4\pi$, defining the local $T^3(x,y,z)$ coordinates as:
{\footnotesize
	\begin{eqnarray}
	\label{}
	&& x=\sqrt{h_2}4^{1/4}\pi^{1/4}g_{s}^{1/4}N^{1/20}\alpha_{\theta_{1}}\phi_{1},\ \ y=\sqrt{h_4}4^{1/4}\pi^{1/4}g_{s}^{1/4}N^{1/20}\alpha_{\theta_{2}}\phi_{2},\ \ z=\sqrt{h_1}4^{1/4}\pi^{1/4}g_{s}^{1/4}N^{1/20}\psi,\nonumber\\
\end{eqnarray}}
$h_{1,2,4}$ are defined in \cite{metrics}, and defining:
{
\begin{eqnarray}
\label{def-A}
& & f \equiv 1 - \frac{r_h^4}{r^4}\nonumber\\
&&A(r)\equiv\frac{1}{4} \left(\frac{3}{\pi }\right)^{2/3} \left(-{N_f} \log \left(9 b^2 r^4 {r_h}^2+r^6\right)+\frac{8 \pi }{{g_s}}-4 N_f \log\left(\frac{\alpha_{\theta_1}\alpha_{\theta_2}}{4\sqrt{\log N}}\right)\right){}^{2/3}
	\nonumber\\
	& &  \times  \left(1-\frac{32 \pi  b {g_s} M^2 {N_f} {r_h}^2 \gamma(1 + \log r_h)}{N \left(9 b^2 {r_h}^2+r^2\right)
		\left(-{N_f} \log \left(9 b^2 r^4 {r_h}^2+r^6\right)+\frac{8 \pi }{{g_s}}-4 N_f \log\left(\frac{\alpha_{\theta_1}\alpha_{\theta_2}}{4\sqrt{\log N}}\right)\right)}\right),\nonumber\\
\end{eqnarray}}
 the M-theory metric components used in Sections {\bf 4.2} - {\bf 4.5}, are given by:
 {
	\begin{eqnarray*}
	&&g^{\cal M}_{00}=-f(r)\frac{A(r)}{\sqrt{h}}\nonumber\\
	&&g^{\cal M}_{ii}=\frac{A(r)}{\sqrt{h}}\ \  i=1,2,3\nonumber\\
	&& g^{\cal M}_{rr}=\sqrt{h}\frac{A(r)}{f(r)}\nonumber\\
	&&g^{\cal M}_{\theta_{1,2}\theta_{1,2}}=0,\ \ g^{\cal M}_{\theta_{1}\theta_{2}}=0,\ \ g^{\cal M}_{xr}=0,\ \ g^{\cal M}_{r\theta_{1,2}}=0\nonumber\\
	&&g^{\cal M}_{11x}=0,\ \ g^{\cal M}_{11r}=0,\ \ g^{\cal M}_{11\theta_{1,2}}=0\nonumber\\
	&&g^{\cal M}_{11,11}=\frac{16 \pi ^{4/3} \left(\frac{64 \pi  b {g_s} M^2 {N_f} {r_h}^2 \gamma(1 + \log r_h)}{N \left(9 b^2 {r_h}^2+r^2\right)
			\left(-{N_f} \log \left(9 b^2 r^4 {r_h}^2+r^6\right)+\frac{8 \pi }{{g_s}}-4 N_f \log\left(\frac{\alpha_{\theta_1}\alpha_{\theta_2}}{4\sqrt{\log N}}\right)\right)}+1\right)}{3 \sqrt[3]{3} \left(-{N_f} \log \left(9 b^2 r^4
		{r_h}^2+r^6\right)+\frac{8 \pi }{{g_s}}-4 N_f \log\left(\frac{\alpha_{\theta_1}\alpha_{\theta_2}}{4\sqrt{\log N}}\right)\right){}^{4/3}}\nonumber\\
		&&g^{\cal M}_{x\theta_{1}}=A(r)\frac{1}{972 \pi ^{5/4} r^2 \alpha _{\theta _1}^3 \alpha _{\theta _2}^2}\Biggl\{{g_s}^{3/4} M N^{7/20} \left(-243 \sqrt{3} \alpha _{\theta _1}^3+4 \sqrt{2} \alpha _{\theta _2}^2+81 \sqrt{2} \sqrt[5]{N} \alpha _{\theta
		_1}^2\right)\nonumber\\
	& & \times \Biggl({g_s} {N_f} \left(3 a^2-r^2\right) \log (N) (2 \log (r)+1)+\log (r) \Biggl(4 {g_s} {N_f} \left(r^2-3 a^2\right) \log
	\left(\frac{1}{4} \alpha _{\theta _1} \alpha _{\theta _2}\right)\nonumber\\
& & -24 \pi  a^2+r^2 (8 \pi -3 {g_s} {N_f})\Biggr)
	\nonumber\\
	& & +2 {g_s} {N_f}
	\left(r^2-3 a^2\right) \log \left(\frac{1}{4} \alpha _{\theta _1} \alpha _{\theta _2}\right)+18 {g_s} {N_f} \left(r^2-3 a^2 (6 r+1)\right) \log
	^2(r)\Biggr)\Biggr\}\nonumber\\
&&g^{\cal M}_{x\theta_{2}}=-A(r)\frac{{g_s}^{7/4} M N^{9/20} {N_f} \log (r) \left(36 a^2 \log (r)+r\right) \left(486 \sqrt{6} \alpha _{\theta _1}^3+11 \alpha _{\theta _2}^2-324
		\sqrt[5]{N} \alpha _{\theta _1}^2\right)}{972 \sqrt{2} \pi ^{5/4} r \alpha _{\theta _1}^2 \alpha _{\theta _2}^3}\nonumber\\
	&&g^{\cal M}_{y\theta_{1}}=A(r)\frac{1}{72 \sqrt{3} \pi ^{5/4} N^{7/20} r^2 \alpha _{\theta _1}
		\alpha _{\theta _2}}\Biggl\{{g_s}^{3/4} M \left(67 \alpha _{\theta _2}^2+81 \sqrt[5]{N} \alpha _{\theta _1}^2\right)\nonumber\\
& & \times \Biggl[{g_s} {N_f} \left(3 a^2-r^2\right) \log
	(N) (2 \log (r)+1)+\nonumber\\
	& & \log (r) \left(4 {g_s} {N_f} \left(r^2-3 a^2\right) \log \left(\frac{1}{4} \alpha _{\theta _1} \alpha _{\theta _2}\right)-24
	\pi  a^2+r^2 (8 \pi -3 {g_s} {N_f})\right)\nonumber\\
	& & +2 {g_s} {N_f} \left(r^2-3 a^2\right) \log \left(\frac{1}{4} \alpha _{\theta _1} \alpha
	_{\theta _2}\right)+18 {g_s} {N_f} \left(r^2-3 a^2 (6 r+1)\right) \log ^2(r)\Biggr]\Biggr\}\nonumber\\
&&g^{\cal M}_{y\theta_{2}}=A(r)\frac{\sqrt{2} \sqrt[4]{\pi } \sqrt[4]{{g_s}} \sqrt[4]{N} \alpha _{\theta _2} \left(3 \sqrt[10]{N} \alpha _{\theta _1}-7 {h5} \alpha _{\theta
			_2}\right)}{27 \alpha _{\theta _1}^3}\nonumber\\
\end{eqnarray*}
\begin{eqnarray}
\label{M-th components}
			&&g^{\cal M}_{z\theta_{1}}=-A(r)\frac{1}{324 \sqrt{2} \pi ^{5/4} \sqrt[20]{N} r^2 \alpha _{\theta
			_1} \alpha _{\theta _2}^2}\Biggl\{{g_s}^{3/4} M \left(49 \alpha _{\theta _2}^2+81 \sqrt[5]{N} \alpha _{\theta _1}^2\right)\nonumber\\
& & \times \Biggl[{g_s} {N_f} \left(3 a^2-r^2\right) \log
	(N) (2 \log (r)+1)\nonumber\\
	& & +\log (r) \left(4 {g_s} {N_f} \left(r^2-3 a^2\right) \log \left(\frac{1}{4} \alpha _{\theta _1} \alpha _{\theta _2}\right)-24
	\pi  a^2+r^2 (8 \pi -3 {g_s} {N_f})\right)\nonumber\\
	& & +2 {g_s} {N_f} \left(r^2-3 a^2\right) \log \left(\frac{1}{4} \alpha _{\theta _1} \alpha
	_{\theta _2}\right)+18 {g_s} {N_f} \left(r^2-3 a^2 (6 r+1)\right) \log ^2(r)\Biggr]\Biggr\}\nonumber\\
	&&g^{\cal M}_{z\theta_{2}}=-A(r)\frac{{g_s}^{7/4} M {N_f} \log (r) \left(36 a^2 \log (r)+r\right) \left(324 \sqrt[4]{N} \alpha _{\theta _1}^2+169 \sqrt[20]{N} \alpha _{\theta
			_2}^2\right)}{648 \sqrt{2} \pi ^{5/4} r \alpha _{\theta _2}^3}\nonumber\\
	&&g^{\cal M}_{xy}=A(r)\Bigg\{\frac{2 \sqrt{\frac{2}{3}} N^{7/10}}{9 \alpha _{\theta _1}^2 \alpha _{\theta _2}}-\frac{\sqrt{\frac{2}{3}} \sqrt{N} \left(243 \sqrt{6} \alpha _{\theta
			_1}^3+118 \alpha _{\theta _2}^2\right)}{729 \alpha _{\theta _1}^4 \alpha _{\theta _2}}\Bigg\}\nonumber\\
	&&g^{\cal M}_{xz}=-A(r)\frac{2 N^{4/5} \left(-243 \sqrt{6} \alpha _{\theta _1}^3+8 \alpha _{\theta _2}^2+162 \sqrt[5]{N} \alpha _{\theta _1}^2\right)}{6561 \alpha _{\theta
			_1}^4 \alpha _{\theta _2}^2}\nonumber\\
	&&g^{\cal M}_{yz}=A(r)\Bigg\{\frac{14 \sqrt{\frac{2}{3}} \sqrt[10]{N} \alpha _{\theta _2}}{243 \alpha _{\theta _1}^2}-\frac{\sqrt{\frac{2}{3}} N^{3/10}}{3 \alpha _{\theta _2}}\Bigg\}
	\end{eqnarray}}

\section{Glueballs equation of motion from M-theory metric perturbation}

The explicit expressions for EOM in Section {\bf 4.2} derived from 11-D action

\noindent$\bullet\delta {\rm R[t,t]}$\\
{\small

\begin{eqnarray}
\label{EOM11}
& & {q_1}''(r)+{q_1}'(r) \Biggl(\frac{2 \left(6 a^2 \log (r)+r^2\right)}{r (2
	\log (r)+1) \left(r^2-3 a^2\right)}\nonumber\\
& &  -\frac{1}{52488 \pi ^{3/2} r^2 (2 \log (r)+1) \alpha _{\theta _1}^4 \alpha _{\theta _2}^3 \left(r^2-a^2\right)}\Biggl\{\left(\frac{1}{N}\right)^{2/5} \Biggl[38416 \pi ^{3/2} r \alpha _{\theta _2}^7
\left(6 a^2 \log (r)+r^2\right)\nonumber\\
& & +177147 \sqrt{6} {g_s}^{3/2} M {N_f} \alpha _{\theta _1}^8 \Biggl(12 a^2 \log ^2(r) \left(-27 a^2+15 r^2+r\right)-3 a^2 r +72 \log ^3(r) \left(9 a^4 -a^2 r^2
\right)\nonumber\\
& &  +\log (r) \left(-216 a^4+72 a^2 r^2 -6 a^2 r +4
r^3\right)+r^3\Biggr)\Biggr]\Biggr\} -\frac{3 a^2 r}{54 a^4+15 a^2 r^2 +r^4}+\frac{4 {r_h}^4}{r^5-r
	{r_h}^4}+\frac{5}{r}\Biggr)\nonumber\\
& &  +{q_1}(r) \Biggl(\frac{4 \pi  {g_s} (K^1)\ ^2 N \left(6 a^2+r^2\right)}{r^4 \left(9 a^2
	+r^2\right)\left(1-\frac{r_h^4}{r^4}\right)}  -\frac{3 {g_s}^3 (K^1)\ ^2 \log N  M^2 {N_f} \log (r) \left(6 a^2+r^2\right)}{4
	\pi  (r^4-{r_h}^4) \left(9 a^2+r^2\right)}\Biggr) =0.
\end{eqnarray}
}
\noindent$\bullet\delta {\rm R[x^1,x^1]}$\\
{\small
\begin{eqnarray*}
& &  {q_5}''(r)+\left(\frac{100 {g_s} N \pi  \left(r^2+6 a^2 \right)}{r^4 \left(r^2+9 a^2 \right)\left(1-\frac{{r_h}^4}{r^4}\right)}-\frac{75 {g_s}^3
	\log N  M^2 {N_f} \left(r^2+6 a^2 \right) \log (r)}{4 \pi  \left(r^4-{r_h}^4\right) \left(r^2+9 a^2 \right)}\right)
{q_1}(r)\nonumber\\\\
& &  +\left(\frac{75 {g_s}^3 \log N  M^2 {N_f} \left(r^2+6 a^2 \right) \log (r)}{2 \pi  \left(r^4-{r_h}^4\right)
	\left(r^2+9 a^2\right)}-\frac{200 {g_s} N \pi  \left(r^2+6 a^2\right)}{\left(r^4-{r_h}^4\right) \left(r^2+9 a^2\right)}\right) {q_4}(r)\nonumber\\\\
& &  +\Biggl(-\frac{12 \sqrt{6} a^4 {g_s}^{3/2} M^3 \sqrt[5]{\frac{1}{N}} \left(r^2+6 a^2\right) \left(\frac{68260644 \left(54 a^2+5\right) \left({r_h}^4-10000\right) \log (10)}{\left(100-3 a^2
		\right)^4}-\frac{30876125 \left(12 a^2 +1\right) \left({r_h}^4-6561\right) \log (9)}{9 \left(a^2
	-27\right)^4}\right) }{5 \log N ^5 M_g^2 \pi ^{3/2} r^2 \left(r^2+9 a^2 \right)\left(1-\frac{{r_h}^4}{r^4}\right)}
	\nonumber\\
	& & -\frac{1}{r^2 \left(r^2-3 a^2\right)
		\left(r^2+6 a^2 \right) \left(r^2+9 a^2 \right) \left(r^4-{r_h}^4\right) (2 \log (r)+1)}\nonumber\\
	& &  \times\Biggl\{2 \Biggl[r^2 \left(648 a^6 r^2-9 a^4 \left(17 r^4-13 {r_h}^4\right)+a^2 \left(27 r^2 {r_h}^4-75 r^6\right)-6 r^8+2 r^4 {r_h}^4\right)\nonumber\\
& &  +2 \log (r) \left(324
   a^6 \left(r^4+{r_h}^4\right)+a^4 \left(99 r^2 {r_h}^4-135 r^6\right)+a^2 \left(3 r^4 {r_h}^4-51 r^8\right)-4 r^{10}\right)\Biggr]\Biggr\}\Biggr)
{q_5}(r)\nonumber\\
& &  +{q_3}(r) \Biggl[\frac{8 {r_h}^4}{r^5 \left(1-\frac{{r_h}^4}{r^4}\right)}-\frac{\left(r^2+9 a^2\right) \left(\frac{2 r}{r^2+9 a^2}-\frac{2 r \left(r^2+6 a^2 \right)}{\left(r^2+9 a^2
		\right)^2}\right)}{r^2+6 a^2 } +\frac{4 \left(r^2+6 a^2  \log (r)\right)}{r \left(r^2-3 a^2
	\right) (2 \log (r)+1)}\nonumber\\
& &   -\frac{1}{13122 \sqrt{2} \pi ^{3/2} r^2 \left(r^2-3 a^2 \right) (2 \log (r)+1) \alpha _{\theta _1}^4 \alpha _{\theta
		_2}^3}\Biggl\{\left(\frac{1}{N}\right)^{2/5} \Biggl(177147 \sqrt{3} {g_s}^{3/2} M {N_f}
	\biggl(r^3-3 a^2  r+72 \left(9 a^4 -a^2 r^2 \right) \log ^3(r)\nonumber\\
	& &  +12 a^2 \left(15 r^2+r-27
	a^2\right) \log ^2(r)+\left(-216 a^4+72 a^2 r^2 -6 a^2 r +4 r^3\right) \log
	(r)\biggr) \alpha _{\theta _1}^8\nonumber\\
	& &  +19208 \sqrt{2} \pi ^{3/2} r \left(r^2+6 a^2  \log (r)\right) \alpha _{\theta
		_2}^7\Biggr)\Biggr\}+\frac{12}{r}\Biggr]+\left(\frac{75 {g_s}^2 \log N  {N_f} M^2}{64 (K^1)\ ^2 N \pi ^2 r}+\frac{25}{(K^1)\ ^2
	r}\right) {q_1}'(r)\nonumber\\
& &  +2 {q_3}'(r)+\Biggl(\frac{25 \left(-\frac{8 {r_h}^4}{r^5-r {r_h}^4}+\frac{6 a^2 r}{r^4+15 a^2 r^2+54 a^4 }-\frac{4 \left(r^2+6 a^2 \log (r)\right)}{r \left(r^2-3 a^2\right) (2 \log (r)+1)}-\frac{16}{r}\right)}{2 (K^1)\ ^2}\nonumber\\
	& &  +\frac{1}{26244 \sqrt{2} (K^1)\ ^2 \pi ^{3/2} r^2 \left(r^2-3 a^2 \right) (2 \log
			(r)+1) \alpha _{\theta _1}^4 \alpha _{\theta _2}^3}\Biggl\{25 \left(\frac{1}{N}\right)^{2/5}
		\nonumber\\
& &  \times		 \Biggl(177147
	\sqrt{3} {g_s}^{3/2} M {N_f} \Biggl[r^3-3 a^2  r+72 \left(9 a^4 -b^2 r^2 {r_h}^2\right) \log
	^3(r)+12 a^2  \left(15 r^2+r-27 a^2 \right) \log ^2(r)\nonumber\\
& & +\left(-216 a^4+72 a^2 r^2 -6
	a^2 r +4 r^3\right) \log (r)\Biggr] \alpha _{\theta _1}^8+19208 \sqrt{2} \pi ^{3/2} r \left(r^2+6 a^2  \log
	(r)\right) \alpha _{\theta _2}^7\Biggr)\Biggr\}\Biggr) {q_4}'(r)\nonumber\\
\end{eqnarray*}}
{\small
\begin{eqnarray}
\label{EOM22}
	& &  +\Biggl(\frac{1}{12} \left(\frac{48 {r_h}^4}{r^5
	\left(1-\frac{{r_h}^4}{r^4}\right)}-\frac{6 \left(r^2+9 b^2 {r_h}^2\right) \left(\frac{2 r}{r^2+9 b^2 {r_h}^2}-\frac{2
		r \left(r^2+6 b^2 {r_h}^2\right)}{\left(r^2+9 b^2 {r_h}^2\right)^2}\right)}{r^2+6 b^2 {r_h}^2}+\frac{24 \left(r^2+6 b^2
	{r_h}^2 \log (r)\right)}{r \left(r^2-3 b^2 {r_h}^2\right) (2 \log
	(r)+1)}+\frac{9}{r}\right)\nonumber\\
& &  -\frac{1}{26244
	\sqrt{2} \pi ^{3/2} r^2 \left(r^2-3 a^2 \right) (2 \log (r)+1) \alpha _{\theta _1}^4 \alpha _{\theta _2}^3}\Biggl\{\left(\frac{1}{N}\right)^{2/5}\nonumber\\
& &  \times \Biggl(177147 \sqrt{3} {g_s}^{3/2} M {N_f} \biggl[r^3-3 a^2 r+72 \left(9 a^4 -a^2 r^2 \right) \log ^3(r)+12 a^2  \left(15 r^2+r-27 a^2
	\right) \log ^2(r)\nonumber\\
	& &  +\left(-216 a^4+72 a^2 r^2 -6 a^2 r +4 r^3\right) \log (r)\biggr]
	\alpha _{\theta _1}^8 +19208 \sqrt{2} \pi ^{3/2} r \left(r^2+6 a^2  \log (r)\right) \alpha _{\theta _2}^7\Biggr)\Biggr\}\Biggr)
{q_5}'(r)\nonumber\\
& & -\frac{25 {q_4}''(r)}{(K^1)\ ^2}  = 0.\nonumber\\
& &
\end{eqnarray}}

\noindent$\bullet\delta {\rm R[x^1,r]}$\\
{\footnotesize
\begin{eqnarray}
\label{EOM25}
& &  {q_1}'(r) \left(\frac{200 \pi  {g_s} N r\left(6 b^2 {r_h}^2+r^2\right)}{21 \left(r^4-{r_h}^4\right) \left(9 a^2 +r^2\right)}-\frac{25
	{g_s}^3 \log N  M^2 {N_f} r(4 \log (r)+1) \left(6 a^2+r^2\right)}{56 \pi  \left(r^4-{r_h}^4\right) \left(9 a^2+r^2\right)}\right)\nonumber\\
& &  +{q_1}(r) \left(\frac{400 \pi  {g_s} N {r_h}^4 \left(6 a^2 +r^2\right)}{21
	\left(r^4-{r_h}^4\right)^2 \left(9 a^2 +r^2\right)}-\frac{25 {g_s}^3 \log N  M^2 {N_f} {r_h}^4 (4 \log
	(r)+1) \left(6 a^2+r^2\right)}{28 \pi   \left(r^4-{r_h}^4\right)^2 \left(9 a^2
	+r^2\right)}\right)\nonumber\\
& &  +{q_4}'(r) \left(\frac{25 {g_s}^3 \log N  M^2 {N_f}r (4 \log (r)+1) \left(6 a^2+r^2\right)}{28 \pi  \left(r^4-{r_h}^4\right) \left(9 a^2 +r^2\right)}-\frac{400 \pi  {g_s} N r\left(6 a^2
	+r^2\right)}{21 \left(r^4-{r_h}^4\right) \left(9 a^2+r^2\right)}\right)\nonumber\\
& &  +{q_3}(r) \Biggl[\frac{8 \sqrt{6} a^4 {g_s}^{3/2}
	M^3 \sqrt[5]{\frac{1}{N}}  \left(6 a^2+r^2\right) \left(\frac{68260644 \left({r_h}^4-10000\right) \log
		(10) \left(54 a^2+5\right)}{\left(100-3 a^2 \right)^4}-\frac{30876125 \left({r_h}^4-6561\right) \log (9)
		\left(12 a^2 +1\right)}{9 \left(a^2-27\right)^4}\right)}{35 \pi ^{3/2} \log N ^5 M_g^2 r \left(9 a^2
	+r^2\right)\left(1-\frac{{r_h}^4}{r^4}\right)}\nonumber\\
& &  +\frac{1}{21 r \left(r^4-{r_h}^4\right) (2 \log (r)+1) \left(r^2-3 a^2\right) \left(6 a^2 +r^2\right) \left(9 a^2+r^2\right)}\nonumber\\
& & \times\Biggl\{-7614 a^6 r^4-17820 a^6 r^4  \log (r)+20412 a^6 {r_h}^{4} \log (r)+8910
	a^6 {r_h}^{4}+144 a^4 r^6 +432 a^4 r^6  \log (r)\nonumber\\
	& &  -216 a^4 r^2 {r_h}^4-576 a^4 r^2 {r_h}^4 \log
	(r)+393 a^2 r^8 +978 a^2 r^8  \log (r)-489 a^2 r^4 {r_h}^4-1170 a^2 r^4 {r_h}^4 \log (r)+39 r^{10}\nonumber\\
	& & +94
	r^{10} \log (r)-47 r^6 {r_h}^4-110 r^6 {r_h}^4 \log (r)\Biggr\}\Biggr] +{q_3}'(r) = 0.
\end{eqnarray}}

\section{Schr\"{o}dinger-Like Potential for the Radial Profile Function for $\rho$ Mesons}

The Schr\"{o}dinger-like equation (\ref{near_Z=0-EOM-redefined_psi1}) satisfied by $g(Z)\equiv\sqrt{{\cal V}_1(Z)}\psi_1(Z)$ will have a potential given by:
{\scriptsize
	\begin{eqnarray}
\label{V-psi1}
	& & \hskip -0.4in V(Z) = \frac{1}{4 \left(-1+e^{4 Z}\right)^2}\Biggl\{\frac{1}{{r_h}^2}\Biggl\{3 \left(-1+e^{4 Z}\right) \Biggl(-4 m_0^2 + \frac{1}{\Delta_3}\nonumber\\
& &\hskip -0.4in \times\Biggl\{e^{-2 Z} \left(6 \left(1+e^{4 Z}\right) {g_s} {N_f} (Z+\log
   ({r_h}))^2-2 \left(4 \pi  \left(1+e^{4 Z}\right)+{g_s} {N_f} \left(e^{4 Z} (\log N -3)+\log N +3\right)\right) (Z+\log
   ({r_h}))-\left(-1+e^{4 Z}\right) ({g_s} {N_f} \log N +4 \pi )\right)\nonumber\\
   & &\hskip -0.4in \times \Biggl[216 e^Z {g_s}^2 {N_f}^2 {r_h}
   (Z+\log ({r_h}))^3-18 {g_s} {N_f} \left(32 e^Z \pi  {r_h}+{g_s} {N_f} \left(4 e^Z {r_h} (2 \log
   (N)+3)-1\right)\right) (Z+\log ({r_h}))^2\nonumber\\
   & &\hskip -0.4in +6 \left({g_s}^2 \left(4 e^Z {r_h} \log ^2 N+\left(24 e^Z {r_h}-2\right) \log
   (N)+3\right) {N_f}^2+8 {g_s} \pi  \left(4 e^Z {r_h} (\log N +3)-1\right) {N_f}+64 e^Z \pi ^2 {r_h}\right) (Z+\log
   ({r_h}))-32 \pi ^2 \left(12 e^Z {r_h}-1\right)\nonumber\\
   & &\hskip -0.4in -8 {g_s} {N_f} \pi  \left(\left(24 e^Z {r_h}-2\right) \log
   (N)+3\right)-{g_s}^2 {N_f}^2 \left(\left(24 e^Z {r_h}-2\right) \log ^2(N)+6 \log N -9\right)\Biggr]\Biggr\}\nonumber\\
   & &\hskip -0.4in +\frac{1}{\Delta_2}\Biggl\{2 e^{-2 Z} \Biggl[\Biggl(-6 \left(-1+e^{4
   Z}\right) {g_s} {N_f} (Z+\log ({r_h}))^2+2 \left(4 \pi  \left(-1+e^{4 Z}\right)+{g_s} {N_f} \left(e^{4 Z} (\log
   (N)-6)-\log N -6\right)\right)\nonumber\\
    & & \hskip -0.4in\times(Z+\log ({r_h}))+{g_s} {N_f} \left(2 \log N +e^{4 Z} (2 \log N -3)+3\right)+8 \left(1+e^{4
   Z}\right) \pi \Biggr)\nonumber\\
   & &\hskip -0.4in \times \left(-72 e^Z {g_s} {N_f} {r_h} (Z+\log ({r_h}))^2+3 \left(32 e^Z \pi  {r_h}+{g_s}
   {N_f} \left(8 e^Z {r_h} \log N -1\right)\right) (Z+\log ({r_h}))+{g_s} {N_f} (\log N -3)+4 \pi \right)\nonumber\\
   & & \hskip -0.4in-({g_s}
   {N_f} \log N -3 {g_s} {N_f} (Z+\log ({r_h}))+4 \pi )\nonumber\\
   & & \hskip -0.4in\times \Biggl[-36 e^Z \left(-9+e^{4 Z}\right) {g_s} {N_f}
   {r_h} (Z+\log ({r_h}))^3+12 \left(4 e^Z \left(-9+e^{4 Z}\right) \pi  {r_h}+{g_s} {N_f} \left(e^{5 Z} {r_h} (\log
   (N)-18)-9 e^Z {r_h} (\log N +6)+2\right)\right) (Z+\log ({r_h}))^2\nonumber\\
   & &\hskip -0.4in +8 \left(-27 e^Z \left(-1+e^{4 Z}\right) {g_s} {N_f}
   {r_h}+4 \pi  \left(18 e^Z {r_h}+6 e^{5 Z} {r_h}-1\right)+{g_s} {N_f} \left(18 e^Z {r_h}+6 e^{5 Z}
   {r_h}-1\right) \log N \right)\nonumber\\
    & & \hskip -0.4in\times(Z+\log ({r_h}))+16 \pi  \left(-6 e^Z {r_h}+6 e^{5 Z} {r_h}+1\right)+{g_s} {N_f}
   \left(4 \left(-6 e^Z {r_h}+6 e^{5 Z} {r_h}+1\right) \log N -3 \left(3+e^{4 Z}\right)\right)\Biggr]\Biggr]\Biggr\}\Biggr) a^2\Biggr\}\nonumber\\
   & & \hskip -0.4in+4 e^{2 Z} \left(-1+e^{4
   Z}\right) m_0^2+\nonumber\\
   & & \hskip -0.4in\frac{\left(-6 \left(1+e^{4 Z}\right) {g_s} {N_f} (Z+\log ({r_h}))^2+2 \left(4 \pi  \left(1+e^{4
   Z}\right)+{g_s} {N_f} \left(e^{4 Z} (\log N -3)+\log N +3\right)\right) (Z+\log ({r_h}))+\left(-1+e^{4 Z}\right)
   ({g_s} {N_f} \log N +4 \pi )\right)^2}{\Delta_1^2}\nonumber\\
   & & \hskip -0.4in-\frac{1}{\Delta_1}\Biggl\{4 \left(-1+e^{4 Z}\right) \Biggl[-6 \left(-1+e^{4 Z}\right) {g_s} {N_f} (Z+\log ({r_h}))^2\nonumber\\
   & & \hskip -0.4in+2
   \left(4 \pi  \left(-1+e^{4 Z}\right)+{g_s} {N_f} \left(e^{4 Z} (\log N -6)-\log N -6\right)\right) (Z+\log
   ({r_h}))+{g_s} {N_f} \left(2 \log N +e^{4 Z} (2 \log N -3)+3\right)+8 \left(1+e^{4 Z}\right) \pi \Biggr]\Biggr\}\Biggr\},\nonumber\\
	& & \hskip -0.4in
	\end{eqnarray}}
where:
\begin{equation*}
\Delta_n\equiv (Z+\log ({r_h}))
   ({g_s} {N_f} \log N -3 {g_s} {N_f} (Z+\log ({r_h}))+4 \pi )^n.
\end{equation*}

\section{Coefficient $c_{i}s$ appearing in the interaction lagrangian}

In this appendix assuming that in (\ref{interaction action full}), $\int_{Z=0}^\infty dZ = \int_{Z=0}^{\log \sqrt{3}b} dZ + \int_{\log \sqrt{3}b}^\infty dZ$, expressions for the coefficients $c_{i}s$, are given as:
{\scriptsize
	\begin{eqnarray*}
\label{interaction coefficients}		
&&  c_{1} =\int dZ\Biggl[ \frac{e^{-2 Z} \phi_0(Z)^2
			\sqrt{-{\cal A}_{\theta_2\theta_2} {g^{\cal M}_{11,11}}  ^2 g^{\cal M}_{yy}  -B_{\theta_2y}^2+g^{\cal M}_{11,11}   {g^{\cal M}_{\theta_2 y}} ^2}\sqrt{{g^{\cal M}_{x^1x^1}} ^3 {g^{\cal M}_{11,11}}  ^{5/2} g^{\cal M}_{rr}   g^{\cal M}_{tt} {r_h}^2 e^{2 Z}} }{g^{\cal M}_{x^1 x^1}  {g^{\cal M}_{11,11}}  ^{7/4} g^{\cal M}_{rr}   {r_h}^2}\nonumber\\
		& &  \times \left(-\frac{
			q_5 (Z){m^2}}{m^2}-q_1  (Z)-q_2  (Z)+3 q_4 (Z)\right)\nonumber\\
		& & + \frac{2 e^{-2 Z}
			\phi_0(Z)^2 (q_2  (Z)-q_4 (Z)-q_6 (Z)) \sqrt{-{\cal A}_{\theta_2\theta_2} {g^{\cal M}_{11,11}}  ^2
				g^{\cal M}_{yy}  -B_{\theta_2y}^2+g^{\cal M}_{11,11}   {g^{\cal M}_{\theta_2 y} }^2} \sqrt{g^{\cal M}_{x^1 x^1}  {g^{\cal M}_{11,11}}  ^{5/2} g^{\cal M}_{x^2 x^2}   g^{\cal M}_{x^3 x^3}   g^{\cal M}_{rr}
				g^{\cal M}_{tt} {r_h}^2 e^{2 Z}}}{g^{\cal M}_{x^1 x^1}  {g^{\cal M}_{11,11}}  ^{7/4} g^{\cal M}_{rr}   {r_h}^2}\Biggr]\nonumber\\
		& & = -\int \frac{dZ}{216 \pi ^2   M_g^2 \alpha _{\theta _1}^3 \alpha
			_{\theta _2}^2}\Biggl\{{g_s} M \sqrt[5]{\frac{1}{N}} {N_f}^2 e^{-4 Z} (e^{4 Z}-1) \phi_0(Z)^2 \left(2 \sqrt[5]{\frac{1}{N}} \alpha _{\theta _2}^2+81 \alpha
		_{\theta _1}^2\right)\nonumber\\
		& & \times (\log(e^Zr_h)) \left(72 a^2 {r_h} e^Z  \log(e^Zr_h)+3 a^2 +2 {r_h}^2 e^{2 Z} \right)
		M_g^2\left( {q_5}(Z) + ({q_1}(Z)-{q_2}(Z)-{q_4}(Z)+2 {q_6}(Z))\right)\Biggr\},\nonumber\\
		& & {\rm which\ for\ }b\sim 0.6\ {\rm yields:}\nonumber\\
		& & \Biggl[-\frac{1.16\times 10^{-7} {{\cal C}^2_{\phi_{0}}} N^{6/5} \alpha _{\theta _1} \alpha _{\theta _2}^2 c_{1_{q4}}}{M {N_f}^2 {r_h}^3}+\frac{15.9759 {{{\cal C}^{UV}_{\phi_0}}}^2 {\log (N)}^5 \sqrt[5]{N} {N_f^{UV}}^2 \log ({r_h}) ({c^{UV}_{2_{ q1}}}-3.01538 {c^{UV}_{2_{ q4}}})}{\sqrt{{g_s}}
   {M^{UV}}^2 {r_h}^6 \alpha _{\theta _1} \alpha _{\theta _2}^2}\Biggr] \nonumber
   \end{eqnarray*}}
   {\scriptsize
   \begin{eqnarray*}
			&&c_{2} = \int dZ \Biggl[ \frac{2 e^{-2 Z} \phi _0(Z)^2 {q_5}(Z) \sqrt{-{\cal A}_{\theta_2\theta_2} {g^{\cal M}_{11\ 11}}^2 g^{\cal M}_{yy}\ -{B^{IIA}_{\theta_2y}}^2+g^{\cal M}_{11\ 11}
				{g^{\cal M}_{\theta_2 y}}^2} \sqrt{g^{\cal M}_{x^1 x^1} {g^{\cal M}_{11\ 11}}^{5/2} g^{\cal M}_{x^2 x^2} g^{\cal M}_{x^3 x^3}   g^{\cal M}_{rr} g^{\cal M}_{tt} {r_h}^2 e^{2 Z}}}{g^{\cal M}_{x^1 x^1}
			{g^{\cal M}_{11\ 11}}^{7/4} g^{\cal M}_{rr} {r_h}^2} \nonumber\\
		& &=\int  \frac{dZ}{108 \pi ^2  \alpha _{\theta _1}^3 \alpha _{\theta _2}^2}\Biggl\{{g_s} M \sqrt[5]{\frac{1}{N}} {N_f}^2 e^{-4 Z} \sqrt{e^{4 Z}-1} \phi_0(Z)^2 {q_5}(Z) \left(2 \sqrt[5]{\frac{1}{N}} \alpha _{\theta _2}^2+81
		\alpha _{\theta _1}^2\right)\nonumber\\
		& &\times (\log(e^Zr_h)) \left(72 a^2   {r_h} e^Z (\log(e^Zr_h))+3 a^2 +2 {r_h}^2 e^{2 Z}\right)\Biggr\}\Biggr]\nonumber\\
		& & {\rm which\ for\ }b\sim 0.6\ {\rm yields:}\nonumber\\
		& &\Biggl[-\frac{2.32\times 10^{-7} {{\cal C}^2_{\phi_{0}}} N^{6/5} \alpha _{\theta _1} \alpha _{\theta _2}^2  c_{1_{{ q4}}}}{M {N_f}^2 {r_h}^3}-\frac{31.9518 {{\cal C}^{UV}_{\phi_{0}}}^2 {\log(N)}^5 \sqrt[5]{N} {N^{UV}_{f}}^2 \log ({r_h}) ({{c_2}^{UV}_{q1}}-3.01538 {{c_2}^{UV}_{q4}})}{\sqrt{{g^{UV}_s}}
   {M^{UV}}^2 {r_h}^6 \alpha _{\theta _1} \alpha _{\theta _2}^2}\Biggr]\nonumber\\
		&&c_{3} = \int dZ \Biggl[ \frac{e^{-2 Z} \psi_1'(Z)^2 \sqrt{-{\cal A}_{\theta_2\theta_2} {g^{\cal M}_{11\ 11}}^2 g^{\cal M}_{yy}\ -{B^{IIA}_{\theta_2y}}^2+g^{\cal M}_{11\ 11}\  {g^{\cal M}_{\theta_2 y}}^2}
		\sqrt{{g^{\cal M}_{x^1x^1}}^3 {g^{\cal M}_{11\ 11}}^{5/2} g^{\cal M}_{rr} g^{\cal M}_{tt} {r_h}^2 e^{2 Z}}}{g^{\cal M}_{x^1x^1} {g^{\cal M}_{11\ 11}}^{7/4} g^{\cal M}_{rr} {r_h}^2}\nonumber\\
	& & \times \left(-\frac{
		{q_5}(Z)m^2}{m^2}-{q_1}(Z)-{q_2}(Z)+3 {q_4}(Z)\right)\nonumber\\
	& & +\frac{2 e^{-2 Z}
			\psi_1'(Z)^2 ({q_2}(Z)-{q_4}(Z)-{q_6}(Z)) \sqrt{-{\cal A}_{\theta_2\theta_2} {g^{\cal M}_{11\ 11}}^2
				g^{\cal M}_{yy}\ -{B^{IIA}_{\theta_2y}}^2+g^{\cal M}_{11\ 11}\  {g^{\cal M}_{\theta_2 y}}^2} \sqrt{g^{\cal M}_{x^1x^1} {g^{\cal M}_{11\ 11}}^{5/2} g^{\cal M}_{x^2x^2} g^{\cal M}_{x^3x^3} g^{\cal M}_{rr}
				g^{\cal M}_{tt} {r_h}^2 e^{2 Z}}}{g^{\cal M}_{x^1x^1} {g^{\cal M}_{11\ 11}}^{7/4} g^{\cal M}_{rr} {r_h}^2}\Biggr]\nonumber\\
		& & = -\int  \frac{dZ}{216 \pi ^2   M_g^2 \alpha _{\theta _1}^3 \alpha
			_{\theta _2}^2}\Biggl\{{g_s} M \sqrt[5]{\frac{1}{N}} {N_f}^2 e^{-4 Z} \left(e^{4 Z}-1\right)
 \left(2 \sqrt[5]{\frac{1}{N}} \alpha _{\theta _2}^2+81 \alpha _{\theta _1}^2\right)\nonumber\\
 \end{eqnarray*}}
 {\scriptsize
 \begin{eqnarray*}
		& & \times
		\psi_1'(Z)^2 (\log(e^Zr_h)) \left(72 a^2 {r_h} e^Z (\log(e^Zr_h))+3 a^2 +2 {r_h}^2 e^{2 Z} \right)
	M_g^2\left( {q_5}(Z) + ({q_1}(Z)-{q_2}(Z)-{q_4}(Z)+2 {q_6}(Z))\right)\Biggr\}\nonumber\\
	& &{\rm which\ for\ }b\sim 0.6\ {\rm yields:}\nonumber\\
	& & \Biggl[\frac{0.68 {g_s}^2 M N^{4/5} {N_f}^2 \Biggl(\upsilon_2+\frac{\upsilon_1 g_s M^{2}(m_0^2-4)\log(r_h)}{N}\Biggr)^{3/2} {r_h} {c_{{\psi_1}}}^2 c_{1_{q4}} \log ^2({r_h})}{\alpha _{\theta _1} \alpha
   _{\theta _2}^2}+\frac{821.55 {\log(N)}^5 \sqrt[5]{N} {N^{UV}_{f}}^2 \log ({r_h}) {c^{UV}_{2_{\psi_{1}}}}^2 (1. {{c^{UV}_2}_{q1}}-3.01
   {{c^{UV}_2}_{q4}})}{\sqrt{{g^{UV}_s}} {M^{UV}}^2 {r_h}^6 \alpha _{\theta _1} \alpha _{\theta _2}^2}\Biggr]\nonumber\\
	&&c_{4} = \int dZ \Biggl[\frac{2 e^{-2 Z} {q_5}(Z) \psi_1'(Z)^2 \sqrt{-{\cal A}_{\theta_2\theta_2} {g^{\cal M}_{11\ 11}}^2 g^{\cal M}_{yy}\ -{B^{IIA}_{\theta_2y}}^2+g^{\cal M}_{11\ 11}\
				{g^{\cal M}_{\theta_2 y}}^2} \sqrt{g^{\cal M}_{x^1x^1} {g^{\cal M}_{11\ 11}}^{5/2} g^{\cal M}_{x^2x^2} g^{\cal M}_{x^3x^3} g^{\cal M}_{rr} g^{\cal M}_{tt} {r_h}^2 e^{2
					Z}}}{{g^{\cal M}_{11\ 11}}^{7/4} g^{\cal M}_{x^2x^2} g^{\cal M}_{rr} {r_h}^2}\Biggr]\nonumber\\
		& & =\int \frac{dZ}{108 \pi ^2  \alpha _{\theta _1}^3 \alpha _{\theta _2}^2}\Biggl\{{g_s} M \sqrt[5]{\frac{1}{N}} {N_f}^2 e^{-4 Z} \left(e^{4 Z}-1\right) {q_5}(Z) \left(2 \sqrt[5]{\frac{1}{N}} \alpha _{\theta _2}^2+81 \alpha _{\theta
			_1}^2\right) \psi_1'(Z)^2 (\log(e^Zr_h))\nonumber\\
		& &  \times \left(3 a^2 +72 a^2 {r_h} e^Z  (\log(e^Zr_h))+2 {r_h}^2 e^{2 Z} \log
		(N)\right)\Biggr\}\nonumber\\
		& &{\rm which\ for\ }b\sim 0.6\ {\rm yields:}\nonumber\\
		& & \Biggl[-\frac{1.36 {g_s}^2 M N^{4/5} {N_f}^2 \Biggl(\upsilon_2+\frac{\upsilon_1 g_s M^{2}(m_0^2-4)\log(r_h)}{N}\Biggr)^{3/2} {r_h} {c_{{\psi_{1}}}}^2 c_{1_{q4}} \log ^2({r_h})}{\alpha _{\theta _1} \alpha
   _{\theta _2}^2}\nonumber\\
   &&-\frac{1643.11 {\log(N)}^5 \sqrt[5]{N} {N^{UV}_{f}}^2 \log ({r_h}) {{c_2}^{UV}_{\psi_{1}}}^2 (1. {{c_2}^{UV}_{q1}}-3.01
   {{c_2}^{UV}_{q4}})}{\sqrt{{g^{UV}_s}} {M^{UV}}^2 {r_h}^6 \alpha _{\theta _1} \alpha _{\theta _2}^2}\Biggr]\nonumber\\
		&&c_{5} =\int dZ\Biggl[ \frac{\psi_1(Z)^2 \sqrt{-{\cal A}_{\theta_2\theta_2} {g^{\cal M}_{11\ 11}}^2 g^{\cal M}_{yy}\ -{B^{IIA}_{\theta_2y}}^2+g^{\cal M}_{11\ 11}\  {g^{\cal M}_{\theta_2 y}}^2}
			\sqrt{{g^{\cal M}_{x^1 x^1}}^3 {g^{\cal M}_{11\ 11}}^{5/2} g^{\cal M}_{rr} g^{\cal M}_{tt} {r_h}^2 e^{2 Z}}}{2 {g^{\cal M}_{x^1 x^1}}^2 {g^{\cal M}_{11\ 11}}^{7/4}}\nonumber\\
& &  \times  \left(-\frac{
				{q_5}(Z)m^2}{m^2}-{q_1}(Z)-{q_2}(Z)+3 {q_4}(Z)\right)\nonumber\\
			& & +\frac{\psi_1(Z)^2 (-2
			{q_4}(Z)-{q_6}(Z)) \sqrt{-{\cal A}_{\theta_2\theta_2} {g^{\cal M}_{11\ 11}}^2 g^{\cal M}_{yy}\ -{B^{IIA}_{\theta_2y}}^2+g^{\cal M}_{11\ 11}\  {g^{\cal M}_{\theta_2 y}}^2}
			\sqrt{g^{\cal M}_{x^1x^1} {g^{\cal M}_{11\ 11}}^{5/2} g^{\cal M}_{x^2x^2} g^{\cal M}_{x^3x^3} g^{\cal M}_{rr} g^{\cal M}_{tt} {r_h}^2 e^{2 Z}}}{{g^{\cal M}_{x^1x^1}}^2 {g^{\cal M}_{11\ 11}}^{7/4}}\Biggr]\nonumber\\
		& &-\int\frac{dZ}{108 \pi  M_g^2 {r_h}^2  \alpha _{\theta _1}^3 \alpha _{\theta _2}^2}\Biggl\{{g_s}^2 M N^{3/5} {N_f}^2 e^{-2 Z}  \psi_1(Z)^2 \left(81 \sqrt[5]{N} \alpha _{\theta _1}^2+2 \alpha _{\theta _2}^2\right) (\log
		({r_h})+Z)\nonumber\\
		& &\times \left(72 a^2 {r_h} e^Z (\log(e^Zr_h))-3 a^2 +2 {r_h}^2 e^{2 Z} \right)
		 M_g^2\left( {q_5}(Z) + ({q_1}(Z)+{q_2}(Z)+{q_4}(Z)+2 {q_6}(Z))\right)\Biggr\}\nonumber\\
		& & {\rm which\ for\ }b\sim 0.6\ {\rm yields:}\nonumber\\
		& &\Biggl[\frac{-55.75 {g_s}^3 M N^{9/5} {N_f}^2 \sqrt{\Biggl(\upsilon_2+\frac{\upsilon_1 g_s M^{2}(m_0^2-4)\log(r_h)}{N}\Biggr)} {c^2_{{\psi_{1}}}} c_{1_{q4}} \log ^2({r_h})}{{r_h} \alpha _{\theta _1} \alpha
   _{\theta _2}^2}\nonumber\\
   & & +\frac{176.96 \sqrt{{g^{UV}_s}} {\log(N)}^5 N^{6/5} {N^{UV}_{f}}^2 \log ({r_h}) {c^{UV}_{2_{\psi_{1}}}}^2 (0.0196 {{c_2}^{UV}_{q4}}-0.006
   {{c_2}^{UV}_{q1}})}{{M^{UV}}^2 {r_h}^8 \alpha _{\theta _1} \alpha _{\theta _2}^2}\Biggr]\nonumber\\
   &&c_{6} = \int dZ \Biggl[ \frac{2 \psi_1(Z)^2 {q_5}(Z) \sqrt{-{\cal A}_{\theta_2\theta_2} {g^{\cal M}_{11\ 11}}^2 g^{\cal M}_{yy}\ -{B^{IIA}_{\theta_2y}}^2+g^{\cal M}_{11\ 11}\  {g^{\cal M}_{\theta_2 y}}^2}
			\sqrt{g^{\cal M}_{x^1x^1} {g^{\cal M}_{11\ 11}}^{5/2} g^{\cal M}_{x^2x^2} g^{\cal M}_{x^3x^3} g^{\cal M}_{rr} g^{\cal M}_{tt} {r_h}^2 e^{2 Z}}}{{g^{\cal M}_{x^1x^1}}^2 {g^{\cal M}_{11\ 11}}^{7/4}}\Biggr]\nonumber\\
		& & =\int  \frac{dZ}{27 \pi
			{r_h}^2  \alpha _{\theta _1}^3 \alpha _{\theta _2}^2}\Biggl\{{g_s}^2 M N^{3/5} {N_f}^2 e^{-2 Z} \psi_1(Z)^2 {q_5}(Z) \left(81 \sqrt[5]{N} \alpha _{\theta _1}^2+2 \alpha _{\theta
			_2}^2\right)\nonumber\\
		& &\times (\log(e^Zr_h)) \left(72 a^2 {r_h} e^Z  (\log(e^Zr_h))-3 a^2 +2 {r_h}^2 e^{2 Z} \right)\Biggr\}\nonumber\\
&&
\end{eqnarray*}}
{\scriptsize
\begin{eqnarray*}
		& &{\rm which for\ }b\sim 0.6\ {\rm yields:}\nonumber\\
		& &\Biggl[\frac{223.007 {g_s}^3 M N^{9/5} {Nf}^2 \sqrt{\Biggl(\upsilon_2+\frac{\upsilon_1 g_s M^{2}(m_0^2-4)\log(r_h)}{N}\Biggr)} {c^2_{{\psi_{1}}}} c_{1_{q4}} \log ^2({r_h})}{{r_h} \alpha _{\theta _1} \alpha
   _{\theta _2}^2}\nonumber\\
   &&+\frac{4.60 \sqrt{{g^{UV}_s}} {\log(N)}^5 N^{6/5} {N^{UV}_{f}}^2 \log ({r_h}) {c^{UV}_{2_{\psi_{1}}}}^2 (1. {{c_2}^{UV}_{q1}}-3.015
   {{c_2}^{UV}_{q4}})}{{M^{UV}}^2 {r_h}^8 \alpha _{\theta _1} \alpha _{\theta _2}^2}\Biggr]\nonumber\\
	&&c_{7} = \int dZ \Biggl[\frac{2 e^{-2 Z} \phi_0(Z)^2 \psi_1(Z) \sqrt{-{\cal A}_{\theta_2\theta_2} {g^{\cal M}_{11\ 11}}^2 g^{\cal M}_{yy}\ -{B^{IIA}_{\theta_2y}}^2+g^{\cal M}_{11\ 11}\
			{g^{\cal M}_{\theta_2 y}}^2} \sqrt{{g^{\cal M}_{x^1 x^1}}^3 {g^{\cal M}_{11\ 11}}^{5/2} g^{\cal M}_{rr} g^{\cal M}_{tt} {r_h}^2 e^{2 Z}} }{g^{\cal M}_{x^1x^1} {g^{\cal M}_{11\ 11}}^{7/4} g^{\cal M}_{rr} {r_h}^2}\nonumber\\
		& & \times \left(-\frac{
			{q_5}(Z)m^2}{m^2}-{q_1}(Z)-{q_2}(Z)+3 {q_4}(Z)\right)\nonumber\\
& &+4 e^{-2 Z}
		\phi_0(Z)^2 \psi_1(Z) ({q_2}(Z)-{q_4}(Z)-{q_6}(Z))\nonumber\\
			&&\times\frac{ \sqrt{-{\cal A}_{\theta_2\theta_2} {g^{\cal M}_{11\ 11}}^2
		g^{\cal M}_{yy}\ -{B^{IIA}_{\theta_2y}}^2+g^{\cal M}_{11\ 11}\  {g^{\cal M}_{\theta_2 y}}^2} \sqrt{g^{\cal M}_{x^1 x^1} {g^{\cal M}_{11\ 11}}^{5/2} g^{\cal M}_{x^2 x^2} g^{\cal M}_{x^3x^3} g^{\cal M}_{rr}
			g^{\cal M}_{tt} {r_h}^2 e^{2 Z}}}{g^{\cal M}_{x^1x^1} {g^{\cal M}_{11\ 11}}^{7/4} g^{\cal M}_{rr} {r_h}^2}\Biggr]\nonumber\\
	& & = -\int \frac{dZ}{108 \pi ^2 M_g^2  \alpha _{\theta _1}^3
		\alpha _{\theta _2}^2}\Biggl\{{g_s} M \sqrt[5]{\frac{1}{N}} {N_f}^2 e^{-2 Z} \left(e^{4 Z}-1\right) \phi_0(Z)^2 \psi_1(Z) \left(2 \sqrt[5]{\frac{1}{N}} \alpha _{\theta
		_2}^2+81 \alpha _{\theta _1}^2\right)\nonumber\\
	& &\times (\log(e^Zr_h)) \left(72 a^2 {r_h} e^Z  (\log(e^Zr_h))+3 a^2 +2 {r_h}^2 e^{2 Z}
	\log (N)\right)\nonumber\\
	& &\times M_g^2\left( {q_5}(Z) + ({q_1}(Z)-{q_2}(Z)-{q_4}(Z)+2 {q_6}(Z))\right)\Biggr\}\nonumber\\
	& &{\rm which\ for\ }b\sim 0.6\ {\rm yields:}\nonumber\\
	& &\Biggl[-\frac{3.28\times 10^{-7} {{\cal C}_{\phi_{0}}}^2 N^{6/5} (\Biggl(\upsilon_2+\frac{\upsilon_1 g_s M^{2}(m_0^2-4)\log(r_h)}{N}\Biggr))^{1/4} \alpha _{\theta _1} \alpha _{\theta _2}^2
   c_{{\psi_{1}}} c_{1_{q4}}}{M {N_f}^2{r_h}^3}\nonumber\\
   & &+\frac{2666.71 {{\cal C}^{UV}_{\phi_{0}}}^2 {\log(N)}^5 \sqrt[5]{N} {N^{UV}_{f}}^2 \log ({r_h}) {{c_2}^{UV}_{\psi_{1}}} (1. {{c_2}^{UV}_{q1}}-3.015
   {{c_2}^{UV}_{q4}})}{\sqrt{{g^{UV}_s}} {M^{UV}}^2 {r_h}^6 \alpha _{\theta _1} \alpha _{\theta _2}^2}\Biggr]\nonumber\\
	&&c_{8} = \int dZ \Biggl[\frac{4 e^{-2 Z} \phi_0(Z)^2 \psi_1(Z) {q_5}(Z) \sqrt{-{\cal A}_{\theta_2\theta_2} {g^{\cal M}_{11\ 11}}^2
			g^{\cal M}_{yy}\ -{B^{IIA}_{\theta_2y}}^2+g^{\cal M}_{11\ 11}\  {g^{\cal M}_{\theta_2 y}}^2} \sqrt{g^{\cal M}_{x^1x^1} {g^{\cal M}_{11\ 11}}^{5/2} g^{\cal M}_{x^2x^2} g^{\cal M}_{x^3x^3} g^{\cal M}_{rr}
			g^{\cal M}_{tt} {r_h}^2 e^{2 Z}}}{g^{\cal M}_{x^1x^1} {g^{\cal M}_{11\ 11}}^{7/4} g^{\cal M}_{rr} {r_h}^2}\Biggr]\nonumber\\
	& &= \int \frac{dZ}{54 \pi ^2 \alpha _{\theta _1}^3 \alpha _{\theta _2}^2}\Biggl\{{g_s} M \sqrt[5]{\frac{1}{N}} {N_f}^2 e^{-4 Z} \left(e^{4 Z}-1\right) \phi_0(Z)^2 \psi_1(Z) {q_5}(Z) \left(2 \sqrt[5]{\frac{1}{N}} \alpha
	_{\theta _2}^2+81 \alpha _{\theta _1}^2\right)\nonumber\\
	& &\times (\log(e^Zr_h)) \left(72 a^2 {r_h} e^Z (\log(e^Zr_h))+3 a^2 +2 {r_h}^2
	e^{2 Z} \right)\Biggr\}\nonumber\\
	& &{\rm which\ for\ }b\sim 0.6\ {\rm yields:}\nonumber\\
	& &\Biggl[\frac{6.57\times 10^{-7} {{\cal C}^2_{\phi_{0}}} N^{6/5} \Biggl(\upsilon_2+\frac{\upsilon_1 g_s M^{2}(m_0^2-4)\log(r_h)}{N}\Biggr)^{1/4} \alpha _{\theta _1} \alpha _{\theta _2}^2
   {c}_{\psi_1} c_{1_{q4}}}{M {N_f}^2 {r_h}^3}\nonumber\\
   & &-\frac{5333.42 {{\cal C}^{UV}_{\phi_{0}}}^2 {\log(N)}^5 \sqrt[5]{N} {N^{UV}_{f}}^2 \log ({r_h}) {{c_2}^{UV}_{\psi_{1}}} (1. {{c_2}^{UV}_{q1}}-3.015
   {{c_2}^{UV}_{q4}})}{\sqrt{{g^{UV}_s}} {M^{UV}}^2 {r_h}^6 \alpha _{\theta _1} \alpha _{\theta _2}^2}\Biggr]\nonumber\\
   &&
   \end{eqnarray*}}
   {\scriptsize
   \begin{eqnarray}
   \label{interaction coefficients}
	&&c_{9} = -\int dZ \Biggl[\frac{4 e^{-Z} \psi_1(Z) {q_3}(Z) \psi_1'(Z) \sqrt{-{\cal A}_{\theta_2\theta_2} {g^{\cal M}_{11\ 11}}^2
			g^{\cal M}_{yy}\ -{B^{IIA}_{\theta_2y}}^2+g^{\cal M}_{11\ 11}\  {g^{\cal M}_{\theta_2 y}}^2} \sqrt{g^{\cal M}_{x^1x^1} {g^{\cal M}_{11\ 11}}^{5/2} g^{\cal M}_{x^2x^2} g^{\cal M}_{x^3x^3} g^{\cal M}_{rr}
			g^{\cal M}_{tt} {r_h}^2 e^{2 Z}}}{g^{\cal M}_{x^1x^1} {g^{\cal M}_{11\ 11}}^{7/4} g^{\cal M}_{rr} {r_h}}\Biggr]\nonumber\\
	& &= -\int \frac{dZ}{54 \pi ^2  \alpha _{\theta _1}^3 \alpha _{\theta _2}^2}\Biggl\{{g_s} M \sqrt[5]{\frac{1}{N}} {N_f}^2 {r_h} e^{-3Z} \left(e^{4 Z}-1\right) \psi_1(Z) {q_3}(Z)
	\left(2 \sqrt[5]{\frac{1}{N}} \alpha _{\theta_2}^2+81 \alpha _{\theta _1}^2\right) \psi_1'(Z) (\log(e^Zr_h))\nonumber\\
	& &\times \left(72 a^2 {r_h} e^Z  (\log(e^Zr_h))+3 a^2+2
	{r_h}^2 e^{2 Z} \right)\Biggr\}\nonumber\\
	& &{\rm which\ for\ }b\sim 0.6\ {\rm yields:}\nonumber\\
	& &\Biggl[-\frac{0.000514915 {g_s}^{3/2} {\log(N)}^5 {m_0}^4 {N_f}^2 \Biggl(\upsilon_2+\frac{\upsilon_1 g_s M^{2}(m_0^2-4)\log(r_h)}{N}\Biggr) {c_{{\psi_{1}}}}^2 c_{1_{q4}} \log ({r_h})}{M^2 {\pi g_s}^2
   {r_h}^2 \alpha _{\theta _1} \alpha _{\theta _2}^2}\nonumber\\
   & & +\frac{2709.66 {g^{UV}_s} {M^{UV}} \sqrt[5]{\frac{1}{N}} {N^{UV}_{f}}^2 {r_h}^3 \log ({r_h}) {c_1}^{UV}_{q3} {{c_2}^{UV}_{\psi_{1}}}^2}{\alpha
   _{\theta _1} \alpha _{\theta _2}^2}\Biggr],	\nonumber\\
   & &
	\end{eqnarray}}
where:
{\footnotesize
\begin{eqnarray}
\label{defs}
& & {\cal A}_{\theta_2\theta_2} \equiv \frac{9 {g_s}^{7/2} M^2 N^{11/10} {N_f}^4 e^{-2 Z} \log ^2\left({r_h} e^Z\right) \left(36 a^2 \log
	\left({r_h} e^Z\right)+{r_h} e^Z\right)^2}{2 \pi ^{5/2} {r_h}^2 \alpha _{\theta _1}^2 \alpha _{\theta _2}^4}.
\end{eqnarray}} 
\chapter{}
\section{Equations of Motion}

This appendix lists the various EOMs corresponding to the first equation of (\ref{eoms}). The discussion has been split into two sub-sections - {\bf D.1.1} on the ${\cal O}(l_p^6 R^4)$ corrections to the ${\cal M}$-Theory uplift of \cite{MQGP} worked out in the $\psi=0, 2\pi, 4\pi$-patches, and {\bf D.1.2} on the same worked out in the $\psi\neq0$-patch. The former has all the EOMs listed out expanding the coefficients of $f^{(n)}_{MN}, n=0,1,2$ near $r=r_h$ and retaining the LO terms in the powers of $(r-r_h)$ in the same, and then performing a large-$N$-large-$|\log r_h|$-$\log N$ expansion the resulting LO terms are written out. Due to a more complicated nature of the EOMs in the $\psi=\psi_0\neq0$-patch (wherein some $g^{\cal M}_{rM}, M\neq r$ and $g^{\cal M}_{x^{10}N}, N\neq x^{10}$ components are non-zero), only the LO-in-$(r-r_h)$ LO-in-$N$ form of the EOMs are written out.

This appendix is rather large as the various EOMs are quite long (which is why the EOMs were relegated to an appendix). The solutions of the EOMs are discussed in Section {\bf 5.2}.

\subsection{$\psi=0, 2\pi, 4\pi$-Coordinate Patches Near $r=r_h$}
Working in the IR,  the EOMs near the $\psi=0\ 2\pi\ 4\pi$-branches near $r=r_h$ as described in Section {\bf 5.2}, can be written as follows:
{\footnotesize
\begin{eqnarray}
\label{IR-psi=2nPi-EOMs}
& & {\rm EOM}_{MN}:\nonumber\\
& &  \sum_{p=0}^2\sum_{i=0}^2a_{MN}^{(p,i)}\left(r_h, a, N, M, N_f, g_s, \alpha_{\theta_{1,2}}\right)(r-r_h)^if_{MN}^{(p)}(r) +
\beta {\cal F}_{MN}\left(r_h, a, N, M, N_f, g_s, \alpha_{\theta_{1,2}}\right)(r-r_h)^{\alpha_{MN}^{\rm LO}} = 0,\nonumber\\
& &
\end{eqnarray}
}
where $M, N$ run over the $D=11$ coordinates,   $f^{(p)}_{MN}\equiv \frac{d^p f_{MN}}{dr^p}, p=0, 1, 2$, $\alpha_{MN}^{\rm LO}=0, 1, 2, 3$ denotes the leading order (LO) terms in powers of $r-r_h$ in the IR when the ${\cal O}(\beta)$-terms are expanded in a Taylor series about $r=r_h$, and ${\cal F}_{\theta_1\theta_2} = {\cal F}_{\theta_2\theta_2} = 0$ (i.e. EOM$_{\theta_1\theta_2}$ and EOM$_{\theta_2\theta_2}$ are homogeneous up to ${\cal O}(\beta)$).
In the EOMs in this appendix and their solutions in Section {\bf 5.2}:
\begin{eqnarray}
\label{Sigma_1-def}
& & \Sigma_1 \equiv 19683
   \sqrt{6} \alpha _{\theta _1}^6+6642 \alpha _{\theta _2}^2 \alpha _{\theta _1}^3-40 \sqrt{6} \alpha _{\theta _2}^4\nonumber\\
   & & \stackrel{\rm Global}{\longrightarrow} N^{\frac{6}{5}}\left(19683
   \sqrt{6} \sin^6\theta_1+6642 \sin^2{\theta _2} \sin^3{\theta _1}-40 \sqrt{6} \sin^4{\theta _2}\right).
\end{eqnarray}

(i) \underline{${\rm EOM}_{tt}$}

{\scriptsize
\begin{eqnarray*}
& & \frac{4 \left(9 b^2+1\right)^3 \left(4374 b^6+1035 b^4+9 b^2-4\right) M \left(\frac{1}{N}\right)^{9/4} (r-{r_h})^2 \beta  \log
   ({r_h})\Sigma_1 b^8}{27 \left(18 b^4-3 b^2-1\right)^5 {g_s} \log N ^2 {N_f} \pi ^2 \alpha _{\theta _2}^3}
\nonumber\\
& &+\frac{243
   \left(\frac{1}{N}\right)^{12/5} (r-{r_h}) {r_h} {f_{xx}}(r) \alpha _{\theta _1}^4 \alpha _{\theta _2}^2}{2 {g_s} \pi
   \log \left({r_h}^6+9 a^2 {r_h}^4\right)}\nonumber\\
& & -\frac{6 (r-{r_h}) {r_h} {f_t}(r) \left(-36 a^4+6 {r_h}^2 a^2+2
   {r_h}^4+\left(-18 a^4+7 {r_h}^2 a^2+{r_h}^4\right) \log \left({r_h}^6+9 a^2 {r_h}^4\right)\right)}{{g_s} N \pi
   \left({r_h}^2-3 a^2\right) \left(6 a^2+{r_h}^2\right) \log \left({r_h}^6+9 a^2 {r_h}^4\right)}\nonumber\\
& & -\frac{245
   \left(\frac{1}{N}\right)^{4/5} \sqrt{\frac{\pi }{6}} (r-{r_h}) {r_h}^5 {f_{xy}}(r) \left(27 \sqrt{6} \alpha _{\theta _1}^3+10
   \alpha _{\theta _2}^2\right)}{41472 {g_s}^{5/2} M {N_f} \left({r_h}^2-3 a^2\right)^2 \log ^3({r_h}) \alpha _{\theta
   _2}}\nonumber\\
   & & +\frac{343 \left(\frac{1}{N}\right)^{4/5} \sqrt{\frac{\pi }{2}} (r-{r_h}) {r_h}^5 {f_{\theta_1x}}(r) \left(81 \sqrt{2} \alpha
   _{\theta _1}^3+10 \sqrt{3} \alpha _{\theta _2}^2\right)}{31104 {g_s}^{5/2} M {N_f} \left({r_h}^2-3 a^2\right)^2 \log
   ^3({r_h}) \alpha _{\theta _2}}\nonumber\\
   & & -\frac{1715 \left(\frac{1}{N}\right)^{4/5} \sqrt{\frac{\pi }{2}} (r-{r_h}) {r_h}^5
   {f_{\theta_2x}}(r) \left(81 \sqrt{2} \alpha _{\theta _1}^3+10 \sqrt{3} \alpha _{\theta _2}^2\right)}{93312 {g_s}^{5/2} M {N_f}
   \left({r_h}^2-3 a^2\right)^2 \log ^3({r_h}) \alpha _{\theta _2}}\nonumber\\
   & & +\frac{3479 \left(\frac{1}{N}\right)^{4/5} \sqrt{\frac{\pi
   }{2}} (r-{r_h}) {r_h}^5 {f_{xz}}(r) \left(81 \sqrt{2} \alpha _{\theta _1}^3+10 \sqrt{3} \alpha _{\theta _2}^2\right)}{373248
   {g_s}^{5/2} M {N_f} \left({r_h}^2-3 a^2\right)^2 \log ^3({r_h}) \alpha _{\theta _2}}\nonumber\\
   & & -\frac{3 (r-{r_h}) {r_h}^2
   \left(9 a^2+{r_h}^2\right) f'(r)}{{g_s} N \pi  \left(6 a^2+{r_h}^2\right)}+\frac{(r-{r_h}) {r_h}^2 \left(9
   a^2+{r_h}^2\right) {f_{zz}}'(r)}{{g_s} N \pi  \left(6 a^2+{r_h}^2\right)}-\frac{(r-{r_h}) {r_h}^2 \left(9
   a^2+{r_h}^2\right) {f_{x^{10} x^{10}}}'(r)}{{g_s} N \pi  \left(6 a^2+{r_h}^2\right)}\nonumber\\
 & & -\frac{(r-{r_h}) {r_h}^2 \left(9
   a^2+{r_h}^2\right) {f_{\theta_1z}}'(r)}{{g_s} N \pi  \left(6 a^2+{r_h}^2\right)}+\frac{6 a^2 (r-{r_h})^2 {r_h} \left(9
   a^2+{r_h}^2\right) {f_{\theta_1x}}'(r)}{{g_s} N \pi  \left({r_h}^2-3 a^2\right) \left(6 a^2+{r_h}^2\right)}\nonumber\\
 & &+\frac{27
   \sqrt{\frac{3}{2}} \sqrt{{g_s}} M \left(\frac{1}{N}\right)^{7/5} {N_f} (r-{r_h}) {r_h}^2 \left(9 a^2+{r_h}^2\right)
   \log ({r_h}) \alpha _{\theta _1}^4 {f_{\theta_1y}}'(r)}{4 \pi ^{5/2} \left(6 a^2+{r_h}^2\right) \alpha _{\theta _2}^3}\nonumber\\
   & & +\frac{27
   \sqrt{\frac{3}{2}} \sqrt{{g_s}} M \left(\frac{1}{N}\right)^{7/5} {N_f} (r-{r_h}) {r_h}^2 \left(9 a^2+{r_h}^2\right)
   \log ({r_h}) \alpha _{\theta _1}^4 {f_{\theta_2z}}'(r)}{4 \pi ^{5/2} \left(6 a^2+{r_h}^2\right) \alpha _{\theta _2}^3}\nonumber\\
      & & -\frac{81
   \sqrt{\frac{3}{2}} a^2 \sqrt{{g_s}} M \left(\frac{1}{N}\right)^{8/5} {N_f} (r-{r_h})^2 {r_h} \left(9
   a^2+{r_h}^2\right) \log ({r_h}) \alpha _{\theta _1}^2 {f_{\theta_2x}}'(r)}{2 \pi ^{5/2} \left({r_h}^2-3 a^2\right) \left(6
   a^2+{r_h}^2\right) \alpha _{\theta _2}}\nonumber\\
   & & -\frac{(r-{r_h}) {r_h}^2 \left(9 a^2+{r_h}^2\right) {f_{\theta_2y}}'(r)}{{g_s} N
   \pi  \left(6 a^2+{r_h}^2\right)}-\frac{(r-{r_h}) {r_h}^2 \left(9 a^2+{r_h}^2\right) {f_{xz}}'(r)}{{g_s} N \pi
   \left(6 a^2+{r_h}^2\right)}\nonumber\\
   & & +\frac{243 \left(\frac{1}{N}\right)^{12/5} (r-{r_h}) {r_h}^2 \left(9 a^2+{r_h}^2\right)
   \alpha _{\theta _1}^4 \alpha _{\theta _2}^2 {f_{xx}}'(r)}{8 {g_s} \pi  \left(6 a^2+{r_h}^2\right)}+\frac{81
   \sqrt{\frac{3}{2}} a^2 \sqrt{{g_s}} M \left(\frac{1}{N}\right)^{7/5} {N_f} (r-{r_h})^2 {r_h} \left(9
   a^2+{r_h}^2\right) \log ({r_h}) \alpha _{\theta _1}^4 {f_{xy}}'(r)}{2 \pi ^{5/2} \left(-18 a^4+3 {r_h}^2
   a^2+{r_h}^4\right) \alpha _{\theta _2}^3}\nonumber\\
   & & +\frac{81 \sqrt{\frac{3}{2}} a^2 \sqrt{{g_s}} M \left(\frac{1}{N}\right)^{7/5}
   {N_f} (r-{r_h})^2 {r_h} \left(9 a^2+{r_h}^2\right) \log ({r_h}) \alpha _{\theta _1}^4 {f_{yz}}'(r)}{2 \pi ^{5/2}
   \left(-18 a^4+3 {r_h}^2 a^2+{r_h}^4\right) \alpha _{\theta _2}^3}\nonumber\\
   & & +\frac{27 a^2 {g_s}^2 M^2 \left(\frac{1}{N}\right)^{12/5}
   {N_f}^2 (r-{r_h})^2 {r_h} \left(9 a^2+{r_h}^2\right) \log ^2({r_h}) \alpha _{\theta _1}^2 \left(2187 \sqrt{2}
   \alpha _{\theta _1}^6+540 \sqrt{3} \alpha _{\theta _2}^2 \alpha _{\theta _1}^3+50 \sqrt{2} \alpha _{\theta _2}^4\right)
   {f_{yy}}'(r)}{128 \sqrt{2} \pi ^4 \left(-18 a^4+3 {r_h}^2 a^2+{r_h}^4\right) \alpha _{\theta _2}^4}
    \end{eqnarray*}}
   {\footnotesize
   \begin{eqnarray*}
   & & +\frac{3 (r-{r_h})^2
   {r_h} \left(6 \left(-18 a^4+3 {r_h}^2 a^2+{r_h}^4\right)+\left(-27 a^4+15 {r_h}^2 a^2+2 {r_h}^4\right) \log
   \left({r_h}^6+9 a^2 {r_h}^4\right)\right) {f_r}'(r)}{{g_s} N \pi  \left({r_h}^2-3 a^2\right) \left(6
   a^2+{r_h}^2\right) \log \left({r_h}^6+9 a^2 {r_h}^4\right)}\nonumber\\
   & & +\frac{3 (r-{r_h})^2 {r_h} \left(-36 a^4+6 {r_h}^2
   a^2+\left(9 a^2+{r_h}^2\right) \log \left({r_h}^6+9 a^2 {r_h}^4\right) a^2+2 {r_h}^4\right) {f_t}'(r)}{{g_s} N
   \pi  \left({r_h}^2-3 a^2\right) \left(6 a^2+{r_h}^2\right) \log \left({r_h}^6+9 a^2 {r_h}^4\right)}
   \nonumber\\
& & -\frac{6
   (r-{r_h})^2 {r_h}^2 \left(9 a^2+{r_h}^2\right) f''(r)}{{g_s} N \pi  \left(6 a^2+{r_h}^2\right)}+\frac{2
   (r-{r_h})^2 {r_h}^2 \left(9 a^2+{r_h}^2\right) {f_{zz}}''(r)}{{g_s} N \pi  \left(6 a^2+{r_h}^2\right)}-\frac{2
   (r-{r_h})^2 {r_h}^2 \left(9 a^2+{r_h}^2\right) {f_{x^{10} x^{10}}}''(r)}{{g_s} N \pi  \left(6 a^2+{r_h}^2\right)}
   \nonumber\\
   & & -\frac{3
   (r-{r_h})^2 {r_h}^2 \left(9 a^2+{r_h}^2\right) {f_{\theta_1z}}''(r)}{{g_s} N \pi  \left(6 a^2+{r_h}^2\right)}-\frac{7
   \sqrt{{g_s}} M \left(\frac{1}{N}\right)^{9/5} {N_f} (r-{r_h})^2 {r_h}^2 \left(9 a^2+{r_h}^2\right) \log ({r_h})
   \left(81 \sqrt{2} \alpha _{\theta _1}^3+10 \sqrt{3} \alpha _{\theta _2}^2\right) {f_{\theta_1x}}''(r)}{8 \sqrt{2} \pi ^{5/2} \left(6
   a^2+{r_h}^2\right) \alpha _{\theta _2}}\nonumber\\
   & &+\frac{81 \sqrt{\frac{3}{2}} \sqrt{{g_s}} M \left(\frac{1}{N}\right)^{7/5} {N_f}
   (r-{r_h})^2 {r_h}^2 \left(9 a^2+{r_h}^2\right) \log ({r_h}) \alpha _{\theta _1}^4 {f_{\theta_1y}}''(r)}{4 \pi ^{5/2} \left(6
   a^2+{r_h}^2\right) \alpha _{\theta _2}^3}\nonumber\\
   & &+\frac{81 \sqrt{\frac{3}{2}} \sqrt{{g_s}} M \left(\frac{1}{N}\right)^{7/5} {N_f}
   (r-{r_h})^2 {r_h}^2 \left(9 a^2+{r_h}^2\right) \log ({r_h}) \alpha _{\theta _1}^4 {f_{\theta_2z}}''(r)}{4 \pi ^{5/2}
   \left(6 a^2+{r_h}^2\right) \alpha _{\theta _2}^3}\nonumber\\
      & & +\frac{7 \sqrt{{g_s}} M \left(\frac{1}{N}\right)^{9/5} {N_f}
   (r-{r_h})^2 {r_h}^2 \left(9 a^2+{r_h}^2\right) \log ({r_h}) \left(81 \sqrt{2} \alpha _{\theta _1}^3+10 \sqrt{3} \alpha
   _{\theta _2}^2\right) {f_{\theta_2x}}''(r)}{8 \sqrt{2} \pi ^{5/2} \left(6 a^2+{r_h}^2\right) \alpha _{\theta _2}}\nonumber\\
   & & -\frac{3
   (r-{r_h})^2 {r_h}^2 \left(9 a^2+{r_h}^2\right) {f_{\theta_2y}}''(r)}{{g_s} N \pi  \left(6 a^2+{r_h}^2\right)}-\frac{3
   (r-{r_h})^2 {r_h}^2 \left(9 a^2+{r_h}^2\right) {f_{xz}}''(r)}{{g_s} N \pi  \left(6 a^2+{r_h}^2\right)}
   \nonumber\\
   & & +\frac{243
   \left(\frac{1}{N}\right)^{12/5} (r-{r_h})^2 {r_h}^2 \left(9 a^2+{r_h}^2\right) \alpha _{\theta _1}^4 \alpha _{\theta _2}^2
   {f_{xx}}''(r)}{4 {g_s} \pi  \left(6 a^2+{r_h}^2\right)}\nonumber\\
   & &+\frac{3 \sqrt{3} \sqrt{{g_s}} M \left(\frac{1}{N}\right)^{9/5}
   {N_f} (r-{r_h})^2 {r_h}^2 \left(9 a^2+{r_h}^2\right) \log ({r_h}) \left(27 \sqrt{3} \alpha _{\theta _1}^3+5
   \sqrt{2} \alpha _{\theta _2}^2\right) {f_{xy}}''(r)}{32 \pi ^{5/2} \left(6 a^2+{r_h}^2\right) \alpha _{\theta _2}}\nonumber\\
   & &-\frac{3
   \sqrt{3} \sqrt{{g_s}} M \left(\frac{1}{N}\right)^{9/5} {N_f} (r-{r_h})^2 {r_h}^2 \left(9 a^2+{r_h}^2\right) \log
   ({r_h}) \left(27 \sqrt{3} \alpha _{\theta _1}^3+5 \sqrt{2} \alpha _{\theta _2}^2\right) {f_{yz}}''(r)}{32 \pi ^{5/2} \left(6
   a^2+{r_h}^2\right) \alpha _{\theta _2}}\nonumber\\
   & &+\frac{3 {g_s}^2 M^2 \left(\frac{1}{N}\right)^{13/5} {N_f}^2 (r-{r_h})^2
   {r_h}^2 \left(9 a^2+{r_h}^2\right) \log ^2({r_h}) \left(2187 \alpha _{\theta _1}^6+270 \sqrt{6} \alpha _{\theta _2}^2
   \alpha _{\theta _1}^3+50 \alpha _{\theta _2}^4\right) {f_{yy}}''(r)}{512 \pi ^4 \left(6 a^2+{r_h}^2\right) \alpha _{\theta
   _2}^2}\nonumber\\
   & & +\frac{9 {g_s} M^2 {N_f} (r-{r_h})^2 {r_h}^2 \left(9 a^2+{r_h}^2\right) \log ^2({r_h}) {f_t}''(r)}{4
   N^2 \pi ^3 \left(6 a^2+{r_h}^2\right)}+\frac{243 {g_s}^3 \log N  M^4 {N_f}^2 (r-{r_h}) {r_h} \left(9
   a^2+{r_h}^2\right) f(r) \log ^4({r_h})}{16 N^3 \pi ^5 \left(6 a^2+{r_h}^2\right) \log \left({r_h}^6+9 a^2
   {r_h}^4\right)}\nonumber\\
   & & +\frac{5 \left(\frac{1}{N}\right)^{3/5} (r-{r_h})^2 {r_h}^4 \left(9 a^2+{r_h}^2\right) \left(108 b^2
   {r_h}^2+1\right)^2 {f_r}(r)}{9 {g_s} \pi  \left({r_h}^2-3 a^2\right)^2 \left(6 a^2+{r_h}^2\right) \log
   ^2({r_h}) \alpha _{\theta _1}^2}+\frac{4 \left(\frac{1}{N}\right)^{2/5} (r-{r_h})^2 {r_h}^4 \left(9 a^2+{r_h}^2\right)
   \left(108 b^2 {r_h}^2+1\right)^2 {f_{yz}}(r)}{3 {g_s} \pi  \left({r_h}^2-3 a^2\right)^2 \left(6 a^2+{r_h}^2\right)
   \log ^2({r_h}) \alpha _{\theta _2}^2}\nonumber\\
   & & -\frac{2 \left(\frac{1}{N}\right)^{2/5} (r-{r_h})^2 {r_h}^4 \left(9
   a^2+{r_h}^2\right) \left(108 b^2 {r_h}^2+1\right)^2 {f_{yy}}(r)}{3 {g_s} \pi  \left({r_h}^2-3 a^2\right)^2 \left(6
   a^2+{r_h}^2\right) \log ^2({r_h}) \alpha _{\theta _2}^2}-\frac{245 \left(\frac{1}{N}\right)^{2/5} \sqrt{\frac{\pi }{6}}
   (r-{r_h}) {r_h}^5 {f_{\theta_1y}}(r) \alpha _{\theta _1}^4}{192 {g_s}^{5/2} M {N_f} \left({r_h}^2-3 a^2\right)^2 \log
   ^3({r_h}) \alpha _{\theta _2}^3}\nonumber\\
    & & -\frac{245 \left(\frac{1}{N}\right)^{2/5} \sqrt{\frac{\pi }{6}} (r-{r_h}) {r_h}^5
   {f_{\theta_2z}}(r) \alpha _{\theta _1}^4}{192 {g_s}^{5/2} M {N_f} \left({r_h}^2-3 a^2\right)^2 \log ^3({r_h}) \alpha
   _{\theta _2}^3}-\frac{49 \pi ^2 (r-{r_h}) {r_h}^5 {f_{zz}}(r)}{1944 {g_s}^4 M^2 {N_f}^2 \left({r_h}^2-3
   a^2\right)^2 \log ^4({r_h})}\nonumber\\
   & & +\frac{49 \pi ^2 (r-{r_h}) {r_h}^5 {f_{x^{10} x^{10}}}(r)}{1296 {g_s}^4 M^2 {N_f}^2
   \left({r_h}^2-3 a^2\right)^2 \log ^4({r_h})}+\frac{49 \pi ^2 (r-{r_h}) {r_h}^5 {f_{\theta_1z}}(r)}{972 {g_s}^4 M^2
   {N_f}^2 \left({r_h}^2-3 a^2\right)^2 \log ^4({r_h})}+\frac{245 \pi ^2 (r-{r_h}) {r_h}^5 {f_{\theta_2y}}(r)}{3888
   {g_s}^4 M^2 {N_f}^2 \left({r_h}^2-3 a^2\right)^2 \log ^4({r_h})} = 0.
\end{eqnarray*}
}

(ii) \underline{${\rm EOM}_{x^ix^i}, i=1, 2, 3$}

{\scriptsize
\begin{eqnarray*}
& & \frac{\left(9 b^2+1\right)^4 \left(39 b^2-4\right) M \left(\frac{1}{N}\right)^{9/4} (r-{r_h}) {r_h} \log ({r_h}) \Sigma_1 b^8}{9 \left(3
   b^2-1\right)^5 \left(6 b^2+1\right)^4 {g_s} \log N ^2 {N_f} \pi ^2 \alpha _{\theta _2}^3}-\frac{243
   \left(\frac{1}{N}\right)^{12/5} {r_h}^2 {f_{xx}}(r) \alpha _{\theta _1}^4 \alpha _{\theta _2}^2}{8 {g_s} \pi  \log
   \left({r_h}^6+9 a^2 {r_h}^4\right)}\nonumber\\
   & & +\frac{3 {r_h}^2 f(r) \left(6 \left(-108 a^6+9 {r_h}^4
   a^2+{r_h}^6\right)+\left({r_h}^6+14 a^2 {r_h}^4+57 a^4 {r_h}^2\right) \log \left({r_h}^6+9 a^2
   {r_h}^4\right)\right)}{2 {g_s} N \pi  \left({r_h}^2-3 a^2\right) \left(6 a^2+{r_h}^2\right)^2 \log \left({r_h}^6+9
   a^2 {r_h}^4\right)}\nonumber\\
   & & +\frac{81 {g_s}^3 M^4 {N_f}^2 {r_h}^2 {f_t}(r) \log ^4({r_h}) \left(16 \left(-108 a^6+9
   {r_h}^4 a^2+{r_h}^6\right)+\left(2 {r_h}^6+27 a^2 {r_h}^4+117 a^4 {r_h}^2\right) \log \left({r_h}^6+9 a^2
   {r_h}^4\right)\right)}{256 N^3 \pi ^5 \left({r_h}^2-3 a^2\right) \left(6 a^2+{r_h}^2\right)^2 \log \left({r_h}^6+9 a^2
   {r_h}^4\right)}\nonumber\\
   & & +\frac{245 \left(\frac{1}{N}\right)^{4/5} \sqrt{\frac{\pi }{6}} {r_h}^6 {f_{xy}}(r) \left(27 \sqrt{6} \alpha
   _{\theta _1}^3+10 \alpha _{\theta _2}^2\right)}{165888 {g_s}^{5/2} M {N_f} \left({r_h}^2-3 a^2\right)^2 \log ^3({r_h})
   \alpha _{\theta _2}}\nonumber\\
   & & -\frac{343 \left(\frac{1}{N}\right)^{4/5} \sqrt{\frac{\pi }{2}} {r_h}^6 {f_{\theta_1x}}(r) \left(81 \sqrt{2} \alpha
   _{\theta _1}^3+10 \sqrt{3} \alpha _{\theta _2}^2\right)}{124416 {g_s}^{5/2} M {N_f} \left({r_h}^2-3 a^2\right)^2 \log
   ^3({r_h}) \alpha _{\theta _2}}\nonumber\\
   & & +\frac{1715 \left(\frac{1}{N}\right)^{4/5} \sqrt{\frac{\pi }{2}} {r_h}^6 {f_{\theta_2x}}(r) \left(81
   \sqrt{2} \alpha _{\theta _1}^3+10 \sqrt{3} \alpha _{\theta _2}^2\right)}{373248 {g_s}^{5/2} M {N_f} \left({r_h}^2-3
   a^2\right)^2 \log ^3({r_h}) \alpha _{\theta _2}}\nonumber\\
   & & -\frac{3479 \left(\frac{1}{N}\right)^{4/5} \sqrt{\frac{\pi }{2}} {r_h}^6
   {f_{xz}}(r) \left(81 \sqrt{2} \alpha _{\theta _1}^3+10 \sqrt{3} \alpha _{\theta _2}^2\right)}{1492992 {g_s}^{5/2} M {N_f}
   \left({r_h}^2-3 a^2\right)^2 \log ^3({r_h}) \alpha _{\theta _2}}\nonumber\\
   & & +\frac{\left({r_h}^5+9 a^2 {r_h}^3\right) f'(r)}{6
   {g_s} N \pi  a^2+{g_s} N \pi  {r_h}^2}-\frac{\left({r_h}^5+9 a^2 {r_h}^3\right) {f_{zz}}'(r)}{12 {g_s} N
   \pi  a^2+2 {g_s} N \pi  {r_h}^2}+\frac{\left({r_h}^5+9 a^2 {r_h}^3\right) {f_{x^{10} x^{10}}}'(r)}{12 {g_s} N \pi  a^2+2
   {g_s} N \pi  {r_h}^2}\nonumber\\
      & & +\frac{\left(3 {r_h}^5+27 a^2 {r_h}^3\right) {f_{\theta_1z}}'(r)}{24 {g_s} N \pi  a^2+4 {g_s}
   N \pi  {r_h}^2}-\frac{3 a^2 (r-{r_h}) {r_h}^2 \left(9 a^2+{r_h}^2\right) {f_{\theta_1x}}'(r)}{2 {g_s} N \pi
   \left({r_h}^2-3 a^2\right) \left(6 a^2+{r_h}^2\right)}-\frac{81 \sqrt{\frac{3}{2}} \sqrt{{g_s}} M
   \left(\frac{1}{N}\right)^{7/5} {N_f} {r_h}^3 \left(9 a^2+{r_h}^2\right) \log ({r_h}) \alpha _{\theta _1}^4
   {f_{\theta_1y}}'(r)}{16 \pi ^{5/2} \left(6 a^2+{r_h}^2\right) \alpha _{\theta _2}^3}\nonumber\\
   & & -\frac{81 \sqrt{\frac{3}{2}} \sqrt{{g_s}} M
   \left(\frac{1}{N}\right)^{7/5} {N_f} {r_h}^3 \left(9 a^2+{r_h}^2\right) \log ({r_h}) \alpha _{\theta _1}^4
   {f_{\theta_2z}}'(r)}{16 \pi ^{5/2} \left(6 a^2+{r_h}^2\right) \alpha _{\theta _2}^3}\nonumber\\
   & & +\frac{81 \sqrt{\frac{3}{2}} a^2 \sqrt{{g_s}} M
   \left(\frac{1}{N}\right)^{8/5} {N_f} (r-{r_h}) {r_h}^2 \left(9 a^2+{r_h}^2\right) \log ({r_h}) \alpha _{\theta
   _1}^2 {f_{\theta_2x}}'(r)}{8 \pi ^{5/2} \left({r_h}^2-3 a^2\right) \left(6 a^2+{r_h}^2\right) \alpha _{\theta _2}}\nonumber\\
   & & +\frac{\left(3
   {r_h}^5+27 a^2 {r_h}^3\right) {f_{\theta_2y}}'(r)}{24 {g_s} N \pi  a^2+4 {g_s} N \pi  {r_h}^2}+\frac{\left(3
   {r_h}^5+27 a^2 {r_h}^3\right) {f_{xz}}'(r)}{24 {g_s} N \pi  a^2+4 {g_s} N \pi  {r_h}^2}-\frac{243
   \left(\frac{1}{N}\right)^{12/5} {r_h}^3 \left(9 a^2+{r_h}^2\right) \alpha _{\theta _1}^4 \alpha _{\theta _2}^2
   {f_{xx}}'(r)}{16 {g_s} \pi  \left(6 a^2+{r_h}^2\right)}\nonumber\\
   & & -\frac{81 \sqrt{\frac{3}{2}} a^2 \sqrt{{g_s}} M
   \left(\frac{1}{N}\right)^{7/5} {N_f} (r-{r_h}) {r_h}^2 \left(9 a^2+{r_h}^2\right) \log ({r_h}) \alpha _{\theta
   _1}^4 {f_{xy}}'(r)}{8 \pi ^{5/2} \left(-18 a^4+3 {r_h}^2 a^2+{r_h}^4\right) \alpha _{\theta _2}^3}\nonumber\\
   & & -\frac{81
   \sqrt{\frac{3}{2}} a^2 \sqrt{{g_s}} M \left(\frac{1}{N}\right)^{7/5} {N_f} (r-{r_h}) {r_h}^2 \left(9
   a^2+{r_h}^2\right) \log ({r_h}) \alpha _{\theta _1}^4 {f_{yz}}'(r)}{8 \pi ^{5/2} \left(-18 a^4+3 {r_h}^2
   a^2+{r_h}^4\right) \alpha _{\theta _2}^3}\nonumber\\
   & & -\frac{27 a^2 {g_s}^2 M^2 \left(\frac{1}{N}\right)^{12/5} {N_f}^2 (r-{r_h})
   {r_h}^2 \left(9 a^2+{r_h}^2\right) \log ^2({r_h}) \alpha _{\theta _1}^2 \left(2187 \sqrt{2} \alpha _{\theta _1}^6+540
   \sqrt{3} \alpha _{\theta _2}^2 \alpha _{\theta _1}^3+50 \sqrt{2} \alpha _{\theta _2}^4\right) {f_{yy}}'(r)}{512 \sqrt{2} \pi ^4
   \left(-18 a^4+3 {r_h}^2 a^2+{r_h}^4\right) \alpha _{\theta _2}^4}\nonumber\\
         & & -\frac{\left({r_h}^5+9 a^2 {r_h}^3\right)
   {f_r}'(r)}{24 {g_s} N \pi  a^2+4 {g_s} N \pi  {r_h}^2}+\frac{\left(3 {r_h}^5+27 a^2 {r_h}^3\right)
   {f_t}'(r)}{24 {g_s} N \pi  a^2+4 {g_s} N \pi  {r_h}^2}+\frac{(r-{r_h}) {r_h}^3 \left(9 a^2+{r_h}^2\right)
   f''(r)}{{g_s} N \pi  \left(6 a^2+{r_h}^2\right)}\nonumber\\
   & & -\frac{(r-{r_h}) {r_h}^3 \left(9 a^2+{r_h}^2\right)
   {f_{zz}}''(r)}{2 {g_s} N \pi  \left(6 a^2+{r_h}^2\right)}+\frac{(r-{r_h}) {r_h}^3 \left(9 a^2+{r_h}^2\right)
   {f_{x^{10} x^{10}}}''(r)}{2 {g_s} N \pi  \left(6 a^2+{r_h}^2\right)}+\frac{3 (r-{r_h}) {r_h}^3 \left(9 a^2+{r_h}^2\right)
   {f_{\theta_1z}}''(r)}{4 {g_s} N \pi  \left(6 a^2+{r_h}^2\right)}\nonumber\\
   & & +\frac{7 \sqrt{{g_s}} M \left(\frac{1}{N}\right)^{9/5}
   {N_f} (r-{r_h}) {r_h}^3 \left(9 a^2+{r_h}^2\right) \log ({r_h}) \left(81 \sqrt{2} \alpha _{\theta _1}^3+10 \sqrt{3}
   \alpha _{\theta _2}^2\right) {f_{\theta_1x}}''(r)}{32 \sqrt{2} \pi ^{5/2} \left(6 a^2+{r_h}^2\right) \alpha _{\theta _2}}\nonumber\\
    & & -\frac{81
   \sqrt{\frac{3}{2}} \sqrt{{g_s}} M \left(\frac{1}{N}\right)^{7/5} {N_f} (r-{r_h}) {r_h}^3 \left(9 a^2+{r_h}^2\right)
   \log ({r_h}) \alpha _{\theta _1}^4 {f_{\theta_1y}}''(r)}{16 \pi ^{5/2} \left(6 a^2+{r_h}^2\right) \alpha _{\theta _2}^3}-\frac{81
   \sqrt{\frac{3}{2}} \sqrt{{g_s}} M \left(\frac{1}{N}\right)^{7/5} {N_f} (r-{r_h}) {r_h}^3 \left(9 a^2+{r_h}^2\right)
   \log ({r_h}) \alpha _{\theta _1}^4 {f_{\theta_2z}}''(r)}{16 \pi ^{5/2} \left(6 a^2+{r_h}^2\right) \alpha _{\theta _2}^3}\nonumber\\
        & & -\frac{7
   \sqrt{{g_s}} M \left(\frac{1}{N}\right)^{9/5} {N_f} (r-{r_h}) {r_h}^3 \left(9 a^2+{r_h}^2\right) \log ({r_h})
   \left(81 \sqrt{2} \alpha _{\theta _1}^3+10 \sqrt{3} \alpha _{\theta _2}^2\right) {f_{\theta_2x}}''(r)}{32 \sqrt{2} \pi ^{5/2} \left(6
   a^2+{r_h}^2\right) \alpha _{\theta _2}}+\frac{3 (r-{r_h}) {r_h}^3 \left(9 a^2+{r_h}^2\right) {f_{\theta_2y}}''(r)}{4
   {g_s} N \pi  \left(6 a^2+{r_h}^2\right)}\nonumber\\
    \end{eqnarray*}
\begin{eqnarray*}
   & & +\frac{3 (r-{r_h}) {r_h}^3 \left(9 a^2+{r_h}^2\right) {f_{xz}}''(r)}{4
   {g_s} N \pi  \left(6 a^2+{r_h}^2\right)}-\frac{243 \left(\frac{1}{N}\right)^{12/5} (r-{r_h}) {r_h}^3 \left(9
   a^2+{r_h}^2\right) \alpha _{\theta _1}^4 \alpha _{\theta _2}^2 {f_{xx}}''(r)}{16 {g_s} \pi  \left(6
   a^2+{r_h}^2\right)}\nonumber\\
   & & -\frac{3 \sqrt{3} \sqrt{{g_s}} M \left(\frac{1}{N}\right)^{9/5} {N_f} (r-{r_h}) {r_h}^3 \left(9
   a^2+{r_h}^2\right) \log ({r_h}) \left(27 \sqrt{3} \alpha _{\theta _1}^3+5 \sqrt{2} \alpha _{\theta _2}^2\right)
   {f_{xy}}''(r)}{128 \pi ^{5/2} \left(6 a^2+{r_h}^2\right) \alpha _{\theta _2}}\nonumber\\
   & & +\frac{3 \sqrt{3} \sqrt{{g_s}} M
   \left(\frac{1}{N}\right)^{9/5} {N_f} (r-{r_h}) {r_h}^3 \left(9 a^2+{r_h}^2\right) \log ({r_h}) \left(27 \sqrt{3}
   \alpha _{\theta _1}^3+5 \sqrt{2} \alpha _{\theta _2}^2\right) {f_{yz}}''(r)}{128 \pi ^{5/2} \left(6 a^2+{r_h}^2\right) \alpha
   _{\theta _2}}\nonumber\\
 & & -\frac{3 {g_s}^2 M^2 \left(\frac{1}{N}\right)^{13/5} {N_f}^2 (r-{r_h}) {r_h}^3 \left(9
   a^2+{r_h}^2\right) \log ^2({r_h}) \left(2187 \alpha _{\theta _1}^6+270 \sqrt{6} \alpha _{\theta _2}^2 \alpha _{\theta _1}^3+50
   \alpha _{\theta _2}^4\right) {f_{yy}}''(r)}{2048 \pi ^4 \left(6 a^2+{r_h}^2\right) \alpha _{\theta _2}^2}
   \nonumber\\
  & & +\frac{(r-{r_h})
   {r_h}^3 \left(9 a^2+{r_h}^2\right) {f_t}''(r)}{2 {g_s} N \pi  \left(6 a^2+{r_h}^2\right)}-\frac{5
   \left(\frac{1}{N}\right)^{3/5} (r-{r_h}) {r_h}^5 \left(9 a^2+{r_h}^2\right) \left(108 b^2 {r_h}^2+1\right)^2
   {f_r}(r)}{36 {g_s} \pi  \left({r_h}^2-3 a^2\right)^2 \left(6 a^2+{r_h}^2\right) \log ^2({r_h}) \alpha _{\theta
   _1}^2}\nonumber\\
   &&-\frac{\left(\frac{1}{N}\right)^{2/5} (r-{r_h}) {r_h}^5 \left(9 a^2+{r_h}^2\right) \left(108 b^2
   {r_h}^2+1\right)^2 {f_{yz}}(r)}{3 {g_s} \pi  \left({r_h}^2-3 a^2\right)^2 \left(6 a^2+{r_h}^2\right) \log
   ^2({r_h}) \alpha _{\theta _2}^2}\nonumber\\
   & & +\frac{\left(\frac{1}{N}\right)^{2/5} (r-{r_h}) {r_h}^5 \left(9 a^2+{r_h}^2\right)
   \left(108 b^2 {r_h}^2+1\right)^2 {f_{yy}}(r)}{6 {g_s} \pi  \left({r_h}^2-3 a^2\right)^2 \left(6 a^2+{r_h}^2\right)
   \log ^2({r_h}) \alpha _{\theta _2}^2}\nonumber\\
   & & +\frac{245 \left(\frac{1}{N}\right)^{2/5} \sqrt{\frac{\pi }{6}} {r_h}^6 {f_{\theta_1y}}(r)
   \alpha _{\theta _1}^4}{768 {g_s}^{5/2} M {N_f} \left({r_h}^2-3 a^2\right)^2 \log ^3({r_h}) \alpha _{\theta
   _2}^3}+\frac{245 \left(\frac{1}{N}\right)^{2/5} \sqrt{\frac{\pi }{6}} {r_h}^6 {f_{\theta_2z}}(r) \alpha _{\theta _1}^4}{768
   {g_s}^{5/2} M {N_f} \left({r_h}^2-3 a^2\right)^2 \log ^3({r_h}) \alpha _{\theta _2}^3}\nonumber\\
   & & +\frac{49 \pi ^2 {r_h}^6
   {f_{zz}}(r)}{7776 {g_s}^4 M^2 {N_f}^2 \left({r_h}^2-3 a^2\right)^2 \log ^4({r_h})}-\frac{49 \pi ^2 {r_h}^6
   {f_{x^{10} x^{10}}}(r)}{5184 {g_s}^4 M^2 {N_f}^2 \left({r_h}^2-3 a^2\right)^2 \log ^4({r_h})}\nonumber\\
   & & -\frac{49 \pi ^2 {r_h}^6
   {f_{\theta_1z}}(r)}{3888 {g_s}^4 M^2 {N_f}^2 \left({r_h}^2-3 a^2\right)^2 \log ^4({r_h})}
    -\frac{245 \pi ^2 {r_h}^6
   {f_{\theta_2y}}(r)}{15552 {g_s}^4 M^2 {N_f}^2 \left({r_h}^2-3 a^2\right)^2 \log ^4({r_h})}=0.
\end{eqnarray*}
}

(iii) ${\rm EOM}_{rr}$

{\scriptsize
\begin{eqnarray*}
& & -\frac{3 \left(9 b^2+1\right)^3 M \left(\frac{1}{N}\right)^{5/4} \beta  \log ({r_h}) \Sigma_1 b^{10}}{\left(3 b^2-1\right)^5 \left(6
   b^2+1\right)^3 \log N ^2 {N_f} \pi  {r_h}^2 \alpha _{\theta _2}^3}-\frac{243 \left(\frac{1}{N}\right)^{7/5} \left(6
   a^2+{r_h}^2\right) {f_{xx}}(r) \alpha _{\theta _1}^4 \alpha _{\theta _2}^2}{8 (r-{r_h}) {r_h} \left(9
   a^2+{r_h}^2\right) \log \left({r_h}^6+9 a^2 {r_h}^4\right)}\nonumber\\
   & &+\frac{81 {g_s}^4 M^4 {N_f}^2 {f_t}(r) \log
   ^4({r_h}) \left(16 \left(-108 a^6+9 {r_h}^4 a^2+{r_h}^6\right)+\left(2 {r_h}^6+27 a^2 {r_h}^4+117 a^4
   {r_h}^2\right) \log \left({r_h}^6+9 a^2 {r_h}^4\right)\right)}{256 N^2 \pi ^4 (r-{r_h}) {r_h} \left({r_h}^2-3
   a^2\right) \left(6 a^2+{r_h}^2\right) \left(9 a^2+{r_h}^2\right) \log \left({r_h}^6+9 a^2 {r_h}^4\right)}
   \nonumber\\
   & & +\frac{245 \pi
   ^{3/2} {r_h}^3 \left(6 a^2+{r_h}^2\right) {f_{xy}}(r) \left(27 \sqrt{6} \alpha _{\theta _1}^3+10 \alpha _{\theta
   _2}^2\right)}{165888 \sqrt{6} {g_s}^{3/2} M \sqrt[5]{\frac{1}{N}} {N_f} (r-{r_h}) \left({r_h}^2-3 a^2\right)^2 \left(9
   a^2+{r_h}^2\right) \log ^3({r_h}) \alpha _{\theta _2}}\nonumber\\
   &&-\frac{343 \pi ^{3/2} {r_h}^3 \left(6 a^2+{r_h}^2\right)
   {f_{\theta_1x}}(r) \left(81 \sqrt{2} \alpha _{\theta _1}^3+10 \sqrt{3} \alpha _{\theta _2}^2\right)}{124416 \sqrt{2} {g_s}^{3/2} M
   \sqrt[5]{\frac{1}{N}} {N_f} (r-{r_h}) \left({r_h}^2-3 a^2\right)^2 \left(9 a^2+{r_h}^2\right) \log ^3({r_h}) \alpha
   _{\theta _2}}\nonumber\\
   & &+\frac{1715 \pi ^{3/2} {r_h}^3 \left(6 a^2+{r_h}^2\right) {f_{\theta_2x}}(r) \left(81 \sqrt{2} \alpha _{\theta _1}^3+10
   \sqrt{3} \alpha _{\theta _2}^2\right)}{373248 \sqrt{2} {g_s}^{3/2} M \sqrt[5]{\frac{1}{N}} {N_f} (r-{r_h})
   \left({r_h}^2-3 a^2\right)^2 \left(9 a^2+{r_h}^2\right) \log ^3({r_h}) \alpha _{\theta _2}}\nonumber\\
   &&-\frac{3479 \pi ^{3/2}
   {r_h}^3 \left(6 a^2+{r_h}^2\right) {f_{xz}}(r) \left(81 \sqrt{2} \alpha _{\theta _1}^3+10 \sqrt{3} \alpha _{\theta
   _2}^2\right)}{1492992 \sqrt{2} {g_s}^{3/2} M \sqrt[5]{\frac{1}{N}} {N_f} (r-{r_h}) \left({r_h}^2-3 a^2\right)^2 \left(9
   a^2+{r_h}^2\right) \log ^3({r_h}) \alpha _{\theta _2}}\nonumber\\
   & & +\frac{3 f'(r)}{4 (r-{r_h})}-\frac{{f_{zz}}'(r)}{4
   (r-{r_h})}+\frac{{f_{x^{10} x^{10}}}'(r)}{4 r-4 {r_h}}+\frac{{f_{\theta_1z}}'(r)}{4 r-4 {r_h}}+\frac{3 a^2 {f_{\theta_1x}}'(r)}{6 a^2
   {r_h}-2 {r_h}^3}\nonumber\\
   & &-\frac{27 \sqrt{\frac{3}{2}} {g_s}^{3/2} M \left(\frac{1}{N}\right)^{2/5} {N_f} \log ({r_h}) \alpha
   _{\theta _1}^4 {f_{\theta_1y}}'(r)}{16 \pi ^{3/2} (r-{r_h}) \alpha _{\theta _2}^3}-\frac{27 \sqrt{\frac{3}{2}} {g_s}^{3/2} M
   \left(\frac{1}{N}\right)^{2/5} {N_f} \log ({r_h}) \alpha _{\theta _1}^4 {f_{\theta_2z}}'(r)}{16 \pi ^{3/2} (r-{r_h}) \alpha
   _{\theta _2}^3}\nonumber\\
     \end{eqnarray*}
\begin{eqnarray*}
   & & +\frac{81 \sqrt{\frac{3}{2}} a^2 {g_s}^{3/2} M \left(\frac{1}{N}\right)^{3/5} {N_f} \log ({r_h}) \alpha
   _{\theta _1}^2 {f_{\theta_2x}}'(r)}{8 \pi ^{3/2} {r_h} \left({r_h}^2-3 a^2\right) \alpha _{\theta _2}}+\frac{{f_{\theta_2y}}'(r)}{4 r-4
   {r_h}}+\frac{{f_{xz}}'(r)}{4 r-4 {r_h}}\nonumber\\
   & & -\frac{243 \left(\frac{1}{N}\right)^{7/5} \alpha _{\theta _1}^4 \alpha _{\theta _2}^2
   {f_{xx}}'(r)}{32 (r-{r_h})}-\frac{81 \sqrt{\frac{3}{2}} a^2 {g_s}^{3/2} M \left(\frac{1}{N}\right)^{2/5} {N_f} \log
   ({r_h}) \alpha _{\theta _1}^4 {f_{xy}}'(r)}{8 \pi ^{3/2} \left({r_h}^3-3 a^2 {r_h}\right) \alpha _{\theta _2}^3}\nonumber\\
   & &-\frac{81
   \sqrt{\frac{3}{2}} a^2 {g_s}^{3/2} M \left(\frac{1}{N}\right)^{2/5} {N_f} \log ({r_h}) \alpha _{\theta _1}^4
   {f_{yz}}'(r)}{8 \pi ^{3/2} \left({r_h}^3-3 a^2 {r_h}\right) \alpha _{\theta _2}^3}-\frac{27 a^2 {g_s}^3 M^2
   \left(\frac{1}{N}\right)^{7/5} {N_f}^2 \log ^2({r_h}) \alpha _{\theta _1}^2 \left(2187 \sqrt{2} \alpha _{\theta _1}^6+540
   \sqrt{3} \alpha _{\theta _2}^2 \alpha _{\theta _1}^3+50 \sqrt{2} \alpha _{\theta _2}^4\right) {f_{yy}}'(r)}{512 \sqrt{2} \pi ^3
   \left({r_h}^3-3 a^2 {r_h}\right) \alpha _{\theta _2}^4}\nonumber\\
   & & +\frac{\left(\frac{6 a^2}{{r_h}^2-3 a^2}+\frac{12 \left(6
   a^2+{r_h}^2\right)}{\left(9 a^2+{r_h}^2\right) \log \left({r_h}^6+9 a^2 {r_h}^4\right)}\right) {f_r}'(r)}{8
   {r_h}}+\frac{3 \left(\frac{6 a^2}{{r_h}^2-3 a^2}+\frac{12 \left(6 a^2+{r_h}^2\right)}{\left(9 a^2+{r_h}^2\right) \log
   \left({r_h}^6+9 a^2 {r_h}^4\right)}+4\right) {f_t}'(r)}{8 {r_h}}+\frac{27 {g_s}^2 M^2 {N_f} \log ^2({r_h})
   f''(r)}{16 N \pi ^2}\nonumber\\
   & & +\frac{49 \sqrt[5]{\frac{1}{N}} \alpha _{\theta _2}^2 {f_{zz}}''(r)}{324 \alpha _{\theta _1}^2}+\frac{9
   {g_s}^2 M^2 {N_f} \log ^2({r_h}) {f_{x^{10} x^{10}}}''(r)}{32 N \pi ^2}-\frac{{f_{\theta_1z}}''(r)}{4}-\frac{7 {g_s}^{3/2} M
   \left(\frac{1}{N}\right)^{4/5} {N_f} \log ({r_h}) \left(81 \sqrt{2} \alpha _{\theta _1}^3+10 \sqrt{3} \alpha _{\theta
   _2}^2\right) {f_{\theta_1x}}''(r)}{96 \sqrt{2} \pi ^{3/2} \alpha _{\theta _2}}\nonumber\\
   & & +\frac{27 \sqrt{\frac{3}{2}} {g_s}^{3/2} M
   \left(\frac{1}{N}\right)^{2/5} {N_f} \log ({r_h}) \alpha _{\theta _1}^4 {f_{\theta_1y}}''(r)}{16 \pi ^{3/2} \alpha _{\theta
   _2}^3}+\frac{27 \sqrt{\frac{3}{2}} {g_s}^{3/2} M \left(\frac{1}{N}\right)^{2/5} {N_f} \log ({r_h}) \alpha _{\theta _1}^4
   {f_{\theta_2z}}''(r)}{16 \pi ^{3/2} \alpha _{\theta _2}^3}\nonumber\\
   & & +\frac{7 {g_s}^{3/2} M \left(\frac{1}{N}\right)^{4/5} {N_f} \log
   ({r_h}) \left(81 \sqrt{2} \alpha _{\theta _1}^3+10 \sqrt{3} \alpha _{\theta _2}^2\right) {f_{\theta_2x}}''(r)}{96 \sqrt{2} \pi ^{3/2}
   \alpha _{\theta _2}}-\frac{{f_{\theta_2y}}''(r)}{4}-\frac{{f_{xz}}''(r)}{4}\nonumber\\
   & & -\frac{3}{32} \left(\frac{1}{N}\right)^{8/5} \alpha _{\theta
   _2}^2 \left(243 \sqrt{6} \alpha _{\theta _1}^5+41 \alpha _{\theta _2}^2 \alpha _{\theta _1}^2\right) {f_{xx}}''(r)+\frac{\sqrt{3}
   {g_s}^{3/2} M \left(\frac{1}{N}\right)^{4/5} {N_f} \log ({r_h}) \left(27 \sqrt{3} \alpha _{\theta _1}^3+5 \sqrt{2} \alpha
   _{\theta _2}^2\right) {f_{xy}}''(r)}{128 \pi ^{3/2} \alpha _{\theta _2}}\nonumber\\
   & & -\frac{\sqrt{3} {g_s}^{3/2} M
   \left(\frac{1}{N}\right)^{4/5} {N_f} \log ({r_h}) \left(27 \sqrt{3} \alpha _{\theta _1}^3+5 \sqrt{2} \alpha _{\theta
   _2}^2\right) {f_{yz}}''(r)}{128 \pi ^{3/2} \alpha _{\theta _2}}\nonumber\\
   & &  +\frac{{g_s}^3 M^2 \left(\frac{1}{N}\right)^{9/5} {N_f}^2
   \log ^2({r_h}) \left(19 \alpha _{\theta _2}^2-81 \sqrt{6} \alpha _{\theta _1}^3\right) \left(2187 \alpha _{\theta _1}^6+270
   \sqrt{6} \alpha _{\theta _2}^2 \alpha _{\theta _1}^3+50 \alpha _{\theta _2}^4\right) {f_{yy}}''(r)}{36864 \pi ^3 \alpha _{\theta
   _1}^2 \alpha _{\theta _2}^2}\nonumber\\
   & & +\frac{9 {g_s}^2 M^2 {N_f} \log ^2({r_h}) {f_t}''(r)}{16 N \pi ^2}-\frac{243 {g_s}^4
   \log N  M^4 {N_f}^2 f(r) \log ^4({r_h})}{64 N^2 \pi ^4 (r-{r_h}) {r_h} \log \left({r_h}^6+9 a^2
   {r_h}^4\right)}-\frac{5 {r_h}^2 {f_r}(r)}{36 \left(\frac{1}{N}\right)^{2/5} \left({r_h}^2-3 a^2\right)^2 \log
   ^2({r_h}) \alpha _{\theta _1}^2}\nonumber\\
   & & +\frac{5 {r_h}^2 {f_{yz}}(r)}{3 \left(\frac{1}{N}\right)^{3/5} \left({r_h}^2-3
   a^2\right)^2 \log ^2({r_h}) \alpha _{\theta _2}^2}-\frac{5 {r_h}^2 {f_{yy}}(r)}{6 \left(\frac{1}{N}\right)^{3/5}
   \left({r_h}^2-3 a^2\right)^2 \log ^2({r_h}) \alpha _{\theta _2}^2}\nonumber\\
   & &+\frac{245 \pi ^{3/2} {r_h}^3 \left(6
   a^2+{r_h}^2\right) {f_{\theta_1y}}(r) \alpha _{\theta _1}^4}{768 \sqrt{6} {g_s}^{3/2} M \left(\frac{1}{N}\right)^{3/5} {N_f}
   (r-{r_h}) \left({r_h}^2-3 a^2\right)^2 \left(9 a^2+{r_h}^2\right) \log ^3({r_h}) \alpha _{\theta _2}^3}\nonumber\\
   & & +\frac{245 \pi
   ^{3/2} {r_h}^3 \left(6 a^2+{r_h}^2\right) {f_{\theta_2z}}(r) \alpha _{\theta _1}^4}{768 \sqrt{6} {g_s}^{3/2} M
   \left(\frac{1}{N}\right)^{3/5} {N_f} (r-{r_h}) \left({r_h}^2-3 a^2\right)^2 \left(9 a^2+{r_h}^2\right) \log
   ^3({r_h}) \alpha _{\theta _2}^3}+\frac{49 N \pi ^3 {r_h}^3 \left(6 a^2+{r_h}^2\right) {f_{zz}}(r)}{7776 {g_s}^3 M^2
   {N_f}^2 (r-{r_h}) \left({r_h}^2-3 a^2\right)^2 \left(9 a^2+{r_h}^2\right) \log ^4({r_h})}\nonumber\\
   & & -\frac{49 N \pi ^3
   {r_h}^3 \left(6 a^2+{r_h}^2\right) {f_{x^{10} x^{10}}}(r)}{5184 {g_s}^3 M^2 {N_f}^2 (r-{r_h}) \left({r_h}^2-3
   a^2\right)^2 \left(9 a^2+{r_h}^2\right) \log ^4({r_h})}-\frac{49 N \pi ^3 {r_h}^3 \left(6 a^2+{r_h}^2\right)
   {f_{\theta_1z}}(r)}{3888 {g_s}^3 M^2 {N_f}^2 (r-{r_h}) \left({r_h}^2-3 a^2\right)^2 \left(9 a^2+{r_h}^2\right) \log
   ^4({r_h})}\nonumber\\
   & & -\frac{245 N \pi ^3 {r_h}^3 \left(6 a^2+{r_h}^2\right) {f_{\theta_2y}}(r)}{15552 {g_s}^3 M^2 {N_f}^2
   (r-{r_h}) \left({r_h}^2-3 a^2\right)^2 \left(9 a^2+{r_h}^2\right) \log ^4({r_h})} = 0.
\end{eqnarray*}
}

(iv) \underline{${\rm EOM}_{\theta_1\theta_1}$}

{\scriptsize
\begin{eqnarray*}
& & -\frac{8 \sqrt{2} \left(9 b^2+1\right)^3 {g_s}^{5/4} M^2 \left(\frac{1}{N}\right)^{17/10} (r-{r_h})^3 \beta  \log ^2({r_h})
   \left(-19683 \alpha _{\theta _1}^6+216 \sqrt{6} \alpha _{\theta _2}^2 \alpha _{\theta _1}^3+530 \alpha _{\theta _2}^4\right)
   \Sigma_1
   b^{12}}{6561 \left(1-3 b^2\right)^4 \log N  \pi ^{11/4} \left(6 {r_h} b^2+{r_h}\right)^3 \alpha _{\theta _1}^6 \alpha
   _{\theta _2}^4}\nonumber\\
   \end{eqnarray*}}
   {\scriptsize
   \begin{eqnarray*}
   & &-\frac{3969 \sqrt{\frac{3}{2}} {g_s}^{3/2} M \left(\frac{1}{N}\right)^{3/5} {N_f} {f_{xz}}(r) \log
   ({r_h}) \alpha _{\theta _1}^2 \left(27 \sqrt{6} \alpha _{\theta _1}^3+10 \alpha _{\theta _2}^2\right)}{16384 \pi ^{3/2} \alpha
   _{\theta _2}^3}+\frac{3969 \sqrt{\frac{3}{2}} {g_s}^{3/2} M \left(\frac{1}{N}\right)^{3/5} {N_f} {f_{xy}}(r) \log ({r_h})
   \alpha _{\theta _1}^2 \left(27 \sqrt{6} \alpha _{\theta _1}^3+10 \alpha _{\theta _2}^2\right)}{16384 \pi ^{3/2} \alpha _{\theta
   _2}^3}\nonumber\\
   & & -\frac{83349 {g_s}^{3/2} M \left(\frac{1}{N}\right)^{12/5} {N_f} {f_{xx}}(r) \log ({r_h}) \alpha _{\theta _1}^2
   \alpha _{\theta _2} \left(27 \sqrt{6} \alpha _{\theta _1}^3+10 \alpha _{\theta _2}^2\right) \left(81 \sqrt{2} \alpha _{\theta _1}^3+10
   \sqrt{3} \alpha _{\theta _2}^2\right)}{262144 \sqrt{2} \pi ^{3/2}}\nonumber\\
   & & -\frac{9261 {g_s}^{3/2} M \left(\frac{1}{N}\right)^{3/5}
   {N_f} {f_{\theta_1x}}(r) \log ({r_h}) \left(81 \sqrt{2} \alpha _{\theta _1}^5+10 \sqrt{3} \alpha _{\theta _2}^2 \alpha _{\theta
   _1}^2\right)}{4096 \sqrt{2} \pi ^{3/2} \alpha _{\theta _2}^3}\nonumber\\
   & & +\frac{9261 {g_s}^{3/2} M \left(\frac{1}{N}\right)^{3/5} {N_f}
   {f_{\theta_2x}}(r) \log ({r_h}) \left(81 \sqrt{2} \alpha _{\theta _1}^5+10 \sqrt{3} \alpha _{\theta _2}^2 \alpha _{\theta
   _1}^2\right)}{4096 \sqrt{2} \pi ^{3/2} \alpha _{\theta _2}^3} \nonumber\\
   & & -\frac{3 {g_s}^3 M^2 \left(\frac{1}{N}\right)^{6/5} {N_f}^2
   (r-{r_h}) \left({r_h}^2-3 a^2\right) \log ^4({r_h}) \left(-36 a^4+6 {r_h}^2 a^2+3 \left(9 a^2+{r_h}^2\right) \log
   \left({r_h}^6+9 a^2 {r_h}^4\right) a^2+2 {r_h}^4\right) \Sigma_1 {f_{\theta_1z}}'(r)}{8 \pi ^3 {r_h}^4 \left(6 a^2+{r_h}^2\right) \log
   \left({r_h}^6+9 a^2 {r_h}^4\right) \alpha _{\theta _1}^2 \alpha _{\theta _2}^2}\nonumber\\
   & & +\frac{3 {g_s}^3 M^2
   \left(\frac{1}{N}\right)^{6/5} {N_f}^2 (r-{r_h}) \left({r_h}^2-3 a^2\right) \log ^4({r_h}) \left(243 \sqrt{6} \alpha
   _{\theta _1}^3-8 \alpha _{\theta _2}^2\right) \left(27 \sqrt{6} \alpha _{\theta _1}^3+10 \alpha _{\theta _2}^2\right) {f_{\theta_1x}}'(r)}{8
   \pi ^3 {r_h}^4 \left(6 a^2+{r_h}^2\right) \log \left({r_h}^6+9 a^2 {r_h}^4\right) \alpha _{\theta _1}^2 \alpha _{\theta
   _2}^2}\nonumber\\
   & & \times \left(-36 a^4+6 {r_h}^2
   a^2+3 \left(9 a^2+{r_h}^2\right) \log \left({r_h}^6+9 a^2 {r_h}^4\right) a^2+2 {r_h}^4\right) \nonumber\\
   & & +\frac{9 \sqrt{3} {g_s}^{9/2} M^3 {N_f}^3 (r-{r_h}) \left({r_h}^2-3 a^2\right) \log ^5({r_h}) \left(-36 a^4+6
   {r_h}^2 a^2+3 \left(9 a^2+{r_h}^2\right) \log \left({r_h}^6+9 a^2 {r_h}^4\right) a^2+2 {r_h}^4\right)  {f_{\theta_1y}}'(r)}{256 N^2 \pi ^{9/2} {r_h}^4 \left(6 a^2+{r_h}^2\right)
   \log \left({r_h}^6+9 a^2 {r_h}^4\right) \alpha _{\theta _1}^2 \alpha _{\theta _2}^3}\nonumber\\
   & &\times \left(354294
   \sqrt{3} \alpha _{\theta _1}^9+89667 \sqrt{2} \alpha _{\theta _2}^2 \alpha _{\theta _1}^6-2160 \sqrt{3} \alpha _{\theta _2}^4 \alpha
   _{\theta _1}^3-950 \sqrt{2} \alpha _{\theta _2}^6\right)\nonumber\\
 & & -\frac{2187 {g_s}^3 M^2
   \left(\frac{1}{N}\right)^{2/5} {N_f}^2 (r-{r_h}) \left(9 a^2+{r_h}^2\right) \left(108 b^2 {r_h}^2+1\right)^2
   {f_r}(r) \log ^2({r_h})}{8 \pi ^3 \left({r_h}^3+6 a^2 {r_h}\right) \alpha _{\theta _2}^2}-\frac{3969 {f_{x^{10} x^{10}}}(r)
   \alpha _{\theta _1}^2}{256 \sqrt[5]{\frac{1}{N}} \alpha _{\theta _2}^2}-\frac{3969 {f_{\theta_2y}}(r) \alpha _{\theta _1}^2}{512
   \sqrt[5]{\frac{1}{N}} \alpha _{\theta _2}^2}\nonumber\\
   & & +\frac{2187 {g_s}^3 M^2 \sqrt[5]{\frac{1}{N}} {N_f}^2 (r-{r_h}) \left(9
   a^2+{r_h}^2\right) \left(108 b^2 {r_h}^2+1\right)^2 {f_{zz}}(r) \log ^2({r_h}) \alpha _{\theta _1}^2}{8 \pi ^3
   \left({r_h}^3+6 a^2 {r_h}\right) \alpha _{\theta _2}^4}\nonumber\\
   & &  -\frac{2187 {g_s}^3 M^2 \sqrt[5]{\frac{1}{N}} {N_f}^2
   (r-{r_h}) \left(9 a^2+{r_h}^2\right) \left(108 b^2 {r_h}^2+1\right)^2 {f_{yz}}(r) \log ^2({r_h}) \alpha _{\theta
   _1}^2}{4 \pi ^3 \left({r_h}^3+6 a^2 {r_h}\right) \alpha _{\theta _2}^4}\nonumber\\
   & &+\frac{2187 {g_s}^3 M^2 \sqrt[5]{\frac{1}{N}}
   {N_f}^2 (r-{r_h}) \left(9 a^2+{r_h}^2\right) \left(108 b^2 {r_h}^2+1\right)^2 {f_{yy}}(r) \log ^2({r_h}) \alpha
   _{\theta _1}^2}{8 \pi ^3 \left({r_h}^3+6 a^2 {r_h}\right) \alpha _{\theta _2}^4}-\frac{107163 \sqrt{\frac{3}{2}}
   {g_s}^{3/2} M \sqrt[5]{\frac{1}{N}} {N_f} {f_{\theta_1z}}(r) \log ({r_h}) \alpha _{\theta _1}^6}{2048 \pi ^{3/2} \alpha
   _{\theta _2}^5}\nonumber\\
   & & +\frac{107163 \sqrt{\frac{3}{2}} {g_s}^{3/2} M \sqrt[5]{\frac{1}{N}} {N_f} {f_{\theta_1y}}(r) \log ({r_h}) \alpha
   _{\theta _1}^6}{2048 \pi ^{3/2} \alpha _{\theta _2}^5}+\frac{107163 \sqrt{\frac{3}{2}} {g_s}^{3/2} M \sqrt[5]{\frac{1}{N}}
   {N_f} {f_{\theta_2z}}(r) \log ({r_h}) \alpha _{\theta _1}^6}{2048 \pi ^{3/2} \alpha _{\theta _2}^5}=0.
\end{eqnarray*}
}
(vi) ${\rm EOM}_{\theta_1\theta_2}$

{\scriptsize
\begin{eqnarray*}
& & -\frac{11907 \sqrt{\frac{3}{2}} {g_s}^{3/2} M \sqrt[10]{\frac{1}{N}} {N_f} {r_h}^2 {f_{\theta_1z}}(r) \alpha _{\theta _1}^7}{2048
   \pi ^{3/2} \left({r_h}^2-3 a^2\right) \alpha _{\theta _2}^6}+\frac{11907 \sqrt{\frac{3}{2}} {g_s}^{3/2} M
   \sqrt[10]{\frac{1}{N}} {N_f} {r_h}^2 {f_{\theta_1y}}(r) \alpha _{\theta _1}^7}{2048 \pi ^{3/2} \left({r_h}^2-3 a^2\right) \alpha
   _{\theta _2}^6}\nonumber\\
 & &   +\frac{11907 \sqrt{\frac{3}{2}} {g_s}^{3/2} M \sqrt[10]{\frac{1}{N}} {N_f} {r_h}^2 {f_{\theta_2z}}(r) \alpha
   _{\theta _1}^7}{2048 \pi ^{3/2} \left({r_h}^2-3 a^2\right) \alpha _{\theta _2}^6}+\frac{441 \sqrt{3} {g_s}^{3/2} M
   \sqrt{\frac{1}{N}} {N_f} {r_h}^2 {f_{zz}}(r) \left(27 \sqrt{3} \alpha _{\theta _1}^3+5 \sqrt{2} \alpha _{\theta
   _2}^2\right) \alpha _{\theta _1}^3}{16384 \pi ^{3/2} \left({r_h}^2-3 a^2\right) \alpha _{\theta _2}^4}\nonumber\\
& &  -\frac{441 \sqrt{3}
   {g_s}^{3/2} M \sqrt{\frac{1}{N}} {N_f} {r_h}^2 {f_{xz}}(r) \left(27 \sqrt{3} \alpha _{\theta _1}^3+5 \sqrt{2} \alpha
   _{\theta _2}^2\right) \alpha _{\theta _1}^3}{16384 \pi ^{3/2} \left({r_h}^2-3 a^2\right) \alpha _{\theta _2}^4}+\frac{441 \sqrt{3}
   {g_s}^{3/2} M \sqrt{\frac{1}{N}} {N_f} {r_h}^2 {f_{xy}}(r) \left(27 \sqrt{3} \alpha _{\theta _1}^3+5 \sqrt{2} \alpha
   _{\theta _2}^2\right) \alpha _{\theta _1}^3}{16384 \pi ^{3/2} \left({r_h}^2-3 a^2\right) \alpha _{\theta _2}^4}
   \nonumber\\
    & & -\frac{441 \sqrt{3}
   {g_s}^{3/2} M \sqrt{\frac{1}{N}} {N_f} {r_h}^2 {f_{yz}}(r) \left(27 \sqrt{3} \alpha _{\theta _1}^3+5 \sqrt{2} \alpha
   _{\theta _2}^2\right) \alpha _{\theta _1}^3}{16384 \pi ^{3/2} \left({r_h}^2-3 a^2\right) \alpha _{\theta _2}^4}
   \nonumber\\
   \end{eqnarray*}}
   {\scriptsize
   \begin{eqnarray*}
   & & -\frac{9261
   {g_s}^{3/2} M \left(\frac{1}{N}\right)^{23/10} {N_f} {r_h}^2 {f_{xx}}(r) \left(27 \sqrt{3} \alpha _{\theta _1}^3+5
   \sqrt{2} \alpha _{\theta _2}^2\right) \left(81 \sqrt{2} \alpha _{\theta _1}^3+10 \sqrt{3} \alpha _{\theta _2}^2\right) \alpha _{\theta
   _1}^3}{262144 \pi ^{3/2} \left({r_h}^2-3 a^2\right)}\nonumber\\
   & & -\frac{1029 {g_s}^{3/2} M \sqrt{\frac{1}{N}} {N_f} {r_h}^2
   {f_{\theta_1x}}(r) \left(81 \sqrt{2} \alpha _{\theta _1}^3+10 \sqrt{3} \alpha _{\theta _2}^2\right) \alpha _{\theta _1}^3}{4096 \sqrt{2} \pi
   ^{3/2} \left({r_h}^2-3 a^2\right) \alpha _{\theta _2}^4} +
   \frac{1029 {g_s}^{3/2} M \sqrt{\frac{1}{N}} {N_f} {r_h}^2
   {f_{\theta_2x}}(r) \left(81 \sqrt{2} \alpha _{\theta _1}^3+10 \sqrt{3} \alpha _{\theta _2}^2\right) \alpha _{\theta _1}^3}{4096 \sqrt{2} \pi
   ^{3/2} \left({r_h}^2-3 a^2\right) \alpha _{\theta _2}^4}\nonumber\\
   & & +\frac{1323 {g_s}^3 M^2 \left(\frac{1}{N}\right)^{13/10} {N_f}^2
   {r_h}^2 {f_{yy}}(r) \log ({r_h}) \left(2187 \alpha _{\theta _1}^6+270 \sqrt{6} \alpha _{\theta _2}^2 \alpha _{\theta _1}^3+50
   \alpha _{\theta _2}^4\right) \alpha _{\theta _1}^3}{524288 \pi ^3 \left({r_h}^2-3 a^2\right) \alpha _{\theta _2}^5}-\frac{441
   {r_h}^2 {f_{x^{10} x^{10}}}(r) \alpha _{\theta _1}^3}{256 \left(\frac{1}{N}\right)^{3/10} \left({r_h}^2-3 a^2\right) \log ({r_h})
   \alpha _{\theta _2}^3}\nonumber\\
   & & -\frac{441 {r_h}^2 {f_{\theta_2y}}(r) \alpha _{\theta _1}^3}{512 \left(\frac{1}{N}\right)^{3/10}
   \left({r_h}^2-3 a^2\right) \log ({r_h}) \alpha _{\theta _2}^3}-\frac{{g_s}^3 M^2 \left(\frac{1}{N}\right)^{11/10}
   {N_f}^2 (r-{r_h}) \left({r_h}^2-3 a^2\right) \log ^3({r_h}) \left(19683 \alpha _{\theta _1}^6-216 \sqrt{6} \alpha
   _{\theta _2}^2 \alpha _{\theta _1}^3-530 \alpha _{\theta _2}^4\right) {f_{\theta_1z}}'(r)}{24 \pi ^3 {r_h}^2 \log \left({r_h}^6+9
   a^2 {r_h}^4\right) \alpha _{\theta _2}^3 \alpha _{\theta _1}}\nonumber\\
   & & +\frac{{g_s}^3 M^2 \left(\frac{1}{N}\right)^{11/10} {N_f}^2
   (r-{r_h}) \left({r_h}^2-3 a^2\right) \log ^3({r_h}) \left(19683 \alpha _{\theta _1}^6+1107 \sqrt{6} \alpha _{\theta _2}^2
   \alpha _{\theta _1}^3-40 \alpha _{\theta _2}^4\right) {f_{\theta_1x}}'(r)}{12 \pi ^3 {r_h}^2 \log \left({r_h}^6+9 a^2
   {r_h}^4\right) \alpha _{\theta _2}^3 \alpha _{\theta _1}}\nonumber\\
   & & +\frac{\sqrt{\frac{3}{2}} {g_s}^{9/2} M^3
   \left(\frac{1}{N}\right)^{19/10} {N_f}^3 (r-{r_h}) \left({r_h}^2-3 a^2\right) \log ^4({r_h}) \left(177147 \sqrt{6}
   \alpha _{\theta _1}^9+89667 \alpha _{\theta _2}^2 \alpha _{\theta _1}^6-1080 \sqrt{6} \alpha _{\theta _2}^4 \alpha _{\theta _1}^3-950
   \alpha _{\theta _2}^6\right) {f_{\theta_1y}}'(r)}{128 \pi ^{9/2} {r_h}^2 \log \left({r_h}^6+9 a^2 {r_h}^4\right) \alpha _{\theta
   _2}^4 \alpha _{\theta _1}}\nonumber\\
   & & \hskip -0.4in-\frac{{g_s}^3 M^2 \left(\frac{1}{N}\right)^{11/10} {N_f}^2 (r-{r_h}) \log ^3({r_h})
   \left(-36 a^4+6 {r_h}^2 a^2+3 \left(9 a^2+{r_h}^2\right) \log \left({r_h}^6+9 a^2 {r_h}^4\right) a^2+2
   {r_h}^4\right) \left(19683 \alpha _{\theta _1}^6-216 \sqrt{6} \alpha _{\theta _2}^2 \alpha _{\theta _1}^3-530 \alpha _{\theta
   _2}^4\right) {f_{\theta_2z}}'(r)}{48 \pi ^3 {r_h}^2 \left(6 a^2+{r_h}^2\right) \log \left({r_h}^6+9 a^2 {r_h}^4\right)
   \alpha _{\theta _2}^3 \alpha _{\theta _1}}\nonumber\\
   & &\hskip -0.4in +\frac{{g_s}^3 M^2 \left(\frac{1}{N}\right)^{11/10} {N_f}^2 (r-{r_h}) \log
   ^3({r_h}) \left(-36 a^4+6 {r_h}^2 a^2+3 \left(9 a^2+{r_h}^2\right) \log \left({r_h}^6+9 a^2 {r_h}^4\right) a^2+2
   {r_h}^4\right) \left(19683 \alpha _{\theta _1}^6+1107 \sqrt{6} \alpha _{\theta _2}^2 \alpha _{\theta _1}^3-40 \alpha _{\theta
   _2}^4\right) {f_{\theta_2x}}'(r)}{24 \pi ^3 {r_h}^2 \left(6 a^2+{r_h}^2\right) \log \left({r_h}^6+9 a^2 {r_h}^4\right)
   \alpha _{\theta _2}^3 \alpha _{\theta _1}}\nonumber\\
   & & -\frac{3 \sqrt{\frac{3}{2}} {g_s}^{3/2} M {N_f} (r-{r_h}) {r_h} \left(9
   a^2+{r_h}^2\right) \left(108 b^2 {r_h}^2+1\right)^2 {f_r}(r)}{\sqrt[10]{\frac{1}{N}} \pi ^{3/2} \left(-18 a^4+3 {r_h}^2
   a^2+{r_h}^4\right) \alpha _{\theta _1}^3}\nonumber\\
   & & +\frac{{g_s}^3 M^2 \left(\frac{1}{N}\right)^{3/2} {N_f}^2 (r-{r_h}) \log
   ^3({r_h}) \left(-36 a^4+6 {r_h}^2 a^2+3 \left(9 a^2+{r_h}^2\right) \log \left({r_h}^6+9 a^2 {r_h}^4\right) a^2+2
   {r_h}^4\right)  {f_{\theta_2y}}'(r)}{1728 \sqrt{2} \pi ^3 {r_h}^2
   \left(6 a^2+{r_h}^2\right) \log \left({r_h}^6+9 a^2 {r_h}^4\right) \alpha _{\theta _2} \alpha _{\theta _1}^5}
   \nonumber\\
   & & \times \left(354294 \sqrt{3} \alpha _{\theta _1}^9+89667 \sqrt{2} \alpha _{\theta _2}^2 \alpha _{\theta _1}^6-2160 \sqrt{3}
   \alpha _{\theta _2}^4 \alpha _{\theta _1}^3-950 \sqrt{2} \alpha _{\theta _2}^6\right)
\end{eqnarray*}
}
(vii) \underline{${\rm EOM}_{\theta_1x}$}

{\scriptsize
\begin{eqnarray*}
& & \frac{16 \sqrt{\frac{2}{3}} \left(9 b^2+1\right)^4 {g_s} M^2 \left(\frac{1}{N}\right)^{17/20} (r-{r_h})^3 \beta  \log ({r_h})
\Sigma_1   b^{12}}{9 \left(3 b^2-1\right)^5 \left(6 b^2+1\right)^4 \log N  \pi ^3 {r_h}^3 \alpha _{\theta _1} \alpha _{\theta
   _2}^7}-\frac{\sqrt{2} \left(9 b^2+1\right)^4 {g_s}^{5/4} M^2 \left(\frac{1}{N}\right)^{6/5} (r-{r_h}) \beta  \log ^2({r_h})
\Sigma_1   b^{10}}{\left(-18 b^4+3 b^2+1\right)^4 \log N  \pi ^{11/4} {r_h} \alpha _{\theta _1} \alpha _{\theta _2}^5}\nonumber\\
& & -\frac{729
   {g_s}^{5/4} M \left(\frac{1}{N}\right)^{27/20} {N_f} \left({r_h}^2-3 a^2\right) {f_{xx}}(r) \log ^2({r_h}) \alpha
   _{\theta _1}^3}{4 \sqrt{2} \pi ^{7/4} {r_h}^2 \log \left({r_h}^6+9 a^2 {r_h}^4\right)}+\frac{245 {r_h}^2 {f_{xy}}(r)
   \left(27 \sqrt{6} \alpha _{\theta _1}^3+10 \alpha _{\theta _2}^2\right)}{55296 \sqrt{3} \sqrt[4]{{g_s}} \sqrt[4]{\frac{1}{N}}
   \sqrt[4]{\pi } \left({r_h}^2-3 a^2\right) \log ({r_h}) \alpha _{\theta _1} \alpha _{\theta _2}^3}\nonumber\\
   & &-\frac{343 {r_h}^2
   {f_{\theta_1x}}(r) \left(81 \sqrt{2} \alpha _{\theta _1}^3+10 \sqrt{3} \alpha _{\theta _2}^2\right)}{41472 \sqrt[4]{{g_s}}
   \sqrt[4]{\frac{1}{N}} \sqrt[4]{\pi } \left({r_h}^2-3 a^2\right) \log ({r_h}) \alpha _{\theta _1} \alpha _{\theta
   _2}^3}+\frac{1715 {r_h}^2 {f_{\theta_2x}}(r) \left(81 \sqrt{2} \alpha _{\theta _1}^3+10 \sqrt{3} \alpha _{\theta _2}^2\right)}{124416
   \sqrt[4]{{g_s}} \sqrt[4]{\frac{1}{N}} \sqrt[4]{\pi } \left({r_h}^2-3 a^2\right) \log ({r_h}) \alpha _{\theta _1} \alpha
   _{\theta _2}^3}\nonumber\\
   & & -\frac{3479 {r_h}^2 {f_{xz}}(r) \left(81 \sqrt{2} \alpha _{\theta _1}^3+10 \sqrt{3} \alpha _{\theta
   _2}^2\right)}{497664 \sqrt[4]{{g_s}} \sqrt[4]{\frac{1}{N}} \sqrt[4]{\pi } \left({r_h}^2-3 a^2\right) \log ({r_h}) \alpha
   _{\theta _1} \alpha _{\theta _2}^3}+\frac{9 {g_s}^{5/4} M {N_f} \left({r_h}^2-3 a^2\right) \left(9 a^2+{r_h}^2\right)
   \log ^2({r_h}) f'(r)}{\sqrt{2} \sqrt[20]{\frac{1}{N}} \pi ^{7/4} {r_h} \left(6 a^2+{r_h}^2\right) \alpha _{\theta _1}
   \alpha _{\theta _2}^2}\nonumber\\
     & & -\frac{3 {g_s}^{5/4} M {N_f} \left({r_h}^2-3 a^2\right) \left(9 a^2+{r_h}^2\right) \log
   ^2({r_h}) {f_{zz}}'(r)}{\sqrt{2} \sqrt[20]{\frac{1}{N}} \pi ^{7/4} {r_h} \left(6 a^2+{r_h}^2\right) \alpha _{\theta _1}
   \alpha _{\theta _2}^2}+\frac{3 {g_s}^{5/4} M {N_f} \left({r_h}^2-3 a^2\right) \left(9 a^2+{r_h}^2\right) \log
   ^2({r_h}) {f_{x^{10} x^{10}}}'(r)}{\sqrt{2} \sqrt[20]{\frac{1}{N}} \pi ^{7/4} {r_h} \left(6 a^2+{r_h}^2\right) \alpha _{\theta _1}
   \alpha _{\theta _2}^2}\nonumber\\
   \end{eqnarray*}}
   {\scriptsize
   \begin{eqnarray*}
   & & +\frac{9 {g_s}^{5/4} M {N_f} \left({r_h}^2-3 a^2\right) \left(9 a^2+{r_h}^2\right) \log
   ^2({r_h}) {f_{\theta_1z}}'(r)}{2 \sqrt{2} \sqrt[20]{\frac{1}{N}} \pi ^{7/4} {r_h} \left(6 a^2+{r_h}^2\right) \alpha _{\theta
   _1} \alpha _{\theta _2}^2}-\frac{3 {g_s}^{5/4} M {N_f} \left({r_h}^2-3 a^2\right) \left(9 a^2+{r_h}^2\right) \log
   ^2({r_h}) {f_{\theta_1x}}'(r)}{\sqrt{2} \sqrt[20]{\frac{1}{N}} \pi ^{7/4} {r_h} \left(6 a^2+{r_h}^2\right) \alpha _{\theta _1}
   \alpha _{\theta _2}^2}\nonumber\\
   & & -\frac{243 \sqrt{3} {g_s}^{11/4} M^2 \left(\frac{1}{N}\right)^{7/20} {N_f}^2 \left({r_h}^2-3
   a^2\right) \left(9 a^2+{r_h}^2\right) \log ^3({r_h}) \alpha _{\theta _1}^3 {f_{\theta_1y}}'(r)}{16 \pi ^{13/4} \left({r_h}^3+6
   a^2 {r_h}\right) \alpha _{\theta _2}^5}\nonumber\\
   & &  -\frac{243 \sqrt{3} {g_s}^{11/4} M^2 \left(\frac{1}{N}\right)^{7/20} {N_f}^2
   \left({r_h}^2-3 a^2\right) \left(9 a^2+{r_h}^2\right) \log ^3({r_h}) \alpha _{\theta _1}^3 {f_{\theta_2z}}'(r)}{16 \pi ^{13/4}
   \left({r_h}^3+6 a^2 {r_h}\right) \alpha _{\theta _2}^5}\nonumber\\
   & & +\frac{243 \sqrt{3} a^2 {g_s}^{11/4} M^2
   \left(\frac{1}{N}\right)^{11/20} {N_f}^2 (r-{r_h}) \left(9 a^2+{r_h}^2\right) \log ^3({r_h}) \alpha _{\theta _1}
   {f_{\theta_2x}}'(r)}{8 \pi ^{13/4} {r_h}^2 \left(6 a^2+{r_h}^2\right) \alpha _{\theta _2}^3}\nonumber\\
   & & +\frac{9 {g_s}^{5/4} M {N_f}
   \left({r_h}^2-3 a^2\right) \left(9 a^2+{r_h}^2\right) \log ^2({r_h}) {f_{\theta_2y}}'(r)}{2 \sqrt{2} \sqrt[20]{\frac{1}{N}} \pi
   ^{7/4} {r_h} \left(6 a^2+{r_h}^2\right) \alpha _{\theta _1} \alpha _{\theta _2}^2}+\frac{9 {g_s}^{5/4} M {N_f}
   \left({r_h}^2-3 a^2\right) \left(9 a^2+{r_h}^2\right) \log ^2({r_h}) {f_{xz}}'(r)}{2 \sqrt{2} \sqrt[20]{\frac{1}{N}} \pi
   ^{7/4} {r_h} \left(6 a^2+{r_h}^2\right) \alpha _{\theta _1} \alpha _{\theta _2}^2}\nonumber\\
   & &  -\frac{729 {g_s}^{5/4} M
   \left(\frac{1}{N}\right)^{27/20} {N_f} \left({r_h}^2-3 a^2\right) \left(9 a^2+{r_h}^2\right) \log ^2({r_h}) \alpha
   _{\theta _1}^3 {f_{xx}}'(r)}{8 \sqrt{2} \pi ^{7/4} \left({r_h}^3+6 a^2 {r_h}\right)}\nonumber\\
      & & -\frac{243 \sqrt{3} a^2 {g_s}^{11/4}
   M^2 \left(\frac{1}{N}\right)^{7/20} {N_f}^2 (r-{r_h}) \left(9 a^2+{r_h}^2\right) \log ^3({r_h}) \alpha _{\theta _1}^3
   {f_{xy}}'(r)}{8 \pi ^{13/4} {r_h}^2 \left(6 a^2+{r_h}^2\right) \alpha _{\theta _2}^5}\nonumber\\
   & & -\frac{243 \sqrt{3} a^2 {g_s}^{11/4}
   M^2 \left(\frac{1}{N}\right)^{7/20} {N_f}^2 (r-{r_h}) \left(9 a^2+{r_h}^2\right) \log ^3({r_h}) \alpha _{\theta _1}^3
   {f_{yz}}'(r)}{8 \pi ^{13/4} {r_h}^2 \left(6 a^2+{r_h}^2\right) \alpha _{\theta _2}^5}\nonumber\\
   & &-\frac{81 a^2 {g_s}^{17/4} M^3
   \left(\frac{1}{N}\right)^{27/20} {N_f}^3 (r-{r_h}) \left(9 a^2+{r_h}^2\right) \log ^4({r_h}) \alpha _{\theta _1}
   \left(2187 \sqrt{2} \alpha _{\theta _1}^6+540 \sqrt{3} \alpha _{\theta _2}^2 \alpha _{\theta _1}^3+50 \sqrt{2} \alpha _{\theta
   _2}^4\right) {f_{yy}}'(r)}{512 \pi ^{19/4} {r_h}^2 \left(6 a^2+{r_h}^2\right) \alpha _{\theta _2}^6}
   \nonumber\\
   & &  -\frac{3 {g_s}^{5/4}
   M {N_f} \left({r_h}^2-3 a^2\right) \left(9 a^2+{r_h}^2\right) \log ^2({r_h}) {f_r}'(r)}{2 \sqrt{2}
   \sqrt[20]{\frac{1}{N}} \pi ^{7/4} {r_h} \left(6 a^2+{r_h}^2\right) \alpha _{\theta _1} \alpha _{\theta _2}^2}\nonumber\\
& & +\frac{9
   {g_s}^{5/4} M {N_f} \left({r_h}^2-3 a^2\right) \left(9 a^2+{r_h}^2\right) \log ^2({r_h}) {f_t}'(r)}{2 \sqrt{2}
   \sqrt[20]{\frac{1}{N}} \pi ^{7/4} {r_h} \left(6 a^2+{r_h}^2\right) \alpha _{\theta _1} \alpha _{\theta _2}^2}+\frac{9
   {g_s}^{5/4} M {N_f} (r-{r_h}) \left({r_h}^2-3 a^2\right) \left(9 a^2+{r_h}^2\right) \log ^2({r_h})
   f''(r)}{\sqrt{2} \sqrt[20]{\frac{1}{N}} \pi ^{7/4} {r_h} \left(6 a^2+{r_h}^2\right) \alpha _{\theta _1} \alpha _{\theta
   _2}^2}\nonumber\\
   & & -\frac{3 {g_s}^{5/4} M {N_f} (r-{r_h}) \left({r_h}^2-3 a^2\right) \left(9 a^2+{r_h}^2\right) \log
   ^2({r_h}) {f_{zz}}''(r)}{\sqrt{2} \sqrt[20]{\frac{1}{N}} \pi ^{7/4} {r_h} \left(6 a^2+{r_h}^2\right) \alpha _{\theta
   _1} \alpha _{\theta _2}^2}+\frac{3 {g_s}^{5/4} M {N_f} (r-{r_h}) \left({r_h}^2-3 a^2\right) \left(9
   a^2+{r_h}^2\right) \log ^2({r_h}) {f_{x^{10} x^{10}}}''(r)}{\sqrt{2} \sqrt[20]{\frac{1}{N}} \pi ^{7/4} {r_h} \left(6
   a^2+{r_h}^2\right) \alpha _{\theta _1} \alpha _{\theta _2}^2}\nonumber\\
   & & +\frac{9 {g_s}^{5/4} M {N_f} (r-{r_h}) \left({r_h}^2-3
   a^2\right) \left(9 a^2+{r_h}^2\right) \log ^2({r_h}) {f_{\theta_1z}}''(r)}{2 \sqrt{2} \sqrt[20]{\frac{1}{N}} \pi ^{7/4} {r_h}
   \left(6 a^2+{r_h}^2\right) \alpha _{\theta _1} \alpha _{\theta _2}^2}-\frac{3 {g_s}^{5/4} M {N_f} (r-{r_h})
   \left({r_h}^2-3 a^2\right) \left(9 a^2+{r_h}^2\right) \log ^2({r_h}) {f_{\theta_1x}}''(r)}{\sqrt{2} \sqrt[20]{\frac{1}{N}} \pi
   ^{7/4} {r_h} \left(6 a^2+{r_h}^2\right) \alpha _{\theta _1} \alpha _{\theta _2}^2}\nonumber\\
      & & -\frac{243 \sqrt{3} {g_s}^{11/4} M^2
   \left(\frac{1}{N}\right)^{7/20} {N_f}^2 (r-{r_h}) \left({r_h}^2-3 a^2\right) \left(9 a^2+{r_h}^2\right) \log
   ^3({r_h}) \alpha _{\theta _1}^3 {f_{\theta_1y}}''(r)}{16 \pi ^{13/4} \left({r_h}^3+6 a^2 {r_h}\right) \alpha _{\theta
   _2}^5}\nonumber\\
   & & -\frac{243 \sqrt{3} {g_s}^{11/4} M^2 \left(\frac{1}{N}\right)^{7/20} {N_f}^2 (r-{r_h}) \left({r_h}^2-3 a^2\right)
   \left(9 a^2+{r_h}^2\right) \log ^3({r_h}) \alpha _{\theta _1}^3 {f_{\theta_2z}}''(r)}{16 \pi ^{13/4} \left({r_h}^3+6 a^2
   {r_h}\right) \alpha _{\theta _2}^5}\nonumber\\
   & & -\frac{21 {g_s}^{11/4} M^2 \left(\frac{1}{N}\right)^{3/4} {N_f}^2 (r-{r_h})
   \left({r_h}^2-3 a^2\right) \left(9 a^2+{r_h}^2\right) \log ^3({r_h}) \left(81 \sqrt{2} \alpha _{\theta _1}^3+10 \sqrt{3}
   \alpha _{\theta _2}^2\right) {f_{\theta_2x}}''(r)}{32 \pi ^{13/4} {r_h} \left(6 a^2+{r_h}^2\right) \alpha _{\theta _1} \alpha
   _{\theta _2}^3}\nonumber\\
   & & +\frac{9 {g_s}^{5/4} M {N_f} (r-{r_h}) \left({r_h}^2-3 a^2\right) \left(9 a^2+{r_h}^2\right) \log
   ^2({r_h}) {f_{\theta_2y}}''(r)}{2 \sqrt{2} \sqrt[20]{\frac{1}{N}} \pi ^{7/4} {r_h} \left(6 a^2+{r_h}^2\right) \alpha _{\theta
   _1} \alpha _{\theta _2}^2}\nonumber\\
    \end{eqnarray*}}
   {\scriptsize
   \begin{eqnarray*}
   & & +\frac{9 {g_s}^{5/4} M {N_f} (r-{r_h}) \left({r_h}^2-3 a^2\right) \left(9
   a^2+{r_h}^2\right) \log ^2({r_h}) {f_{xz}}''(r)}{2 \sqrt{2} \sqrt[20]{\frac{1}{N}} \pi ^{7/4} {r_h} \left(6
   a^2+{r_h}^2\right) \alpha _{\theta _1} \alpha _{\theta _2}^2}\nonumber\\
   & &-\frac{729 {g_s}^{5/4} M \left(\frac{1}{N}\right)^{27/20}
   {N_f} (r-{r_h}) \left({r_h}^2-3 a^2\right) \left(9 a^2+{r_h}^2\right) \log ^2({r_h}) \alpha _{\theta _1}^3
   {f_{xx}}''(r)}{8 \sqrt{2} \pi ^{7/4} \left({r_h}^3+6 a^2 {r_h}\right)}\nonumber\\
   & &-\frac{9 \sqrt{\frac{3}{2}} {g_s}^{11/4} M^2
   \left(\frac{1}{N}\right)^{3/4} {N_f}^2 (r-{r_h}) \left({r_h}^2-3 a^2\right) \left(9 a^2+{r_h}^2\right) \log
   ^3({r_h}) \left(27 \sqrt{3} \alpha _{\theta _1}^3+5 \sqrt{2} \alpha _{\theta _2}^2\right) {f_{xy}}''(r)}{64 \pi ^{13/4} {r_h}
   \left(6 a^2+{r_h}^2\right) \alpha _{\theta _1} \alpha _{\theta _2}^3}\nonumber\\
   & & +\frac{9 \sqrt{\frac{3}{2}} {g_s}^{11/4} M^2
   \left(\frac{1}{N}\right)^{3/4} {N_f}^2 (r-{r_h}) \left({r_h}^2-3 a^2\right) \left(9 a^2+{r_h}^2\right) \log
   ^3({r_h}) \left(27 \sqrt{3} \alpha _{\theta _1}^3+5 \sqrt{2} \alpha _{\theta _2}^2\right) {f_{yz}}''(r)}{64 \pi ^{13/4}
   {r_h} \left(6 a^2+{r_h}^2\right) \alpha _{\theta _1} \alpha _{\theta _2}^3}\nonumber\\
   & & -\frac{9 {g_s}^{17/4} M^3
   \left(\frac{1}{N}\right)^{31/20} {N_f}^3 (r-{r_h}) \left({r_h}^2-3 a^2\right) \left(9 a^2+{r_h}^2\right) \log
   ^4({r_h}) \left(2187 \alpha _{\theta _1}^6+270 \sqrt{6} \alpha _{\theta _2}^2 \alpha _{\theta _1}^3+50 \alpha _{\theta _2}^4\right)
   {f_{yy}}''(r)}{1024 \sqrt{2} \pi ^{19/4} {r_h} \left(6 a^2+{r_h}^2\right) \alpha _{\theta _1} \alpha _{\theta _2}^4}\nonumber\\
   & & +\frac{3
   {g_s}^{5/4} M {N_f} (r-{r_h}) \left({r_h}^2-3 a^2\right) \left(9 a^2+{r_h}^2\right) \log ^2({r_h})
   {f_t}''(r)}{\sqrt{2} \sqrt[20]{\frac{1}{N}} \pi ^{7/4} {r_h} \left(6 a^2+{r_h}^2\right) \alpha _{\theta _1} \alpha _{\theta
   _2}^2}\nonumber\\
   & &+\frac{243 {g_s}^{21/4} M^5 \left(\frac{1}{N}\right)^{39/20} {N_f}^3 {f_t}(r) \log ^6({r_h}) \left(16 \left(-108
   a^6+9 {r_h}^4 a^2+{r_h}^6\right)+\left(2 {r_h}^6+27 a^2 {r_h}^4+117 a^4 {r_h}^2\right) \log \left({r_h}^6+9 a^2
   {r_h}^4\right)\right)}{128 \sqrt{2} \pi ^{23/4} {r_h}^2 \left(6 a^2+{r_h}^2\right)^2 \log \left({r_h}^6+9 a^2
   {r_h}^4\right) \alpha _{\theta _1} \alpha _{\theta _2}^2}\nonumber\\
   & & -\frac{729 {g_s}^{21/4} \log N  M^5
   \left(\frac{1}{N}\right)^{39/20} {N_f}^3 \left({r_h}^2-3 a^2\right) \left(9 a^2+{r_h}^2\right) f(r) \log ^6({r_h})}{32
   \sqrt{2} \pi ^{23/4} {r_h}^2 \left(6 a^2+{r_h}^2\right) \log \left({r_h}^6+9 a^2 {r_h}^4\right) \alpha _{\theta _1}
   \alpha _{\theta _2}^2}\nonumber\\
   & & +\frac{49 \pi ^{5/4} {r_h}^2 {f_{zz}}(r)}{1296 \sqrt{2} {g_s}^{7/4} M
   \left(\frac{1}{N}\right)^{21/20} {N_f} \left({r_h}^2-3 a^2\right) \log ^2({r_h}) \alpha _{\theta _1} \alpha _{\theta
   _2}^2}-\frac{49 \pi ^{5/4} {r_h}^2 {f_{x^{10} x^{10}}}(r)}{864 \sqrt{2} {g_s}^{7/4} M \left(\frac{1}{N}\right)^{21/20} {N_f}
   \left({r_h}^2-3 a^2\right) \log ^2({r_h}) \alpha _{\theta _1} \alpha _{\theta _2}^2}\nonumber\\
   & & -\frac{49 \pi ^{5/4} {r_h}^2
   {f_{\theta_1z}}(r)}{648 \sqrt{2} {g_s}^{7/4} M \left(\frac{1}{N}\right)^{21/20} {N_f} \left({r_h}^2-3 a^2\right) \log
   ^2({r_h}) \alpha _{\theta _1} \alpha _{\theta _2}^2}-\frac{245 \pi ^{5/4} {r_h}^2 {f_{\theta_2y}}(r)}{2592 \sqrt{2} {g_s}^{7/4}
   M \left(\frac{1}{N}\right)^{21/20} {N_f} \left({r_h}^2-3 a^2\right) \log ^2({r_h}) \alpha _{\theta _1} \alpha _{\theta
   _2}^2}\nonumber\\
   & & -\frac{5 {g_s}^{5/4} M {N_f} (r-{r_h}) {r_h} \left(9 a^2+{r_h}^2\right) \left(108 b^2 {r_h}^2+1\right)^2
   {f_r}(r)}{6 \sqrt{2} \left(\frac{1}{N}\right)^{9/20} \pi ^{7/4} \left(-18 a^4+3 {r_h}^2 a^2+{r_h}^4\right) \alpha _{\theta
   _1}^3 \alpha _{\theta _2}^2}-\frac{\sqrt{2} {g_s}^{5/4} M {N_f} (r-{r_h}) {r_h} \left(9 a^2+{r_h}^2\right)
   \left(108 b^2 {r_h}^2+1\right)^2 {f_{yz}}(r)}{\left(\frac{1}{N}\right)^{13/20} \pi ^{7/4} \left({r_h}^2-3 a^2\right) \left(6
   a^2+{r_h}^2\right) \alpha _{\theta _1} \alpha _{\theta _2}^4}\nonumber\\
   & & +\frac{{g_s}^{5/4} M {N_f} (r-{r_h}) {r_h} \left(9
   a^2+{r_h}^2\right) \left(108 b^2 {r_h}^2+1\right)^2 {f_{yy}}(r)}{\sqrt{2} \left(\frac{1}{N}\right)^{13/20} \pi ^{7/4}
   \left({r_h}^2-3 a^2\right) \left(6 a^2+{r_h}^2\right) \alpha _{\theta _1} \alpha _{\theta _2}^4}+\frac{245 {r_h}^2
   {f_{\theta_1y}}(r) \alpha _{\theta _1}^3}{256 \sqrt{3} \sqrt[4]{{g_s}} \left(\frac{1}{N}\right)^{13/20} \sqrt[4]{\pi }
   \left({r_h}^2-3 a^2\right) \log ({r_h}) \alpha _{\theta _2}^5}\nonumber\\
   & & +\frac{245 {r_h}^2 {f_{\theta_2z}}(r) \alpha _{\theta _1}^3}{256
   \sqrt{3} \sqrt[4]{{g_s}} \left(\frac{1}{N}\right)^{13/20} \sqrt[4]{\pi } \left({r_h}^2-3 a^2\right) \log ({r_h}) \alpha
   _{\theta _2}^5}
\end{eqnarray*}
}
(viii) \underline{${\rm EOM}_{\theta_1y}$}

{\scriptsize
\begin{eqnarray*}
& & -\frac{32 \sqrt{\frac{2}{3}} \left(9 b^2+1\right)^4 M \left(\frac{1}{N}\right)^{23/20} (r-{r_h})^2 \beta  \log ^2({r_h})
\Sigma_1   b^{12}}{\left(3 b^2-1\right)^5 \left(6 b^2+1\right)^4 \sqrt{{g_s}} \log N ^2 {N_f} \pi ^{3/2} {r_h}^2 \alpha _{\theta
   _1}^3 \alpha _{\theta _2}^3}\nonumber\\
   & & -\frac{9 \sqrt{3} \left(9 b^2+1\right)^4 {g_s}^{5/4} M^2 \left(\frac{1}{N}\right)^{19/10} (r-{r_h})
   \beta  \log ^2({r_h}) \alpha _{\theta _1} \left(19683 \sqrt{6} \alpha _{\theta _1}^6+6642 \alpha _{\theta _2}^2 \alpha _{\theta
   _1}^3-40 \sqrt{6} \alpha _{\theta _2}^4\right) b^{10}}{2 \left(-18 b^4+3 b^2+1\right)^4 \log N  \pi ^{11/4} {r_h} \alpha
   _{\theta _2}^4}\nonumber\\
   & & -\frac{6561 \sqrt{3} {g_s}^{5/4} M \left(\frac{1}{N}\right)^{41/20} {N_f} \left({r_h}^2-3 a^2\right)
   {f_{xx}}(r) \log ^2({r_h}) \alpha _{\theta _1}^5 \alpha _{\theta _2}}{16 \pi ^{7/4} {r_h}^2 \log \left({r_h}^6+9 a^2
   {r_h}^4\right)}-\frac{343 \left(\frac{1}{N}\right)^{9/20} {r_h}^2 {f_{\theta_1x}}(r) \alpha _{\theta _1} \left(81 \sqrt{2} \alpha
   _{\theta _1}^3+10 \sqrt{3} \alpha _{\theta _2}^2\right)}{3072 \sqrt{6} \sqrt[4]{{g_s}} \sqrt[4]{\pi } \left({r_h}^2-3
   a^2\right) \log ({r_h}) \alpha _{\theta _2}^2}\nonumber\\
    \end{eqnarray*}}
   {\scriptsize
   \begin{eqnarray*}
   & & +\frac{1715 \left(\frac{1}{N}\right)^{9/20} {r_h}^2 {f_{\theta_2x}}(r) \alpha _{\theta
   _1} \left(81 \sqrt{2} \alpha _{\theta _1}^3+10 \sqrt{3} \alpha _{\theta _2}^2\right)}{9216 \sqrt{6} \sqrt[4]{{g_s}} \sqrt[4]{\pi }
   \left({r_h}^2-3 a^2\right) \log ({r_h}) \alpha _{\theta _2}^2}+\frac{21217 \left(\frac{1}{N}\right)^{9/20} {r_h}^2
   {f_{xz}}(r) \alpha _{\theta _1} \left(81 \sqrt{2} \alpha _{\theta _1}^3+10 \sqrt{3} \alpha _{\theta _2}^2\right)}{36864 \sqrt{6}
   \sqrt[4]{{g_s}} \sqrt[4]{\pi } \left({r_h}^2-3 a^2\right) \log ({r_h}) \alpha _{\theta _2}^2}\nonumber\\
   & & -\frac{7987
   \left(\frac{1}{N}\right)^{9/20} {r_h}^2 {f_{xy}}(r) \alpha _{\theta _1} \left(81 \sqrt{2} \alpha _{\theta _1}^3+10 \sqrt{3}
   \alpha _{\theta _2}^2\right)}{12288 \sqrt{6} \sqrt[4]{{g_s}} \sqrt[4]{\pi } \left({r_h}^2-3 a^2\right) \log ({r_h}) \alpha
   _{\theta _2}^2}\nonumber\\
   & & +\frac{81 \sqrt{3} {g_s}^{5/4} M \left(\frac{1}{N}\right)^{13/20} {N_f} \left({r_h}^2-3 a^2\right) \left(9
   a^2+{r_h}^2\right) \log ^2({r_h}) \alpha _{\theta _1} f'(r)}{4 \pi ^{7/4} {r_h} \left(6 a^2+{r_h}^2\right) \alpha
   _{\theta _2}}\nonumber\\
   &&-\frac{27 \sqrt{3} {g_s}^{5/4} M \left(\frac{1}{N}\right)^{13/20} {N_f} \left({r_h}^2-3 a^2\right) \left(9
   a^2+{r_h}^2\right) \log ^2({r_h}) \alpha _{\theta _1} {f_{zz}}'(r)}{4 \pi ^{7/4} {r_h} \left(6 a^2+{r_h}^2\right)
   \alpha _{\theta _2}}\nonumber\\
   & &+\frac{27 \sqrt{3} {g_s}^{5/4} M \left(\frac{1}{N}\right)^{13/20} {N_f} \left({r_h}^2-3 a^2\right)
   \left(9 a^2+{r_h}^2\right) \log ^2({r_h}) \alpha _{\theta _1} {f_{x^{10} x^{10}}}'(r)}{4 \pi ^{7/4} {r_h} \left(6
   a^2+{r_h}^2\right) \alpha _{\theta _2}}\nonumber\\
   &&+\frac{81 \sqrt{3} {g_s}^{5/4} M \left(\frac{1}{N}\right)^{13/20} {N_f}
   \left({r_h}^2-3 a^2\right) \left(9 a^2+{r_h}^2\right) \log ^2({r_h}) \alpha _{\theta _1} {f_{\theta_1z}}'(r)}{8 \pi ^{7/4}
   {r_h} \left(6 a^2+{r_h}^2\right) \alpha _{\theta _2}}\nonumber\\
   & &-\frac{81 \sqrt{3} a^2 {g_s}^{5/4} M \left(\frac{1}{N}\right)^{13/20}
   {N_f} (r-{r_h}) \left(9 a^2+{r_h}^2\right) \log ^2({r_h}) \alpha _{\theta _1} {f_{\theta_1x}}'(r)}{4 \pi ^{7/4} {r_h}^2
   \left(6 a^2+{r_h}^2\right) \alpha _{\theta _2}}\nonumber\\
   &&-\frac{27 \sqrt{3} {g_s}^{5/4} M \left(\frac{1}{N}\right)^{13/20} {N_f}
   \left({r_h}^2-3 a^2\right) \left(9 a^2+{r_h}^2\right) \log ^2({r_h}) \alpha _{\theta _1} {f_{\theta_1y}}'(r)}{4 \pi ^{7/4}
   {r_h} \left(6 a^2+{r_h}^2\right) \alpha _{\theta _2}}\nonumber\\
      & & -\frac{6561 {g_s}^{11/4} M^2 \left(\frac{1}{N}\right)^{21/20}
   {N_f}^2 \left({r_h}^2-3 a^2\right) \left(9 a^2+{r_h}^2\right) \log ^3({r_h}) \alpha _{\theta _1}^5 {f_{\theta_2z}}'(r)}{32
   \sqrt{2} \pi ^{13/4} \left({r_h}^3+6 a^2 {r_h}\right) \alpha _{\theta _2}^4}\nonumber\\
   &&+\frac{6561 a^2 {g_s}^{11/4} M^2
   \left(\frac{1}{N}\right)^{5/4} {N_f}^2 (r-{r_h}) \left(9 a^2+{r_h}^2\right) \log ^3({r_h}) \alpha _{\theta _1}^3
   {f_{\theta_2x}}'(r)}{16 \sqrt{2} \pi ^{13/4} {r_h}^2 \left(6 a^2+{r_h}^2\right) \alpha _{\theta _2}^2}\nonumber\\
   & &+\frac{81 \sqrt{3}
   {g_s}^{5/4} M \left(\frac{1}{N}\right)^{13/20} {N_f} \left({r_h}^2-3 a^2\right) \left(9 a^2+{r_h}^2\right) \log
   ^2({r_h}) \alpha _{\theta _1} {f_{\theta_2y}}'(r)}{8 \pi ^{7/4} {r_h} \left(6 a^2+{r_h}^2\right) \alpha _{\theta _2}}\nonumber\\
   &&+\frac{81\sqrt{3} {g_s}^{5/4} M \left(\frac{1}{N}\right)^{13/20} {N_f} \left({r_h}^2-3 a^2\right) \left(9 a^2+{r_h}^2\right)
   \log ^2({r_h}) \alpha _{\theta _1} {f_{xz}}'(r)}{8 \pi ^{7/4} {r_h} \left(6 a^2+{r_h}^2\right) \alpha _{\theta
   _2}}\nonumber\\
   & & -\frac{6561 \sqrt{3} {g_s}^{5/4} M \left(\frac{1}{N}\right)^{41/20} {N_f} \left({r_h}^2-3 a^2\right) \left(9
   a^2+{r_h}^2\right) \log ^2({r_h}) \alpha _{\theta _1}^5 \alpha _{\theta _2} {f_{xx}}'(r)}{32 \pi ^{7/4} \left({r_h}^3+6
   a^2 {r_h}\right)}\nonumber\\
   & & -\frac{6561 a^2 {g_s}^{11/4} M^2 \left(\frac{1}{N}\right)^{21/20} {N_f}^2 (r-{r_h}) \left(9
   a^2+{r_h}^2\right) \log ^3({r_h}) \alpha _{\theta _1}^5 {f_{xy}}'(r)}{16 \sqrt{2} \pi ^{13/4} {r_h}^2 \left(6
   a^2+{r_h}^2\right) \alpha _{\theta _2}^4}\nonumber\\
   & & -\frac{6561 a^2 {g_s}^{11/4} M^2 \left(\frac{1}{N}\right)^{21/20} {N_f}^2
   (r-{r_h}) \left(9 a^2+{r_h}^2\right) \log ^3({r_h}) \alpha _{\theta _1}^5 {f_{yz}}'(r)}{16 \sqrt{2} \pi ^{13/4}
   {r_h}^2 \left(6 a^2+{r_h}^2\right) \alpha _{\theta _2}^4}\nonumber\\
   & &  +\frac{{g_s}^{11/4} M^2 \left(\frac{1}{N}\right)^{37/20}
   {N_f}^2 (r-{r_h}) \log ^3({r_h}) \left(-36 a^4+6 {r_h}^2 a^2+3 \left(9 a^2+{r_h}^2\right) \log \left({r_h}^6+9
   a^2 {r_h}^4\right) a^2+2 {r_h}^4\right){f_{yy}}'(r)}{384
   \pi ^{13/4} {r_h}^2 \left(6 a^2+{r_h}^2\right) \log \left({r_h}^6+9 a^2 {r_h}^4\right) \alpha _{\theta _1}^3 \alpha
   _{\theta _2}^2}\nonumber\\
   & &  \times  \left(354294 \sqrt{3} \alpha _{\theta _1}^9+89667 \sqrt{2} \alpha _{\theta _2}^2 \alpha
   _{\theta _1}^6-2160 \sqrt{3} \alpha _{\theta _2}^4 \alpha _{\theta _1}^3-950 \sqrt{2} \alpha _{\theta _2}^6\right) \nonumber\\
   & & -\frac{27 \sqrt{3} {g_s}^{5/4} M \left(\frac{1}{N}\right)^{13/20} {N_f} \left({r_h}^2-3 a^2\right) \left(9
   a^2+{r_h}^2\right) \log ^2({r_h}) \alpha _{\theta _1} {f_r}'(r)}{8 \pi ^{7/4} {r_h} \left(6 a^2+{r_h}^2\right)
   \alpha _{\theta _2}}\nonumber\\
   &&+\frac{81 \sqrt{3} {g_s}^{5/4} M \left(\frac{1}{N}\right)^{13/20} {N_f} \left({r_h}^2-3 a^2\right)
   \left(9 a^2+{r_h}^2\right) \log ^2({r_h}) \alpha _{\theta _1} {f_t}'(r)}{8 \pi ^{7/4} {r_h} \left(6
   a^2+{r_h}^2\right) \alpha _{\theta _2}}\nonumber\\
   & &+\frac{81 \sqrt{3} {g_s}^{5/4} M \left(\frac{1}{N}\right)^{13/20} {N_f}
   (r-{r_h}) \left({r_h}^2-3 a^2\right) \left(9 a^2+{r_h}^2\right) \log ^2({r_h}) \alpha _{\theta _1} f''(r)}{4 \pi ^{7/4}
   {r_h} \left(6 a^2+{r_h}^2\right) \alpha _{\theta _2}}\nonumber\\
   & & -\frac{27 \sqrt{3} {g_s}^{5/4} M \left(\frac{1}{N}\right)^{13/20}
   {N_f} (r-{r_h}) \left({r_h}^2-3 a^2\right) \left(9 a^2+{r_h}^2\right) \log ^2({r_h}) \alpha _{\theta _1}
   {f_{zz}}''(r)}{4 \pi ^{7/4} {r_h} \left(6 a^2+{r_h}^2\right) \alpha _{\theta _2}}\nonumber\\
\end{eqnarray*}
\begin{eqnarray*}
& &  +\frac{27 \sqrt{3} {g_s}^{5/4} M
   \left(\frac{1}{N}\right)^{13/20} {N_f} (r-{r_h}) \left({r_h}^2-3 a^2\right) \left(9 a^2+{r_h}^2\right) \log
   ^2({r_h}) \alpha _{\theta _1} {f_{x^{10} x^{10}}}''(r)}{4 \pi ^{7/4} {r_h} \left(6 a^2+{r_h}^2\right) \alpha _{\theta
   _2}}\nonumber\\
   & & +\frac{81 \sqrt{3} {g_s}^{5/4} M \left(\frac{1}{N}\right)^{13/20} {N_f} (r-{r_h}) \left({r_h}^2-3 a^2\right)
   \left(9 a^2+{r_h}^2\right) \log ^2({r_h}) \alpha _{\theta _1} {f_{\theta_1z}}''(r)}{8 \pi ^{7/4} {r_h} \left(6
   a^2+{r_h}^2\right) \alpha _{\theta _2}}\nonumber\\
& &+\frac{189 \sqrt{\frac{3}{2}} {g_s}^{11/4} M^2 \left(\frac{1}{N}\right)^{29/20}
   {N_f}^2 (r-{r_h}) \left({r_h}^2-3 a^2\right) \left(9 a^2+{r_h}^2\right) \log ^3({r_h}) \alpha _{\theta _1} \left(81
   \sqrt{2} \alpha _{\theta _1}^3+10 \sqrt{3} \alpha _{\theta _2}^2\right) {f_{\theta_1x}}''(r)}{64 \pi ^{13/4} \left({r_h}^3+6 a^2
   {r_h}\right) \alpha _{\theta _2}^2}\nonumber\\
   & & -\frac{27 \sqrt{3} {g_s}^{5/4} M \left(\frac{1}{N}\right)^{13/20} {N_f} (r-{r_h})
   \left({r_h}^2-3 a^2\right) \left(9 a^2+{r_h}^2\right) \log ^2({r_h}) \alpha _{\theta _1} {f_{\theta_1y}}''(r)}{4 \pi ^{7/4}
   {r_h} \left(6 a^2+{r_h}^2\right) \alpha _{\theta _2}}\nonumber\\
     & & -\frac{6561 {g_s}^{11/4} M^2 \left(\frac{1}{N}\right)^{21/20}
   {N_f}^2 (r-{r_h}) \left({r_h}^2-3 a^2\right) \left(9 a^2+{r_h}^2\right) \log ^3({r_h}) \alpha _{\theta _1}^5
   {f_{\theta_2z}}''(r)}{32 \sqrt{2} \pi ^{13/4} \left({r_h}^3+6 a^2 {r_h}\right) \alpha _{\theta _2}^4}\nonumber\\
   & & -\frac{189 \sqrt{\frac{3}{2}}
   {g_s}^{11/4} M^2 \left(\frac{1}{N}\right)^{29/20} {N_f}^2 (r-{r_h}) \left({r_h}^2-3 a^2\right) \left(9
   a^2+{r_h}^2\right) \log ^3({r_h}) \alpha _{\theta _1} \left(81 \sqrt{2} \alpha _{\theta _1}^3+10 \sqrt{3} \alpha _{\theta
   _2}^2\right) {f_{\theta_2x}}''(r)}{64 \pi ^{13/4} \left({r_h}^3+6 a^2 {r_h}\right) \alpha _{\theta _2}^2}\nonumber\\
   & & +\frac{81 \sqrt{3}
   {g_s}^{5/4} M \left(\frac{1}{N}\right)^{13/20} {N_f} (r-{r_h}) \left({r_h}^2-3 a^2\right) \left(9
   a^2+{r_h}^2\right) \log ^2({r_h}) \alpha _{\theta _1} {f_{\theta_2y}}''(r)}{8 \pi ^{7/4} {r_h} \left(6 a^2+{r_h}^2\right)
   \alpha _{\theta _2}}\nonumber\\
   & &+\frac{81 \sqrt{3} {g_s}^{5/4} M \left(\frac{1}{N}\right)^{13/20} {N_f} (r-{r_h}) \left({r_h}^2-3
   a^2\right) \left(9 a^2+{r_h}^2\right) \log ^2({r_h}) \alpha _{\theta _1} {f_{xz}}''(r)}{8 \pi ^{7/4} {r_h} \left(6
   a^2+{r_h}^2\right) \alpha _{\theta _2}}\nonumber\\
& &   -\frac{6561 \sqrt{3} {g_s}^{5/4} M \left(\frac{1}{N}\right)^{41/20} {N_f}
   (r-{r_h}) \left({r_h}^2-3 a^2\right) \left(9 a^2+{r_h}^2\right) \log ^2({r_h}) \alpha _{\theta _1}^5 \alpha _{\theta
   _2} {f_{xx}}''(r)}{32 \pi ^{7/4} \left({r_h}^3+6 a^2 {r_h}\right)}\nonumber\\
   & & -\frac{243 {g_s}^{11/4} M^2
   \left(\frac{1}{N}\right)^{29/20} {N_f}^2 (r-{r_h}) \left({r_h}^2-3 a^2\right) \left(9 a^2+{r_h}^2\right) \log
   ^3({r_h}) \alpha _{\theta _1} \left(27 \sqrt{3} \alpha _{\theta _1}^3+5 \sqrt{2} \alpha _{\theta _2}^2\right) {f_{xy}}''(r)}{256
   \pi ^{13/4} \left({r_h}^3+6 a^2 {r_h}\right) \alpha _{\theta _2}^2}\nonumber\\
   & & +\frac{243 {g_s}^{11/4} M^2
   \left(\frac{1}{N}\right)^{29/20} {N_f}^2 (r-{r_h}) \left({r_h}^2-3 a^2\right) \left(9 a^2+{r_h}^2\right) \log
   ^3({r_h}) \alpha _{\theta _1} \left(27 \sqrt{3} \alpha _{\theta _1}^3+5 \sqrt{2} \alpha _{\theta _2}^2\right) {f_{yz}}''(r)}{256
   \pi ^{13/4} \left({r_h}^3+6 a^2 {r_h}\right) \alpha _{\theta _2}^2}\nonumber\\
   & & -\frac{81 \sqrt{3} {g_s}^{17/4} M^3
   \left(\frac{1}{N}\right)^{9/4} {N_f}^3 (r-{r_h}) \left({r_h}^2-3 a^2\right) \left(9 a^2+{r_h}^2\right) \log
   ^4({r_h}) \alpha _{\theta _1} \left(2187 \alpha _{\theta _1}^6+270 \sqrt{6} \alpha _{\theta _2}^2 \alpha _{\theta _1}^3+50 \alpha
   _{\theta _2}^4\right) {f_{yy}}''(r)}{4096 \pi ^{19/4} \left({r_h}^3+6 a^2 {r_h}\right) \alpha _{\theta _2}^3}\nonumber\\
& &    +\frac{27
   \sqrt{3} {g_s}^{5/4} M \left(\frac{1}{N}\right)^{13/20} {N_f} (r-{r_h}) \left({r_h}^2-3 a^2\right) \left(9
   a^2+{r_h}^2\right) \log ^2({r_h}) \alpha _{\theta _1} {f_t}''(r)}{4 \pi ^{7/4} {r_h} \left(6 a^2+{r_h}^2\right)
   \alpha _{\theta _2}}\nonumber\\
   & &  -\frac{2187 \sqrt{3} {g_s}^{21/4} M^5 \left(\frac{1}{N}\right)^{53/20} {N_f}^3 {f_t}(r) \log
   ^6({r_h}) \left(16 \left(108 a^6-9 {r_h}^4 a^2-{r_h}^6\right)-\left(2 {r_h}^6+27 a^2 {r_h}^4+117 a^4
   {r_h}^2\right) \log \left({r_h}^6+9 a^2 {r_h}^4\right)\right) \alpha _{\theta _1}}{512 \pi ^{23/4} \left({r_h}^3+6 a^2
   {r_h}\right)^2 \log \left({r_h}^6+9 a^2 {r_h}^4\right) \alpha _{\theta _2}}\nonumber\\
   & & -\frac{6561 \sqrt{3} {g_s}^{21/4}
   \log N  M^5 \left(\frac{1}{N}\right)^{53/20} {N_f}^3 \left({r_h}^2-3 a^2\right) \left(9 a^2+{r_h}^2\right) f(r) \log
   ^6({r_h}) \alpha _{\theta _1}}{128 \pi ^{23/4} {r_h}^2 \left(6 a^2+{r_h}^2\right) \log \left({r_h}^6+9 a^2
   {r_h}^4\right) \alpha _{\theta _2}}+\frac{49 \pi ^{5/4} {r_h}^2 {f_{zz}}(r) \alpha _{\theta _1}}{192 \sqrt{3}
   {g_s}^{7/4} M \left(\frac{1}{N}\right)^{7/20} {N_f} \left({r_h}^2-3 a^2\right) \log ^2({r_h}) \alpha _{\theta
   _2}}\nonumber\\
   & & -\frac{49 \pi ^{5/4} {r_h}^2 {f_{x^{10} x^{10}}}(r) \alpha _{\theta _1}}{128 \sqrt{3} {g_s}^{7/4} M
   \left(\frac{1}{N}\right)^{7/20} {N_f} \left({r_h}^2-3 a^2\right) \log ^2({r_h}) \alpha _{\theta _2}}-\frac{49 \pi ^{5/4}
   {r_h}^2 {f_{\theta_1z}}(r) \alpha _{\theta _1}}{96 \sqrt{3} {g_s}^{7/4} M \left(\frac{1}{N}\right)^{7/20} {N_f}
   \left({r_h}^2-3 a^2\right) \log ^2({r_h}) \alpha _{\theta _2}}\nonumber\\
   & & -\frac{245 \pi ^{5/4} {r_h}^2 {f_{\theta_2y}}(r) \alpha _{\theta
   _1}}{384 \sqrt{3} {g_s}^{7/4} M \left(\frac{1}{N}\right)^{7/20} {N_f} \left({r_h}^2-3 a^2\right) \log ^2({r_h}) \alpha
   _{\theta _2}}\nonumber\\
   & & -\frac{15 \sqrt{3} {g_s}^{5/4} M \sqrt[4]{\frac{1}{N}} {N_f} (r-{r_h}) {r_h} \left(9
   a^2+{r_h}^2\right) \left(108 b^2 {r_h}^2+1\right)^2 {f_r}(r)}{8 \pi ^{7/4} \left({r_h}^2-3 a^2\right) \left(6
   a^2+{r_h}^2\right) \alpha _{\theta _1} \alpha _{\theta _2}}+\frac{2205 \sqrt[20]{\frac{1}{N}} {r_h}^2 {f_{\theta_1y}}(r) \alpha
   _{\theta _1}^5}{512 \sqrt{2} \sqrt[4]{{g_s}} \sqrt[4]{\pi } \left({r_h}^2-3 a^2\right) \log ({r_h}) \alpha _{\theta
   _2}^4}\nonumber\\
   & & +\frac{2205 \sqrt[20]{\frac{1}{N}} {r_h}^2 {f_{\theta_2z}}(r) \alpha _{\theta _1}^5}{512 \sqrt{2} \sqrt[4]{{g_s}}
   \sqrt[4]{\pi } \left({r_h}^2-3 a^2\right) \log ({r_h}) \alpha _{\theta _2}^4}-\frac{3969 \sqrt[20]{\frac{1}{N}} {r_h}^2
   {f_{yz}}(r) \alpha _{\theta _1}^5}{256 \sqrt{2} \sqrt[4]{{g_s}} \sqrt[4]{\pi } \left({r_h}^2-3 a^2\right) \log ({r_h})
   \alpha _{\theta _2}^4} + \frac{3969 \sqrt[20]{\frac{1}{N}} {r_h}^2 {f_{yy}}(r) \alpha _{\theta _1}^5}{512 \sqrt{2}
   \sqrt[4]{{g_s}} \sqrt[4]{\pi } \left({r_h}^2-3 a^2\right) \log ({r_h}) \alpha _{\theta _2}^4} = 0.
\end{eqnarray*}
}

(ix) \underline{${\rm EOM}_{\theta_1z}$}

{\scriptsize
\begin{eqnarray*}
& & -\frac{8 \sqrt{\frac{2}{3}} \left(9 b^2+1\right)^4 {g_s} M^2 \left(\frac{1}{N}\right)^{5/4} (r-{r_h})^3 \beta  \log ({r_h})
   \alpha _{\theta _1} \left(19683 \sqrt{6} \alpha _{\theta _1}^6+6642 \alpha _{\theta _2}^2 \alpha _{\theta _1}^3-40 \sqrt{6} \alpha
   _{\theta _2}^4\right) b^{12}}{3 \left(3 b^2-1\right)^5 \left(6 b^2+1\right)^4 \log N  \pi ^3 {r_h}^3 \alpha _{\theta
   _2}^7}\nonumber\\
   & & +\frac{3 \left(9 b^2+1\right)^4 {g_s}^{5/4} M^2 \left(\frac{1}{N}\right)^{8/5} (r-{r_h}) \beta  \log ^2({r_h}) \alpha
   _{\theta _1}\Sigma_1 b^{10}}{\sqrt{2} \left(-18 b^4+3 b^2+1\right)^4 \log N  \pi ^{11/4} {r_h} \alpha _{\theta _2}^5}\nonumber\\
   & & +\frac{2187
   {g_s}^{5/4} M \left(\frac{1}{N}\right)^{7/4} {N_f} \left({r_h}^2-3 a^2\right) {f_{xx}}(r) \log ^2({r_h}) \alpha
   _{\theta _1}^5}{8 \sqrt{2} \pi ^{7/4} {r_h}^2 \log \left({r_h}^6+9 a^2 {r_h}^4\right)}-\frac{245
   \left(\frac{1}{N}\right)^{3/20} {r_h}^2 {f_{xy}}(r) \alpha _{\theta _1} \left(27 \sqrt{6} \alpha _{\theta _1}^3+10 \alpha
   _{\theta _2}^2\right)}{36864 \sqrt{3} \sqrt[4]{{g_s}} \sqrt[4]{\pi } \left({r_h}^2-3 a^2\right) \log ({r_h}) \alpha
   _{\theta _2}^3}\nonumber\\
   & & +\frac{343 \left(\frac{1}{N}\right)^{3/20} {r_h}^2 {f_{\theta_1x}}(r) \alpha _{\theta _1} \left(81 \sqrt{2} \alpha
   _{\theta _1}^3+10 \sqrt{3} \alpha _{\theta _2}^2\right)}{27648 \sqrt[4]{{g_s}} \sqrt[4]{\pi } \left({r_h}^2-3 a^2\right) \log
   ({r_h}) \alpha _{\theta _2}^3}-\frac{1715 \left(\frac{1}{N}\right)^{3/20} {r_h}^2 {f_{\theta_2x}}(r) \alpha _{\theta _1} \left(81
   \sqrt{2} \alpha _{\theta _1}^3+10 \sqrt{3} \alpha _{\theta _2}^2\right)}{82944 \sqrt[4]{{g_s}} \sqrt[4]{\pi } \left({r_h}^2-3
   a^2\right) \log ({r_h}) \alpha _{\theta _2}^3}\nonumber\\
   & & +\frac{3479 \left(\frac{1}{N}\right)^{3/20} {r_h}^2 {f_{xz}}(r) \alpha
   _{\theta _1} \left(81 \sqrt{2} \alpha _{\theta _1}^3+10 \sqrt{3} \alpha _{\theta _2}^2\right)}{331776 \sqrt[4]{{g_s}} \sqrt[4]{\pi
   } \left({r_h}^2-3 a^2\right) \log ({r_h}) \alpha _{\theta _2}^3}-\frac{27 {g_s}^{5/4} M \left(\frac{1}{N}\right)^{7/20}
   {N_f} \left({r_h}^2-3 a^2\right) \left(9 a^2+{r_h}^2\right) \log ^2({r_h}) \alpha _{\theta _1} f'(r)}{2 \sqrt{2} \pi
   ^{7/4} \left({r_h}^3+6 a^2 {r_h}\right) \alpha _{\theta _2}^2}\nonumber\\
   & & +\frac{9 {g_s}^{5/4} M \left(\frac{1}{N}\right)^{7/20}
   {N_f} \left({r_h}^2-3 a^2\right) \left(9 a^2+{r_h}^2\right) \log ^2({r_h}) \alpha _{\theta _1} {f_{zz}}'(r)}{2
   \sqrt{2} \pi ^{7/4} \left({r_h}^3+6 a^2 {r_h}\right) \alpha _{\theta _2}^2}-\frac{9 {g_s}^{5/4} M
   \left(\frac{1}{N}\right)^{7/20} {N_f} \left({r_h}^2-3 a^2\right) \left(9 a^2+{r_h}^2\right) \log ^2({r_h}) \alpha
   _{\theta _1} {f_{x^{10} x^{10}}}'(r)}{2 \sqrt{2} \pi ^{7/4} \left({r_h}^3+6 a^2 {r_h}\right) \alpha _{\theta _2}^2}
   \nonumber\\
   & &-\frac{9
   {g_s}^{5/4} M \left(\frac{1}{N}\right)^{7/20} {N_f} \left({r_h}^2-3 a^2\right) \left(9 a^2+{r_h}^2\right) \log
   ^2({r_h}) \alpha _{\theta _1} {f_{\theta_1z}}'(r)}{4 \sqrt{2} \pi ^{7/4} \left({r_h}^3+6 a^2 {r_h}\right) \alpha _{\theta
   _2}^2}+\frac{27 a^2 {g_s}^{5/4} M \left(\frac{1}{N}\right)^{7/20} {N_f} (r-{r_h}) \left(9 a^2+{r_h}^2\right) \log
   ^2({r_h}) \alpha _{\theta _1} {f_{\theta_1x}}'(r)}{2 \sqrt{2} \pi ^{7/4} {r_h}^2 \left(6 a^2+{r_h}^2\right) \alpha _{\theta
   _2}^2}\nonumber\\
   & & +\frac{729 \sqrt{3} {g_s}^{11/4} M^2 \left(\frac{1}{N}\right)^{3/4} {N_f}^2 \left({r_h}^2-3 a^2\right) \left(9
   a^2+{r_h}^2\right) \log ^3({r_h}) \alpha _{\theta _1}^5 {f_{\theta_1y}}'(r)}{32 \pi ^{13/4} \left({r_h}^3+6 a^2 {r_h}\right)
   \alpha _{\theta _2}^5}\nonumber\\
   & & +\frac{729 \sqrt{3} {g_s}^{11/4} M^2 \left(\frac{1}{N}\right)^{3/4} {N_f}^2 \left({r_h}^2-3
   a^2\right) \left(9 a^2+{r_h}^2\right) \log ^3({r_h}) \alpha _{\theta _1}^5 {f_{\theta_2z}}'(r)}{32 \pi ^{13/4} \left({r_h}^3+6
   a^2 {r_h}\right) \alpha _{\theta _2}^5}\nonumber\\
   & & -\frac{729 \sqrt{3} a^2 {g_s}^{11/4} M^2 \left(\frac{1}{N}\right)^{19/20} {N_f}^2
   (r-{r_h}) \left(9 a^2+{r_h}^2\right) \log ^3({r_h}) \alpha _{\theta _1}^3 {f_{\theta_2x}}'(r)}{16 \pi ^{13/4} {r_h}^2
   \left(6 a^2+{r_h}^2\right) \alpha _{\theta _2}^3}\nonumber\\
   &&-\frac{27 {g_s}^{5/4} M \left(\frac{1}{N}\right)^{7/20} {N_f}
   \left({r_h}^2-3 a^2\right) \left(9 a^2+{r_h}^2\right) \log ^2({r_h}) \alpha _{\theta _1} {f_{\theta_2y}}'(r)}{4 \sqrt{2} \pi
   ^{7/4} \left({r_h}^3+6 a^2 {r_h}\right) \alpha _{\theta _2}^2}\nonumber\\
   & & -\frac{27 {g_s}^{5/4} M \left(\frac{1}{N}\right)^{7/20}
   {N_f} \left({r_h}^2-3 a^2\right) \left(9 a^2+{r_h}^2\right) \log ^2({r_h}) \alpha _{\theta _1} {f_{xz}}'(r)}{4
   \sqrt{2} \pi ^{7/4} \left({r_h}^3+6 a^2 {r_h}\right) \alpha _{\theta _2}^2}+\frac{2187 {g_s}^{5/4} M
   \left(\frac{1}{N}\right)^{7/4} {N_f} \left({r_h}^2-3 a^2\right) \left(9 a^2+{r_h}^2\right) \log ^2({r_h}) \alpha
   _{\theta _1}^5 {f_{xx}}'(r)}{16 \sqrt{2} \pi ^{7/4} \left({r_h}^3+6 a^2 {r_h}\right)}\nonumber\\
   & & +\frac{729 \sqrt{3} a^2 {g_s}^{11/4}
   M^2 \left(\frac{1}{N}\right)^{3/4} {N_f}^2 (r-{r_h}) \left(9 a^2+{r_h}^2\right) \log ^3({r_h}) \alpha _{\theta _1}^5
   {f_{xy}}'(r)}{16 \pi ^{13/4} {r_h}^2 \left(6 a^2+{r_h}^2\right) \alpha _{\theta _2}^5}\nonumber\\
   &&+\frac{729 \sqrt{3} a^2
   {g_s}^{11/4} M^2 \left(\frac{1}{N}\right)^{3/4} {N_f}^2 (r-{r_h}) \left(9 a^2+{r_h}^2\right) \log ^3({r_h}) \alpha
   _{\theta _1}^5 {f_{yz}}'(r)}{16 \pi ^{13/4} {r_h}^2 \left(6 a^2+{r_h}^2\right) \alpha _{\theta _2}^5}
\nonumber\\
& &   +\frac{243 a^2
   {g_s}^{17/4} M^3 \left(\frac{1}{N}\right)^{7/4} {N_f}^3 (r-{r_h}) \left(9 a^2+{r_h}^2\right) \log ^4({r_h}) \alpha
   _{\theta _1}^3 \left(2187 \sqrt{2} \alpha _{\theta _1}^6+540 \sqrt{3} \alpha _{\theta _2}^2 \alpha _{\theta _1}^3+50 \sqrt{2} \alpha
   _{\theta _2}^4\right) {f_{yy}}'(r)}{1024 \pi ^{19/4} {r_h}^2 \left(6 a^2+{r_h}^2\right) \alpha _{\theta _2}^6}\nonumber\\
   & & +\frac{9
   {g_s}^{5/4} M \left(\frac{1}{N}\right)^{7/20} {N_f} \left({r_h}^2-3 a^2\right) \left(9 a^2+{r_h}^2\right) \log
   ^2({r_h}) \alpha _{\theta _1} {f_r}'(r)}{4 \sqrt{2} \pi ^{7/4} \left({r_h}^3+6 a^2 {r_h}\right) \alpha _{\theta
   _2}^2}\nonumber\\
& &  -\frac{27 {g_s}^{5/4} M \left(\frac{1}{N}\right)^{7/20} {N_f} \left({r_h}^2-3 a^2\right) \left(9
   a^2+{r_h}^2\right) \log ^2({r_h}) \alpha _{\theta _1} {f_t}'(r)}{4 \sqrt{2} \pi ^{7/4} \left({r_h}^3+6 a^2
   {r_h}\right) \alpha _{\theta _2}^2}\nonumber\\
   &&-\frac{27 {g_s}^{5/4} M \left(\frac{1}{N}\right)^{7/20} {N_f} (r-{r_h})
   \left({r_h}^2-3 a^2\right) \left(9 a^2+{r_h}^2\right) \log ^2({r_h}) \alpha _{\theta _1} f''(r)}{2 \sqrt{2} \pi ^{7/4}
   \left({r_h}^3+6 a^2 {r_h}\right) \alpha _{\theta _2}^2}\nonumber\\
    \end{eqnarray*}}
   {\scriptsize
   \begin{eqnarray*}
   & & +\frac{9 {g_s}^{5/4} M \left(\frac{1}{N}\right)^{7/20} {N_f}
   (r-{r_h}) \left({r_h}^2-3 a^2\right) \left(9 a^2+{r_h}^2\right) \log ^2({r_h}) \alpha _{\theta _1} {f_{zz}}''(r)}{2
   \sqrt{2} \pi ^{7/4} \left({r_h}^3+6 a^2 {r_h}\right) \alpha _{\theta _2}^2}\nonumber\\
   & & -\frac{9 {g_s}^{5/4} M
   \left(\frac{1}{N}\right)^{7/20} {N_f} (r-{r_h}) \left({r_h}^2-3 a^2\right) \left(9 a^2+{r_h}^2\right) \log
   ^2({r_h}) \alpha _{\theta _1} {f_{x^{10} x^{10}}}''(r)}{2 \sqrt{2} \pi ^{7/4} \left({r_h}^3+6 a^2 {r_h}\right) \alpha _{\theta
   _2}^2}\nonumber\\
 & & -\frac{9 {g_s}^{5/4} M \left(\frac{1}{N}\right)^{7/20} {N_f} (r-{r_h}) \left({r_h}^2-3 a^2\right) \left(9
   a^2+{r_h}^2\right) \log ^2({r_h}) \alpha _{\theta _1} {f_{\theta_1z}}''(r)}{4 \sqrt{2} \pi ^{7/4} \left({r_h}^3+6 a^2
   {r_h}\right) \alpha _{\theta _2}^2}\nonumber\\
   & & -\frac{63 {g_s}^{11/4} M^2 \left(\frac{1}{N}\right)^{23/20} {N_f}^2 (r-{r_h})
   \left({r_h}^2-3 a^2\right) \left(9 a^2+{r_h}^2\right) \log ^3({r_h}) \alpha _{\theta _1} \left(81 \sqrt{2} \alpha _{\theta
   _1}^3+10 \sqrt{3} \alpha _{\theta _2}^2\right) {f_{\theta_1x}}''(r)}{64 \pi ^{13/4} \left({r_h}^3+6 a^2 {r_h}\right) \alpha _{\theta
   _2}^3}\nonumber\\
   & &  +\frac{729 \sqrt{3} {g_s}^{11/4} M^2 \left(\frac{1}{N}\right)^{3/4} {N_f}^2 (r-{r_h}) \left({r_h}^2-3 a^2\right)
   \left(9 a^2+{r_h}^2\right) \log ^3({r_h}) \alpha _{\theta _1}^5 {f_{\theta_1y}}''(r)}{32 \pi ^{13/4} \left({r_h}^3+6 a^2
   {r_h}\right) \alpha _{\theta _2}^5}\nonumber\\
   & & +\frac{729 \sqrt{3} {g_s}^{11/4} M^2 \left(\frac{1}{N}\right)^{3/4} {N_f}^2
   (r-{r_h}) \left({r_h}^2-3 a^2\right) \left(9 a^2+{r_h}^2\right) \log ^3({r_h}) \alpha _{\theta _1}^5
   {f_{\theta_2z}}''(r)}{32 \pi ^{13/4} \left({r_h}^3+6 a^2 {r_h}\right) \alpha _{\theta _2}^5}\nonumber\\
   & & +\frac{63 {g_s}^{11/4} M^2
   \left(\frac{1}{N}\right)^{23/20} {N_f}^2 (r-{r_h}) \left({r_h}^2-3 a^2\right) \left(9 a^2+{r_h}^2\right) \log
   ^3({r_h}) \alpha _{\theta _1} \left(81 \sqrt{2} \alpha _{\theta _1}^3+10 \sqrt{3} \alpha _{\theta _2}^2\right) {f_{\theta_2x}}''(r)}{64
   \pi ^{13/4} \left({r_h}^3+6 a^2 {r_h}\right) \alpha _{\theta _2}^3}\nonumber\\
   & & -\frac{27 {g_s}^{5/4} M \left(\frac{1}{N}\right)^{7/20}
   {N_f} (r-{r_h}) \left({r_h}^2-3 a^2\right) \left(9 a^2+{r_h}^2\right) \log ^2({r_h}) \alpha _{\theta _1}
   {f_{\theta_2y}}''(r)}{4 \sqrt{2} \pi ^{7/4} \left({r_h}^3+6 a^2 {r_h}\right) \alpha _{\theta _2}^2}\nonumber\\
   & & -\frac{27 {g_s}^{5/4} M
   \left(\frac{1}{N}\right)^{7/20} {N_f} (r-{r_h}) \left({r_h}^2-3 a^2\right) \left(9 a^2+{r_h}^2\right) \log
   ^2({r_h}) \alpha _{\theta _1} {f_{xz}}''(r)}{4 \sqrt{2} \pi ^{7/4} \left({r_h}^3+6 a^2 {r_h}\right) \alpha _{\theta
   _2}^2}\nonumber\\
   & &  +\frac{2187 {g_s}^{5/4} M \left(\frac{1}{N}\right)^{7/4} {N_f} (r-{r_h}) \left({r_h}^2-3 a^2\right) \left(9
   a^2+{r_h}^2\right) \log ^2({r_h}) \alpha _{\theta _1}^5 {f_{xx}}''(r)}{16 \sqrt{2} \pi ^{7/4} \left({r_h}^3+6 a^2
   {r_h}\right)}\nonumber\\
   & & +\frac{27 \sqrt{\frac{3}{2}} {g_s}^{11/4} M^2 \left(\frac{1}{N}\right)^{23/20} {N_f}^2 (r-{r_h})
   \left({r_h}^2-3 a^2\right) \left(9 a^2+{r_h}^2\right) \log ^3({r_h}) \alpha _{\theta _1} \left(27 \sqrt{3} \alpha _{\theta
   _1}^3+5 \sqrt{2} \alpha _{\theta _2}^2\right) {f_{xy}}''(r)}{128 \pi ^{13/4} \left({r_h}^3+6 a^2 {r_h}\right) \alpha _{\theta
   _2}^3}\nonumber\\
   & & -\frac{27 \sqrt{\frac{3}{2}} {g_s}^{11/4} M^2 \left(\frac{1}{N}\right)^{23/20} {N_f}^2 (r-{r_h}) \left({r_h}^2-3
   a^2\right) \left(9 a^2+{r_h}^2\right) \log ^3({r_h}) \alpha _{\theta _1} \left(27 \sqrt{3} \alpha _{\theta _1}^3+5 \sqrt{2}
   \alpha _{\theta _2}^2\right) {f_{yz}}''(r)}{128 \pi ^{13/4} \left({r_h}^3+6 a^2 {r_h}\right) \alpha _{\theta _2}^3}\nonumber\\
  & &  +\frac{27
   {g_s}^{17/4} M^3 \left(\frac{1}{N}\right)^{39/20} {N_f}^3 (r-{r_h}) \left({r_h}^2-3 a^2\right) \left(9
   a^2+{r_h}^2\right) \log ^4({r_h}) \alpha _{\theta _1} \left(2187 \alpha _{\theta _1}^6+270 \sqrt{6} \alpha _{\theta _2}^2
   \alpha _{\theta _1}^3+50 \alpha _{\theta _2}^4\right) {f_{yy}}''(r)}{2048 \sqrt{2} \pi ^{19/4} \left({r_h}^3+6 a^2
   {r_h}\right) \alpha _{\theta _2}^4}\nonumber\\
   & & -\frac{9 {g_s}^{5/4} M \left(\frac{1}{N}\right)^{7/20} {N_f} (r-{r_h})
   \left({r_h}^2-3 a^2\right) \left(9 a^2+{r_h}^2\right) \log ^2({r_h}) \alpha _{\theta _1} {f_t}''(r)}{2 \sqrt{2} \pi
   ^{7/4} \left({r_h}^3+6 a^2 {r_h}\right) \alpha _{\theta _2}^2}\nonumber\\
   & & -\frac{729 {g_s}^{21/4} M^5 \left(\frac{1}{N}\right)^{47/20}
   {N_f}^3 {f_t}(r) \log ^6({r_h}) \left(16 \left(-108 a^6+9 {r_h}^4 a^2+{r_h}^6\right)+\left(2 {r_h}^6+27 a^2
   {r_h}^4+117 a^4 {r_h}^2\right) \log \left({r_h}^6+9 a^2 {r_h}^4\right)\right) \alpha _{\theta _1}}{256 \sqrt{2} \pi
   ^{23/4} {r_h}^2 \left(6 a^2+{r_h}^2\right)^2 \log \left({r_h}^6+9 a^2 {r_h}^4\right) \alpha _{\theta _2}^2}\nonumber\\
   & &  +\frac{2187
   {g_s}^{21/4} \log N  M^5 \left(\frac{1}{N}\right)^{47/20} {N_f}^3 \left({r_h}^2-3 a^2\right) \left(9
   a^2+{r_h}^2\right) f(r) \log ^6({r_h}) \alpha _{\theta _1}}{64 \sqrt{2} \pi ^{23/4} {r_h}^2 \left(6 a^2+{r_h}^2\right)
   \log \left({r_h}^6+9 a^2 {r_h}^4\right) \alpha _{\theta _2}^2}\nonumber\\
   & & -\frac{49 \pi ^{5/4} {r_h}^2 {f_{zz}}(r) \alpha _{\theta
   _1}}{864 \sqrt{2} {g_s}^{7/4} M \left(\frac{1}{N}\right)^{13/20} {N_f} \left({r_h}^2-3 a^2\right) \log ^2({r_h}) \alpha
   _{\theta _2}^2}+\frac{49 \pi ^{5/4} {r_h}^2 {f_{x^{10} x^{10}}}(r) \alpha _{\theta _1}}{576 \sqrt{2} {g_s}^{7/4} M
   \left(\frac{1}{N}\right)^{13/20} {N_f} \left({r_h}^2-3 a^2\right) \log ^2({r_h}) \alpha _{\theta _2}^2}\nonumber\\
   & & +\frac{49 \pi ^{5/4}
   {r_h}^2 {f_{\theta_1z}}(r) \alpha _{\theta _1}}{432 \sqrt{2} {g_s}^{7/4} M \left(\frac{1}{N}\right)^{13/20} {N_f}
   \left({r_h}^2-3 a^2\right) \log ^2({r_h}) \alpha _{\theta _2}^2}+\frac{245 \pi ^{5/4} {r_h}^2 {f_{\theta_2y}}(r) \alpha _{\theta
   _1}}{1728 \sqrt{2} {g_s}^{7/4} M \left(\frac{1}{N}\right)^{13/20} {N_f} \left({r_h}^2-3 a^2\right) \log ^2({r_h})
   \alpha _{\theta _2}^2}\nonumber\\
   \end{eqnarray*}}
{\scriptsize
\begin{eqnarray*}
   & & +\frac{5 {g_s}^{5/4} M {N_f} (r-{r_h}) {r_h} \left(9 a^2+{r_h}^2\right) \left(108 b^2
   {r_h}^2+1\right)^2 {f_r}(r)}{4 \sqrt{2} \sqrt[20]{\frac{1}{N}} \pi ^{7/4} \left({r_h}^2-3 a^2\right) \left(6
   a^2+{r_h}^2\right) \alpha _{\theta _1} \alpha _{\theta _2}^2}+\frac{3 {g_s}^{5/4} M {N_f} (r-{r_h}) {r_h} \left(9
   a^2+{r_h}^2\right) \left(108 b^2 {r_h}^2+1\right)^2 {f_{yz}}(r) \alpha _{\theta _1}}{\sqrt{2} \sqrt[4]{\frac{1}{N}} \pi
   ^{7/4} \left(-18 a^4+3 {r_h}^2 a^2+{r_h}^4\right) \alpha _{\theta _2}^4}\nonumber\\
   & & -\frac{3 {g_s}^{5/4} M {N_f} (r-{r_h})
   {r_h} \left(9 a^2+{r_h}^2\right) \left(108 b^2 {r_h}^2+1\right)^2 {f_{yy}}(r) \alpha _{\theta _1}}{2 \sqrt{2}
   \sqrt[4]{\frac{1}{N}} \pi ^{7/4} \left(-18 a^4+3 {r_h}^2 a^2+{r_h}^4\right) \alpha _{\theta _2}^4}-\frac{245 \sqrt{3}
   {r_h}^2 {f_{\theta_1y}}(r) \alpha _{\theta _1}^5}{512 \sqrt[4]{{g_s}} \sqrt[4]{\frac{1}{N}} \sqrt[4]{\pi } \left({r_h}^2-3
   a^2\right) \log ({r_h}) \alpha _{\theta _2}^5}\nonumber\\
   & & -\frac{245 \sqrt{3} {r_h}^2 {f_{\theta_2z}}(r) \alpha _{\theta _1}^5}{512
   \sqrt[4]{{g_s}} \sqrt[4]{\frac{1}{N}} \sqrt[4]{\pi } \left({r_h}^2-3 a^2\right) \log ({r_h}) \alpha _{\theta _2}^5} = 0.
\end{eqnarray*}
}
(x) \underline{${\rm EOM}_{\theta_2\theta_2}$}

{\scriptsize
\begin{eqnarray*}
& & -\frac{49 \left(\frac{1}{N}\right)^{9/5} {f_{xx}}(r) \left(27 \sqrt{3} \alpha _{\theta _2} \alpha _{\theta _1}^3+5 \sqrt{2} \alpha
   _{\theta _2}^3\right){}^2 {r_h}^4}{196608 \left({r_h}^2-3 a^2\right)^2 \log ^2({r_h})}-\frac{49 {f_{\theta_1x}}(r) \left(27
   \sqrt{6} \alpha _{\theta _1}^3+10 \alpha _{\theta _2}^2\right) {r_h}^4}{36864 \left({r_h}^2-3 a^2\right)^2 \log ^2({r_h})
   \alpha _{\theta _2}^2}\nonumber\\
   & & +\frac{49 {f_{\theta_2x}}(r) \left(27 \sqrt{6} \alpha _{\theta _1}^3+10 \alpha _{\theta _2}^2\right)
   {r_h}^4}{27648 \left({r_h}^2-3 a^2\right)^2 \log ^2({r_h}) \alpha _{\theta _2}^2}-\frac{49 {f_{xz}}(r) \left(27 \sqrt{6}
   \alpha _{\theta _1}^3+10 \alpha _{\theta _2}^2\right) {r_h}^4}{110592 \left({r_h}^2-3 a^2\right)^2 \log ^2({r_h}) \alpha
   _{\theta _2}^2}\nonumber\\
   & & +\frac{49 \sqrt{\frac{3}{2}} {g_s}^{3/2} M \left(\frac{1}{N}\right)^{2/5} {N_f} {f_{xy}}(r) \alpha _{\theta
   _1}^4 \left(27 \sqrt{6} \alpha _{\theta _1}^3+10 \alpha _{\theta _2}^2\right) {r_h}^4}{16384 \pi ^{3/2} \left({r_h}^2-3
   a^2\right)^2 \log ({r_h}) \alpha _{\theta _2}^5}+\frac{49 {f_{\theta_2z}}(r) \alpha _{\theta _1}^4 {r_h}^4}{512
   \left(\frac{1}{N}\right)^{2/5} \left({r_h}^2-3 a^2\right)^2 \log ^2({r_h}) \alpha _{\theta _2}^4}\nonumber\\
   & & -\frac{49 {f_{\theta_2y}}(r) \alpha
   _{\theta _1}^4 {r_h}^4}{512 \left(\frac{1}{N}\right)^{2/5} \left({r_h}^2-3 a^2\right)^2 \log ^2({r_h}) \alpha _{\theta
   _2}^4}+\frac{\sqrt{\frac{2}{3}} {g_s}^{3/2} M {N_f} (r-{r_h}) \left(9 a^2+{r_h}^2\right) \left(108 b^2
   {r_h}^2+1\right)^2 {f_{\theta_1y}}(r) {r_h}^3}{\left(\frac{1}{N}\right)^{2/5} \pi ^{3/2} \left({r_h}^2-3 a^2\right)^2 \left(6
   a^2+{r_h}^2\right) \log ({r_h}) \alpha _{\theta _2}^3}\nonumber\\
   & & -\frac{\sqrt{\frac{2}{3}} {g_s}^{3/2} M {N_f} (r-{r_h})
   \left(9 a^2+{r_h}^2\right) \left(108 b^2 {r_h}^2+1\right)^2 {f_{yz}}(r) {r_h}^3}{\left(\frac{1}{N}\right)^{2/5} \pi
   ^{3/2} \left({r_h}^2-3 a^2\right)^2 \left(6 a^2+{r_h}^2\right) \log ({r_h}) \alpha _{\theta _2}^3}+\frac{4 (r-{r_h})
   \left(9 a^2+{r_h}^2\right) \left(108 b^2 {r_h}^2+1\right)^2 {f_{zz}}(r) {r_h}^3}{81 \left(\frac{1}{N}\right)^{4/5}
   \left({r_h}^2-3 a^2\right)^2 \left(6 a^2+{r_h}^2\right) \log ^2({r_h}) \alpha _{\theta _1}^4}\nonumber\\
   & & -\frac{4 (r-{r_h}) \left(9
   a^2+{r_h}^2\right) \left(108 b^2 {r_h}^2+1\right)^2 {f_{x^{10} x^{10}}}(r) {r_h}^3}{81 \left(\frac{1}{N}\right)^{4/5}
   \left({r_h}^2-3 a^2\right)^2 \left(6 a^2+{r_h}^2\right) \log ^2({r_h}) \alpha _{\theta _1}^4}-\frac{8 (r-{r_h}) \left(9
   a^2+{r_h}^2\right) \left(108 b^2 {r_h}^2+1\right)^2 {f_{\theta_1z}}(r) {r_h}^3}{81 \left(\frac{1}{N}\right)^{4/5}
   \left({r_h}^2-3 a^2\right)^2 \left(6 a^2+{r_h}^2\right) \log ^2({r_h}) \alpha _{\theta _1}^4}\nonumber\\
   & & -\frac{4 (r-{r_h}) \left(9
   a^2+{r_h}^2\right) \left(108 b^2 {r_h}^2+1\right)^2 {f_r}(r) {r_h}^3}{81 \left(\frac{1}{N}\right)^{4/5}
   \left({r_h}^2-3 a^2\right)^2 \left(6 a^2+{r_h}^2\right) \log ^2({r_h}) \alpha _{\theta _1}^4}+\frac{27 {g_s}^3 M^2
   {N_f}^2 (r-{r_h}) \left(9 a^2+{r_h}^2\right) \left(108 b^2 {r_h}^2+1\right)^2 {f_{yy}}(r) \alpha _{\theta _1}^4
   {r_h}^3}{8 \pi ^3 \left({r_h}^2-3 a^2\right)^2 \left(6 a^2+{r_h}^2\right) \alpha _{\theta _2}^6}\nonumber\\
   & & +\frac{{g_s}^3 M^2
   {N_f}^2 (r-{r_h}) \log ^2({r_h}) \left(-19683 \alpha _{\theta _1}^6+216 \sqrt{6} \alpha _{\theta _2}^2 \alpha _{\theta
   _1}^3+530 \alpha _{\theta _2}^4\right) {f_{\theta_2z}}'(r)}{108 N \pi ^3 \log \left({r_h}^6+9 a^2 {r_h}^4\right) \alpha _{\theta
   _2}^4}\nonumber\\
 & & +\frac{{g_s}^3 M^2 {N_f}^2 (r-{r_h}) \log ^2({r_h}) \left(19683 \alpha _{\theta _1}^6+1107 \sqrt{6} \alpha
   _{\theta _2}^2 \alpha _{\theta _1}^3-40 \alpha _{\theta _2}^4\right) {f_{\theta_2x}}'(r)}{54 N \pi ^3 \log \left({r_h}^6+9 a^2
   {r_h}^4\right) \alpha _{\theta _2}^4}\nonumber\\
   & & +\frac{{g_s}^3 M^2 \left(\frac{1}{N}\right)^{7/5} {N_f}^2 (r-{r_h}) \log
   ^2({r_h}) \left(177147 \sqrt{6} \alpha _{\theta _1}^9+89667 \alpha _{\theta _2}^2 \alpha _{\theta _1}^6-1080 \sqrt{6} \alpha
   _{\theta _2}^4 \alpha _{\theta _1}^3-950 \alpha _{\theta _2}^6\right) {f_{\theta_2y}}'(r)}{3888 \pi ^3 \log \left({r_h}^6+9 a^2
   {r_h}^4\right) \alpha _{\theta _1}^4 \alpha _{\theta _2}^2}=0
\end{eqnarray*}
}

(xi) \underline{${\rm EOM}_{\theta_2x}$}

{\scriptsize
\begin{eqnarray*}
& &-\frac{512 \left(9 b^2+1\right)^4 M \left(\frac{1}{N}\right)^{7/20} (r-{r_h})^3 \beta  \log ({r_h}) \Sigma_1 b^{14}}{243 \left(3
   b^2-1\right)^7 \left(6 b^2+1\right)^4 \sqrt{{g_s}} \log N ^2 {N_f} \pi ^{3/2} {r_h}^3 \alpha _{\theta _1}^4 \alpha
   _{\theta _2}^5}-\frac{\sqrt{2} \left(9 b^2+1\right)^4 {g_s}^{5/4} M^2 \left(\frac{1}{N}\right)^{11/10} (r-{r_h}) \beta  \log
   ^2({r_h})\Sigma_1 b^{10}}{\left(3 b^2-1\right)^5 \left(6 b^2+1\right)^4 \log N ^2 \pi ^{11/4} {r_h} \alpha _{\theta _2}^6}\nonumber\\
   & & +\frac{245
   {r_h}^4 {f_{xy}}(r) \left(27 \sqrt{6} \alpha _{\theta _1}^3+10 \alpha _{\theta _2}^2\right)}{497664 \sqrt{3} \sqrt[4]{{g_s}}
   \left(\frac{1}{N}\right)^{7/20} \sqrt[4]{\pi } \left({r_h}^2-3 a^2\right)^2 \log ^2({r_h}) \alpha _{\theta _2}^4}-\frac{343
   {r_h}^4 {f_{\theta_1x}}(r) \left(81 \sqrt{2} \alpha _{\theta _1}^3+10 \sqrt{3} \alpha _{\theta _2}^2\right)}{373248 \sqrt[4]{{g_s}}
   \left(\frac{1}{N}\right)^{7/20} \sqrt[4]{\pi } \left({r_h}^2-3 a^2\right)^2 \log ^2({r_h}) \alpha _{\theta _2}^4}
\nonumber\\
& &  +\frac{1715
   {r_h}^4 {f_{\theta_2x}}(r) \left(81 \sqrt{2} \alpha _{\theta _1}^3+10 \sqrt{3} \alpha _{\theta _2}^2\right)}{1119744 \sqrt[4]{{g_s}}
   \left(\frac{1}{N}\right)^{7/20} \sqrt[4]{\pi } \left({r_h}^2-3 a^2\right)^2 \log ^2({r_h}) \alpha _{\theta _2}^4}-\frac{3479
   {r_h}^4 {f_{xz}}(r) \left(81 \sqrt{2} \alpha _{\theta _1}^3+10 \sqrt{3} \alpha _{\theta _2}^2\right)}{4478976
   \sqrt[4]{{g_s}} \left(\frac{1}{N}\right)^{7/20} \sqrt[4]{\pi } \left({r_h}^2-3 a^2\right)^2 \log ^2({r_h}) \alpha _{\theta
   _2}^4}\nonumber\\
   & & +\frac{{g_s}^{5/4} M {N_f} {r_h} \left(9 a^2+{r_h}^2\right) \log ({r_h}) f'(r)}{\sqrt{2}
   \left(\frac{1}{N}\right)^{3/20} \pi ^{7/4} \left(6 a^2+{r_h}^2\right) \alpha _{\theta _2}^3}-\frac{{g_s}^{5/4} M {N_f}
   {r_h} \left(9 a^2+{r_h}^2\right) \log ({r_h}) {f_{zz}}'(r)}{3 \sqrt{2} \left(\frac{1}{N}\right)^{3/20} \pi ^{7/4}
   \left(6 a^2+{r_h}^2\right) \alpha _{\theta _2}^3}\nonumber\\
   & & +\frac{{g_s}^{5/4} M {N_f} {r_h} \left(9 a^2+{r_h}^2\right) \log
   ({r_h}) {f_{x^{10} x^{10}}}'(r)}{3 \sqrt{2} \left(\frac{1}{N}\right)^{3/20} \pi ^{7/4} \left(6 a^2+{r_h}^2\right) \alpha _{\theta
   _2}^3}+\frac{{g_s}^{5/4} M {N_f} {r_h} \left(9 a^2+{r_h}^2\right) \log ({r_h}) {f_{\theta_1z}}'(r)}{2 \sqrt{2}
   \left(\frac{1}{N}\right)^{3/20} \pi ^{7/4} \left(6 a^2+{r_h}^2\right) \alpha _{\theta _2}^3}\nonumber\\
   & & +\frac{a^2 {g_s}^{5/4} M {N_f}
   (r-{r_h}) \left(9 a^2+{r_h}^2\right) \log ({r_h}) {f_{\theta_1x}}'(r)}{\sqrt{2} \left(\frac{1}{N}\right)^{3/20} \pi ^{7/4}
   \left(3 a^2-{r_h}^2\right) \left(6 a^2+{r_h}^2\right) \alpha _{\theta _2}^3}-\frac{27 \sqrt{3} {g_s}^{11/4} M^2
   \sqrt[4]{\frac{1}{N}} {N_f}^2 {r_h} \left(9 a^2+{r_h}^2\right) \log ^2({r_h}) \alpha _{\theta _1}^4 {f_{\theta_1y}}'(r)}{16
   \pi ^{13/4} \left(6 a^2+{r_h}^2\right) \alpha _{\theta _2}^6}\nonumber\\
& & +\frac{\sqrt{2} a^2 {g_s}^{5/4} M {N_f} (r-{r_h}) \left(9
   a^2+{r_h}^2\right) \log ({r_h}) {f_{\theta_2z}}'(r)}{\left(\frac{1}{N}\right)^{3/20} \pi ^{7/4} \left(-18 a^4+3 {r_h}^2
   a^2+{r_h}^4\right) \alpha _{\theta _2}^3}-\frac{{g_s}^{5/4} M {N_f} {r_h} \left(9 a^2+{r_h}^2\right) \log
   ({r_h}) {f_{\theta_2x}}'(r)}{3 \sqrt{2} \left(\frac{1}{N}\right)^{3/20} \pi ^{7/4} \left(6 a^2+{r_h}^2\right) \alpha _{\theta
   _2}^3}\nonumber\\
   & & +\frac{{g_s}^{5/4} M {N_f} {r_h} \left(9 a^2+{r_h}^2\right) \log ({r_h}) {f_{\theta_2y}}'(r)}{2 \sqrt{2}
   \left(\frac{1}{N}\right)^{3/20} \pi ^{7/4} \left(6 a^2+{r_h}^2\right) \alpha _{\theta _2}^3}+\frac{{g_s}^{5/4} M {N_f}
   {r_h} \left(9 a^2+{r_h}^2\right) \log ({r_h}) {f_{xz}}'(r)}{2 \sqrt{2} \left(\frac{1}{N}\right)^{3/20} \pi ^{7/4}
   \left(6 a^2+{r_h}^2\right) \alpha _{\theta _2}^3}\nonumber\\
   & & -\frac{81 {g_s}^{5/4} M \left(\frac{1}{N}\right)^{5/4} {N_f} {r_h}
   \left(9 a^2+{r_h}^2\right) \log ({r_h}) \alpha _{\theta _1}^4 {f_{xx}}'(r)}{8 \sqrt{2} \pi ^{7/4} \left(6
   a^2+{r_h}^2\right) \alpha _{\theta _2}}-\frac{27 \sqrt{3} a^2 {g_s}^{11/4} M^2 \sqrt[4]{\frac{1}{N}} {N_f}^2 (r-{r_h})
   \left(9 a^2+{r_h}^2\right) \log ^2({r_h}) \alpha _{\theta _1}^4 {f_{xy}}'(r)}{8 \pi ^{13/4} \left(-18 a^4+3 {r_h}^2
   a^2+{r_h}^4\right) \alpha _{\theta _2}^6}\nonumber\\
   & &-\frac{27 \sqrt{3} a^2 {g_s}^{11/4} M^2 \sqrt[4]{\frac{1}{N}} {N_f}^2
   (r-{r_h}) \left(9 a^2+{r_h}^2\right) \log ^2({r_h}) \alpha _{\theta _1}^4 {f_{yz}}'(r)}{8 \pi ^{13/4} \left(-18 a^4+3
   {r_h}^2 a^2+{r_h}^4\right) \alpha _{\theta _2}^6}\nonumber\\
   & & -\frac{9 a^2 {g_s}^{17/4} M^3 \left(\frac{1}{N}\right)^{5/4} {N_f}^3
   (r-{r_h}) \left(9 a^2+{r_h}^2\right) \log ^3({r_h}) \alpha _{\theta _1}^2 \left(2187 \sqrt{2} \alpha _{\theta _1}^6+540
   \sqrt{3} \alpha _{\theta _2}^2 \alpha _{\theta _1}^3+50 \sqrt{2} \alpha _{\theta _2}^4\right) {f_{yy}}'(r)}{512 \pi ^{19/4} \left(-18
   a^4+3 {r_h}^2 a^2+{r_h}^4\right) \alpha _{\theta _2}^7}\nonumber\\
   & & -\frac{{g_s}^{5/4} M {N_f} {r_h} \left(9
   a^2+{r_h}^2\right) \log ({r_h}) {f_r}'(r)}{6 \sqrt{2} \left(\frac{1}{N}\right)^{3/20} \pi ^{7/4} \left(6
   a^2+{r_h}^2\right) \alpha _{\theta _2}^3}+\frac{{g_s}^{5/4} M {N_f} {r_h} \left(9 a^2+{r_h}^2\right) \log
   ({r_h}) {f_t}'(r)}{2 \sqrt{2} \left(\frac{1}{N}\right)^{3/20} \pi ^{7/4} \left(6 a^2+{r_h}^2\right) \alpha _{\theta
   _2}^3}\nonumber\\
   & & +\frac{{g_s}^{5/4} M {N_f} (r-{r_h}) {r_h} \left(9 a^2+{r_h}^2\right) \log ({r_h}) f''(r)}{\sqrt{2}
   \left(\frac{1}{N}\right)^{3/20} \pi ^{7/4} \left(6 a^2+{r_h}^2\right) \alpha _{\theta _2}^3}-\frac{{g_s}^{5/4} M {N_f}
   (r-{r_h}) {r_h} \left(9 a^2+{r_h}^2\right) \log ({r_h}) {f_{zz}}''(r)}{3 \sqrt{2} \left(\frac{1}{N}\right)^{3/20}
   \pi ^{7/4} \left(6 a^2+{r_h}^2\right) \alpha _{\theta _2}^3}
  \nonumber\\
  & & +\frac{{g_s}^{5/4} M {N_f} (r-{r_h}) {r_h} \left(9
   a^2+{r_h}^2\right) \log ({r_h}) {f_{x^{10} x^{10}}}''(r)}{3 \sqrt{2} \left(\frac{1}{N}\right)^{3/20} \pi ^{7/4} \left(6
   a^2+{r_h}^2\right) \alpha _{\theta _2}^3}+\frac{{g_s}^{5/4} M {N_f} (r-{r_h}) {r_h} \left(9 a^2+{r_h}^2\right)
   \log ({r_h}) {f_{\theta_1z}}''(r)}{2 \sqrt{2} \left(\frac{1}{N}\right)^{3/20} \pi ^{7/4} \left(6 a^2+{r_h}^2\right) \alpha _{\theta
   _2}^3}\nonumber\\
   & &+\frac{7 {g_s}^{11/4} M^2 \left(\frac{1}{N}\right)^{13/20} {N_f}^2 (r-{r_h}) {r_h} \left(9 a^2+{r_h}^2\right)
   \log ^2({r_h}) \left(81 \sqrt{2} \alpha _{\theta _1}^3+10 \sqrt{3} \alpha _{\theta _2}^2\right) {f_{\theta_1x}}''(r)}{96 \pi ^{13/4}
   \left(6 a^2+{r_h}^2\right) \alpha _{\theta _2}^4}\nonumber\\
   & & -\frac{27 \sqrt{3} {g_s}^{11/4} M^2 \sqrt[4]{\frac{1}{N}} {N_f}^2
   (r-{r_h}) {r_h} \left(9 a^2+{r_h}^2\right) \log ^2({r_h}) \alpha _{\theta _1}^4 {f_{\theta_1y}}''(r)}{16 \pi ^{13/4} \left(6
   a^2+{r_h}^2\right) \alpha _{\theta _2}^6}\nonumber\\
   & &    -\frac{27 \sqrt{3} {g_s}^{11/4} M^2 \sqrt[4]{\frac{1}{N}} {N_f}^2 (r-{r_h})
   {r_h} \left(9 a^2+{r_h}^2\right) \log ^2({r_h}) \alpha _{\theta _1}^4 {f_{\theta_2z}}''(r)}{16 \pi ^{13/4} \left(6
   a^2+{r_h}^2\right) \alpha _{\theta _2}^6}-\frac{{g_s}^{5/4} M {N_f} (r-{r_h}) {r_h} \left(9 a^2+{r_h}^2\right)
   \log ({r_h}) {f_{\theta_2x}}''(r)}{3 \sqrt{2} \left(\frac{1}{N}\right)^{3/20} \pi ^{7/4} \left(6 a^2+{r_h}^2\right) \alpha _{\theta
   _2}^3}\nonumber\\
   & &+\frac{{g_s}^{5/4} M {N_f} (r-{r_h}) {r_h} \left(9 a^2+{r_h}^2\right) \log ({r_h}) {f_{\theta_2y}}''(r)}{2
   \sqrt{2} \left(\frac{1}{N}\right)^{3/20} \pi ^{7/4} \left(6 a^2+{r_h}^2\right) \alpha _{\theta _2}^3}+\frac{{g_s}^{5/4} M
   {N_f} (r-{r_h}) {r_h} \left(9 a^2+{r_h}^2\right) \log ({r_h}) {f_{xz}}''(r)}{2 \sqrt{2}
   \left(\frac{1}{N}\right)^{3/20} \pi ^{7/4} \left(6 a^2+{r_h}^2\right) \alpha _{\theta _2}^3}\nonumber\\
    \end{eqnarray*}}
   {\scriptsize
\begin{eqnarray*}
   & & -\frac{81 {g_s}^{5/4} M
   \left(\frac{1}{N}\right)^{5/4} {N_f} (r-{r_h}) {r_h} \left(9 a^2+{r_h}^2\right) \log ({r_h}) \alpha _{\theta _1}^4
   {f_{xx}}''(r)}{8 \sqrt{2} \pi ^{7/4} \left(6 a^2+{r_h}^2\right) \alpha _{\theta _2}}\nonumber\\
   & &-\frac{\sqrt{\frac{3}{2}} {g_s}^{11/4}
   M^2 \left(\frac{1}{N}\right)^{13/20} {N_f}^2 (r-{r_h}) {r_h} \left(9 a^2+{r_h}^2\right) \log ^2({r_h}) \left(27
   \sqrt{3} \alpha _{\theta _1}^3+5 \sqrt{2} \alpha _{\theta _2}^2\right) {f_{xy}}''(r)}{64 \pi ^{13/4} \left(6 a^2+{r_h}^2\right)
   \alpha _{\theta _2}^4}\nonumber\\
   & &+\frac{\sqrt{\frac{3}{2}} {g_s}^{11/4} M^2 \left(\frac{1}{N}\right)^{13/20} {N_f}^2 (r-{r_h})
   {r_h} \left(9 a^2+{r_h}^2\right) \log ^2({r_h}) \left(27 \sqrt{3} \alpha _{\theta _1}^3+5 \sqrt{2} \alpha _{\theta
   _2}^2\right) {f_{yz}}''(r)}{64 \pi ^{13/4} \left(6 a^2+{r_h}^2\right) \alpha _{\theta _2}^4}
   \nonumber\\
   & &  -\frac{{g_s}^{17/4} M^3
   \left(\frac{1}{N}\right)^{29/20} {N_f}^3 (r-{r_h}) {r_h} \left(9 a^2+{r_h}^2\right) \log ^3({r_h}) \left(2187
   \alpha _{\theta _1}^6+270 \sqrt{6} \alpha _{\theta _2}^2 \alpha _{\theta _1}^3+50 \alpha _{\theta _2}^4\right) {f_{yy}}''(r)}{1024
   \sqrt{2} \pi ^{19/4} \left(6 a^2+{r_h}^2\right) \alpha _{\theta _2}^5}\nonumber\\
& &  +\frac{{g_s}^{5/4} M {N_f} (r-{r_h}) {r_h}
   \left(9 a^2+{r_h}^2\right) \log ({r_h}) {f_t}''(r)}{3 \sqrt{2} \left(\frac{1}{N}\right)^{3/20} \pi ^{7/4} \left(6
   a^2+{r_h}^2\right) \alpha _{\theta _2}^3}-\frac{81 {g_s}^{5/4} M \left(\frac{1}{N}\right)^{5/4} {N_f} {f_{xx}}(r) \log
   ({r_h}) \alpha _{\theta _1}^4}{4 \sqrt{2} \pi ^{7/4} \log \left({r_h}^6+9 a^2 {r_h}^4\right) \alpha _{\theta _2}}\nonumber\\
   & &  +\frac{27
   {g_s}^{21/4} M^5 \left(\frac{1}{N}\right)^{37/20} {N_f}^3 {f_t}(r) \log ^5({r_h}) \left(16 \left(108 a^6-9 {r_h}^4
   a^2-{r_h}^6\right)-\left(2 {r_h}^6+27 a^2 {r_h}^4+117 a^4 {r_h}^2\right) \log \left({r_h}^6+9 a^2
   {r_h}^4\right)\right)}{128 \sqrt{2} \pi ^{23/4} \left(3 a^2-{r_h}^2\right) \left(6 a^2+{r_h}^2\right)^2 \log
   \left({r_h}^6+9 a^2 {r_h}^4\right) \alpha _{\theta _2}^3}\nonumber\\
& & -\frac{81 {g_s}^{21/4} \log N  M^5
   \left(\frac{1}{N}\right)^{37/20} {N_f}^3 \left(9 a^2+{r_h}^2\right) f(r) \log ^5({r_h})}{32 \sqrt{2} \pi ^{23/4} \left(6
   a^2+{r_h}^2\right) \log \left({r_h}^6+9 a^2 {r_h}^4\right) \alpha _{\theta _2}^3}\nonumber\\
   & &  -\frac{5 {g_s}^{5/4} M {N_f}
   (r-{r_h}) {r_h}^3 \left(9 a^2+{r_h}^2\right) \left(108 b^2 {r_h}^2+1\right)^2 {f_r}(r)}{54 \sqrt{2}
   \left(\frac{1}{N}\right)^{11/20} \pi ^{7/4} \left({r_h}^2-3 a^2\right)^2 \left(6 a^2+{r_h}^2\right) \log ({r_h}) \alpha
   _{\theta _1}^2 \alpha _{\theta _2}^3}+\frac{49 \pi ^{5/4} {r_h}^4 {f_{zz}}(r)}{11664 \sqrt{2} {g_s}^{7/4} M
   \left(\frac{1}{N}\right)^{23/20} {N_f} \left({r_h}^2-3 a^2\right)^2 \log ^3({r_h}) \alpha _{\theta _2}^3}\nonumber\\
      & & -\frac{49 \pi
   ^{5/4} {r_h}^4 {f_{x^{10} x^{10}}}(r)}{7776 \sqrt{2} {g_s}^{7/4} M \left(\frac{1}{N}\right)^{23/20} {N_f} \left({r_h}^2-3
   a^2\right)^2 \log ^3({r_h}) \alpha _{\theta _2}^3}-\frac{49 \pi ^{5/4} {r_h}^4 {f_{\theta_1z}}(r)}{5832 \sqrt{2} {g_s}^{7/4} M
   \left(\frac{1}{N}\right)^{23/20} {N_f} \left({r_h}^2-3 a^2\right)^2 \log ^3({r_h}) \alpha _{\theta _2}^3}\nonumber\\
   & & -\frac{245 \pi
   ^{5/4} {r_h}^4 {f_{\theta_2y}}(r)}{23328 \sqrt{2} {g_s}^{7/4} M \left(\frac{1}{N}\right)^{23/20} {N_f} \left({r_h}^2-3
   a^2\right)^2 \log ^3({r_h}) \alpha _{\theta _2}^3}-\frac{\sqrt{2} {g_s}^{5/4} M {N_f} (r-{r_h}) {r_h}^3 \left(9
   a^2+{r_h}^2\right) \left(108 b^2 {r_h}^2+1\right)^2 {f_{yz}}(r)}{9 \left(\frac{1}{N}\right)^{3/4} \pi ^{7/4}
   \left({r_h}^2-3 a^2\right)^2 \left(6 a^2+{r_h}^2\right) \log ({r_h}) \alpha _{\theta _2}^5}\nonumber\\
   & & +\frac{{g_s}^{5/4} M
   {N_f} (r-{r_h}) {r_h}^3 \left(9 a^2+{r_h}^2\right) \left(108 b^2 {r_h}^2+1\right)^2 {f_{yy}}(r)}{9 \sqrt{2}
   \left(\frac{1}{N}\right)^{3/4} \pi ^{7/4} \left({r_h}^2-3 a^2\right)^2 \left(6 a^2+{r_h}^2\right) \log ({r_h}) \alpha
   _{\theta _2}^5}+\frac{245 {r_h}^4 {f_{\theta_1y}}(r) \alpha _{\theta _1}^4}{2304 \sqrt{3} \sqrt[4]{{g_s}}
   \left(\frac{1}{N}\right)^{3/4} \sqrt[4]{\pi } \left({r_h}^2-3 a^2\right)^2 \log ^2({r_h}) \alpha _{\theta _2}^6}\nonumber\\
   & & +\frac{245
   {r_h}^4 {f_{\theta_2z}}(r) \alpha _{\theta _1}^4}{2304 \sqrt{3} \sqrt[4]{{g_s}} \left(\frac{1}{N}\right)^{3/4} \sqrt[4]{\pi }
   \left({r_h}^2-3 a^2\right)^2 \log ^2({r_h}) \alpha _{\theta _2}^6} = 0.
\end{eqnarray*}
}

(xii) \underline{${\rm EOM}_{\theta_2y}$}
{\scriptsize
\begin{eqnarray*}
& &  \frac{1280 \left(9 b^2+1\right)^4 \left(\frac{1}{N}\right)^{13/20} (r-{r_h})^2 \beta  \Sigma_1 b^{12}}{2187 \left(1-3 b^2\right)^6 \left(6
   b^2+1\right)^4 {g_s}^2 \log N ^2 {N_f}^2 {r_h}^2 \alpha _{\theta _1}^6 \alpha _{\theta _2}}-\frac{2 \sqrt{2} \left(9
   b^2+1\right)^4 M \left(\frac{1}{N}\right)^{7/5} (r-{r_h}) \beta  \log ({r_h}) \Sigma_1 b^{10}}{3 \left(3 b^2-1\right)^5 \left(6
   b^2+1\right)^4 \sqrt[4]{{g_s}} \log N ^2 {N_f} \pi ^{5/4} {r_h} \alpha _{\theta _1}^2 \alpha _{\theta _2}^2}
   \nonumber\\
   & & -\frac{27
   \left(\frac{1}{N}\right)^{31/20} {f_{xx}}(r) \alpha _{\theta _1}^2 \alpha _{\theta _2}^3}{2 \sqrt{2} \sqrt[4]{{g_s}} \sqrt[4]{\pi
   } \log \left({r_h}^6+9 a^2 {r_h}^4\right)}\nonumber\\
   & & +\frac{9 {g_s}^{15/4} M^4 \left(\frac{1}{N}\right)^{43/20} {N_f}^2
   {f_t}(r) \log ^4({r_h}) \left(16 \left(108 a^6-9 {r_h}^4 a^2-{r_h}^6\right)-\left(2 {r_h}^6+27 a^2 {r_h}^4+117
   a^4 {r_h}^2\right) \log \left({r_h}^6+9 a^2 {r_h}^4\right)\right) \alpha _{\theta _2}}{64 \sqrt{2} \pi ^{17/4} \left(3
   a^2-{r_h}^2\right) \left(6 a^2+{r_h}^2\right)^2 \log \left({r_h}^6+9 a^2 {r_h}^4\right) \alpha _{\theta _1}^2}
   \nonumber\\
    \end{eqnarray*}}
   {\scriptsize
   \begin{eqnarray*}
   & & -\frac{27
   {g_s}^{15/4} \log N  M^4 \left(\frac{1}{N}\right)^{43/20} {N_f}^2 \left(9 a^2+{r_h}^2\right) f(r) \log ^4({r_h})
   \alpha _{\theta _2}}{16 \sqrt{2} \pi ^{17/4} \left(6 a^2+{r_h}^2\right) \log \left({r_h}^6+9 a^2 {r_h}^4\right) \alpha
   _{\theta _1}^2}-\frac{343 \pi ^{11/4} {r_h}^4 {f_{zz}}(r) \alpha _{\theta _2}}{34992 \sqrt{2} {g_s}^{13/4} M^2
   \left(\frac{1}{N}\right)^{17/20} {N_f}^2 \left({r_h}^2-3 a^2\right)^2 \log ^4({r_h}) \alpha _{\theta _1}^2}\nonumber\\
   & &+\frac{245 \pi
   ^{11/4} {r_h}^4 {f_{x^{10} x^{10}}}(r) \alpha _{\theta _2}}{11664 \sqrt{2} {g_s}^{13/4} M^2 \left(\frac{1}{N}\right)^{17/20}
   {N_f}^2 \left({r_h}^2-3 a^2\right)^2 \log ^4({r_h}) \alpha _{\theta _1}^2}+\frac{343 \pi ^{11/4} {r_h}^4 {f_{\theta_1z}}(r)
   \alpha _{\theta _2}}{17496 \sqrt{2} {g_s}^{13/4} M^2 \left(\frac{1}{N}\right)^{17/20} {N_f}^2 \left({r_h}^2-3 a^2\right)^2
   \log ^4({r_h}) \alpha _{\theta _1}^2}\nonumber\\
   & & -\frac{245 \pi ^{11/4} {r_h}^4 {f_{\theta_2y}}(r) \alpha _{\theta _2}}{34992 \sqrt{2}
   {g_s}^{13/4} M^2 \left(\frac{1}{N}\right)^{17/20} {N_f}^2 \left({r_h}^2-3 a^2\right)^2 \log ^4({r_h}) \alpha _{\theta
   _1}^2}-\frac{5 (r-{r_h}) {r_h}^3 \left(9 a^2+{r_h}^2\right) \left(108 b^2 {r_h}^2+1\right)^2 {f_r}(r) \alpha
   _{\theta _2}}{81 \sqrt{2} \sqrt[4]{{g_s}} \sqrt[4]{\frac{1}{N}} \sqrt[4]{\pi } \left({r_h}^2-3 a^2\right)^2 \left(6
   a^2+{r_h}^2\right) \log ^2({r_h}) \alpha _{\theta _1}^4}\nonumber\\
   & & +\frac{245 \pi ^{5/4} {r_h}^4 {f_{xy}}(r) \left(27 \sqrt{6}
   \alpha _{\theta _1}^3+10 \alpha _{\theta _2}^2\right)}{746496 \sqrt{3} {g_s}^{7/4} M \sqrt[20]{\frac{1}{N}} {N_f}
   \left({r_h}^2-3 a^2\right)^2 \log ^3({r_h}) \alpha _{\theta _1}^2}-\frac{343 \pi ^{5/4} {r_h}^4 {f_{\theta_1x}}(r) \left(81
   \sqrt{2} \alpha _{\theta _1}^3+10 \sqrt{3} \alpha _{\theta _2}^2\right)}{559872 {g_s}^{7/4} M \sqrt[20]{\frac{1}{N}} {N_f}
   \left({r_h}^2-3 a^2\right)^2 \log ^3({r_h}) \alpha _{\theta _1}^2}\nonumber\\
   & & +\frac{1715 \pi ^{5/4} {r_h}^4 {f_{\theta_2x}}(r) \left(81
   \sqrt{2} \alpha _{\theta _1}^3+10 \sqrt{3} \alpha _{\theta _2}^2\right)}{1679616 {g_s}^{7/4} M \sqrt[20]{\frac{1}{N}} {N_f}
   \left({r_h}^2-3 a^2\right)^2 \log ^3({r_h}) \alpha _{\theta _1}^2}-\frac{3479 \pi ^{5/4} {r_h}^4 {f_{xz}}(r) \left(81
   \sqrt{2} \alpha _{\theta _1}^3+10 \sqrt{3} \alpha _{\theta _2}^2\right)}{6718464 {g_s}^{7/4} M \sqrt[20]{\frac{1}{N}} {N_f}
   \left({r_h}^2-3 a^2\right)^2 \log ^3({r_h}) \alpha _{\theta _1}^2}\nonumber\\
   & & +\frac{\sqrt{2} \left(\frac{1}{N}\right)^{3/20} {r_h}
   \left(9 a^2+{r_h}^2\right) \alpha _{\theta _2} f'(r)}{3 \sqrt[4]{{g_s}} \sqrt[4]{\pi } \left(6 a^2+{r_h}^2\right) \alpha
   _{\theta _1}^2}-\frac{\sqrt{2} \left(\frac{1}{N}\right)^{3/20} {r_h} \left(9 a^2+{r_h}^2\right) \alpha _{\theta _2}
   {f_{zz}}'(r)}{9 \sqrt[4]{{g_s}} \sqrt[4]{\pi } \left(6 a^2+{r_h}^2\right) \alpha _{\theta _1}^2}\nonumber\\
   & & +\frac{\sqrt{2}
   \left(\frac{1}{N}\right)^{3/20} {r_h} \left(9 a^2+{r_h}^2\right) \alpha _{\theta _2} {f_{x^{10} x^{10}}}'(r)}{9 \sqrt[4]{{g_s}}
   \sqrt[4]{\pi } \left(6 a^2+{r_h}^2\right) \alpha _{\theta _1}^2}+\frac{\left(\frac{1}{N}\right)^{3/20} {r_h} \left(9
   a^2+{r_h}^2\right) \alpha _{\theta _2} {f_{\theta_1z}}'(r)}{3 \sqrt{2} \sqrt[4]{{g_s}} \sqrt[4]{\pi } \left(6
   a^2+{r_h}^2\right) \alpha _{\theta _1}^2}\nonumber\\
   & & -\frac{\sqrt{2} a^2 \left(\frac{1}{N}\right)^{3/20} (r-{r_h}) \left(9
   a^2+{r_h}^2\right) \alpha _{\theta _2} {f_{\theta_1x}}'(r)}{3 \sqrt[4]{{g_s}} \sqrt[4]{\pi } \left(-18 a^4+3 {r_h}^2
   a^2+{r_h}^4\right) \alpha _{\theta _1}^2}-\frac{9 \sqrt{3} {g_s}^{5/4} M \left(\frac{1}{N}\right)^{11/20} {N_f} {r_h}
   \left(9 a^2+{r_h}^2\right) \log ({r_h}) \alpha _{\theta _1}^2 {f_{\theta_1y}}'(r)}{8 \pi ^{7/4} \left(6 a^2+{r_h}^2\right)
   \alpha _{\theta _2}^2}\nonumber\\
   & & -\frac{9 \sqrt{3} {g_s}^{5/4} M \left(\frac{1}{N}\right)^{11/20} {N_f} {r_h} \left(9
   a^2+{r_h}^2\right) \log ({r_h}) \alpha _{\theta _1}^2 {f_{\theta_2z}}'(r)}{8 \pi ^{7/4} \left(6 a^2+{r_h}^2\right) \alpha
   _{\theta _2}^2}+\frac{9 \sqrt{3} a^2 {g_s}^{5/4} M \left(\frac{1}{N}\right)^{3/4} {N_f} (r-{r_h}) \left(9
   a^2+{r_h}^2\right) \log ({r_h}) {f_{\theta_2x}}'(r)}{4 \pi ^{7/4} \left(-18 a^4+3 {r_h}^2
   a^2+{r_h}^4\right)}\nonumber\\
 & & +\frac{\left(\frac{1}{N}\right)^{3/20} {r_h} \left(9 a^2+{r_h}^2\right) \alpha _{\theta _2}
   {f_{\theta_2y}}'(r)}{9 \sqrt{2} \sqrt[4]{{g_s}} \sqrt[4]{\pi } \left(6 a^2+{r_h}^2\right) \alpha _{\theta
   _1}^2}+\frac{\left(\frac{1}{N}\right)^{3/20} {r_h} \left(9 a^2+{r_h}^2\right) \alpha _{\theta _2} {f_{xz}}'(r)}{3 \sqrt{2}
   \sqrt[4]{{g_s}} \sqrt[4]{\pi } \left(6 a^2+{r_h}^2\right) \alpha _{\theta _1}^2}\nonumber\\
   & & -\frac{27 \left(\frac{1}{N}\right)^{31/20}
   {r_h} \left(9 a^2+{r_h}^2\right) \alpha _{\theta _1}^2 \alpha _{\theta _2}^3 {f_{xx}}'(r)}{4 \sqrt{2} \sqrt[4]{{g_s}}
   \sqrt[4]{\pi } \left(6 a^2+{r_h}^2\right)}-\frac{9 \sqrt{3} a^2 {g_s}^{5/4} M \left(\frac{1}{N}\right)^{11/20} {N_f}
   (r-{r_h}) \left(9 a^2+{r_h}^2\right) \log ({r_h}) \alpha _{\theta _1}^2 {f_{xy}}'(r)}{4 \pi ^{7/4} \left(-18 a^4+3
   {r_h}^2 a^2+{r_h}^4\right) \alpha _{\theta _2}^2}\nonumber\\
     & & -\frac{9 \sqrt{3} a^2 {g_s}^{5/4} M \left(\frac{1}{N}\right)^{11/20}
   {N_f} (r-{r_h}) \left(9 a^2+{r_h}^2\right) \log ({r_h}) \alpha _{\theta _1}^2 {f_{yz}}'(r)}{4 \pi ^{7/4} \left(-18
   a^4+3 {r_h}^2 a^2+{r_h}^4\right) \alpha _{\theta _2}^2}\nonumber\\
   & & -\frac{3 a^2 {g_s}^{11/4} M^2 \left(\frac{1}{N}\right)^{31/20}
   {N_f}^2 (r-{r_h}) \left(9 a^2+{r_h}^2\right) \log ^2({r_h}) \left(2187 \sqrt{2} \alpha _{\theta _1}^6+540 \sqrt{3}
   \alpha _{\theta _2}^2 \alpha _{\theta _1}^3+50 \sqrt{2} \alpha _{\theta _2}^4\right) {f_{yy}}'(r)}{256 \pi ^{13/4} \left(-18 a^4+3
   {r_h}^2 a^2+{r_h}^4\right) \alpha _{\theta _2}^3}  \nonumber\\
    & & -\frac{\left(\frac{1}{N}\right)^{3/20} {r_h} \left(9
   a^2+{r_h}^2\right) \alpha _{\theta _2} {f_r}'(r)}{9 \sqrt{2} \sqrt[4]{{g_s}} \sqrt[4]{\pi } \left(6 a^2+{r_h}^2\right)
   \alpha _{\theta _1}^2}+\frac{\left(\frac{1}{N}\right)^{3/20} {r_h} \left(9 a^2+{r_h}^2\right) \alpha _{\theta _2}
   {f_t}'(r)}{3 \sqrt{2} \sqrt[4]{{g_s}} \sqrt[4]{\pi } \left(6 a^2+{r_h}^2\right) \alpha _{\theta _1}^2}\nonumber\\
   & & +\frac{\sqrt{2}
   \left(\frac{1}{N}\right)^{3/20} (r-{r_h}) {r_h} \left(9 a^2+{r_h}^2\right) \alpha _{\theta _2} f''(r)}{3
   \sqrt[4]{{g_s}} \sqrt[4]{\pi } \left(6 a^2+{r_h}^2\right) \alpha _{\theta _1}^2}-\frac{\sqrt{2} \left(\frac{1}{N}\right)^{3/20}
   (r-{r_h}) {r_h} \left(9 a^2+{r_h}^2\right) \alpha _{\theta _2} {f_{zz}}''(r)}{9 \sqrt[4]{{g_s}} \sqrt[4]{\pi }
   \left(6 a^2+{r_h}^2\right) \alpha _{\theta _1}^2}\nonumber\\
   & & +\frac{\sqrt{2} \left(\frac{1}{N}\right)^{3/20} (r-{r_h}) {r_h} \left(9
   a^2+{r_h}^2\right) \alpha _{\theta _2} {f_{x^{10} x^{10}}}''(r)}{9 \sqrt[4]{{g_s}} \sqrt[4]{\pi } \left(6 a^2+{r_h}^2\right)
   \alpha _{\theta _1}^2}+\frac{\left(\frac{1}{N}\right)^{3/20} (r-{r_h}) {r_h} \left(9 a^2+{r_h}^2\right) \alpha _{\theta _2}
   {f_{\theta_1z}}''(r)}{3 \sqrt{2} \sqrt[4]{{g_s}} \sqrt[4]{\pi } \left(6 a^2+{r_h}^2\right) \alpha _{\theta _1}^2}\nonumber\\
   & & +\frac{7
   {g_s}^{5/4} M \left(\frac{1}{N}\right)^{19/20} {N_f} (r-{r_h}) {r_h} \left(9 a^2+{r_h}^2\right) \log ({r_h})
   \left(81 \sqrt{2} \alpha _{\theta _1}^3+10 \sqrt{3} \alpha _{\theta _2}^2\right) {f_{\theta_1x}}''(r)}{144 \pi ^{7/4} \left(6
   a^2+{r_h}^2\right) \alpha _{\theta _1}^2}\nonumber\\
   & & -\frac{9 \sqrt{3} {g_s}^{5/4} M \left(\frac{1}{N}\right)^{11/20} {N_f}
   (r-{r_h}) {r_h} \left(9 a^2+{r_h}^2\right) \log ({r_h}) \alpha _{\theta _1}^2 {f_{\theta_1y}}''(r)}{8 \pi ^{7/4} \left(6
   a^2+{r_h}^2\right) \alpha _{\theta _2}^2}\nonumber\\
   & & -\frac{9 \sqrt{3} {g_s}^{5/4} M \left(\frac{1}{N}\right)^{11/20} {N_f}
   (r-{r_h}) {r_h} \left(9 a^2+{r_h}^2\right) \log ({r_h}) \alpha _{\theta _1}^2 {f_{\theta_2z}}''(r)}{8 \pi ^{7/4} \left(6
   a^2+{r_h}^2\right) \alpha _{\theta _2}^2}\nonumber\\
   \end{eqnarray*}}
{\scriptsize
\begin{eqnarray*}
   & & -\frac{7 {g_s}^{5/4} M \left(\frac{1}{N}\right)^{19/20} {N_f} (r-{r_h})
   {r_h} \left(9 a^2+{r_h}^2\right) \log ({r_h}) \left(81 \sqrt{2} \alpha _{\theta _1}^3+10 \sqrt{3} \alpha _{\theta
   _2}^2\right) {f_{\theta_2x}}''(r)}{144 \pi ^{7/4} \left(6 a^2+{r_h}^2\right) \alpha _{\theta
   _1}^2}\nonumber\\
   & & +\frac{\left(\frac{1}{N}\right)^{3/20} (r-{r_h}) {r_h} \left(9 a^2+{r_h}^2\right) \alpha _{\theta _2}
   {f_{\theta_2y}}''(r)}{9 \sqrt{2} \sqrt[4]{{g_s}} \sqrt[4]{\pi } \left(6 a^2+{r_h}^2\right) \alpha _{\theta
   _1}^2}+\frac{\left(\frac{1}{N}\right)^{3/20} (r-{r_h}) {r_h} \left(9 a^2+{r_h}^2\right) \alpha _{\theta _2}
   {f_{xz}}''(r)}{3 \sqrt{2} \sqrt[4]{{g_s}} \sqrt[4]{\pi } \left(6 a^2+{r_h}^2\right) \alpha _{\theta _1}^2}\nonumber\\
 & & -\frac{27
   \left(\frac{1}{N}\right)^{31/20} (r-{r_h}) {r_h} \left(9 a^2+{r_h}^2\right) \alpha _{\theta _1}^2 \alpha _{\theta _2}^3
   {f_{xx}}''(r)}{4 \sqrt{2} \sqrt[4]{{g_s}} \sqrt[4]{\pi } \left(6 a^2+{r_h}^2\right)}-\frac{{g_s}^{5/4} M
   \left(\frac{1}{N}\right)^{19/20} {N_f} (r-{r_h}) {r_h} \left(9 a^2+{r_h}^2\right) \log ({r_h}) \left(27 \sqrt{3}
   \alpha _{\theta _1}^3+5 \sqrt{2} \alpha _{\theta _2}^2\right) {f_{xy}}''(r)}{32 \sqrt{6} \pi ^{7/4} \left(6 a^2+{r_h}^2\right)
   \alpha _{\theta _1}^2}\nonumber\\
      & & +\frac{{g_s}^{5/4} M \left(\frac{1}{N}\right)^{19/20} {N_f} (r-{r_h}) {r_h} \left(9
   a^2+{r_h}^2\right) \log ({r_h}) \left(27 \sqrt{3} \alpha _{\theta _1}^3+5 \sqrt{2} \alpha _{\theta _2}^2\right)
   {f_{yz}}''(r)}{32 \sqrt{6} \pi ^{7/4} \left(6 a^2+{r_h}^2\right) \alpha _{\theta _1}^2}\nonumber\\
   & & -\frac{{g_s}^{11/4} M^2
   \left(\frac{1}{N}\right)^{7/4} {N_f}^2 (r-{r_h}) {r_h} \left(9 a^2+{r_h}^2\right) \log ^2({r_h}) \left(2187 \alpha
   _{\theta _1}^6+270 \sqrt{6} \alpha _{\theta _2}^2 \alpha _{\theta _1}^3+50 \alpha _{\theta _2}^4\right) {f_{yy}}''(r)}{1536 \sqrt{2}
   \pi ^{13/4} \left(6 a^2+{r_h}^2\right) \alpha _{\theta _1}^2 \alpha _{\theta _2}}\nonumber\\
   & & +\frac{\sqrt{2} \left(\frac{1}{N}\right)^{3/20}
   (r-{r_h}) {r_h} \left(9 a^2+{r_h}^2\right) \alpha _{\theta _2} {f_t}''(r)}{9 \sqrt[4]{{g_s}} \sqrt[4]{\pi } \left(6
   a^2+{r_h}^2\right) \alpha _{\theta _1}^2}+\frac{\sqrt{2} (r-{r_h}) {r_h}^3 \left(9 a^2+{r_h}^2\right) \left(108 b^2
   {r_h}^2+1\right)^2 {f_{yy}}(r)}{27 \sqrt[4]{{g_s}} \left(\frac{1}{N}\right)^{9/20} \sqrt[4]{\pi } \left({r_h}^2-3
   a^2\right)^2 \left(6 a^2+{r_h}^2\right) \log ^2({r_h}) \alpha _{\theta _1}^2 \alpha _{\theta _2}}\nonumber\\
   & & -\frac{49 \pi ^{5/4}
   {r_h}^4 {f_{\theta_1y}}(r) \alpha _{\theta _1}^2}{864 \sqrt{3} {g_s}^{7/4} M \left(\frac{1}{N}\right)^{9/20} {N_f}
   \left({r_h}^2-3 a^2\right)^2 \log ^3({r_h}) \alpha _{\theta _2}^2}+\frac{245 \pi ^{5/4} {r_h}^4 {f_{\theta_2z}}(r) \alpha
   _{\theta _1}^2}{3456 \sqrt{3} {g_s}^{7/4} M \left(\frac{1}{N}\right)^{9/20} {N_f} \left({r_h}^2-3 a^2\right)^2 \log
   ^3({r_h}) \alpha _{\theta _2}^2}\nonumber\\
   & & +\frac{49 \pi ^{5/4} {r_h}^4 {f_{yz}}(r) \alpha _{\theta _1}^2}{384 \sqrt{3} {g_s}^{7/4}
   M \left(\frac{1}{N}\right)^{9/20} {N_f} \left({r_h}^2-3 a^2\right)^2 \log ^3({r_h}) \alpha _{\theta _2}^2} = 0
\end{eqnarray*}
}

(xiii) \underline{${\rm EOM}_{\theta_2z}$}
{\scriptsize
\begin{eqnarray*}
& &  \frac{256 \left(9 b^2+1\right)^4 M \left(\frac{1}{N}\right)^{3/4} (r-{r_h})^3 \beta  \log ({r_h}) \Sigma_1 b^{14}}{81 \left(3
   b^2-1\right)^7 \left(6 b^2+1\right)^4 \sqrt{{g_s}} \log N ^2 {N_f} \pi ^{3/2} {r_h}^3 \alpha _{\theta _1}^2 \alpha
   _{\theta _2}^5}\nonumber\\
   &&+\frac{3 \left(9 b^2+1\right)^4 {g_s}^{5/4} M^2 \left(\frac{1}{N}\right)^{3/2} (r-{r_h}) \beta  \log
   ^2({r_h}) \alpha _{\theta _1}^2 \left(19683 \sqrt{6} \alpha _{\theta _1}^6+6642 \alpha _{\theta _2}^2 \alpha _{\theta _1}^3-40
   \sqrt{6} \alpha _{\theta _2}^4\right) b^{10}}{\sqrt{2} \left(3 b^2-1\right)^5 \left(6 b^2+1\right)^4 \log N ^2 \pi ^{11/4}
   {r_h} \alpha _{\theta _2}^6}\nonumber\\
   & & -\frac{245 \sqrt[20]{\frac{1}{N}} {r_h}^4 {f_{xy}}(r) \alpha _{\theta _1}^2 \left(27 \sqrt{6}
   \alpha _{\theta _1}^3+10 \alpha _{\theta _2}^2\right)}{331776 \sqrt{3} \sqrt[4]{{g_s}} \sqrt[4]{\pi } \left({r_h}^2-3
   a^2\right)^2 \log ^2({r_h}) \alpha _{\theta _2}^4}+\frac{343 \sqrt[20]{\frac{1}{N}} {r_h}^4 {f_{\theta_1x}}(r) \alpha _{\theta _1}^2
   \left(81 \sqrt{2} \alpha _{\theta _1}^3+10 \sqrt{3} \alpha _{\theta _2}^2\right)}{248832 \sqrt[4]{{g_s}} \sqrt[4]{\pi }
   \left({r_h}^2-3 a^2\right)^2 \log ^2({r_h}) \alpha _{\theta _2}^4}\nonumber\\
   & & -\frac{1715 \sqrt[20]{\frac{1}{N}} {r_h}^4 {f_{\theta_2x}}(r)
   \alpha _{\theta _1}^2 \left(81 \sqrt{2} \alpha _{\theta _1}^3+10 \sqrt{3} \alpha _{\theta _2}^2\right)}{746496 \sqrt[4]{{g_s}}
   \sqrt[4]{\pi } \left({r_h}^2-3 a^2\right)^2 \log ^2({r_h}) \alpha _{\theta _2}^4} +\frac{3479 \sqrt[20]{\frac{1}{N}} {r_h}^4
   {f_{xz}}(r) \alpha _{\theta _1}^2 \left(81 \sqrt{2} \alpha _{\theta _1}^3+10 \sqrt{3} \alpha _{\theta _2}^2\right)}{2985984
   \sqrt[4]{{g_s}} \sqrt[4]{\pi } \left({r_h}^2-3 a^2\right)^2 \log ^2({r_h}) \alpha _{\theta _2}^4}\nonumber\\
   & & -\frac{3 {g_s}^{5/4} M
   \sqrt[4]{\frac{1}{N}} {N_f} {r_h} \left(9 a^2+{r_h}^2\right) \log ({r_h}) \alpha _{\theta _1}^2 f'(r)}{2 \sqrt{2} \pi
   ^{7/4} \left(6 a^2+{r_h}^2\right) \alpha _{\theta _2}^3}+\frac{{g_s}^{5/4} M \sqrt[4]{\frac{1}{N}} {N_f} {r_h} \left(9
   a^2+{r_h}^2\right) \log ({r_h}) \alpha _{\theta _1}^2 {f_{zz}}'(r)}{2 \sqrt{2} \pi ^{7/4} \left(6 a^2+{r_h}^2\right)
   \alpha _{\theta _2}^3}\nonumber\\
   & & -\frac{{g_s}^{5/4} M \sqrt[4]{\frac{1}{N}} {N_f} {r_h} \left(9 a^2+{r_h}^2\right) \log
   ({r_h}) \alpha _{\theta _1}^2 {f_{x^{10} x^{10}}}'(r)}{2 \sqrt{2} \pi ^{7/4} \left(6 a^2+{r_h}^2\right) \alpha _{\theta _2}^3}\nonumber\\
   & & -\frac{3
   {g_s}^{5/4} M \sqrt[4]{\frac{1}{N}} {N_f} {r_h} \left(9 a^2+{r_h}^2\right) \log ({r_h}) \alpha _{\theta _1}^2
   {f_{\theta_1z}}'(r)}{4 \sqrt{2} \pi ^{7/4} \left(6 a^2+{r_h}^2\right) \alpha _{\theta _2}^3}+\frac{3 a^2 {g_s}^{5/4} M
   \sqrt[4]{\frac{1}{N}} {N_f} (r-{r_h}) \left(9 a^2+{r_h}^2\right) \log ({r_h}) \alpha _{\theta _1}^2 {f_{\theta_1x}}'(r)}{2
   \sqrt{2} \pi ^{7/4} \left(-18 a^4+3 {r_h}^2 a^2+{r_h}^4\right) \alpha _{\theta _2}^3}\nonumber\\
    \end{eqnarray*}}
{\scriptsize
\begin{eqnarray*}
   & & +\frac{81 \sqrt{3} {g_s}^{11/4} M^2
   \left(\frac{1}{N}\right)^{13/20} {N_f}^2 {r_h} \left(9 a^2+{r_h}^2\right) \log ^2({r_h}) \alpha _{\theta _1}^6
   {f_{\theta_1y}}'(r)}{32 \pi ^{13/4} \left(6 a^2+{r_h}^2\right) \alpha _{\theta _2}^6}+\frac{{g_s}^{5/4} M \sqrt[4]{\frac{1}{N}}
   {N_f} {r_h} \left(9 a^2+{r_h}^2\right) \log ({r_h}) \alpha _{\theta _1}^2 {f_{\theta_2z}}'(r)}{2 \sqrt{2} \pi ^{7/4}
   \left(6 a^2+{r_h}^2\right) \alpha _{\theta _2}^3}\nonumber\\
      & & -\frac{\sqrt{2} {g_s}^{5/4} M \left(\frac{1}{N}\right)^{13/20} {N_f}
   (r-{r_h}) \log ({r_h}) \left(19683 \alpha _{\theta _1}^6+1107 \sqrt{6} \alpha _{\theta _2}^2 \alpha _{\theta _1}^3-40 \alpha
   _{\theta _2}^4\right) {f_{\theta_2x}}'(r)}{729 \pi ^{7/4} \log \left({r_h}^6+9 a^2 {r_h}^4\right) \alpha _{\theta _1}^2 \alpha
   _{\theta _2}^3}\nonumber\\
   & & -\frac{3 {g_s}^{5/4} M \sqrt[4]{\frac{1}{N}} {N_f} {r_h} \left(9 a^2+{r_h}^2\right) \log ({r_h})
   \alpha _{\theta _1}^2 {f_{\theta_2y}}'(r)}{4 \sqrt{2} \pi ^{7/4} \left(6 a^2+{r_h}^2\right) \alpha _{\theta _2}^3}\nonumber\\
   & & -\frac{3
   {g_s}^{5/4} M \sqrt[4]{\frac{1}{N}} {N_f} {r_h} \left(9 a^2+{r_h}^2\right) \log ({r_h}) \alpha _{\theta _1}^2
   {f_{xz}}'(r)}{4 \sqrt{2} \pi ^{7/4} \left(6 a^2+{r_h}^2\right) \alpha _{\theta _2}^3}
   \nonumber\\
   & & +\frac{243 {g_s}^{5/4} M
   \left(\frac{1}{N}\right)^{33/20} {N_f} {r_h} \left(9 a^2+{r_h}^2\right) \log ({r_h}) \alpha _{\theta _1}^6
   {f_{xx}}'(r)}{16 \sqrt{2} \pi ^{7/4} \left(6 a^2+{r_h}^2\right) \alpha _{\theta _2}}\nonumber\\
   & & +\frac{81 \sqrt{3} a^2 {g_s}^{11/4} M^2
   \left(\frac{1}{N}\right)^{13/20} {N_f}^2 (r-{r_h}) \left(9 a^2+{r_h}^2\right) \log ^2({r_h}) \alpha _{\theta _1}^6
   {f_{xy}}'(r)}{16 \pi ^{13/4} \left(-18 a^4+3 {r_h}^2 a^2+{r_h}^4\right) \alpha _{\theta _2}^6}
   \nonumber\\
   & & +\frac{81 \sqrt{3} a^2
   {g_s}^{11/4} M^2 \left(\frac{1}{N}\right)^{13/20} {N_f}^2 (r-{r_h}) \left(9 a^2+{r_h}^2\right) \log ^2({r_h})
   \alpha _{\theta _1}^6 {f_{yz}}'(r)}{16 \pi ^{13/4} \left(-18 a^4+3 {r_h}^2 a^2+{r_h}^4\right) \alpha _{\theta
   _2}^6}\nonumber\\
   & & +\frac{27 a^2 {g_s}^{17/4} M^3 \left(\frac{1}{N}\right)^{33/20} {N_f}^3 (r-{r_h}) \left(9 a^2+{r_h}^2\right) \log
   ^3({r_h}) \alpha _{\theta _1}^4 \left(2187 \sqrt{2} \alpha _{\theta _1}^6+540 \sqrt{3} \alpha _{\theta _2}^2 \alpha _{\theta
   _1}^3+50 \sqrt{2} \alpha _{\theta _2}^4\right) {f_{yy}}'(r)}{1024 \pi ^{19/4} \left(-18 a^4+3 {r_h}^2 a^2+{r_h}^4\right)
   \alpha _{\theta _2}^7}\nonumber\\
   & & +\frac{{g_s}^{5/4} M \sqrt[4]{\frac{1}{N}} {N_f} {r_h} \left(9 a^2+{r_h}^2\right) \log
   ({r_h}) \alpha _{\theta _1}^2 {f_r}'(r)}{4 \sqrt{2} \pi ^{7/4} \left(6 a^2+{r_h}^2\right) \alpha _{\theta _2}^3}\nonumber\\
   & & -\frac{3
   {g_s}^{5/4} M \sqrt[4]{\frac{1}{N}} {N_f} {r_h} \left(9 a^2+{r_h}^2\right) \log ({r_h}) \alpha _{\theta _1}^2
   {f_t}'(r)}{4 \sqrt{2} \pi ^{7/4} \left(6 a^2+{r_h}^2\right) \alpha _{\theta _2}^3}-\frac{3 {g_s}^{5/4} M
   \sqrt[4]{\frac{1}{N}} {N_f} (r-{r_h}) {r_h} \left(9 a^2+{r_h}^2\right) \log ({r_h}) \alpha _{\theta _1}^2 f''(r)}{2
   \sqrt{2} \pi ^{7/4} \left(6 a^2+{r_h}^2\right) \alpha _{\theta _2}^3}\nonumber\\
& & +\frac{{g_s}^{5/4} M \sqrt[4]{\frac{1}{N}} {N_f}
   (r-{r_h}) {r_h} \left(9 a^2+{r_h}^2\right) \log ({r_h}) \alpha _{\theta _1}^2 {f_{zz}}''(r)}{2 \sqrt{2} \pi ^{7/4}
   \left(6 a^2+{r_h}^2\right) \alpha _{\theta _2}^3}-\frac{{g_s}^{5/4} M \sqrt[4]{\frac{1}{N}} {N_f} (r-{r_h}) {r_h}
   \left(9 a^2+{r_h}^2\right) \log ({r_h}) \alpha _{\theta _1}^2 {f_{x^{10} x^{10}}}''(r)}{2 \sqrt{2} \pi ^{7/4} \left(6
   a^2+{r_h}^2\right) \alpha _{\theta _2}^3}\nonumber\\
   & & -\frac{3 {g_s}^{5/4} M \sqrt[4]{\frac{1}{N}} {N_f} (r-{r_h}) {r_h} \left(9
   a^2+{r_h}^2\right) \log ({r_h}) \alpha _{\theta _1}^2 {f_{\theta_1z}}''(r)}{4 \sqrt{2} \pi ^{7/4} \left(6 a^2+{r_h}^2\right)
   \alpha _{\theta _2}^3}\nonumber\\
      & &  -\frac{7 {g_s}^{11/4} M^2 \left(\frac{1}{N}\right)^{21/20} {N_f}^2 (r-{r_h}) {r_h} \left(9
   a^2+{r_h}^2\right) \log ^2({r_h}) \alpha _{\theta _1}^2 \left(81 \sqrt{2} \alpha _{\theta _1}^3+10 \sqrt{3} \alpha _{\theta
   _2}^2\right) {f_{\theta_1x}}''(r)}{64 \pi ^{13/4} \left(6 a^2+{r_h}^2\right) \alpha _{\theta _2}^4}\nonumber\\
   & & +\frac{81 \sqrt{3} {g_s}^{11/4}
   M^2 \left(\frac{1}{N}\right)^{13/20} {N_f}^2 (r-{r_h}) {r_h} \left(9 a^2+{r_h}^2\right) \log ^2({r_h}) \alpha
   _{\theta _1}^6 {f_{\theta_1y}}''(r)}{32 \pi ^{13/4} \left(6 a^2+{r_h}^2\right) \alpha _{\theta _2}^6}+\frac{{g_s}^{5/4} M
   \sqrt[4]{\frac{1}{N}} {N_f} (r-{r_h}) {r_h} \left(9 a^2+{r_h}^2\right) \log ({r_h}) \alpha _{\theta _1}^2
   {f_{\theta_2z}}''(r)}{2 \sqrt{2} \pi ^{7/4} \left(6 a^2+{r_h}^2\right) \alpha _{\theta _2}^3}\nonumber\\
   & & +\frac{7 {g_s}^{11/4} M^2
   \left(\frac{1}{N}\right)^{21/20} {N_f}^2 (r-{r_h}) {r_h} \left(9 a^2+{r_h}^2\right) \log ^2({r_h}) \alpha _{\theta
   _1}^2 \left(81 \sqrt{2} \alpha _{\theta _1}^3+10 \sqrt{3} \alpha _{\theta _2}^2\right) {f_{\theta_2x}}''(r)}{64 \pi ^{13/4} \left(6
   a^2+{r_h}^2\right) \alpha _{\theta _2}^4}\nonumber\\
   & & -\frac{3 {g_s}^{5/4} M \sqrt[4]{\frac{1}{N}} {N_f} (r-{r_h}) {r_h} \left(9
   a^2+{r_h}^2\right) \log ({r_h}) \alpha _{\theta _1}^2 {f_{\theta_2y}}''(r)}{4 \sqrt{2} \pi ^{7/4} \left(6 a^2+{r_h}^2\right)
   \alpha _{\theta _2}^3}\nonumber\\
   & & -\frac{3 {g_s}^{5/4} M \sqrt[4]{\frac{1}{N}} {N_f} (r-{r_h}) {r_h} \left(9 a^2+{r_h}^2\right)
   \log ({r_h}) \alpha _{\theta _1}^2 {f_{xz}}''(r)}{4 \sqrt{2} \pi ^{7/4} \left(6 a^2+{r_h}^2\right) \alpha _{\theta
   _2}^3}+\frac{243 {g_s}^{5/4} M \left(\frac{1}{N}\right)^{33/20} {N_f} (r-{r_h}) {r_h} \left(9 a^2+{r_h}^2\right)
   \log ({r_h}) \alpha _{\theta _1}^6 {f_{xx}}''(r)}{16 \sqrt{2} \pi ^{7/4} \left(6 a^2+{r_h}^2\right) \alpha _{\theta
   _2}}\nonumber\\
   & & +\frac{3 \sqrt{\frac{3}{2}} {g_s}^{11/4} M^2 \left(\frac{1}{N}\right)^{21/20} {N_f}^2 (r-{r_h}) {r_h} \left(9
   a^2+{r_h}^2\right) \log ^2({r_h}) \alpha _{\theta _1}^2 \left(27 \sqrt{3} \alpha _{\theta _1}^3+5 \sqrt{2} \alpha _{\theta
   _2}^2\right) {f_{xy}}''(r)}{128 \pi ^{13/4} \left(6 a^2+{r_h}^2\right) \alpha _{\theta _2}^4}
   \nonumber\\
    \end{eqnarray*}}
{\scriptsize
\begin{eqnarray*}
   & &
   -\frac{3 \sqrt{\frac{3}{2}}
   {g_s}^{11/4} M^2 \left(\frac{1}{N}\right)^{21/20} {N_f}^2 (r-{r_h}) {r_h} \left(9 a^2+{r_h}^2\right) \log
   ^2({r_h}) \alpha _{\theta _1}^2 \left(27 \sqrt{3} \alpha _{\theta _1}^3+5 \sqrt{2} \alpha _{\theta _2}^2\right)
   {f_{yz}}''(r)}{128 \pi ^{13/4} \left(6 a^2+{r_h}^2\right) \alpha _{\theta _2}^4}\nonumber\\
   & & +\frac{3 {g_s}^{17/4} M^3
   \left(\frac{1}{N}\right)^{37/20} {N_f}^3 (r-{r_h}) {r_h} \left(9 a^2+{r_h}^2\right) \log ^3({r_h}) \alpha _{\theta
   _1}^2 \left(2187 \alpha _{\theta _1}^6+270 \sqrt{6} \alpha _{\theta _2}^2 \alpha _{\theta _1}^3+50 \alpha _{\theta _2}^4\right)
   {f_{yy}}''(r)}{2048 \sqrt{2} \pi ^{19/4} \left(6 a^2+{r_h}^2\right) \alpha _{\theta _2}^5}\nonumber\\
   & & -\frac{{g_s}^{5/4} M
   \sqrt[4]{\frac{1}{N}} {N_f} (r-{r_h}) {r_h} \left(9 a^2+{r_h}^2\right) \log ({r_h}) \alpha _{\theta _1}^2
   {f_t}''(r)}{2 \sqrt{2} \pi ^{7/4} \left(6 a^2+{r_h}^2\right) \alpha _{\theta _2}^3} +\frac{243 {g_s}^{5/4} M
   \left(\frac{1}{N}\right)^{33/20} {N_f} {f_{xx}}(r) \log ({r_h}) \alpha _{\theta _1}^6}{8 \sqrt{2} \pi ^{7/4} \log
   \left({r_h}^6+9 a^2 {r_h}^4\right) \alpha _{\theta _2}}\nonumber\\
   & & -\frac{81 {g_s}^{21/4} M^5 \left(\frac{1}{N}\right)^{9/4}
   {N_f}^3 {f_t}(r) \log ^5({r_h}) \left(16 \left(108 a^6-9 {r_h}^4 a^2-{r_h}^6\right)-\left(2 {r_h}^6+27 a^2
   {r_h}^4+117 a^4 {r_h}^2\right) \log \left({r_h}^6+9 a^2 {r_h}^4\right)\right) \alpha _{\theta _1}^2}{256 \sqrt{2} \pi
   ^{23/4} \left(3 a^2-{r_h}^2\right) \left(6 a^2+{r_h}^2\right)^2 \log \left({r_h}^6+9 a^2 {r_h}^4\right) \alpha _{\theta
   _2}^3}\nonumber\\
& &    +\frac{243 {g_s}^{21/4} \log N  M^5 \left(\frac{1}{N}\right)^{9/4} {N_f}^3 \left(9 a^2+{r_h}^2\right) f(r) \log
   ^5({r_h}) \alpha _{\theta _1}^2}{64 \sqrt{2} \pi ^{23/4} \left(6 a^2+{r_h}^2\right) \log \left({r_h}^6+9 a^2
   {r_h}^4\right) \alpha _{\theta _2}^3}-\frac{49 \pi ^{5/4} {r_h}^4 {f_{zz}}(r) \alpha _{\theta _1}^2}{7776 \sqrt{2}
   {g_s}^{7/4} M \left(\frac{1}{N}\right)^{3/4} {N_f} \left({r_h}^2-3 a^2\right)^2 \log ^3({r_h}) \alpha _{\theta
   _2}^3}\nonumber\\
   & &   +\frac{49 \pi ^{5/4} {r_h}^4 {f_{x^{10} x^{10}}}(r) \alpha _{\theta _1}^2}{5184 \sqrt{2} {g_s}^{7/4} M
   \left(\frac{1}{N}\right)^{3/4} {N_f} \left({r_h}^2-3 a^2\right)^2 \log ^3({r_h}) \alpha _{\theta _2}^3}+\frac{49 \pi ^{5/4}
   {r_h}^4 {f_{\theta_1z}}(r) \alpha _{\theta _1}^2}{3888 \sqrt{2} {g_s}^{7/4} M \left(\frac{1}{N}\right)^{3/4} {N_f}
   \left({r_h}^2-3 a^2\right)^2 \log ^3({r_h}) \alpha _{\theta _2}^3}\nonumber\\
   & &   +\frac{245 \pi ^{5/4} {r_h}^4 {f_{\theta_2y}}(r) \alpha
   _{\theta _1}^2}{15552 \sqrt{2} {g_s}^{7/4} M \left(\frac{1}{N}\right)^{3/4} {N_f} \left({r_h}^2-3 a^2\right)^2 \log
   ^3({r_h}) \alpha _{\theta _2}^3}+\frac{5 {g_s}^{5/4} M {N_f} (r-{r_h}) {r_h}^3 \left(9 a^2+{r_h}^2\right)
   \left(108 b^2 {r_h}^2+1\right)^2 {f_r}(r)}{36 \sqrt{2} \left(\frac{1}{N}\right)^{3/20} \pi ^{7/4} \left({r_h}^2-3
   a^2\right)^2 \left(6 a^2+{r_h}^2\right) \log ({r_h}) \alpha _{\theta _2}^3}\nonumber\\
   & &   +\frac{{g_s}^{5/4} M {N_f} (r-{r_h})
   {r_h}^3 \left(9 a^2+{r_h}^2\right) \left(108 b^2 {r_h}^2+1\right)^2 {f_{yz}}(r) \alpha _{\theta _1}^2}{3 \sqrt{2}
   \left(\frac{1}{N}\right)^{7/20} \pi ^{7/4} \left({r_h}^2-3 a^2\right)^2 \left(6 a^2+{r_h}^2\right) \log ({r_h}) \alpha
   _{\theta _2}^5}-\frac{{g_s}^{5/4} M {N_f} (r-{r_h}) {r_h}^3 \left(9 a^2+{r_h}^2\right) \left(108 b^2
   {r_h}^2+1\right)^2 {f_{yy}}(r) \alpha _{\theta _1}^2}{6 \sqrt{2} \left(\frac{1}{N}\right)^{7/20} \pi ^{7/4} \left({r_h}^2-3
   a^2\right)^2 \left(6 a^2+{r_h}^2\right) \log ({r_h}) \alpha _{\theta _2}^5}\nonumber\\
   & &   -\frac{245 {r_h}^4 {f_{\theta_1y}}(r) \alpha _{\theta
   _1}^6}{1536 \sqrt{3} \sqrt[4]{{g_s}} \left(\frac{1}{N}\right)^{7/20} \sqrt[4]{\pi } \left({r_h}^2-3 a^2\right)^2 \log
   ^2({r_h}) \alpha _{\theta _2}^6}-\frac{245 {r_h}^4 {f_{\theta_2z}}(r) \alpha _{\theta _1}^6}{1536 \sqrt{3} \sqrt[4]{{g_s}}
   \left(\frac{1}{N}\right)^{7/20} \sqrt[4]{\pi } \left({r_h}^2-3 a^2\right)^2 \log ^2({r_h}) \alpha _{\theta _2}^6} = 0.
\end{eqnarray*}
}
(xiv) ${\rm EOM}_{xx}$

{\scriptsize
\begin{eqnarray*}
& & -\frac{6 \left(9 b^2+1\right)^4 M \left(\frac{1}{N}\right)^{7/4} (r-{r_h}) \beta  \log ({r_h}) \Sigma_1 b^{10}}{\left(3 b^2-1\right)^5
   \left(6 b^2+1\right)^4 \sqrt{{g_s}} \log N ^2 {N_f} \pi ^{3/2} {r_h} \alpha _{\theta _2}^3}+\frac{2048 \left(9
   b^2+1\right)^4 \left(9 b^4-5 b^2+1\right) \sqrt[4]{\frac{1}{N}} (r-{r_h})^2 \beta  b^{10}}{9 \sqrt{3} \left(3 b^2-1\right)^7
   \left(6 b^2+1\right)^4 {g_s}^{9/4} \log N ^3 {N_f}^2 \sqrt[4]{\pi } {r_h}^2 \alpha _{\theta _1}^4 \alpha _{\theta
   _2}^4}\nonumber\\
   & & +\frac{32 a^2 (r-{r_h}) {f_{xz}}(r) \left(243 \sqrt{6} \alpha _{\theta _1}^3-8 \alpha _{\theta _2}^2\right){}^2}{531441
   \sqrt{{g_s}} \sqrt{\frac{1}{N}} \sqrt{\pi } {r_h} \left({r_h}^2-3 a^2\right) \log \left({r_h}^6+9 a^2
   {r_h}^4\right) \alpha _{\theta _1}^8 \alpha _{\theta _2}^2}\nonumber\\
   & & +\frac{\sqrt{\frac{1}{N}} {f_{xx}}(r) \left(18 \left(108 a^6-9
   {r_h}^4 a^2-{r_h}^6\right)+\left(324 a^6-189 {r_h}^2 a^4-66 {r_h}^4 a^2-5 {r_h}^6\right) \log \left({r_h}^6+9
   a^2 {r_h}^4\right)\right)}{\sqrt{{g_s}} \sqrt{\pi } \left(3 a^2-{r_h}^2\right) \left(6 a^2+{r_h}^2\right)^2 \log
   \left({r_h}^6+9 a^2 {r_h}^4\right)}\nonumber\\
& & +\frac{81 {g_s}^{7/2} M^4 \left(\frac{1}{N}\right)^{5/2} {N_f}^2 {f_t}(r) \log
   ^4({r_h}) \left(16 \left(108 a^6-9 {r_h}^4 a^2-{r_h}^6\right)-\left(2 {r_h}^6+27 a^2 {r_h}^4+117 a^4
   {r_h}^2\right) \log \left({r_h}^6+9 a^2 {r_h}^4\right)\right)}{128 \pi ^{9/2} \left(3 a^2-{r_h}^2\right) \left(6
   a^2+{r_h}^2\right)^2 \log \left({r_h}^6+9 a^2 {r_h}^4\right)}\nonumber\\
   & & +\frac{8 \sqrt{2} a^2 {g_s} M {N_f} (r-{r_h})
   {f_{xy}}(r) \log ({r_h}) \left(-177147 \sqrt{3} \alpha _{\theta _1}^6+5832 \sqrt{2} \alpha _{\theta _2}^2 \alpha _{\theta
   _1}^3-32 \sqrt{3} \alpha _{\theta _2}^4\right)}{19683 \sqrt[10]{\frac{1}{N}} \pi ^2 {r_h} \left({r_h}^2-3 a^2\right) \log
   \left({r_h}^6+9 a^2 {r_h}^4\right) \alpha _{\theta _1}^4 \alpha _{\theta _2}^5}\nonumber\\
   & & +\frac{3 \sqrt{\frac{1}{N}} {r_h} \left(9
   a^2+{r_h}^2\right) f'(r)}{\sqrt{{g_s}} \sqrt{\pi } \left(6 a^2+{r_h}^2\right)}-\frac{\sqrt{\frac{1}{N}} {r_h} \left(9
   a^2+{r_h}^2\right) {f_{zz}}'(r)}{\sqrt{{g_s}} \sqrt{\pi } \left(6 a^2+{r_h}^2\right)}+\frac{\sqrt{\frac{1}{N}}
   {r_h} \left(9 a^2+{r_h}^2\right) {f_{x^{10} x^{10}}}'(r)}{\sqrt{{g_s}} \sqrt{\pi } \left(6 a^2+{r_h}^2\right)}\nonumber\\
    \end{eqnarray*}}
{\scriptsize
\begin{eqnarray*}
   & & +\frac{3
   \sqrt{\frac{1}{N}} {r_h} \left(9 a^2+{r_h}^2\right) {f_{\theta_1z}}'(r)}{2 \sqrt{{g_s}} \sqrt{\pi } \left(6
   a^2+{r_h}^2\right)}-\frac{32 a^2 (r-{r_h}) \left(9 a^2+{r_h}^2\right) {f_{\theta_1x}}'(r)}{81 \sqrt{{g_s}}
   \left(\frac{1}{N}\right)^{9/10} \sqrt{\pi } \left(-18 a^4+3 {r_h}^2 a^2+{r_h}^4\right) \alpha _{\theta _1}^4 \alpha _{\theta
   _2}^2}\nonumber\\
   & & -\frac{81 \sqrt{\frac{3}{2}} {g_s} M \left(\frac{1}{N}\right)^{9/10} {N_f} {r_h} \left(9 a^2+{r_h}^2\right) \log
   ({r_h}) \alpha _{\theta _1}^4 {f_{\theta_1y}}'(r)}{8 \pi ^2 \left(6 a^2+{r_h}^2\right) \alpha _{\theta _2}^3}-\frac{81
   \sqrt{\frac{3}{2}} {g_s} M \left(\frac{1}{N}\right)^{9/10} {N_f} {r_h} \left(9 a^2+{r_h}^2\right) \log ({r_h})
   \alpha _{\theta _1}^4 {f_{\theta_2z}}'(r)}{8 \pi ^2 \left(6 a^2+{r_h}^2\right) \alpha _{\theta _2}^3}\nonumber\\
   & & +\frac{4 \sqrt{\frac{2}{3}} a^2
   {g_s} M {N_f} (r-{r_h}) \left(9 a^2+{r_h}^2\right) \log ({r_h}) {f_{\theta_2x}}'(r)}{\left(\frac{1}{N}\right)^{3/10} \pi
   ^2 \left(-18 a^4+3 {r_h}^2 a^2+{r_h}^4\right) \alpha _{\theta _1}^2 \alpha _{\theta _2}^3}+\frac{3 \sqrt{\frac{1}{N}} {r_h}
   \left(9 a^2+{r_h}^2\right) {f_{\theta_2y}}'(r)}{2 \sqrt{{g_s}} \sqrt{\pi } \left(6 a^2+{r_h}^2\right)}\nonumber\\
   & & +\frac{32 a^2
   (r-{r_h}) \left(9 a^2+{r_h}^2\right) {f_{xz}}'(r)}{81 \sqrt{{g_s}} \left(\frac{1}{N}\right)^{9/10} \sqrt{\pi } \left(-18
   a^4+3 {r_h}^2 a^2+{r_h}^4\right) \alpha _{\theta _1}^4 \alpha _{\theta _2}^2}-\frac{\sqrt{\frac{1}{N}} {r_h} \left(9
   a^2+{r_h}^2\right) {f_{xx}}'(r)}{\sqrt{{g_s}} \sqrt{\pi } \left(6 a^2+{r_h}^2\right)}\nonumber\\
   & & -\frac{4 \sqrt{\frac{2}{3}} a^2
   {g_s} M {N_f} (r-{r_h}) \left(9 a^2+{r_h}^2\right) \log ({r_h}) {f_{xy}}'(r)}{\sqrt{\frac{1}{N}} \pi ^2 \left(-18
   a^4+3 {r_h}^2 a^2+{r_h}^4\right) \alpha _{\theta _2}^5}-\frac{81 \sqrt{\frac{3}{2}} a^2 {g_s} M
   \left(\frac{1}{N}\right)^{9/10} {N_f} (r-{r_h}) \left(9 a^2+{r_h}^2\right) \log ({r_h}) \alpha _{\theta _1}^4
   {f_{yz}}'(r)}{4 \pi ^2 \left(-18 a^4+3 {r_h}^2 a^2+{r_h}^4\right) \alpha _{\theta _2}^3}\nonumber\\
   & & -\frac{27 a^2 {g_s}^{5/2} M^2
   \left(\frac{1}{N}\right)^{19/10} {N_f}^2 (r-{r_h}) \left(9 a^2+{r_h}^2\right) \log ^2({r_h}) \alpha _{\theta _1}^2
   \left(2187 \sqrt{2} \alpha _{\theta _1}^6+540 \sqrt{3} \alpha _{\theta _2}^2 \alpha _{\theta _1}^3+50 \sqrt{2} \alpha _{\theta
   _2}^4\right) {f_{yy}}'(r)}{256 \sqrt{2} \pi ^{7/2} \left(-18 a^4+3 {r_h}^2 a^2+{r_h}^4\right) \alpha _{\theta
   _2}^4}\nonumber\\
   & & -\frac{\sqrt{\frac{1}{N}} {r_h} \left(9 a^2+{r_h}^2\right) {f_r}'(r)}{2 \sqrt{{g_s}} \sqrt{\pi } \left(6
   a^2+{r_h}^2\right)}+\frac{3 \sqrt{\frac{1}{N}} {r_h} \left(9 a^2+{r_h}^2\right) {f_t}'(r)}{2 \sqrt{{g_s}} \sqrt{\pi
   } \left(6 a^2+{r_h}^2\right)}+\frac{3 \sqrt{\frac{1}{N}} (r-{r_h}) {r_h} \left(9 a^2+{r_h}^2\right)
   f''(r)}{\sqrt{{g_s}} \sqrt{\pi } \left(6 a^2+{r_h}^2\right)}\nonumber\\
     & &  -\frac{\sqrt{\frac{1}{N}} (r-{r_h}) {r_h} \left(9
   a^2+{r_h}^2\right) {f_{zz}}''(r)}{\sqrt{{g_s}} \sqrt{\pi } \left(6 a^2+{r_h}^2\right)}+\frac{\sqrt{\frac{1}{N}}
   (r-{r_h}) {r_h} \left(9 a^2+{r_h}^2\right) {f_{x^{10} x^{10}}}''(r)}{\sqrt{{g_s}} \sqrt{\pi } \left(6
   a^2+{r_h}^2\right)}\nonumber\\
   & & +\frac{3 \sqrt{\frac{1}{N}} (r-{r_h}) {r_h} \left(9 a^2+{r_h}^2\right) {f_{\theta_1z}}''(r)}{2
   \sqrt{{g_s}} \sqrt{\pi } \left(6 a^2+{r_h}^2\right)}+\frac{7 {g_s} M \left(\frac{1}{N}\right)^{13/10} {N_f}
   (r-{r_h}) {r_h} \left(9 a^2+{r_h}^2\right) \log ({r_h}) \left(81 \sqrt{2} \alpha _{\theta _1}^3+10 \sqrt{3} \alpha
   _{\theta _2}^2\right) {f_{\theta_1x}}''(r)}{16 \sqrt{2} \pi ^2 \left(6 a^2+{r_h}^2\right) \alpha _{\theta _2}}\nonumber\\
   & & -\frac{81
   \sqrt{\frac{3}{2}} {g_s} M \left(\frac{1}{N}\right)^{9/10} {N_f} (r-{r_h}) {r_h} \left(9 a^2+{r_h}^2\right) \log
   ({r_h}) \alpha _{\theta _1}^4 {f_{\theta_1y}}''(r)}{8 \pi ^2 \left(6 a^2+{r_h}^2\right) \alpha _{\theta _2}^3}-\frac{81
   \sqrt{\frac{3}{2}} {g_s} M \left(\frac{1}{N}\right)^{9/10} {N_f} (r-{r_h}) {r_h} \left(9 a^2+{r_h}^2\right) \log
   ({r_h}) \alpha _{\theta _1}^4 {f_{\theta_2z}}''(r)}{8 \pi ^2 \left(6 a^2+{r_h}^2\right) \alpha _{\theta _2}^3}
   \nonumber\\
   & &  -\frac{7 {g_s} M
   \left(\frac{1}{N}\right)^{13/10} {N_f} (r-{r_h}) {r_h} \left(9 a^2+{r_h}^2\right) \log ({r_h}) \left(81 \sqrt{2}
   \alpha _{\theta _1}^3+10 \sqrt{3} \alpha _{\theta _2}^2\right) {f_{\theta_2x}}''(r)}{16 \sqrt{2} \pi ^2 \left(6 a^2+{r_h}^2\right)
   \alpha _{\theta _2}}\nonumber\\
   & & +\frac{3 \sqrt{\frac{1}{N}} (r-{r_h}) {r_h} \left(9 a^2+{r_h}^2\right) {f_{\theta_2y}}''(r)}{2
   \sqrt{{g_s}} \sqrt{\pi } \left(6 a^2+{r_h}^2\right)}+\frac{3 \sqrt{\frac{1}{N}} (r-{r_h}) {r_h} \left(9
   a^2+{r_h}^2\right) {f_{xz}}''(r)}{2 \sqrt{{g_s}} \sqrt{\pi } \left(6 a^2+{r_h}^2\right)}\nonumber\\
   & &    -\frac{\sqrt{\frac{1}{N}}
   (r-{r_h}) {r_h} \left(9 a^2+{r_h}^2\right) {f_{xx}}''(r)}{\sqrt{{g_s}} \sqrt{\pi } \left(6
   a^2+{r_h}^2\right)}-\frac{3 \sqrt{3} {g_s} M \left(\frac{1}{N}\right)^{13/10} {N_f} (r-{r_h}) {r_h} \left(9
   a^2+{r_h}^2\right) \log ({r_h}) \left(27 \sqrt{3} \alpha _{\theta _1}^3+5 \sqrt{2} \alpha _{\theta _2}^2\right)
   {f_{xy}}''(r)}{64 \pi ^2 \left(6 a^2+{r_h}^2\right) \alpha _{\theta _2}}
 \nonumber\\
& &    +\frac{3 \sqrt{3} {g_s} M
   \left(\frac{1}{N}\right)^{13/10} {N_f} (r-{r_h}) {r_h} \left(9 a^2+{r_h}^2\right) \log ({r_h}) \left(27 \sqrt{3}
   \alpha _{\theta _1}^3+5 \sqrt{2} \alpha _{\theta _2}^2\right) {f_{yz}}''(r)}{64 \pi ^2 \left(6 a^2+{r_h}^2\right) \alpha
   _{\theta _2}}\nonumber\\
   & &  -\frac{3 {g_s}^{5/2} M^2 \left(\frac{1}{N}\right)^{21/10} {N_f}^2 (r-{r_h}) {r_h} \left(9
   a^2+{r_h}^2\right) \log ^2({r_h}) \left(2187 \alpha _{\theta _1}^6+270 \sqrt{6} \alpha _{\theta _2}^2 \alpha _{\theta _1}^3+50
   \alpha _{\theta _2}^4\right) {f_{yy}}''(r)}{1024 \pi ^{7/2} \left(6 a^2+{r_h}^2\right) \alpha _{\theta
   _2}^2}\nonumber\\
   & & +\frac{\sqrt{\frac{1}{N}} (r-{r_h}) {r_h} \left(9 a^2+{r_h}^2\right) {f_t}''(r)}{\sqrt{{g_s}} \sqrt{\pi }
   \left(6 a^2+{r_h}^2\right)}-\frac{243 {g_s}^{7/2} \log N  M^4 \left(\frac{1}{N}\right)^{5/2} {N_f}^2 \left(9
   a^2+{r_h}^2\right) f(r) \log ^4({r_h})}{32 \pi ^{9/2} \left(6 a^2+{r_h}^2\right) \log \left({r_h}^6+9 a^2
   {r_h}^4\right)}\nonumber\\
   & & -\frac{32 a^4 (r-{r_h}) \left(9 a^2+{r_h}^2\right) {f_{zz}}(r)}{27 \sqrt{{g_s}}
   \left(\frac{1}{N}\right)^{9/10} \sqrt{\pi } \left({r_h}^2-3 a^2\right)^2 \left({r_h}^3+6 a^2 {r_h}\right) \alpha _{\theta
   _1}^4 \alpha _{\theta _2}^2}+\frac{64 a^4 (r-{r_h}) \left(9 a^2+{r_h}^2\right) {f_{\theta_1z}}(r)}{27 \sqrt{{g_s}}
   \left(\frac{1}{N}\right)^{9/10} \sqrt{\pi } \left({r_h}^2-3 a^2\right)^2 \left({r_h}^3+6 a^2 {r_h}\right) \alpha _{\theta
   _1}^4 \alpha _{\theta _2}^2}\nonumber\\
   & & -\frac{64 a^4 (r-{r_h}) \left(9 a^2+{r_h}^2\right) {f_{\theta_1x}}(r)}{27 \sqrt{{g_s}}
   \left(\frac{1}{N}\right)^{9/10} \sqrt{\pi } \left({r_h}^2-3 a^2\right)^2 \left({r_h}^3+6 a^2 {r_h}\right) \alpha _{\theta
   _1}^4 \alpha _{\theta _2}^2}+\frac{32 a^4 (r-{r_h}) \left(9 a^2+{r_h}^2\right) {f_r}(r)}{27 \sqrt{{g_s}}
   \left(\frac{1}{N}\right)^{9/10} \sqrt{\pi } \left({r_h}^2-3 a^2\right)^2 \left({r_h}^3+6 a^2 {r_h}\right) \alpha _{\theta
   _1}^4 \alpha _{\theta _2}^2}\nonumber\\
    \end{eqnarray*}}
{\scriptsize
\begin{eqnarray*}
   & & +\frac{4 \sqrt{\frac{2}{3}} a^2 {g_s} M {N_f} (r-{r_h}) \left(9 a^2+{r_h}^2\right)
   {f_{\theta_2x}}(r)}{\left(\frac{1}{N}\right)^{3/10} \pi ^2 \left({r_h}^5+3 a^2 {r_h}^3-18 a^4 {r_h}\right) \alpha _{\theta _1}^2
   \alpha _{\theta _2}^3}-\frac{8 \sqrt{6} a^4 {g_s} M {N_f} (r-{r_h}) \left(9 a^2+{r_h}^2\right) {f_{\theta_2z}}(r) \log
   ({r_h})}{\left(\frac{1}{N}\right)^{3/10} \pi ^2 \left({r_h}^2-3 a^2\right)^2 \left({r_h}^3+6 a^2 {r_h}\right) \alpha
   _{\theta _1}^2 \alpha _{\theta _2}^3}\nonumber\\
& &    -\frac{49 \pi ^{5/2} {r_h}^4 {f_{x^{10} x^{10}}}(r)}{2592 {g_s}^{7/2} M^2 \sqrt{\frac{1}{N}}
   {N_f}^2 \left({r_h}^2-3 a^2\right)^2 \log ^4({r_h})}-\frac{245 \pi ^{5/2} {r_h}^4 {f_{\theta_2y}}(r)}{7776 {g_s}^{7/2}
   M^2 \sqrt{\frac{1}{N}} {N_f}^2 \left({r_h}^2-3 a^2\right)^2 \log ^4({r_h})}\nonumber\\
   & & -\frac{8 \sqrt{6} a^4 {g_s} M {N_f}
   (r-{r_h}) \left(9 a^2+{r_h}^2\right) {f_{\theta_1y}}(r) \log ({r_h})}{\sqrt{\frac{1}{N}} \pi ^2 \left({r_h}^2-3 a^2\right)^2
   \left({r_h}^3+6 a^2 {r_h}\right) \alpha _{\theta _2}^5}+\frac{8 \sqrt{6} a^4 {g_s} M {N_f} (r-{r_h}) \left(9
   a^2+{r_h}^2\right) {f_{yz}}(r) \log ({r_h})}{\sqrt{\frac{1}{N}} \pi ^2 \left({r_h}^2-3 a^2\right)^2 \left({r_h}^3+6
   a^2 {r_h}\right) \alpha _{\theta _2}^5}\nonumber\\
   & & -\frac{81 a^4 {g_s}^{5/2} M^2 {N_f}^2 (r-{r_h}) \left(9 a^2+{r_h}^2\right)
   {f_{yy}}(r) \log ^2({r_h}) \alpha _{\theta _1}^4}{\sqrt[10]{\frac{1}{N}} \pi ^{7/2} \left({r_h}^2-3 a^2\right)^2
   \left({r_h}^3+6 a^2 {r_h}\right) \alpha _{\theta _2}^8} = 0.
\end{eqnarray*}
}

(xv) \underline{${\rm EOM}_{xy}$}

{\scriptsize
\begin{eqnarray*}
& & -\frac{512 \left(9 b^2+1\right)^4 \left(9 b^4-5 b^2+1\right) \left(\frac{1}{N}\right)^{3/10} (r-{r_h})^3 \beta  \Sigma_1 b^{12}}{243 \sqrt{3}
   \left(1-3 b^2\right)^8 \left(6 b^2+1\right)^4 {g_s}^{9/4} \log N ^3 {N_f}^2 \sqrt[4]{\pi } {r_h}^3 \alpha _{\theta
   _1}^6 \alpha _{\theta _2}^3}\nonumber\\
& & -\frac{4 \sqrt{\frac{2}{3}} \left(9 b^2+1\right)^4 M \left(\frac{1}{N}\right)^{21/20} (r-{r_h}) \beta
   \log ({r_h}) \left(19683 \sqrt{6} \alpha _{\theta _1}^6+6642 \alpha _{\theta _2}^2 \alpha _{\theta _1}^3-40 \sqrt{6} \alpha
   _{\theta _2}^4\right) b^{10}}{3 \left(3 b^2-1\right)^5 \left(6 b^2+1\right)^4 \sqrt{{g_s}} \log N ^2 {N_f} \pi ^{3/2}
   {r_h} \alpha _{\theta _1}^2 \alpha _{\theta _2}^4}\nonumber\\
   & & -\frac{9 \left(\frac{1}{N}\right)^{6/5} \sqrt{\frac{3}{2 \pi }} {f_{xx}}(r)
   \alpha _{\theta _1}^2 \alpha _{\theta _2}}{\sqrt{{g_s}} \log \left({r_h}^6+9 a^2 {r_h}^4\right)}+\frac{245 \pi  {r_h}^4
   {f_{xy}}(r) \left(27 \sqrt{6} \alpha _{\theta _1}^3+10 \alpha _{\theta _2}^2\right)}{1119744 {g_s}^2 M
   \left(\frac{1}{N}\right)^{2/5} {N_f} \left({r_h}^2-3 a^2\right)^2 \log ^3({r_h}) \alpha _{\theta _1}^2 \alpha _{\theta
   _2}^2}\nonumber\\
   & & -\frac{343 \pi  {r_h}^4 {f_{\theta_1x}}(r) \left(81 \sqrt{2} \alpha _{\theta _1}^3+10 \sqrt{3} \alpha _{\theta
   _2}^2\right)}{279936 \sqrt{3} {g_s}^2 M \left(\frac{1}{N}\right)^{2/5} {N_f} \left({r_h}^2-3 a^2\right)^2 \log
   ^3({r_h}) \alpha _{\theta _1}^2 \alpha _{\theta _2}^2}+\frac{1715 \pi  {r_h}^4 {f_{\theta_2x}}(r) \left(81 \sqrt{2} \alpha _{\theta
   _1}^3+10 \sqrt{3} \alpha _{\theta _2}^2\right)}{839808 \sqrt{3} {g_s}^2 M \left(\frac{1}{N}\right)^{2/5} {N_f}
   \left({r_h}^2-3 a^2\right)^2 \log ^3({r_h}) \alpha _{\theta _1}^2 \alpha _{\theta _2}^2}\nonumber\\
   & & -\frac{3479 \pi  {r_h}^4
   {f_{xz}}(r) \left(81 \sqrt{2} \alpha _{\theta _1}^3+10 \sqrt{3} \alpha _{\theta _2}^2\right)}{3359232 \sqrt{3} {g_s}^2 M
   \left(\frac{1}{N}\right)^{2/5} {N_f} \left({r_h}^2-3 a^2\right)^2 \log ^3({r_h}) \alpha _{\theta _1}^2 \alpha _{\theta
   _2}^2}+\frac{2 \sqrt{\frac{2}{3 \pi }} {r_h} \left(9 a^2+{r_h}^2\right) f'(r)}{3 \sqrt{{g_s}} \sqrt[5]{\frac{1}{N}} \left(6
   a^2+{r_h}^2\right) \alpha _{\theta _1}^2 \alpha _{\theta _2}}\nonumber\\
   & & -\frac{2 \sqrt{\frac{2}{3 \pi }} {r_h} \left(9
   a^2+{r_h}^2\right) {f_{zz}}'(r)}{9 \sqrt{{g_s}} \sqrt[5]{\frac{1}{N}} \left(6 a^2+{r_h}^2\right) \alpha _{\theta _1}^2
   \alpha _{\theta _2}}+\frac{2 \sqrt{\frac{2}{3 \pi }} {r_h} \left(9 a^2+{r_h}^2\right) {f_{x^{10} x^{10}}}'(r)}{9 \sqrt{{g_s}}
   \sqrt[5]{\frac{1}{N}} \left(6 a^2+{r_h}^2\right) \alpha _{\theta _1}^2 \alpha _{\theta _2}}\nonumber\\
   & & +\frac{\sqrt{\frac{2}{3 \pi }} {r_h}
   \left(9 a^2+{r_h}^2\right) {f_{\theta_1z}}'(r)}{3 \sqrt{{g_s}} \sqrt[5]{\frac{1}{N}} \left(6 a^2+{r_h}^2\right) \alpha _{\theta
   _1}^2 \alpha _{\theta _2}}-\frac{2 a^2 \sqrt{\frac{2}{3 \pi }} (r-{r_h}) \left(9 a^2+{r_h}^2\right)
   {f_{\theta_1x}}'(r)}{\sqrt{{g_s}} \sqrt[5]{\frac{1}{N}} \left({r_h}^2-3 a^2\right) \left(6 a^2+{r_h}^2\right) \alpha _{\theta
   _1}^2 \alpha _{\theta _2}}\nonumber\\
   \end{eqnarray*}
   \begin{eqnarray*}
   & & -\frac{4 a^2 \sqrt{\frac{2}{3 \pi }} (r-{r_h}) \left(9 a^2+{r_h}^2\right) {f_{\theta_1y}}'(r)}{3
   \sqrt{{g_s}} \sqrt[5]{\frac{1}{N}} \left({r_h}^2-3 a^2\right) \left(6 a^2+{r_h}^2\right) \alpha _{\theta _1}^2 \alpha
   _{\theta _2}}-\frac{9 {g_s} M \sqrt[5]{\frac{1}{N}} {N_f} {r_h} \left(9 a^2+{r_h}^2\right) \log ({r_h}) \alpha
   _{\theta _1}^2 {f_{\theta_2z}}'(r)}{4 \pi ^2 \left(6 a^2+{r_h}^2\right) \alpha _{\theta _2}^4}\nonumber\\
   & & +\frac{27 a^2 {g_s} M
   \left(\frac{1}{N}\right)^{2/5} {N_f} (r-{r_h}) \left(9 a^2+{r_h}^2\right) \log ({r_h}) {f_{\theta_2x}}'(r)}{2 \pi ^2
   \left(-18 a^4+3 {r_h}^2 a^2+{r_h}^4\right) \alpha _{\theta _2}^2}+\frac{\sqrt{\frac{2}{3 \pi }} {r_h} \left(9
   a^2+{r_h}^2\right) {f_{\theta_2y}}'(r)}{3 \sqrt{{g_s}} \sqrt[5]{\frac{1}{N}} \left(6 a^2+{r_h}^2\right) \alpha _{\theta _1}^2
   \alpha _{\theta _2}}\nonumber\\
   & & +\frac{\sqrt{\frac{2}{3 \pi }} {r_h} \left(9 a^2+{r_h}^2\right) {f_{xz}}'(r)}{3 \sqrt{{g_s}}
   \sqrt[5]{\frac{1}{N}} \left(6 a^2+{r_h}^2\right) \alpha _{\theta _1}^2 \alpha _{\theta _2}}-\frac{9 \left(\frac{1}{N}\right)^{6/5}
   \sqrt{\frac{3}{2 \pi }} {r_h} \left(9 a^2+{r_h}^2\right) \alpha _{\theta _1}^2 \alpha _{\theta _2} {f_{xx}}'(r)}{2
   \sqrt{{g_s}} \left(6 a^2+{r_h}^2\right)}\nonumber\\
   & & -\frac{2 \sqrt{\frac{2}{3 \pi }} {r_h} \left(9 a^2+{r_h}^2\right)
   {f_{xy}}'(r)}{9 \sqrt{{g_s}} \sqrt[5]{\frac{1}{N}} \left(6 a^2+{r_h}^2\right) \alpha _{\theta _1}^2 \alpha _{\theta
   _2}}+\frac{4 a^2 \sqrt{\frac{2}{3 \pi }} (r-{r_h}) \left(9 a^2+{r_h}^2\right) {f_{yz}}'(r)}{3 \sqrt{{g_s}}
   \sqrt[5]{\frac{1}{N}} \left({r_h}^2-3 a^2\right) \left(6 a^2+{r_h}^2\right) \alpha _{\theta _1}^2 \alpha _{\theta _2}}\nonumber\\
   & & -\frac{9
   a^2 {g_s} M \sqrt[5]{\frac{1}{N}} {N_f} (r-{r_h}) \left(9 a^2+{r_h}^2\right) \log ({r_h}) \alpha _{\theta _1}^2
   {f_{yy}}'(r)}{\pi ^2 \left(-18 a^4+3 {r_h}^2 a^2+{r_h}^4\right) \alpha _{\theta _2}^4}-\frac{\sqrt{\frac{2}{3 \pi }}
   {r_h} \left(9 a^2+{r_h}^2\right) {f_r}'(r)}{9 \sqrt{{g_s}} \sqrt[5]{\frac{1}{N}} \left(6 a^2+{r_h}^2\right) \alpha
   _{\theta _1}^2 \alpha _{\theta _2}}\nonumber\\
   & & +\frac{\sqrt{\frac{2}{3 \pi }} {r_h} \left(9 a^2+{r_h}^2\right) {f_t}'(r)}{3
   \sqrt{{g_s}} \sqrt[5]{\frac{1}{N}} \left(6 a^2+{r_h}^2\right) \alpha _{\theta _1}^2 \alpha _{\theta _2}}+\frac{2
   \sqrt{\frac{2}{3 \pi }} (r-{r_h}) {r_h} \left(9 a^2+{r_h}^2\right) f''(r)}{3 \sqrt{{g_s}} \sqrt[5]{\frac{1}{N}} \left(6
   a^2+{r_h}^2\right) \alpha _{\theta _1}^2 \alpha _{\theta _2}}\nonumber\\
   & & -\frac{2 \sqrt{\frac{2}{3 \pi }} (r-{r_h}) {r_h} \left(9
   a^2+{r_h}^2\right) {f_{zz}}''(r)}{9 \sqrt{{g_s}} \sqrt[5]{\frac{1}{N}} \left(6 a^2+{r_h}^2\right) \alpha _{\theta _1}^2
   \alpha _{\theta _2}}+\frac{2 \sqrt{\frac{2}{3 \pi }} (r-{r_h}) {r_h} \left(9 a^2+{r_h}^2\right) {f_{x^{10} x^{10}}}''(r)}{9
   \sqrt{{g_s}} \sqrt[5]{\frac{1}{N}} \left(6 a^2+{r_h}^2\right) \alpha _{\theta _1}^2 \alpha _{\theta _2}}+\frac{\sqrt{\frac{2}{3
   \pi }} (r-{r_h}) {r_h} \left(9 a^2+{r_h}^2\right) {f_{\theta_1z}}''(r)}{3 \sqrt{{g_s}} \sqrt[5]{\frac{1}{N}} \left(6
   a^2+{r_h}^2\right) \alpha _{\theta _1}^2 \alpha _{\theta _2}}\nonumber\\
   & & +\frac{7 {g_s} M \left(\frac{1}{N}\right)^{3/5} {N_f}
   (r-{r_h}) {r_h} \left(9 a^2+{r_h}^2\right) \log ({r_h}) \left(81 \sqrt{2} \alpha _{\theta _1}^3+10 \sqrt{3} \alpha
   _{\theta _2}^2\right) {f_{\theta_1x}}''(r)}{72 \sqrt{3} \pi ^2 \left(6 a^2+{r_h}^2\right) \alpha _{\theta _1}^2 \alpha _{\theta
   _2}^2}\nonumber\\
   & & -\frac{9 {g_s} M \sqrt[5]{\frac{1}{N}} {N_f} (r-{r_h}) {r_h} \left(9 a^2+{r_h}^2\right) \log ({r_h})
   \alpha _{\theta _1}^2 {f_{\theta_1y}}''(r)}{4 \pi ^2 \left(6 a^2+{r_h}^2\right) \alpha _{\theta _2}^4}-\frac{9 {g_s} M
   \sqrt[5]{\frac{1}{N}} {N_f} (r-{r_h}) {r_h} \left(9 a^2+{r_h}^2\right) \log ({r_h}) \alpha _{\theta _1}^2
   {f_{\theta_2z}}''(r)}{4 \pi ^2 \left(6 a^2+{r_h}^2\right) \alpha _{\theta _2}^4}\nonumber\\
   & & -\frac{7 {g_s} M \left(\frac{1}{N}\right)^{3/5}
   {N_f} (r-{r_h}) {r_h} \left(9 a^2+{r_h}^2\right) \log ({r_h}) \left(81 \sqrt{2} \alpha _{\theta _1}^3+10 \sqrt{3}
   \alpha _{\theta _2}^2\right) {f_{\theta_2x}}''(r)}{72 \sqrt{3} \pi ^2 \left(6 a^2+{r_h}^2\right) \alpha _{\theta _1}^2 \alpha _{\theta
   _2}^2}\nonumber\\
   & & +\frac{\sqrt{\frac{2}{3 \pi }} (r-{r_h}) {r_h} \left(9 a^2+{r_h}^2\right) {f_{\theta_2y}}''(r)}{3 \sqrt{{g_s}}
   \sqrt[5]{\frac{1}{N}} \left(6 a^2+{r_h}^2\right) \alpha _{\theta _1}^2 \alpha _{\theta _2}}+\frac{\sqrt{\frac{2}{3 \pi }}
   (r-{r_h}) {r_h} \left(9 a^2+{r_h}^2\right) {f_{xz}}''(r)}{3 \sqrt{{g_s}} \sqrt[5]{\frac{1}{N}} \left(6
   a^2+{r_h}^2\right) \alpha _{\theta _1}^2 \alpha _{\theta _2}}\nonumber\\
   & & -\frac{9 \left(\frac{1}{N}\right)^{6/5} \sqrt{\frac{3}{2 \pi }}
   (r-{r_h}) {r_h} \left(9 a^2+{r_h}^2\right) \alpha _{\theta _1}^2 \alpha _{\theta _2} {f_{xx}}''(r)}{2 \sqrt{{g_s}}
   \left(6 a^2+{r_h}^2\right)}-\frac{2 \sqrt{\frac{2}{3 \pi }} (r-{r_h}) {r_h} \left(9 a^2+{r_h}^2\right)
   {f_{xy}}''(r)}{9 \sqrt{{g_s}} \sqrt[5]{\frac{1}{N}} \left(6 a^2+{r_h}^2\right) \alpha _{\theta _1}^2 \alpha _{\theta
   _2}}\nonumber\\
   & & +\frac{{g_s} M \left(\frac{1}{N}\right)^{3/5} {N_f} (r-{r_h}) {r_h} \left(9 a^2+{r_h}^2\right) \log ({r_h})
   \left(27 \sqrt{3} \alpha _{\theta _1}^3+5 \sqrt{2} \alpha _{\theta _2}^2\right) {f_{yz}}''(r)}{48 \sqrt{2} \pi ^2 \left(6
   a^2+{r_h}^2\right) \alpha _{\theta _1}^2 \alpha _{\theta _2}^2}\nonumber\\
   & & -\frac{{g_s}^{5/2} M^2 \left(\frac{1}{N}\right)^{7/5}
   {N_f}^2 (r-{r_h}) {r_h} \left(9 a^2+{r_h}^2\right) \log ^2({r_h}) \left(2187 \alpha _{\theta _1}^6+270 \sqrt{6}
   \alpha _{\theta _2}^2 \alpha _{\theta _1}^3+50 \alpha _{\theta _2}^4\right) {f_{yy}}''(r)}{768 \sqrt{6} \pi ^{7/2} \left(6
   a^2+{r_h}^2\right) \alpha _{\theta _1}^2 \alpha _{\theta _2}^3}\nonumber\\
   \nonumber\\
& & +\frac{2 \sqrt{\frac{2}{3 \pi }} (r-{r_h}) {r_h} \left(9
   a^2+{r_h}^2\right) {f_t}''(r)}{9 \sqrt{{g_s}} \sqrt[5]{\frac{1}{N}} \left(6 a^2+{r_h}^2\right) \alpha _{\theta _1}^2
   \alpha _{\theta _2}}\nonumber\\
   & & +\frac{3 \sqrt{\frac{3}{2}} {g_s}^{7/2} M^4 \left(\frac{1}{N}\right)^{9/5} {N_f}^2 {f_t}(r) \log
   ^4({r_h}) \left(16 \left(108 a^6-9 {r_h}^4 a^2-{r_h}^6\right)-\left(2 {r_h}^6+27 a^2 {r_h}^4+117 a^4
   {r_h}^2\right) \log \left({r_h}^6+9 a^2 {r_h}^4\right)\right)}{32 \pi ^{9/2} \left(3 a^2-{r_h}^2\right) \left(6
   a^2+{r_h}^2\right)^2 \log \left({r_h}^6+9 a^2 {r_h}^4\right) \alpha _{\theta _1}^2 \alpha _{\theta _2}}\nonumber\\
   & & -\frac{9
   \sqrt{\frac{3}{2}} {g_s}^{7/2} \log N  M^4 \left(\frac{1}{N}\right)^{9/5} {N_f}^2 \left(9 a^2+{r_h}^2\right) f(r) \log
   ^4({r_h})}{8 \pi ^{9/2} \left(6 a^2+{r_h}^2\right) \log \left({r_h}^6+9 a^2 {r_h}^4\right) \alpha _{\theta _1}^2 \alpha
   _{\theta _2}}+\frac{49 \pi ^{5/2} {r_h}^4 {f_{zz}}(r)}{8748 \sqrt{6} {g_s}^{7/2} M^2 \left(\frac{1}{N}\right)^{6/5}
   {N_f}^2 \left({r_h}^2-3 a^2\right)^2 \log ^4({r_h}) \alpha _{\theta _1}^2 \alpha _{\theta _2}}\nonumber\\
   & & -\frac{49 \pi ^{5/2}
   {r_h}^4 {f_{x^{10} x^{10}}}(r)}{5832 \sqrt{6} {g_s}^{7/2} M^2 \left(\frac{1}{N}\right)^{6/5} {N_f}^2 \left({r_h}^2-3
   a^2\right)^2 \log ^4({r_h}) \alpha _{\theta _1}^2 \alpha _{\theta _2}}-\frac{49 \pi ^{5/2} {r_h}^4 {f_{\theta_1z}}(r)}{4374
   \sqrt{6} {g_s}^{7/2} M^2 \left(\frac{1}{N}\right)^{6/5} {N_f}^2 \left({r_h}^2-3 a^2\right)^2 \log ^4({r_h}) \alpha
   _{\theta _1}^2 \alpha _{\theta _2}}\nonumber\\
   \end{eqnarray*}
   \begin{eqnarray*}
      & & -\frac{245 \pi ^{5/2} {r_h}^4 {f_{\theta_2y}}(r)}{17496 \sqrt{6} {g_s}^{7/2} M^2
   \left(\frac{1}{N}\right)^{6/5} {N_f}^2 \left({r_h}^2-3 a^2\right)^2 \log ^4({r_h}) \alpha _{\theta _1}^2 \alpha _{\theta
   _2}}-\frac{5 \sqrt{\frac{2}{3 \pi }} (r-{r_h}) {r_h}^3 \left(9 a^2+{r_h}^2\right) \left(108 b^2 {r_h}^2+1\right)^2
   {f_r}(r)}{81 \sqrt{{g_s}} \left(\frac{1}{N}\right)^{3/5} \left({r_h}^2-3 a^2\right)^2 \left(6 a^2+{r_h}^2\right) \log
   ^2({r_h}) \alpha _{\theta _1}^4 \alpha _{\theta _2}}\nonumber\\
   & & -\frac{4 \sqrt{\frac{2}{3 \pi }} (r-{r_h}) {r_h}^3 \left(9
   a^2+{r_h}^2\right) \left(108 b^2 {r_h}^2+1\right)^2 {f_{yz}}(r)}{27 \sqrt{{g_s}} \left(\frac{1}{N}\right)^{4/5}
   \left({r_h}^2-3 a^2\right)^2 \left(6 a^2+{r_h}^2\right) \log ^2({r_h}) \alpha _{\theta _1}^2 \alpha _{\theta _2}^3}+\frac{2
   \sqrt{\frac{2}{3 \pi }} (r-{r_h}) {r_h}^3 \left(9 a^2+{r_h}^2\right) \left(108 b^2 {r_h}^2+1\right)^2 {f_{yy}}(r)}{27
   \sqrt{{g_s}} \left(\frac{1}{N}\right)^{4/5} \left({r_h}^2-3 a^2\right)^2 \left(6 a^2+{r_h}^2\right) \log ^2({r_h})
   \alpha _{\theta _1}^2 \alpha _{\theta _2}^3}\nonumber\\
   & & +\frac{245 \pi  {r_h}^4 {f_{\theta_1y}}(r) \alpha _{\theta _1}^2}{5184 {g_s}^2 M
   \left(\frac{1}{N}\right)^{4/5} {N_f} \left({r_h}^2-3 a^2\right)^2 \log ^3({r_h}) \alpha _{\theta _2}^4}+\frac{245 \pi
   {r_h}^4 {f_{\theta_2z}}(r) \alpha _{\theta _1}^2}{5184 {g_s}^2 M \left(\frac{1}{N}\right)^{4/5} {N_f} \left({r_h}^2-3
   a^2\right)^2 \log ^3({r_h}) \alpha _{\theta _2}^4}
 \end{eqnarray*}
}

(xvi) \underline{${\rm EOM}_{xz}$}

{\scriptsize
\begin{eqnarray*}
& & \frac{8 \left(9 b^2+1\right)^4 M \left(\frac{1}{N}\right)^{3/4} (r-{r_h}) \beta  \log ({r_h}) \Sigma_1 b^{10}}{27 \left(3 b^2-1\right)^5
   \left(6 b^2+1\right)^4 \sqrt{{g_s}} \log N ^2 {N_f} \pi ^{3/2} {r_h} \alpha _{\theta _1}^2 \alpha _{\theta
   _2}^5}\nonumber\\
   & & -\frac{1024 \left(9 b^2+1\right)^3 \left(\frac{1}{N}\right)^{13/20} (r-{r_h})^2 \beta  \left(-{g_s} \log \left(\left(9
   b^2+1\right) {r_h}^6\right) {N_f}-4 {g_s} \log \left(\alpha _{\theta _1}\right) {N_f}-4 {g_s} \log \left(\alpha
   _{\theta _2}\right) {N_f}+4 {g_s} \log (4) {N_f}+8 \pi \right) b^{10}}{27 \sqrt{3} \left(3 b^2-1\right)^5 {g_s}^{13/4}
   \log N ^4 \left(6 {N_f} b^2+{N_f}\right)^3 \sqrt[4]{\pi } {r_h}^2 \alpha _{\theta _1}^2 \alpha _{\theta _2}^4}
   \nonumber\\
   & & +\frac{3
   \left(\frac{1}{N}\right)^{9/10} {f_{xx}}(r) \alpha _{\theta _1}^2}{\sqrt{{g_s}} \sqrt{\pi } \log \left({r_h}^6+9 a^2
   {r_h}^4\right)}-\frac{245 \pi  {r_h}^4 {f_{xy}}(r) \left(27 \sqrt{6} \alpha _{\theta _1}^3+10 \alpha _{\theta
   _2}^2\right)}{1679616 \sqrt{6} {g_s}^2 M \left(\frac{1}{N}\right)^{7/10} {N_f} \left({r_h}^2-3 a^2\right)^2 \log
   ^3({r_h}) \alpha _{\theta _1}^2 \alpha _{\theta _2}^3}\nonumber\\
   & & +\frac{343 \pi  {r_h}^4 {f_{\theta_1x}}(r) \left(81 \sqrt{2} \alpha _{\theta
   _1}^3+10 \sqrt{3} \alpha _{\theta _2}^2\right)}{1259712 \sqrt{2} {g_s}^2 M \left(\frac{1}{N}\right)^{7/10} {N_f}
   \left({r_h}^2-3 a^2\right)^2 \log ^3({r_h}) \alpha _{\theta _1}^2 \alpha _{\theta _2}^3}-\frac{1715 \pi  {r_h}^4
   {f_{\theta_2x}}(r) \left(81 \sqrt{2} \alpha _{\theta _1}^3+10 \sqrt{3} \alpha _{\theta _2}^2\right)}{3779136 \sqrt{2} {g_s}^2 M
   \left(\frac{1}{N}\right)^{7/10} {N_f} \left({r_h}^2-3 a^2\right)^2 \log ^3({r_h}) \alpha _{\theta _1}^2 \alpha _{\theta
   _2}^3}\nonumber\\
   & & +\frac{3479 \pi  {r_h}^4 {f_{xz}}(r) \left(81 \sqrt{2} \alpha _{\theta _1}^3+10 \sqrt{3} \alpha _{\theta
   _2}^2\right)}{15116544 \sqrt{2} {g_s}^2 M \left(\frac{1}{N}\right)^{7/10} {N_f} \left({r_h}^2-3 a^2\right)^2 \log
   ^3({r_h}) \alpha _{\theta _1}^2 \alpha _{\theta _2}^3}-\frac{4 {r_h} \left(9 a^2+{r_h}^2\right) f'(r)}{27 \sqrt{{g_s}}
   \sqrt{\frac{1}{N}} \sqrt{\pi } \left(6 a^2+{r_h}^2\right) \alpha _{\theta _1}^2 \alpha _{\theta _2}^2}\nonumber\\
   & & +\frac{4 {r_h} \left(9
   a^2+{r_h}^2\right) {f_{zz}}'(r)}{81 \sqrt{{g_s}} \sqrt{\frac{1}{N}} \sqrt{\pi } \left(6 a^2+{r_h}^2\right) \alpha
   _{\theta _1}^2 \alpha _{\theta _2}^2}-\frac{4 {r_h} \left(9 a^2+{r_h}^2\right) {f_{x^{10} x^{10}}}'(r)}{81 \sqrt{{g_s}}
   \sqrt{\frac{1}{N}} \sqrt{\pi } \left(6 a^2+{r_h}^2\right) \alpha _{\theta _1}^2 \alpha _{\theta _2}^2}\nonumber\\
   & & -\frac{2 {r_h} \left(9
   a^2+{r_h}^2\right) {f_{\theta_1z}}'(r)}{27 \sqrt{{g_s}} \sqrt{\frac{1}{N}} \sqrt{\pi } \left(6 a^2+{r_h}^2\right) \alpha
   _{\theta _1}^2 \alpha _{\theta _2}^2}+\frac{4 a^2 (r-{r_h}) \left(9 a^2+{r_h}^2\right) {f_{\theta_1x}}'(r)}{9 \sqrt{{g_s}}
   \sqrt{\frac{1}{N}} \sqrt{\pi } \left(-18 a^4+3 {r_h}^2 a^2+{r_h}^4\right) \alpha _{\theta _1}^2 \alpha _{\theta
   _2}^2}\nonumber\\
   & & +\frac{\sqrt{\frac{3}{2}} {g_s} M {N_f} {r_h} \left(9 a^2+{r_h}^2\right) \log ({r_h}) \alpha _{\theta _1}^2
   {f_{\theta_1y}}'(r)}{2 \sqrt[10]{\frac{1}{N}} \pi ^2 \left(6 a^2+{r_h}^2\right) \alpha _{\theta _2}^5}+\frac{\sqrt{\frac{3}{2}}
   {g_s} M {N_f} {r_h} \left(9 a^2+{r_h}^2\right) \log ({r_h}) \alpha _{\theta _1}^2 {f_{\theta_2z}}'(r)}{2
   \sqrt[10]{\frac{1}{N}} \pi ^2 \left(6 a^2+{r_h}^2\right) \alpha _{\theta _2}^5}\nonumber\\
   & & -\frac{3 \sqrt{\frac{3}{2}} a^2 {g_s} M
   \sqrt[10]{\frac{1}{N}} {N_f} (r-{r_h}) \left(9 a^2+{r_h}^2\right) \log ({r_h}) {f_{\theta_2x}}'(r)}{\pi ^2 \left(-18 a^4+3
   {r_h}^2 a^2+{r_h}^4\right) \alpha _{\theta _2}^3}-\frac{2 {r_h} \left(9 a^2+{r_h}^2\right) {f_{\theta_2y}}'(r)}{27
   \sqrt{{g_s}} \sqrt{\frac{1}{N}} \sqrt{\pi } \left(6 a^2+{r_h}^2\right) \alpha _{\theta _1}^2 \alpha _{\theta _2}^2}\nonumber\\
 & & -\frac{2
   {r_h} \left(9 a^2+{r_h}^2\right) {f_{xz}}'(r)}{81 \sqrt{{g_s}} \sqrt{\frac{1}{N}} \sqrt{\pi } \left(6
   a^2+{r_h}^2\right) \alpha _{\theta _1}^2 \alpha _{\theta _2}^2}+\frac{3 \left(\frac{1}{N}\right)^{9/10} {r_h} \left(9
   a^2+{r_h}^2\right) \alpha _{\theta _1}^2 {f_{xx}}'(r)}{2 \sqrt{{g_s}} \sqrt{\pi } \left(6 a^2+{r_h}^2\right)}\nonumber\\
   & & +\frac{3
   \sqrt{\frac{3}{2}} a^2 {g_s} M {N_f} (r-{r_h}) \left(9 a^2+{r_h}^2\right) \log ({r_h}) \alpha _{\theta _1}^2
   {f_{xy}}'(r)}{\sqrt[10]{\frac{1}{N}} \pi ^2 \left(-18 a^4+3 {r_h}^2 a^2+{r_h}^4\right) \alpha _{\theta _2}^5}+\frac{3
   \sqrt{\frac{3}{2}} a^2 {g_s} M {N_f} (r-{r_h}) \left(9 a^2+{r_h}^2\right) \log ({r_h}) \alpha _{\theta _1}^2
   {f_{yz}}'(r)}{\sqrt[10]{\frac{1}{N}} \pi ^2 \left(-18 a^4+3 {r_h}^2 a^2+{r_h}^4\right) \alpha _{\theta _2}^5}\nonumber\\
   & & +\frac{a^2
   {g_s}^{5/2} M^2 \left(\frac{1}{N}\right)^{9/10} {N_f}^2 (r-{r_h}) \left(9 a^2+{r_h}^2\right) \log ^2({r_h})
   \left(2187 \sqrt{2} \alpha _{\theta _1}^6+540 \sqrt{3} \alpha _{\theta _2}^2 \alpha _{\theta _1}^3+50 \sqrt{2} \alpha _{\theta
   _2}^4\right) {f_{yy}}'(r)}{192 \sqrt{2} \pi ^{7/2} \left(-18 a^4+3 {r_h}^2 a^2+{r_h}^4\right) \alpha _{\theta _2}^6}\nonumber\\
   & & +\frac{2
   {r_h} \left(9 a^2+{r_h}^2\right) {f_r}'(r)}{81 \sqrt{{g_s}} \sqrt{\frac{1}{N}} \sqrt{\pi } \left(6
   a^2+{r_h}^2\right) \alpha _{\theta _1}^2 \alpha _{\theta _2}^2}-\frac{2 {r_h} \left(9 a^2+{r_h}^2\right) {f_t}'(r)}{27
   \sqrt{{g_s}} \sqrt{\frac{1}{N}} \sqrt{\pi } \left(6 a^2+{r_h}^2\right) \alpha _{\theta _1}^2 \alpha _{\theta _2}^2}\nonumber\\
      \end{eqnarray*}
   \begin{eqnarray*}
& & -\frac{4
   (r-{r_h}) {r_h} \left(9 a^2+{r_h}^2\right) f''(r)}{27 \sqrt{{g_s}} \sqrt{\frac{1}{N}} \sqrt{\pi } \left(6
   a^2+{r_h}^2\right) \alpha _{\theta _1}^2 \alpha _{\theta _2}^2}+\frac{4 (r-{r_h}) {r_h} \left(9 a^2+{r_h}^2\right)
   {f_{zz}}''(r)}{81 \sqrt{{g_s}} \sqrt{\frac{1}{N}} \sqrt{\pi } \left(6 a^2+{r_h}^2\right) \alpha _{\theta _1}^2 \alpha
   _{\theta _2}^2}\nonumber\\
        & & -\frac{4 (r-{r_h}) {r_h} \left(9 a^2+{r_h}^2\right) {f_{x^{10} x^{10}}}''(r)}{81 \sqrt{{g_s}}
   \sqrt{\frac{1}{N}} \sqrt{\pi } \left(6 a^2+{r_h}^2\right) \alpha _{\theta _1}^2 \alpha _{\theta _2}^2}-\frac{2 (r-{r_h})
   {r_h} \left(9 a^2+{r_h}^2\right) {f_{\theta_1z}}''(r)}{27 \sqrt{{g_s}} \sqrt{\frac{1}{N}} \sqrt{\pi } \left(6
   a^2+{r_h}^2\right) \alpha _{\theta _1}^2 \alpha _{\theta _2}^2}\nonumber\\
   & & -\frac{7 {g_s} M \left(\frac{1}{N}\right)^{3/10} {N_f}
   (r-{r_h}) {r_h} \left(9 a^2+{r_h}^2\right) \log ({r_h}) \left(81 \sqrt{2} \alpha _{\theta _1}^3+10 \sqrt{3} \alpha
   _{\theta _2}^2\right) {f_{\theta_1x}}''(r)}{324 \sqrt{2} \pi ^2 \left(6 a^2+{r_h}^2\right) \alpha _{\theta _1}^2 \alpha _{\theta
   _2}^3}\nonumber\\
& &    +\frac{\sqrt{\frac{3}{2}} {g_s} M {N_f} (r-{r_h}) {r_h} \left(9 a^2+{r_h}^2\right) \log ({r_h}) \alpha
   _{\theta _1}^2 {f_{\theta_1y}}''(r)}{2 \sqrt[10]{\frac{1}{N}} \pi ^2 \left(6 a^2+{r_h}^2\right) \alpha _{\theta
   _2}^5}+\frac{\sqrt{\frac{3}{2}} {g_s} M {N_f} (r-{r_h}) {r_h} \left(9 a^2+{r_h}^2\right) \log ({r_h}) \alpha
   _{\theta _1}^2 {f_{\theta_2z}}''(r)}{2 \sqrt[10]{\frac{1}{N}} \pi ^2 \left(6 a^2+{r_h}^2\right) \alpha _{\theta _2}^5}\nonumber\\
   & & +\frac{7
   {g_s} M \left(\frac{1}{N}\right)^{3/10} {N_f} (r-{r_h}) {r_h} \left(9 a^2+{r_h}^2\right) \log ({r_h}) \left(81
   \sqrt{2} \alpha _{\theta _1}^3+10 \sqrt{3} \alpha _{\theta _2}^2\right) {f_{\theta_2x}}''(r)}{324 \sqrt{2} \pi ^2 \left(6
   a^2+{r_h}^2\right) \alpha _{\theta _1}^2 \alpha _{\theta _2}^3}-\frac{2 (r-{r_h}) {r_h} \left(9 a^2+{r_h}^2\right)
   {f_{\theta_2y}}''(r)}{27 \sqrt{{g_s}} \sqrt{\frac{1}{N}} \sqrt{\pi } \left(6 a^2+{r_h}^2\right) \alpha _{\theta _1}^2 \alpha
   _{\theta _2}^2}\nonumber\\
   & & -\frac{2 (r-{r_h}) {r_h} \left(9 a^2+{r_h}^2\right) {f_{xz}}''(r)}{81 \sqrt{{g_s}} \sqrt{\frac{1}{N}}
   \sqrt{\pi } \left(6 a^2+{r_h}^2\right) \alpha _{\theta _1}^2 \alpha _{\theta _2}^2}+\frac{3 \left(\frac{1}{N}\right)^{9/10}
   (r-{r_h}) {r_h} \left(9 a^2+{r_h}^2\right) \alpha _{\theta _1}^2 {f_{xx}}''(r)}{2 \sqrt{{g_s}} \sqrt{\pi } \left(6
   a^2+{r_h}^2\right)}\nonumber\\
   & & +\frac{{g_s} M \left(\frac{1}{N}\right)^{3/10} {N_f} (r-{r_h}) {r_h} \left(9
   a^2+{r_h}^2\right) \log ({r_h}) \left(27 \sqrt{3} \alpha _{\theta _1}^3+5 \sqrt{2} \alpha _{\theta _2}^2\right)
   {f_{xy}}''(r)}{144 \sqrt{3} \pi ^2 \left(6 a^2+{r_h}^2\right) \alpha _{\theta _1}^2 \alpha _{\theta _2}^3}\nonumber\\
& & -\frac{{g_s} M
   \left(\frac{1}{N}\right)^{3/10} {N_f} (r-{r_h}) {r_h} \left(9 a^2+{r_h}^2\right) \log ({r_h}) \left(27 \sqrt{3}
   \alpha _{\theta _1}^3+5 \sqrt{2} \alpha _{\theta _2}^2\right) {f_{yz}}''(r)}{144 \sqrt{3} \pi ^2 \left(6 a^2+{r_h}^2\right)
   \alpha _{\theta _1}^2 \alpha _{\theta _2}^3}\nonumber\\
   & & +\frac{{g_s}^{5/2} M^2 \left(\frac{1}{N}\right)^{11/10} {N_f}^2 (r-{r_h})
   {r_h} \left(9 a^2+{r_h}^2\right) \log ^2({r_h}) \left(2187 \alpha _{\theta _1}^6+270 \sqrt{6} \alpha _{\theta _2}^2 \alpha
   _{\theta _1}^3+50 \alpha _{\theta _2}^4\right) {f_{yy}}''(r)}{6912 \pi ^{7/2} \left(6 a^2+{r_h}^2\right) \alpha _{\theta _1}^2
   \alpha _{\theta _2}^4}\nonumber\\
   & & -\frac{4 (r-{r_h}) {r_h} \left(9 a^2+{r_h}^2\right) {f_t}''(r)}{81 \sqrt{{g_s}}
   \sqrt{\frac{1}{N}} \sqrt{\pi } \left(6 a^2+{r_h}^2\right) \alpha _{\theta _1}^2 \alpha _{\theta _2}^2}\nonumber\\
   & & -\frac{{g_s}^{7/2} M^4
   \left(\frac{1}{N}\right)^{3/2} {N_f}^2 {f_t}(r) \log ^4({r_h}) \left(16 \left(108 a^6-9 {r_h}^4
   a^2-{r_h}^6\right)-\left(2 {r_h}^6+27 a^2 {r_h}^4+117 a^4 {r_h}^2\right) \log \left({r_h}^6+9 a^2
   {r_h}^4\right)\right)}{32 \pi ^{9/2} \left(3 a^2-{r_h}^2\right) \left(6 a^2+{r_h}^2\right)^2 \log \left({r_h}^6+9 a^2
   {r_h}^4\right) \alpha _{\theta _1}^2 \alpha _{\theta _2}^2}\nonumber\\
   & & +\frac{3 {g_s}^{7/2} \log N  M^4 \left(\frac{1}{N}\right)^{3/2}
   {N_f}^2 \left(9 a^2+{r_h}^2\right) f(r) \log ^4({r_h})}{8 \pi ^{9/2} \left(6 a^2+{r_h}^2\right) \log
   \left({r_h}^6+9 a^2 {r_h}^4\right) \alpha _{\theta _1}^2 \alpha _{\theta _2}^2}-\frac{49 \pi ^{5/2} {r_h}^4
   {f_{zz}}(r)}{78732 {g_s}^{7/2} M^2 \left(\frac{1}{N}\right)^{3/2} {N_f}^2 \left({r_h}^2-3 a^2\right)^2 \log
   ^4({r_h}) \alpha _{\theta _1}^2 \alpha _{\theta _2}^2}\nonumber\\
   & & +\frac{49 \pi ^{5/2} {r_h}^4 {f_{x^{10} x^{10}}}(r)}{52488 {g_s}^{7/2} M^2
   \left(\frac{1}{N}\right)^{3/2} {N_f}^2 \left({r_h}^2-3 a^2\right)^2 \log ^4({r_h}) \alpha _{\theta _1}^2 \alpha _{\theta
   _2}^2}+\frac{49 \pi ^{5/2} {r_h}^4 {f_{\theta_1z}}(r)}{39366 {g_s}^{7/2} M^2 \left(\frac{1}{N}\right)^{3/2} {N_f}^2
   \left({r_h}^2-3 a^2\right)^2 \log ^4({r_h}) \alpha _{\theta _1}^2 \alpha _{\theta _2}^2}\nonumber\\
   & & +\frac{245 \pi ^{5/2} {r_h}^4
   {f_{\theta_2y}}(r)}{157464 {g_s}^{7/2} M^2 \left(\frac{1}{N}\right)^{3/2} {N_f}^2 \left({r_h}^2-3 a^2\right)^2 \log
   ^4({r_h}) \alpha _{\theta _1}^2 \alpha _{\theta _2}^2}+\frac{10 (r-{r_h}) {r_h}^3 \left(9 a^2+{r_h}^2\right) \left(108
   b^2 {r_h}^2+1\right)^2 {f_r}(r)}{729 \sqrt{{g_s}} \left(\frac{1}{N}\right)^{9/10} \sqrt{\pi } \left({r_h}^2-3
   a^2\right)^2 \left(6 a^2+{r_h}^2\right) \log ^2({r_h}) \alpha _{\theta _1}^4 \alpha _{\theta _2}^2}\nonumber\\
   & & +\frac{8 (r-{r_h})
   {r_h}^3 \left(9 a^2+{r_h}^2\right) \left(108 b^2 {r_h}^2+1\right)^2 {f_{yz}}(r)}{243 \sqrt{{g_s}}
   \left(\frac{1}{N}\right)^{11/10} \sqrt{\pi } \left({r_h}^2-3 a^2\right)^2 \left(6 a^2+{r_h}^2\right) \log ^2({r_h}) \alpha
   _{\theta _1}^2 \alpha _{\theta _2}^4}-\frac{4 (r-{r_h}) {r_h}^3 \left(9 a^2+{r_h}^2\right) \left(108 b^2
   {r_h}^2+1\right)^2 {f_{yy}}(r)}{243 \sqrt{{g_s}} \left(\frac{1}{N}\right)^{11/10} \sqrt{\pi } \left({r_h}^2-3
   a^2\right)^2 \left(6 a^2+{r_h}^2\right) \log ^2({r_h}) \alpha _{\theta _1}^2 \alpha _{\theta _2}^4}\nonumber\\
   & & -\frac{245 \pi  {r_h}^4
   {f_{\theta_1y}}(r) \alpha _{\theta _1}^2}{7776 \sqrt{6} {g_s}^2 M \left(\frac{1}{N}\right)^{11/10} {N_f} \left({r_h}^2-3
   a^2\right)^2 \log ^3({r_h}) \alpha _{\theta _2}^5}-\frac{245 \pi  {r_h}^4 {f_{\theta_2z}}(r) \alpha _{\theta _1}^2}{7776 \sqrt{6}
   {g_s}^2 M \left(\frac{1}{N}\right)^{11/10} {N_f} \left({r_h}^2-3 a^2\right)^2 \log ^3({r_h}) \alpha _{\theta _2}^5} = 0
\end{eqnarray*}
}
(xvii) \underline{${\rm EOM}_{yy}$}\\
{\scriptsize
\begin{eqnarray*}
& & -\frac{128 \sqrt{2} \left(9 b^2+1\right)^4 \left(9 b^4-5 b^2+1\right) (r-{r_h})^3 \beta  \Sigma_1 b^{12}}{27 \left(1-3 b^2\right)^8
   \left(6 b^2+1\right)^4 {g_s}^{9/4} \log N ^3 N {N_f}^2 \sqrt[4]{\pi } {r_h}^3 \alpha _{\theta _1}^4 \alpha _{\theta
   _2}^2}-\frac{6 \left(9 b^2+1\right)^4 M \left(\frac{1}{N}\right)^{7/4} (r-{r_h}) \beta  \log ({r_h}) \Sigma_1 b^{10}}{\left(3
   b^2-1\right)^5 \left(6 b^2+1\right)^4 \sqrt{{g_s}} \log N ^2 {N_f} \pi ^{3/2} {r_h} \alpha _{\theta _2}^3}\nonumber\\
   & & -\frac{49 \pi
   ^{5/2} {r_h}^4 {f_{\theta_1y}}(r) \alpha _{\theta _2}^2}{432 {g_s}^{7/2} M^2 \left(\frac{1}{N}\right)^{3/10} {N_f}^2
   \left({r_h}^2-3 a^2\right)^2 \log ^4({r_h}) \alpha _{\theta _1}^2}-\frac{2401 \left(\frac{1}{N}\right)^{17/10} \pi ^{5/2}
   {r_h}^4 {f_{xx}}(r) \alpha _{\theta _2}^6 \left(81 \sqrt{2} \alpha _{\theta _1}^3+10 \sqrt{3} \alpha _{\theta
   _2}^2\right){}^2}{1119744 {g_s}^{7/2} M^2 {N_f}^2 \left({r_h}^2-3 a^2\right)^2 \log ^4({r_h}) \alpha _{\theta
   _1}^4}\nonumber\\
   & & +\frac{81 {g_s}^{7/2} M^4 \left(\frac{1}{N}\right)^{5/2} {N_f}^2 {f_t}(r) \log ^4({r_h}) \left(16 \left(108 a^6-9
   {r_h}^4 a^2-{r_h}^6\right)-\left(2 {r_h}^6+27 a^2 {r_h}^4+117 a^4 {r_h}^2\right) \log \left({r_h}^6+9 a^2
   {r_h}^4\right)\right)}{128 \pi ^{9/2} \left(3 a^2-{r_h}^2\right) \left(6 a^2+{r_h}^2\right)^2 \log \left({r_h}^6+9 a^2
   {r_h}^4\right)}\nonumber\\
   & & +\frac{343 \sqrt[10]{\frac{1}{N}} \pi ^{5/2} {r_h}^4 {f_{\theta_1x}}(r) \alpha _{\theta _2}^4 \left(27 \sqrt{6}
   \alpha _{\theta _1}^3+10 \alpha _{\theta _2}^2\right)}{69984 {g_s}^{7/2} M^2 {N_f}^2 \left({r_h}^2-3 a^2\right)^2 \log
   ^4({r_h}) \alpha _{\theta _1}^6}\nonumber\\
   & & -\frac{343 \pi ^{5/2} {r_h}^4 {f_{xz}}(r) \alpha _{\theta _2}^2 \left(81 \sqrt{2} \alpha
   _{\theta _1}^3+10 \sqrt{3} \alpha _{\theta _2}^2\right)}{34992 \sqrt{3} {g_s}^{7/2} M^2 \sqrt[10]{\frac{1}{N}} {N_f}^2
   \left({r_h}^2-3 a^2\right)^2 \log ^4({r_h}) \alpha _{\theta _1}^4}+\frac{343 \pi ^{5/2} {r_h}^4 {f_{xy}}(r) \alpha
   _{\theta _2}^2 \left(81 \sqrt{2} \alpha _{\theta _1}^3+10 \sqrt{3} \alpha _{\theta _2}^2\right)}{34992 \sqrt{3} {g_s}^{7/2} M^2
   \sqrt[10]{\frac{1}{N}} {N_f}^2 \left({r_h}^2-3 a^2\right)^2 \log ^4({r_h}) \alpha _{\theta _1}^4}\nonumber\\
   & & +\frac{1715
   \left(\frac{1}{N}\right)^{3/10} \pi  {r_h}^4 {f_{\theta_2x}}(r) \left(81 \sqrt{2} \alpha _{\theta _1}^3+10 \sqrt{3} \alpha _{\theta
   _2}^2\right)}{186624 \sqrt{2} {g_s}^2 M {N_f} \left({r_h}^2-3 a^2\right)^2 \log ^3({r_h}) \alpha _{\theta _2}}+\frac{3
   \sqrt{\frac{1}{N}} {r_h} \left(9 a^2+{r_h}^2\right) f'(r)}{\sqrt{{g_s}} \sqrt{\pi } \left(6
   a^2+{r_h}^2\right)}\nonumber\\
   & & -\frac{\sqrt{\frac{1}{N}} {r_h} \left(9 a^2+{r_h}^2\right) {f_{zz}}'(r)}{\sqrt{{g_s}} \sqrt{\pi
   } \left(6 a^2+{r_h}^2\right)}+\frac{\sqrt{\frac{1}{N}} {r_h} \left(9 a^2+{r_h}^2\right) {f_{x^{10} x^{10}}}'(r)}{\sqrt{{g_s}}
   \sqrt{\pi } \left(6 a^2+{r_h}^2\right)}\nonumber\\
   & & +\frac{3 \sqrt{\frac{1}{N}} {r_h} \left(9 a^2+{r_h}^2\right) {f_{\theta_1z}}'(r)}{2
   \sqrt{{g_s}} \sqrt{\pi } \left(6 a^2+{r_h}^2\right)}-\frac{3 a^2 \sqrt{\frac{1}{N}} (r-{r_h}) \left(9
   a^2+{r_h}^2\right) {f_{\theta_1x}}'(r)}{\sqrt{{g_s}} \sqrt{\pi } \left(-18 a^4+3 {r_h}^2 a^2+{r_h}^4\right)}
   \nonumber\\
& &    -\frac{12 a^2
   \sqrt{\frac{1}{N}} (r-{r_h}) \left(9 a^2+{r_h}^2\right) {f_{\theta_1y}}'(r)}{\sqrt{{g_s}} \sqrt{\pi } \left(-18 a^4+3
   {r_h}^2 a^2+{r_h}^4\right)}-\frac{81 \sqrt{\frac{3}{2}} {g_s} M \left(\frac{1}{N}\right)^{9/10} {N_f} {r_h} \left(9
   a^2+{r_h}^2\right) \log ({r_h}) \alpha _{\theta _1}^4 {f_{\theta_2z}}'(r)}{8 \pi ^2 \left(6 a^2+{r_h}^2\right) \alpha _{\theta
   _2}^3}\nonumber\\
   & & +\frac{81 \sqrt{\frac{3}{2}} a^2 {g_s} M \left(\frac{1}{N}\right)^{11/10} {N_f} (r-{r_h}) \left(9
   a^2+{r_h}^2\right) \log ({r_h}) \alpha _{\theta _1}^2 {f_{\theta_2x}}'(r)}{4 \pi ^2 \left({r_h}^2-3 a^2\right) \left(6
   a^2+{r_h}^2\right) \alpha _{\theta _2}}+\frac{3 \sqrt{\frac{1}{N}} {r_h} \left(9 a^2+{r_h}^2\right) {f_{\theta_2y}}'(r)}{2
   \sqrt{{g_s}} \sqrt{\pi } \left(6 a^2+{r_h}^2\right)}\nonumber\\
   & & +\frac{3 \sqrt{\frac{1}{N}} {r_h} \left(9 a^2+{r_h}^2\right)
   {f_{xz}}'(r)}{2 \sqrt{{g_s}} \sqrt{\pi } \left(6 a^2+{r_h}^2\right)}-\frac{243 \left(\frac{1}{N}\right)^{19/10} {r_h}
   \left(9 a^2+{r_h}^2\right) \alpha _{\theta _1}^4 \alpha _{\theta _2}^2 {f_{xx}}'(r)}{8 \sqrt{{g_s}} \sqrt{\pi } \left(6
   a^2+{r_h}^2\right)}\nonumber\\
   & & -\frac{81 \sqrt{\frac{3}{2}} a^2 {g_s} M \left(\frac{1}{N}\right)^{9/10} {N_f} (r-{r_h}) \left(9
   a^2+{r_h}^2\right) \log ({r_h}) \alpha _{\theta _1}^4 {f_{xy}}'(r)}{4 \pi ^2 \left(-18 a^4+3 {r_h}^2
   a^2+{r_h}^4\right) \alpha _{\theta _2}^3}+\frac{12 a^2 \sqrt{\frac{1}{N}} (r-{r_h}) \left(9 a^2+{r_h}^2\right)
   {f_{yz}}'(r)}{\sqrt{{g_s}} \sqrt{\pi } \left(-18 a^4+3 {r_h}^2 a^2+{r_h}^4\right)}\nonumber\\
   & & -\frac{\sqrt{\frac{1}{N}} {r_h}
   \left(9 a^2+{r_h}^2\right) {f_{yy}}'(r)}{\sqrt{{g_s}} \sqrt{\pi } \left(6 a^2+{r_h}^2\right)}-\frac{\sqrt{\frac{1}{N}}
   {r_h} \left(9 a^2+{r_h}^2\right) {f_r}'(r)}{2 \sqrt{{g_s}} \sqrt{\pi } \left(6 a^2+{r_h}^2\right)}\nonumber\\
   & & +\frac{3
   \sqrt{\frac{1}{N}} {r_h} \left(9 a^2+{r_h}^2\right) {f_t}'(r)}{2 \sqrt{{g_s}} \sqrt{\pi } \left(6
   a^2+{r_h}^2\right)}+\frac{3 \sqrt{\frac{1}{N}} (r-{r_h}) {r_h} \left(9 a^2+{r_h}^2\right) f''(r)}{\sqrt{{g_s}}
   \sqrt{\pi } \left(6 a^2+{r_h}^2\right)}\nonumber\\
   & & -\frac{\sqrt{\frac{1}{N}} (r-{r_h}) {r_h} \left(9 a^2+{r_h}^2\right)
   {f_{zz}}''(r)}{\sqrt{{g_s}} \sqrt{\pi } \left(6 a^2+{r_h}^2\right)}+\frac{\sqrt{\frac{1}{N}} (r-{r_h}) {r_h}
   \left(9 a^2+{r_h}^2\right) {f_{x^{10} x^{10}}}''(r)}{\sqrt{{g_s}} \sqrt{\pi } \left(6 a^2+{r_h}^2\right)}\nonumber\\
   & & +\frac{3
   \sqrt{\frac{1}{N}} (r-{r_h}) {r_h} \left(9 a^2+{r_h}^2\right) {f_{\theta_1z}}''(r)}{2 \sqrt{{g_s}} \sqrt{\pi } \left(6
   a^2+{r_h}^2\right)}+\frac{7 {g_s} M \left(\frac{1}{N}\right)^{13/10} {N_f} (r-{r_h}) {r_h} \left(9
   a^2+{r_h}^2\right) \log ({r_h}) \left(81 \sqrt{2} \alpha _{\theta _1}^3+10 \sqrt{3} \alpha _{\theta _2}^2\right)
   {f_{\theta_1x}}''(r)}{16 \sqrt{2} \pi ^2 \left(6 a^2+{r_h}^2\right) \alpha _{\theta _2}}\nonumber\\
   & & -\frac{81 \sqrt{\frac{3}{2}} {g_s} M
   \left(\frac{1}{N}\right)^{9/10} {N_f} (r-{r_h}) {r_h} \left(9 a^2+{r_h}^2\right) \log ({r_h}) \alpha _{\theta _1}^4
   {f_{\theta_1y}}''(r)}{8 \pi ^2 \left(6 a^2+{r_h}^2\right) \alpha _{\theta _2}^3}\nonumber\\
   & & -\frac{81 \sqrt{\frac{3}{2}} {g_s} M
   \left(\frac{1}{N}\right)^{9/10} {N_f} (r-{r_h}) {r_h} \left(9 a^2+{r_h}^2\right) \log ({r_h}) \alpha _{\theta _1}^4
   {f_{\theta_2z}}''(r)}{8 \pi ^2 \left(6 a^2+{r_h}^2\right) \alpha _{\theta _2}^3}\nonumber\\
   & & -\frac{7 {g_s} M \left(\frac{1}{N}\right)^{13/10}
   {N_f} (r-{r_h}) {r_h} \left(9 a^2+{r_h}^2\right) \log ({r_h}) \left(81 \sqrt{2} \alpha _{\theta _1}^3+10 \sqrt{3}
   \alpha _{\theta _2}^2\right) {f_{\theta_2x}}''(r)}{16 \sqrt{2} \pi ^2 \left(6 a^2+{r_h}^2\right) \alpha _{\theta _2}}
      \end{eqnarray*}

   \begin{eqnarray*}
   & & +\frac{3
   \sqrt{\frac{1}{N}} (r-{r_h}) {r_h} \left(9 a^2+{r_h}^2\right) {f_{\theta_2y}}''(r)}{2 \sqrt{{g_s}} \sqrt{\pi } \left(6
   a^2+{r_h}^2\right)}+\frac{3 \sqrt{\frac{1}{N}} (r-{r_h}) {r_h} \left(9 a^2+{r_h}^2\right) {f_{xz}}''(r)}{2
   \sqrt{{g_s}} \sqrt{\pi } \left(6 a^2+{r_h}^2\right)}\nonumber\\
   & & -\frac{243 \left(\frac{1}{N}\right)^{19/10} (r-{r_h}) {r_h} \left(9
   a^2+{r_h}^2\right) \alpha _{\theta _1}^4 \alpha _{\theta _2}^2 {f_{xx}}''(r)}{8 \sqrt{{g_s}} \sqrt{\pi } \left(6
   a^2+{r_h}^2\right)}\nonumber\\
   & & -\frac{3 \sqrt{3} {g_s} M \left(\frac{1}{N}\right)^{13/10} {N_f} (r-{r_h}) {r_h} \left(9
   a^2+{r_h}^2\right) \log ({r_h}) \left(27 \sqrt{3} \alpha _{\theta _1}^3+5 \sqrt{2} \alpha _{\theta _2}^2\right)
   {f_{xy}}''(r)}{64 \pi ^2 \left(6 a^2+{r_h}^2\right) \alpha _{\theta _2}}\nonumber\\
   & & +\frac{3 \sqrt{3} {g_s} M
   \left(\frac{1}{N}\right)^{13/10} {N_f} (r-{r_h}) {r_h} \left(9 a^2+{r_h}^2\right) \log ({r_h}) \left(27 \sqrt{3}
   \alpha _{\theta _1}^3+5 \sqrt{2} \alpha _{\theta _2}^2\right) {f_{yz}}''(r)}{64 \pi ^2 \left(6 a^2+{r_h}^2\right) \alpha
   _{\theta _2}}\nonumber\\
   & & -\frac{\sqrt{\frac{1}{N}} (r-{r_h}) {r_h} \left(9 a^2+{r_h}^2\right) {f_{yy}}''(r)}{\sqrt{{g_s}}
   \sqrt{\pi } \left(6 a^2+{r_h}^2\right)}+\frac{\sqrt{\frac{1}{N}} (r-{r_h}) {r_h} \left(9 a^2+{r_h}^2\right)
   {f_t}''(r)}{\sqrt{{g_s}} \sqrt{\pi } \left(6 a^2+{r_h}^2\right)}\nonumber\\
   & & -\frac{243 {g_s}^{7/2} \log N  M^4
   \left(\frac{1}{N}\right)^{5/2} {N_f}^2 \left(9 a^2+{r_h}^2\right) f(r) \log ^4({r_h})}{32 \pi ^{9/2} \left(6
   a^2+{r_h}^2\right) \log \left({r_h}^6+9 a^2 {r_h}^4\right)}-\frac{5 \sqrt[10]{\frac{1}{N}} (r-{r_h}) {r_h}^3
   \left(9 a^2+{r_h}^2\right) \left(108 b^2 {r_h}^2+1\right)^2 {f_r}(r)}{18 \sqrt{{g_s}} \sqrt{\pi } \left({r_h}^2-3
   a^2\right)^2 \left(6 a^2+{r_h}^2\right) \log ^2({r_h}) \alpha _{\theta _1}^2}\nonumber\\
   & & +\frac{245 \pi  {r_h}^4 {f_{\theta_2z}}(r) \alpha
   _{\theta _1}^4}{384 \sqrt{6} {g_s}^2 M \sqrt[10]{\frac{1}{N}} {N_f} \left({r_h}^2-3 a^2\right)^2 \log ^3({r_h}) \alpha
   _{\theta _2}^3}-\frac{49 \pi ^{5/2} {r_h}^4 {f_{zz}}(r)}{486 {g_s}^{7/2} M^2 \sqrt{\frac{1}{N}} {N_f}^2
   \left({r_h}^2-3 a^2\right)^2 \log ^4({r_h})}\nonumber\\
   & & -\frac{49 \pi ^{5/2} {r_h}^4 {f_{x^{10} x^{10}}}(r)}{2592 {g_s}^{7/2} M^2
   \sqrt{\frac{1}{N}} {N_f}^2 \left({r_h}^2-3 a^2\right)^2 \log ^4({r_h})}-\frac{49 \pi ^{5/2} {r_h}^4
   {f_{\theta_1z}}(r)}{1944 {g_s}^{7/2} M^2 \sqrt{\frac{1}{N}} {N_f}^2 \left({r_h}^2-3 a^2\right)^2 \log ^4({r_h})}
   \nonumber\\
   & & -\frac{245
   \pi ^{5/2} {r_h}^4 {f_{\theta_2y}}(r)}{7776 {g_s}^{7/2} M^2 \sqrt{\frac{1}{N}} {N_f}^2 \left({r_h}^2-3 a^2\right)^2 \log
   ^4({r_h})}+\frac{49 \pi ^{5/2} {r_h}^4 {f_{yz}}(r)}{216 {g_s}^{7/2} M^2 \sqrt{\frac{1}{N}} {N_f}^2
   \left({r_h}^2-3 a^2\right)^2 \log ^4({r_h})}\nonumber\\
   & & -\frac{49 \pi ^{5/2} {r_h}^4 {f_{yy}}(r)}{432 {g_s}^{7/2} M^2
   \sqrt{\frac{1}{N}} {N_f}^2 \left({r_h}^2-3 a^2\right)^2 \log ^4({r_h})} = 0.
\end{eqnarray*}
}

(xviii) \underline{${\rm EOM}_{yz}$}

{\scriptsize
\begin{eqnarray*}
& &  -\frac{512 \left(9 b^2+1\right)^3 \left(\frac{1}{N}\right)^{21/20} (r-{r_h})^3 \beta  \Sigma_1 b^{12}}{729 \left(1-3 b^2\right)^6 \left(6
   b^2+1\right)^3 {g_s}^2 \log N ^3 {N_f}^2 {r_h}^3 \alpha _{\theta _1}^4 \alpha _{\theta _2}}+\frac{2 \sqrt{\frac{2}{3}}
   \left(9 b^2+1\right)^4 M \left(\frac{1}{N}\right)^{29/20} (r-{r_h}) \beta  \log ({r_h}) \Sigma_1 b^{10}}{\left(3 b^2-1\right)^5 \left(6
   b^2+1\right)^4 \sqrt{{g_s}} \log N ^2 {N_f} \pi ^{3/2} {r_h} \alpha _{\theta _2}^4}\nonumber\\
& &  -\frac{27
   \left(\frac{1}{N}\right)^{8/5} \sqrt{\frac{3}{2 \pi }} {f_{xx}}(r) \alpha _{\theta _1}^4 \alpha _{\theta _2}}{4 \sqrt{{g_s}}
   \log N }-\frac{245 \pi  {r_h}^4 {f_{xy}}(r) \left(27 \sqrt{6} \alpha _{\theta _1}^3+10 \alpha _{\theta _2}^2\right)}{746496
   {g_s}^2 M {N_f} \left({r_h}^2-3 a^2\right)^2 \log ^3({r_h}) \alpha _{\theta _2}^2}\nonumber\\
& &  +\frac{343 \pi  {r_h}^4
   {f_{\theta_1x}}(r) \left(81 \sqrt{2} \alpha _{\theta _1}^3+10 \sqrt{3} \alpha _{\theta _2}^2\right)}{186624 \sqrt{3} {g_s}^2 M {N_f}
   \left({r_h}^2-3 a^2\right)^2 \log ^3({r_h}) \alpha _{\theta _2}^2}-\frac{1715 \pi  {r_h}^4 {f_{\theta_2x}}(r) \left(81 \sqrt{2}
   \alpha _{\theta _1}^3+10 \sqrt{3} \alpha _{\theta _2}^2\right)}{559872 \sqrt{3} {g_s}^2 M {N_f} \left({r_h}^2-3
   a^2\right)^2 \log ^3({r_h}) \alpha _{\theta _2}^2}\nonumber\\
& & +\frac{3479 \pi  {r_h}^4 {f_{xz}}(r) \left(81 \sqrt{2} \alpha _{\theta
   _1}^3+10 \sqrt{3} \alpha _{\theta _2}^2\right)}{2239488 \sqrt{3} {g_s}^2 M {N_f} \left({r_h}^2-3 a^2\right)^2 \log
   ^3({r_h}) \alpha _{\theta _2}^2}-\frac{\sqrt[5]{\frac{1}{N}} \sqrt{\frac{2}{3 \pi }} {r_h} \left(9 a^2+{r_h}^2\right)
   f'(r)}{\sqrt{{g_s}} \left(6 a^2+{r_h}^2\right) \alpha _{\theta _2}}\nonumber\\
& &  +\frac{\sqrt[5]{\frac{1}{N}} \sqrt{\frac{2}{3 \pi }}
   {r_h} \left(9 a^2+{r_h}^2\right) {f_{zz}}'(r)}{3 \sqrt{{g_s}} \left(6 a^2+{r_h}^2\right) \alpha _{\theta
   _2}}-\frac{\sqrt[5]{\frac{1}{N}} \sqrt{\frac{2}{3 \pi }} {r_h} \left(9 a^2+{r_h}^2\right) {f_{x^{10} x^{10}}}'(r)}{3 \sqrt{{g_s}}
   \left(6 a^2+{r_h}^2\right) \alpha _{\theta _2}}\nonumber\\
   & &  -\frac{\sqrt[5]{\frac{1}{N}} {r_h} \left(9 a^2+{r_h}^2\right)
   {f_{\theta_1z}}'(r)}{\sqrt{{g_s}} \sqrt{6 \pi } \left(6 a^2+{r_h}^2\right) \alpha _{\theta _2}}+\frac{a^2 \sqrt[5]{\frac{1}{N}}
   \sqrt{\frac{2}{3 \pi }} (r-{r_h}) \left(9 a^2+{r_h}^2\right) {f_{\theta_1x}}'(r)}{\sqrt{{g_s}} \left({r_h}^2-3 a^2\right)
   \left(6 a^2+{r_h}^2\right) \alpha _{\theta _2}}\nonumber\\
& &  +\frac{2 a^2 \sqrt[5]{\frac{1}{N}} \sqrt{\frac{2}{3 \pi }} (r-{r_h}) \left(9
   a^2+{r_h}^2\right) {f_{\theta_1y}}'(r)}{\sqrt{{g_s}} \left({r_h}^2-3 a^2\right) \left(6 a^2+{r_h}^2\right) \alpha _{\theta
   _2}}+\frac{27 {g_s} M \left(\frac{1}{N}\right)^{3/5} {N_f} {r_h} \left(9 a^2+{r_h}^2\right) \log ({r_h}) \alpha
   _{\theta _1}^4 {f_{\theta_2z}}'(r)}{8 \pi ^2 \left(6 a^2+{r_h}^2\right) \alpha _{\theta _2}^4}\nonumber\\
   \end{eqnarray*}
   \begin{eqnarray*}
& &  -\frac{27 a^2 {g_s} M
   \left(\frac{1}{N}\right)^{4/5} {N_f} (r-{r_h}) \left(9 a^2+{r_h}^2\right) \log ({r_h}) \alpha _{\theta _1}^2
   {f_{\theta_2x}}'(r)}{4 \pi ^2 \left(-18 a^4+3 {r_h}^2 a^2+{r_h}^4\right) \alpha _{\theta _2}^2}-\frac{\sqrt[5]{\frac{1}{N}}
   {r_h} \left(9 a^2+{r_h}^2\right) {f_{\theta_2y}}'(r)}{\sqrt{{g_s}} \sqrt{6 \pi } \left(6 a^2+{r_h}^2\right) \alpha _{\theta
   _2}}\nonumber\\
& &  -\frac{\sqrt[5]{\frac{1}{N}} {r_h} \left(9 a^2+{r_h}^2\right) {f_{xz}}'(r)}{\sqrt{{g_s}} \sqrt{6 \pi } \left(6
   a^2+{r_h}^2\right) \alpha _{\theta _2}}+\frac{27 \left(\frac{1}{N}\right)^{8/5} \sqrt{\frac{3}{2 \pi }} {r_h} \left(9
   a^2+{r_h}^2\right) \alpha _{\theta _1}^4 \alpha _{\theta _2} {f_{xx}}'(r)}{4 \sqrt{{g_s}} \left(6
   a^2+{r_h}^2\right)}\nonumber\\
& &  +\frac{27 a^2 {g_s} M \left(\frac{1}{N}\right)^{3/5} {N_f} (r-{r_h}) \left(9 a^2+{r_h}^2\right)
   \log ({r_h}) \alpha _{\theta _1}^4 {f_{xy}}'(r)}{4 \pi ^2 \left(-18 a^4+3 {r_h}^2 a^2+{r_h}^4\right) \alpha _{\theta
   _2}^4}+\frac{\sqrt[5]{\frac{1}{N}} \sqrt{\frac{2}{3 \pi }} {r_h} \left(9 a^2+{r_h}^2\right) {f_{yz}}'(r)}{3 \sqrt{{g_s}}
   \left(6 a^2+{r_h}^2\right) \alpha _{\theta _2}}\nonumber\\
& &  +\frac{27 a^2 {g_s} M \left(\frac{1}{N}\right)^{3/5} {N_f} (r-{r_h})
   \left(9 a^2+{r_h}^2\right) \log ({r_h}) \alpha _{\theta _1}^4 {f_{yy}}'(r)}{2 \pi ^2 \left(-18 a^4+3 {r_h}^2
   a^2+{r_h}^4\right) \alpha _{\theta _2}^4}+\frac{\sqrt[5]{\frac{1}{N}} {r_h} \left(9 a^2+{r_h}^2\right) {f_r}'(r)}{3
   \sqrt{{g_s}} \sqrt{6 \pi } \left(6 a^2+{r_h}^2\right) \alpha _{\theta _2}}\nonumber\\
& &  -\frac{\sqrt[5]{\frac{1}{N}} {r_h} \left(9
   a^2+{r_h}^2\right) {f_t}'(r)}{\sqrt{{g_s}} \sqrt{6 \pi } \left(6 a^2+{r_h}^2\right) \alpha _{\theta
   _2}}-\frac{\sqrt[5]{\frac{1}{N}} \sqrt{\frac{2}{3 \pi }} (r-{r_h}) {r_h} \left(9 a^2+{r_h}^2\right)
   f''(r)}{\sqrt{{g_s}} \left(6 a^2+{r_h}^2\right) \alpha _{\theta _2}}\nonumber\\
& &  +\frac{\sqrt[5]{\frac{1}{N}} \sqrt{\frac{2}{3 \pi }}
   (r-{r_h}) {r_h} \left(9 a^2+{r_h}^2\right) {f_{zz}}''(r)}{3 \sqrt{{g_s}} \left(6 a^2+{r_h}^2\right) \alpha
   _{\theta _2}}-\frac{\sqrt[5]{\frac{1}{N}} \sqrt{\frac{2}{3 \pi }} (r-{r_h}) {r_h} \left(9 a^2+{r_h}^2\right)
   {f_{x^{10} x^{10}}}''(r)}{3 \sqrt{{g_s}} \left(6 a^2+{r_h}^2\right) \alpha _{\theta _2}}\nonumber\\
& &  -\frac{\sqrt[5]{\frac{1}{N}} (r-{r_h})
   {r_h} \left(9 a^2+{r_h}^2\right) {f_{\theta_1z}}''(r)}{\sqrt{{g_s}} \sqrt{6 \pi } \left(6 a^2+{r_h}^2\right) \alpha
   _{\theta _2}}-\frac{7 {g_s} M {N_f} (r-{r_h}) {r_h} \left(9 a^2+{r_h}^2\right) \log ({r_h}) \left(81 \sqrt{2}
   \alpha _{\theta _1}^3+10 \sqrt{3} \alpha _{\theta _2}^2\right) {f_{\theta_1x}}''(r)}{48 \sqrt{3} N \pi ^2 \left(6 a^2+{r_h}^2\right)
   \alpha _{\theta _2}^2}\nonumber\\
& &  +\frac{27 {g_s} M \left(\frac{1}{N}\right)^{3/5} {N_f} (r-{r_h}) {r_h} \left(9
   a^2+{r_h}^2\right) \log ({r_h}) \alpha _{\theta _1}^4 {f_{\theta_1y}}''(r)}{8 \pi ^2 \left(6 a^2+{r_h}^2\right) \alpha _{\theta
   _2}^4}+\frac{27 {g_s} M \left(\frac{1}{N}\right)^{3/5} {N_f} (r-{r_h}) {r_h} \left(9 a^2+{r_h}^2\right) \log
   ({r_h}) \alpha _{\theta _1}^4 {f_{\theta_2z}}''(r)}{8 \pi ^2 \left(6 a^2+{r_h}^2\right) \alpha _{\theta _2}^4}\nonumber\\
& &  +\frac{7 {g_s} M
   {N_f} (r-{r_h}) {r_h} \left(9 a^2+{r_h}^2\right) \log ({r_h}) \left(81 \sqrt{2} \alpha _{\theta _1}^3+10 \sqrt{3}
   \alpha _{\theta _2}^2\right) {f_{\theta_2x}}''(r)}{48 \sqrt{3} N \pi ^2 \left(6 a^2+{r_h}^2\right) \alpha _{\theta
   _2}^2}\nonumber\\
& &  -\frac{\sqrt[5]{\frac{1}{N}} (r-{r_h}) {r_h} \left(9 a^2+{r_h}^2\right) {f_{\theta_2y}}''(r)}{\sqrt{{g_s}} \sqrt{6 \pi
   } \left(6 a^2+{r_h}^2\right) \alpha _{\theta _2}}-\frac{\sqrt[5]{\frac{1}{N}} (r-{r_h}) {r_h} \left(9
   a^2+{r_h}^2\right) {f_{xz}}''(r)}{\sqrt{{g_s}} \sqrt{6 \pi } \left(6 a^2+{r_h}^2\right) \alpha _{\theta _2}}\nonumber\\
& &  +\frac{27
   \left(\frac{1}{N}\right)^{8/5} \sqrt{\frac{3}{2 \pi }} (r-{r_h}) {r_h} \left(9 a^2+{r_h}^2\right) \alpha _{\theta _1}^4
   \alpha _{\theta _2} {f_{xx}}''(r)}{4 \sqrt{{g_s}} \left(6 a^2+{r_h}^2\right)}+\frac{{g_s} M {N_f} (r-{r_h})
   {r_h} \left(9 a^2+{r_h}^2\right) \log ({r_h}) \left(27 \sqrt{3} \alpha _{\theta _1}^3+5 \sqrt{2} \alpha _{\theta
   _2}^2\right) {f_{xy}}''(r)}{32 \sqrt{2} N \pi ^2 \left(6 a^2+{r_h}^2\right) \alpha _{\theta _2}^2}
\nonumber\\
& &  +\frac{\sqrt[5]{\frac{1}{N}}
   \sqrt{\frac{2}{3 \pi }} (r-{r_h}) {r_h} \left(9 a^2+{r_h}^2\right) {f_{yz}}''(r)}{3 \sqrt{{g_s}} \left(6
   a^2+{r_h}^2\right) \alpha _{\theta _2}}\nonumber\\
   &&+\frac{{g_s}^{5/2} M^2 \left(\frac{1}{N}\right)^{9/5} {N_f}^2 (r-{r_h})
   {r_h} \left(9 a^2+{r_h}^2\right) \log ^2({r_h}) \left(2187 \alpha _{\theta _1}^6+270 \sqrt{6} \alpha _{\theta _2}^2 \alpha
   _{\theta _1}^3+50 \alpha _{\theta _2}^4\right) {f_{\phi_{2}\phi_{2}}}''(r)}{512 \sqrt{6} \pi ^{7/2} \left(6 a^2+{r_h}^2\right) \alpha _{\theta
   _2}^3}\nonumber\\
& &  -\frac{\sqrt[5]{\frac{1}{N}} \sqrt{\frac{2}{3 \pi }} (r-{r_h}) {r_h} \left(9 a^2+{r_h}^2\right) {f_t}''(r)}{3
   \sqrt{{g_s}} \left(6 a^2+{r_h}^2\right) \alpha _{\theta _2}}\nonumber\\
& &  -\frac{9 \sqrt{\frac{3}{2}} {g_s}^{7/2} M^4
   \left(\frac{1}{N}\right)^{11/5} {N_f}^2 {f_t}(r) \log ^4({r_h}) \left(16 \left(108 a^6-9 {r_h}^4
   a^2-{r_h}^6\right)-\left(2 {r_h}^6+27 a^2 {r_h}^4+117 a^4 {r_h}^2\right) \log \left({r_h}^6+9 a^2
   {r_h}^4\right)\right)}{64 \pi ^{9/2} \left(3 a^2-{r_h}^2\right) \left(6 a^2+{r_h}^2\right)^2 \log \left({r_h}^6+9 a^2
   {r_h}^4\right) \alpha _{\theta _2}}\nonumber\\
& &  +\frac{27 \sqrt{\frac{3}{2}} {g_s}^{7/2} \log N  M^4 \left(\frac{1}{N}\right)^{11/5}
   {N_f}^2 \left(9 a^2+{r_h}^2\right) f(r) \log ^4({r_h})}{16 \pi ^{9/2} \left(6 a^2+{r_h}^2\right) \log
   \left({r_h}^6+9 a^2 {r_h}^4\right) \alpha _{\theta _2}}+\frac{5 (r-{r_h}) {r_h}^3 \left(9 a^2+{r_h}^2\right)
   \left(108 b^2 {r_h}^2+1\right)^2 {f_r}(r)}{27 \sqrt{{g_s}} \sqrt[5]{\frac{1}{N}} \sqrt{6 \pi } \left({r_h}^2-3
   a^2\right)^2 \left(6 a^2+{r_h}^2\right) \log ^2({r_h}) \alpha _{\theta _1}^2 \alpha _{\theta _2}}\nonumber\\
& &  -\frac{49 \pi ^{5/2}
   {r_h}^4 {f_{zz}}(r)}{5832 \sqrt{6} {g_s}^{7/2} M^2 \left(\frac{1}{N}\right)^{4/5} {N_f}^2 \left({r_h}^2-3
   a^2\right)^2 \log ^4({r_h}) \alpha _{\theta _2}}+\frac{49 \pi ^{5/2} {r_h}^4 {f_{x^{10} x^{10}}}(r)}{3888 \sqrt{6} {g_s}^{7/2} M^2
   \left(\frac{1}{N}\right)^{4/5} {N_f}^2 \left({r_h}^2-3 a^2\right)^2 \log ^4({r_h}) \alpha _{\theta _2}}\nonumber\\
& &  +\frac{49 \pi ^{5/2}
   {r_h}^4 {f_{\theta_1z}}(r)}{2916 \sqrt{6} {g_s}^{7/2} M^2 \left(\frac{1}{N}\right)^{4/5} {N_f}^2 \left({r_h}^2-3
   a^2\right)^2 \log ^4({r_h}) \alpha _{\theta _2}}+\frac{245 \pi ^{5/2} {r_h}^4 {f_{\theta_2y}}(r)}{11664 \sqrt{6} {g_s}^{7/2} M^2
   \left(\frac{1}{N}\right)^{4/5} {N_f}^2 \left({r_h}^2-3 a^2\right)^2 \log ^4({r_h}) \alpha _{\theta _2}}\nonumber\\
& &  +\frac{2
   \sqrt{\frac{2}{3 \pi }} (r-{r_h}) {r_h}^3 \left(9 a^2+{r_h}^2\right) \left(108 b^2 {r_h}^2+1\right)^2 {f_{yz}}(r)}{9
   \sqrt{{g_s}} \left(\frac{1}{N}\right)^{2/5} \left({r_h}^2-3 a^2\right)^2 \left(6 a^2+{r_h}^2\right) \log ^2({r_h})
   \alpha _{\theta _2}^3}-\frac{\sqrt{\frac{2}{3 \pi }} (r-{r_h}) {r_h}^3 \left(9 a^2+{r_h}^2\right) \left(108 b^2
   {r_h}^2+1\right)^2 {f_{yy}}(r)}{9 \sqrt{{g_s}} \left(\frac{1}{N}\right)^{2/5} \left({r_h}^2-3 a^2\right)^2 \left(6
   a^2+{r_h}^2\right) \log ^2({r_h}) \alpha _{\theta _2}^3}\nonumber\\
& &  -\frac{245 \pi  {r_h}^4 {f_{\theta_1y}}(r) \alpha _{\theta _1}^4}{3456
   {g_s}^2 M \left(\frac{1}{N}\right)^{2/5} {N_f} \left({r_h}^2-3 a^2\right)^2 \log ^3({r_h}) \alpha _{\theta
   _2}^4}-\frac{245 \pi  {r_h}^4 {f_{\theta_2z}}(r) \alpha _{\theta _1}^4}{3456 {g_s}^2 M \left(\frac{1}{N}\right)^{2/5} {N_f}
   \left({r_h}^2-3 a^2\right)^2 \log ^3({r_h}) \alpha _{\theta _2}^4} = 0.
\end{eqnarray*}
}

(xix) \underline{${\rm EOM}_{zz}$}

{\scriptsize
\begin{eqnarray*}
& & \frac{512 \sqrt{2} \left(9 b^2+1\right)^3 \left(\frac{1}{N}\right)^{2/5} (r-{r_h})^3 \beta  \Sigma_1 b^{12}}{6561 \left(1-3 b^2\right)^6
   \left(6 b^2+1\right)^3 {g_s}^{9/4} \log N ^3 {N_f}^2 \sqrt[4]{\pi } {r_h}^3 \alpha _{\theta _1}^4 \alpha _{\theta
   _2}^4}+\frac{4 \left(9 b^2+1\right)^4 M \left(\frac{1}{N}\right)^{23/20} (r-{r_h}) \beta  \log ({r_h}) (-\Sigma_1)b^{10}}{9 \left(3
   b^2-1\right)^5 \left(6 b^2+1\right)^4 \sqrt{{g_s}} \log N ^2 {N_f} \pi ^{3/2} {r_h} \alpha _{\theta _2}^5}\nonumber\\
& &    -\frac{9
   \left(\frac{1}{N}\right)^{13/10} {f_{xx}}(r) \alpha _{\theta _1}^4}{2 \sqrt{{g_s}} \sqrt{\pi } \log \left({r_h}^6+9 a^2
   {r_h}^4\right)}+\frac{245 \pi  {r_h}^4 {f_{xy}}(r) \left(27 \sqrt{6} \alpha _{\theta _1}^3+10 \alpha _{\theta
   _2}^2\right)}{1119744 \sqrt{6} {g_s}^2 M \left(\frac{1}{N}\right)^{3/10} {N_f} \left({r_h}^2-3 a^2\right)^2 \log
   ^3({r_h}) \alpha _{\theta _2}^3}\nonumber\\
   & & -\frac{343 \pi  {r_h}^4 {f_{\theta_1x}}(r) \left(81 \sqrt{2} \alpha _{\theta _1}^3+10 \sqrt{3}
   \alpha _{\theta _2}^2\right)}{839808 \sqrt{2} {g_s}^2 M \left(\frac{1}{N}\right)^{3/10} {N_f} \left({r_h}^2-3 a^2\right)^2
   \log ^3({r_h}) \alpha _{\theta _2}^3}+\frac{1715 \pi  {r_h}^4 {f_{\theta_2x}}(r) \left(81 \sqrt{2} \alpha _{\theta _1}^3+10 \sqrt{3}
   \alpha _{\theta _2}^2\right)}{2519424 \sqrt{2} {g_s}^2 M \left(\frac{1}{N}\right)^{3/10} {N_f} \left({r_h}^2-3 a^2\right)^2
   \log ^3({r_h}) \alpha _{\theta _2}^3}\nonumber\\
   & & -\frac{3479 \pi  {r_h}^4 {f_{xz}}(r) \left(81 \sqrt{2} \alpha _{\theta _1}^3+10
   \sqrt{3} \alpha _{\theta _2}^2\right)}{10077696 \sqrt{2} {g_s}^2 M \left(\frac{1}{N}\right)^{3/10} {N_f} \left({r_h}^2-3
   a^2\right)^2 \log ^3({r_h}) \alpha _{\theta _2}^3}+\frac{2 {r_h} \left(9 a^2+{r_h}^2\right) f'(r)}{9 \sqrt{{g_s}}
   \sqrt[10]{\frac{1}{N}} \sqrt{\pi } \left(6 a^2+{r_h}^2\right) \alpha _{\theta _2}^2}\nonumber\\
   & & -\frac{4 {r_h} \left(9
   a^2+{r_h}^2\right) {f_{zz}}'(r)}{27 \sqrt{{g_s}} \sqrt[10]{\frac{1}{N}} \sqrt{\pi } \left(6 a^2+{r_h}^2\right) \alpha
   _{\theta _2}^2}+\frac{2 {r_h} \left(9 a^2+{r_h}^2\right) {f_{x^{10} x^{10}}}'(r)}{27 \sqrt{{g_s}} \sqrt[10]{\frac{1}{N}} \sqrt{\pi
   } \left(6 a^2+{r_h}^2\right) \alpha _{\theta _2}^2}\nonumber\\
   & & +\frac{{r_h} \left(9 a^2+{r_h}^2\right) {f_{\theta_1z}}'(r)}{9
   \sqrt{{g_s}} \sqrt[10]{\frac{1}{N}} \sqrt{\pi } \left(6 a^2+{r_h}^2\right) \alpha _{\theta _2}^2}-\frac{2 a^2 (r-{r_h})
   \left(9 a^2+{r_h}^2\right) {f_{\theta_1x}}'(r)}{9 \sqrt{{g_s}} \sqrt[10]{\frac{1}{N}} \sqrt{\pi } \left(-18 a^4+3 {r_h}^2
   a^2+{r_h}^4\right) \alpha _{\theta _2}^2}\nonumber\\
   & & -\frac{3 \sqrt{\frac{3}{2}} {g_s} M \left(\frac{1}{N}\right)^{3/10} {N_f}
   {r_h} \left(9 a^2+{r_h}^2\right) \log ({r_h}) \alpha _{\theta _1}^4 {f_{\theta_1y}}'(r)}{4 \pi ^2 \left(6 a^2+{r_h}^2\right)
   \alpha _{\theta _2}^5}-\frac{3 \sqrt{\frac{3}{2}} {g_s} M \left(\frac{1}{N}\right)^{3/10} {N_f} {r_h} \left(9
   a^2+{r_h}^2\right) \log ({r_h}) \alpha _{\theta _1}^4 {f_{\theta_2z}}'(r)}{4 \pi ^2 \left(6 a^2+{r_h}^2\right) \alpha _{\theta
   _2}^5}\nonumber\\
   & & +\frac{3 \sqrt{\frac{3}{2}} a^2 {g_s} M \sqrt{\frac{1}{N}} {N_f} (r-{r_h}) \left(9 a^2+{r_h}^2\right) \log
   ({r_h}) \alpha _{\theta _1}^2 {f_{\theta_2x}}'(r)}{2 \pi ^2 \left(-18 a^4+3 {r_h}^2 a^2+{r_h}^4\right) \alpha _{\theta
   _2}^3}+\frac{{r_h} \left(9 a^2+{r_h}^2\right) {f_{\theta_2y}}'(r)}{9 \sqrt{{g_s}} \sqrt[10]{\frac{1}{N}} \sqrt{\pi } \left(6
   a^2+{r_h}^2\right) \alpha _{\theta _2}^2}\nonumber\\
& &    +\frac{{r_h} \left(9 a^2+{r_h}^2\right) {f_{xz}}'(r)}{9 \sqrt{{g_s}}
   \sqrt[10]{\frac{1}{N}} \sqrt{\pi } \left(6 a^2+{r_h}^2\right) \alpha _{\theta _2}^2}-\frac{9 \left(\frac{1}{N}\right)^{13/10}
   {r_h} \left(9 a^2+{r_h}^2\right) \alpha _{\theta _1}^4 {f_{xx}}'(r)}{4 \sqrt{{g_s}} \sqrt{\pi } \left(6
   a^2+{r_h}^2\right)}\nonumber\\
   & & -\frac{3 \sqrt{\frac{3}{2}} a^2 {g_s} M \left(\frac{1}{N}\right)^{3/10} {N_f} (r-{r_h}) \left(9
   a^2+{r_h}^2\right) \log ({r_h}) \alpha _{\theta _1}^4 {f_{xy}}'(r)}{2 \pi ^2 \left(-18 a^4+3 {r_h}^2
   a^2+{r_h}^4\right) \alpha _{\theta _2}^5}-\frac{15 \sqrt{\frac{3}{2}} a^2 {g_s} M \left(\frac{1}{N}\right)^{3/10} {N_f}
   (r-{r_h}) \left(9 a^2+{r_h}^2\right) \log ({r_h}) \alpha _{\theta _1}^4 {f_{yz}}'(r)}{2 \pi ^2 \left(-18 a^4+3
   {r_h}^2 a^2+{r_h}^4\right) \alpha _{\theta _2}^5}\nonumber\\
   & & -\frac{a^2 {g_s}^{5/2} M^2 \left(\frac{1}{N}\right)^{13/10} {N_f}^2
   (r-{r_h}) \left(9 a^2+{r_h}^2\right) \log ^2({r_h}) \alpha _{\theta _1}^2 \left(2187 \sqrt{2} \alpha _{\theta _1}^6+540
   \sqrt{3} \alpha _{\theta _2}^2 \alpha _{\theta _1}^3+50 \sqrt{2} \alpha _{\theta _2}^4\right) {f_{yy}}'(r)}{128 \sqrt{2} \pi ^{7/2}
   \left(-18 a^4+3 {r_h}^2 a^2+{r_h}^4\right) \alpha _{\theta _2}^6} \nonumber\\
& &    -\frac{{r_h} \left(9 a^2+{r_h}^2\right)
   {f_r}'(r)}{27 \sqrt{{g_s}} \sqrt[10]{\frac{1}{N}} \sqrt{\pi } \left(6 a^2+{r_h}^2\right) \alpha _{\theta
   _2}^2}+\frac{{r_h} \left(9 a^2+{r_h}^2\right) {f_t}'(r)}{9 \sqrt{{g_s}} \sqrt[10]{\frac{1}{N}} \sqrt{\pi } \left(6
   a^2+{r_h}^2\right) \alpha _{\theta _2}^2}+\frac{2 (r-{r_h}) {r_h} \left(9 a^2+{r_h}^2\right) f''(r)}{9 \sqrt{{g_s}}
   \sqrt[10]{\frac{1}{N}} \sqrt{\pi } \left(6 a^2+{r_h}^2\right) \alpha _{\theta _2}^2}\nonumber\\
   & & -\frac{4 (r-{r_h}) {r_h} \left(9
   a^2+{r_h}^2\right) {f_{zz}}''(r)}{27 \sqrt{{g_s}} \sqrt[10]{\frac{1}{N}} \sqrt{\pi } \left(6 a^2+{r_h}^2\right) \alpha
   _{\theta _2}^2}+\frac{2 (r-{r_h}) {r_h} \left(9 a^2+{r_h}^2\right) {f_{x^{10} x^{10}}}''(r)}{27 \sqrt{{g_s}}
   \sqrt[10]{\frac{1}{N}} \sqrt{\pi } \left(6 a^2+{r_h}^2\right) \alpha _{\theta _2}^2}\nonumber\\
   & & +\frac{(r-{r_h}) {r_h} \left(9
   a^2+{r_h}^2\right) {f_{\theta_1z}}''(r)}{9 \sqrt{{g_s}} \sqrt[10]{\frac{1}{N}} \sqrt{\pi } \left(6 a^2+{r_h}^2\right) \alpha
   _{\theta _2}^2}+\frac{7 {g_s} M \left(\frac{1}{N}\right)^{7/10} {N_f} (r-{r_h}) {r_h} \left(9 a^2+{r_h}^2\right)
   \log ({r_h}) \left(81 \sqrt{2} \alpha _{\theta _1}^3+10 \sqrt{3} \alpha _{\theta _2}^2\right) {f_{\theta_1x}}''(r)}{216 \sqrt{2} \pi ^2
   \left(6 a^2+{r_h}^2\right) \alpha _{\theta _2}^3}\nonumber\\
  & & -\frac{3 \sqrt{\frac{3}{2}} {g_s} M \left(\frac{1}{N}\right)^{3/10} {N_f}
   (r-{r_h}) {r_h} \left(9 a^2+{r_h}^2\right) \log ({r_h}) \alpha _{\theta _1}^4 {f_{\theta_1y}}''(r)}{4 \pi ^2 \left(6
   a^2+{r_h}^2\right) \alpha _{\theta _2}^5}-\frac{3 \sqrt{\frac{3}{2}} {g_s} M \left(\frac{1}{N}\right)^{3/10} {N_f}
   (r-{r_h}) {r_h} \left(9 a^2+{r_h}^2\right) \log ({r_h}) \alpha _{\theta _1}^4 {f_{\theta_2z}}''(r)}{4 \pi ^2 \left(6
   a^2+{r_h}^2\right) \alpha _{\theta _2}^5}\nonumber\\
   & & -\frac{7 {g_s} M \left(\frac{1}{N}\right)^{7/10} {N_f} (r-{r_h}) {r_h}
   \left(9 a^2+{r_h}^2\right) \log ({r_h}) \left(81 \sqrt{2} \alpha _{\theta _1}^3+10 \sqrt{3} \alpha _{\theta _2}^2\right)
   {f_{\theta_2x}}''(r)}{216 \sqrt{2} \pi ^2 \left(6 a^2+{r_h}^2\right) \alpha _{\theta _2}^3}\nonumber\\
   & & +\frac{(r-{r_h}) {r_h} \left(9
   a^2+{r_h}^2\right) {f_{\theta_2y}}''(r)}{9 \sqrt{{g_s}} \sqrt[10]{\frac{1}{N}} \sqrt{\pi } \left(6 a^2+{r_h}^2\right) \alpha
   _{\theta _2}^2}+\frac{(r-{r_h}) {r_h} \left(9 a^2+{r_h}^2\right) {f_{xz}}''(r)}{9 \sqrt{{g_s}}
   \sqrt[10]{\frac{1}{N}} \sqrt{\pi } \left(6 a^2+{r_h}^2\right) \alpha _{\theta _2}^2}\nonumber\\
    \end{eqnarray*}}
{\scriptsize
\begin{eqnarray*}
   & & -\frac{9 \left(\frac{1}{N}\right)^{13/10}
   (r-{r_h}) {r_h} \left(9 a^2+{r_h}^2\right) \alpha _{\theta _1}^4 {f_{xx}}''(r)}{4 \sqrt{{g_s}} \sqrt{\pi } \left(6
   a^2+{r_h}^2\right)}-\frac{{g_s} M \left(\frac{1}{N}\right)^{7/10} {N_f} (r-{r_h}) {r_h} \left(9
   a^2+{r_h}^2\right) \log ({r_h}) \left(27 \sqrt{3} \alpha _{\theta _1}^3+5 \sqrt{2} \alpha _{\theta _2}^2\right)
   {f_{xy}}''(r)}{96 \sqrt{3} \pi ^2 \left(6 a^2+{r_h}^2\right) \alpha _{\theta _2}^3}\nonumber\\
& & +\frac{{g_s} M
   \left(\frac{1}{N}\right)^{7/10} {N_f} (r-{r_h}) {r_h} \left(9 a^2+{r_h}^2\right) \log ({r_h}) \left(27 \sqrt{3}
   \alpha _{\theta _1}^3+5 \sqrt{2} \alpha _{\theta _2}^2\right) {f_{yz}}''(r)}{96 \sqrt{3} \pi ^2 \left(6 a^2+{r_h}^2\right)
   \alpha _{\theta _2}^3}\nonumber\\
   & & -\frac{{g_s}^{5/2} M^2 \left(\frac{1}{N}\right)^{3/2} {N_f}^2 (r-{r_h}) {r_h} \left(9
   a^2+{r_h}^2\right) \log ^2({r_h}) \left(2187 \alpha _{\theta _1}^6+270 \sqrt{6} \alpha _{\theta _2}^2 \alpha _{\theta _1}^3+50
   \alpha _{\theta _2}^4\right) {f_{yy}}''(r)}{4608 \pi ^{7/2} \left(6 a^2+{r_h}^2\right) \alpha _{\theta _2}^4}\nonumber\\
   & & +\frac{2
   (r-{r_h}) {r_h} \left(9 a^2+{r_h}^2\right) {f_t}''(r)}{27 \sqrt{{g_s}} \sqrt[10]{\frac{1}{N}} \sqrt{\pi } \left(6
   a^2+{r_h}^2\right) \alpha _{\theta _2}^2}\nonumber\\
   & & +\frac{3 {g_s}^{7/2} M^4 \left(\frac{1}{N}\right)^{19/10} {N_f}^2 {f_t}(r)
   \log ^4({r_h}) \left(16 \left(108 a^6-9 {r_h}^4 a^2-{r_h}^6\right)-\left(2 {r_h}^6+27 a^2 {r_h}^4+117 a^4
   {r_h}^2\right) \log \left({r_h}^6+9 a^2 {r_h}^4\right)\right)}{64 \pi ^{9/2} \left(3 a^2-{r_h}^2\right) \left(6
   a^2+{r_h}^2\right)^2 \log \left({r_h}^6+9 a^2 {r_h}^4\right) \alpha _{\theta _2}^2}
   \nonumber\\
& &    -\frac{9 {g_s}^{7/2} \log N  M^4
   \left(\frac{1}{N}\right)^{19/10} {N_f}^2 \left(9 a^2+{r_h}^2\right) f(r) \log ^4({r_h})}{16 \pi ^{9/2} \left(6
   a^2+{r_h}^2\right) \log \left({r_h}^6+9 a^2 {r_h}^4\right) \alpha _{\theta _2}^2}-\frac{5 (r-{r_h}) {r_h}^3 \left(9
   a^2+{r_h}^2\right) \left(108 b^2 {r_h}^2+1\right)^2 {f_r}(r)}{243 \sqrt{{g_s}} \sqrt{\frac{1}{N}} \sqrt{\pi }
   \left({r_h}^2-3 a^2\right)^2 \left(6 a^2+{r_h}^2\right) \log ^2({r_h}) \alpha _{\theta _1}^2 \alpha _{\theta
   _2}^2}\nonumber\\
   & &    +\frac{49 \pi ^{5/2} {r_h}^4 {f_{zz}}(r)}{52488 {g_s}^{7/2} M^2 \left(\frac{1}{N}\right)^{11/10} {N_f}^2
   \left({r_h}^2-3 a^2\right)^2 \log ^4({r_h}) \alpha _{\theta _2}^2}-\frac{49 \pi ^{5/2} {r_h}^4 {f_{x^{10} x^{10}}}(r)}{34992
   {g_s}^{7/2} M^2 \left(\frac{1}{N}\right)^{11/10} {N_f}^2 \left({r_h}^2-3 a^2\right)^2 \log ^4({r_h}) \alpha _{\theta
   _2}^2}\nonumber\\
   & & -\frac{49 \pi ^{5/2} {r_h}^4 {f_{\theta_1z}}(r)}{26244 {g_s}^{7/2} M^2 \left(\frac{1}{N}\right)^{11/10} {N_f}^2
   \left({r_h}^2-3 a^2\right)^2 \log ^4({r_h}) \alpha _{\theta _2}^2}-\frac{245 \pi ^{5/2} {r_h}^4 {f_{\theta_2y}}(r)}{104976
   {g_s}^{7/2} M^2 \left(\frac{1}{N}\right)^{11/10} {N_f}^2 \left({r_h}^2-3 a^2\right)^2 \log ^4({r_h}) \alpha _{\theta
   _2}^2}\nonumber\\
   & & -\frac{4 (r-{r_h}) {r_h}^3 \left(9 a^2+{r_h}^2\right) \left(108 b^2 {r_h}^2+1\right)^2 {f_{yz}}(r)}{81
   \sqrt{{g_s}} \left(\frac{1}{N}\right)^{7/10} \sqrt{\pi } \left({r_h}^2-3 a^2\right)^2 \left(6 a^2+{r_h}^2\right) \log
   ^2({r_h}) \alpha _{\theta _2}^4}+\frac{2 (r-{r_h}) {r_h}^3 \left(9 a^2+{r_h}^2\right) \left(108 b^2
   {r_h}^2+1\right)^2 {f_{yy}}(r)}{81 \sqrt{{g_s}} \left(\frac{1}{N}\right)^{7/10} \sqrt{\pi } \left({r_h}^2-3 a^2\right)^2
   \left(6 a^2+{r_h}^2\right) \log ^2({r_h}) \alpha _{\theta _2}^4}\nonumber\\
   & & +\frac{245 \pi  {r_h}^4 {f_{\theta_1y}}(r) \alpha _{\theta
   _1}^4}{5184 \sqrt{6} {g_s}^2 M \left(\frac{1}{N}\right)^{7/10} {N_f} \left({r_h}^2-3 a^2\right)^2 \log ^3({r_h}) \alpha
   _{\theta _2}^5}+\frac{245 \pi  {r_h}^4 {f_{\theta_2z}}(r) \alpha _{\theta _1}^4}{5184 \sqrt{6} {g_s}^2 M
   \left(\frac{1}{N}\right)^{7/10} {N_f} \left({r_h}^2-3 a^2\right)^2 \log ^3({r_h}) \alpha _{\theta _2}^5}
\end{eqnarray*}
}

(xx) \underline{${\rm EOM}_{x^{10}x^{10}}$}

{\scriptsize
\begin{eqnarray*}
& & -\frac{32 \left(9 b^2+1\right)^4 M \left(\frac{1}{N}\right)^{7/4} \sqrt{\pi } (r-{r_h}) \beta  \log ({r_h}) \Sigma_1 b^{10}}{3 \left(3
   b^2-1\right)^5 \sqrt{{g_s}} \left(6 \log N  b^2+\log N \right)^4 {N_f}^3 {r_h} \alpha _{\theta _2}^3}+\frac{256
   \left(9 b^2+1\right)^3 M \left(\frac{1}{N}\right)^{7/4} \sqrt{\pi } (r-{r_h}) \beta  \log ({r_h}) (-\Sigma_1) b^8}{81 \left(1-3 b^2\right)^4
   \left(6 b^2+1\right)^3 \sqrt{{g_s}} \log N ^5 {N_f}^3 {r_h} \alpha _{\theta _2}^3}\nonumber\\
   & & -\frac{432
   \left(\frac{1}{N}\right)^{19/10} \pi ^{3/2} {f_{xx}}(r) \alpha _{\theta _1}^4 \alpha _{\theta _2}^2}{\sqrt{{g_s}} {N_f}^2
   \log ^3\left({r_h}^6+9 a^2 {r_h}^4\right)}\nonumber\\
   & & +\frac{9 {g_s}^{7/2} M^4 \left(\frac{1}{N}\right)^{5/2} {f_t}(r) \log
   ^4({r_h}) \left(16 \left(108 a^6-9 {r_h}^4 a^2-{r_h}^6\right)-\left(2 {r_h}^6+27 a^2 {r_h}^4+117 a^4
   {r_h}^2\right) \log \left({r_h}^6+9 a^2 {r_h}^4\right)\right)}{2 \pi ^{5/2} \left(3 a^2-{r_h}^2\right) \left(6
   a^2+{r_h}^2\right)^2 \log ^3\left({r_h}^6+9 a^2 {r_h}^4\right)}\nonumber\\
   & & -\frac{637 \left(\frac{1}{N}\right)^{3/10} \pi ^3
   {r_h}^4 {f_{xy}}(r) \left(27 \sqrt{6} \alpha _{\theta _1}^3+10 \alpha _{\theta _2}^2\right)}{11664 \sqrt{6} {g_s}^2 M
   {N_f}^3 \left({r_h}^2-3 a^2\right)^2 \log ^3({r_h}) \log ^2\left({r_h}^6+9 a^2 {r_h}^4\right) \alpha _{\theta
   _2}}\nonumber\\
   & & +\frac{1715 \left(\frac{1}{N}\right)^{3/10} \pi ^3 {r_h}^4 {f_{\theta_1x}}(r) \left(81 \sqrt{2} \alpha _{\theta _1}^3+10 \sqrt{3}
   \alpha _{\theta _2}^2\right)}{8748 \sqrt{2} {g_s}^2 M {N_f}^3 \left({r_h}^2-3 a^2\right)^2 \log ^3({r_h}) \log
   ^2\left({r_h}^6+9 a^2 {r_h}^4\right) \alpha _{\theta _2}}-\frac{4459 \left(\frac{1}{N}\right)^{3/10} \pi ^3 {r_h}^4
   {f_{\theta_2x}}(r) \left(81 \sqrt{2} \alpha _{\theta _1}^3+10 \sqrt{3} \alpha _{\theta _2}^2\right)}{26244 \sqrt{2} {g_s}^2 M
   {N_f}^3 \left({r_h}^2-3 a^2\right)^2 \log ^3({r_h}) \log ^2\left({r_h}^6+9 a^2 {r_h}^4\right) \alpha _{\theta
   _2}}\nonumber\\
    \end{eqnarray*}}
   {\scriptsize
   \begin{eqnarray*}
   & & -\frac{833 \left(\frac{1}{N}\right)^{3/10} \pi ^3 {r_h}^4 {f_{xz}}(r) \left(81 \sqrt{2} \alpha _{\theta _1}^3+10 \sqrt{3}
   \alpha _{\theta _2}^2\right)}{104976 \sqrt{2} {g_s}^2 M {N_f}^3 \left({r_h}^2-3 a^2\right)^2 \log ^3({r_h}) \log
   ^2\left({r_h}^6+9 a^2 {r_h}^4\right) \alpha _{\theta _2}}+\frac{64 \sqrt{\frac{1}{N}} \pi ^{3/2} {r_h} \left(9
   a^2+{r_h}^2\right) f'(r)}{3 \sqrt{{g_s}} {N_f}^2 \left(6 a^2+{r_h}^2\right) \log ^2\left({r_h}^6+9 a^2
   {r_h}^4\right)}\nonumber\\
   & & -\frac{64 \sqrt{\frac{1}{N}} \pi ^{3/2} {r_h} \left(9 a^2+{r_h}^2\right) {f_{zz}}'(r)}{9
   \sqrt{{g_s}} {N_f}^2 \left(6 a^2+{r_h}^2\right) \log ^2\left({r_h}^6+9 a^2 {r_h}^4\right)}\nonumber\\
   & & -\frac{32
   \sqrt{\frac{1}{N}} \pi ^{3/2} (r-{r_h}) \left(-36 a^4+6 {r_h}^2 a^2+\left(9 a^2+{r_h}^2\right) \log \left({r_h}^6+9 a^2
   {r_h}^4\right) a^2+2 {r_h}^4\right) {f_{x^{10} x^{10}}}'(r)}{3 \sqrt{{g_s}} {N_f}^2 \left(-18 a^4+3 {r_h}^2
   a^2+{r_h}^4\right) \log ^3\left({r_h}^6+9 a^2 {r_h}^4\right)}+\frac{32 \sqrt{\frac{1}{N}} \pi ^{3/2} {r_h} \left(9
   a^2+{r_h}^2\right) {f_{\theta_1z}}'(r)}{3 \sqrt{{g_s}} {N_f}^2 \left(6 a^2+{r_h}^2\right) \log ^2\left({r_h}^6+9 a^2
   {r_h}^4\right)}\nonumber\\
   & & -\frac{64 a^2 \sqrt{\frac{1}{N}} \pi ^{3/2} (r-{r_h}) \left(9 a^2+{r_h}^2\right) {f_{\theta_1x}}'(r)}{3
   \sqrt{{g_s}} {N_f}^2 \left(-18 a^4+3 {r_h}^2 a^2+{r_h}^4\right) \log ^2\left({r_h}^6+9 a^2
   {r_h}^4\right)}-\frac{36 \sqrt{6} {g_s} M \left(\frac{1}{N}\right)^{9/10} {r_h} \left(9 a^2+{r_h}^2\right) \log
   ({r_h}) \alpha _{\theta _1}^4 {f_{\theta_1y}}'(r)}{{N_f} \left(6 a^2+{r_h}^2\right) \log ^2\left({r_h}^6+9 a^2
   {r_h}^4\right) \alpha _{\theta _2}^3}\nonumber\\
  & &  -\frac{36 \sqrt{6} {g_s} M \left(\frac{1}{N}\right)^{9/10} {r_h} \left(9
   a^2+{r_h}^2\right) \log ({r_h}) \alpha _{\theta _1}^4 {f_{\theta_2z}}'(r)}{{N_f} \left(6 a^2+{r_h}^2\right) \log
   ^2\left({r_h}^6+9 a^2 {r_h}^4\right) \alpha _{\theta _2}^3}+\frac{72 \sqrt{6} a^2 {g_s} M \left(\frac{1}{N}\right)^{11/10}
   (r-{r_h}) \left(9 a^2+{r_h}^2\right) \log ({r_h}) \alpha _{\theta _1}^2 {f_{\theta_2x}}'(r)}{{N_f} \left({r_h}^2-3
   a^2\right) \left(6 a^2+{r_h}^2\right) \log ^2\left({r_h}^6+9 a^2 {r_h}^4\right) \alpha _{\theta _2}}\nonumber\\
   & & +\frac{32
   \sqrt{\frac{1}{N}} \pi ^{3/2} {r_h} \left(9 a^2+{r_h}^2\right) {f_{\theta_2y}}'(r)}{3 \sqrt{{g_s}} {N_f}^2 \left(6
   a^2+{r_h}^2\right) \log ^2\left({r_h}^6+9 a^2 {r_h}^4\right)}\nonumber\\
   & & +\frac{32 \sqrt{\frac{1}{N}} \pi ^{3/2} {r_h} \left(9
   a^2+{r_h}^2\right) {f_{xz}}'(r)}{3 \sqrt{{g_s}} {N_f}^2 \left(6 a^2+{r_h}^2\right) \log ^2\left({r_h}^6+9 a^2
   {r_h}^4\right)}-\frac{216 \left(\frac{1}{N}\right)^{19/10} \pi ^{3/2} {r_h} \left(9 a^2+{r_h}^2\right) \alpha _{\theta
   _1}^4 \alpha _{\theta _2}^2 {f_{xx}}'(r)}{\sqrt{{g_s}} {N_f}^2 \left(6 a^2+{r_h}^2\right) \log ^2\left({r_h}^6+9 a^2
   {r_h}^4\right)}\nonumber\\
   & & -\frac{72 \sqrt{6} a^2 {g_s} M \left(\frac{1}{N}\right)^{9/10} (r-{r_h}) \left(9 a^2+{r_h}^2\right) \log
   ({r_h}) \alpha _{\theta _1}^4 {f_{xy}}'(r)}{{N_f} \left({r_h}^2-3 a^2\right) \left(6 a^2+{r_h}^2\right) \log
   ^2\left({r_h}^6+9 a^2 {r_h}^4\right) \alpha _{\theta _2}^3}-\frac{72 \sqrt{6} a^2 {g_s} M \left(\frac{1}{N}\right)^{9/10}
   (r-{r_h}) \left(9 a^2+{r_h}^2\right) \log ({r_h}) \alpha _{\theta _1}^4 {f_{yz}}'(r)}{{N_f} \left({r_h}^2-3
   a^2\right) \left(6 a^2+{r_h}^2\right) \log ^2\left({r_h}^6+9 a^2 {r_h}^4\right) \alpha _{\theta _2}^3}\nonumber\\
   & & -\frac{3 a^2
   {g_s}^{5/2} M^2 \left(\frac{1}{N}\right)^{19/10} (r-{r_h}) \left(9 a^2+{r_h}^2\right) \log ^2({r_h}) \alpha _{\theta
   _1}^2 \left(2187 \sqrt{2} \alpha _{\theta _1}^6+540 \sqrt{3} \alpha _{\theta _2}^2 \alpha _{\theta _1}^3+50 \sqrt{2} \alpha _{\theta
   _2}^4\right) {f_{yy}}'(r)}{4 \sqrt{2} \pi ^{3/2} \left(-18 a^4+3 {r_h}^2 a^2+{r_h}^4\right) \log ^2\left({r_h}^6+9 a^2
   {r_h}^4\right) \alpha _{\theta _2}^4}\nonumber\\
    &&-\frac{32 \sqrt{\frac{1}{N}} \pi ^{3/2} {r_h} \left(9 a^2+{r_h}^2\right)
   {f_r}'(r)}{9 \sqrt{{g_s}} {N_f}^2 \left(6 a^2+{r_h}^2\right) \log ^2\left({r_h}^6+9 a^2
   {r_h}^4\right)}\nonumber\\
   & & +\frac{32 \sqrt{\frac{1}{N}} \pi ^{3/2} {r_h} \left(9 a^2+{r_h}^2\right) {f_t}'(r)}{3 \sqrt{{g_s}}
   {N_f}^2 \left(6 a^2+{r_h}^2\right) \log ^2\left({r_h}^6+9 a^2 {r_h}^4\right)}+\frac{64 \sqrt{\frac{1}{N}} \pi ^{3/2}
   (r-{r_h}) {r_h} \left(9 a^2+{r_h}^2\right) f''(r)}{3 \sqrt{{g_s}} {N_f}^2 \left(6 a^2+{r_h}^2\right) \log
   ^2\left({r_h}^6+9 a^2 {r_h}^4\right)}-\frac{64 \sqrt{\frac{1}{N}} \pi ^{3/2} (r-{r_h}) {r_h} \left(9
   a^2+{r_h}^2\right) {f_{zz}}''(r)}{9 \sqrt{{g_s}} {N_f}^2 \left(6 a^2+{r_h}^2\right) \log ^2\left({r_h}^6+9 a^2
   {r_h}^4\right)}\nonumber\\
   & & -\frac{4 {g_s}^{3/2} M^2 \left(\frac{1}{N}\right)^{3/2} (r-{r_h}) {r_h} \left(9 a^2+{r_h}^2\right)
   \log ^2({r_h}) {f_{x^{10} x^{10}}}''(r)}{{N_f} \sqrt{\pi } \left(6 a^2+{r_h}^2\right) \log ^2\left({r_h}^6+9 a^2
   {r_h}^4\right)}+\frac{32 \sqrt{\frac{1}{N}} \pi ^{3/2} (r-{r_h}) {r_h} \left(9 a^2+{r_h}^2\right) {f_{\theta_1z}}''(r)}{3
   \sqrt{{g_s}} {N_f}^2 \left(6 a^2+{r_h}^2\right) \log ^2\left({r_h}^6+9 a^2 {r_h}^4\right)}\nonumber\\
& &  +\frac{14 \sqrt{2}
   {g_s} M \left(\frac{1}{N}\right)^{13/10} (r-{r_h}) {r_h} \left(9 a^2+{r_h}^2\right) \log ({r_h}) \left(81 \sqrt{2}
   \alpha _{\theta _1}^3+10 \sqrt{3} \alpha _{\theta _2}^2\right) {f_{\theta_1x}}''(r)}{9 {N_f} \left(6 a^2+{r_h}^2\right) \log
   ^2\left({r_h}^6+9 a^2 {r_h}^4\right) \alpha _{\theta _2}}\nonumber\\
   & & -\frac{36 \sqrt{6} {g_s} M \left(\frac{1}{N}\right)^{9/10}
   (r-{r_h}) {r_h} \left(9 a^2+{r_h}^2\right) \log ({r_h}) \alpha _{\theta _1}^4 {f_{\theta_1y}}''(r)}{{N_f} \left(6
   a^2+{r_h}^2\right) \log ^2\left({r_h}^6+9 a^2 {r_h}^4\right) \alpha _{\theta _2}^3}-\frac{36 \sqrt{6} {g_s} M
   \left(\frac{1}{N}\right)^{9/10} (r-{r_h}) {r_h} \left(9 a^2+{r_h}^2\right) \log ({r_h}) \alpha _{\theta _1}^4
   {f_{\theta_2z}}''(r)}{{N_f} \left(6 a^2+{r_h}^2\right) \log ^2\left({r_h}^6+9 a^2 {r_h}^4\right) \alpha _{\theta
   _2}^3}\nonumber\\
   & & -\frac{14 \sqrt{2} {g_s} M \left(\frac{1}{N}\right)^{13/10} (r-{r_h}) {r_h} \left(9 a^2+{r_h}^2\right) \log
   ({r_h}) \left(81 \sqrt{2} \alpha _{\theta _1}^3+10 \sqrt{3} \alpha _{\theta _2}^2\right) {f_{\theta_2x}}''(r)}{9 {N_f} \left(6
   a^2+{r_h}^2\right) \log ^2\left({r_h}^6+9 a^2 {r_h}^4\right) \alpha _{\theta _2}}+\frac{32 \sqrt{\frac{1}{N}} \pi ^{3/2}
   (r-{r_h}) {r_h} \left(9 a^2+{r_h}^2\right) {f_{\theta_2y}}''(r)}{3 \sqrt{{g_s}} {N_f}^2 \left(6 a^2+{r_h}^2\right)
   \log ^2\left({r_h}^6+9 a^2 {r_h}^4\right)}\nonumber\\
& &  +\frac{32 \sqrt{\frac{1}{N}} \pi ^{3/2} (r-{r_h}) {r_h} \left(9
   a^2+{r_h}^2\right) {f_{xz}}''(r)}{3 \sqrt{{g_s}} {N_f}^2 \left(6 a^2+{r_h}^2\right) \log ^2\left({r_h}^6+9 a^2
   {r_h}^4\right)}\nonumber\\
   & & -\frac{216 \left(\frac{1}{N}\right)^{19/10} \pi ^{3/2} (r-{r_h}) {r_h} \left(9 a^2+{r_h}^2\right) \alpha
   _{\theta _1}^4 \alpha _{\theta _2}^2 {f_{\phi_{10}\phi_{1}}}''(r)}{\sqrt{{g_s}} {N_f}^2 \left(6 a^2+{r_h}^2\right) \log
   ^2\left({r_h}^6+9 a^2 {r_h}^4\right)}-\frac{{g_s} M \left(\frac{1}{N}\right)^{13/10} (r-{r_h}) {r_h} \left(9
   a^2+{r_h}^2\right) \log ({r_h}) \left(27 \sqrt{3} \alpha _{\theta _1}^3+5 \sqrt{2} \alpha _{\theta _2}^2\right)
   {f_{xy}}''(r)}{\sqrt{3} {N_f} \left(6 a^2+{r_h}^2\right) \log ^2\left({r_h}^6+9 a^2 {r_h}^4\right) \alpha _{\theta
   _2}}\nonumber\\
& &  +\frac{{g_s} M \left(\frac{1}{N}\right)^{13/10} (r-{r_h}) {r_h} \left(9 a^2+{r_h}^2\right) \log ({r_h})
   \left(27 \sqrt{3} \alpha _{\theta _1}^3+5 \sqrt{2} \alpha _{\theta _2}^2\right) {f_{yz}}''(r)}{\sqrt{3} {N_f} \left(6
   a^2+{r_h}^2\right) \log ^2\left({r_h}^6+9 a^2 {r_h}^4\right) \alpha _{\theta _2}}\nonumber\\
    \end{eqnarray*}}
   {\scriptsize
   \begin{eqnarray*}
   & & -\frac{{g_s}^{5/2} M^2
   \left(\frac{1}{N}\right)^{21/10} (r-{r_h}) {r_h} \left(9 a^2+{r_h}^2\right) \log ^2({r_h}) \left(2187 \alpha _{\theta
   _1}^6+270 \sqrt{6} \alpha _{\theta _2}^2 \alpha _{\theta _1}^3+50 \alpha _{\theta _2}^4\right) {f_{\phi_{20}\phi_{2}}}''(r)}{48 \pi ^{3/2} \left(6
   a^2+{r_h}^2\right) \log ^2\left({r_h}^6+9 a^2 {r_h}^4\right) \alpha _{\theta _2}^2}\nonumber\\
   &&+\frac{64 \sqrt{\frac{1}{N}} \pi ^{3/2}
   (r-{r_h}) {r_h} \left(9 a^2+{r_h}^2\right) {f_t}''(r)}{9 \sqrt{{g_s}} {N_f}^2 \left(6 a^2+{r_h}^2\right)
   \log ^2\left({r_h}^6+9 a^2 {r_h}^4\right)}\nonumber\\
& &  -\frac{1568 \pi ^{9/2} {r_h}^4 {f_{zz}}(r)}{2187 {g_s}^{7/2} M^2
   \sqrt{\frac{1}{N}} {N_f}^4 \left({r_h}^2-3 a^2\right)^2 \log ^4({r_h}) \log ^2\left({r_h}^6+9 a^2
   {r_h}^4\right)}\nonumber\\
   & & -\frac{98 \pi ^{9/2} {r_h}^4 {f_{x^{10} x^{10}}}(r)}{729 {g_s}^{7/2} M^2 \sqrt{\frac{1}{N}} {N_f}^4
   \left({r_h}^2-3 a^2\right)^2 \log ^4({r_h}) \log ^2\left({r_h}^6+9 a^2 {r_h}^4\right)}+\frac{3136 \pi ^{9/2}
   {r_h}^4 {f_{\theta_1z}}(r)}{2187 {g_s}^{7/2} M^2 \sqrt{\frac{1}{N}} {N_f}^4 \left({r_h}^2-3 a^2\right)^2 \log ^4({r_h})
   \log ^2\left({r_h}^6+9 a^2 {r_h}^4\right)}\nonumber\\
& &  +\frac{1274 \pi ^{9/2} {r_h}^4 {f_{\theta_2y}}(r)}{2187 {g_s}^{7/2} M^2
   \sqrt{\frac{1}{N}} {N_f}^4 \left({r_h}^2-3 a^2\right)^2 \log ^4({r_h}) \log ^2\left({r_h}^6+9 a^2
   {r_h}^4\right)}\nonumber\\
   & & +\frac{992 \sqrt[10]{\frac{1}{N}} \pi ^{3/2} (r-{r_h}) {r_h}^3 \left(9 a^2+{r_h}^2\right) \left(108 b^2
   {r_h}^2+1\right)^2 {f_r}(r)}{81 \sqrt{{g_s}} {N_f}^2 \left({r_h}^2-3 a^2\right)^2 \left(6 a^2+{r_h}^2\right)
   \log ^2({r_h}) \log ^2\left({r_h}^6+9 a^2 {r_h}^4\right) \alpha _{\theta _1}^2}\nonumber\\
   &&-\frac{320 \pi ^{3/2} (r-{r_h})
   {r_h}^3 \left(9 a^2+{r_h}^2\right) \left(108 b^2 {r_h}^2+1\right)^2 {f_{yy}}(r)}{27 \sqrt{{g_s}}
   \sqrt[10]{\frac{1}{N}} {N_f}^2 \left({r_h}^2-3 a^2\right)^2 \left(6 a^2+{r_h}^2\right) \log ^2({r_h}) \log
   ^2\left({r_h}^6+9 a^2 {r_h}^4\right) \alpha _{\theta _2}^2}\nonumber\\
   & & -\frac{54 {g_s}^{7/2} \log N  M^4
   \left(\frac{1}{N}\right)^{5/2} \left(9 a^2+{r_h}^2\right) f(r) \log ^4({r_h})}{\pi ^{5/2} \left(6 a^2+{r_h}^2\right) \log
   ^3\left({r_h}^6+9 a^2 {r_h}^4\right)}-\frac{1519 \pi ^3 {r_h}^4 {f_{\theta_1y}}(r) \alpha _{\theta _1}^4}{54 \sqrt{6}
   {g_s}^2 M \sqrt[10]{\frac{1}{N}} {N_f}^3 \left({r_h}^2-3 a^2\right)^2 \log ^3({r_h}) \log ^2\left({r_h}^6+9 a^2
   {r_h}^4\right) \alpha _{\theta _2}^3}\nonumber\\
   & & -\frac{637 \pi ^3 {r_h}^4 {f_{\theta_2z}}(r) \alpha _{\theta _1}^4}{54 \sqrt{6} {g_s}^2 M
   \sqrt[10]{\frac{1}{N}} {N_f}^3 \left({r_h}^2-3 a^2\right)^2 \log ^3({r_h}) \log ^2\left({r_h}^6+9 a^2
   {r_h}^4\right) \alpha _{\theta _2}^3}\nonumber\\
   &&+\frac{49 \pi ^3 {r_h}^4 {f_{yz}}(r) \alpha _{\theta _1}^4}{3 \sqrt{6} {g_s}^2 M
   \sqrt[10]{\frac{1}{N}} {N_f}^3 \left({r_h}^2-3 a^2\right)^2 \log ^3({r_h}) \log ^2\left({r_h}^6+9 a^2
   {r_h}^4\right) \alpha _{\theta _2}^3}=0
\end{eqnarray*}
}

\subsection{$\psi\neq0$ near $r=r_h$}

Working in the IR,  the EOMs  near $r=r_h$ and up to LO in $N$, can be written as follows:
{\footnotesize
\begin{eqnarray}
\label{IR-psi=2nPi-EOMs}
& & {\rm EOM}_{MN}:\nonumber\\
& &  \sum_{p=0}^2\sum_{i=0}^2b_{MN}^{(p,i)}\left(r_h, a, N, M, N_f, g_s, \alpha_{\theta_{1,2}}\right)(r-r_h)^if_{MN}^{(p)}(r) +
\beta \frac{{\cal H}_{MN}\left(r_h, a, N, M, N_f, g_s, \alpha_{\theta_{1,2}}\right)}{(r-r_h)^{\gamma_{MN}^{\rm LO}}} = 0,\nonumber\\
& &
\end{eqnarray}
}
$M, N$ run over the $D=11$ coordinates,   $f^{(p)}_{MN}\equiv \frac{d^p f_{MN}}{dr^p}, p=0, 1, 2$, $\gamma_{MN}^{\rm LO}=1, 2$ denotes the leading order (LO) terms in powers of $r-r_h$ in the IR when the ${\cal O}(\beta)$-terms are Laurent-expanded about $r=r_h$.
where:
{\footnotesize
\begin{eqnarray}
\label{Sigma_2-def}
& & \hskip -0.9in\Sigma_2 \equiv \frac{ \left(40 ({g_s}-1) {g_s}^3 M^2 {N_f}^2 \sin^2\left(\frac{\psi_0}{2}\right) \alpha _{\theta _2}^2+108 \sqrt{6} ({g_s}-1) {g_s}^3 M^2 {N_f}^2 \sin^2\left(\frac{\psi_0}{2}\right)
   \alpha _{\theta _1}^3+8 \pi ^3 \alpha _{\theta _1}^2 \alpha _{\theta _2}^4\right)}{ \left(-4 ({g_s}-1) {g_s}^3 M^2 {N_f}^2 \psi
   ^2 \alpha _{\theta _2}^2+108 \sqrt{6} ({g_s}-1) {g_s}^3 M^2 {N_f}^2 \sin^2\left(\frac{\psi_0}{2}\right) \alpha _{\theta _1}^3+8 \pi ^3 \alpha _{\theta _1}^2
   \alpha _{\theta _2}^4\right)}
\end{eqnarray}
}

(i) ${\rm EOM}_{tt}$:

{\footnotesize
\begin{eqnarray}
& & -\frac{16\sin^4\left(\frac{\psi_0}{2}\right)  {r_h} \left(9 a^2+ {r_h}^2\right) (r- {r_h})^2  {f_{x^{10} x^{10}}}'(r)}{72 \pi   {g_s} \sqrt[5]{\frac{1}{N}}  \sin^4\phi_{20}  \left(6
   a^2+ {r_h}^2\right) \log ( {r_h}) \alpha _{\theta _2}^4}+\frac{16\sin^4\left(\frac{\psi_0}{2}\right)  {r_h} \left(9 a^2+ {r_h}^2\right) (r- {r_h})^2  {f_{\theta_1x^{10}}}'(r)}{36
   \pi   {g_s} \sqrt[5]{\frac{1}{N}}  \sin^4\phi_{20}  \left(6 a^2+ {r_h}^2\right) \log ( {r_h}) \alpha _{\theta _2}^4}\nonumber\\
   & & -\frac{16\sin^4\left(\frac{\psi_0}{2}\right)  {r_h} \left(9
   a^2+ {r_h}^2\right) (r- {r_h})^2  {f_{\theta_1\theta_1}}'(r)}{72 \pi   {g_s} \sqrt[5]{\frac{1}{N}}  \sin^4\phi_{20}  \left(6 a^2+ {r_h}^2\right) \log ( {r_h})
   \alpha _{\theta _2}^4}+\frac{12 a^2 \left(\frac{1}{N}\right)^{2/5} 4\sin^2\left(\frac{\psi_0}{2}\right) \left(9 a^2+ {r_h}^2\right) (r- {r_h})^2 \log ( {r_h})
    {f_{\theta_1\theta_2}}'(r)}{\pi  ( {g_s}-1)  {g_s}  \sin^2\phi_{20}  \left(6 a^2+ {r_h}^2\right) \alpha _{\theta _2}^2}
 \nonumber\\
&   & +\frac{8192 \pi ^{9/2} a^2
   \sqrt[10]{\frac{1}{N}} 16\sin^4\left(\frac{\psi_0}{2}\right) \beta  \left(9 a^2+ {r_h}^2\right)^2 \left(\left(9 a^2+ {r_h}^2\right) \log \left(9 a^2
    {r_h}^4+ {r_h}^6\right)-8 \left(6 a^2+ {r_h}^2\right) \log ( {r_h})\right)^2}{729 ( {g_s}-1)  {g_s}^{3/2}  {N_f}^6  \sin^4\phi_{20}
    {r_h}^2 \left(6 a^2+ {r_h}^2\right)^4 (r- {r_h}) \log ^2( {r_h}) \alpha _{\theta _2}^4 \log ^8\left(9 a^2  {r_h}^4+ {r_h}^6\right)}=0.\nonumber\\
& &
   \end{eqnarray}
}

(ii) ${\rm EOM}_{x^ix^i}, i=1, 2, 3$:

{\footnotesize
\begin{eqnarray}
& &  \frac{16\sin^4\left(\frac{\psi_0}{2}\right)  {r_h}^2 \left(9 a^2+ {r_h}^2\right) (r- {r_h})  {f_{x^{10} x^{10}}}'(r)}{288 \pi   {g_s} \sqrt[5]{\frac{1}{N}}
    \sin^4\phi_{20}  \left(6 a^2+ {r_h}^2\right) \log ( {r_h}) \alpha _{\theta _2}^4}-\frac{16\sin^4\left(\frac{\psi_0}{2}\right)  {r_h}^2 \left(9
   a^2+ {r_h}^2\right) (r- {r_h})  {f_{\theta_1x^{10}}}'(r)}{144 \pi   {g_s} \sqrt[5]{\frac{1}{N}}  \sin^4\phi_{20}  \left(6
   a^2+ {r_h}^2\right) \log ( {r_h}) \alpha _{\theta _2}^4}\nonumber\\
   & &  +\frac{16\sin^4\left(\frac{\psi_0}{2}\right)  {r_h}^2 \left(9 a^2+ {r_h}^2\right)
   (r- {r_h})  {f_{\theta_1\theta_1}}'(r)}{288 \pi   {g_s} \sqrt[5]{\frac{1}{N}}  \sin^4\phi_{20}  \left(6 a^2+ {r_h}^2\right) \log
   ( {r_h}) \alpha _{\theta _2}^4}\nonumber\\
   & &  + \frac{\sqrt{\frac{3}{2}} \left(\frac{1}{N}\right)^{9/10} \psi  \left( {r_h}^2-3 a^2\right)^2
   \left(9 a^2+ {r_h}^2\right)  {f_{\theta_1y}}(r) (r- {r_h}) \alpha _{\theta _1}^4 \log ^2\left(9 a^2  {r_h}^4+ {r_h}^6\right)
  }{128 \pi  ( {g_s}-1)  {g_s}^4 M^2  {N_f}^2  \sin^5\phi_{20}
    {r_h}^3 \left(6 a^2+ {r_h}^2\right) \log ^2( {r_h}) \alpha _{\theta _2}^3}\nonumber\\
    & &  \times  \left(\pi ^3 \alpha _{\theta _1}^2 \alpha _{\theta _2}^2 \log ^2\left(9 a^2  {r_h}^4+ {r_h}^6\right)+49 ( {g_s}-1)
    {g_s}^3 M^2  {N_f}^2 4\sin^2\left(\frac{\psi_0}{2}\right) \log ^2( {r_h})\right)\nonumber\\
    & &  +\frac{2048 \pi ^{9/2} a^2
   \sqrt[10]{\frac{1}{N}} 16\sin^4\left(\frac{\psi_0}{2}\right) \beta  \left(9 a^2+ {r_h}^2\right)^2 (\log ( {r_h})-1) \left(\left(9 a^2+ {r_h}^2\right)
   \log \left(9 a^2  {r_h}^4+ {r_h}^6\right)-8 \left(6 a^2+ {r_h}^2\right) \log ( {r_h})\right)^2}{729 ( {g_s}-1)
    {g_s}^{3/2}  {N_f}^6  \sin^4\phi_{20}   {r_h} \left(6 a^2+ {r_h}^2\right)^4 (r- {r_h})^2 \log ^3( {r_h}) \alpha
   _{\theta _2}^4 \log ^8\left(9 a^2  {r_h}^4+ {r_h}^6\right)}=0.\nonumber\\
& &
\end{eqnarray}
}

(iii) ${\rm EOM}_{rr}\sim{\rm EOM}_{x^1x^1}$

(iv) ${\rm EOM}_{r\theta_1}$:

{\footnotesize
\begin{eqnarray}
& &  -\frac{7 \pi ^3  \log N  \left(108 a^2+ {r_h}\right) \left( {r_h}^2-3 a^2\right)  {f_{\theta_1y}}(r) \alpha _{\theta _1}^7 \log ^4\left(9
   a^2  {r_h}^4+ {r_h}^6\right)}{1024 ( {g_s}-1)  {g_s}^3 M^2 \sqrt[5]{\frac{1}{N}}  {N_f}^2  \sin^4\phi_{20}   {r_h}^4 \log
   ^2( {r_h}) \alpha _{\theta _2}^2}\nonumber\\
& & +\frac{7 \pi ^3 4\sin^2\left(\frac{\psi_0}{2}\right) \left(108 a^2+ {r_h}\right)  {f_{xy}}(r) \alpha _{\theta _1}^6 \log
   ^2\left(9 a^2  {r_h}^4+ {r_h}^6\right)}{64 \sqrt{6}  {g_s}^3 M^2 \left(\frac{1}{N}\right)^{2/5}  {N_f}^2  \sin^4\phi_{20}
    {r_h}^2 \log ^2( {r_h}) \alpha _{\theta _2}^2}\nonumber\\
    & & -\frac{7 \pi ^3 4\sin^2\left(\frac{\psi_0}{2}\right) \left(108 a^2+ {r_h}\right)  {f_{yz}}(r) \alpha _{\theta
   _1}^6 \log ^2\left(9 a^2  {r_h}^4+ {r_h}^6\right)}{64 \sqrt{6}  {g_s}^3 M^2 \left(\frac{1}{N}\right)^{2/5}  {N_f}^2
    \sin^4\phi_{20}   {r_h}^2 \log ^2( {r_h}) \alpha _{\theta _2}^2}\nonumber\\
    & & \hskip -0.2in-\frac{163840 \pi ^{15/4} a^2  {g_s}^{3/4} M N^{13/20}  \sin\phi_{10}
   32\sin^5\left(\frac{\psi_0}{2}\right) \beta  \left(9 a^2+ {r_h}^2\right)^2 \left(\left(9 a^2+ {r_h}^2\right) \log \left(9 a^2  {r_h}^4+ {r_h}^6\right)-8
   \left(6 a^2+ {r_h}^2\right) \log ( {r_h})\right)^2}{243 \sqrt{3} ( {g_s}-1)  {N_f}^5  \sin^4\phi_{20}   {r_h}^3 \left(6
   a^2+ {r_h}^2\right)^4 (r- {r_h})^2 \log ( {r_h}) \alpha _{\theta _2}^5 \alpha _{\theta _1} \log ^{10}\left(9 a^2
    {r_h}^4+ {r_h}^6\right)}\nonumber\\
& &  =0.
\end{eqnarray}
}

(v) ${\rm EOM}_{r\theta_2}\sim$${\rm EOM}_{r\theta_1}$

(vi) ${\rm EOM}_{r\phi_1}$:

{\footnotesize
\begin{eqnarray}
& &  -\frac{10  \sin\phi_{10}  32\sin^5\left(\frac{\psi_0}{2}\right) (r- {r_h}) \alpha _{\theta _1}  {f_{x^{10} x^{10}}}'(r) \left(24 ( {g_s}-1)^2 \left(6 a^2+ {r_h}^2\right) \log ( {r_h})+(2
    {g_s}-3) \left(9 a^2+ {r_h}^2\right) \log \left(9 a^2  {r_h}^4+ {r_h}^6\right)\right)}{27 \sqrt{\pi } \sqrt{ {g_s}}
   \left(\frac{1}{N}\right)^{4/5}  \sin^4\phi_{20}  \left(6 a^2+ {r_h}^2\right) \log ( {r_h}) \alpha _{\theta _2}^5 \log ^3\left(9 a^2
    {r_h}^4+ {r_h}^6\right)}\nonumber\\
    & &  +\frac{20  \sin\phi_{10}  32\sin^5\left(\frac{\psi_0}{2}\right) (r- {r_h}) \alpha _{\theta _1}  {f_{\theta_1x^{10}}}'(r) \left(24 ( {g_s}-1)^2 \left(6
   a^2+ {r_h}^2\right) \log ( {r_h})+(2  {g_s}-3) \left(9 a^2+ {r_h}^2\right) \log \left(9 a^2  {r_h}^4+ {r_h}^6\right)\right)}{27 \sqrt{\pi }
   \sqrt{ {g_s}} \left(\frac{1}{N}\right)^{4/5}  \sin^4\phi_{20}  \left(6 a^2+ {r_h}^2\right) \log ( {r_h}) \alpha _{\theta _2}^5 \log ^3\left(9 a^2
    {r_h}^4+ {r_h}^6\right)}\nonumber\\
    & &  -\frac{10  \sin\phi_{10}  32\sin^5\left(\frac{\psi_0}{2}\right) (r- {r_h}) \alpha _{\theta _1}  {f_{\theta_1\theta_1}}'(r) \left(24 ( {g_s}-1)^2 \left(6
   a^2+ {r_h}^2\right) \log ( {r_h})+(2  {g_s}-3) \left(9 a^2+ {r_h}^2\right) \log \left(9 a^2  {r_h}^4+ {r_h}^6\right)\right)}{27 \sqrt{\pi }
   \sqrt{ {g_s}} \left(\frac{1}{N}\right)^{4/5}  \sin^4\phi_{20}  \left(6 a^2+ {r_h}^2\right) \log ( {r_h}) \alpha _{\theta _2}^5 \log ^3\left(9 a^2
    {r_h}^4+ {r_h}^6\right)}\nonumber\\
    & &  +\frac{5 ( {g_s}-1)  \log N   \sin\phi_{10}  2 \sin\left(\frac{\psi_0}{2}\right)  \left( {r_h}^2-3 a^2\right) \left(9 a^2+ {r_h}^2\right)
   (r- {r_h}) (2 \log ( {r_h})+1) \alpha _{\theta _1}^2  {f_{xy}}''(r)}{3 \sqrt{6 \pi } \sqrt{ {g_s}}  \sin^2\phi_{20}   {r_h} \left(6
   a^2+ {r_h}^2\right) \log ( {r_h}) \alpha _{\theta _2}^3}\nonumber\\
   & &  -\frac{5 ( {g_s}-1)  \log N   \sin\phi_{10}  4\sin^2\left(\frac{\psi_0}{2}\right) \left( {r_h}^2-3 a^2\right) \left(9
   a^2+ {r_h}^2\right) (r- {r_h}) (2 \log ( {r_h})+1) \alpha _{\theta _1}^2  {f_{xy}}'(r)}{18 \sqrt{\pi } \sqrt{ {g_s}}
   \left(\frac{1}{N}\right)^{3/10}  \sin^3\phi_{20}  {r_h}^2 \left(6 a^2+ {r_h}^2\right) \log ^2( {r_h}) \alpha _{\theta _2}^4}
   \nonumber\\
   & & \hskip -0.2in +\frac{655360 \pi ^5 a^2
   \sqrt{N}  \sin\phi_{10}  32\sin^5\left(\frac{\psi_0}{2}\right) \beta  \left(9 a^2+ {r_h}^2\right)^2 (\log ( {r_h})-1) \alpha _{\theta _1} \left(\left(9 a^2+ {r_h}^2\right) \log
   \left(9 a^2  {r_h}^4+ {r_h}^6\right)-8 \left(6 a^2+ {r_h}^2\right) \log ( {r_h})\right)^2}{2187 ( {g_s}-1)  {g_s}  {N_f}^6  \sin^4\phi_{20}
    {r_h}^3 \left(6 a^2+ {r_h}^2\right)^4 (r- {r_h})^2 \log ^3( {r_h}) \alpha _{\theta _2}^5 \log ^{10}\left(9 a^2  {r_h}^4+ {r_h}^6\right)}
    \nonumber\\
    & &   =0.
\end{eqnarray}
}

(vii) ${\rm EOM}_{r\phi_2}$:

{\footnotesize
\begin{eqnarray}
& &  \frac{28 \pi ^{17/4} \sin^2\left(\frac{\psi_0}{2}\right) \left(108 a^2+ {r_h}\right)  {f_{x^{10} x^{10}}}(r) \alpha _{\theta _1}^4 \log ^4\left(9 a^2
    {r_h}^4+ {r_h}^6\right)}{4608 \sqrt{3} ( {g_s}-1)  {g_s}^{19/4} M^3 \left(\frac{1}{N}\right)^{3/4}  {N_f}^3  \sin^6\phi_{20}
    {r_h}^2 \log ^3( {r_h}) \alpha _{\theta _2}}\nonumber\\
& & -\frac{28 \pi ^{17/4} \sin^2\left(\frac{\psi_0}{2}\right) \left(108 a^2+ {r_h}\right)  {f_{\theta_1x^{10}}}(r) \alpha _{\theta
   _1}^4 \log ^4\left(9 a^2  {r_h}^4+ {r_h}^6\right)}{2304 \sqrt{3} ( {g_s}-1)  {g_s}^{19/4} M^3 \left(\frac{1}{N}\right)^{3/4}
    {N_f}^3  \sin^6\phi_{20}   {r_h}^2 \log ^3( {r_h}) \alpha _{\theta _2}}\nonumber\\
    & & +\frac{28 \pi ^{17/4} \sin^2\left(\frac{\psi_0}{2}\right) \left(108 a^2+ {r_h}\right)
    {f_{\theta_1\theta_1}}(r) \alpha _{\theta _1}^4 \log ^4\left(9 a^2  {r_h}^4+ {r_h}^6\right)}{4608 \sqrt{3} ( {g_s}-1)  {g_s}^{19/4} M^3
   \left(\frac{1}{N}\right)^{3/4}  {N_f}^3  \sin^6\phi_{20}   {r_h}^2 \log ^3( {r_h}) \alpha _{\theta _2}}\nonumber\\
   &&+\frac{7 \pi ^{17/4} \left(108
   a^2+ {r_h}\right)  {f_{\theta_1\theta_2}}(r) \alpha _{\theta _2} \alpha _{\theta _1}^4 \log ^4\left(9 a^2  {r_h}^4+ {r_h}^6\right)}{768 \sqrt{3}
   ( {g_s}-1)^2  {g_s}^{19/4} M^3 \left(\frac{1}{N}\right)^{3/20}  {N_f}^3  \sin^4\phi_{20}   {r_h}^2 \log ^3( {r_h})}\nonumber\\
   & & +\frac{7
   \sqrt{3} \pi ^{17/4}  \log N ^2 \left(\frac{1}{N}\right)^{9/20} \left(108 a^2+ {r_h}\right) \left( {r_h}^2-3 a^2\right)^2
    {f_{yy}}(r) (2 \log ( {r_h})+1)^2 \alpha _{\theta _1}^{10} \log ^6\left(9 a^2  {r_h}^4+ {r_h}^6\right)}{65536 ( {g_s}-1)
    {g_s}^{19/4} M^3  {N_f}^3  \sin^6\phi_{20}   {r_h}^6 \log ^5( {r_h}) \alpha _{\theta _2}}=0
    \nonumber\\
    & &
\end{eqnarray}
}

(viii) ${\rm EOM}_{\theta_1\theta_1}$:

{\scriptsize
\begin{eqnarray}
& &  -\frac{ {g_s}^3 M^2  {N_f}^2 64\sin^6\left(\frac{\psi_0}{2}\right) (r- {r_h}) \log ( {r_h}) \left(8 \left(6 a^2+ {r_h}^2\right) \log ( {r_h})-3 \left(9
   a^2+ {r_h}^2\right) \log \left(9 a^2  {r_h}^4+ {r_h}^6\right)\right)  {f_{x^{10} x^{10}}}'(r)}{72 \pi ^3 \left(\frac{1}{N}\right)^{8/5}
    \sin^4\phi_{20}  \left(6 a^2+ {r_h}^2\right) \alpha _{\theta _1}^4 \alpha _{\theta _2}^6 \log ^3\left(9 a^2
    {r_h}^4+ {r_h}^6\right)}\nonumber\\
    & &  -\frac{ {g_s}^3 M^2  {N_f}^2 256\sin^8\left(\frac{\psi_0}{2}\right) \left(9 a^2+ {r_h}^2\right)  {f_{x^{10} x^{10}}}(r)
   (r- {r_h})}{216 \pi ^3 \left(\frac{1}{N}\right)^{11/5}  \sin^6\phi_{20}  \left(6 a^2  {r_h}+ {r_h}^3\right) \alpha _{\theta _1}^4
   \alpha _{\theta _2}^8 \log ^2\left(9 a^2  {r_h}^4+ {r_h}^6\right)}\nonumber\\
   & &  +\frac{ {g_s}^3 M^2  {N_f}^2 64\sin^6\left(\frac{\psi_0}{2}\right) (r- {r_h}) \log
   ( {r_h}) \left(8 \left(6 a^2+ {r_h}^2\right) \log ( {r_h})-\left(9 a^2+ {r_h}^2\right) \log \left(9 a^2
    {r_h}^4+ {r_h}^6\right)\right)  {f_{\theta_1x^{10}}}'(r)}{36 \pi ^3 \left(\frac{1}{N}\right)^{8/5}  \sin^4\phi_{20}  \left(6
   a^2+ {r_h}^2\right) \alpha _{\theta _1}^4 \alpha _{\theta _2}^6 \log ^3\left(9 a^2  {r_h}^4+ {r_h}^6\right)}
   \nonumber\\
   & &  +\frac{ {g_s}^3
   M^2  {N_f}^2 256\sin^8\left(\frac{\psi_0}{2}\right) \left(9 a^2+ {r_h}^2\right)  {f_{\theta_1x^{10}}}(r) (r- {r_h})}{108 \pi ^3 \left(\frac{1}{N}\right)^{11/5}
    \sin^6\phi_{20}  \left(6 a^2  {r_h}+ {r_h}^3\right) \alpha _{\theta _1}^4 \alpha _{\theta _2}^8 \log ^2\left(9 a^2
    {r_h}^4+ {r_h}^6\right)}\nonumber\\
    & &  -\frac{ {g_s}^3 M^2  {N_f}^2 64\sin^6\left(\frac{\psi_0}{2}\right) (r- {r_h}) \log ( {r_h}) \left(8 \left(6
   a^2+ {r_h}^2\right) \log ( {r_h})+\left(9 a^2+ {r_h}^2\right) \log \left(9 a^2  {r_h}^4+ {r_h}^6\right)\right)
    {f_{\theta_1\theta_1}}'(r)}{72 \pi ^3 \left(\frac{1}{N}\right)^{8/5}  \sin^4\phi_{20}  \left(6 a^2+ {r_h}^2\right) \alpha _{\theta _1}^4 \alpha
   _{\theta _2}^6 \log ^3\left(9 a^2  {r_h}^4+ {r_h}^6\right)}\nonumber\\
   & &  -\frac{ {g_s}^3 M^2  {N_f}^2 256\sin^8\left(\frac{\psi_0}{2}\right) \left(9 a^2+ {r_h}^2\right)
    {f_{\theta_1\theta_1}}(r) (r- {r_h})}{216 \pi ^3 \left(\frac{1}{N}\right)^{11/5}  \sin^6\phi_{20}  \left(6 a^2  {r_h}+ {r_h}^3\right) \alpha
   _{\theta _1}^4 \alpha _{\theta _2}^8 \log ^2\left(9 a^2  {r_h}^4+ {r_h}^6\right)}\nonumber\\
   & &  -\frac{8192 \pi ^{5/2} a^2  {g_s}^{5/2} M^2
   N^{13/10} 64\sin^6\left(\frac{\psi_0}{2}\right) \beta  \left(9 a^2+ {r_h}^2\right)^2 (\log ( {r_h})-1) \left(\left(9 a^2+ {r_h}^2\right) \log \left(9 a^2
    {r_h}^4+ {r_h}^6\right)-8 \left(6 a^2+ {r_h}^2\right) \log ( {r_h})\right)^2}{729 ( {g_s}-1)  {N_f}^4  \sin^4\phi_{20}
    {r_h}^3 \left(6 a^2+ {r_h}^2\right)^4 (r- {r_h})^2 \log ( {r_h}) \alpha _{\theta _1}^4 \alpha _{\theta _2}^6 \log
   ^{10}\left(9 a^2  {r_h}^4+ {r_h}^6\right)} \nonumber\\
   & &  =0
\end{eqnarray}
}

(ix) EOM$\ {\theta_1\theta_2}=$${\rm EOM}_{\theta_1\theta_1}$:

(x) ${\rm EOM}_{\theta_1x}$:

{\scriptsize
\begin{eqnarray*}
& &  \frac{ 256{g_s}^{5/4} M  {N_f} (r- {r_h}) \left(9 a^2+ {r_h}^2\right)  {f_{x^{10} x^{10}}}(r) \sin^8\left(\frac{\psi_0}{2}\right)}{162 \sqrt{3}
   \left(\frac{1}{N}\right)^{41/20}  \sin^6\phi_{20}  \pi ^{7/4}  {r_h} \left(6 a^2+ {r_h}^2\right) \log ( {r_h}) \log
   ^2\left( {r_h}^6+9 a^2  {r_h}^4\right) \alpha _{\theta _1}^2 \alpha _{\theta _2}^8}\nonumber\\
   & &  -\frac{ 256{g_s}^{5/4} M  {N_f} (r- {r_h})
   \left(9 a^2+ {r_h}^2\right)  {f_{\theta_1x^{10}}}(r) \sin^8\left(\frac{\psi_0}{2}\right)}{81 \sqrt{3} \left(\frac{1}{N}\right)^{41/20}  \sin^6\phi_{20}  \pi ^{7/4}  {r_h}
   \left(6 a^2+ {r_h}^2\right) \log ( {r_h}) \log ^2\left( {r_h}^6+9 a^2  {r_h}^4\right) \alpha _{\theta _1}^2 \alpha _{\theta
   _2}^8}\nonumber\\
   & &  +\frac{ 256{g_s}^{5/4} M  {N_f} (r- {r_h}) \left(9 a^2+ {r_h}^2\right)  {f_{\theta_1\theta_1}}(r) \sin^8\left(\frac{\psi_0}{2}\right)}{162 \sqrt{3}
   \left(\frac{1}{N}\right)^{41/20}  \sin^6\phi_{20}  \pi ^{7/4}  {r_h} \left(6 a^2+ {r_h}^2\right) \log ( {r_h}) \log
   ^2\left( {r_h}^6+9 a^2  {r_h}^4\right) \alpha _{\theta _1}^2 \alpha _{\theta _2}^8}\nonumber\\
   & &  +\frac{ {g_s}^{5/4} M  {N_f} (r- {r_h})
   \left(8 \left(6 a^2+ {r_h}^2\right) \log ( {r_h})-3 \left(9 a^2+ {r_h}^2\right) \log \left( {r_h}^6+9 a^2
    {r_h}^4\right)\right)  {f_{x^{10} x^{10}}}'(r) 64\sin^6\left(\frac{\psi_0}{2}\right)}{54 \sqrt{3} \left(\frac{1}{N}\right)^{29/20}  \sin^4\phi_{20}  \pi ^{7/4} \left(6
   a^2+ {r_h}^2\right) \log ^3\left( {r_h}^6+9 a^2  {r_h}^4\right) \alpha _{\theta _1}^2 \alpha _{\theta _2}^6}
   \nonumber\\
   & &  +\frac{ {g_s}^{5/4} M
    {N_f} (r- {r_h}) \left(\left(9 a^2+ {r_h}^2\right) \log \left( {r_h}^6+9 a^2  {r_h}^4\right)-8 \left(6 a^2+ {r_h}^2\right)
   \log ( {r_h})\right)  {f_{\theta_1x^{10}}}'(r) 64\sin^6\left(\frac{\psi_0}{2}\right)}{27 \sqrt{3} \left(\frac{1}{N}\right)^{29/20}  \sin^4\phi_{20}  \pi ^{7/4} \left(6
   a^2+ {r_h}^2\right) \log ^3\left( {r_h}^6+9 a^2  {r_h}^4\right) \alpha _{\theta _1}^2 \alpha _{\theta _2}^6}
   \nonumber\\
    \end{eqnarray*}}
   {\scriptsize
   \begin{eqnarray*}
   & &  +\frac{ {g_s}^{5/4} M
    {N_f} (r- {r_h}) \left(8 \left(6 a^2+ {r_h}^2\right) \log ( {r_h})+\left(9 a^2+ {r_h}^2\right) \log \left( {r_h}^6+9 a^2
    {r_h}^4\right)\right)  {f_{\theta_1\theta_1}}'(r) 64\sin^6\left(\frac{\psi_0}{2}\right)}{54 \sqrt{3} \left(\frac{1}{N}\right)^{29/20}  \sin^4\phi_{20}  \pi ^{7/4} \left(6
   a^2+ {r_h}^2\right) \log ^3\left( {r_h}^6+9 a^2  {r_h}^4\right) \alpha _{\theta _1}^2 \alpha _{\theta _2}^6}
   \nonumber\\
   & &  +\frac{32768 a^2
    {g_s}^{3/4} M N^{23/20} \pi ^{15/4} \left(9 a^2+ {r_h}^2\right)^2 \beta  (\log ( {r_h})-1) \left(\left(9 a^2+ {r_h}^2\right)
   \log \left( {r_h}^6+9 a^2  {r_h}^4\right)-8 \left(6 a^2+ {r_h}^2\right) \log ( {r_h})\right)^2 64\sin^6\left(\frac{\psi_0}{2}\right)}{2187 \sqrt{3}
   ( {g_s}-1)  {N_f}^5  \sin^4\phi_{20}  (r- {r_h})^2  {r_h}^3 \left(6 a^2+ {r_h}^2\right)^4 \log ^2( {r_h}) \log
   ^{10}\left( {r_h}^6+9 a^2  {r_h}^4\right) \alpha _{\theta _1}^2 \alpha _{\theta _2}^6}\nonumber\\
   & &  -\frac{ {g_s}^{5/4}  \log N  M  {N_f}
   (r- {r_h}) \left( {r_h}^2-3 a^2\right) \left(9 a^2+ {r_h}^2\right) (2 \log ( {r_h})+1)  {f_{xy}}'(r) 4\sin^2\left(\frac{\psi_0}{2}\right)}{36 \sqrt{2}
   \left(\frac{1}{N}\right)^{13/20}  \sin^2\phi_{20}  \pi ^{7/4}  {r_h}^2 \left(6 a^2+ {r_h}^2\right) \log ( {r_h}) \alpha _{\theta _1}
   \alpha _{\theta _2}^4}=0.
\end{eqnarray*}
}

(xi) {\footnotesize ${\rm EOM}_{\theta_1y}={\rm EOM}_{\theta_1\theta_1}={\rm EOM}_{\theta_1z}={\rm EOM}_{\theta_1x^{10}}={\rm EOM}_{\theta_2\theta_2}={\rm EOM}_{\theta_2x}={\rm EOM}_{\theta_2y}={\rm EOM}_{\theta_2z}={\rm EOM}_{\theta_2x^{10}}={\rm EOM}_{xy}$
=${\rm EOM}_{xz}$=
${\rm EOM}_{x x^{10}}={\rm EOM}_{yy}={\rm EOM}_{yz}={\rm EOM}_{zz}={\rm EOM}_{x^{10}x^{10}}$}

(xii) ${\rm EOM}_{xx}$:

{\scriptsize
\begin{eqnarray}
& &  -\frac{32  {r_h} \left(9 a^2+ {r_h}^2\right) (r- {r_h}) \alpha _{\theta _2}^2  {f_{xz}}'(r)
  }{177147 \sqrt{\pi } ( {g_s}-1)
    {g_s}^{7/2} M^2 \left(\frac{1}{N}\right)^{7/10}  {N_f}^2  \sin^2\phi_{20}  4\sin^2\left(\frac{\psi_0}{2}\right) \left(6
   a^2+ {r_h}^2\right) \log ^2( {r_h}) \alpha _{\theta _1}^6 \log ^3\left(9 a^2
    {r_h}^4+ {r_h}^6\right)}\nonumber\\
    & &  \times  \Biggl(\frac{63 ( {g_s}-1)  {g_s}^3 M^2  {N_f}^2  \sin^2\phi_{20}  4\sin^2\left(\frac{\psi_0}{2}\right) \left(6 a^2+ {r_h}^2\right)
   \log ^2( {r_h}) \left(4 \alpha _{\theta _2}^2-27 \sqrt{6} \alpha _{\theta _1}^3\right) \log ^2\left(9 a^2
    {r_h}^4+ {r_h}^6\right)}{\left(9 a^2  {r_h}+ {r_h}^3\right) \alpha _{\theta _2}^4}\nonumber\\
    & &  -\frac{14 \pi ^3
    \sin^2\phi_{20}  \left(6 a^2+ {r_h}^2\right) \alpha _{\theta _1}^2 \log ^4\left(9 a^2
    {r_h}^4+ {r_h}^6\right)}{9 a^2  {r_h}+ {r_h}^3}\Biggr)\nonumber\\
    & &  -\frac{224 (r- {r_h})  {f_{xy}}'(r) }{177147 \sqrt{\pi } ( {g_s}-1)
    {g_s}^{7/2} M^2 \left(\frac{1}{N}\right)^{7/10}  {N_f}^2 4\sin^2\left(\frac{\psi_0}{2}\right) \log ^2( {r_h}) \alpha _{\theta _1}^6
   \alpha _{\theta _2}^2 \log \left(9 a^2  {r_h}^4+ {r_h}^6\right)}\nonumber\\
& & \times \left(2 \pi ^3 \alpha _{\theta _1}^2
   \alpha _{\theta _2}^4 \log ^2\left(9 a^2  {r_h}^4+ {r_h}^6\right)+90 ( {g_s}-1)  {g_s}^3 M^2
    {N_f}^2 4\sin^2\left(\frac{\psi_0}{2}\right) \log ^2( {r_h}) \alpha _{\theta _2}^2+243 \sqrt{6} ( {g_s}-1)  {g_s}^3 M^2
    {N_f}^2 4\sin^2\left(\frac{\psi_0}{2}\right) \log ^2( {r_h}) \alpha _{\theta _1}^3\right)   =0\nonumber\\
    & &
\end{eqnarray}
}

(xiii) ${\rm EOM}_{y x^{10}}$:

{\footnotesize
 \begin{eqnarray*}
 & & \frac{\sqrt{\frac{\pi }{6}}  \log N  \sqrt{\frac{1}{N}} 2 \sin \left(\frac{\psi_0}{2}\right)  \left(9 a^2+ {r_h}^2\right) (r- {r_h}) \alpha _{\theta _1}^3
   \left(6 a^2 \log ( {r_h})+ {r_h}^2\right)  {f_{x^{10} x^{10}}}'(r)}{\sqrt{ {g_s}}  {N_f}  \sin^2\phi_{20}   {r_h}^2 \left(6
   a^2+ {r_h}^2\right) \log ( {r_h}) \alpha _{\theta _2}^2}\nonumber\\
   & & -\frac{8\sqrt{\frac{\pi }{6}}  \log N  \sin^3\left(\frac{\psi_0}{2}\right) \left(9
   a^2+ {r_h}^2\right)  {f_{x^{10} x^{10}}}(r) (r- {r_h}) \alpha _{\theta _1}^3 \left(6 a^2 \log ( {r_h})+ {r_h}^2\right)}{6
   \sqrt{ {g_s}} \sqrt[10]{\frac{1}{N}}  {N_f}  \sin^4\phi_{20}   {r_h}^3 \left(6 a^2+ {r_h}^2\right) \log ^2( {r_h})
   \alpha _{\theta _2}^4}\nonumber\\
   & & +\frac{49 \pi ^{13/2}  \log N  \left(\frac{1}{N}\right)^{3/10} \left( {r_h}^2-3 a^2\right)
    {f_{\theta_1z}}(r) (2 \log ( {r_h})+1) \alpha _{\theta _1}^9 \log ^4\left(9 a^2  {r_h}^4+ {r_h}^6\right)}{9216 \sqrt{6}
   ( {g_s}-1)  {g_s}^{13/2} M^4  {N_f}^5  \sin^4\phi_{20}  \sin\left(\frac{\psi_0}{2}\right)   {r_h}^2 \log ^5( {r_h})}\nonumber\\
   & & -\frac{\sqrt{\frac{\pi }{6}}
    \log N  \sqrt{\frac{1}{N}} \sin\left(\frac{\psi_0}{2}\right)  \left(9 a^2+ {r_h}^2\right) (r- {r_h}) \alpha _{\theta _1}^3 \left(6 a^2 \log
   ( {r_h})+ {r_h}^2\right)  {f_{\theta_1x^{10}}}'(r)}{\sqrt{ {g_s}}  {N_f}  \sin^2\phi_{20}   {r_h}^2 \left(6
   a^2+ {r_h}^2\right) \log ( {r_h}) \alpha _{\theta _2}^2}\nonumber\\
   & & +\frac{8\sqrt{\frac{\pi }{6}}  \log N  \sin^3\left(\frac{\psi_0}{2}\right) \left(9
   a^2+ {r_h}^2\right)  {f_{\theta_1x^{10}}}(r) (r- {r_h}) \alpha _{\theta _1}^3 \left(6 a^2 \log ( {r_h})+ {r_h}^2\right)}{3
   \sqrt{ {g_s}} \sqrt[10]{\frac{1}{N}}  {N_f}  \sin^4\phi_{20}   {r_h}^3 \left(6 a^2+ {r_h}^2\right) \log ^2( {r_h})
   \alpha _{\theta _2}^4}\nonumber\\
   & & -\frac{8\sqrt{\frac{\pi }{6}}  \log N  \sin^3\left(\frac{\psi_0}{2}\right) \left(9 a^2+ {r_h}^2\right)  {f_{\theta_1\theta_1}}(r)
   (r- {r_h}) \alpha _{\theta _1}^3 \left(6 a^2 \log ( {r_h})+ {r_h}^2\right)}{6 \sqrt{ {g_s}} \sqrt[10]{\frac{1}{N}}
    {N_f}  \sin^4\phi_{20}   {r_h}^3 \left(6 a^2+ {r_h}^2\right) \log ^2( {r_h}) \alpha _{\theta _2}^4}\nonumber\\
    & & -\frac{49 \pi
   ^{13/2}  \log N  \left(\frac{1}{N}\right)^{3/10} \left( {r_h}^2-3 a^2\right)  {f_{\theta_1y}}(r) (2 \log ( {r_h})+1) \alpha
   _{\theta _1}^9 \log ^4\left(9 a^2  {r_h}^4+ {r_h}^6\right)}{9216 \sqrt{6} ( {g_s}-1)  {g_s}^{13/2} M^4  {N_f}^5
    \sin^4\phi_{20}  \sin\left(\frac{\psi_0}{2}\right)  {r_h}^2 \log ^5( {r_h})}\nonumber\\
    & & +\frac{49 \pi ^{13/2}  \log N  \left(\frac{1}{N}\right)^{3/10}
   \left( {r_h}^2-3 a^2\right)  {f_{yy}}(r) (2 \log ( {r_h})+1) \alpha _{\theta _1}^9 \log ^4\left(9 a^2
    {r_h}^4+ {r_h}^6\right)}{9216 \sqrt{6} ( {g_s}-1)  {g_s}^{13/2} M^4  {N_f}^5  \sin^4\phi_{20}  \psi   {r_h}^2
   \log ^5( {r_h})}=0.
 \end{eqnarray*}
}

 (xiv) ${\rm EOM}_{z x^{10}}$: \\
{\scriptsize
 \begin{eqnarray}
 & & -\frac{\sqrt{\pi }  \log N  \sqrt[5]{\frac{1}{N}} \sin\left(\frac{\psi_0}{2}\right)  \left(9 a^2+ {r_h}^2\right) (r- {r_h})
   \alpha _{\theta _1}^3 \left(6 a^2 \log ( {r_h})+ {r_h}^2\right)  {f_{x^{10} x^{10}}}'(r)}{9 \sqrt{ {g_s}}
    {N_f}  \sin^2\phi_{20}   {r_h}^2 \left(6 a^2+ {r_h}^2\right) \log ( {r_h}) \alpha _{\theta
   _2}^3}\nonumber\\
   & & +\frac{8\sqrt{\pi }  \log N  \sin^3\left(\frac{\psi_0}{2}\right) \left(9 a^2+ {r_h}^2\right)  {f_{x^{10} x^{10}}}(r) (r- {r_h})
   \alpha _{\theta _1}^3 \left(6 a^2 \log ( {r_h})+ {r_h}^2\right)}{54 \sqrt{ {g_s}}
   \left(\frac{1}{N}\right)^{2/5}  {N_f}  \sin^4\phi_{20}   {r_h}^3 \left(6 a^2+ {r_h}^2\right) \log
   ^2( {r_h}) \alpha _{\theta _2}^5}\nonumber\\
   & & +\frac{\sqrt{\pi }  \log N  \sqrt[5]{\frac{1}{N}} \sin\left(\frac{\psi_0}{2}\right)  \left(9
   a^2+ {r_h}^2\right) (r- {r_h}) \alpha _{\theta _1}^3 \left(6 a^2 \log ( {r_h})+ {r_h}^2\right)
    {f_{\theta_1x^{10}}}'(r)}{9 \sqrt{ {g_s}}  {N_f}  \sin^2\phi_{20}   {r_h}^2 \left(6 a^2+ {r_h}^2\right)
   \log ( {r_h}) \alpha _{\theta _2}^3}\nonumber\\
   & & -\frac{8\sqrt{\pi }  \log N  \sin^3\left(\frac{\psi_0}{2}\right) \left(9
   a^2+ {r_h}^2\right)  {f_{\theta_1x^{10}}}(r) (r- {r_h}) \alpha _{\theta _1}^3 \left(6 a^2 \log
   ( {r_h})+ {r_h}^2\right)}{27 \sqrt{ {g_s}} \left(\frac{1}{N}\right)^{2/5}  {N_f}  \sin^4\phi_{20}
    {r_h}^3 \left(6 a^2+ {r_h}^2\right) \log ^2( {r_h}) \alpha _{\theta _2}^5}\nonumber\\
    & & +\frac{8\sqrt{\pi }
    \log N  \sin^3\left(\frac{\psi_0}{2}\right) \left(9 a^2+ {r_h}^2\right)  {f_{\theta_1\theta_1}}(r) (r- {r_h}) \alpha _{\theta _1}^3
   \left(6 a^2 \log ( {r_h})+ {r_h}^2\right)}{54 \sqrt{ {g_s}} \left(\frac{1}{N}\right)^{2/5}
    {N_f}  \sin^4\phi_{20}   {r_h}^3 \left(6 a^2+ {r_h}^2\right) \log ^2( {r_h}) \alpha _{\theta
   _2}^5}=0.
\end{eqnarray}
}

\section{${\cal T}^{(3)}_{PQ}$ \label{seceqbb}}

In this appendix, we list the non-zero components of ${\cal T}^{(3)}_{MN}$ that figure in the EOMs via (\ref{TMN-first-order-in-beta-1}) - (\ref{TMN-first-order-in-beta-2}) near $\psi=0, 2\pi, 4\pi$-coordinate patches in {\bf D.2.1} and away from the same in {\bf D.2.2}
\subsection{$\psi=0, 2\pi, 4\pi$-Coordinate Patches }
We begin by listing out the non-zero components of $T^{(3)}_{MN}$ near the $\psi=0, 2\pi, 4\pi$-coordinate patches:
{\footnotesize
\begin{eqnarray*}
& & {\cal T}^{(3)}_{rr} = \frac{15 {g_s}^{7/2} M^2 N^{2/5} {N_f}^2 \log ^2(r) \left(108 b^2
   {r_h}^2+r\right)^2}{\pi ^{5/2} r^4 \ b_1(r)^2 \alpha _{\theta
   _1}^4 \alpha _{\theta _2}^4},\nonumber\\
   & & {\cal T}^{(3)}_{r\theta_1} = -\frac{63 {g_s}^{7/4} M \sqrt[5]{N} {N_f} \log (r) \left(108 b^2
   {r_h}^2+r\right)}{2 \sqrt{2} \pi ^{5/4} r^2 \ b_1(r) \alpha
   _{\theta _1}^2 \alpha _{\theta _2}^2},\nonumber\\
      & & {\cal T}^{(3)}_{r\theta_2} = -\frac{7 \sqrt{\frac{2}{3}} {g_s}^2 M N^{9/10} {N_f} \log (r)
   \left(108 b^2 {r_h}^2+r\right)}{9 \pi  r^2 \ b_1(r)^2 \alpha
   _{\theta _1}^4 \alpha _{\theta _2}^4},\nonumber\\
   & & {\cal T}^{(3)}_{r\phi_1} = \frac{5 {g_s}^{15/4} \log N  M^3 \sqrt[20]{N} {N_f}^2 \log
   ^2(r) \left(108 b^2 {r_h}^2+r\right)}{18 \sqrt{2} \pi ^{13/4} r^2
   \ b_1(r)^2 \alpha _{\theta _1}^6 \alpha _{\theta _2}^4},\nonumber\\
   & & {\cal T}^{(3)}_{r\phi_2} = -\frac{35 {g_s}^{7/4} M N^{7/20} {N_f} \log (r) \left(108 b^2
   {r_h}^2+r\right)}{4 \sqrt{3} \pi ^{5/4} r^2 \ b_1(r)^2 \alpha
   _{\theta _1}^4 \alpha _{\theta _2}^3},\nonumber\\
   & & {\cal T}^{(3)}_{r\psi} = -\frac{5 {g_s}^{15/4} M^3 \left(\frac{1}{N}\right)^{7/20} {N_f}^2
   \log ^2(r) \left(108 b^2 {r_h}^2+r\right)}{12 \sqrt{2} \pi ^{13/4}
   r^2 \ b_1(r)^2 \alpha _{\theta _1}^4 \alpha _{\theta _2}^4}\nonumber\\
   & & {\cal T}^{(3)}_{rx^{10}} = \frac{49 {g_s}^{5/4}  \log N  M^2
   \left(\frac{1}{N}\right)^{17/20} {N_f}}{17280 \sqrt{2} \pi ^{7/4}
   r \ a_1(r)^2 \ b_1(r)^2 \alpha _{\theta _1}^2 \alpha _{\theta
   _2}^2}\nonumber\\
   & & {\cal T}^{(3)}_{\theta_1\theta_1} = -\frac{1323 \sqrt[5]{N} \alpha _{\theta _1}^2}{32 \alpha _{\theta _2}^2},\nonumber\\
   & & {\cal T}^{(3)}_{\theta_1\theta_2} = -\frac{1323 N^{3/10} \ b_2(r) \alpha _{\theta _1}^2}{32 \ b_1(r)
   \alpha _{\theta _2}^2}\nonumber\\
   & & {\cal T}^{(3)}_{\theta_1x} = -\frac{35 {g_s}^2 \log N  M^2 \left(\frac{1}{N}\right)^{3/20}
   {N_f} \log (r)}{96 \pi ^2 \ b_1(r) \alpha _{\theta _1}^4
   \alpha _{\theta _2}^2},\nonumber\\
   & & {\cal T}^{(3)}_{\theta_1y} = \frac{245 \sqrt{3} N^{3/20}}{32 \sqrt{2} \ b_1(r) \alpha _{\theta
   _1}^2 \alpha _{\theta _2}},\nonumber\\
   & & {\cal T}^{(3)}_{\theta_1z} = \frac{35 {g_s}^2 M^2 \left(\frac{1}{N}\right)^{11/20} {N_f} \log
   (r)}{64 \pi ^2 \ b_1(r) \alpha _{\theta _1}^2 \alpha _{\theta
   _2}^2},\nonumber\\
   \end{eqnarray*}}
   {\scriptsize
   \begin{eqnarray}
   & & {\cal T}^{(3)}_{\theta_1x^{10}} = \frac{27 {g_s}^{5/2}  \log N  M^3
   \left(\frac{1}{N}\right)^{29/20} {N_f}^2 \left(1-\frac{r_h^4}{r^4}\right) \log (r)
   \left(108 b^2 {r_h}^2+r\right)}{1280 \pi ^{7/2} r \ a_1(r)^2
   \ a_4(r) \ b_1(r) \alpha _{\theta _2}^2},\nonumber\\
    & & {\cal T}^{(3)}_{\theta_2\theta_2} = -\frac{2 {g_s}^{7/2} M^2 N^{4/5} {N_f}^2 \left(1-\frac{r_h^4}{r^4}\right) \log ^2(r)
   \left(108 b^2 {r_h}^2+r\right)^2}{9 \pi ^{5/2} r^2 \ a_4(r)
   \ b_1(r)^2 \alpha _{\theta _1}^6 \alpha _{\theta _2}^4},\nonumber\\
   & & {\cal T}^{(3)}_{\theta_2x} = -\frac{35 {g_s}^{9/4} \log N  M^2 N^{11/20} {N_f} \log
   (r)}{1944 \sqrt{3} \pi ^{7/4} \ b_1(r)^2 \alpha _{\theta _1}^6
   \alpha _{\theta _2}^4},\nonumber\\
   & & {\cal T}^{(3)}_{\theta_2y} = \frac{245 \sqrt[4]{\pi } \sqrt[4]{{g_s}} N^{17/20}}{648 \sqrt{2}
   \ b_1(r)^2 \alpha _{\theta _1}^4 \alpha _{\theta _2}^3},\nonumber\\
& & {\cal T}^{(3)}_{\theta_2z} = \frac{35 {g_s}^{9/4} M^2 N^{3/20} {N_f} \log (r)}{1296 \sqrt{3}
   \pi ^{7/4} \ b_1(r)^2 \alpha _{\theta _1}^4 \alpha _{\theta _2}^4},\nonumber\\
   & & {\cal T}^{(3)}_{\theta_2x^{10}} =  \frac{7 {g_s}^{11/4}  \log N  M^3
   \left(\frac{1}{N}\right)^{19/20} {N_f}^2 \left(1-\frac{r_h^4}{r^4}\right) \log (r)
   \left(108 b^2 {r_h}^2+r\right)}{8640 \sqrt{3} \pi ^{13/4} r
   \ a_1(r)^2 \ a_4(r) \ b_1(r)^2 \alpha _{\theta _1}^4
   \alpha _{\theta _2}^2}, \nonumber\\
 & & {\cal T}^{(3)}_{xx} = \frac{{g_s}^4 \log N ^2 M^4 \left(\frac{1}{N}\right)^{3/10}
   {N_f}^2 \log ^2(r)}{324 \pi ^4 \ b_1(r)^2 \alpha _{\theta
   _1}^8 \alpha _{\theta _2}^4},\nonumber\\
& & T^{(3)}_{xy} = -\frac{7 {g_s}^2 \log N  M^2 {N_f} \log (r)}{36 \sqrt{6} \pi
   ^2 \ b_1(r)^2 \alpha _{\theta _1}^6 \alpha _{\theta _2}^3},\nonumber\\
& & T^{(3)}_{xz} = -\frac{{g_s}^4 \log N  M^4 \left(\frac{1}{N}\right)^{7/10}
   {N_f}^2 \log ^2(r)}{216 \pi ^4 \ b_1(r)^2 \alpha _{\theta
   _1}^6 \alpha _{\theta _2}^4},\nonumber\\
& & T^{(3)}_{x x^{10}} = \frac{{g_s}^{5/2}  \log N  M^3
   \left(\frac{1}{N}\right)^{3/5} {N_f}^2 \left(1-\frac{r_h^4}{r^4}\right) \log (r)
   \left(108 b^2 {r_h}^2+r\right)}{720 \pi ^{7/2} r \ a_1(r)^2
   \ a_4(r) \ b_1(r)^2 \alpha _{\theta _1}^4 \alpha _{\theta
   _2}^4},\nonumber\\
& & T^{(3)}_{yy} = \frac{49 N^{3/10}}{24 \ b_1(r)^2 \alpha _{\theta _1}^4 \alpha
   _{\theta _2}^2},\nonumber\\
& & T^{(3)}_{yz} = \frac{7 {g_s}^2 M^2 \left(\frac{1}{N}\right)^{2/5} {N_f} \log
   (r)}{24 \sqrt{6} \pi ^2 \ b_1(r)^2 \alpha _{\theta _1}^4 \alpha
   _{\theta _2}^3},\nonumber\\
& & T^{(3)}_{y x^{10}} = \frac{\sqrt{\frac{3}{2}} {g_s}^{5/2}
   \log N  M^3 \left(\frac{1}{N}\right)^{13/10} {N_f}^2
   \left(1-\frac{r_h^4}{r^4}\right) \log (r) \left(108 b^2 {r_h}^2+r\right)}{320 \pi
   ^{7/2} r \ a_1(r)^2 \ a_4(r) \ b_1(r)^2 \alpha _{\theta
   _1}^2 \alpha _{\theta _2}^3},\nonumber\\
& & T^{(3)}_{zz} = \frac{{g_s}^4 M^4 \left(\frac{1}{N}\right)^{11/10} {N_f}^2 \log
   ^2(r)}{144 \pi ^4 \ b_1(r)^2 \alpha _{\theta _1}^4 \alpha _{\theta
   _2}^4},\nonumber\\
& & T^{(3)}_{z x^{10}} = \frac{{g_s}^{5/2}  \log N  M^3 {N_f}^2
   \left(1-\frac{r_h^4}{r^4}\right) \log (r) \left(108 b^2 {r_h}^2+r\right)}{960 \pi
   ^{7/2} N r \ a_1(r)^2 \ a_4(r) \ b_1(r)^2 \alpha _{\theta
   _1}^2 \alpha _{\theta _2}^4},\nonumber\\
& & T^{(3)}_{x^{10}x^{10}} = \frac{49 \sqrt{N}}{24 \ a_1(r)^2 \ b_1(r)^2 \alpha _{\theta _1}^2 \alpha _{\theta
   _2}^4},
\end{eqnarray}
}
where:
{\footnotesize
\begin{eqnarray}
\label{a1+a4+b1}
& & \ a_1(r) = \frac{3 \left(-{N_f} \log \left(9 a^2 r^4+r^6\right)+\frac{8 \pi }{{g_s}}+2 \log N  {N_f}-4 {N_f} \log \left(\alpha
   _{\theta _1}\right)-4 {N_f} \log \left(\alpha _{\theta _2}\right)+4 {N_f} \log (4)\right)}{8 \pi },
   \nonumber\\
& &a_4(r) = \frac{6 a^2+r^2}{9 a^2+r^2},\nonumber\\
& &  b_1(r) =  \frac{1}{6 \sqrt{2} \pi ^{5/4} r^2 \alpha _{\theta _1} \alpha
   _{\theta _2}^2}\nonumber\\
& & \Biggl\{{g_s}^{3/4} M \Biggl(-{g_s} {N_f} \left(r^2-3 a^2\right) \log (N) (2 \log (r)+1)+\log (r) \left(4 {g_s} {N_f}
   \left(r^2-3 a^2\right) \log \left(\frac{1}{4} \alpha _{\theta _1} \alpha _{\theta _2}\right)+8 \pi  \left(r^2-3 a^2\right)-3 {g_s}
   {N_f} r^2\right)\nonumber\\
& &+2 {g_s} {N_f} \left(r^2-3 a^2\right) \log \left(\frac{1}{4} \alpha _{\theta _1} \alpha _{\theta
   _2}\right)+18 {g_s} {N_f} \left(r^2-3 a^2 (6 r+1)\right) \log ^2(r)\Biggr)\Biggr\},\nonumber\\
   & & b_2(r) = \frac{{g_s}^{7/4} M {N_f} \log (r) \left(36 a^2 \log (r)+r\right)}{3 \sqrt{2} \pi ^{5/4} r \alpha _{\theta _2}^3}.
\end{eqnarray}
}

\subsection{$\psi\neq0$-Coordinate Patch}

We now list the non-zero components of $T^{(3)}_{MN}$ in the $\psi=\psi_0\neq0$-coordinate patch (wherein some $g^{\cal M}_{rM}, M\neq r$ and $g^{\cal M}_{x^{10}N}, N\neq x^{10}$ components are non-zero) that figure in the EOMs via (\ref{TMN-first-order-in-beta-1}) - (\ref{TMN-first-order-in-beta-2}):

{\footnotesize
\begin{eqnarray*}
& & \hskip -0.6in {\cal T}^{(3)}_{rr} = \frac{2 {g_s}^4 M^2 \left(\frac{1}{N}\right)^{2/5} {N_f}^2 \left(108 a^2+r\right)^2  a_1(r)^4 \log ^2(r) \alpha _{\theta _1}^4}{81\times4
   ({g_s}-1)^2 \sin^2\phi_2 \sin^2\left(\frac{\psi_0}{2}\right) r^4 b_2(r)^2 b_3(r)^2 \alpha _{\theta _2}^4},\nonumber\\
& & \hskip -0.6in {\cal T}^{(3)}_{r\theta_1} = -\frac{245 \pi ^{3/2} \sqrt{{g_s}} \sqrt[20]{\frac{1}{N}} \sin\phi_1  a_1(r)^2 \alpha _{\theta _1}^5}{648 \sqrt{3} ({g_s}-1)
   \sin^2\phi_2 \sin\left(\frac{\psi_0}{2}\right)  b_2(r)^2 b_3(r) \alpha _{\theta _2}^5},\nonumber\\
& & \hskip -0.6in {\cal T}^{(3)}_{r\theta_2} = -\frac{7 \pi ^{3/2} {g_s}^{5/2} M {N_f} \left(108 a^2+r\right)  a_1(r)^4 \log (r) \alpha _{\theta _1}^4}{2187\times4 \sqrt{6} ({g_s}-1)^2
   \sqrt[10]{\frac{1}{N}} \sin^2\phi_2 \sin^2\left(\frac{\psi_0}{2}\right) r^2 b_2(r)^2 b_3(r)^2 \alpha _{\theta _2}^4},\nonumber\\
& &  \hskip -0.6in{\cal T}^{(3)}_{r\phi_1} = -\frac{245 \sqrt{\pi } \sqrt{{g_s}} \sqrt[5]{\frac{1}{N}} {N_f}^2 \sin\phi_1  a_1(r)^2 \alpha _{\theta _1}^7}{11664 ({g_s}-1)
   \sin^2\phi_2 \sin\left(\frac{\psi_0}{2}\right)  b_2(r)^2 b_3(r)^2 \alpha _{\theta _2}^5},\nonumber\\
& & \hskip -0.6in{\cal T}^{(3)}_{r\phi_2} = -\frac{7 \pi ^{5/4} {g_s}^{9/4} M \left(\frac{1}{N}\right)^{9/20} {N_f} \left(108 a^2+r\right)  a_1(r)^4 \log (r) \alpha _{\theta
   _1}^4}{486\times4 \sqrt{3} ({g_s}-1)^2 \sin^2\phi_2 \sin^2\left(\frac{\psi_0}{2}\right) r^2 b_2(r)^2 b_3(r)^2 \alpha _{\theta _2}^3},\nonumber\\
   \end{eqnarray*}}
   {\scriptsize
   \begin{eqnarray*}
& & {\cal T}^{(3)}_{r\psi} = -\frac{{g_s}^{17/4} M^3 \left(\frac{1}{N}\right)^{23/20} {N_f}^2 \left(108 a^2+r\right)  a_1(r)^4 \log ^2(r) \alpha _{\theta
   _1}^4}{1458\times4 \sqrt{2} \pi ^{3/4} ({g_s}-1)^2 \sin^2\phi_2 \sin^2\left(\frac{\psi_0}{2}\right) r^2 b_2(r)^2 b_3(r)^2 \alpha _{\theta _2}^4},\nonumber\\
& & {\cal T}^{(3)}_{rx^{10}} = \frac{245 \pi ^{3/2} \sqrt{{g_s}} \sqrt{\frac{1}{N}} {N_f} \sin\phi_1  a_1(r)^2 \alpha _{\theta _1}^7}{1944\times4 \sqrt{6} ({g_s}-1)
   \sin^2\phi_2 \sin^2\left(\frac{\psi_0}{2}\right) b_2(r)^2 b_3(r)^2 \alpha _{\theta _2}^4},\nonumber\\
& & {\cal T}^{(3)}_{\theta_1\theta_1} = -\frac{49 \pi ^{5/2} \sqrt{{g_s}}  a_1(r)^2 \alpha _{\theta _1}^2}{972 ({g_s}-1) \left(\frac{1}{N}\right)^{3/5} {N_f}^2
   \sin^2\phi_2 b_2(r)^2 \alpha _{\theta _2}^6},\nonumber\\
& &  {\cal T}^{(3)}_{\theta_1\theta_2} = \frac{49 \pi ^{3/2} \sqrt{{g_s}}  a_1(r)^2 \alpha _{\theta _1}^3}{1296 \sqrt{2} ({g_s}-1) \left(\frac{1}{N}\right)^{7/10}
   \sin^2\phi_2 b_2(r) b_3(r) \alpha _{\theta _2}^4},\nonumber\\
& & {\cal T}^{(3)}_{\theta_1x} = -\frac{49 \pi ^{3/2} \sqrt{{g_s}}  a_1(r)^2 \alpha _{\theta _1}^4}{5832 \sqrt{3} ({g_s}-1) \left(\frac{1}{N}\right)^{9/20}
   \sin^2\phi_2 b_2(r)^2 b_3(r) \alpha _{\theta _2}^6},\nonumber\\
& & {\cal T}^{(3)}_{\theta_1y} = \frac{1}{264479053824\times4
   \sqrt{\pi } ({g_s}-1)^2 {N_f}^2 \sin^6\phi_2 \sin^2\left(\frac{\psi_0}{2}\right) b_2(r)^4 b_3(r)^2 \alpha _{\theta _2}^7}\Biggl\{392\sqrt{{g_s}} \sqrt[4]{\frac{1}{N}} \nonumber\\
  &&  \Bigg[81 \pi  ({g_s}-1)^2 {N_f}^2 \sin^3\left(\frac{\psi_0}{2}\right) a_1(r)^2 b_2(r)^2 \alpha _{\theta
   _2}^2 \nonumber\\
   & &  \times\biggl(224 \pi  \sin^2\phi_2 b_3(r) \biggl(27 \sqrt{3} (243 \sin\phi_2+131 \sin\left(\frac{\psi_0}{2}\right) ) \alpha _{\theta _2}^2 \alpha _{\theta _1}^3+5
   \sqrt{2} (243\sin\phi_2+41 \sin\left(\frac{\psi_0}{2}\right) ) \alpha _{\theta _2}^4\nonumber\\
   &&+19683 \sqrt{2} \sin\left(\frac{\psi_0}{2}\right)  \alpha _{\theta _1}^6\biggr) \nonumber\\
& &  -59049\times8 \sqrt{6} ({g_s}-1)
   {N_f}^2 \sin^3\left(\frac{\psi_0}{2}\right) {b_1}(r) \alpha _{\theta _1}^{10}\biggr) \nonumber\\
   & & +224 \pi ^3 ({g_s}-1) \sin^2\phi_2 \psi ^2 a_1(r)^4 \alpha _{\theta
   _1}^2 \biggl[32 \pi ^{3/2} \sqrt{{g_s}} b_3(r) \left(27 \sqrt{3} \alpha _{\theta _1}^3+5 \sqrt{2} \alpha _{\theta _2}^2\right)-15309
   \sqrt{6} {N_f}^2 {b_1}(r) b_2(r)^2 \alpha _{\theta _1}^2 \alpha _{\theta _2}^6\biggr]\nonumber\\
   & &  -27648 \sqrt{6} \pi ^{11/2} \sqrt{{g_s}}
   \sin^2\phi_2 a_1(r)^6 b_1(r) \alpha _{\theta _1}^6 \alpha _{\theta _2}^2 +1240029\times 2^8 ({g_s}-1)^4 {N_f}^4 \sin^8\left(\frac{\psi_0}{2}\right)\nonumber\\
   && b_2(r)^2 b_3(r) \alpha _{\theta _1}^6 \left(27 \sqrt{3} \alpha _{\theta _1}^3+5 \sqrt{2} \alpha _{\theta _2}^2\right)\Biggr]\Biggr\},\nonumber\\
& &{\cal T}^{(3)}_{\theta_1z} = \frac{1}{1586874322944\times4 \pi ^{7/2} ({g_s}-1)^2 {N_f} \sin\phi_2^6 \sin^2\left(\frac{\psi_0}{2}\right) r^2 b_2(r)^4 b_3(r)^2
   \alpha _{\theta _2}^8}\nonumber\\
   & & \times\Biggl\{{g_s}^{5/2} M^2 \left(\frac{1}{N}\right)^{19/20} \log (r) \Biggl(162\times4 ({g_s}-1) {N_f} \sin^2\left(\frac{\psi_0}{2}\right) a_1(r)^2 b_2(r)^2 \alpha
   _{\theta _2}^2 \Biggl[16 {N_f} \sin^2\phi_2 b_3(r)\nonumber\\
   & & \times \Biggl[354294 {g_s}^{5/4} \log N  M {N_f} \sin^2\phi_2
  \left(108
   a^2+r\right) \left(1-\frac{r_h^4}{r^4}\right) \kappa (r) \alpha _{\theta _1} \left(12 \sqrt{2} a^2 {g_s}^{7/4} M {N_f} \log ^2(r)-\pi ^{5/4} r b_2(r)
   \alpha _{\theta _2}^3\right)\nonumber\\
   & & +49 \pi ^3 ({g_s}-1) \sin\left(\frac{\psi_0}{2}\right)  r^2 \Biggl(81 \sqrt{2} (243 \sin\phi_2+131\sin\left(\frac{\psi_0}{2}\right) ) \alpha _{\theta _2}^2 \alpha
   _{\theta _1}^3+10 \sqrt{3} (243 \sin\phi_2+41 \sin\left(\frac{\psi_0}{2}\right) ) \alpha _{\theta _2}^4\nonumber\\
   &&+39366 \sqrt{3} \sin\left(\frac{\psi_0}{2}\right)  \alpha _{\theta _1}^6\Biggr)\biggr]\nonumber\\
   & & +7558272
   \sqrt{2} \pi ^{9/4} {g_s}^{5/4} \log N  M \sin\phi_2^4 r \left(108 a^2+r\right) b_3(r)^2 \left(1-\frac{r_h^4}{r^4}\right) \kappa (r) \alpha _{\theta
   _1}-1240029 \pi ^2 ({g_s}-1)^2 {N_f}^3 \psi ^4 r^2 b_1(r) \alpha _{\theta _1}^{10}\Biggr]\nonumber\\
   & & +3136 \pi ^4 ({g_s}-1) \sin^2\phi_2
   \psi ^2 r^2 a_1(r)^4 \alpha _{\theta _1}^2\left(16 \pi ^{3/2} \sqrt{{g_s}} b_3(r) \left(81 \sqrt{2} \alpha _{\theta _1}^3+10
   \sqrt{3} \alpha _{\theta _2}^2\right)-45927 {N_f}^2 b_1(r) b_2(r)^2 \alpha _{\theta _1}^2 \alpha _{\theta _2}^6\right)\nonumber\\
   & & -1161216
   \pi ^{13/2} \sqrt{{g_s}} \sin^2\phi_2 r^2 a_1(r)^6 b_1(r) \alpha _{\theta _1}^6 \alpha _{\theta _2}^2+8680203\times2^8 \pi
   ({g_s}-1)^4 {N_f}^4 \nonumber\\
   &&\sin^8\left(\frac{\psi_0}{2}\right) r^2 b_2(r)^2 b_3(r) \alpha _{\theta _1}^6 \left(81 \sqrt{2} \alpha _{\theta _1}^3+10 \sqrt{3}
   \alpha _{\theta _2}^2\right)\Biggr)\Biggr\}\nonumber\\
    &&
   \end{eqnarray*}}
   {\footnotesize
   \begin{eqnarray*}
& &  {\cal T}^{(3)}_{\theta_1x^{10}} = \frac{49 \pi ^{5/2} \sqrt{{g_s}} a_1(r)^2 \alpha _{\theta _1}^4}{2916\times2 \sqrt{2} ({g_s}-1) \left(\frac{1}{N}\right)^{3/20} {N_f}
   \sin^2\phi_2 \sin\left(\frac{\psi_0}{2}\right)  b_2(r)^2 b_3(r) \alpha _{\theta _2}^5},\nonumber\\
& &  {\cal T}^{(3)}_{\theta_2\theta_2} = \frac{49 \pi ^3 {g_s} a_1(r)^4 \alpha _{\theta _1}^4}{708588 \times4({g_s}-1)^2 \left(\frac{1}{N}\right)^{3/5} \sin^2\phi_2 \sin^2\left(\frac{\psi_0}{2}\right)
   b_2(r)^2 b_3(r)^2 \alpha _{\theta _2}^4},\nonumber\\
& &  {\cal T}^{(3)}_{\theta_2x} = \frac{49 \sqrt{\pi } \sqrt{{g_s}} {N_f}^2 a_1(r)^2 \alpha _{\theta _1}^5}{7776 \sqrt{6} ({g_s}-1)
   \left(\frac{1}{N}\right)^{11/20} \sin^2\phi_2 b_2(r) b_3(r)^2 \alpha _{\theta _2}^4},\nonumber\\
& & {\cal T}^{(3)}_{\theta_2y} = \frac{49 \pi ^{11/4} {g_s}^{3/4} a_1(r)^4 \alpha _{\theta _1}^4}{78732 \sqrt{2} ({g_s}-1)^2 \sqrt[20]{\frac{1}{N}} \sin^2\phi_2 \sin\left(\frac{\psi_0}{2}\right)
   ^2 b_2(r)^2 b_3(r)^2 \alpha _{\theta _2}^3},\nonumber\\
& & {\cal T}^{(3)}_{\theta_2z} = \frac{7 \pi ^{3/4} {g_s}^{11/4} M^2 \left(\frac{1}{N}\right)^{13/20} {N_f} a_1(r)^4 \log (r) \alpha _{\theta _1}^4}{157464\times4 \sqrt{3}
   ({g_s}-1)^2 \sin^2\phi_2 \sin^2\left(\frac{\psi_0}{2}\right) b_2(r)^2 b_3(r)^2 \alpha _{\theta _2}^4},\nonumber\\
& & {\cal T}^{(3)}_{\theta_2x^{10}} = -\frac{49 \pi ^{3/2} \sqrt{{g_s}} {N_f} a_1(r)^2 \alpha _{\theta _1}^5}{7776 ({g_s}-1) \sqrt[4]{\frac{1}{N}} \sin^2\phi_2 \sin\left(\frac{\psi_0}{2}\right)
   b_2(r) b_3(r)^2 \alpha _{\theta _2}^3},\nonumber\\
& & {\cal T}^{(3)}_{xx} = -\frac{49 \sqrt{\pi } \sqrt{{g_s}} {N_f}^2 a_1(r)^2 \alpha _{\theta _1}^6}{104976 ({g_s}-1) \left(\frac{1}{N}\right)^{3/10}
   \sin^2\phi_2 b_2(r)^2 b_3(r)^2 \alpha _{\theta _2}^6},\nonumber\\
\label{T3PQ-psi-neq-0}
& &  {\cal T}^{(3)}_{xy} = \frac{1}{1586874322944\times4 \sqrt{3} \pi ^{3/2} ({g_s}-1)^2 \sin\phi_2^6 \sin^2\left(\frac{\psi_0}{2}\right) b_2(r)^4 b_3(r)^3 \alpha _{\theta
   _2}^7}\Biggl\{49 \sqrt{{g_s}} \left(\frac{1}{N}\right)^{2/5} \alpha _{\theta _1}^2\nonumber\\
   &&\Biggl(81 \pi  ({g_s}-1)^2 {N_f}^2 \psi ^3 a_1(r)^2
   b_2(r)^2 \alpha _{\theta _2}^2\nonumber\\
   & & \times \biggl[224 \pi  \sin^2\phi_2 b_3(r) \left(27 \sqrt{3} (243 \sin\phi_2+131 \sin\left(\frac{\psi_0}{2}\right) ) \alpha _{\theta
   _2}^2 \alpha _{\theta _1}^3+5 \sqrt{2} (243 \sin\phi_2+41 \sin\left(\frac{\psi_0}{2}\right) ) \alpha _{\theta _2}^4+19683 \sqrt{2} \psi  \alpha _{\theta _1}^6\right)\nonumber\\
& & -59049
   \sqrt{6} ({g_s}-1) {N_f}^2 \psi ^3 {b_1}(r) \alpha _{\theta _1}^{10}\biggr]\nonumber\\
   & &+224\times4 \pi ^3 ({g_s}-1) \sin^2\phi_2 \sin^2\left(\frac{\psi_0}{2}\right)
   a_1(r)^4 \alpha _{\theta _1}^2 \Biggl(32 \pi ^{3/2} \sqrt{{g_s}} b_3(r) \left(27 \sqrt{3} \alpha _{\theta _1}^3+5 \sqrt{2} \alpha
   _{\theta _2}^2\right)\nonumber\\
   &&-15309 \sqrt{6} {N_f}^2 b_1(r) b_2(r)^2 \alpha _{\theta _1}^2 \alpha _{\theta _2}^6\Biggr)\nonumber\\
   & &-27648 \sqrt{6}
   \pi ^{11/2} \sqrt{{g_s}} \sin^2\phi_2 a_1(r)^6 b_1(r) \alpha _{\theta _1}^6 \alpha _{\theta _2}^2+1240029\times2^8 ({g_s}-1)^4
   {N_f}^4\nonumber\\
    &&\sin^8\left(\frac{\psi_0}{2}\right) b_2(r)^2 b_3(r) \alpha _{\theta _1}^6 \left(27 \sqrt{3} \alpha _{\theta _1}^3+5 \sqrt{2} \alpha _{\theta
   _2}^2\right)\Biggr)\Biggr\},\nonumber
   \end{eqnarray*}}
   {\footnotesize
   \begin{eqnarray*}
& &  {\cal T}^{(3)}_{xz}\nonumber\\
& & = \frac{1}{9521245937664 \pi ^{9/2} \times4({g_s}-1)^2 \sin\phi_2^6
   \sin^2\left(\frac{\psi_0}{2}\right) r^2 b_2(r)^4 b_3(r)^3 \alpha _{\theta _2}^8}\nonumber\\
& & \Biggl\{{g_s}^{5/2} M^2 \left(\frac{1}{N}\right)^{11/10} {N_f} \log (r) \alpha _{\theta _1}^2 \Biggl(162\times4 ({g_s}-1) {N_f} \sin^2\left(\frac{\psi_0}{2}\right)
   a_1(r)^2 b_2(r)^2 \alpha _{\theta _2}^2\nonumber\\
   & &\times\biggl(16 {N_f} \sin^2\phi_2 b_3(r) \biggl[118098 \sqrt{3} {g_s}^{5/4}
   \log N  M {N_f} \sin^2\phi_2 \left(108 a^2+r\right) \left(1-\frac{r_h^4}{r^4}\right) \kappa (r) \alpha _{\theta _1} \nonumber\\
    &&\left(12 \sqrt{2} a^2 {g_s}^{7/4} M
   {N_f} \log ^2(r)-\pi ^{5/4} r b_2(r) \alpha _{\theta _2}^3\right)\nonumber\\
   & &+49 \pi ^3 ({g_s}-1) \sin\left(\frac{\psi_0}{2}\right) r^2 \left(27 \sqrt{6} (243
   \sin\phi_2+131 \psi ) \alpha _{\theta _2}^2 \alpha _{\theta _1}^3+10 (243 \sin\phi_2+41 \sin\left(\frac{\psi_0}{2}\right) ) \alpha _{\theta _2}^4+39366 \psi  \alpha
   _{\theta _1}^6\right)\biggr]\nonumber\\
   & &+2519424 \sqrt{6} \pi ^{9/4} {g_s}^{5/4} \log N  M \sin\phi_2^4 r \left(108 a^2+r\right) b_3(r)^2
   \left(1-\frac{r_h^4}{r^4}\right) \kappa (r) \alpha _{\theta _1}\nonumber\\
   &&-413343\times2^4 \sqrt{3} \pi ^2 ({g_s}-1)^2 {N_f}^3 \sin^4\left(\frac{\psi_0}{2}\right)  r^2 b_1(r) \alpha _{\theta
   _1}^{10}\biggr)\nonumber\\
& & +3136\times4 \pi ^4 ({g_s}-1) \sin^2\phi_2 \sin^2\left(\frac{\psi_0}{2}\right) r^2 a_1(r)^4 \alpha _{\theta _1}^2 \Biggl(16 \pi ^{3/2} \sqrt{{g_s}}
   b_3(r) \left(27 \sqrt{6} \alpha _{\theta _1}^3+10 \alpha _{\theta _2}^2\right)\nonumber\\
   &&-15309 \sqrt{3} {N_f}^2 b_1(r) b_2(r)^2
   \alpha _{\theta _1}^2 \alpha _{\theta _2}^6\Biggr)\nonumber\\
   & & -387072 \sqrt{3} \pi ^{13/2} \sqrt{{g_s}} \sin^2\phi_2 r^2 a_1(r)^6 b_1(r)
   \alpha _{\theta _1}^6 \alpha _{\theta _2}^2+8680203\times2^8 \pi  ({g_s}-1)^4 {N_f}^4\nonumber\\
    &&\sin^8\left(\frac{\psi_0}{2}\right) r^2 b_2(r)^2 b_3(r) \alpha _{\theta
   _1}^6 \left(27 \sqrt{6} \alpha _{\theta _1}^3+10 \alpha _{\theta _2}^2\right)\Biggr)\Biggr\},\nonumber\\
& &  {\cal T}^{(3)}_{x x^{10}} = \frac{49 \pi ^{3/2} \sqrt{{g_s}} {N_f} a_1(r)^2 \alpha _{\theta _1}^6}{17496 \sqrt{6} ({g_s}-1) \sin^2\phi_2 \psi
   b_2(r)^2 b_3(r)^2 \alpha _{\theta _2}^5},\nonumber\\
& &  {\cal T}^{(3)}_{yy} = \frac{49 \pi ^{5/2} \sqrt{{g_s}} \sqrt{\frac{1}{N}} a_1(r)^4 \alpha _{\theta _1}^4}{17496 \times4({g_s}-1)^2 \sin^2\phi_2 \sin^2\left(\frac{\psi_0}{2}\right)
   b_2(r)^2 b_3(r)^2 \alpha _{\theta _2}^2},\nonumber\\
& & {\cal T}^{(3)}_{yz} = \frac{7 \sqrt{\frac{\pi }{6}} {g_s}^{5/2} M^2 \left(\frac{1}{N}\right)^{6/5} {N_f} a_1(r)^4 \log (r) \alpha _{\theta _1}^4}{17496
   \times4({g_s}-1)^2 \sin^2\phi_2 \sin^2\left(\frac{\psi_0}{2}\right) b_2(r)^2 b_3(r)^2 \alpha _{\theta _2}^3},\nonumber\\
 & & {\cal T}^{(3)}_{y x^{10}} = -\frac{1}{1586874322944 \sqrt{\pi } ({g_s}-1)^2 {N_f} \sin\phi_2^6 \psi ^3 b_2(r)^4 b_3(r)^3 \alpha _{\theta
   _2}^6}\Biggl\{49 \sqrt{{g_s}} \left(\frac{1}{N}\right)^{7/10} \alpha _{\theta _1}^2 \nonumber\\
   &&\biggl(162 \pi  ({g_s}-1)^2 {N_f}^2 \psi ^3 a_1(r)^2
   b_2(r)^2 \alpha _{\theta _2}^2 \nonumber\\
   & & \times \biggl(112 \pi  \sin^2\phi_2 b_3(r)\biggl[27 \sqrt{6} (243 \sin\phi_2+262 \sin\left(\frac{\psi_0}{2}\right) ) \alpha _{\theta
   _2}^2 \alpha _{\theta _1}^3\nonumber\\
& & +10 (243 \sin\phi_2+41 \sin\left(\frac{\psi_0}{2}\right) ) \alpha _{\theta _2}^4+39366 \psi  \alpha _{\theta _1}^6\biggr]-59049 \sqrt{3}
   ({g_s}-1) {N_f}^2 \psi ^3 b_1(r) \alpha _{\theta _1}^{10}\biggr)\nonumber\\
   & &  +448 \pi ^3 \times4({g_s}-1) \sin^2\phi_2 \sin^2\left(\frac{\psi_0}{2}\right) a_1(r)^4
   \alpha _{\theta _1}^2 \left(16 \pi ^{3/2} \sqrt{{g_s}} b_3(r) \left(27 \sqrt{6} \alpha _{\theta _1}^3+10 \alpha _{\theta
   _2}^2\right)-15309 \sqrt{3} {N_f}^2 b_1(r) b_2(r)^2 \alpha _{\theta _1}^2 \alpha _{\theta _2}^6\right)\nonumber\\
   & & -55296 \sqrt{3} \pi
   ^{11/2} \sqrt{{g_s}} \sin^2\phi_2 a_1(r)^6 b_1(r) \alpha _{\theta _1}^6 \alpha _{\theta _2}^2+1240029 ({g_s}-1)^4
   {N_f}^4 \sin\left(\frac{\psi_0}{2}\right) ^8 b_2(r)^2 b_3(r) \alpha _{\theta _1}^6 \left(27 \sqrt{6} \alpha _{\theta _1}^3+10 \alpha _{\theta
   _2}^2\right)\biggr)\Biggr\},\nonumber\\
    &&
   \end{eqnarray*}}
   {\footnotesize
   \begin{eqnarray}
   & &{\cal T}^{(3)}_{zz} = \frac{1}{6718464\times4 \pi ^{13/2} ({g_s}-1)^2 \sin^2\phi_2 \sin^2\left(\frac{\psi_0}{2}\right) r^2 b_2(r)^2
   b_3(r)^2 \alpha _{\theta _2}^6}
   \Biggl\{{g_s}^4 M^4 \left(\frac{1}{N}\right)^{19/10} a_1(r)^2 \alpha _{\theta _1}^2\nonumber\\
   & & \Biggl(\sqrt{{g_s}} {N_f}^2 \log ^2(r) \Biggl(64
   \pi ^5 r^2 a_1(r)^2 \alpha _{\theta _1}^2 \alpha _{\theta _2}^2-243 a^2 ({g_s}-1) {g_s}^{5/4} \log N ^2 M {N_f}^3 \psi ^2
   \left(1-\frac{r_h^4}{r^4}\right) \kappa (r) \nonumber\\
   &&\left(12 a^2 {g_s}^{7/4} M {N_f} \log ^2(r)-\sqrt{2} \pi ^{5/4} r b_2(r) \alpha _{\theta
   _2}^3\right)\Biggr)\nonumber\\
   & & +27 \pi ^{9/4} \times4({g_s}-1) \log N ^2 {N_f}^2 \sin^2\left(\frac{\psi_0}{2}\right) r b_3(r) \left(1-\frac{r_h^4}{r^4}\right) \kappa (r) \left(\sqrt{2} \pi
   ^{5/4} r b_2(r) \alpha _{\theta _2}^3-24 a^2 {g_s}^{7/4} M {N_f} \log ^2(r)\right)\nonumber\\
   & &-36 \pi ^{9/2} \times4({g_s}-1) \log N ^2 \sin^2\left(\frac{\psi_0}{2}\right) r^2 b_3(r)^2 \left(1-\frac{r_h^4}{r^4}\right) \kappa (r)\Biggr)\Biggr\},\nonumber\\
   & &{\cal T}^{(3)}_{z x^{10}} = -\frac{1}{4760622968832 \pi ^{7/2}\times2^3 ({g_s}-1)^2 \sin\phi_2^6 \sin^3\left(\frac{\psi_0}{2}\right) r^2 b_2(r)^4
   b_3(r)^3 \alpha _{\theta _2}^7}\nonumber\\
& &  \times\Biggl\{{g_s}^{5/2} M^2 \left(\frac{1}{N}\right)^{7/5} \log (r) \alpha _{\theta _1}^2 \Biggl(81\times4 ({g_s}-1) {N_f} \sin^2\left(\frac{\psi_0}{2}\right) a_1(r)^2
   b_2(r)^2 \alpha _{\theta _2}^2\nonumber\\
   & & \times \biggl[32 {N_f} \sin^2\phi_2 b_3(r) \biggl(177147 {g_s}^{5/4} \log N  M {N_f}
   \sin^2\phi_2 \left(108 a^2+r\right) \left(1-\frac{r_h^4}{r^4}\right) \kappa (r) \alpha _{\theta _1} \nonumber\\
   &&\left(24 a^2 {g_s}^{7/4} M {N_f} \log ^2(r)-\sqrt{2}
   \pi ^{5/4} r b_2(r) \alpha _{\theta _2}^3\right)\nonumber\\
& & +49 \pi ^3 ({g_s}-1)\sin\left(\frac{\psi_0}{2}\right) r^2   \Biggl(81 (243 \sin\phi_2+131\times2 \sin\left(\frac{\psi_0}{2}\right)) \alpha _{\theta
   _2}^2 \alpha _{\theta _1}^3+5 \sqrt{6} (243 \sin\phi_2+41 \psi ) \alpha _{\theta _2}^4\nonumber\\
   &&+19683\times2 \sqrt{6} \sin\left(\frac{\psi_0}{2}\right)  \alpha _{\theta_1}^6\Biggr)\biggr)\nonumber\\
   & & +15116544 \pi ^{9/4} {g_s}^{5/4} \log N  M \sin\phi_2^4 r \left(108 a^2+r\right) b_3(r)^2 \left(1-\frac{r_h^4}{r^4}\right) \kappa
   (r) \alpha _{\theta _1}\nonumber\\
   &&-1240029\times2^4 \sqrt{2} \pi ^2 ({g_s}-1)^2 {N_f}^3 \sin^4\left(\frac{\psi_0}{2}\right) r^2 b_1(r) \alpha _{\theta _1}^{10}\biggr]\nonumber\\
   & & +1568 \pi ^4
   4({g_s}-1) \sin^2\phi_2 \sin^2\left(\frac{\psi_0}{2}\right) r^2 a_1(r)^4 \alpha _{\theta _1}^2 \left(32 \pi ^{3/2} \sqrt{{g_s}} b_3(r) \left(81 \alpha
   _{\theta _1}^3+5 \sqrt{6} \alpha _{\theta _2}^2\right)-45927 \sqrt{2} {N_f}^2 b_1(r) b_2(r)^2 \alpha _{\theta _1}^2 \alpha
   _{\theta _2}^6\right)\nonumber\\
   & & -580608 \sqrt{2} \pi ^{13/2} \sqrt{{g_s}} \sin^2\phi_2 r^2 a_1(r)^6 b_1(r) \alpha _{\theta _1}^6 \alpha
   _{\theta _2}^2+8680203\times2^8 \pi  ({g_s}-1)^4 {N_f}^4 \nonumber\\
   &&\sin^8\left(\frac{\psi_0}{2}\right) r^2 b_2(r)^2 b_3(r) \alpha _{\theta _1}^6 \left(81 \alpha _{\theta
   _1}^3+5 \sqrt{6} \alpha _{\theta _2}^2\right)\Biggr)\Biggr\},\nonumber\\
   & & {\cal T}^{(3)}_{x^{10}x^{10}} = -\frac{49 \pi ^{5/2} \sqrt{{g_s}} \left(\frac{1}{N}\right)^{3/10} a_1(r)^2 \alpha _{\theta _1}^6}{17496\times2 ({g_s}-1) \sin^2\phi_2 \sin\left(\frac{\psi_0}{2}\right)
   ^2 b_2(r)^2 b_3(r)^2 \alpha _{\theta _2}^4},
\end{eqnarray}
}
where:
\begin{equation}
\label{b3}
b_3(r) = \frac{g_s^{\frac{3}{4}}MN_f^3\log r\left(r + 108 a^2 \log r\right)}{8\pi^{\frac{9}{4}}r}.
\end{equation} 
\chapter{}

\section{Equations of Motion for $f_{MN} $s, their large-$N$ large-$r$ (UV) Limit and Proof of $f_{MN}$s to be Vanishingly Small in the UV}

In this appendix, we will show that $f_{MN}$ is vanishingly small in the UV  inclusive of the ${\cal O}(l_p^6 R^4)$ corrections to the ${\cal M}$-Theory uplift of large-$N$ thermal QCD at low temperatures (i.e. below the deconfinement temperature) near the $\psi=2n\pi, n=0, 1, 2$-patches. In the following:
{\footnotesize
\begin{eqnarray}
\label{g_2+a_i+b_j-defs}
& & \hskip -0.6in g_2(r) \equiv 1 - \frac{r_0^4}{r^4},\nonumber\\
& & \hskip -0.6in{a_1}(r)=\frac{3 \left(-{N_f^{\rm UV}}\  \log \left(9 a^2 r^4+r^6\right)+\frac{8 \pi }{{g_s}}+2 \log N  {N_f^{\rm UV}}\ -4 {N_f^{\rm UV}}\
   \log \left(\alpha _{\theta _1}\right)-4 {N_f^{\rm UV}}\  \log \left(\alpha _{\theta _2}\right)+4 {N_f^{\rm UV}}\  \log (4)\right)}{8 \pi }\nonumber\\
& & \hskip -0.6in{a_2}(r)=\frac{12 a^2 {g_s} M_{\rm UV}^2 {N_f^{\rm UV}}\  ({c_1}+{c_2} \log ({r_0}))}{9 a^2+r^2}\nonumber\\
& &\hskip -0.6in {a_3}(r)=\frac{r^2 {a_2}(r)}{2 {N_f^{\rm UV}}\  \left(6 a^2+r^2\right)} \nonumber\\
& &\hskip -0.6in {a_4}(r)=\frac{6 a^2+r^2}{9 a^2+r^2} \nonumber\\
& &\hskip -0.6in B(r)=\frac{3 {g_s} M_{\rm UV}^2 \log (r) \left(-{g_s} \log N  {N_f^{\rm UV}}\ +2 {g_s} {N_f^{\rm UV}}\  \log \left(\alpha _{\theta
   _1}\right)+2 {g_s} {N_f^{\rm UV}}\  \log \left(\alpha _{\theta _2}\right)+12 {g_s} {N_f^{\rm UV}}\  \log (r)+6 {g_s} {N_f^{\rm UV}}\ -2
   {g_s} {N_f^{\rm UV}}\  \log (4)+8 \pi \right)}{32 \pi ^2}\nonumber\\
& &\hskip -0.6in {b_1}(r)=\frac{1}{6
   \sqrt{2} \pi ^{5/4} r^2 \alpha _{\theta _1} \alpha _{\theta _2}^2}\Biggl\{{g_s}^{3/4} M_{\rm UV}\nonumber\\
& & \hskip -0.6in \times\Biggl[-{g_s} {N_f^{\rm UV}}\  \left(r^2-3 a^2\right) \log N  (2 \log (r)+1)+\log (r) \left(4
   {g_s} {N_f^{\rm UV}}\  \left(r^2-3 a^2\right) \log \left(\frac{1}{4} \alpha _{\theta _1} \alpha _{\theta _2}\right)+8 \pi
   \left(r^2-3 a^2\right)-3 {g_s} {N_f^{\rm UV}}\  r^2\right)\nonumber\\
& & \hskip -0.6in+2 {g_s} {N_f^{\rm UV}}\  \left(r^2-3 a^2\right) \log \left(\frac{1}{4}
   \alpha _{\theta _1} \alpha _{\theta _2}\right)+18 {g_s} {N_f^{\rm UV}}\  \left(r^2-3 a^2 (6 r+1)\right) \log ^2(r)\Biggr]\Biggr\}\nonumber\\
& & \hskip -0.6in {b_2}(r)=\frac{{g_s}^{7/4} M_{\rm UV} {N_f^{\rm UV}}\  \log (r) \left(36 a^2 \log (r)+r\right)}{3 \sqrt{2} \pi ^{5/4} r \alpha _{\theta
   _2}^3} ,\nonumber\\
& &
\end{eqnarray}
}
\noindent wherein $M_{\rm UV}\equiv M_{\rm eff}(r>{\cal R}_{D5/\overline{D5}})$ and similarly $N_f^{\rm UV}\equiv N_f^{\rm eff}(r>{\cal R}_{D5/\overline{D5}})$. We should keep in  mind that near the $\psi=\psi_0\neq2n\pi, n=0, 1, 2$-patch, some $g^{\cal M}_{rM}, M\neq r$ and $g^{\cal M}_{x^{10}N}, N\neq x^{10}$ components are non-zero, making this exercise much more non-trivial. As shown in \cite{OR4-Yadav+Misra}, the contributions from $E_8$ is sub-dominant as compared to the contributions from $J_0$ terms.

As the EOMs are too long, they have not been explicitly typed but their forms have been written out. The explicit forms of ${\cal F}_{MN}$ have been given.
Using (\ref{g_2+a_i+b_j-defs}), one obtains the following EOMs in the UV.:
{\footnotesize
\begin{eqnarray}
\label{IR-psi=2nPi-EOMs}
& & {\rm EOM}_{MN}:\nonumber\\
& & \sum_{{\cal M},{\cal N}=0}^{x^{10}} \sum_{p=0}^2{\cal H}^{{\cal M}{\cal N}\ (p)}_{MN}\left(r; r_0, a, N, M^{\rm UV}, N_f^{\rm UV}, g_s, \alpha_{\theta_{1,2}}\right)f_{{\cal M}{\cal N}}^{(p)}(r) +
\beta {\cal F}_{MN}\left(r; r_0, a, N, M, N_f, g_s, \alpha_{\theta_{1,2}}\right) = 0,\nonumber\\
& &
\end{eqnarray}
}
where $M, N$ run over the $D=11$ coordinates,   $f^{(p)}_{MN}\equiv \frac{d^p f_{MN}}{dr^p}, p=0, 1, 2$.
In the UV, and in the large-$N$ limit, one can show that EOMs are algebraic and are given as under:
\begin{eqnarray*}
&  & \hskip -0.7in EOM_{x^3x^3} : -\frac{49 r^2 {g_2} (r) (2 {f_{zz} } (r)-3 {f_{x^{10}x^{10}}} (r)-4 {f_{\theta_1z} } (r)-5 {f_{\theta_2y} } (r))}{3456 \sqrt{\pi } \sqrt{{g_s}} \alpha
   _{\theta _1}^2 \alpha _{\theta _2}^4 {b_1}(r)^2} -
\frac{7 a^8 \beta  M_{\rm UV} \left(\frac{1}{N}\right)^{9/4}\Sigma_1 \log \left({r_0}\right)}{216 \pi ^2 {g_s} \log N  ^2 {N_f^{\rm UV}}\  r^6 \alpha
   _{\theta _2}^3} = 0\nonumber\\
& & \nonumber\\
& & \hskip -0.7in EOM_{\mathbb{R}^{1,2}(t,x^{1,2})} :  \frac{49 r^2 (2 {f_{zz} } (r)-3 {f_{x^{10}x^{10}}} (r)-4 {f_{\theta_1z} } (r)-5 {f_{\theta_2y} } (r))}{3456 \sqrt{\pi } \sqrt{{g_s}} \alpha
   _{\theta _1}^2 \alpha _{\theta _2}^4 {b_1}(r)^2}  -\frac{7 a^8 \beta  M_{\rm UV} \left(\frac{1}{N}\right)^{9/4} \Sigma_1 \log (r)}{216 \pi ^2 {g_s} \log N  ^2 {N_f^{\rm UV}}\  r^6
   \alpha _{\theta _2}^3} = 0\nonumber\\
& & \nonumber\\
& & \hskip -0.7in  EOM_{rr} :  \frac{49 \sqrt{\pi } \sqrt{{g_s}} N {a_4}(r) (2 {f_{zz} } (r)-3 {f_{x^{10}x^{10}}} (r)-4
   {f_{\theta_1z} } (r)-5 {f_{\theta_2y} } (r))}{864 r^2 \alpha _{\theta _1}^2 \alpha _{\theta _2}^4 {b_1}(r)^2
   {g_2} (r)} -\frac{2 a^{12} \beta  M_{\rm UV} \left(\frac{1}{N}\right)^{9/10} \Sigma_1}{3 \sqrt{3} \pi
   ^{5/4} \sqrt[4]{{g_s}} \log N  ^2 {N_f^{\rm UV}}\  r^{14} \alpha _{\theta _2}^5} = 0\nonumber\\
& & \hskip -0.7in EOM_{\theta_1\theta_1} : -\frac{3969 \sqrt[5]{N} \alpha _{\theta _1}^2 (2 {f_{x^{10}x^{10}}} (r)+{f_{\theta_2y} } (r))}{512 \alpha _{\theta _2}^2}  + \frac{a^{12} \beta  {g_s}^{5/4} M_{\rm UV}^2 \left(\frac{1}{N}\right)^{17/10} (-\Sigma_1) \log ^2\left({r_0}\right)}{26244 \sqrt{2} \pi ^{11/4} \log N   r^{12} \alpha _{\theta
   _1}^6 \alpha _{\theta _2}^4} = 0\nonumber\\
& & \nonumber\\
& & \hskip -0.7in   EOM_{\theta_1\theta_2} : -\frac{3969 N^{3/10} \alpha _{\theta _1}^2 {b_2}(r) (2 {f_{x^{10}x^{10}}} (r)+{f_{\theta_2y} } (r))}{512 \alpha _{\theta
   _2}^2 {b_1}(r)}=0 \nonumber\\
& & \nonumber\\
& & \hskip -0.7in   EOM_{\theta_1x} : \frac{49 N^{21/20} (2 {f_{zz} } (r)-3 {f_{x^{10}x^{10}}} (r)-4 {f_{\theta_1z} } (r)-5 {f_{\theta_2y} } (r))}{1728 \alpha _{\theta _1}^2 \alpha
   _{\theta _2}^4 {b_1}(r)} \nonumber\\
& & +\frac{a^{12} \beta  {g_s} M_{\rm UV}^2 \left(\frac{1}{N}\right)^{17/20} \left(19683 \sqrt{6} \alpha _{\theta _1}^6+6642 \alpha
   _{\theta _2}^2 \alpha _{\theta _1}^3-40 \sqrt{6} \alpha _{\theta _2}^4\right) \log \left({r_0}\right)}{18 \sqrt{6} \pi ^3
   \log N   r^{12} \alpha _{\theta _1} \alpha _{\theta _2}^7} = 0\nonumber\\
& & \nonumber\\
& & \hskip -0.7in EOM_{\theta_1y} : \frac{49 N^{7/20} (2 {f_{zz} } (r)-3 {f_{x^{10}x^{10}}} (r)-4 {f_{\theta_1z} } (r)-5 {f_{\theta_2y} } (r))}{128 \sqrt{6} \alpha _{\theta
   _2}^3 {b_1}(r)}  - \frac{11 \sqrt{\frac{2}{3}} a^{12} \beta  M_{\rm UV} \left(\frac{1}{N}\right)^{23/20} (-\Sigma_1) \log
   ^2\left({r_0}\right)}{3 \pi ^{3/2} \sqrt{{g_s}} \log N  ^2 {N_f^{\rm UV}}\  r^{12} \alpha _{\theta _1}^3 \alpha
   _{\theta _2}^3} = 0 \nonumber\\
& & \hskip -0.7in EOM_{\theta_1z} : -\frac{49 N^{13/20} (2 {f_{zz} } (r)-3 {f_{x^{10}x^{10}}} (r)-4 {f_{\theta_1z} } (r)-5 {f_{\theta_2y} } (r))}{1152 \alpha _{\theta _2}^4
   {b_1}(r)} + \frac{a^{12} \beta  {g_s} M_{\rm UV}^2 \left(\frac{1}{N}\right)^{5/4} \alpha _{\theta _1}\Sigma_1 \log
   \left({r_0}\right)}{12 \sqrt{6} \pi ^3 \log N   r^{12} \alpha _{\theta _2}^7} = 0
\nonumber\\
& & \nonumber\\
\end{eqnarray*}
\begin{eqnarray}
\label{EOMs}
& &  \hskip -0.7in EOM_{\theta_2\theta_2} : {f_{zz} } (r) = {f_{x^{10}x^{10}}} (r)\nonumber\\
& & \nonumber\\
& & \hskip -0.7in EOM_{\theta_2x} : \frac{49 N^{23/20} {b_2}(r) (2 {f_{zz} } (r)-3 {f_{x^{10}x^{10}}} (r)-4 {f_{\theta_1z} } (r)-5 {f_{\theta_2y} } (r))}{1728 \alpha _{\theta _1}^2 \alpha _{\theta _2}^4
   {b_1}(r)^2} +
\frac{8 a^{14} \beta  M_{\rm UV} \left(\frac{1}{N}\right)^{7/20}\Sigma_1 \log \left({r_0}\right)}{243 \pi ^{3/2} \sqrt{{g_s}} \log N  ^2 {N_f^{\rm UV}}\  r^{14} \alpha _{\theta
   _1}^4 \alpha _{\theta _2}^5} = 0\nonumber\\
&& \nonumber\\
& & \hskip -0.7in EOM_{\theta_2y}: -\frac{49 N^{17/20} \left(7 \sqrt{2} \sqrt[4]{\pi } \sqrt[4]{{g_s}} {f_{zz} } (r)-15 \sqrt{2} \sqrt[4]{\pi } \sqrt[4]{{g_s}}
   {f_{x^{10}x^{10}}} (r)-14 \sqrt{2} \sqrt[4]{\pi } \sqrt[4]{{g_s}} {f_{\theta_1z} } (r)+5 \sqrt{2} \sqrt[4]{\pi } \sqrt[4]{{g_s}}
   {f_{\theta_2y} } (r)\right)}{15552 \alpha _{\theta _1}^4 \alpha _{\theta _2}^3 {b_1}(r)^2} \nonumber\\
& & +
\frac{8 a^{12} \beta  \left(\frac{1}{N}\right)^{13/20} \left(19683 \sqrt{6} \alpha _{\theta _1}^6+6642 \alpha _{\theta _2}^2 \alpha _{\theta
   _1}^3-40 \sqrt{6} \alpha _{\theta _2}^4\right)}{243 {g_s}^2 \log N  ^2 {N_f^{\rm UV}}\ ^2 r^{12} \alpha _{\theta _1}^6 \alpha _{\theta
   _2}} = 0\nonumber\\
& & \hskip -0.7 in EOM_{\theta_2z} :  -\frac{49 N^{3/4} {b_2}(r) (2 {f_{zz} } (r)-3 {f_{x^{10}x^{10}}} (r)-4 {f_{\theta_1z} } (r)-5 {f_{\theta_2y} } (r))}{1152 \alpha _{\theta
   _2}^4 {b_1}(r)^2}   - \frac{4 a^{14} \beta  M_{\rm UV} \left(\frac{1}{N}\right)^{3/4}(-\Sigma_1) \log (r)}{81 \pi ^{3/2} \sqrt{{g_s}} \log N  ^2
   {N_f^{\rm UV}}\  r^{14} \alpha _{\theta _1}^2 \alpha _{\theta _2}^5} = 0\nonumber\\
& & \hskip -0.7in  EOM_{xx} : \frac{144 \sqrt{3} \sqrt[4]{\pi } a^{10} \beta  \sqrt[4]{\frac{1}{N}}}{\log N  ^3 {N_f^{\rm UV}}\ ^2
   r^{10}} +\frac{{g_s}^{7/4} N^{9/10} \alpha _{\theta _2}^2 (\log N  -18 \log r )^2 ({f_{zz} } (r)-2
   {f_{\theta_1z} } (r)+2 {f_{\theta_1x} } (r)-{f_r}(r))}{\log r ^2 (\log N  -9 \log r )^2}=0\nonumber\\
& & \hskip -0.7in EOM_{xy} : \frac{49 N^{6/5} (2 {f_{zz} } (r)-3 {f_{x^{10}x^{10}}} (r)-4 {f_{\theta_1z} } (r)-5 {f_{\theta_2y} } (r))}{3888 \sqrt{6} \alpha _{\theta _1}^4
   \alpha _{\theta _2}^5 {b_1}(r)^2}  + \frac{8 a^{12} \beta  \left(\frac{1}{N}\right)^{3/10} \left(-\Sigma_1\right)}{243 \sqrt{3} \sqrt[4]{\pi } {g_s}^{9/4} \log N  ^3
   {N_f^{\rm UV}}\ ^2 r^{12} \alpha _{\theta _1}^6 \alpha _{\theta _2}^3} = 0\nonumber\\
& & \nonumber\\
& & \hskip -0.7in EOM_{xz} : -\frac{49 N^{3/2} (2 {f_{zz} } (r)-3 {f_{x^{10}x^{10}}} (r)-4 {f_{\theta_1z} } (r)-5 {f_{\theta_2y} } (r))}{34992 \alpha _{\theta _1}^4
   \alpha _{\theta _2}^6 {b_1}(r)^2} \nonumber\\
   & & \hskip -0.3in + \frac{16 a^{10} \beta  \left(\frac{1}{N}\right)^{13/20} \left(-4 {g_s} {N_f^{\rm UV}}\  \log \left(\alpha _{\theta
   _1}\right)-4 {g_s} {N_f^{\rm UV}}\  \log \left(\alpha _{\theta _2}\right)+6 {g_s} {N_f^{\rm UV}}\  \log
   \left(\frac{1}{r_0}\right)+4 {g_s} {N_f^{\rm UV}}\  \log (4)+8 \pi \right)}{27 \sqrt{3} \sqrt[4]{\pi } {g_s}^{13/4}
   \log N  ^4 {N_f^{\rm UV}}\ ^3 r^{10} \alpha _{\theta _1}^2 \alpha _{\theta _2}^4}\nonumber\\
& &  = 0\nonumber\\
& & \hskip -0.7in EOM_{yy} : -\frac{49 \sqrt{N} (16 {f_{zz} } (r)+3 {f_{x^{10}x^{10}}} (r)+4 {f_{\theta_1z} } (r)+5 {f_{\theta_2y} } (r)-36 {f_{yz} }(r)+18
   {f_{yy}}(r))}{1728 \alpha _{\theta _1}^2 \alpha _{\theta _2}^4 {b_1}(r)^2}\nonumber\\
& &  +
\frac{2 \sqrt{2} a^{12} \beta (-\Sigma_1)}{27 \sqrt[4]{\pi } {g_s}^{9/4} \log N  ^3 N {N_f^{\rm UV}}\ ^2 r^{12} \alpha _{\theta
   _1}^4 \alpha _{\theta _2}^2}=0\nonumber\\
& & \nonumber\\
& & \hskip -0.7in EOM_{yz} :  -\frac{49 N^{4/5} (2 {f_{zz} } (r)-3 {f_{x^{10}x^{10}}} (r)-4 {f_{\theta_1z} } (r)-5 {f_{\theta_2y} } (r))}{2592 \sqrt{6} \alpha _{\theta _1}^2
   \alpha _{\theta _2}^5 {b_1}(r)^2}
- \frac{8 a^{12} \beta  \left(\frac{1}{N}\right)^{21/20} \Sigma_1}{729 {g_s}^2 \log N  ^3 {N_f^{\rm UV}}\ ^2 r^{12} \alpha
   _{\theta _1}^4 \alpha _{\theta _2}} = 0\nonumber\\
& & \nonumber\\
& & \hskip -0.7in EOM_{zz} : \frac{49 N^{11/10} (2 {f_{zz} } (r)-3 {f_{x^{10}x^{10}}} (r)-4 {f_{\theta_1z} } (r)-5 {f_{\theta_2y} } (r))}{23328 \alpha _{\theta _1}^2
   \alpha _{\theta _2}^6 {b_1}(r)^2}
-\frac{8 \sqrt{2} a^{12} \beta  \left(\frac{1}{N}\right)^{2/5} (-\Sigma_1)}{6561 \sqrt[4]{\pi } {g_s}^{9/4}
   \log N  ^3 {N_f^{\rm UV}}\ ^2 r^{12} \alpha _{\theta _1}^4 \alpha _{\theta _2}^4} = 0\nonumber\\
& & \hskip -0.7in EOM_{x^{10}x^{10}} :  -\frac{49 \sqrt{N} (16 {f_{zz} } (r)+3 {f_{x^{10}x^{10}}} (r)-32 {f_{\theta_1z} } (r)-13 {f_{\theta_2y} } (r))}{1728 \alpha _{\theta _1}^2
   \alpha _{\theta _2}^4 {a_1}(r)^2 {b_1}(r)^2} \nonumber\\
& &  +
\frac{8 \sqrt{\pi } a^{10} \beta  (27 \log N  +14) M_{\rm UV} \left(\frac{1}{N}\right)^{7/4}\Sigma_1 \log (r)}{81
   \sqrt{{g_s}} \log N  ^5 {N_f^{\rm UV}}\ ^3 r^{10} \alpha _{\theta _2}^3} = 0\nonumber\\
& &
\end{eqnarray}
where:
\begin{equation}
\label{Sigma_1}
\Sigma_1 \equiv  \left(-19683 \sqrt{6} \alpha _{\theta
   _1}^6-6642 \alpha _{\theta _2}^2 \alpha _{\theta _1}^3+40 \sqrt{6} \alpha _{\theta _2}^4\right).
\end{equation}

From (\ref{EOMs}), one sees that the ${\cal O}(\beta)$-terms are further large-$N$-large-$r$-suppressed. Hence, in the UV one sees that one can consistently assume $f_{MN}$ to be vanishingly small.
\section{Coupling Constants $y_{1,3,5,7}$ and $z_{1,....,8}$}
In this appendix, we work out the IR and UV contributions to the coupling constants $y_{1,3,5,7}$ and $z_{1,2,...,8}$.
In the following, $\psi_0(Z) = \int_0^Z \phi_0(Z)dZ$. In the following, we will be splitting $\int dZ$ into $\int_{\rm IR} + \int_{\rm UV}$. Now, one can argue
that $\int_{\rm UV}\sim \left({\cal C}_{\phi_0}^{\rm UV}\right)^m\left({\cal C}_{\psi_1}^{\rm UV}\right)^n, m, n \in \mathbb{Z}^+$, and we can self-consistently set ${\cal C}_{\phi_0}^{\rm UV} = {\cal C}_{\psi_1}^{\rm UV} = 0$, and one can hence argue that one can disregard $\int_{\rm UV}$.

In the following calculations, one will need the value of ${\cal V}_2$ in the IR, which can be shown to be given by:
\begin{eqnarray}
\label{V2-IR}
& & \hskip -0.6in {\cal V}_2(Z\in{\rm IR}) = -\frac{3 \left(3 b^2-2\right)
   {g_s}^2 M N^{4/5} {N_f}^2 \log ({r_0}) (\log N -3 \log ({r_0}))}{2 \pi  \log N  \alpha _{\theta _1} \alpha _{\theta
   _2}^2}\nonumber\\
& & \hskip -0.6in \frac{3 \left(2-3 b^2\right) \beta  {g_s}^2 \log r_0  M N^{4/5} {N_f}^2 Z ({\cal C}_{zz}^{(1)}-2   {\cal C}_{\theta_1z}^{(1)}+2   {\cal C}_{\theta_1x}^{(1)})
  }{4 \pi  \alpha _{\theta _1} \alpha _{\theta _2}^2}  \left(\frac{9 b^2 {g_s}^2 M N^{4/5} {N_f}^2 \log ({r_0}) (\log N -3 \log ({r_0}))}{\pi
   \log N  \alpha _{\theta _1} \alpha _{\theta _2}^2}\right)\nonumber\\
\end{eqnarray}
Therefore, using (\ref{y_i+z_i}) and using $b=\frac{1}{\sqrt{3}} + \epsilon$ \cite{OR4-Yadav+Misra} and setting $r_0 = N^{-\frac{f_{r_0}}{3}}$ \cite{Bulk-Viscosity}, one obtains the following results for $y_{1,3,5,7}$:
\begin{eqnarray}
\label{y_i}
& & \hskip -0.5in \bullet\  y_1 = \int {\cal V}_2(Z)  \left(1 + \psi_1(Z) - \psi_0^2(Z)\right)^2\nonumber\\
& & \hskip -0.5in  = \frac{177147 \pi ^6 \beta  {\cal C}_{\phi_0}^{\rm IR}\ ^4 \epsilon ^6 {f_{r_0}} ({f_{r_0}}+1)^3 \log r_0  \alpha _{\theta _1}^6
   N^{4 {f_{r_0}}+\frac{14}{5}} \left({\cal C}_{zz}^{(1)} - 2 {\cal C}_{\theta_1z}^{(1)} + 2 {\cal C}_{\theta_1x}^{(1)}\right)}{8192 ({f_{r_0}}-1)^6 {g_s}^4
   \left(\log N\right) ^7 M^2 N_f ^4}\nonumber\\
& & \hskip -0.5in -\frac{4782969 \sqrt{3} \pi ^7 {\cal C}_{\phi_0}^{\rm IR}\ ^4 \epsilon ^5 {f_{r_0}} ({f_{r_0}}+1)^4
   \alpha _{\theta _1}^9 N^{4 {f_{r_0}}+\frac{11}{5}}}{20480 ({f_{r_0}}-1)^7 {g_s}^6 \left(\log N\right) ^7 M^3 N_f ^6}\nonumber\\
& & \hskip -0.5in \nonumber\\
& & \hskip -0.5in \bullet\  y_3 = \int dZ {\cal V}_2\psi_1^2(Z)\left(1 + \psi_1(Z)\right)^2\nonumber\\
& & \hskip -0.5in  = -\frac{21 {g_s}^2 \log r_0  M N^{4/5} {N_f}^2 {\cal C}_{\psi_1}^{\rm IR}\ ^4 (\log N -3 \log r_0 )}{8 \pi
   \log N  \alpha _{\theta _1} \alpha _{\theta _2}^2}\nonumber\\
& &  \hskip -0.5in+ \frac{3^{9/8} 7^{3/4} \sqrt[4]{\epsilon } {g_s}^2 \log r_0  M N^{4/5} {N_f}^2 {\cal C}_{\psi_1}^{\rm IR}\ ^3
   (\log N -3 \log r_0 )}{2 \sqrt{2} \pi  \log N  \alpha _{\theta _1} \alpha _{\theta _2}^2}\nonumber\\
& & \hskip -0.5in + 945 \sqrt[3]{3} \sqrt[6]{\frac{\pi }{2}} \beta  \epsilon  {g_s}^{3/2} \left(\frac{1}{\log N }\right)^{2/3}
   \log r_0  M N^{3/10} {N_f}^{4/3} {r_0}^2 {\cal C}_{\psi_1}^{\rm IR}\ ^4 ({\cal C}_{zz}^{(1)}-2 {\cal C}_{\theta_1z}^{(1)}+2{\cal C}_{\theta_1x}^{(1)})\nonumber\\
& & \hskip -0.5in \nonumber\\
& &  \hskip -0.5in \bullet\  y_5 = \int dZ {\cal V}_2\psi_0^2(Z)\left(\psi_1(Z)\right)^2) =\frac{6561 \sqrt[4]{3} \sqrt{7} \pi ^3 {\cal C}_{\phi_0}^{\rm IR}\ ^2 \epsilon ^{5/2} {f_{r_0}} ({f_{r_0}}+1)^3 \alpha _{\theta _1}^5 {\cal C}_{\psi_1}^{\rm IR}\ ^2 N^{2
   {f_{r_0}}+\frac{8}{5}}}{256 ({f_{r_0}}-1)^4 {g_s}^2 \left(\log N\right) ^3 M N_f ^2 \alpha _{\theta _2}^2}
\nonumber\\
& & \hskip -0.5in -\frac{63 \sqrt[4]{3} \sqrt{7} \pi ^3 \beta  {\cal C}_{\phi_0}^{\rm IR}\ ^2 \epsilon ^{3/2} {f_{r_0}} ({f_{r_0}}+1) \alpha _{\theta _1}^3 {\cal C}_{\psi_1}^{\rm IR}\ ^2 N^{2
   {f_{r_0}}-\frac{3}{5}} \left({\cal C}_{zz}^{(1)} - 2 {\cal C}_{\theta_1z}^{(1)} + 2 {\cal C}_{\theta_1x}^{(1)}\right) \left(6 \epsilon  N+{f_{r_0}} \gamma  {g_s} \log N  M^2\right)^2}{32768
   ({f_{r_0}}-1)^3 {g_s}^2 \left(\log N\right) ^3 M N_f ^2}\nonumber\\
& &  \hskip -0.5in\nonumber\\
& & \hskip -0.5in \bullet\  y_7= \int dZ {\cal V}_2\psi_1(Z) (1 + \psi_1(Z)) (1 + \psi_1(Z) - \psi_0(Z)^2)^2\nonumber\\
& &  \hskip -0.5in = \frac{4782969 \sqrt[4]{3} \sqrt{7} \pi ^7 {\cal C}_{\phi_0}^{\rm IR}\ ^4 \epsilon ^{9/2} {f_{r_0}} ({f_{r_0}}+1)^4 \alpha _{\theta _1}^9
   {\cal C}_{\psi_1}^{\rm IR}\ ^2 N^{4 {f_{r_0}}+\frac{11}{5}}}{40960 ({f_{r_0}}-1)^7 {g_s}^6 \left(\log N\right) ^7 M^3 N_f ^6}\nonumber\\
& & \hskip -0.5in -\frac{3720087
   \sqrt[4]{3} \sqrt{7} \pi ^7 \beta  {\cal C}_{\phi_0}^{\rm IR}\ ^4 \epsilon ^{9/2} {f_{r_0}} ({f_{r_0}}+1)^3 \alpha _{\theta _1}^9
   {\cal C}_{\psi_1}^{\rm IR}\ ^2 N^{4 {f_{r_0}}+\frac{11}{5}} \left({\cal C}_{zz}^{(1)} - 2 {\cal C}_{\theta_1z}^{(1)} + 2 {\cal C}_{\theta_1x}^{(1)}\right)}{262144 ({f_{r_0}}-1)^7
   {g_s}^6 \left(\log N\right) ^7 M^3 N_f ^6}.
\end{eqnarray}
Similarly, one obtains the following expressions for $z_{1,...,8}$:
{\footnotesize
\begin{eqnarray}
\label{z_i}
& &\hskip -0.3in \bullet\  z_1 = \int dZ {\cal V}_2 \left(1 + \psi_1(Z)\right)^2= \frac{3 \sqrt[4]{3} \sqrt{7} \sqrt{\epsilon } {g_s}^2 M N^{4/5} {N_f}^2 {\cal C}_{\psi_1}^{\rm IR}\ ^2 \log ({r_0})
   (\log N -3 \log ({r_0}))}{16 \pi  \alpha _{\theta _1} \alpha _{\theta _2}^2 \log N } \nonumber\\
& & \hskip -0.3in + \frac{7 \sqrt{7} \beta  \sqrt{\epsilon } {g_s}^2 \log r_0  M N^{4/5} {N_f}^2 {\cal C}_{\psi_1}^{\rm IR}\ ^2
   ({\cal C}_{zz}^{(1)}-2 {\cal C}_{\theta_1z}^{(1)}+2 {\cal C}_{\theta_1x}^{(1)}) \left(12 \log N +\left(-36+6 \log ^2(3)+\log (9) \log (27)-\log (27)
   \log (81)\right) \log ({r_0})\right)}{2048\ 3^{3/4} \pi  \alpha _{\theta _1} \alpha _{\theta _2}^2 (\log N -3 \log
   ({r_0}))}\nonumber\\
& &\hskip -0.3in \nonumber\\
& &\hskip -0.3in \bullet\  z_2 = \int dZ {\cal V}_2\psi_0^2(Z)  =\frac{81 \sqrt{3} \pi ^3 \beta  {\cal C}_{\phi_0}^{\rm IR}\ ^2 \epsilon ^3 ({f_{r_0}}+1)^2 \alpha _{\theta _1}^3 N^{\frac{5 {f_{r_0}}}{3}+\frac{7}{5}}
   \left({\cal C}_{zz}^{(1)} - 2 {\cal C}_{\theta_1z}^{(1)} + 2 {\cal C}_{\theta_1x}^{(1)}\right)}{128 ({f_{r_0}}-1)^2 {g_s} \left(\log N\right) ^2 M N_f ^2}\nonumber\\
& & \hskip -0.3in -\frac{81 \sqrt{3} \pi ^3
   {\cal C}_{\phi_0}^{\rm IR}\ ^2 \epsilon ^3 {f_{r_0}} ({f_{r_0}}+1)^2 \alpha _{\theta _1}^3 N^{2 {f_{r_0}}+\frac{7}{5}}}{256 ({f_{r_0}}-1)^3
   {g_s}^2 \left(\log N\right) ^3 M N_f ^2}\nonumber\\
& &\hskip -0.3in \nonumber\\
& &\hskip -0.3in \bullet\  z_3 = \int dZ {\cal V}_2\psi_1(1 + \psi_1)(Z) = \frac{3 \sqrt[4]{3} \sqrt{7} \sqrt{\epsilon } {g_s}^2 M N^{4/5} {N_f}^2 {\cal C}_{\psi_1}^{\rm IR}\ ^2 \log ({r_0})
   (\log N -3 \log ({r_0}))}{8 \pi  \alpha _{\theta _1} \alpha _{\theta _2}^2 \log N }\nonumber\\
& &\hskip -0.3in + \frac{7 \sqrt[4]{3} \sqrt{7} \beta  \sqrt{\epsilon } {g_s}^2 M N^{4/5} {N_f}^2 {\cal C}_{\psi_1}^{\rm IR}\ ^2 \log
   ({r_0}) ({\cal C}_{zz}^{(1)}-2 {\cal C}_{\theta_1z}^{(1)}+2 {\cal C}_{\theta_1x}^{(1)})}{512 \pi  \alpha _{\theta _1} \alpha _{\theta _2}^2}\nonumber\\
& &\hskip -0.3in \nonumber\\
& &\hskip -0.3in \bullet\  z_4 = \int dZ {\cal V}_2\psi_1(1 + \psi_1 - \phi_0^2)(Z)  = \frac{189\ 3^{3/8} \sqrt[4]{7} \pi ^3 \beta  {\cal C}_{\phi_0}^{\rm IR}\ ^2 \epsilon ^{3/4} {f_{r_0}} ({f_{r_0}}+1) \alpha _{\theta _1}^3
   {\cal C}_{\psi_1}^{\rm IR}\  N^{2 {f_{r_0}}+\frac{7}{5}} \left({\cal C}_{zz}^{(1)} - 2 {\cal C}_{\theta_1z}^{(1)} + 2 {\cal C}_{\theta_1x}^{(1)}\right)}{16384 \sqrt{2} ({f_{r_0}}-1)^3
   {g_s}^2 \left(\log N\right) ^3 M N_f ^2}\nonumber\\
& &\hskip -0.3in +\frac{81\ 3^{3/8} \sqrt[4]{7} \pi ^3 {\cal C}_{\phi_0}^{\rm IR}\ ^2 \epsilon ^{11/4} {f_{r_0}} ({f_{r_0}}+1)^2
   \alpha _{\theta _1}^3 {\cal C}_{\psi_1}^{\rm IR}\  N^{2 {f_{r_0}}+\frac{7}{5}}}{256 \sqrt{2} ({f_{r_0}}-1)^3 {g_s}^2 \left(\log N\right) ^3 M
   N_f ^2}\nonumber\\
& &\hskip -0.3in \nonumber\\
& &\hskip -0.3in \bullet\  z_5 = \int dZ {\cal V}_2\psi_1^2(1 + \psi_1)(Z),
  = -\frac{3 \sqrt[8]{3} 7^{3/4} \sqrt[4]{\epsilon } {g_s}^2 M N^{4/5} {N_f}^2 {\cal C}_{\psi_1}^{\rm IR}\ ^3 \log ({r_0}) (\log N -3 \log
   ({r_0}))}{4 \sqrt{2} \pi  \alpha _{\theta _1} \alpha _{\theta _2}^2 \log N }
\nonumber\\
& & \hskip -0.3in -\frac{21 \sqrt[8]{3} 7^{3/4} \beta  \sqrt[4]{\epsilon } {g_s}^2 M N^{4/5} {N_f}^2 {\cal C}_{\psi_1}^{\rm IR}\ ^3 \log ({r_0})
   ({\cal C}_{zz}^{(1)}-2 {\cal C}_{\theta_1z}^{(1)}+2 {\cal C}_{\theta_1x}^{(1)})}{256 \sqrt{2} \pi  \alpha _{\theta _1} \alpha _{\theta _2}^2}\nonumber\\
& &  \nonumber\\
& &\hskip -0.3in \bullet\  z_6 = \int dZ {\cal V}_2(1 + \psi_1)(1 + \psi_1 - \psi_0^2)(Z)  = \frac{567\ 3^{3/8} \sqrt[4]{7} \pi ^3 {\cal C}_{\phi_0}^{\rm IR}\ ^2 \epsilon ^{11/4} {f_{r_0}} ({f_{r_0}}+1) \alpha _{\theta _1}^3 {\cal C}_{\psi_1}^{\rm IR}\
   N^{2 {f_{r_0}}+\frac{7}{5}} \left({\cal C}_{zz}^{(1)} - 2 {\cal C}_{\theta_1z}^{(1)} + 2 {\cal C}_{\theta_1x}^{(1)}\right)}{8192 \sqrt{2} ({f_{r_0}}-1)^3 {g_s}^2 \left(\log N\right) ^3 M
   N_f ^2}\nonumber\\
& &\hskip -0.3in +\frac{81\ 3^{3/8} \sqrt[4]{7} \pi ^3 {\cal C}_{\phi_0}^{\rm IR}\ ^2 \epsilon ^{11/4} {f_{r_0}} ({f_{r_0}}+1)^2 \alpha _{\theta _1}^3
   {\cal C}_{\psi_1}^{\rm IR}\  N^{2 {f_{r_0}}+\frac{7}{5}}}{64 \sqrt{2} ({f_{r_0}}-1)^3 {g_s}^2 \left(\log N\right) ^3 M N_f ^2}\nonumber\\
& & \hskip -0.3in= \frac{567\ 3^{3/8} \sqrt[4]{7} \pi ^3 {\cal C}_{\phi_0}^{\rm IR}\ ^2 \epsilon ^{11/4} {f_{r_0}} ({f_{r_0}}+1) \alpha _{\theta _1}^3 {\cal C}_{\psi_1}^{\rm IR}\
   N^{2 {f_{r_0}}+\frac{7}{5}} \left({\cal C}_{zz}^{(1)} - 2 {\cal C}_{\theta_1z}^{(1)} + 2 {\cal C}_{\theta_1x}^{(1)}\right)}{8192 \sqrt{2} ({f_{r_0}}-1)^3 {g_s}^2 \left(\log N\right) ^3 M
   N_f ^2}\nonumber\\
& &\hskip -0.3in +\frac{81\ 3^{3/8} \sqrt[4]{7} \pi ^3 {\cal C}_{\phi_0}^{\rm IR}\ ^2 \epsilon ^{11/4} {f_{r_0}} ({f_{r_0}}+1)^2 \alpha _{\theta _1}^3
   {\cal C}_{\psi_1}^{\rm IR}\  N^{2 {f_{r_0}}+\frac{7}{5}}}{64 \sqrt{2} ({f_{r_0}}-1)^3 {g_s}^2 \left(\log N\right) ^3 M N_f ^2}\nonumber\\
& & \nonumber\\
& & \hskip -0.3in \bullet\  z_7 = \int dZ {\cal V}_2\psi_1(1 + \psi_1)^2(Z)  =
\frac{3 \sqrt[8]{3} 7^{3/4} \sqrt[4]{\epsilon } g_s ^2 M N^{4/5} {N_f}^2 {\cal C}_{\psi_1}^{\rm IR}\ ^3 \log (r_0 )
   (\log N -3 \log (r_0 ))}{4 \sqrt{2} \pi  \alpha _{\theta _1} \alpha _{\theta _2}^2 \log N }\nonumber\\
& &\hskip -0.3in -\frac{21 \sqrt[8]{3} 7^{3/4} \beta  \sqrt[4]{\epsilon } g_s ^2 M N^{4/5} {N_f}^2 {\cal C}_{\psi_1}^{\rm IR}\ ^3 \log
   (r_0 ) ( {\cal C}_{zz}^{(1)} - 2 {\cal C}_{\theta_1z}^{(1)} + 2 {\cal C}_{\theta_1x})}{256 \sqrt{2} \pi  \alpha _{\theta _1} \alpha _{\theta
   _2}^2} +\frac{48\ 3^{3/4} \sqrt{\frac{1}{\epsilon }} g_s ^2 M N^{4/5} {N_f}^2 r_0  \log N  \log ^2(r_0 )
   {\cal C}_{\psi_1}^{\rm UV}\ }{\pi  \log N  \alpha _{\theta _1} \alpha _{\theta _2}^2}\nonumber\\
& & \nonumber\\
& & \hskip -0.3in \bullet\  z_8 = \int dZ {\cal V}_2\psi_0^2\psi_1(Z) =  \frac{567\ 3^{3/8} \sqrt[4]{7} \pi ^3 \beta  {\cal C}_{\phi_0}^{\rm IR}\ ^2 \epsilon ^{11/4} {f_{r_0}} ({f_{r_0}}+1) \alpha _{\theta
   _1}^3 {\cal C}_{\psi_1}^{\rm IR}\  N^{2 {f_{r_0}}+\frac{7}{5}} \left({\cal C}_{zz}^{(1)} - 2 {\cal C}_{\theta_1z}^{(1)} + 2 {\cal C}_{\theta_1x}^{(1)}\right)}{8192
   \sqrt{2} ({f_{r_0}}-1)^3 {g_s}^2 \left(\log N\right) ^3 M N_f ^2}\nonumber\\
& & \hskip -0.3in +\frac{81\ 3^{3/8} \sqrt[4]{7} \pi ^3
   {\cal C}_{\phi_0}^{\rm IR}\ ^2 \epsilon ^{11/4} {f_{r_0}} ({f_{r_0}}+1)^2 \alpha _{\theta _1}^3 {\cal C}_{\psi_1}^{\rm IR}\  N^{2
   {f_{r_0}}+\frac{7}{5}}}{64 \sqrt{2} ({f_{r_0}}-1)^3 {g_s}^2 \left(\log N\right) ^3 M N_f ^2}.
\end{eqnarray}
}

\section{$G_{\rm global} \times H_{\rm local}$ HLS Formalism and Obtaining GL's Mesonic $\chi$PT Lagrangian up to  ${\cal O}(p^4)$ After Integrating Out the Vector Mesons from the HLS Lagrangian \cite{HLS-Physics-Reports}}

In this appendix, we summarize the HLS formalism and the arguments of how to obtain the $SU(3)$ $\chi$PT Lagrangian of \cite{GL} up to (NLO in momentum below the chiral symmetry breaking scale) ${\cal O}(p^4)$ by integrating out the vector mesons from the HLS Lagrangian (both as discussed in detail in  \cite{HLS-Physics-Reports})

\subsection{HLS Formalism}
The HLS formalism describes a model based on $G_{\rm global} \times H_{\rm local}$ symmetry, where
$G = \mbox{SU($N_f$)}_{\rm L} \times
\mbox{SU($N_f$)}_{\rm R}$  is the
global chiral symmetry and
$H = \mbox{SU($N_f$)}_{\rm V}$ is the H(idden) L(ocal) S(ymmetry). The building blocks of
this model are  SU($N_f$)-matrix valued variables $\xi_{\rm L}$ and
$\xi_{\rm R}$ which are introduced by splitting $U$ in the ChPT as
\begin{equation*}
\label{split-U-xiLR}
U = \xi_{\rm L}^\dagger \xi_{\rm R} \ .
\end{equation*}
Now, under $G_{\rm global} \times H_{\rm local}$ $\xi_{\rm L,R}(x)$ transform as follows:
\begin{equation*}
\label{xiLR-trans}
\xi_{\rm L,R}(x) \rightarrow \xi_{\rm L,R}^{\prime}(x) =
h(x) \cdot \xi_{\rm L,R}(x) \cdot g^{\dag}_{\rm L,R}
\ ,
\end{equation*}
where $h(x) \in H_{\rm local},\
g_{\rm L,R} \in G_{\rm global}$.  These variables are parameterized as
\begin{eqnarray*}
\label{def:xiLR}
&& \xi_{\rm L,R} = e^{i\sigma/F_\sigma} e^{\mp i\pi/F_\pi} \ ,
\quad
\left[ \, \pi = \pi^a T_a \,,\, \sigma = \sigma^a T_a \right] \ ,
\end{eqnarray*}
where the matrix-valued $\pi$ denotes the Nambu-Goldstone (NG) bosons associated with
the spontaneous breaking of $G$ chiral symmetry and $\sigma$ denotes the NG bosons absorbed into the gauge bosons. Further, $F_\pi$ and $F_\sigma$ denote the relevant decay constants, and
the parameter $a$ is defined via: $a \equiv \frac{F_\sigma^2}{F_\pi^2}$.

From $\xi_{\rm L}$ and $\xi_{\rm R}$ the following are constructed:
\begin{eqnarray*}
\alpha_{\perp\mu} &=&
\left(
  \partial_\mu \xi_{\rm R} \cdot \xi_{\rm R}^\dag -
  \partial_\mu \xi_{\rm L} \cdot \xi_{\rm L}^\dag
\right)
/ (2i) \ ,
\label{def-alpha perp}
\\
\alpha_{\parallel\mu} &=&
\left(
  \partial_\mu \xi_{\rm R} \cdot \xi_{\rm R}^\dag +
  \partial_\mu \xi_{\rm L} \cdot \xi_{\rm L}^\dag
\right)
/ (2i) \ ,
\label{def-alpha parallel}
\end{eqnarray*}
which transform under $H_{\rm local}$ as
\begin{eqnarray*}
&& \alpha_{\perp\mu} \rightarrow
  h(x) \cdot \alpha_{\perp\mu} \cdot h^\dag (x) \ ,
\\
&& \alpha_{\parallel\mu} \rightarrow
  h(x) \cdot \alpha_{\perp\mu} \cdot h^\dag (x)
  -i  \partial_\mu h(x) \cdot h^\dag(x) \ .
\end{eqnarray*}
The covariant derivatives of $\xi_{\rm L}$ and $\xi_{\rm R}$ can be obtained from the transformation properties of $\xi_{\rm L,R}(x)$ as:
\begin{equation*}
D_\mu \xi_{\rm L,R} = \partial_\mu \xi_{\rm L,R} - i V_\mu
\xi_{\rm L,R} \ ,
\end{equation*}
where $V_\mu = V_\mu^a T_a$ are the gauge fields corresponding to $H_{\rm local}$.
Now, $V_\mu$ transforms under $H_{\rm local}$ as:
\begin{equation*}
V_\mu \rightarrow
 h(x) \cdot V_\mu \cdot h^\dag (x)
  - i \partial_\mu h(x) \cdot h^\dag(x)
\ .
\end{equation*}
The ``covariantized 1-forms" defined as
\begin{eqnarray*}
\label{def:al hat perp 0}
& & \widehat{\alpha}_{\perp\mu} =
\frac{1}{2i}
\left(
  D_\mu \xi_{\rm R} \cdot \xi_{\rm R}^\dag -
  D_\mu \xi_{\rm L} \cdot \xi_{\rm L}^\dag
\right) = \alpha_{\perp\mu}\ ,
\nonumber\\
& & \widehat{\alpha}_{\parallel\mu} =
\frac{1}{2i}
\left(
  D_\mu \xi_{\rm R} \cdot \xi_{\rm R}^\dag +
  D_\mu \xi_{\rm L} \cdot \xi_{\rm L}^\dag
\right)= \alpha_{\parallel\mu}
- V_\mu \  ,
\end{eqnarray*}
transform homogeneously under $H_{\rm local}$:
\begin{equation*}
\widehat{\alpha}_{\perp,\parallel}^\mu \rightarrow
  h(x) \cdot \widehat{\alpha}_{\perp,\parallel}^\mu \cdot h^\dag (x) \ .
\end{equation*}

Thus, one can construct the following two invariants:
\begin{eqnarray*}
&& {\cal L}_{\rm A} \equiv
F_\pi^2 \, \mbox{tr}
\left[ \widehat{\alpha}_{\perp\mu} \widehat{\alpha}_{\perp}^\mu
\right]
\ ,
\label{def:LA 0}
\\
&& a {\cal L}_{\rm V} \equiv
F_\sigma^2 \, \mbox{tr}
\left[
  \widehat{\alpha}_{\parallel\mu} \widehat{\alpha}_{\parallel}^\mu
\right]
=
F_\sigma^2 \, \mbox{tr}
\left[
  \left(
    V_\mu - \alpha_{\parallel\mu}
  \right)^2
\right]
\ .
\label{def:LV 0}
\end{eqnarray*}
Therefore, the most general Lagrangian
made out of $\xi_{\rm L,R}$ and $D_\mu\xi_{\rm L,R}$ with the lowest number of
derivatives, i.e., at ${\cal O}(p^2)$ (see \cite{GLF, GL, HLS-Physics-Reports} for power counting arguments), is  given by:
\begin{equation*}
\label{L-Op2}
{\cal L} = {\cal L}_{\rm A} + a {\cal L}_{\rm V}.
\end{equation*}
Using the EOM for $V_\mu$ at ${\cal O}(p^2)$, one obtains:
\begin{equation*}
\label{Vmu-EOM-Op2}
V_\mu = \alpha_{\parallel\mu} \ .
\end{equation*}
Further, with the relation
\begin{equation*}
\widehat{\alpha}_{\perp\mu}
= \frac{1}{2i}\,\xi_{\rm L} \cdot \partial_\mu U \cdot
  \xi_{\rm R}^\dag
= \frac{i}{2} \,\xi_{\rm R} \cdot \partial_\mu
  U^\dag \cdot \xi_{\rm L}^\dag
\label{perp U0}
\end{equation*}
substituted into (\ref{L-Op2}), one obtains the following ${\cal O}(p^2)$ term in the $\chi$PT Lagrangian:
\begin{equation*}
{\cal L} = {\cal L}_{\rm A} =
\frac{F_\pi^2}{4} \mbox{tr}
\left[ \partial_\mu U^\dag \partial^\mu U \right]
\ .
\end{equation*}

In the unitary gauge, $\sigma=0$\footnote{This unitary gauge is not preserved though under the $G_{\rm global}$
transformation, which in general has the following form
\begin{eqnarray*}
\xi \rightarrow \xi^{\prime}
&=& \xi \cdot g_{\rm R}^\dag = g_{\rm L} \cdot \xi
\nonumber\\
&=&
\exp\left[i\sigma^\prime( \pi, g_{\rm R}, g_{\rm L} )/F_\sigma \right]
\exp\left[i\pi^\prime/F_\pi \right]
\nonumber\\
&=&
\exp\left[i\pi^\prime/F_\pi\right]
\exp\left[- i\sigma^\prime(\pi, g_{\rm R}, g_{\rm L})/F_\sigma\right]
\ .
\end{eqnarray*}
However, $\exp\left[i\sigma^\prime( \pi, g_{\rm R}, g_{\rm L} )/F_\sigma
\right]$
can be eliminated if we simultaneously perform the $H_{\rm local}$
gauge transformation:
\begin{eqnarray*}
h =
\exp\left[i\sigma^\prime( \pi, g_{\rm R}, g_{\rm L} )/F_\sigma
\right]
\equiv
h \left( \pi, g_{\rm R}, g_{\rm L}\right) \ .
\end{eqnarray*}

Therefore there is a global symmetry $G =
\mbox{SU($N_f$)}_{\rm L} \times
\mbox{SU($N_f$)}_{\rm R}$ under the following combined transformation:
\begin{equation*}
G \ : \ \xi \rightarrow h \left( \pi, g_{\rm R}, g_{\rm L}\right)
\cdot \xi \cdot g_{\rm R}^\dag
= g_{\rm L} \cdot \xi \cdot
h^\dag \left( \pi, g_{\rm R}, g_{\rm L}\right)
\ .
\label{com trans}
\end{equation*}}, two
$SU(N_f)$-matrix valued variables $\xi_{\rm L}$ and
$\xi_{\rm R}$ are related via:
\begin{equation*}
\xi_{\rm L}^\dag = \xi_{\rm R} \equiv \xi = e^{i\pi/F_\pi}
\ .
\end{equation*}


\subsection{Obtaining the ${\cal O}(p^2)$ $SU(3)$ $\chi$PT Lagrangian by Integrating out the Vector Mesons from the HLS Lagrangian}

First, we introduce the external gauge fields ${\cal L}_\mu$
and ${\cal R}_\mu$ which include $W$ boson, $Z$-boson and photon
fields.
This is done by gauging the $G_{\rm global}$ symmetry.
The transformation properties of ${\cal L}_\mu$
and ${\cal R}_\mu$, ${\cal L}_\mu\rightarrow g_L {\cal L}_\mu g_L^\dagger - i\partial_\mu g_L . g_L^\dagger$ and ${\cal R}_\mu\rightarrow g_R {\cal R}_\mu g_R^\dagger - i\partial_\mu g_R . g_R^\dagger$ are used to define covariant derivatives of $\xi_{\rm L,R}$:
\begin{eqnarray}
D_\mu \xi_{\rm L} &=&
\partial_\mu \xi_{\rm L} - i V_\mu \xi_{\rm L}
+ i \xi_{\rm L} {\cal L}_\mu \ ,
\nonumber\\
D_\mu \xi_{\rm R} &=&
\partial_\mu \xi_{\rm R} - i V_\mu \xi_{\rm R}
+ i \xi_{\rm R} {\cal R}_\mu \ .
\label{covder}
\end{eqnarray}
It should be noticed that in the HLS these external gauge fields are
included without assuming the vector dominance.
It is outstanding feature of the HLS model that $\xi_{\rm L,R}$ have
two independent source charges and hence two independent gauge bosons
are automatically introduced in the HLS model.
Both the vector meson fields and external gauge fields are
simultaneously incorporated into the Lagrangian fully consistent with
the chiral symmetry.
By using the above covariant derivatives two
Maurer-Cartan 1-forms are constructed as
\begin{eqnarray}
\widehat{\alpha}_{\perp\mu} &=&
\left(
  D_\mu \xi_{\rm R} \cdot \xi_{\rm R}^\dag -
  D_\mu \xi_{\rm L} \cdot \xi_{\rm L}^\dag
\right)
/ (2i) \ ,
\nonumber\\
\widehat{\alpha}_{\parallel\mu} &=&
\left(
  D_\mu \xi_{\rm R} \cdot \xi_{\rm R}^\dag +
  D_\mu \xi_{\rm L} \cdot \xi_{\rm L}^\dag
\right)
/ (2i) \ .
\label{def:al hat}
\end{eqnarray}
These 1-forms upon expansion in a power series in $\pi$ yield:
\begin{eqnarray}
\label{al perp exp}
\widehat{\alpha}_{\perp\mu} &=&
\frac{1}{F_\pi} \partial_\mu \pi
+ {\cal A}_\mu
- \frac{i}{F_\pi} \left[ {\cal V}_\mu \,,\, \pi \right]
- \frac{1}{6F_\pi^3}
  \Bigl[ \bigl[ \partial_\mu \pi \,,\, \pi \bigr] \,,\, \pi \Bigr]
+ \cdots
\ ,
\\
\label{al para exp}
\widehat{\alpha}_{\parallel\mu} &=&
\frac{1}{F_\sigma} \partial_\mu \sigma - V_\mu
+ {\cal V}_\mu
- \frac{i}{2F_\pi^2} \bigl[ \partial_\mu \pi \,,\, \pi \bigr]
- \frac{i}{F_\pi} \left[ {\cal A}_\mu \,,\, \pi \right]
+ \cdots
\ ,
\end{eqnarray}
where
${\cal V}_\mu = \left({\cal R}_\mu + {\cal L}_\mu\right)/2$ and
${\cal A}_\mu = \left({\cal R}_\mu - {\cal L}_\mu\right)/2$.

The covariantized 1-forms in Eqs.~(\ref{def:al hat})
transform homogeneously:
\begin{equation}
\widehat{\alpha}_{\parallel,\perp}^\mu \rightarrow
h(x) \cdot \widehat{\alpha}_{\parallel,\perp}^\mu \cdot h^\dag(x) \ .
\label{hat al:trans}
\end{equation}
Then we can construct two independent terms with lowest derivatives
which are invariant under the full
$G_{\rm global}\times H_{\rm local}$ symmetry as
\begin{eqnarray}
&& {\cal L}_{\rm A} \equiv
F_\pi^2 \, \mbox{tr}
\left[ \hat{\alpha}_{\perp\mu} \hat{\alpha}_{\perp}^\mu \right]
=
\mbox{tr} \left[ \partial_\mu \pi \partial^\mu \pi \right]
+ \cdot
\ ,
\label{def:LA}
\\
&& a {\cal L}_{\rm V} \equiv
F_\sigma^2 \, \mbox{tr}
\left[
  \hat{\alpha}_{\parallel\mu} \hat{\alpha}_{\parallel}^\mu
\right]
=
\mbox{tr} \biggl[
\left( \partial_\mu \sigma - F_\sigma V_\mu \right)
\left( \partial^\mu \sigma - F_\sigma V^\mu \right)
\biggr]
+ \cdots
\ ,
\label{def:LV}
\end{eqnarray}
where the expansions of the covariantized 1-forms in
(\ref{al perp exp}) and (\ref{al para exp}) were substituted to
obtain the second expressions.
These expansions imply that ${\cal L}_A$ generates the kinetic term
of pseudoscalar meson, while
${\cal L}_V$ generates the kinetic term of the would-be NG boson
$\sigma$ in addition to the mass term of the vector meson.

The HLS gauge boson field strength defined by $
V_{\mu\nu} \equiv \partial_\mu V_\nu - \partial_\nu
V_\mu - i [ V_\mu , V_\nu ] $,
which also transforms homogeneously:
\begin{equation}
V_{\mu\nu} \rightarrow h(x) \cdot V_{\mu\nu} \cdot h^\dag(x) \ ,
\end{equation}
is another building block of the $\chi$PT Lagrangian. Then the simplest term with $V_{\mu\nu}$ is the kinetic term of the
gauge boson:
\begin{equation}
{\cal L}_{\rm kin}(V_\mu) = - \frac{1}{2g^2} \, \mbox{tr}
\left[ V_{\mu\nu} V^{\mu\nu} \right] \ ,
\label{HLS gauge kinetic}
\end{equation}
where
$g$ is the HLS gauge coupling constant.

Now the Lagrangian with lowest derivatives
is given by:
\begin{eqnarray}
{\cal L} &=&
{\cal L}_{\rm A} + a {\cal L}_{\rm V} +
{\cal L}_{\rm kin}(V_\mu)
\nonumber\\
&=&
F_\pi^2 \, \mbox{tr}
\left[ \hat{\alpha}_{\perp\mu} \hat{\alpha}_{\perp}^\mu \right]
+ F_\sigma^2 \, \mbox{tr}
\left[
  \hat{\alpha}_{\parallel\mu} \hat{\alpha}_{\parallel}^\mu
\right]
- \frac{1}{2g^2} \, \mbox{tr}
\left[ V_{\mu\nu} V^{\mu\nu} \right]
\ .
\label{leading Lagrangian 0}
\end{eqnarray}

\subsection{Obtaining the ${\cal O}(p^4)$ $SU(3)$ $\chi$PT Lagrangian by Integrating out the Vector Mesons from the HLS Lagrangian}
Integrating out the vector mesons in the Lagrangian of the
HLS given in (\ref{leading Lagrangian 0})
we obtain the Lagrangian for pseudoscalar mesons.
The resultant Lagrangian includes ${\cal O}(p^4)$ terms of the ChPT
in addition to
${\cal O}(p^2)$ terms.
To perform this it is convenient to introduce the following quantities:
\begin{eqnarray}
\alpha_{\perp\mu} &=&
\left(
  {\cal D}_\mu \xi_{\rm R} \cdot \xi_{\rm R}^\dag -
  {\cal D}_\mu \xi_{\rm L} \cdot \xi_{\rm L}^\dag
\right)
/ (2i) \ ,
\nonumber\\
\alpha_{\parallel\mu} &=&
\left(
  {\cal D}_\mu \xi_{\rm R} \cdot \xi_{\rm R}^\dag +
  {\cal D}_\mu \xi_{\rm L} \cdot \xi_{\rm L}^\dag
\right)
/ (2i) \ ,
\end{eqnarray}
where ${\cal D}_\mu\xi_{\rm L}$ and ${\cal D}_\mu\xi_{\rm L}$ are
defined by
\begin{eqnarray}
&& {\cal D}_\mu\xi_{\rm L} =
  \partial_\mu \xi_{\rm L} + i \xi_{\rm L} {\cal L}_\mu \ ,
\nonumber\\
&& {\cal D}_\mu\xi_{\rm R} =
  \partial_\mu \xi_{\rm R} + i \xi_{\rm R} {\cal R}_\mu
\ .
\label{covder:2}
\end{eqnarray}
The relations of these $\alpha_{\perp\mu}$ and $\alpha_{\parallel\mu}$
with $\widehat{\alpha}_{\perp\mu}$ and
$\widehat{\alpha}_{\parallel\mu}$ in (\ref{def:al hat})
are given by
\begin{eqnarray}
&& \widehat{\alpha}_{\perp\mu} = \alpha_{\perp\mu} \ , \nonumber\\
&& \widehat{\alpha}_{\parallel\mu} = \alpha_{\parallel\mu} - V_\mu \ .
\label{rel: al perp para}
\end{eqnarray}

From the Lagrangian in (\ref{leading Lagrangian 0})
the equation of motion for the vector meson is given by
\begin{equation}
F_\sigma^2 \left( V_\mu - \alpha_{\parallel\mu} \right)
- \frac{1}{g^2} \left( \partial^\nu V_{\mu\nu}
- i \left[ V^\nu \,,\, V_{\mu\nu} \right] \right) = 0 \ .
\label{para0}
\end{equation}
In the leading order of the derivative expansion the solution of
(\ref{para0}) is given by
\begin{equation}
\label{EOM V}
V_\mu = \alpha_{\parallel\mu}
+ \frac{1}{M_\rho^2} {\cal O}(p^3) \ ,
\end{equation}
consistent with (\ref{Vmu-EOM-Op2}). Substituting this into the field strength of the HLS gauge boson and
performing the derivative expansion one can show that one obtains \cite{HLS-Physics-Reports}:
\begin{eqnarray}
V_{\mu\nu} &=& \xi_{\rm R} \left( {\cal R}_{\mu\nu} + U^\dag {\cal L}_{\mu\nu} U
+ \frac{i}{4} \nabla_\mu U^\dag \cdot \nabla_\nu U
- \frac{i}{4} \nabla_\nu U^\dag \cdot \nabla_\mu U
\right) \xi_{\rm R}^\dag
+ \frac{1}{M_\rho^2} {\cal O}(p^4)
\ ,
\label{V0}
\end{eqnarray}
where,
\begin{equation}
\label{perpU}
\hat{\alpha}_{\perp\mu}
= \frac{i}{2} \xi_{\rm L} \cdot \nabla_\mu U \cdot \xi_{\rm R}^\dag
= \frac{1}{2i} \xi_{\rm R} \cdot \nabla_\mu
  U^\dag \cdot \xi_{\rm L}^\dag
\ ,
\end{equation}
has been used. By substituting (\ref{perpU})
into the HLS Lagrangian,
the first term in the HLS Lagrangian
(\ref{leading Lagrangian 0}) becomes
the first term in
the leading order ChPT Lagrangian:
\begin{equation}
\left. {\cal L}_{(2)}^{\rm ChPT} \right\vert_{\chi=0}
=
\frac{F_\pi^2}{4} \mbox{tr}
\left[ \nabla_\mu U^\dag \nabla^\mu U \right]
\ .
\label{leading ChPT:2}
\end{equation}
In addition,
the second term in (\ref{leading Lagrangian 0})
with (\ref{para0}) substituted
becomes of ${\cal O}(p^6)$ in the ChPT
and the third term (the kinetic term of the HLS gauge boson)
with (\ref{V0}) becomes
of ${\cal O}(p^4)$ in the ChPT:
\begin{eqnarray}
{\cal L}_4^V &=&
\frac{1}{32g^2}
\left(
\mbox{tr}\left[ \nabla_\mu U \nabla^\mu U^\dag \right] \right)^2
+ \frac{1}{16g^2}
\,\mbox{tr}\left[ \nabla_\mu U \nabla_\nu U^\dag \right]
\mbox{tr} \left[ \nabla^\mu U \nabla^\nu U^\dag \right]
\nonumber\\
&&
{} -  \frac{3}{16g^2} \, \mbox{tr} \left[
  \nabla_\mu U \nabla^\mu U^\dag \nabla_\nu U \nabla^\nu U^\dag
\right]
-  \frac{i}{4g^2} \, \mbox{tr}\left[
  {\cal L}_{\mu\nu} \nabla^\mu U \nabla^\nu U^\dag
  + {\cal R}_{\mu\nu} \nabla^\mu U^\dag \nabla^\nu U
\right]
\nonumber\\
&&
- \frac{1}{4g^2} \,\mbox{tr}
\left[ {\cal L}_{\mu\nu} U {\cal R}^{\mu\nu} U^\dag \right] - \frac{1}{8g^2} \left[ {\cal L}_{\mu\nu} {\cal L}^{\mu\nu}
 + {\cal R}_{\mu\nu} {\cal R}^{\mu\nu} \right]
\ ,
\end{eqnarray}
where we fixed $N_f=3$.
Comparing this with the ${\cal O}(p^4)$ terms of the ChPT Lagrangian
given in (\ref{ChPT-Op4}),
we obtain the contributions of vector mesons to the low-energy
parameters of the ChPT:
\begin{equation}
\label{GL-LEC-integrate-rho-out}
\begin{array}{ccc}
\displaystyle L_1^V = \frac{1}{32g^2} \ , \quad
& \displaystyle L_2^V = \frac{1}{16g^2} \ , \quad
& \displaystyle L_3^V = - \frac{3}{16g^2} \ ,
\\
& & \\
\displaystyle L_9^V = \frac{1}{4g^2} \ , \quad
& \displaystyle L_{10}^V = - \frac{1}{4g^2} \ . \quad
&
\end{array}
\end{equation}

\section{Inclusion of ${\cal O}(R^4)$ Corrections in the $D6$-Brane DBI Action}
Inclusive of the ${\cal O}(R^4)$-corrections indicated by a $\tilde{}$ (e.g., the  ${\cal M}$-theory  metric: $\tilde{g}_{MN}^{\cal M}=g^{\rm MQGP}_{MN}\left(1+f_{MN}\right)$ \cite{OR4-Yadav+Misra}) in (\ref{V_12}), one sees that
\begin{eqnarray}
\label{V_12}
& & {\cal V}_1 = 2\sqrt{h} e^{-\tilde{\phi}_{\rm IIA}}\sqrt{-\tilde{g}_{7\times7}^{\rm IIA}}\tilde{g}^{ZZ}_{\rm IIA},\nonumber\\
& & {\cal V}_2 = h e^{-\tilde{\phi}_{\rm IIA}}\sqrt{-\tilde{g}^{\rm IIA}_{7\times7}},
\end{eqnarray}
wherein using (\ref{TypeIIA-from-M-theory-Witten-prescription}):
\begin{eqnarray}
\label{O4-corrections}
& & e^{-\tilde{\phi}_{\rm IIA}}\ ^{{\cal O}(R^4)} = \tilde{g}^{\cal M}_{x^{10}x^{10}}\ ^{-\frac{3}{4}}
= \left(g^{\cal M}_{x^{10}x^{10}} + {\cal F}_{x^{10}x^{10}}\right)^{-\frac{3}{4}} \equiv e^{-{\phi}_{\rm IIA}} + {\cal E}_{x^{10}x^{10}}{\cal F}_{x^{10}x^{10}},\nonumber\\
& & {\cal E}_{x^{10}x^{10}} = -\frac{3}{4}\frac{1}{\left(g^{\cal M}_{x^{10}x^{10}}\right)^{\frac{7}{4}}}\nonumber\\
& & \tilde{G}_{\cal M}^{ZZ} = \frac{e^{-2 Z}}{\sqrt{\tilde{g}^{\cal M}_{x^{10}x^{10}}} g_{\cal M}^{ZZ} {r_0}^2}-\frac{ e^{-2 Z} ({\cal F}_{x^{10}x^{10}}
  g_{\cal M}^{ZZ}+2 {\cal F}_{ZZ} \tilde{g}^{\cal M}_{x^{10}x^{10}})}{2 \tilde{g}^{\cal M}_{x^{10}x^{10}}\ ^{3/2} g_{\cal M}^{rr}\ ^2r_0^2},\nonumber\\
& & \sqrt{-\tilde{g}_{7\times7}} = \sqrt{-g_{7\times7}} + {\rm Tr}({\cal C}{\cal F}).
\end{eqnarray}
In (\ref{O4-corrections}),
{\footnotesize
\begin{eqnarray*}
& & \hskip -0.8in {\cal F}_{MN} \equiv f_{MN} g^{\cal M}_{MN};\nonumber\\
& & \hskip -0.8in {\cal C}_{MN} \equiv \delta^t_M\delta^t_N {\cal C}_{tt} + \delta^{x^1}_M\delta^{x^1}_N {\cal C}_{x^1x^1}
+ \delta^{x^2}_M\delta^{x^2}_N {\cal C}_{x^2x^2} + \delta^{x^3}_M\delta^{x^3}_N {\cal C}_{x^3x^3} \nonumber\\
& &  \hskip -0.8in + \delta^{r_0}_M\delta^{r_0}_N {\cal C}_{rr} + \delta^{\theta_2}_M\delta^{\theta_2}_N {\cal C}_{\theta_2\theta_2}
+ \delta^{\theta_2}_M\delta^{\phi_2}_N {\cal C}_{\theta_2y} + \delta^{\phi_2}_M\delta^{\phi_2}_N {\cal C}_{yy} + \delta^{x^{11}}_M\delta^{x^{11}}_N {\cal C}_{x^{10}x^{10}},\nonumber\\
& & \hskip -0.8in {\cal C}_{tt} =\nonumber\\
& & \hskip -0.8in  -\frac{\sqrt{-g} \left(-4 {\cal A}^{\cal M}_{\theta_2\theta_2} {g^{\cal M}_{x^1x^1}} {g^{\cal M}_{x^{10}x^{10}}}^3 {g^{\cal M}_{x^2x^2}} {g^{\cal M}_{x^3x^3}} {g^{\cal M}_{rr}}
   {g^{\cal M}_{yy}}-4 B_{r\theta_2}^2 {g^{\cal M}_{x^1x^1}} {g^{\cal M}_{x^{10}x^{10}}} {g^{\cal M}_{x^2x^2}} {g^{\cal M}_{x^3x^3}} {g^{\cal M}_{rr}}+4 {g^{\cal M}_{x^1x^1}}
   {g^{\cal M}_{x^{10}x^{10}}}^2 {g^{\cal M}_{x^2x^2}} {g^{\cal M}_{x^3x^3}} {g^{\cal M}_{rr}} {g^{\cal M}_{\theta_2y}}^2\right)}{\Delta}\nonumber\\
& & \hskip -0.8in = \frac{2 \sqrt[6]{\pi } {g_s}^{3/2} M N^{3/10} {N_f} \log \left({r_0} e^Z\right) \left(72 a^2 {r_0} e^Z \log
   \left({r_0} e^Z\right)-3 a^2+2 {r_0}^2 e^{2 Z}\right)}{3^{2/3} \log N  \left(1-e^{-4 Z}\right) \alpha _{\theta _1}
   \alpha _{\theta _2}^2 {\cal G}{}^{2/3}}\nonumber\\
& & \hskip -0.8in {\cal C}_{x^1x^1} = \nonumber\\
& & \hskip -0.8in -\frac{\sqrt{-g} \left(-4 {\cal A}^{\cal M}_{\theta_2\theta_2} {g^{\cal M}_{x^{10}x^{10}}}^3 {g^{\cal M}_{x^2x^2}} {g^{\cal M}_{x^3x^3}} {g^{\cal M}_{rr}} {g^{\cal M}_{tt}}
   {g^{\cal M}_{yy}}-4 B_{r\theta_2}^2 {g^{\cal M}_{x^{10}x^{10}}} {g^{\cal M}_{x^2x^2}} {g^{\cal M}_{x^3x^3}} {g^{\cal M}_{rr}} {g^{\cal M}_{tt}}+4 {g^{\cal M}_{x^{10}x^{10}}}^2
   {g^{\cal M}_{x^2x^2}} {g^{\cal M}_{x^3x^3}} {g^{\cal M}_{rr}} {g^{\cal M}_{\theta_2y}}^2 {g^{\cal M}_{tt}}\right)}{\Delta}\nonumber\\
& & \hskip -0.8in= \frac{2 \sqrt[6]{\pi } {g_s}^{3/2} M N^{3/10} {N_f} \log \left({r_0} e^Z\right) \left(72 a^2 {r_0} e^Z \log
   \left({r_0} e^Z\right)-3 a^2+2 {r_0}^2 e^{2 Z}\right)}{3^{2/3} \log N  \alpha _{\theta _1} \alpha _{\theta _2}^2
  {\cal G}{}^{2/3}},\nonumber\\
& & \hskip -0.8in {\cal C}_{x^2x^2} = {\cal C}_{x^3x^3} =\nonumber\\
& & \hskip -0.8in -\frac{\sqrt{-g} \left(-4 {\cal A}^{\cal M}_{\theta_2\theta_2} {g^{\cal M}_{x^1x^1}} {g^{\cal M}_{x^{10}x^{10}}}^3 {g^{\cal M}_{x^3x^3}} {g^{\cal M}_{rr}} {g^{\cal M}_{tt}}
   {g^{\cal M}_{yy}}-4 B_{r\theta_2}^2 {g^{\cal M}_{x^1x^1}} {g^{\cal M}_{x^{10}x^{10}}} {g^{\cal M}_{x^3x^3}} {g^{\cal M}_{rr}} {g^{\cal M}_{tt}}+4 {g^{\cal M}_{x^1x^1}}
   {g^{\cal M}_{x^{10}x^{10}}}^2 {g^{\cal M}_{x^3x^3}} {g^{\cal M}_{rr}} {g^{\cal M}_{\theta_2y}}^2 {g^{\cal M}_{tt}}\right)}{\Delta}\nonumber\\
& & \hskip -0.8in = \frac{2 \sqrt[6]{\pi } {g_s}^{3/2} M N^{3/10} {N_f} \log \left({r_0} e^Z\right) \left(72 a^2 {r_0} e^Z \log
   \left({r_0} e^Z\right)-3 a^2+2 {r_0}^2 e^{2 Z}\right)}{3^{2/3} \log N  \alpha _{\theta _1} \alpha _{\theta _2}^2
  {\cal G}{}^{2/3}},\nonumber\\
\end{eqnarray*}
\begin{eqnarray*}
& & \hskip -0.8in {\cal C}_{rr} =\nonumber\\
& & \hskip -0.8in  -\frac{\sqrt{-g} \left(-4 {\cal A}^{\cal M}_{\theta_2\theta_2} {g^{\cal M}_{x^1x^1}} {g^{\cal M}_{x^{10}x^{10}}}^3 {g^{\cal M}_{x^2x^2}} {g^{\cal M}_{x^3x^3}} {g^{\cal M}_{tt}}
   {g^{\cal M}_{yy}}-4 B_{r\theta_2}^2 {g^{\cal M}_{x^1x^1}} {g^{\cal M}_{x^{10}x^{10}}} {g^{\cal M}_{x^2x^2}} {g^{\cal M}_{x^3x^3}} {g^{\cal M}_{tt}}+4 {g^{\cal M}_{x^1x^1}}
   {g^{\cal M}_{x^{10}x^{10}}}^2 {g^{\cal M}_{x^2x^2}} {g^{\cal M}_{x^3x^3}} {g^{\cal M}_{\theta_2y}}^2 {g^{\cal M}_{tt}}\right)}{\Delta}\nonumber\\
& & \hskip -0.8in= \frac{\sqrt{{g_s}} M \left(\frac{1}{N}\right)^{7/10} {N_f} {r_0}^4 \left(e^{4 Z}-1\right) \left(9 a^2+{r_0}^2 e^{2
   Z}\right) \log \left({r_0} e^Z\right) \left(72 a^2 {r_0} e^Z \log \left({r_0} e^Z\right)-3 a^2+2 {r_0}^2 e^{2
   Z}\right)}{2\ 3^{2/3} \pi ^{5/6} \log N  \alpha _{\theta _1} \alpha _{\theta _2}^2 \left(6 a^2+{r_0}^2 e^{2 Z}\right)
  {\cal G}{}^{2/3}},\nonumber\\
& &\hskip -0.8in {\cal C}_{\theta_2\theta_2} = \frac{{g^{\cal M}_{x^{10}x^{10}}} {g^{\cal M}_{yy}} \sqrt{-g}}{2 \left({\cal A}^{\cal M}_{\theta_2\theta_2} {g^{\cal M}_{x^{10}x^{10}}}^2
   {g^{\cal M}_{yy}}+B_{r\theta_2}^2-{g^{\cal M}_{x^{10}x^{10}}} {g^{\cal M}_{\theta_2y}}^2\right)} = \frac{\sqrt[6]{\pi } \left(\frac{1}{N}\right)^{13/10} {r_0}^4 e^{4 Z} \alpha _{\theta _1} \alpha _{\theta _2}^2 \left(72 a^2
   {r_0} e^Z \log \left({r_0} e^Z\right)-3 a^2+2 {r_0}^2 e^{2 Z}\right){\cal G}{}^{4/3}}{32\ 3^{2/3} {g_s}^{5/2}
   \log N  M {N_f}^3 \log \left({r_0} e^Z\right) \left(36 a^2 \log \left({r_0} e^Z\right)+{r_0} e^Z\right)^2}\nonumber\\
& &\hskip -0.8in {\cal C}_{\theta_2y} = -\frac{{g^{\cal M}_{x^{10}x^{10}}} {g^{\cal M}_{\theta_2y}} \sqrt{-g}}{{\cal A}^{\cal M}_{\theta_2\theta_2} {g^{\cal M}_{x^{10}x^{10}}}^2 {g^{\cal M}_{yy}}+B_{r\theta_2}^2-{g^{\cal M}_{x^{10}x^{10}}}
   {g^{\cal M}_{\theta_2y}}^2} = -\frac{\pi ^{5/12} \left(\frac{1}{N}\right)^{19/20} {r_0}^4 e^{4 Z} \alpha _{\theta _2}^3 \left(72 a^2 {r_0} e^Z \log
   \left({r_0} e^Z\right)-3 a^2+2 {r_0}^2 e^{2 Z}\right){\cal G}{}^{4/3}}{72 \sqrt{2} 3^{2/3} {g_s}^{9/4} \log N  M
   {N_f}^3 \alpha _{\theta _1} \log \left({r_0} e^Z\right) \left(36 a^2 \log \left({r_0} e^Z\right)+{r_0}
   e^Z\right)^2}\nonumber\\
\label{C_{MN}}
& &\hskip -0.8in {\cal C}_{yy} = \frac{{\cal A}^{\cal M}_{\theta_2\theta_2} {g^{\cal M}_{x^{10}x^{10}}}^2 \sqrt{-g}}{2 \left({\cal A}^{\cal M}_{\theta_2\theta_2} {g^{\cal M}_{x^{10}x^{10}}}^2
   {g^{\cal M}_{yy}}+B_{r\theta_2}^2-{g^{\cal M}_{x^{10}x^{10}}} {g^{\cal M}_{\theta_2y}}^2\right)}\nonumber\\
& & \hskip -0.8in= \frac{{g_s} M \sqrt[5]{\frac{1}{N}} {N_f} {r_0}^2 e^{2 Z} \log \left({r_0} e^Z\right) \left(72 a^2 {r_0} e^Z
   \log \left({r_0} e^Z\right)-3 a^2+2 {r_0}^2 e^{2 Z}\right)}{3^{2/3} \sqrt[3]{\pi } \log N  \alpha _{\theta _1}
   \alpha _{\theta _2}^2 {\cal G}{}^{2/3}},\nonumber\\
& &\hskip -0.8in {\cal C}_{x^{10}x^{10}} = \nonumber\\
& & \hskip -0.8in  -\frac{\sqrt{-g} \left(-18 {\cal A}^{\cal M}_{\theta_2\theta_2} {g^{\cal M}_{x^1x^1}} {g^{\cal M}_{x^{10}x^{10}}}^2 {g^{\cal M}_{x^2x^2}} {g^{\cal M}_{x^3x^3}} {g^{\cal M}_{rr}} {g^{\cal M}_{tt}}
   {g^{\cal M}_{yy}}-10 B_{r\theta_2}^2 {g^{\cal M}_{x^1x^1}} {g^{\cal M}_{x^2x^2}} {g^{\cal M}_{x^3x^3}} {g^{\cal M}_{rr}} {g^{\cal M}_{tt}}+14 {g^{\cal M}_{x^1x^1}} {g^{\cal M}_{x^{10}x^{10}}} {g^{\cal M}_{x^2x^2}}
   {g^{\cal M}_{x^3x^3}} {g^{\cal M}_{rr}} {g^{\cal M}_{\theta_2y}}^2 {g^{\cal M}_{tt}}\right)}{\Delta}\nonumber\\
& & \hskip -0.8in = \frac{27 \sqrt[3]{3} {g_s} M \sqrt[5]{\frac{1}{N}} {N_f} {r_0}^2 e^{2 Z} \log \left({r_0} e^Z\right) \left(72 a^2
   {r_0} e^Z \log \left({r_0} e^Z\right)-3 a^2+2 {r_0}^2 e^{2 Z}\right) {\cal G}{}^{4/3}}{128 \pi ^{7/3} \log N
   \alpha _{\theta _1} \alpha _{\theta _2}^2};\nonumber\\
& & \hskip -0.8in \Delta = 8 {g^{\cal M}_{x^1x^1}} {g^{\cal M}_{x^{10}x^{10}}}
   {g^{\cal M}_{x^2x^2}} {g^{\cal M}_{x^3x^3}} {g^{\cal M}_{rr}} {g^{\cal M}_{tt}} \left({\cal A}^{\cal M}_{\theta_2\theta_2} {g^{\cal M}_{x^{10}x^{10}}}^2
   {g^{\cal M}_{yy}}+B_{r\theta_2}^2-{g^{\cal M}_{x^{10}x^{10}}} {g^{\cal M}_{\theta_2y}}^2\right);
\nonumber\\
\end{eqnarray*}
\begin{eqnarray}
\label{CMN}
& & \hskip -0.8in {\cal A}^{\cal M}_{\theta_2\theta_2} = \left(x \left(\widetilde{\widetilde{\widetilde{F_3^{IIB}}}}\right)_{x\theta_2} + y \left(\widetilde{\widetilde{\widetilde{F_3^{IIB}}}}\right)_{y\theta_2} + z \left(\widetilde{\widetilde{\widetilde{F_3^{IIB}}}}\right)_{z\theta_2} + r \left(\widetilde{\widetilde{\widetilde{*F_5^{IIB}}}}\right)_{r\theta_2}\right)^2 \sim \left(z \left(\widetilde{\widetilde{\widetilde{F_3^{IIB}}}}\right)_{z\theta_2} + r \left(\widetilde{\widetilde{\widetilde{*F_5^{IIB}}}}\right)_{r\theta_2}\right)^2\nonumber\\
& & \hskip -0.8in = \frac{9 {g_s}^{7/2} M^2 N^{11/10} {N_f}^4 e^{-2 Z} \log ^2\left({r_0} e^Z\right) \left(36 a^2 \log \left({r_0}
   e^Z\right)+{r_0} e^Z\right)^2}{2 \pi ^{5/2} {r_0}^2 \alpha _{\theta _1}^2 \alpha _{\theta _2}^4},
\end{eqnarray}
}
where in the last line use has been made of that in the IR:
\begin{eqnarray}
& & x \left(\widetilde{\widetilde{\widetilde{F_3^{IIB}}}}\right)_{x\theta_2}\sim N^{\frac{3}{20}}\log r,\  y\left(\widetilde{\widetilde{\widetilde{F_3^{IIB}}}}\right)_{y\theta_2}\sim N^{\frac{1}{4}}\log r,\nonumber\\
& & \   z \left(\widetilde{\widetilde{\widetilde{F_3^{IIB}}}}\right)_{z\theta_2}\sim N^{\frac{11}{20}}\log r,\ \left(\widetilde{\widetilde{\widetilde{*F_5^{IIB}}}}\right)_{r\theta_2}\sim N^{\frac{1}{20}}\left(\frac{\log r}{r_0}\right)^2 ;
\end{eqnarray}
 the triple tildes $\hskip 0.1in\widetilde{\widetilde{\widetilde{} }}\hskip 0.1in$ imply a tripe T-dual of the type IIB background of \cite{metrics}; $x, y, z$ are the delocalized $T^3$ coordinates using for effecting SYZ mirror symmetry via a triple T dual in \cite{MQGP}. In (\ref{C_{MN}}), ${\cal G} \equiv  8\pi e^{-\phi^{\rm IIB}}$.

One hence obtains:
\begin{eqnarray}
\label{E1010F1010}
& & \hskip -0.5in {\cal E}_{x^{10}x^{10}}{\cal F}_{x^{10}x^{10}} \nonumber\\
& & \hskip -0.5in  = \frac{243 b^{10} \left(9 b^2+1\right)^4 \beta  M \left(\frac{1}{N}\right)^{5/4} e^Z \left(e^Z-2\right) \left(-19683 \sqrt{6} \alpha _{\theta _1}^6-6642 \alpha _{\theta _2}^2
   \alpha _{\theta _1}^3+40 \sqrt{6} \alpha _{\theta _2}^4\right) \left(r_0 ^2-3 a^2\right) \log ^3(r_0 )}{16 \pi ^2 \left(3 b^2-1\right)^5  \left(\log N\right) ^3 \alpha
   _{\theta _2}^3 \left(6 b^2 r_0 +r_0 \right)^4},\nonumber\\
& & \nonumber\\
\end{eqnarray}
relevant to obtaining the ${\cal O}(R^4)$-corrected type IIA dilaton via (\ref{O4-corrections}).


\begin{spacing}{1}
\end{spacing}


\clearpage
\addcontentsline{toc}{chapter}{Bibliography}
\begin{thebibliography}{99}
\bibitem{metrics}  M.~Mia, K.~Dasgupta, C.~Gale and S.~Jeon, {\it Five Easy Pieces: The Dynamics of Quarks in Strongly Coupled Plasmas}, Nucl.\ Phys.\ B {\bf 839}, 187 (2010) [arXiv:hep-th/0902.1540].
\bibitem{MQGP}M.~Dhuria and A.~Misra, {\it Towards MQGP}, JHEP 1311 (2013) 001, [arXiv:hep-th/1306.4339].
\bibitem{Sil+Yadav+Misra-glueball} K.~Sil, V.~Yadav and A.~Misra,{\it Top-down holographic G-structure glueball spectroscopy at (N)LO in $N$ and finite coupling}, Eur.\ Phys.\ J.\ C {\bf 77}, no. 6, 381 (2017) [arXiv:1703.01306 [hep-th]].
\bibitem{Teper98} M.~J.~Teper, {\it SU(N) gauge theories in (2+1)-dimensions}, Phys.\ Rev.\ D {\bf 59}, 014512 (1999)
  doi:10.1103/PhysRevD.59.014512 [hep-lat/9804008].
\bibitem{Chen_et_al_latest_lattice_2006}Y.~Chen {\it et al.}, {\it Glueball spectrum and matrix elements on anisotropic lattices},
  Phys.\ Rev.\ D {\bf 73}, 014516 (2006) doi:10.1103/PhysRevD.73.014516 [hep-lat/0510074].
\bibitem{Yadav+Misra+Sil-Mesons}V.~Yadav, A.~Misra and K.~Sil, {\it Delocalized SYZ Mirrors and Confronting Top-Down $SU(3)$-Structure Holographic Meson Masses at Finite $g$ and $N_c$ with P(article) D(ata) G(roup) Values}, Eur. Phys. J. C bf{77}, no.10, 656 (2017) [arXiv:1707.02818 [hep-th]].
\bibitem{mesons_0E++-to-mesons-decays} V.~Yadav and A.~Misra, {\it ${\cal M}$-Theory Exotic Scalar Glueball Decays to Mesons at Finite Coupling},  JHEP {\bf 1809}, 133 (2018)  [arXiv:1808.01182 [hep-th]].
\bibitem{OR4-Yadav+Misra} V.~Yadav and A.~Misra, {\it Towards Thermal QCD from M theory at Intermediate 't Hooft Coupling (and IR $SU(3)/G_2/SU(4),Spin(7)$-Structure Torsion Classes of Six-/Seven-/Eight-Folds)} [arXiv:2004.07259 [hep-th]].
\bibitem{Vikas+Gopal+Aalok} V.~Yadav, G.~Yadav and A.~Misra,
{\it (Phenomenology/Lattice-Compatible) $SU(3)$ M$\chi$PT HD up to ${\cal O}(p^4)$ and the ${\cal O}\left(R^4\right)$-Large-$N$ Connection}, to appear in JHEP [arXiv:2011.04660 [hep-th]].
\bibitem{Ecker-2015} G.~Ecker, {\it Status of Chiral Perturbation Theory for Light Mesons}, PoS CD15 (2015) 011 [arXiv:1510.01634[hep-ph]].
\bibitem{HLS-Physics-Reports}M.~Harada and K.~Yamawaki,
{\it Hidden local symmetry at loop: A New perspective of composite gauge boson and chiral phase transition},
Phys. Rept. \textbf{381}, 1-233 (2003)[arXiv:hep-ph/0302103 [hep-ph]].
\bibitem{Minahan} J.~A.~Minahan,{\it Glueball mass spectra and other issues for supergravity duals of QCD models}  JHEP {\bf 9901}, 020 (1999) doi:10.1088/1126-6708/1999/01/020 [hep-th/9811156].
\bibitem{Ouguri et al} C.~Csaki, H.~Ouguri, Y.~Oz, J.~Terning,{\it Glueball mass spectrum from supergravity},
   JHEP {\bf 9901}, 017 (1999) doi:10.1088/1126-6708/1999/01/017 [hep-th/9806021].
\bibitem{Maldacena} Juan M.~Maldacena, {\it The Large N limit of Superconformal Field Theories and Supergravity}, Adv.Theor.Math.Phys.2 231252 (1998), [arXiv:hep-th/9711200].
\bibitem{Minwalla}, {\it Applications of the AdS/CFT correspondence}, Pramana \textbf{79}, 1075(2012).
\bibitem{S.Roy-1}K.~Nayek and S.~Roy, {\it Decoupling limit and throat geometry of non-susy D3 brane}, Phys.\ Lett.\ B {\bf 766}, 192 (2017)   [arXiv:1608.05036 [hep-th]].
\bibitem{S.Roy-2} S.~Chakraborty, K.~Nayek and S.~Roy, {\it Wilson loop calculation in QGP using non-supersymmetric AdS/CFT}, arXiv:1710.08631 [hep-th].
\bibitem{S.Roy-3} S.~Chakraborty, N.~Haque and S.~Roy, {\it Wilson loops in noncommutative Yang-Mills theory using gauge/gravity duality},   Nucl.\ Phys.\ B {\bf 862}, 650 (2012)[arXiv:1201.0129 [hep-th]].
\bibitem{S.Roy-4} S.~Chakraborty and S.~Roy, {\it Calculating the jet quenching parameter in the plasma of NCYM theory from gauge/gravity duality},  Phys.\ Rev.\ D {\bf 85}, 046006 (2012)  [arXiv:1105.3384 [hep-th]].
\bibitem{S.Roy-5}  S.~Chakraborty and S.~Roy, {\it Wilson loops in (p+1)-dimensional Yang-Mills theories using gravity/gauge theory correspondence},   Nucl.\ Phys.\ B {\bf 850}, 463 (2011)  [arXiv:1103.1248 [hep-th]].
\bibitem{S.Roy-6} K.~L.~Panigrahi and S.~Roy,{\it Drag force in a hot non-relativistic, non-commutative Yang-Mills plasma}, JHEP {\bf 1004}, 003 (2010)  [arXiv:1001.2904 [hep-th]].
\bibitem{S.Roy-7}  S.~Roy, {\it Holography and drag force in thermal plasma of non-commutative Yang-Mills theories in diverse dimensions},  Phys.\ Lett.\ B {\bf 682}, 93 (2009)  [arXiv:0907.0333 [hep-th]].
\bibitem{KW} Igor R. Klebanov and Edward Witten, {\it Superconformal Field Theory on Threebranes at a Calabi-Yau Singularity},  Nucl. Phys. B 536, 199 (1998)[arXiv:hep-th/9807080].
\bibitem{KS} I.~R.~Klebanov and  M.~J.~Strassler, {\it Supergravity and a Confining Gauge Theory: Duality Cascades and $X$SB-Resolution of Naked Singularities}, JHEP 0008:052,2000 [arXiv:hep-th/0007191].
\bibitem{Nunez-6} J.~M.~Maldacena and C.~Nunez, {\it Towards the large N limit of pure N=1 superYang-Mills}, Phys.\ Rev.\ Lett.\  {\bf 86}, 588 (2001) [hep-th/0008001].
\bibitem{Kruczenski:2003uq} M.~Kruczenski, D.~Mateos, R.~C.~Myers and D.~J.~Winters, {\it Towards a holographic dual of large N(c) QCD},
JHEP \textbf{05}, 041 (2004) doi:10.1088/1126-6708/2004/05/041 [arXiv:hep-th/0311270 [hep-th]].
\bibitem{Bergman:2001rw} A.~Bergman, K.~Dasgupta, O.~J.~Ganor, J.~L.~Karczmarek and G.~Rajesh, {\it Nonlocal field theories and their gravity duals},
Phys. Rev. D \textbf{65}, 066005 (2002) doi:10.1103/PhysRevD.65.066005 [arXiv:hep-th/0103090 [hep-th]].
\bibitem{Ballon-Bayona:2018ddm} A.~Ballon-Bayona, H.~Boschi-Filho, L.~A.~H.~Mamani, A.~S.~Miranda and V.~T.~Zanchin,
{\it An effective holographic approach to QCD}, [arXiv:1804.01579 [hep-th]].
\bibitem{KT} I.R. Klebanov and A. Tseytlin, {\it Gravity Duals of Supersymmetric $SU(N)\times SU(M+N)$ Gauge Theories}, [hep-th/0002159].
\bibitem{Zayas+Tseytlin}L.~A.~Pando Zayas and A.~A.~Tseytlin, {\it 3-branes on resolved conifold},  JHEP {\bf 0011}, 028 (2000) [hep-th/0010088].
\bibitem{KK} A.~Karch and E.~Katz, {\it Adding flavor to AdS/CFT JHEP 06(2002), 043}; [hep-th/0205236].
\bibitem{ouyang} P.~Ouyang, {\it Holomorphic D7-Branes and Flavored N=1 Gauge Theories}, Nucl.Phys.B 699:207-225 (2004), [arXiv:hep-th/0311084].
\bibitem{Buchel} A.~Buchel, {\it Finite temperature resolution of the Klebanov-Tseytlin singularity}, Nucl.\ Phys.\ B {\bf 600}, 219 (2001) [hep-th/0011146].
\bibitem{Gubser-et-al-finitetemp}  S.~S.~Gubser, C.~P.~Herzog, I.~R.~Klebanov and A.~A.~Tseytlin, {\it Restoration of chiral symmetry: A Supergravity perspective},  JHEP {\bf 0105}, 028 (2001) [hep-th/0102172]; A.~Buchel, C.~P.~Herzog, I.~R.~Klebanov, L.~A.~Pando Zayas and A.~A.~Tseytlin, {\it Nonextremal gravity duals for fractional D-3 branes on the conifold},   JHEP {\bf 0104}, 033 (2001)[hep-th/0102105].
\bibitem{Leo-ii} M.~Mahato, L.~A.~Pando Zayas and C.~A.~Terrero-Escalante, {\it Black Holes in Cascading Theories: Confinement/Deconfinement Transition and other Thermal Properties}, JHEP {\bf 0709}, 083 (2007)  [arXiv:0707.2737 [hep-th]].
\bibitem{Nunez-2} C.~Nunez, A.~Paredes and A.~V.~Ramallo,{\it Unquenched Flavor in the Gauge/Gravity Correspondence}, Adv.\ High Energy Phys.\  {\bf 2010}, 196714 (2010)  [arXiv:1002.1088 [hep-th]].
\bibitem{Nunez-3} F.~Benini, F.~Canoura, S.~Cremonesi, C.~Nunez and A.~V.~Ramallo, {\it Backreacting flavors in the Klebanov-Strassler background},  JHEP {\bf 0709}, 109 (2007)  [arXiv:0706.1238 [hep-th]].
\bibitem{Nunez-4} F.~Benini, F.~Canoura, S.~Cremonesi, C.~Nunez and A.~V.~Ramallo, {\it Unquenched flavors in the Klebanov-Witten model}, JHEP {\bf 0702}, 090 (2007)  [hep-th/0612118].
\bibitem{Ganor:2014wua} O.~J.~Ganor, Y.~P.~Hong, N.~Moore, H.~Y.~Sun, H.~S.~Tan and N.~R.~Torres-Chicon,
{\it Q-balls of quasi-particles in a (2, 0)-theory model of the fractional quantum Hall effect}, JHEP \textbf{09}, 181 (2015) doi:10.1007/JHEP09(2015)181
[arXiv:1410.3575 [hep-th]].
\bibitem{Panda:2009ji} S.~Panda, M.~Sami and I.~Thongkool, {\it Reheating the D-brane universe via instant preheating}, Phys. Rev. D \textbf{81}, 103506 (2010) doi:10.1103/PhysRevD.81.103506 [arXiv:0905.2284 [hep-th]].
\bibitem{Ali:2008ij}
A.~Ali, R.~Chingangbam, S.~Panda and M.~Sami, {\it Prospects of inflation with perturbed throat geometry}, Phys. Lett. B \textbf{674}, 131-136 (2009) doi:10.1016/j.physletb.2009.03.007 [arXiv:0809.4941 [hep-th]].
 \bibitem{K. Dasgupta  et al [2012]} M.~Mia, F.~Chen, K.~Dasgupta, P.~Franche and S.~Vaidya, {\it Non-Extremality, Chemical Potential and the Infrared limit of Large N Thermal QCD}, Phys.\ Rev.\ D {\bf 86}, 086002 (2012)[arXiv:1202.5321 [hep-th]].
\bibitem{arnabkundu0709.1547} T. Albash, V. G. Filev, C. V. Johnson and A. Kundu, {\it Finite temperature large N gauge theory with quarks in an external magnetic field}, JHEP 0807,080 (2008) [arXiv:0709.1547 [hep-th]]
\bibitem{arnabkundu0709.1554} T. Albash, V. G. Filev, C. V. Johnson, and A. Kundu, {\it Quarks in an External Electric Field in Finite Temperature Large N Gauge Theory}, JHEP 08, 092 (2008), arXiv:0709.1554[hep-th].
\bibitem{Minwalla-3} S.~Minwalla, {\it Black holes in large N gauge theories}, Class.\ Quant.\ Grav.\  {\bf 23}, S927 (2006), Lectures from the European RTN Winter School on Strings, Supergravity and Gauge Theories, CERN, January 2006.
\bibitem{Obers:2008pj} N.~A.~Obers, {\it Black Holes in Higher-Dimensional Gravity}, Lect. Notes Phys. \textbf{769}, 211-258 (2009)
doi:10.1007/978-3-540-88460-6\_6 [arXiv:0802.0519 [hep-th]].
\bibitem{A.Sinha-4}A.~Buchel, R.~C.~Myers, M.~F.~Paulos and A.~Sinha, {\it Universal holographic hydrodynamics at finite coupling},   Phys.\ Lett.\ B {\bf 669}, 364 (2008)[arXiv:0808.1837 [hep-th]].
\bibitem{Sakai-Sugimoto-1} T.~Sakai and S.~Sugimoto, {\it Low energy hadron physics in holographic QCD},
  Prog.\ Theor.\ Phys.\  {\bf 113}, 843 (2005) doi:10.1143/PTP.113.843 [hep-th/0412141].
\bibitem{Dasgupta_et_al_Mesons}K.~Dasgupta, C.~Gale, M.~Mia, M.~Richard and O.~Trottier,
{\it Infrared Dynamics of a Large N QCD Model, the Massless String Sector and Mesonic Spectra}, JHEP {\bf 1507}, 122 (2015)
doi:10.1007/JHEP07(2015)122 [arXiv:1409.0559 [hep-th]].
\bibitem{Ni_Kochelev} N.~Kochelev and D.~P.~Min, {\it Role of glueballs in non-perturbative quark-gluon plasma}, Physics Letters B, vol. 650, no. 4, 2007, doi:10.1016/j.physletb.2007.05.019.
\bibitem{Hernandez_glueball} P.~Hernandez and F.~R.~Lopez, {\it The large Nc limit of QCD on the lattice}, Eur. Phys. J. A 57, 52 (2021), https://doi.org/10.1140/epja/s10050-021-00374-2.
\bibitem{Czaki_et_al-0-+}C.~Csaki, Y.~Oz, J.~Russo and J.~Terning, {\it Large N QCD from rotating branes},
  Phys.\ Rev.\ D {\bf 59}, 065012 (1999) doi:10.1103/PhysRevD.59.065012 [hep-th/9810186].
\bibitem{Mathur_et_al}  R.~C.~Brower, S.~D.~Mathur and C.~I.~Tan, {\it Glueball spectrum for QCD from AdS supergravity duality},  Nucl.\ Phys.\ B {\bf 587}, 249 (2000) doi:10.1016/S0550-3213(00)00435-1 [hep-th/0003115].
\bibitem{BH_all_T} P.~Colangelo, F.~Giannuzzi and S.~Nicotri, {\it Holographic Approach to Finite Temperature QCD: The Case of Scalar Glueballs and Scalar Mesons}, Phys.\ Rev.\ D {\bf 80}, 094019 (2009) [arXiv:0909.1534 [hep-ph]].
\bibitem{Nicotri} S.~Nicotri, {\it Scalar glueball in a holographic model of QCD}, Nuovo Cim.\ B {bf 123},796 (2008)
  [Nuovo Cim.\ B {\bf 123}, 851 (2008)] doi:10.1393/ncb/i2008-10579-5, 10.1393/ncb/i2008-10580-0 [arXiv:0807.4377 [hep-ph]].
\bibitem{Forkel} H.~Forkel, {\it Glueball correlators as holograms}, arXiv:0808.0304 [hep-ph]];PoS (CONFINEMENT8) 184 (2008).
\bibitem{ForkelStructure} H.~Forkel, {\it Holograhic glueball structure}, Phys.\ Rev.\ D {\bf 78}, 025001 (2008)
  doi:10.1103/PhysRevD.78.025001 [arXiv:071101179 [hep-ph]].
\bibitem{FolcoCapossoli} E.~Folco Capossoli and H.~Boschi-Filho, {\it  Renormalized AdS$_5$ Mass for even Spin Glueball and Pomeron Trajectory from a Holographic softwall model}, arXiv:1611.09817[hep-ph].
\bibitem{Jugeau} F.~Jugeau, {\it Holographic description of glueballs in a deformed AdS-dilaton background},
  AIP Conf.\ Proc.\ {\bf 964} (2007) 151 doi:10.1063/1.2823842 [arXiv:0709.1093 [hep-ph]].
\bibitem{Colangelolight} P.~Colangelo, F.~De Fazio, F.~jugeau and S.~Nicotri, {\it On the light glueball spectrum in a holographic description of QCD},
  Phys.\ Lett.\ B {\bf 652}, 73 (2007) doi:10.1016/j.physletb.2007.06.072 [hep-th/0703316].
\bibitem{NPB}K.~Sil and A.~Misra, {\it On Aspects of Holographic Thermal QCD at Finite Coupling},
  Nucl.\ Phys.\ B {\bf 910}, 754 (2016) [arXiv:1507.02692 [hep-th]].
\bibitem{transport-coefficients}M.~Dhuria and A.~Misra, {\it Transport Coefficients of Black MQGP M3-Branes},
  Eur.\ Phys.\ J.\ C {\bf 75}, no. 1, 16 (2015)  [arXiv:1406.6076 [hep-th]].
\bibitem{Gubser-Tpq}  S.~S.~Gubser, {\it Einstein manifolds and conformal field theories},
Phys.\ Rev.\ D {\bf 59}, 025006 (1999)doi:10.1103/PhysRevD.59.025006 [hep-th/9807164].
\bibitem{Mathur_et_al-0++-Mtheory}R.~C.~Brower, S.~D.~Mathur and C.~I.~Tan,
{\it Discrete spectrum of the graviton in the AdS(5) black hole background},Nucl.\ Phys.\ B {\bf 574}, 219 (2000)
doi:10.1016/S0550-3213(99)00802-0 [hep-th/9908196].
\bibitem{WKB-i} A.~S.~Miranda, C.~A.~Ballon Bayona, H.~Boschi-Filho and N.~R.~F.~Braga,
  {\it Black-hole quasinormal modes and scalar glueballs in a finite-temperature AdS/QCD model}, JHEP {\bf 0911}, 119 (2009) [arXiv:0909.17904 [hep-th]].
\bibitem{EPJC-2}K.~Sil and A.~Misra,
{\it New Insights into Properties of Large-N Holographic Thermal QCD at Finite Gauge Coupling at (the Non-Conformal/Next-to) Leading Order in N}, Eur.\ Phys.\ J.\ C {\bf 76}, no. 11, 618 (2016) doi:10.1140/epjc/s10052-016-4444-7 [arXiv:1606.04949 [hep-th]
\bibitem{Sommer-r0}R.~Sommer,
{\it A New way to set the energy scale in lattice gauge theories and its applications to the static force and alpha-s in SU(2) Yang-Mills theory,}
  Nucl.\ Phys.\ B {\bf 411}, 839 (1994) doi:10.1016/0550-3213(94)90473-1 [hep-lat/9310022].
\bibitem{Bellantuono:2014lra} L.~Bellantuono, P.~Colangelo and F.~Giannuzzi,{\it Exotic $J^{PC}=1^{-+}$ mesons in a holographic model of QCD},
  Eur.\ Phys.\ J.\ C {\bf 74}, no. 4, 2830 (2014) doi:10.1140/epjc/s10052-014-2830-6 [arXiv:1402.5308 [hep-ph]].
\bibitem{Cui:2013xva} L.~X.~Cui, z.~Fang and Y.~L.~Wu, {\it Infrared-Improved Soft-wall AdS/QCD Model for Mesons},
  Eur.\ Phys.\ J.\ C {\bf 76}, no. 1, 22 (2016) doi:10.1140/epjc/s10052-015-3866-y [arXiv:1310.6487 [hep-ph]].
\bibitem{Alvares:2011wb} R.~Alvares, C.~Hoyos and A.~Karch, {\it An improved model of vector mesons in holographic QCD},
  Phys.\ Rev.\ D {\bf 84}, 095020 (2011) doi:10.1103/PhysRevD.84.095020 [arXiv:1108.1191 [hep-ph]].
\bibitem{Domokos:2007kt} S.~K.~Domokos and J.~A.~Harvey, {\it Baryon number-induced Chern-Simons couplings of vector and axial-vector mesons in holographic QCD}, Phys.\ Rev.\ Lett.\  {\bf 99}, 141602 (2007) doi:10.1103/PhysRevLett.99.141602 [arXiv:0704.1604 [hep-ph]].
\bibitem{Kim:2009bp} H.~C.~Kim, Y.~Kim and U.~Yakhshiev, {\it Mesons and nucleons from holographic QCD in a unified approach},
  JHEP {\bf 0911}, 034 (2009) doi:10.1088/1126-6708/2009/11/034 [arXiv:0908.3406 [hep-ph]].
\bibitem{Li:2013oda} D.~Li and M.~Huang, {\it Dynamical holographic QCD model for glueball and light meson spectra},
  JHEP {\bf 1311}, 088 (2013) doi:10.1007/JHEP11(2013)088 [arXiv:1303.6929 [hep-ph]].
\bibitem{Imoto:2010ef} T.~Imoto, T.~Sakai and S.~Sugimoto, {\it Mesons as Open Strings in a Holographic Dual of QCD}
  Prog.\ Theor.\ Phys.\  {\bf 124}, 263 (2010) doi:10.1143/PTP.124.263 [arXiv:1005.0655 [hep-th]].
\bibitem{Kruczenski:2004me} M.~Kruczenski, L.~A.~Pando zayas, J.~Sonnenschein and D.~Vaman,
  {\it Regge trajectories for mesons in the holographic dual of large-N(c) QCD}, JHEP {\bf 0506}, 046 (2005) doi:10.1088/1126-6708/2005/06/046 [hep-th/0410035].
\bibitem{PDG} K.~A.~Olive et al. (Particle Data Group), Chin. Phys. C, 38, 090001 (2014) and 2015 update.
\bibitem{Misra+Gale_Conformal_Anomaly} A.~Misra and C.~Gale, {\it The QCD Trace Anomaly at Strong Coupling from ${\cal M}$-Theory}, Eur. Phys. J. C {\bf 80}, no.7, 620 (2020) [arXiv:1909.04062 [hep-th]].
\bibitem{Herzog_Tc} C.~P.~Herzog, {\it A Holographic Prediction of the Deconfinement Temperature},
  Phys.\ Rev.\ Lett.\  {\bf 98}, 091601 (2007)  [hep-th/0608151].
\bibitem{Mathieu:2008me} V.~Mathieu, N.~Kochelev and V.~Vento,
{\it The Physics of Glueballs}, Int. J. Mod. Phys. E \textbf{18} (2009), 1-49
doi:10.1142/S0218301309012124[arXiv:0810.4453 [hep-ph]].
\bibitem{PDG-2018} M.~Tanabashi et al. (Particle Data Group), Phys. Rev. D 98, 030001 (2018).
\bibitem{Hashimoto-glueball} K.~Hashimoto, C.~I.~Tan and S.~Terashima, {\it Glueball decay in holographic QCD},  Phys.\ Rev.\ D {\bf 77}, 086001 (2008)
  doi:10.1103/PhysRevD.77.086001  [arXiv:0709.2208 [hep-th]].
\bibitem{Myers} N.~R.~Constable and R.~C.~Myers, {\it Spin two glueballs, positive energy theorems and the AdS/CFT correspondence},
  JHEP {\bf 9910}, 037 (1999) doi:10.1088/1126-6708/1999/10/037 [hep-th/9908175].
\bibitem{Brunner_Hashimoto-results}F.~Brunner, D.~Parganlija and A.~Rebhan, {\it Glueball Decay Rates in the Witten-Sakai-Sugimoto Model},
Phys.\ Rev.\ D {\bf 91}, no. 10, 106002 (2015) Erratum: [Phys.\ Rev.\ D {\bf 93}, no. 10, 109903 (2016)]
doi:10.1103/PhysRevD.93.109903, 10.1103/PhysRevD.91.106002 [arXiv:1501.07906 [hep-ph]].
\bibitem{witten} E. Witten, Anti-de Sitter space, {\it thermal phase transition and confinement in gauge theories},
Adv. Theor. Math. Phys. 2 (1998) 505 [hep-th/9803131].
\bibitem{Savage-Phase-Space-Lecture-Notes}M.J.Savage, {\it Introduction to Nuclear and Particle Physics},
\begin{verbatim}
 www.int.washington.edu/users/mjs5/class_cmu/part_cmu/lecture0395.ps
 \end{verbatim}
\bibitem{Murayama-Phase-Space} H.~Murayama, {\it Notes on Phase Space},
\begin{verbatim}
http://hitoshi.berkeley.edu/233B/phasespace.pdf
\end{verbatim}
\bibitem{Krishna-Rajagopal-et-al-book} J.~Casalderrey-Solana, H.~Liu, 2 D.~Mateos, K.~Rajagopal and U.~A.~Wiedemann, {\it Gauge/String Duality, Hot QCD and Heavy Ion Collisions}, Cambridge, UK: Cambridge University Press, 2014 [arXiv:1101.0618 [hep-th]].
\bibitem{O(R^4)}  M. J. Duff, J. T. Liu and R. Minasian, {\it Eleven-dimensional origin of string-string duality:
A One loop test}, Nucl. Phys. B 452, 261 (1995) [hep-th/9506126];  M. B. Green and P. Vanhove, {\it D instantons, strings and ${\cal M}$ theory}, Phys. Lett. B 408,
122 (1997) [hep-th/9704145];  M. B. Green, M. Gutperle and P. Vanhove, {\it One loop in eleven-dimensions}, Phys. Lett. B 409, 177 (1997) [hep-th/9706175];
E. Kiritsis and B. Pioline, {\it On R**4 threshold corrections in IIb string theory and (p, q)
string instantons}, Nucl. Phys. B 508, 509 (1997) [hep-th/9707018]; J. G. Russo and A. A. Tseytlin, {\it One loop four graviton amplitude in eleven-dimensional
supergravity}, Nucl. Phys. B 508, 245 (1997) [hep-th/9707134]; I. Antoniadis, S. Ferrara, R. Minasian and K. S. Narain, {\it R**4 couplings in M and type
II theories on Calabi-Yau spaces}, Nucl. Phys. B 507, 571 (1997) [hep-th/9707013]; A. A. Tseytlin, {\it R**4 terms in 11 dimensions and conformal anomaly of (2,0) theory}, Nucl. Phys. B 584, 233 (2000) [hep-th/0005072].
\bibitem{O(R^3G^2)} J. T. Liu and R. Minasian, {\it Higher-derivative couplings in string theory: dualities and the
B-field}, arXiv:1304.3137 [hep-th].
\bibitem{previous-higher-ders} S.~Grozdanov and A.~O.~Starinets, {\it Second-order transport, quasinormal modes and zero-viscosity limit in the Gauss-Bonnet holographic fluid},
JHEP \textbf{03}, 166 (2017) [arXiv:1611.07053 [hep-th]]; J.~Casalderrey-Solana, S.~Grozdanov and A.~O.~Starinets, {\it Transport Peak in the Thermal Spectral Function of $\mathcal N=4$ Supersymmetric Yang-Mills Plasma at Intermediate Coupling}, Phys. Rev. Lett. \textbf{121}, no.19, 191603 (2018) [arXiv:1806.10997 [hep-th]].
\bibitem{Becker-sisters-O(R^4)}K.~Becker and M.~Becker, {\it Supersymmetry breaking, ${\cal M}$ theory and fluxes}, JHEP bf{07}, 038 (2001)
doi:10.1088/1126-6708/2001/07/038 [arXiv:hep-th/0107044 [hep-th]].
\bibitem{Tseytlin-epsilonD^2R^4-kroneckerdeltaR^4}
A.~A.~Tseytlin, {\it R**4 terms in 11 dimensions and conformal anomaly of (2,0) theory},Nucl.\ Phys.\ B bf{584}, 233-250 (2000)
doi:10.1016/S0550-3213(00)00380-1 [arXiv:hep-th/0005072 [hep-th]].
\bibitem{SYZ 3 Ts} M.~Becker, K.~Dasgupta, A.~Knauf and R.~Tatar, {\it Geometric transitions, flops and nonKahler manifolds. I.},
  Nucl.\ Phys.\ B {\bf 702}, 207 (2004) [arXiv:hep-th/0403288].
\bibitem{Bulk-Viscosity}A.~Czajka, K.~Dasgupta, C.~Gale, S.~Jeon, A.~Misra, M.~Richard and K.~Sil, {\it Bulk Viscosity at Extreme Limits: From Kinetic Theory to Strings},  JHEP {\bf 1907}, 145 (2019) [arXiv:1807.04713 [hep-th]].
\bibitem{HARADA}  M.~Harada, S.~Matsuzaki and K.~Yamawaki, {\it Holographic QCD Integrated back to Hidden Local Symmetry}, Phys.Rev.D82:076010,2010 [arXiv:hep-th/1007.4715].
\bibitem{MG} A.~Manohar and H.Georgi, {\it Chiral Quarks and the Nonrelativistic Quark Model}, Nucl. Phys. B 234, 189 (1984).
\bibitem{GL} J.~Gasser and H.~Leutwyler, {\it Chiral Perturbation Theory:Expansions in the Mass of the Strange Quark}, Nucl. Phys. B250(1985) 465.
\bibitem{GLF} J.~Gasser and H.~Leutwyler, {\it Chiral perturbation theory to one loop}, Annals of Physics  158, 142 (1984).
\bibitem{MILC} C. ~Bernard et al., {\it Low energy constants from the MILC Collaboration} [arxiv:hep-lat/0611024].
\bibitem{Pich} A.~Pich, {\it Chiral Perturbation Theory}, Rept.Prog.Phys.58:563-610(1995), [arXiv:hep-th/9502366].
\bibitem{EJP} G.~Ecker, J. ~Gasser, A. ~Pich and E. de Rafael, {\it The role of resonances in chiral perturbation theory}, Nucl. Phys. B 321, 311 (1989).
\bibitem{KMMW} M.~Kruczenski, D. ~Mateos, R. C. ~Myers and D. J. ~Winters, {\it Towards a holographic dual of large-$N_c$ QCD}, JHEP 0405 (2004) 041 [arXiv:hep-th/0311270].
\bibitem{SS2} T.~Sakai, S.~Sugimoto, {\it More on a holographic dual of QCD}, Prog.Theor.Phys.114:1083-1118,2005 [arXiv:hep-th/0507073].
\bibitem{HMM} M.~Harada, Y.-Liang~Ma and S.~Matsuzaki, {\it Chiral effective theories from holographic QCD with scalars}, Phys. Rev. D 89, 115012 (2014) [arXiv:hep-th/1404.4532].
\bibitem{Armoni et al-2020} R.~Argurio, A.~Armoni, M.~Bertolini, F.~Mignosa and P.~Niro,
{\it Vacuum structure of large $N$ $QCD_{3}$ from holography}, JHEP bf{07}, 134 (2020).
\bibitem{Knauf-thesis} A.~Knauf, {\it Geometric Transitions on non-Kaehler Manifolds}, Fortsch. Phys. {\bf 55}, 5-107 (2007)
[arXiv:hep-th/0605283 [hep-th]].
\bibitem{theta0-theta} K.~Dasgupta, M.~Grisaru, R.~Gwyn, S.~H.~Katz, A.~Knauf and R.~Tatar, {\it Gauge-Gravity Dualities, Dipoles and New Non-Kahler Manifolds}, Nucl. Phys. B {\bf 755}, 21-78 (2006) [arXiv:hep-th/0605201 [hep-th]].
\bibitem{Moore+Saremi} G.~D.~Moore and O.~Saremi, {\it Bulk viscosity and spectral functions in QCD}, JHEP \textbf{09}, 015 (2008) [arXiv:0805.4201 [hep-ph]].
%
%
%
%
%
%
%
%
%
%
%
%
%
%
%
%
%
%
%
%
%
%
%
%
%
%
%
%
%
%
%
%
%
%
%
%
%
%
%
%
%
%
%
%
%
%
%


\end{thebibliography}

\end{document}